%% file: thesis.tex
\documentstyle[uqthesis,twoside,graphicx]{report}

\begin{document}

\title{Adaptive phase measurements}
\author{Dominic William Berry}
\dept{Physics}
\beforepreface

\signaturepage
\vspace{10cm}

\prefacesection{Acknowledgments}
Primary acknowledgments go to my PhD supervisor, Dr.~Howard Wiseman, who took
me on as a student and continued to come to the university for weekly
discussions, despite having moved to Griffith University. He originated the
topic of this project, as well as directing me to the most fruitful areas for
study. He also originated the general ideas behind some of the approaches I used
in this project. I am grateful for his untiring help with the drafts and
revisions of my papers, as well as this thesis. Also his advice on publishing
and applying for postdoctoral work was invaluable.

I am grateful to Zhong-Xi Zhang and John Breslin, who contributed valuable
insight to the method of solution of some of the problems considered here.
Thanks also go to the Australian Research Council, that supported this research
financially through the Australian Postgraduate Awards. I give thanks to our
system administrator, Ian Mortimer, for enabling me to use the department's
computing resources, as well as Prof.~Peter Drummond for allowing me some
computing time on the quantum optics cluster.

\newcommand{\beq}{\begin{equation}}
\newcommand{\eeq}{\end{equation}}
\newcommand{\bqa}{\begin{eqnarray}}
\newcommand{\eqa}{\end{eqnarray}}
\newcommand{\nn}{\nonumber}
\newcommand{\dg}{^\dagger}
\newcommand{\smallfrac}[2]{\mbox{$\frac{#1}{#2}$}}
\newcommand{\la}{\langle}
\newcommand{\ra}{\rangle}
\newcommand{\ket}[1]{| {#1} \ra}
\newcommand{\bra}[1]{\la {#1} |}
\newcommand{\half}{\smallfrac{1}{2}}
\newcommand{\braket}[2]{\langle{#1}|{#2}\rangle}
\newcommand{\ip}[1]{\left\langle{#1}\right\rangle}
\newcommand{\lr}[1]{\langle{#1}\rangle}
\newcommand{\st}[1]{\left| {#1} \right|}
\newcommand{\atanh}{{\rm atanh}}
\newcommand{\erf}{{\rm erf}}
\newcommand{\erfc}{{\rm erfc}}
\newcommand{\nb}{\overline{n}}
\newcommand{\rt}[1]{\sqrt{#1}\,}
\newcommand{\binom}[2]{\left({#1}\above 0pt {#2}\right)}

\include{pubs}

\prefacesection{Abstract}
In this thesis I consider the general problem of how to make the best possible
phase measurements using feedback. Both the optimum input state and optimum
feedback are considered for both single-mode dyne\footnote[1]{Dyne
measurements are those based on continuous measurement of field quadratures,
including heterodyne, homodyne and more general adaptive measurements.}
measurements and two-mode
interferometric measurements. I derive the optimum input states under general
dyne measurements when the mean photon number is fixed, both for general states
and squeezed states. I propose a new feedback scheme that introduces far less
phase uncertainty than mark II feedback, and is very close to the theoretical
limit. I also derive results for the phase variance when there is a time delay
in the feedback loop, showing that there is a lower limit to the introduced
phase variance, and this is approached quite accurately under some conditions.
I derive the optimum input states for interferometry, showing that the phase
uncertainty scales as $N^{-1}$ for all the common measures of uncertainty. This
is contrasted with the $\ket{j0}_z$ state, which does not scale as $N^{-1}$ for
all measures of phase uncertainty. I introduce an adaptive feedback scheme that
is very close to optimum, and can give scaling very close to $N^{-1}$ for the
uncertainty. Lastly I consider the case of continuous measurements, for both
the dyne and interferometric cases.

\include{symbols}

\prefacesection{List of Abbreviations}

\begin{tabular}{ll}
POM&Probability operator measure\\
POVM&Positive-operator-valued measure\\
BS&Beam splitter\\
PD&Photodetector\\
EOM&Electro-optic phase modulator\\
SDE&Stochastic differential equation
\end{tabular}

\afterpreface



\include{intro}
\include{dynesta}
\include{dyne}
\include{delay}
\include{intersta}
\include{interfer}
\include{continue}
\include{conc}

\appendix
\include{derivs}


\include{thebib}
\end{document}

%% file: pubs.tex
\prefacesection{List of Publications}

In this chronological list of publications, those on which this thesis is based are asterisked.

\begin{enumerate}
	\item  D. W. Berry, N. R. Heckenberg and H. Rubinsztein-Dunlop,
``Effects associated with bubble formation in optical trapping''
J. Mod. Opt. {\bf 47}, 1575 (2000).

	\item  * D. W. Berry, H. M. Wiseman and Zhong-Xi Zhang,
``Heterodyne and adaptive phase measurements on states of fixed mean photon
number'' Phys. Rev. A {\bf 60}, 2458 (1999).

	\item  * D. W. Berry and H. M. Wiseman,
``Phase measurements at the theoretical limit''
Phys. Rev. A {\bf 63}, 013813 (2000).

	\item  * D. W. Berry and H. M. Wiseman,
``The effects of time delays in adaptive phase measurements''
J. Mod. Opt. {\bf 48}, 797  (2001).

	\item  * D. W. Berry and H. M. Wiseman,
``Optimal states and almost optimal adaptive measurements for quantum
interferometry'' Phys. Rev. Lett. {\bf 85}, 5098 (2000).

	\item  * D. W. Berry, H. M. Wiseman and J. K. Breslin,
``Optimal input states and feedback for interferometric phase estimation''
Phys. Rev. A {\bf 63}, 053804 (2001).

	\item  * D. W. Berry and H. M. Wiseman,
``Adaptive quantum measurements of a continuously-varying phase''
Phys. Rev. A (to be published).

\end{enumerate}

%% file: symbols.tex
\prefacesection{List of Symbols}

The following list is neither exhaustive nor exclusive, but may be helpful.
It generally contains only those symbols used frequently or in more than one
chapter.

\begin{tabular}{ll}
$a$ & {annihilation operator for an optical mode} \\
$a\dg$ & {creation operator for an optical mode} \\
$\ket n$ & {a Fock number state with $n$ photons} \\
$\varphi$ & {the system phase} \\
$\phi$ & {a phase estimate} \\
$F(\ldots)$ & {the POM for $\ldots$} \\
$H_{nm}$ & {the matrix giving the POM as in Eq.~(\ref{genFphi})} \\
$V(\phi)$ & {the Holevo phase variance} \\
$V_{\varphi}(\phi)$ & {a modified Holevo phase variance given by Eq.~(\ref{modHol})} \\
$V'(\phi)$ & {an alternative measure of phase variance given by Eq.~(\ref{altvardef})} \\
$h(n)$ & {$1-H_{n,n+1}$} \\
$\Phi$ & {the local oscillator phase} \\
$\hat X_\Phi$ & {the $\Phi$ quadrature operator} \\
$t$ & {unscaled time, in the interval $[0,\infty)$} \\
$I(t)$ & {the unscaled difference photocurrent} \\
$v$ & {scaled time, in the interval $[0,1)$} \\
$I(v)$ & {the scaled difference photocurrent} \\
$A_v$ & {defined in Eq.~(\ref{defA})} \\
$B_v$ & {defined in Eq.~(\ref{defB})} \\
$C_v$ & {defined in Eq.~(\ref{defC})} \\
$A,B,C$ & {the values of these variables at $v=1$} \\
$dW$ & {a Wiener stochastic increment} \\
$\hat\varphi$ & {an intermediate phase estimate used for feedback (for dyne measurements)} \\
$\hat\varphi$ & {the optimal phase estimate (for interferometry)} \\
$c,p$ & {the constants in the asymptotic scaling for $h(n)$} \\
$\alpha$ & {coherent amplitude} \\
$\eta$ & {photodetector efficiency} \\
$\nb$ & {mean photon number} \\
$z_1$ & {the first zero of the Airy function, about -2.338} \\
$\zeta$ & {the squeezing parameter for squeezed states} \\
$n_0$ & {given by $\nb e^{2\zeta}$} \\
$\Delta$ & {a constant for optimised squeezed states, about 2.427} \\
$\Delta'$ & {an alternative constant, about 0.927} \\
$\alpha^{\rm P}$ & {the coherent amplitude for the squeezed state in the POM} \\
$\zeta^{\rm P}$ & {the squeezing parameter for the squeezed state in the POM} \\
$\xi^{\rm P}$ & {the squeezing parameter relative to the phase of $\alpha^{\rm P}$} \\
$\nb^{\rm P}$ & {the mean photon number for the squeezed state in the POM} \\
$Q(A,B)$ & {the ostensible probability distribution for $A$ and $B$} \\
$\varepsilon$ & {defined in Eq.~(\ref{phasestimate})} \\
$\gamma$ & {the local oscillator amplitude} \\
$A_t^{\rm S},B_t^{\rm S}$ & {squeezing parameters defined analogously to $A_v$ and $B_v$ by Eq.~(\ref{defABS})}
\end{tabular}

\begin{tabular}{ll}
$\alpha_v^{\rm P}$ & {an intermediate value of $\alpha^{\rm P}$, defined in Eq.~(\ref{defalvP})} \\
$\zeta_v^{\rm P}$ & {an intermediate value of $\zeta^{\rm P}$, defined in Eq.~(\ref{defzevP})} \\
$\xi_v^{\rm P}$ & {the intermediate squeezing parameter relative to the phase of $\alpha_v^{\rm P}$} \\
$\lambda$ & {the variable controlling when the corrections for dyne measurements are used} \\
$\alpha^f$ & {the value of $\alpha$ found by fitting to the data} \\
$\tau$ & {the feedback time delay} \\
$\hat J_x,\hat J_y,\hat J_z,\hat J^2$ & {operators in the Schwinger representation} \\
$\ket{j\mu}_z$ & {the common eigenstate of $\hat J_x$ and $\hat J^2$} \\
$\ket{j\mu}_y$ & {the common eigenstate of $\hat J_y$ and $\hat J^2$} \\
$I_{\mu,\nu}^j(\pi/2)$ & {interferometer matrix elements} \\
$P_n^{(\alpha,\beta)}$ & {Jacobi polynomials} \\
$H_n$ & {Hermite polynomials} \\
$V_{\pi}(\phi)$ & {the modified Holevo variance for phase modulo $\pi$} \\
$\Delta \phi$ & {the square root of the standard variance} \\
$\Delta \phi_{\rm H}$ & {the square root of the Holevo variance} \\
$L_{\rm H}$ & {the entropic length} \\
$L_{\rm F}$ & {the Fisher length} \\
$L_C$ & {the $C \times 100\%$ confidence interval} \\
$L_{\rm rp}$ & {the reciprocal-of-peak-value} \\
$L_{\rm S}$ & {the S\"ussman measure} \\
$u_m$ & {the result of the $m$th detection for interferometry} \\
$n_m$ & {the sequence of $m$ detection results $u_m \ldots u_2u_1$} \\
$\Phi_m$ & {the feedback phase before the $m$th detection for interferometry} \\
$\psi_{\mu,m,k}(n_m)$ & {the coefficients for the system state after $m$ detections} \\
$\theta(t)$ & {the phase estimate based on the infinitesimal data in the interval $[t,t+dt)$} \\
$\Theta(t)$ & {the phase estimate based on the data up till time t} \\
$\kappa$ & {the diffusion coefficient for the system phase} \\
$\chi$ & {the scaling constant for the exponential weighting} \\
K, X & {the scaled values of $\kappa$ and $\chi$}
\end{tabular}

%% file: intro.tex
\setcounter{chapter}{0}
\chapter{Introduction}

In this study I consider the general problem of how to most efficiently measure
phase. This is a very important subject, as many high precision measurements
are based upon measurements of phase. In particular, the current search for
gravitational waves requires extremely accurate phase measurements, where new
approaches that surpass the standard quantum limit may be necessary to obtain
useful results \cite{Caves}.

Unlike most other quantities that we
would wish to measure, it is not possible to measure phase directly. We must
measure phase indirectly, and this almost always introduces an extra uncertainty
beyond the intrinsic uncertainty in the phase of the mode.
In general, it is possible to improve measurements by introducing an
auxiliary phase shift. In the case of homodyne measurements, this phase shift
would be based upon the previous knowledge about the phase. In this study I
consider the case that the phase is unknown, and instead the auxiliary phase is
adjusted based on the data obtained during the measurement. I
investigate schemes for adjusting the auxiliary phase so as to introduce the
minimum possible phase uncertainty.

Another aspect of the problem of efficiently measuring the phase is the state
itself. Every state with finite energy will have some intrinsic uncertainty in
the phase. We would wish to optimise the state so that it has the minimum
possible intrinsic phase uncertainty, or alternatively so that it gives the
minimum phase uncertainty for some specific phase measurement scheme.

Before I discuss these problems, I will briefly review the theory behind the
description of phase and phase measurements.

\section{The Description of Phase}
\label{firstsec}
Classically there is no ambiguity in the definition of phase. A general
propagating sinusoidal wave can be described by
\beq
\psi(x,t) = A \sin(kx-\omega t+\phi),
\eeq
where $A$ is the amplitude and $\phi$ is the phase. The phase is not a
quantity that can be measured directly; however, it is just a real number and
can be determined unambiguously from the variation of $\psi(x,t)$ (provided the
amplitude is constant and nonzero).

For a monochromatic electromagnetic field propagating in the $\mathbf{k}$
direction, we can express the electric field as
\beq
{\mathbf E}({\mathbf R},t) = {\mathbf E}_0 \sin( {\mathbf k} \cdot {\mathbf R}
-\omega t+\phi).
\eeq
In quantising the electric field in the Heisenberg picture, the operator for
the electric field is
\beq
\label{elecfield}
\hat{\mathbf E}({\mathbf R},t) = \sqrt{\frac{\hbar \omega}{2\epsilon_0 V}}
\left( {\mathbf \varepsilon} e^{i({\mathbf k} \cdot {\mathbf R}-\omega t)}
a+{\mathbf \varepsilon}^* e^{-i({\mathbf k} \cdot {\mathbf R}-\omega t)}
a\dg \right),
\eeq
where $\hbar$ is Planck's constant divided by $2\pi$, $\epsilon_0$ is the
dielectric permittivity for a vacuum, $V$ is the quantisation volume,
${\mathbf \varepsilon}$ is the polarisation vector and $a$ and $a\dg$ are the
annihilation and creation operators respectively. The variation of the electric
field in time and space, as well as the polarisation direction are contained in
this field operator, but the amplitude and phase of the field are contained in
the state.

The annihilation and creation operators have the commutation relations
\bqa
\left[ a , a \dg \right] = 1, \\
\left[ a , a \right] = \left[ a \dg , a \dg \right] = 0.
\eqa
I will generally use the hat over variables to indicate operators when they are
used in both an operator and non-operator sense (like $\mathbf E$), but not when
the variable is used only as an operator (like $a$).

The quantum mechanical decription of phase was first considered by London
\cite{Lon26,Lon27} and Dirac~\cite{dirac}. In general, in quantum mechanics we
wish to represent physical quantities by Hermitian operators. Unfortunately,
this approach produces difficulties when applied to phase. For example, consider
the result for a coherent state
\beq
\ip{a} = e^{i\phi} \sqrt N.
\eeq
From this, an obvious way of defining an operator for the phase is by
\beq
a = e^{i\hat\phi} \sqrt{\hat N},
\eeq
where $\hat N$ is the usual operator for the photon number,
\beq
\hat N = a\dg a.
\eeq
The phase operator defined in this way is equivalent to that considered by
Dirac~\cite{dirac}. It is easily seen that the phase operator defined in this
way is not Hermitian.

This definition of the phase operator produces other problems. For example,
Dirac derives the commutation relation
\beq
\label{phcom}
[\hat N,\hat \phi]=i.
\eeq
This would seem to imply the uncertainty relation \cite{robber}
\beq
\Delta N \Delta \phi \ge \half.
\eeq
This uncertainty relation does not make sense, because the uncertainty in $N$
can become arbitrarily small, whereas the uncertainty in $\phi$ can not be over
$\pi$. In fact, this uncertainty relation is not necessarily implied by
Eq.~(\ref{phcom}), because the phase operator is not Hermitian. Another problem
is that, for a number state, the expectation value of the commutator should be
zero \cite{louise}, rather than $i$.

One resolution of this problem is that although it is not possible to define a
Hermitian phase operator, it is possible to define Hermitian sine and cosine
operators that give valid uncertainty relations \cite{louise,SG}. These
Susskind-Glogower operators satisfy the commutation relations
\bqa
\left[ \widehat{\cos\phi},\hat N \right] \!\!\!\! &=& \!\!\!\!
i\,\widehat{\sin\phi} \\
\left[ \widehat{\sin\phi},\hat N \right] \!\!\!\! &=& \!\!\!\!
-i\,\widehat{\cos\phi}.
\eqa
The corresponding uncertainty relations are
\bqa
\Delta N \Delta \cos \phi \!\!\!\!&\ge&\!\!\!\!
 \half\st{\ip{\widehat{\sin\phi}}} \\
\Delta N \Delta \sin \phi \!\!\!\!&\ge&\!\!\!\!
 \half\st{\ip{\widehat{\cos\phi}}}.
\eqa
These uncertainty relations make sense, because in the limit of small number
uncertainty the expectation values on the right hand side also become small,
so these uncertainty relations do not imply ridiculously large values of
$\Delta\cos\phi$ or $\Delta\sin\phi$.

Unfortunately there are also problems with these operators. For example,
we find that for the vacuum state $\lr{\cos^2\theta}=\smallfrac 14$
\cite{carruthers}. As the vacuum state should have a uniform phase
distribution, we would expect that $\lr{\cos^2\theta}$ is equal to $\half$.
In addition, using these operators we find $\lr{\exp(ip\theta)}=0$ for all
integers $p>0$. This implies a uniform phase distribution, which is what we
would expect, but is inconsistent with the result obtained for
$\lr{\cos^2\theta}$.

Another approach to finding a Hermitian phase operator is the Pegg-Barnett
formalism \cite{pegbarn1,pegbarn2,pegbarn3}. The basis of this formalism is
to put an upper limit $s$ on the photon number, then take the limit as $s$
tends to infinity. The reference phase states are taken to be
\beq
\ket{\theta_m}_s = (s+1)^{-1/2} \sum_{n=0}^s \exp(in\theta_m)\ket n ,
\eeq
where
\beq
\theta_m = \theta_0 + \frac{2m\pi}{s+1}, \;\;\;\; m=0, 1, \ldots, s
\eeq
and $\theta_0$ is an arbitrary constant.

Then a Hermitian phase operator is defined by
\beq
\hat \phi_s = \sum_{m=0}^s \theta_m \ket{\theta_m}_{ss}\bra{\theta_m}.
\eeq
Note that for this operator the phase states $\ket{\theta_m}_s$ are clearly
eigenstates with eigenvalues of $\theta_m$. In terms of the number basis, the
operator is
\beq
\label{phaseoper}
\hat \phi_s = \theta_0 + \frac{s\pi}{s+1} + \frac{2\pi}{s+1} \sum_{j\ne k}^s
\frac{\exp\left[i(j-k)\theta_0 \right]\ket j \bra k}{\exp\left[ i(j-k)2\pi /
(s+1)\right]-1}.
\eeq
In the Pegg-Barnett formalism, the limit of $s \to \infty$ is taken after
expectation values have been determined.

It is also possible to take the limit $s \to \infty$ of Eq.~(\ref{phaseoper}),
which gives
\beq
\label{operlim}
\hat \phi_{\infty} = \theta_0 + \pi + \sum_{j\ne k}^{\infty} \frac{\exp\left[
i(j-k)\theta_0 \right]\ket j \bra k}{i(j-k)}.
\eeq
This operator was also considered before the Pegg-Barnett formalism was
developed \cite{Garrison,Popov,PopJMO}. This would appear to be a Hermitian
operator that can be used to describe phase; however, there are problems with
this operator. This operator leads to different expectation values than given by
the Pegg-Barnett formalism (where the expectation values are taken before the
limit $s\to\infty$).

The problem is that (\ref{phaseoper}) converges to (\ref{operlim}) only weakly
\cite{weak}. The weak limits of operators do not preserve the operator algebra,
for example, the weak limit of $\hat \phi_s^2$ is not $\hat \phi_{\infty}^2$.
One result of this is that, for the vacuum state, $\lr{\hat\phi_{\infty}^2}=
\pi^2/6$~\cite{Barn92,Gantsog}. For a uniform phase distribution, the result
should be $\pi^2/3$, which is that obtained from the limit of
$\lr{\hat\phi_s^2}$.

Fortunately, in this study we do not require an explicit phase operator, and
usually all we need to know is the phase variance, or at most the
probability distribution. For this what we want is the probability operator
measure, or POM.

\section{Probability Operator Measures}

In quantum mechanical systems, the most general way of obtaining the probability
of some measurement result $E$ is by the expectation value of an operator
$F(E)$, i.e.
\beq
\label{POMdef}
P(E)={\rm Tr}[\rho F(E)],
\eeq
where $\rho$ is the state matrix for the system. If the set of all possible
measurement results is $\Omega$, it is evident that $P(\Omega)=1$ for all
$\rho$, which implies that $F(\Omega)=1$. Thus $F(E)$ can be called a
probability operator, and the mapping $E \mapsto F$ defines a probability
operator measure (POM), sometimes also called a positive-operator-valued measure
(POVM), on $\Omega$ \cite{Dav76,Hel76}. This method does not require a specific
operator to represent the quantity that is being measured.

This is quite different to the simple method for pure states, where the
probability is given by the square of the inner product between the initial and
final states. To see the similarity between the two methods, recall that in the
simple method we represent the physical quantity being measured by a Hermitian
operator, for example $\hat R$.  This has associated eigenvalues and
eigenvectors $r$ and $\ket{\psi _r }$. After a measurement yielding the result
$r$ the system is in state $\ket{\psi _r }$, and the probability for this
result is $\st{\braket{\psi_r}{\psi}}^2$.

Alternatively we can define the projection operators
\beq
\Pi _r  = \ket{\psi _r} \bra{\psi _r}.
\eeq
The probability of obtaining the result $r$ is then
\bqa
P_r  \!\!\!\! &=& \!\!\!\! \st{\braket{\psi_r}{\psi}}^2 \nn \\
\!\!\!\! &=& \!\!\!\!\braket{\psi}{\psi_r}\braket{\psi_r}{\psi} \nn \\
\label{projpro}
\!\!\!\! &=& \!\!\!\!\bra{\psi}\Pi_r\ket{\psi}.
\eqa
The normalised state after the measurement is then
\beq
\ket{\psi_r} = \frac{\Pi_r \ket{\psi}}{\sqrt{P_r}}.
\eeq

For more general measurements, we can replace the projection operator with a
more general operator $\Omega_r$, called the measurement operator. Then the
probability is given by
\beq
\label{measop}
P_r = \bra{\psi}\Omega_r\dg \Omega_r \ket{\psi}.
\eeq
Note that in this case $\Omega_r\dg \Omega_r$ does not necessarily simplify to
$\Omega_r$, as is the case for projection operators. The normalised state after
the measurement is given by
\beq
\ket{\psi_r} = \frac{\Omega_r \ket{\psi}}{\sqrt{P_r}}.
\eeq

For the most general case we replace $\Omega_r\dg \Omega_r$ in
Eq.~(\ref{measop}) [or $\Pi_r$ in Eq.~(\ref{projpro})] with a general operator
$F_r$. The probability is then determined using
\beq
P_r=\bra{\psi} F_r \ket{\psi}.
\eeq
Clearly, the generalisation to this for mixed states is Eq.~(\ref{POMdef}).

For phase measurements, the probability distribution for the measurement result
can be determined using the POM $F(\phi)$. This approach was first considered
be Helstrom~\cite{Hel76}, and is also considered in Refs \cite{SSW,SS}. If the
phase measurement treats all phases equally, it should be invariant under a
phase translation
\beq
R(\theta)F(\phi)R(-\theta)=F(\phi+\theta)
\eeq	
where $R(\theta) = \exp ( ia\dg a\theta )$ is the phase translation operator.
Now $F(\phi)$ has the general expansion
\beq
F(\phi) = \sum_{n,m = 0}^\infty \ket{n} \bra{m} F_{nm}(\phi).
\eeq
The phase shift invariance condition gives
\bqa
F(\phi+\theta) \!\!\!\! &=& \!\!\!\! \sum_{n,m = 0}^\infty
e^{i a\dg a\theta}\ket{n} \bra{m} e^{-ia\dg a\theta} F_{nm}(\phi) \\
\sum_{n,m = 0}^\infty \ket{n} \bra{m} F_{nm}(\phi+\theta) \!\!\!\! &=& \!\!\!\!
\sum_{n,m = 0}^\infty e^{in\theta} \ket{n} \bra{m} e^{-im\theta} F_{nm}(\phi)
\eqa		
This means that we must have
\beq
F_{nm}(\phi+\theta)=e^{i(n-m)\theta}F_{nm}(\phi).
\eeq	
This implies that $F_{nm}(\phi)$ must have the form
\beq
F_{mn}(\phi)=\frac 1{2\pi}e^{i(n-m)\phi}H_{nm}.
\eeq
A factor of $1/(2\pi)$ has been added for normalisation. Therefore the general
form of $F(\phi)$ for a shift invariant phase measurement is
\beq
\label{genFphi}
F(\phi) = \frac 1{2\pi}\sum_{n,m = 0}^\infty \ket{n}\bra{m}e^{i(n-m)\phi}
H_{nm}.
\eeq	
For the integral of the probabilities to equal 1, we must have
\beq
\int\limits_{-\pi}^{\pi} F(\phi)d\phi=1.
\eeq	
Applying this to Eq.~(\ref{genFphi}) above we find that
\beq
\sum_{n=0}^\infty \ket{n} \bra{n} H_{nn} =1.
\eeq
This means that the diagonal elements $H_{nn}$ must all be equal to 1.

In addition there is the condition that the probability given by
${\rm Tr} [\rho F(\phi)]$ always be real and positive. This, together with the
above result means that all of the $H_{nm}$ must have absolute values between 0
and 1. Usually these are all assumed to be real and positive; however, we will
see in Sec.~\ref{optimal} that they need not be.

In Ref.~\cite{LeoVacBohPau95} it is shown that the additional condition that
a number shifter does not alter the phase distribution gives $H_{nm}=1$,
corresponding to the POM
\beq
\label{canPOM}
F^{\rm can}(\phi) = \frac 1{2\pi} \sum_{n,m=0}^{\infty} e^{i(n-m)\phi}
\ket n \bra m.
\eeq
This POM corresponds to a canonical phase measurement \cite{LeoVacBohPau95}.
Ref.~\cite{SSW} also derives this POM using the maximum likelihood approach.
In general, real measurements will give smaller values of $H_{nm}$, and the
closer these are to 1 the better the phase measurement is. This POM is the best
possible for the main states that I will be considering in thesis, but not for
every possible input state. The best POM as derived in \cite{Hel76} is actually
dependent on the input state; this is discussed further in Sec.~\ref{optimal}.

Note that we may express the POM (\ref{canPOM}) in the form
\beq
F^{\rm can}(\phi) = \ket{\phi}\bra{\phi},
\eeq
where
\beq
\ket{\phi} = \frac 1{(2\pi)^{1/2}} \sum_{n=0}^{\infty} e^{in\phi}\ket n.
\eeq
These states are eigenstates of the Susskind-Glogower operators 
$\widehat{\cos \phi}$ and $\widehat{\sin \phi}$, and may be interpreted as
phase states. We therefore see that the POM of Eq.~(\ref{canPOM}) is consistent
with the Susskind-Glogower formalism. It can also be shown \cite{Hall91} that
identical results are obtained using this POM as using the Pegg-Barnett
formalism. In addition, London's treatment of phase \cite{Lon26,Lon27} is also
equivalent to this.

The fact that these different approaches to phase give equivalent results is a
compelling reason to consider this to be an accurate description of phase.
Nevertheless, there is another class of descriptions of phase which is not
equivalent to this: those based on an `operational' approach
\cite{oper0,oper1,oper2,oper3,oper4,oper5}. Here the phase is defined as the
quantity measured by a particular experiment. The disadvantage of this approach
is that the description of phase is dependent on the experiment. For further
discussion of the problems involved in the description of phase, see
Refs~\cite{review,Nieto}.

\section{Phase Variance}

\begin{figure}
\centering
\includegraphics[width=0.7\textwidth]{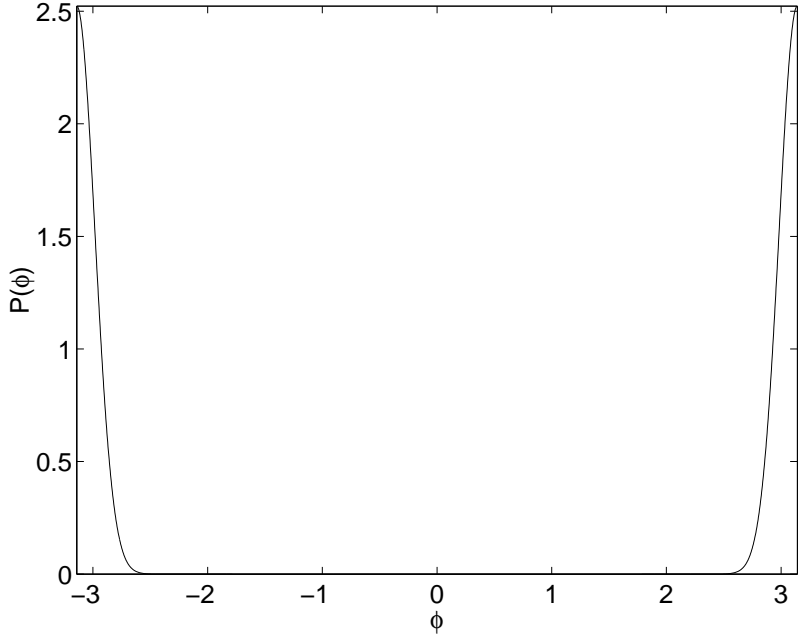}
\caption{A phase probability distribution that is narrowly peaked at $\pm \pi$.}
\label{narrow}
\end{figure}

In this work we are primarily interested in the phase variance, and not the
total phase distribution. Since phase is a cyclic variable the usual definition
of phase variance, as given by
\beq
\label{standard}
{\rm var}(\phi)=\ip{\phi^2}-\ip{\phi}^2,
\eeq
does not work well. For example, we would usually expect the variance to go to
infinity in the limit of a flat distribution. For phase, as the distribution is
limited to a region of length $2\pi$, the variance will be finite for a flat
distribution. This means that it is not possible to give an uncertainty relation
in the usual way. In addition, if the mean of the distribution is at one bound
of the phase, the phase variance obtained from this definition will be
artificially large. For example, the distribution shown in Fig.~\ref{narrow} is
narrowly peaked, but Eq.~(\ref{standard}) will give a very large variance.

These problems can be solved by using the Holevo phase variance \cite{Hol84}
\beq
\label{defholevo}
V(\phi)=\st{\ip{e^{i\phi}}}^{-2}-1.
\eeq
This definition is naturally modulo $2\pi$, and in addition if the distribution
is flat then $\ip{e^{i\phi}}$ will be zero, so the variance will be infinite.
For this definition of the variance, there is the uncertainty relation
\beq
V(\phi)\Delta N^2 \ge \smallfrac 14.
\eeq
In contrast, using Eq.~(\ref{standard}) for the phase variance there is no
uncertainty relation of this form because, for example, a number state has zero
number variance but a finite variance under Eq.~(\ref{standard}).

Also, for a distribution that is sharply peaked, we find
\bqa
V(\phi)\!\!\!\! &=& \!\!\!\!\st{\ip{e^{i\phi}}}^{-2}-1 \nn \\
\!\!\!\! &=& \!\!\!\!\st{e^{i\ip{\phi}}\ip{e^{i(\phi-\ip{\phi})}}}^{-2}-1 \nn \\
\!\!\!\! &=& \!\!\!\!\st{\ip{e^{i(\phi-\ip{\phi})}}}^{-2}-1 \nn \\
\!\!\!\! & \approx & \!\!\!\!\st{\ip{1+i(\phi-\ip{\phi})-\half
(\phi-\ip{\phi})^2 }}^{-2}-1 \nn \\
\!\!\!\! &=& \!\!\!\!\st{1-\half \ip{(\phi-\ip{\phi})^2}}^{-2}-1 \nn \\
\!\!\!\! &=& \!\!\!\!\left[ 1-\half \ip{(\phi-\ip{\phi})^2}\right]^{-2}-1 \nn \\
\!\!\!\! & \approx & \!\!\!\! 1+\ip{(\phi-\ip{\phi})^2}-1 \nn \\
\!\!\!\! &=& \!\!\!\!\ip{(\phi-\ip{\phi})^2}.
\eqa
Therefore this measure of the phase variance is approximately the same as the
usual estimate of the variance for sharply peaked distributions.

The definition of the Holevo variance (\ref{defholevo}) can be used when the
exact phase distribution is known. In obtaining numerical results, it is is
generally the case that the exact distribution is unknown, so the Holevo phase
variance must be estimated from a set of samples. In the case of the standard
variance, it is a standard result in statistics that the variance calculated
from the samples by
\beq
\frac 1M \sum_{i=1}^M (\phi_i-\ip \phi)^2,
\eeq
where $M$ is the number of samples, is a biased estimator for the variance of
the distribution. The unbiased estimator is where the dividing factor is $M-1$,
i.e.\ the unbiased estimator is
\beq
\frac 1{M-1} \sum_{i=1}^M (\phi_i-\ip \phi)^2.
\eeq
This is because taking the average of the data removes one degree of freedom.
If the mean of the distribution is known, then an unbiased estimator for the
variance is
\beq
\frac 1M \sum_{i=1}^M (\phi_i-\bar \phi)^2,
\eeq
where $\bar \phi$ is the known mean for the distribution. If the measurements
are unbiased, then the mean for the distribution will be the same as the actual
phase, $\varphi$. If we use the actual phase, then the estimator becomes
\beq
\label{altstandard}
\frac 1M \sum_{i=1}^M {(\phi_i-\varphi)^2}.
\eeq

The situation is analogous for the Holevo phase variance. If the measurements
are unbiased, we can estimate the Holevo variance from the samples by
\beq
\label{altholevo}
\left[{\rm Re}\left(\frac 1M \sum_{i=1}^M {e^{i(\phi_i-\varphi)}}\right)
\right]^{-2}-1.
\eeq
It is easy to see that when the phase distribution is narrowly peaked, this
simplifies to Eq.~(\ref{altstandard}). On the other hand, if the measurements
are {\it biased}, then this will not be an estimator for the Holevo variance,
and Eq.~(\ref{altstandard}) will not be an estimator for the standard variance.

For measurements that may be biased the variance is not a
good measure of the accuracy of the measurement. An arbitrary bias may be added
to the distribution without altering the variance, as the deviation is measured
from the average of the distribution, rather than the actual phase. For a
biased measurement scheme, it is more appropriate to define the standard
variance as
\beq
{\rm var}_{\varphi}(\phi) = \ip{(\phi-\varphi)^2}.
\eeq
Here the subscript $\varphi$ has been used to distinguish this variance from the
usual definition. For this definition, biased distributions will give larger
variances. Also Eq.~(\ref{altstandard}) will be an unbiased estimator for this
modified variance.

The analogous definition for the Holevo variance is
\beq
\label{modHol}
V_{\varphi}(\phi)=\left[{\rm Re}\ip{e^{i(\phi-\varphi)}}\right]^{-2}-1.
\eeq
In the case that $\varphi=0$, or $\phi$ is the deviation from the system phase,
this definition of the variance simplifies to
\beq
V_0(\phi)=\left[{\rm Re}{\ip{e^{i\phi}}}\right]^{-2}-1.
\eeq
Similarly to the case for the standard variance, this definition will give
larger variances for biased distributions, whereas the standard definition does
not distinguish biased and unbiased distributions. If the measurements are
biased, then Eq.~(\ref{altholevo}) will be an estimator for this variance, but
not the Holevo variance as given by Eq.~(\ref{defholevo}).

In this thesis I usually consider phase distributions that are unbiased, so this
modified definition of the Holevo phase variance is not required. For the
problem of finding optimal phase estimates, however, this altered definition is
necessary in order to eliminate the possibility of biased phase estimates.

Now if we consider arbitrary measurements on an arbitrary pure state
$\ket{\psi}$, the probability distribution for the phase is given by
\bqa
P(\phi)\!\!\!\! &=& \!\!\!\!{\rm Tr}[\rho F(\phi)] \nn \\
\!\!\!\! &=& \!\!\!\!\frac 1{2\pi}\sum_{n,m=0}^\infty \braket{\psi}{n}
\braket{m}{\psi}e^{i(n-m)\phi}H_{nm}.
\eqa
In order to determine the Holevo phase variance we must determine
$\ip{e^{i\phi}}$. Evaluating this gives
\bqa
\label{expectgen}
\ip{e^{i\phi}}\!\!\!\! &=& \!\!\!\!\int\limits_{-\pi}^{\pi}P(\phi)e^{i\phi}
d\phi\nn\\
\!\!\!\! &=& \!\!\!\!\int\limits_{-\pi}^{\pi}\frac 1{2\pi}\sum_{n,m=0}^\infty
\braket{\psi}{n}\braket{m}{\psi}e^{i(n+1-m)\phi}H_{nm}d\phi \nn \\
\!\!\!\! &=& \!\!\!\!\sum_{n,m=0}^\infty \braket{\psi}{n}\braket{m}{\psi}
\delta_{n+1,m}H_{nm} \nn \\
\!\!\!\! &=& \!\!\!\!\sum_{n=0}^\infty \braket{\psi}{n}\braket{n+1}{\psi}
H_{n,n+1}.
\eqa
Therefore the phase variance does not depend on all the elements
$H_{nm}$, but only on the off-diagonal elements $H_{n,n+1}$. As we are often
only interested in the phase variance resulting from a measurement scheme, this
greatly reduces the number of variables required to describe the measurement.
Note that the condition $|H_{nm}|\le 1$ means that this expression gives
$|\lr{e^{i\phi}}|\le 1$, as we would expect.

For canonical measurements we have $H_{nm}^{\rm can}=1$, so this simplifies to
\beq
\ip{e^{i\phi}}_{\rm can}=\sum_{n=0}^{\infty}\braket{\psi}{n}\braket{n+1}{\psi}.
\eeq
For more arbitrary measurements, we will have $H_{n,n+1}\le 1$, so it is
convenient to define the vector $h(n)$ by
\beq
h(n)=1-H_{n,n+1}.
\eeq
In terms of this we obtain
\bqa
\label{varesult}
\ip{e^{i\phi}}\!\!\!\! &=& \!\!\!\!\sum_{n=0}^{\infty}\braket{\psi}{n}
\braket{n+1}{\psi}(1-h(n)) \nn \\
\!\!\!\! &=& \!\!\!\!\sum_{n=0}^{\infty}\braket{\psi}{n}\braket{n+1}{\psi}-
\sum_{n=0}^{\infty}\braket{\psi}{n}\braket{n+1}{\psi}h(n)
\nn \\
\!\!\!\! &=& \!\!\!\!\ip{e^{i\phi}}_{\rm can}-
\sum_{n=0}^{\infty}\braket{\psi}{n}\braket{n+1}{\psi}h(n).
\eqa
If the photon number distribution is reasonably sharply peaked then we can
typically replace $h(n)$ by its value for the average photon number. Then we
find
\beq
\ip{e^{i\phi}} \approx \ip{e^{i\phi}}_{\rm can}(1-h(\bar n)).
\eeq
For large mean photon number, we find that $h(\nb)\ll 1$ for most measurement
schemes. Using this approximation, the Holevo phase variance is
\bqa
\label{additvar}
V(\phi)\!\!\!\!&\approx &\!\!\!\!\st{\ip{e^{i\phi}}_{\rm can}(1-h(\bar n))}^{-2}-1\nn\\
\!\!\!\! &=& \!\!\!\!\st{\ip{e^{i\phi}}_{\rm can}}^{-2}(1-h(\bar n))^{-2}-1\nn\\
\!\!\!\!&\approx&\!\!\!\!\st{\ip{e^{i\phi}}_{\rm can}}^{-2}(1+2h(\bar n))-1\nn\\
\!\!\!\!&\approx&\!\!\!\!\st{\ip{e^{i\phi}}_{\rm can}}^{-2}+2h(\bar n)-1 \nn \\
\!\!\!\! &=& \!\!\!\!V_{\rm can}(\phi)+2h(\bar n).
\eqa
Thus we see that to a first approximation the phase variance is equal to the
intrinsic phase variance plus twice the value of $h(n)$ for the average photon
number. This means that the introduced phase variance is fairly independent of
the input state (apart from the photon number).

For many types of real measurements, the value of $h(n)$ decreases
as some power of $n$ for large $n$. We can therefore approximate it by
\beq
h(n) \approx cn^{-p}.
\eeq
When this is the case, we can describe the measurement by just the two
variables $c$ and $p$, rather than the entire vector $h(n)$, or the matrix
$H_{nm}$.

\section{Real Measurements}

In practice, it is not possible to directly measure phase. What we must do is
infer the phase from measurements of some other quantity. The electric field,
with the operator given by Eq.~(\ref{elecfield}), is directly measurable.
Ignoring the field direction, this operator is proportional to
\beq
\hat X_{\Phi} = a e^{-i\Phi} + a\dg e^{i\Phi},
\eeq
where
\beq
\Phi = \omega t - {\mathbf k} \cdot {\mathbf R}.
\eeq
This is called the $\Phi$ quadrature operator. For high
frequency light $\Phi$ varies far too rapidly for convenience, so we wish to
measure the quadrature in a more convenient way.

\begin{figure}
\centering
\includegraphics[width=0.45\textwidth]{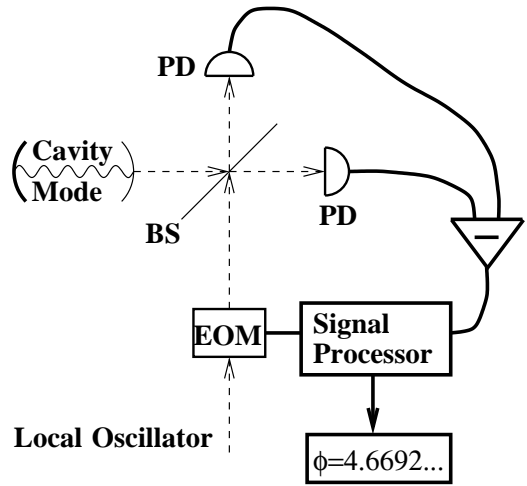}
\caption{Diagram of the apparatus for making a dyne phase measurement.
The signal from the cavity and the local oscillator field are combined at a
50/50 beam splitter (BS) and the outputs are detected by photodetectors (PD).
For adaptive measurements the signals from these photodetectors are processed
by the digital signal processor, which determines a phase estimate and adjusts
the electro-optic phase modulator (EOM) accordingly.}
\label{dynefig}
\end{figure}

This can be done by combining the light with a strong local oscillator field at
a beam splitter, as in Fig.~\ref{dynefig}. The local oscillator field provides
a reference phase. The difference between the photon numbers at the two
photodetectors gives a measurement of the quadrature $\hat X_{\Phi}$, where
$\Phi$ is the phase of the local oscillator field.

In this study I consider measurements that take an appreciable time, so rather
than just considering the total photon counts at the detectors, it is possible
to consider the instantaneous photocurrent. For this extended measurement we can
alter the phase of the local oscillator during the measurement based on
the data obtained so far.

The standard technique for measuring a completely unknown phase is heterodyne
detection, where the phase of the local oscillator is varied linearly. This is
simple to do, as it simply means that the local oscillator has a slightly
different frequency than the signal. This method suffers from the drawback that
it introduces a fairly large phase uncertainty. It is possible to obtain more
accurate measurements if we use a constant feedback phase $\Phi$, that is close
to the actual phase plus $\pi/2$. This is called the homodyne technique.

The major problem with the homodyne technique is that it requires knowledge of
the system phase beforehand. The guiding principle behind adaptive phase
measurements is that we wish to approximate homodyne phase measurements by
varying the local oscillator phase during the measurement based on data
obtained so far. We will see that it is not quite so simple, however, as we do
not use the best estimate of the phase, and in rare cases will not use a
phase estimate at all. I will use the general term ``dyne'' for heterodyne and
homodyne phase measurements, as well as these more complicated adaptive phase
measurements.

In order to describe the adaptive measurements I will first introduce the
notation. The numbers of detections at the two photodetectors in the time
interval $\delta t$ are denoted $\delta N_+$ and $\delta N_-$. The complex
amplitude of the local oscillator will be denoted as $\gamma$. The difference
photocurrent $I(t)$ is then defined in terms of the noncommuting limits
\beq
I(t) = \lim_{\delta t \to 0} \lim_{\st{\gamma} \to \infty } 
\frac{\delta N_+ - \delta N_-} {\st{\gamma} \delta t}.
\eeq
This simplifies to
\beq
I(t) dt = 2{\rm Re} (\ip a e^{-i\Phi}) dt + dW(t),
\eeq
where $dW(t)$ is an infinitesimal Wiener increment with variance $dt$. For a
constant amplitude coherent state we have $\ip a = \alpha$. Note that
$\lr{I(t)} = \lr{\hat X_{\Phi}}$, so this is essentially a measurement of the
$\Phi$ quadrature.

For more general states, $\ip a$ may have a systematic variation given by
\beq
\ip a = \alpha_t \propto \sqrt{u(t)},
\eeq
where $u(t)$ is a mode function. In this case we alter the definition of $I(t)$,
and scale $\alpha_t$ and the time to obtain
\beq
\label{photocur}
I(v) dv = 2{\rm Re} (\alpha_v e^{-i\Phi}) dv + dW(v),
\eeq
similarly to the case for constant amplitude. The detail of how this is done is
given in Ch.~\ref{optdyne}.

We then define the quantities $A_v$ and $B_v$ by
\bqa
\label{defA}
A_v \!\!\!\! &=& \!\!\!\! \int\limits_0^v e^{i\Phi}I(u)du, \\
\label{defB}
B_v \!\!\!\! &=& \!\!\!\! -\int\limits_0^v e^{2i\Phi} du,
\eqa
The time is scaled to the unit interval, and for the values of $A_v$ and $B_v$
at $v=1$ the subscripts are omitted. It is shown in Ref.~\cite{wise96} that the
only relevant information from the measurement record is contained in these two
variables. ($B_v$ does not explicitly depend on the measurement record, but it
does depend on it implicitly if the feedback phase is varied based on the
measurement record.)

To see why this is so, firstly recall that the way the normalised state varies
for measurement operator $\Omega_r$ is
\beq
\ket{\psi_r} = \frac{\Omega_r \ket{\psi}}{\sqrt{P_r}}.
\eeq
Alternatively we can consider an unnormalised state $\ket{\tilde \psi_r}$,
given by
\beq
\ket{\tilde \psi_r} = \frac{\Omega_r \ket{\psi}}{\sqrt{\Lambda_r}},
\eeq
where the $\Lambda_r$ are called the ostensible probabilities. The actual
probability is then given by
\beq
P_r = \Lambda_r \braket{\tilde \psi_r}{\tilde \psi_r}.
\eeq
This method is used in Ref.~\cite{wise96} to consider the evolution of the
signal state under dyne measurements, with the ostensible probabilites chosen
as those for a vacuum.

It is found that the evolution of the unnormalised state depends on the
photocurrent record and feedback phases only through the variables $A_v$ and
$B_v$. This means that the probability distribution for the photocurrent record
up to time $v$, ${\bf I}_{[0,v)}$, is given by
\beq
P({\bf I}_{[0,v)}) = P_0({\bf I}_{[0,v)}) \braket{\tilde \psi_v(A_v,B_v)}
{\tilde \psi_v(A_v,B_v)},
\eeq
where $P_0({\bf I}_{[0,v)})$ is the ostensible probability distribution.
Note that this does not prove that the probability distribution for the
photocurrent record only depends on $A_v$ and $B_v$, as the ostensible
probability distribution may depend on the photocurrent record in some more
complicated way. The ostensible distribution, however, is for a vacuum, and
does not contain any information about the state. This means that all the
information about the state from the photocurrent record ${\bf I}_{[0,v)}$ is
contained in $A_v$ and $B_v$.

To show this in a more rigorous way, consider some arbitrary parameter of the
input state, $x$. This could, for example, be the phase or the photon number.
Using Bayes' theorem we find that the probability distribution for this
quantity given the photocurrent record ${\bf I}_{[0,v)}$ is
\beq
P(x|{\bf I}_{[0,v)}) = \frac{P(x)P({\bf I}_{[0,v)}|x)}{P({\bf I}_{[0,v)})}.
\eeq
Here $P(x)$ is the probability distribution for this quantity at the start of
the measurement. It will be flat, as we are assuming that there is no knowledge
about this parameter before the start of the measurement. The probability in
the denominator is independent of $x$, so the probability distribution is
therefore
\beq
\label{genbayes}
P(x|{\bf I}_{[0,v)}) \propto P({\bf I}_{[0,v)}|x).
\eeq
The ostensible probability distribution $P_0({\bf I}_{[0,v)})$ contains no
information about the state, and will therefore be independent of $x$. Thus we
find
\beq
P(x|{\bf I}_{[0,v)}) \propto \braket{\tilde \psi_v(A_v,B_v)}
{\tilde \psi_v(A_v,B_v)}.
\eeq

This means that the probability distribution for any parameter of the signal
state only depends on the photocurrent record through $A_v$ and $B_v$. This
means, for example, that estimators for the phase should be functions of
$A_v$ and $B_v$. They can be functions of other variables, for example the
initial coherent amplitude, if that is known, but they should depend on the
photocurrent record only through $A_v$ and $B_v$.

\section{Adaptive Phase Measurements}
\label{introadapt}
For a coherent state with $\alpha = \st{\alpha} e^{i\varphi}$,
Eq.~(\ref{photocur}) becomes
\beq
I(v) dv = 2 \st{\alpha}\cos(\varphi-\Phi) dv + dW(v).
\eeq
In order for the measurement to be close to an ideal measurement of the phase,
the $\cos$ function should be as close as possible to zero. This implies that
its argument should be close to $\pm \pi/2$. Therefore it is best to take the
local oscillator phase to be $\hat \varphi +\pi/2$, where $\hat \varphi$ is
some estimate of the system phase. In that case
\beq
I(v) dv = 2 \st{\alpha}\sin(\varphi-\hat\varphi) dv + dW(v).
\eeq
In the limit that $|\varphi-\hat\varphi| \ll 1$, this becomes
\beq
I(v) dv \approx 2 \st{\alpha}(\varphi-\hat\varphi) dv + dW(v),
\eeq
so the measurement is very close to an ideal measurement of the phase. The
reason why there is always an excess phase uncertainty is because the $\sin$
function is not completely linear, so this is not exactly equivalent to a
direct measurement of the phase.

The basis of homodyne measurements is to use an estimate of the phase that is
known before the measurement begins. For adaptive measurements, we only have
information about the phase once the measurement has begun, and we use a phase
estimate based on the data obtained so far. To see what we can use as a phase
estimator, we can expand the variable $A_v$ to give
\beq
\label{expandA0}
A_v = v\alpha - \alpha^* B_v + i\sigma_v,
\eeq
where
\beq
\sigma_v = \int\limits_0^v e^{i\Phi(u)-i\pi/2} dW(u).
\eeq
From this result it is simple to show that
\beq
v A_v + B_v A_v^* = \alpha (v^2 - \st{B_v}^2) + i(v \sigma_v - B_v \sigma_v^*).
\eeq
Taking the expectation value gives
\beq
\ip{v A_v + B_v A_v^*} \approx \alpha \ip{v^2 - \st{B_v}^2}.
\eeq
If the local oscillator phase is independent of the photocurrent record, then
this is exact. In the case of feedback, $B_v$ may be correlated with $\sigma_v$,
but this result should still be approximately true.
This means that the phase of $v A_v + B_v A_v^*$ should be close to the phase
of the signal. It is convenient to define the new variable
\beq
\label{defC}
C_v = v A_v + B_v A_v^*.
\eeq
This should not be confused with the variable $C$ used in Ref.~\cite{fullquan},
which is defined differently.

For heterodyne measurements, the local oscillator phase $\Phi(v)$ is varying
linearly. This means that the average value of $e^{2i\Phi(v)}$ will be zero,
so $B_v$ will be close to zero. This means that $C_v \approx A_v$, so
$\arg A_v$ will also be a good estimator of the phase. It is simple to show
from this \cite{semiclass} that the phase variance for heterodyne measurements
on coherent states is approximately $\half \st{\alpha}^{-2}$. As the intrinsic
phase variance for coherent states is $\smallfrac 14 \st{\alpha}^{-2}$, this
means that there is an introduced phase variance of
$\smallfrac 14 \nb^{-1}$.

For more general measurements where $B_v \ne 0$, $\arg A_v$ will not be as good
an estimator of the phase; however, it can still be used as an estimator. If
$\hat\varphi = \arg A_v$ is used as the estimator for the phase in feedback to
the local oscillator phase, we have a measurement scheme that can be analysed
analytically. In the scheme considered in Ref.~\cite{Wis95c}, $\arg A$ was the
phase estimate used at the end of the measurement as well.

It was shown in Ref.~\cite{Wis95c} that these measurements give the canonical
result if the system has at most one photon. For systems with large numbers of
photons, however, the introduced phase variance scales as
$\smallfrac 14 \nb^{-1/2}$. This means that for large photon numbers
this measurement scheme gives far higher phase variances than heterodyne
measurements.

It is possible to greatly improve on this result by simply using the best phase
estimate at the end of the measurement, $\arg C$. This is called the mark II
adaptive phase measurement scheme, whereas the scheme where $\arg A$ is used at
the end of the measurement is called mark I. This measurement scheme introduces
a phase uncertainty of $\smallfrac 18 \nb^{-3/2}$ \cite{semiclass}. This is a
vast improvement over the mark I scheme, and also improves on the heterodyne
scheme.

Recall from Eq.~(\ref{additvar}) that the introduced phase variance is
approximately $2h(\nb)$. This means that the variation of $h(n)$ for the
heterodyne, mark I and mark II measurement schemes is approximately
\bqa
h_{\rm het}(n) \!\!\!\! &\approx& \!\!\!\! \smallfrac 18 n^{-1}, \\
h_{\rm I}(n) \!\!\!\! &\approx& \!\!\!\! \smallfrac 18 n^{-0.5}, \\
h_{\rm II}(n) \!\!\!\! &\approx& \!\!\!\! \smallfrac 1{16} n^{-1.5}.
\eqa
To summarise this, the values of $c$ and $p$ for these three measurement
schemes are as given in Table \ref{table:meascp}.
\begin{table}[tbp]
\centering
\caption{The values of $c$ and $p$ for heterodyne, mark I and mark II
measurements.}
\begin{tabular}{||c|c|c||} \hline \hline
           & $c$      & $p$ \\ \hline
heterodyne & $1/8$    & 1   \\ \hline
mark I     & $1/8$    & 0.5 \\ \hline
mark II    & $1/16$   & 1.5 \\ \hline \hline
\end{tabular}
\label{table:meascp}
\end{table}

The mark II measurement scheme still leaves questions, however. It has been
shown \cite{SumPeg90} that optimum states have intrinsic phase variances that
scale as $1.89\times\nb^{-2}$. For these states the phase uncertainty that is
introduced by mark II measurements will be far greater than the intrinsic phase
uncertainty of the state. In addition it has been shown \cite{fullquan} that
the theoretical limit to the variance that is introduced by dyne phase
measurements is $\smallfrac 14 \log\nb \times \nb^{-2}$. This is not quite as
good as the scaling for the intrinsic phase variance, but it is far better than
the scaling for mark II measurements.

Another issue is that the intermediate phase estimates used for mark II
measurements are still $\arg A_v$. As evidenced by the poor scaling for mark I
measurements, these phase estimates are far worse than $\arg C_v$. This raises
the question of whether phase measurements can be made closer to the
theoretical limit by using better intermediate phase estimates. This is
considered in Ch.~\ref{optdyne}, and it is shown that measurements can be
made very close to the theoretical limit.

Another factor in attempting to make the most accurate possible phase
measurements is the input state. Rather than just considering states that are
optimised for minimum intrinsic phase variance, the fairest way to evaluate the
various measurement schemes is to consider states that are optimised for
minimum phase variance under that particular phase measurement scheme. This
is considered in Ch.~\ref{dynesta}.

It is necessary to have a constraint on the state being optimised in order to
avoid obtaining a state with indefinitely large photon number. The two main
ways of constraining the state that is being optimised are to put an upper
limit on the photon number and to specify the mean photon number. A third
alternative is to only consider squeezed states. These states are far more
convenient to work with, both theoretically and experimentally, and in addition
they give results that are extremely close to those for the general
optimisation problem for general measurements. All these alternatives are
considered in Ch.~\ref{dynesta}.

\section{Interferometry}
The major alternative to dyne measurements is interferometric phase
measurements. Here, rather than measuring the phase of a single mode, and
assuming that the local oscillator field is sufficiently intense that it can be
treated classically, we are measuring the phase difference between two modes,
both of which are treated quantum mechanically.

\begin{figure}
\centering
\includegraphics[width=0.45\textwidth]{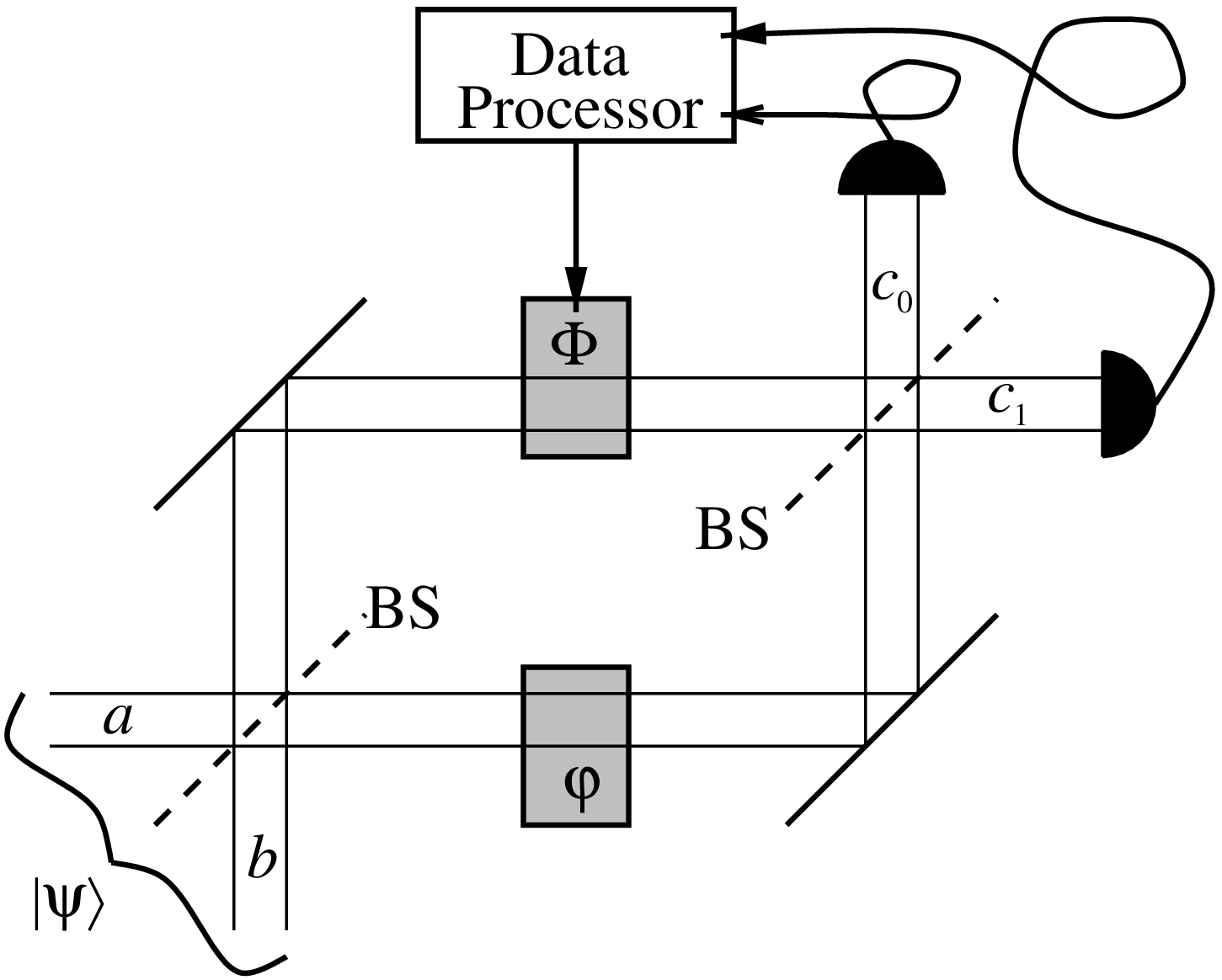}
\caption{The Mach-Zehnder interferometer, with the addition of a controllable
phase $\Phi$ in one arm. The unknown phase to be estimated is $\varphi$. Both
beam splitters (BS) are 50/50.}
\label{diag0}
\end{figure}

The most convenient way of considering this is via a Mach-Zehnder
interferometer, as in Fig.~\ref{diag0}. Two input modes are combined at a
beam splitter, after which each of the modes is subjected to a phase shift, and
the two modes are recombined at a second beam splitter. Usually we would
consider measurement of the phase difference between the two arms. Here for
simplicity I consider the phase shift to be measured, $\varphi$, to
be in one arm, and add a controllable phase shift, $\Phi$, in the second arm.
This allows us to make adaptive measurements analogous to the case for dyne
measurements.

Note that if the first beam splitter is omitted, the configuration is identical
to the configuration used to make dyne measurements of a single mode. The arm of
the interferometer subjected to the phase shift $\Phi$ is equivalent to the
local oscillator mode, and the arm subjected to the phase shift $\varphi$ is
equivalent to the signal mode in the dyne case. The difference here is that we
are now treating both modes quantum mechanically.

The main role of the first beam splitter is that it creates quantum
correlations between the two modes, which generally improves the phase
properties of the state. For example, it is not possible to measure a phase
difference between a number state and a vacuum state, but the two mode state
produced after the beam splitter has good phase properties.

In general it is much easier to produce input states without quantum
correlations between the modes, so the initial beam splitter is very useful.
In this study, I consider two-mode input states with arbitrary
correlations between the modes. In this case, the initial beam splitter is
fairly superfluous, as it merely transforms the input state into another
correlated state that could have been considered directly. I still include
the initial beam splitter in this study, for consistency with previous work.

It is well known that if we feed a state with a mean photon number of $\nb$
into one arm, it is possible to obtain a phase variance scaling as $\nb^{-1}$.
This is analogous to the case of coherent states for dyne measurements, which
are the easiest to produce and also give $\nb^{-1}$ scaling. There have been
several proposals for reducing the phase variance to $\nb^{-2}$. The first of
these is that by Caves \cite{Caves}.

Caves considers an interferometer with a coherent state in one arm, but in the
other arm, rather than a vacuum state, a squeezed vacuum is used. The
measurements considered are equivalent to homodyne measurements, where the phase
difference is very close to $0$. For the right squeezing parameter these
measurements have a variance scaling as $\nb^{-2}$. Squeezed states were also
considered by Bondurant and Shapiro \cite{BondShap}.

Yurke {\it et al.} \cite{Yurke} consider an input state that is a combination
of input number states,
\beq
\ket \psi = \frac 1{\sqrt 2} \left( \ket{j}_a \ket{j}_b + \ket{j+1}_a
\ket{j-1}_b \right),
\eeq
where the subscripts $a$ and $b$ indicate number states in ports $a$ and $b$
respectively. According to their analysis this state should give a phase
uncertainty scaling close to $N^{-1}$, where $N=2j$ is the fixed total photon
number. Similarly to Ref.~\cite{Caves} the phase must be very close to $0$ to
obtain this scaling. Yurke {\it et al.} also consider active interferometers,
characterised by SU(1,1) rather than SU(2), as is the conventional Mach-Zehnder
interferometer considered in this study.

Holland and Burnett \cite{Holland} also considered the case where the total
photon number is fixed. They considered the state with equal photon numbers in
both input ports, $\ket{j}_a \ket{j}_b$. For this state the phase uncertainty
scales as $N^{-1}$, but again only for phases very close to $0$. This state also
has the additional problem that it gives results at $0$ and $\pm \pi$ with equal
probability, and therefore must be considered modulo $\pi$.

Sanders and Milburn \cite{SandMil95,SandMil97} considered ``optimal''
measurements, for which the phase uncertainty is independent of the system
phase. Unfortunately these measurements are derived from projections onto the
phase states, rather than a physical measurement scheme involving
photodetectors. In Ch.~\ref{interfere} it is shown that it is not possible to
implement these measurements using photodetectors, even allowing feedback.
Therefore I will generally call these measurements ideal or canonical rather
than optimal, and reserve the term optimal for the best possible physically
realisable scheme.

Sanders and Milburn considered the same input state as Holland and Burnett,
$\ket{j}_a \ket{j}_b$. In Ref.~\cite{SandMil97} they show that the phase
uncertainty for this state scales as $N^{-1}$ according to two common measures
of the uncertainty. In Ch.~\ref{adaptiveinter} I show that this state has very
poor scaling if we consider the phase variance. The optimal input states as
evaluated using the Holevo phase variance are derived, and it is shown that all
the common measures of the phase uncertainty scale as $N^{-1}$ for these
states. Both the $\ket{j}_a \ket{j}_b$ and $\frac 1{\sqrt 2} \left( \ket{j}_a
\ket{j}_b + \ket{j+1}_a \ket{j-1}_b \right)$ states are rough approximations of
this state.

\section{Experimental Imperfections}
The majority of the work considered in this thesis ignores all experimental
imperfections. This is because the main motivation for this work is to lay a
theoretical foundation for how far phase measurements can, in principle, be
improved, rather than to examine the limits to phase measurements using current
technology.

Firstly there are a number of problems that are very specific to the apparatus.
For example an inaccurate calibration in the phase shifter producing the phase
$\Phi$ will result in a corresponding error in the measured phase. These types
of problems are relatively simple to analyse, and as they are very specific to
the equipment used they will not be discussed here.

One problem that is common to any phase measurement is inefficient
photodetectors. No current photodetectors will register every photon. There are
two main types of photodetectors with different efficiencies. Firstly there are
large amplitude photodetectors that do not need to distinguish the exact photon
number. These are used for dyne measurements, where the photon counts are
treated in the continuous approximation. These can be made very efficient,
and the best current photodetectors have efficiencies around 98\%
\cite{polzik}.

The case of dyne measurements with inefficient detectors was considered in
\cite{semiclass}. For the case of coherent states, the analysis is simple, as
an efficiency $\eta$ can be treated by changing the mean photon number from
$\nb$ to $\eta\nb$. This means that, for mark II measurements, the phase
variance to first order is
\bqa
V(\phi) \!\!\!\! &=& \!\!\!\! \frac 1{4\eta\alpha^2} + O(\alpha^{-3}) \nn \\
\!\!\!\! &=& \!\!\!\! \frac 1{4\alpha^2} + \frac {1-\eta}{4\eta\alpha^2} +
O(\alpha^{-3})
\eqa
We will get the same result (to first order) for any other reasonably good
phase feedback scheme where the introduced phase variance is of higher order
than the intrinsic phase variance. The phase variance is in the form of the
intrinsic phase variance plus an extra term due to the inefficient
photodetectors. As explained previously, the introduced phase variance is
generally independent of the input state (to first order). This means that
\beq
\Delta V(\phi) = \frac {1-\eta}{4\eta\nb}
\eeq
will be the introduced phase variance due to the inefficient photodetectors
for other states with reduced phase uncertainty.

The second type of photodetectors is photon counters, that count photons one by
one. This is the type required for interferometric measurements, where we wish
to alter the feedback phase after every detection. These generally have far
lower efficiencies, less than 90\% \cite{SingleEff}. This is a serious problem,
as the analysis breaks down if even a single photon is missed. Taking account
of a fixed probability of missing a photon would greatly complicate the
analysis, and was not attempted in this study.

Another problem, that is unique to phase measurements involving feedback, is
the time delay in the feedback loop. This is arguably an even more fundamental
problem than inefficient photodetectors, because the feedback loop cannot
operate faster than the time it takes light to reach the phase modulator. This
is a very serious problem for short pulses. For example, if the pulse is
shorter than the distance between the phase modulator and the photodetector it
is not possible to perform any sort of adaptive measurement. Adaptive
measurements rely on the pulse being sufficiently long that the feedback phase
can be altered during the passage of the signal.

Even if the time delay is a small fraction of the length of the signal, there
will be an appreciable increase in the phase variance. This was investigated
by a highly simplified theory in Ref.~\cite{semiclass}, and in
Ch.~\ref{delays} the introduced phase variance is estimated in a far more
rigorous way.

\section{Structure of the Thesis}
This thesis is structured around the problems of determining the optimum input
states and measurement schemes for dyne measurements and interferometry. The
optimum input states for dyne measurements are derived in Ch.~\ref{dynesta}, and
the problem of performing optimum dyne measurements is discussed in
Ch.~\ref{optdyne}. Optimum input states for interferometry are discussed in
Ch.~\ref{adaptiveinter}, and optimum interferometric measurements are discussed
in Ch.~\ref{interfere}. These chapters form the main theme of this thesis. Some
additional problems considered are time delays in Ch.~\ref{delays} and
continuous (rather than pulsed) measurements in Ch.~\ref{continue}.

As mentioned above there are three main alternatives when considering optimum
input states for dyne measurements: \\
1. An upper limit is placed on the photon number. \\
2. The mean photon number is fixed, but the state is otherwise arbitrary. \\
3. The state is a squeezed state, and the mean photon number is fixed. \\
These alternatives are considered in Secs \ref{upper}, \ref{general} and
\ref{squeezed} respectively of Ch.~\ref{dynesta}. In each case both canonical
measurements and more general dyne measurements where $h(n) \approx cn^{-p}$
are considered.

The analytic results for canonical measurements and for general dyne
measurements with an upper limit on the photon number were previously derived
in Refs \cite{SumPeg90,semiclass,collett}. These derivations are summarised
here, as the method involved sheds light on the method of solution for the more
complicated cases of general dyne measurements for fixed mean photon number and
for squeezed states. The accuracy of all these results is evaluated by
extensive numerical calculations.

The problem of performing dyne phase measurements at the theoretical limit is
too complicated to solve analytically, and Ch.~\ref{optdyne} therefore relies
upon numerical methods. The various feedback schemes are evaluated numerically
by solving the stochastic differential equations (SDEs) for a large number of
samples. This is made simpler by using squeezed states, for which only the two
squeezing parameters need be kept track of. The problem of deriving the SDEs
for the squeezing parameters, as well as much of the background theory of the
evolution of the state, is described in Sec.~\ref{method}.

A series of different phase feedback schemes that give results progressively
closer to the theoretical limit are described in Secs~\ref{naive} to
\ref{timedepeps}. The last of these is a corrected feedback scheme that gives
variances within about 5\% of the theoretical limit. Lastly in this chapter,
the possibilities of surpassing the theoretical limit are considered.

The perturbation approach for time delays that is considered in
Ref.~\cite{semiclass} is repeated in a more rigorous way in Ch.~\ref{delays}.
It is shown that the same result is obtained for the final value of the
intermediate phase estimate, but this approach fails when it is performed for
mark II measurements. In Sec.~\ref{thmin} an alternative approach based on the
POM is considered that gives a result that is valid not only for the mark II
measurements, but for all other feedback schemes. These results are all backed
by extensive numerical calculations in Sec.~\ref{result}.

In Ch.~\ref{adaptiveinter} the optimum input states for interferometry are
derived, as evaluated using the Holevo variance for ideal measurements. The
phase variance for these states scales as $N^{-2}$, so the uncertainty scales as
$N^{-1}$. The Holevo variance for three alternative input states is calculated,
and it is shown that all of these have poorer scaling. The optimum input states
are then evaluated under several alternative measures of phase uncertainty, and
are shown to have $N^{-1}$ scaling for all of these.

In Ch.~\ref{interfere} a feedback scheme is introduced that approximates
the ideal measurements very closely. For optimal input states, the scaling in
the phase variance under this measurement scheme is very close to $N^{-2}$.
For photon numbers above 5, this measurement scheme is not exactly optimal, but
it is possible to solve numerically for the optimal feedback scheme (for small
photon numbers). Although it is not always possible to obtain variances as small
as canonical, using this feedback scheme it is possible to obtain variances
smaller than canonical for some states. In Ch.~\ref{interfere} the resolution
of this apparent contradiction is discussed.

The last area that is considered in this thesis is that of continuous
measurements. If we wish to transmit information via the phase, a pulsed signal
with a single phase is not very useful, as only a single real number is
transmitted. In order to transmit a significant amount of information we can
either send a whole series of pulses with different phases, or produce a
constant beam with a fluctuating phase. This is the alternative that is
considered in Ch.~\ref{continue}.

In this chapter both the cases of dyne measurements and interferometry are
considered. Unfortunately the non-classical states with reduced phase
uncertainty do not necessarily have equivalents in the continuous case. This is
because the reduced phase uncertainty is due to the back-action of the
measurement on the state. If the state does not change, then it is not possible
to get the reduced variance.

In the case of dyne measurements it is possible to consider squeezed states by
altering the statistics for the detections in time $dt$, but keeping $\alpha$
constant. Therefore both the coherent and squeezed state cases are considered
in Ch.~\ref{continue}. It does not seem to be possible to use this method for
interferometry, as we cannot alter the statistics for individual detections.
Therefore only the case with all photons in one port is considered in this
chapter.

%% file: dynesta.tex
\setcounter{chapter}{1}

\chapter{Optimal Input States for Dyne Measurements}
\label{dynesta}

There are, in general, two areas for improvement in the use of dyne
measurements for phase estimation. Firstly there is the input state, and
secondly there is the measurement technique that is used. In this chapter I will
be focusing on optimising the input states for minimum phase variance under
various types of dyne measurements, and in the next chapter I discuss
how to perform the best possible dyne measurements.

In optimising the input states, there must be some constraint placed on the
state, because we can reduce the phase variance indefinitely by using larger
and larger photon numbers. The limit of this is phase eigenstates, which have
zero intrinsic phase variance, but infinite mean photon number. There are two
main ways in which we can constrain the input states in order to avoid this
problem. One way is to place an upper limit on the photon number that the state
can have contributions from, and another way is to fix the mean photon number,
but allow the state to have contributions from indefinitely large photon
numbers.

A third alternative is to consider squeezed states rather than arbitrary states.
For these states we can optimise the squeezing parameter while keeping the
mean photon number fixed. The reason for considering this case is that it is
rather more realistic than considering arbitrary states, as it is not possible
to produce an arbitrary state experimentally, whereas it is possible to produce
a squeezed state. In addition squeezed states are easily treated numerically, as
only the two squeezing parameters need be considered, rather than the full
state. A third reason is that squeezed states give results extremely close to
general optimised states for realistic dyne measurements.

I will firstly consider the case where there is an upper limit placed on the
photon number.

\section{Upper Limit on Photon Number}
\label{upper}
This case is far simpler than the case of a fixed mean photon number, and the
analytic results were derived before this study commenced. I will summarise the
derivations here, however, as they are necessary to understand the numerical
results that will be presented. In addition, these derivations cast light on
how to derive the more complicated results with fixed mean photon number. 

\subsection{Canonical Measurements}
\label{simplest}
Firstly I will consider the optimum states for canonical measurements. This is
the simplest case, and in fact the only case which is exactly soluble. It is
solved, for example in Refs~\cite{SumPeg90} and \cite{semiclass}. As discussed
in the introduction, the measure of the phase variance that will be used
throughout this thesis is the Holevo phase variance,
\beq
\label{simpleholevo}
V(\phi) = \st {\ip {e^{i\phi}}}^{-2} -1.
\eeq
For simplicity, the mean phase of the optimised state can be taken to be zero.
This means that $\ip {e^{i\phi}}$ is real. Therefore, rather than using
Eq.~(\ref{simpleholevo}), we can use
\bqa
V(\phi) \!\!\!\! &=& \!\!\!\! \st {\half \left(\ip {e^{i\phi}}+\ip {e^{-i\phi}}
\right)}^{-2} -1 \nn \\
\!\!\!\! &=& \!\!\!\!\ip {\cos \phi}^{-2} -1.
\eqa
This means that minimising the Holevo phase variance is equivalent to maximising
$\ip{\cos \phi}$. Thus minimising the Holevo phase variance is equivalent to
minimising the measure of the variance
\beq
\label{altvardef}
V'(\phi) = 2(1-\ip{\cos \phi}).
\eeq
In addition, this measure of the phase variance has the same value in the limit
of small variance as the Holevo phase variance. This measure is not exactly
equal to the Holevo phase variance, and as is shown later it differs in the
higher order terms. This is significant for general dyne measurements where we
wish to obtain higher order terms in the expressions for the phase variance.

Now recall from Eq.~(\ref{expectgen}) that
\beq
\ip{e^{i\phi}}=\sum_{n=0}^\infty \braket{\psi}{n}\braket{n+1}{\psi}H_{n,n+1}.
\eeq
This means that we can represent $e^{i\phi}$ as the operator
\beq
\widehat{\exp(i\phi)}=\sum_{n=0}^\infty \ket{n}\bra{n+1} H_{n,n+1},
\eeq
and we can represent $\cos \phi$ by the operator
\beq
2 \widehat{\cos \phi}=\sum_{n=0}^\infty \left[ \ket{n}\bra{n+1} +\ket{n+1}
\bra{n}\right] H_{n,n+1}.
\eeq
When we put an upper limit of $N$ on the photon number, we can replace this with
\beq
\label{gencos}
2 \widehat{\cos \phi}=\sum_{n=0}^{N-1} \left[ \ket{n}\bra{n+1} + \ket{n+1}
\bra{n}\right] H_{n,n+1}.
\eeq
For canonical measurements $H_{n,n+1}=1$, so this becomes
\beq
2 \widehat{\cos \phi}=\sum_{n=0}^{N-1} \left[ \ket{n}\bra{n+1}+\ket{n+1}\bra{n}
\right].
\eeq

In general, when we wish to maximise (or minimise) the expectation value of some
Hermitian operator $\hat A$ while keeping the expectation values of other
Hermitian operators $\hat B$, $\hat C \ldots$, constant, we use the method of
undetermined multipliers. For this, we require that the matrix elements of all
of these operators, $\bra n \hat X \ket{m}$, be real, so the corresponding
matrices are symmetric.

The state will be expressed as $\ket \psi = \sum_n \psi_n \ket n$, where the
coefficients $\psi_n$ are real. It is not possible to obtain any smaller phase
variance using complex $\psi_n$. To see this, note that Eq.~(\ref{gencos})
gives
\bqa
2 \ip{\widehat{\cos \phi}}\!\!\!\! &=& \!\!\!\!\sum_{n=0}^{N-1} \left[ \psi_n^*
\psi_{n+1} + \psi_{n+1}^* \psi_n \right] H_{n,n+1} \nn \\
\!\!\!\! &=& \!\!\!\! \sum_{n=0}^{N-1} 2{\rm Re}\left[ \psi_n^*
\psi_{n+1} \right] H_{n,n+1}. 
\eqa
From this it is clear that the maximal value of $\lr{\widehat{\cos \phi}}$
is obtained for real $\psi_n^* \psi_{n+1}$. This in turn implies that the minimum
phase variance will be when $\arg (\psi_n)$ is independent of $n$. Without loss of
generality $\arg (\psi_n)$ can be taken to be zero.

Therefore, taking the $\psi_n$ and increments $d \psi_n$ to be real, we obtain
\bqa
d \bra \psi \hat X \ket \psi \!\!\!\! &=& \!\!\!\! \left(d\bra \psi \right)
\hat X \ket \psi + \bra \psi \hat X \left(d\ket \psi \right) \nn \\
\!\!\!\! &=& \!\!\!\!\sum_n d\psi_n \left( \bra n \hat X \ket \psi + \bra \psi
\hat X \ket n\right) \nn \\
\!\!\!\! &=& \!\!\!\!2\left( d\bra \psi \right) \hat X \ket \psi.
\eqa
If we have a maximum of $\lr{\hat A}$ while $\lr{\hat B}$, $\lr{\hat C} \ldots$
are kept constant, then for any increment $(d\ket\psi)$ which does not change
the expectation values of $\hat B$, $\hat C \ldots$, the increment in
$\lr{\hat A}$ must be zero also. This means that if
\beq
(d\bra\psi)\hat X \ket\psi = 0,
\eeq
for $\hat X=\hat B$, $\hat C \ldots$, then $(d\bra\psi)\hat A \ket\psi$ must be
equal to zero. This means that $\hat A \ket\psi$ cannot be linearly independent
of $\hat B \ket\psi$, $\hat C \ket\psi \ldots$, so it must be possible to write
\beq
\left( \alpha \hat A + \beta \hat B + \gamma \hat C +\ldots \right) \ket \psi=0,
\eeq
for some combination of constants $\alpha$, $\beta$, $\gamma \ldots$.

We wish to maximise the expectation value of the above operator with the single
constraint that the state is normalised. Using the method of undetermined
multipliers gives
\beq
\left[\alpha\left( 2\widehat{\cos\phi} \right)+\beta \hat 1 \right]\ket\psi =0.
\eeq
Rearranging this gives the eigenvalue equation
\beq
\label{eigeqcos}
\left( 2 \widehat{\cos \phi} \right) \ket \psi = \nu \ket \psi.
\eeq

When we expand the state in terms of the photon number states
$\ket \psi = \sum_{n=0}^N \psi_n \ket n$, we find
\beq
\sum_{n=0}^{N-1} \left(\psi_{n+1} \ket n + \psi_n \ket{n+1}\right) 
= \sum_{n=0}^N \nu \psi_n \ket n .
\eeq
Rearranging this gives
\beq
\label{rearranged}
\sum_{n=0}^{N-1} \psi_{n+1} \ket n + \sum_{n=1}^N \psi_{n-1} \ket{n}
= \sum_{n=0}^N \nu \psi_n \ket n .
\eeq
This equation is equivalent to the recurrence relation
\beq
\label{recurrence}
\psi_{n+1} = \nu\psi_n - \psi_{n-1},
\eeq
for $0<n<N$, with the boundary conditions
\bqa
\psi_1 \!\!\!\! &=& \!\!\!\! \nu\psi_0, \nn \\
\psi_{N-1} \!\!\!\! &=& \!\!\!\! \nu\psi_N.
\eqa

The recurrence relation (\ref{recurrence}) is satisfied by exponentials of the
form
\beq
\psi_n=e^{\pm i n{\rm acos} \frac \nu 2}.
\eeq
To obtain real solutions, we can use
\beq
\psi_n= {\rm Re}
 \left( A e^{i n {\rm acos} \frac \nu 2} \right).
\eeq
The boundary conditions can be more conveniently expressed by extending the
range of the recurrence relation to $0 \le n \le N$, and taking
\bqa
\psi_{-1}\!\!\!\! &=& \!\!\!\!0, \nn \\
\psi_{N+1}\!\!\!\! &=& \!\!\!\!0.
\eqa
The first boundary condition then gives the phase of $A$ as
\beq
\arg A={\rm acos} \frac \nu 2 -\frac {\pi}2.
\eeq
This boundary condition does not give the magnitude, which must be found by
normalisation. The last boundary condition gives
\beq
(N+2){\rm acos} \frac \nu 2 = k\pi,
\eeq
where $k$ is an arbitrary integer. The eigenvalues are therefore
\beq
\nu_k = 2 \cos \left( \frac {k\pi}{N+2}\right),
\eeq
and the coefficients for the corresponding (unnormalised) eigenvectors are
\bqa
\psi_n\!\!\!\! &=& \!\!\!\! {\rm Re} \left( -i e^{\frac{i (n+1) k\pi}{N+2}}
\right) \nn \\ \!\!\!\! &=& \!\!\!\!\sin\left( \frac{(n+1) k\pi}{N+2} \right).
\eqa

It is clear from Eq.~(\ref{eigeqcos}) that
\beq
\ip{2 \widehat{\cos \phi}} = \nu,
\eeq
so
\beq
V(\phi) = (\nu/2)^{-2}-1.
\eeq
Therefore the eigenvalue that minimises the Holevo phase variance is the
maximum eigenvalue,
\beq
\nu_1 = 2 \cos \left( \frac {\pi}{N+2}\right).
\eeq
For this eigenvalue the exact Holevo phase variance is
\bqa
V(\phi) \!\!\!\! &=& \!\!\!\! \left(\cos \left( \frac {\pi}{N+2}\right)\right)
^{-2}-1 \nn \\ \!\!\!\! &=& \!\!\!\! \tan ^2 \left( \frac {\pi}{N+2}\right).
\eqa
The state is given by
\beq
\ket \psi = \frac 1 {\sqrt{N/2+1}}
\sum_{n=0}^N \sin\left( \frac{(n+1) \pi}{N+2} \right) \ket n .
\eeq
Here the factor of $1/\sqrt{N/2+1}$ is required for the state to be normalised.

Note that for large photon number we have
\bqa
V(\phi) \!\!\!\! &=& \!\!\!\! \tan ^2 \left( \frac {\pi}{N+2}\right) \nn \\
\!\!\!\! & \approx & \!\!\!\! \left( \frac {\pi}{N+2}\right)^2 \nn \\
\!\!\!\! & \approx & \!\!\!\! \frac {\pi^2}{N^2}.
\eqa
This means that the minimum intrinsic phase variance scales down as
rapidly as $N^{-2}$.

\subsection{General Dyne Measurements}

The method of solution for canonical measurements can be generalised to the case
of more general dyne measurements, where instead of $H_{n,n+1}=1$, we have
\bqa
H_{n,n+1} \!\!\!\! &=& \!\!\!\! 1-h(n) \nn \\
\!\!\!\! & \approx & \!\!\!\! 1-c n^{-p}.
\eqa
This case was also considered in Ref.~\cite{semiclass}. Again the eigenvalue
equation is
\beq
\label{neweigeq}
\left( 2 \widehat{\cos \phi} \right) \ket \psi = \nu \ket \psi,
\eeq
except this time
\bqa
2 \widehat{\cos \phi} \!\!\!\! &=& \!\!\!\! \sum_{n=0}^{N-1} \left[ \ket{n}
\bra{n+1}+\ket{n+1}\bra{n}\right] H_{n,n+1} \nn \\
\!\!\!\! &=& \!\!\!\! \sum_{n=0}^{N-1} \left[ \ket{n}\bra{n+1}+\ket{n+1}\bra{n}
\right] (1-h(n) ) .
\eqa

Expanding this out gives
\bqa
\sum_{n=0}^{N-1} \psi_{n+1} \ket n + \sum_{n=1}^N \psi_{n-1} \ket{n}
-\sum_{n=0}^{N-1} \left[h(n)\psi_{n+1} \ket n + h(n) \psi_n \ket{n+1}\right]
\!\!\!\! &=& \!\!\!\! \sum_{n=0}^N \nu \psi_n \ket n , \nn \\
\sum_{n=0}^{N-1} \psi_{n+1} \ket n + \sum_{n=1}^N \psi_{n-1} \ket{n}
-\sum_{n=0}^{N-1} h(n)\psi_{n+1}\ket n-\sum_{n=1}^N h(n-1)\psi_{n-1} \ket{n}
\!\!\!\! &=& \!\!\!\! \sum_{n=0}^N \nu \psi_n \ket n.
\eqa
This gives the recursion relation for $0<n<N$,
\beq
\psi_{n+1}+\psi_{n-1}-h(n)\psi_{n+1}-h(n-1)\psi_{n-1}=\nu \psi_n.
\eeq

In order to solve this, in Ref.~\cite{semiclass} the photon number $n$ is
treated as a continuous variable, and the coefficients $\psi_n$ are replaced
with the function $\psi (n)$, which is assumed to be twice differentiable. Then
the approximations are made
\bqa
\psi(n+1)+\psi(n-1) \!\!\!\! & \approx & \!\!\!\! \left[2+\frac{\partial^2}
{\partial n^2} \right]\psi(n), \\
\label{secondapprox}
h(n)\psi(n+1)+h(n-1)\psi(n-1) \!\!\!\! & \approx & \!\!\!\! 2h(n) \psi(n).
\eqa
The second approximation (\ref{secondapprox}) is based upon using
$h(n)\approx h(n-1)$ and $\psi(n+1)+\psi(n-1)\approx 2\psi(n)$. The second
derivative is not used, because this term is already much smaller than
$\psi(n)$.

With these approximations, the eigenvalue equation becomes
\beq
\label{contineival}
\left(2+\frac{\partial^2}{\partial n^2}-2cn^{-p}\right)\psi(n)=\nu \psi(n).
\eeq
In Ref.~\cite{semiclass}, the term $2cn^{-p}$ is linearised about $n=N$, and the
variables are changed to $y=1-N^{-1}n$. The equation is then
\beq
\label{maxNdeq}
\left( -\frac{\partial^2}{\partial y^2}+by \right) \psi(y) = a_k \psi(y),
\eeq
where
\bqa
a_k \!\!\!\! &=& \!\!\!\! N^2(2-\nu_k-2cN^{-p}), \nn \\
b \!\!\!\! &=& \!\!\!\! 2cpN^{2-p}.
\eqa
The boundary condition is $\psi(0)=0$, and the boundary condition at $y=1$ is
ignored.

Note that, similarly to the case for canonical measurements, the exact boundary
conditions should be $\psi_{-1}=\psi_{N+1}=0$. In terms of $y$, this means we
should have $\psi(y)=0$ for $y=-1/N$ and $1+1/N$. The $1/N$ terms in these
boundary conditions are ignored, as they merely give higher order corrections
to the results.

Using these approximations, the solutions are
\beq
\label{solnupper}
\psi_k(y) \propto {\rm Ai} (z_k +b^{1/3}y),
\eeq
where Ai is the Airy function and $z_k$ is the $k$th real zero of the Airy
function satisfying $0>z_1>z_2>\ldots$. The eigenvalues are
\beq
\nu_k = 2-\left( 2cN^{-p} + \st{z_k}(2cp)^{2/3}N^{-2(1+p)/3}\right),
\eeq
and so it is obvious that the maximum eigenvalue is for $k=1$. The corresponding
Holevo phase variance is
\bqa
\label{asymexp}
V(\phi) \!\!\!\! & = & \!\!\!\! (\nu_k/2)^{-2}-1 \nn \\
\!\!\!\! & = & \!\!\!\! \left[1-\left( cN^{-p} + \half \st{z_1}
(2cp)^{2/3}N^{-2(1+p)/3}\right)\right]^{-2}-1 \nn \\
\!\!\!\! & \approx & \!\!\!\! \left[1+\left(2cN^{-p}+\st{z_1}
(2cp)^{2/3}N^{-2(1+p)/3}\right)+3(cN^{-p})^2\right]-1 \nn \\
\!\!\!\! & = & \!\!\!\! 2cN^{-p}+\st{z_1}
(2cp)^{2/3}N^{-2(1+p)/3}+3c^2 N^{-2p}.
\eqa
This method should be used rather than using the approximation $V(\phi)\approx
2-v_k$ (as is used in \cite{semiclass}), as the third term here will be of the
same order as the second for mark I measurements.

The value of the first zero of the Airy function is $z_1\approx
-2.33810741045976$. Using these results, and the values of $c$ and $p$ for
heterodyne, mark I and mark II measurements given in Table \ref{table:meascp},
the variances for these measurements should be given by
\bqa
V(\phi_{\rm het}) \!\!\!\! & \approx & \!\!\!\! \smallfrac 14 N^{-1} +
0.927879 \times N^{-4/3}, \\
\label{markIasym}
V(\phi_{\rm I}) \!\!\!\! & \approx & \!\!\!\! \smallfrac 14 N^{-1/2} +
0.631402 \times N^{-1}, \\
V(\phi_{\rm II}) \!\!\!\! & \approx & \!\!\!\! \smallfrac 18 N^{-3/2} +
0.765947 \times N^{-5/3}.
\eqa
The second term for mark I measurements here includes the third term from
Eq.~(\ref{asymexp}). This term has been omitted in the other two cases, as it
is of higher order. As is discussed in the next section, $h_{\rm I}(n)$ has an
extra term of order $N^{-1}$. This term would also need to be taken into account
in order for the second term for mark I measurements to be accurate.

\subsection{Numerical Results}
\label{maxnumer}

These analytic results have been verified numerically by calculating the
optimised states for heterodyne measurements and adaptive mark I and II
measurements. For moderate maximum photon numbers the calculations were exact.
For heterodyne measurements the vector $h_{\rm het}(n)$ is given exactly by
\cite{fullquan}
\beq
\label{exacthet}
h_{\rm het}(n)=1-\frac{\Gamma\left(n+3/2\right)}{\sqrt{\Gamma\left(
n+1\right)\Gamma\left(n+2\right)}}.
\eeq
This form of the equation can not be used for large $n$ due to roundoff error.
For $n \ge 10$ the first twelve terms of an asymptotic expansion were used, giving
results as accurate as or more accurate than the exact expression (\ref{exacthet}).
These terms were determined from the asymptotic expansion for
$\log\Gamma\left(n\right)$ \cite{Erdelyi53}, and are given below:
\bqa
&&\!\!\!\!\!\!\!\! h_{\rm het}(n) \sim \frac 1{8(n+1)}-\frac 1{2^7(n+1)^2}
- \frac 5{2^{10}(n+1)^3} + \frac {21}{2^{15}(n+1)^4} + 
\frac{399}{2^{18}(n+1)^5} - \frac{869}{2^{22}(n+1)^6} \nn \\
&&\!\!\!\!\!\!\!\! - \frac{39325}{2^{25}(n+1)^7} + \frac{334477}{2^{31}(n+1)^8}
+ \frac{28717403}{2^{34}(n+1)^9} - \frac{59697183}{2^{38}(n+1)^{10}}
- \frac{8400372435}{2^{41}(n+1)^{11}} + \frac{34429291905}{2^{46}(n+1)^{12}}
+\ldots
\nn \\
\eqa

The exact expression for mark I measurements is slightly more complicated. From
\cite{fullquan} the exact expression for $H_{mn}$ for mark I measurements is
\beq
H_{mn}^{\rm I} = \sum_{p=0}^{\lfloor m/2 \rfloor} \sum_{q=0}^{\lfloor n/2
\rfloor} \gamma_{mp} \gamma_{nq} M^{p,q},
\eeq
where $\lfloor m/2 \rfloor$ is the integer part of $m/2$,
\beq
\gamma_{mp} = \frac{\sqrt{m!}}{2^p (m-2p)! p!},
\eeq
and $M^{p,q}$ are calculated using the recursion relation
\beq
M^{n,m}=\frac{n M^{n-1,m}+m M^{n,m-1}}{2(n-m)^2+n+m},
\eeq
with the boundary values
\beq
M^{n,0}=M^{0,n}=\frac 1{(2n+1)(2n-1) \ldots 1}=\frac 1{(2n+1)!!}.
\eeq

The values of $h_{\rm I}(n)=1-H_{n,n+1}^{\rm I}$ were calculated up to $n=3000$
using this expression. Further values were extrapolated by fitting an asymptotic
expansion to the results below 3000. The first term, $1/(8\sqrt n)$, was
assumed, and three further terms were obtained by fitting techniques. The terms
found were
\beq
\label{aprxhI}
h_{\rm I}(n) \approx \frac{1}{8 n^{1/2}}-\frac{0.101561734071163}{n}-
\frac{0.0508542807551548}{n^{3/2}}+\frac{0.0959147066823866}{n^2}.
\eeq
These coefficients were the exact values used in calculations, but the digits
given do not reflect the accuracy of the fit. (I have given all digits used
so that it is possible to accurately reproduce the results.)

The exact expression for mark II measurements is even more complicated than that
for mark I measurements. From \cite{fullquan} the expression is
\beq
\label{exactm2}
H_{mn}^{\rm II} = \sum_{p=0}^{\lfloor m/2 \rfloor} \sum_{q=0}^{\lfloor n/2
\rfloor} \gamma_{mp} \gamma_{nq} \ip{\left(\frac{1+D}{1+D^*}\right)
^{(n-m)/2}D^p(D^*)^q}_Q.
\eeq
where $D=BA^*/A$, and the subscript $Q$ indicates that the expectation value is
for the ``ostensible'' or vacuum distribution. This notation differs slightly
from that in \cite{fullquan}, where the symbol $C$ was used, rather than $D$.
In Ref.~\cite{fullquan} it is shown that
\beq
\ip{D^n {D^*}^m}_Q=M^{n,m},
\eeq
where $M^{n,m}$ is calculated as above. In order to obtain $h_{\rm II}(n)$ we
require $H_{n,n+1}^{\rm II}$. In order to calculate these we can expand
Eq.~(\ref{exactm2}) to obtain
\beq
\label{exactmp1}
H_{n,n+1}^{\rm II} = \sum_{p=0}^{\lfloor n/2 \rfloor} \sum_{q=0}^{\lfloor
(n+1)/2 \rfloor} \gamma_{np} \gamma_{n+1,q} \sum_{k=0}^\infty \sum_{l=0}^\infty 
\frac{\left(3/2-k\right)_k}{k!} \frac{\left(1/2-l\right)_l}{l!}
 M^{k+p,l+q},
\eeq
where
\beq
(\alpha)_n = \alpha (\alpha+1)\ldots (\alpha+n-1).
\eeq

The infinite sum in (\ref{exactmp1}) can be determined to within double
precision (15 digits) by summing about the first 100 terms. The values of
$h_{\rm II}(n)$ were calculated up to $n=1000$ in this way. I attempted to
obtain the higher order terms using this data set in a similar way as for the mark I
case, but unfortunately it was not found to be possible to consistently obtain
any terms beyond $\smallfrac 1{16}n^{-1.5}$. Therefore all elements beyond
$n=1000$ were determined using $h_{\rm II}(n) \approx \smallfrac 1{16}n^{-1.5}$.

Using the above methods for calculating $h(n)$, the minimum phase variance was
determined by numerically determining the eigenvalues for heterodyne, mark I and
mark II measurements, for photon numbers up to $2^{18}$. These calculations were
exact except for the above approximations to $h(n)$ for mark I and II
measurements.

For larger photon numbers it was not feasible to solve the exact eigenvalue
problem, but an approximate solution was obtained by using the continuous
approximation of the eigenvalue problem and discretising it. In order to reduce
the number of intervals required in the discretised equation, the equation was
solved for three different numbers (512, 1024 and 2048) of intervals. The result
for the continuous case was then estimated by projecting to zero step size
assuming the error is quadratic in the step size. The approximations that were
{\it not} made (that were made in order to derive the analytic result) were the
linear approximation of $h(n)$ and the omission of the boundary condition at
$y=1$.

\begin{figure}
\centering
\includegraphics[width=0.7\textwidth]{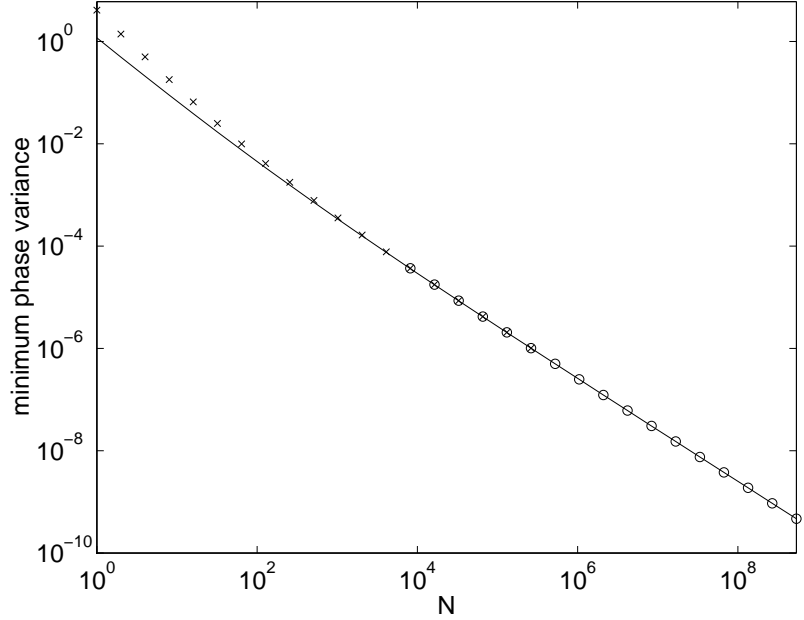}
\caption{The minimum phase variance for heterodyne measurements on states with
an upper limit $N$ on the photon number. The exact calculations are shown as the
crosses, the continuous approximation as the circles, and the asymptotic
analytic expression as the continuous line.}
\label{hetlog}
\end{figure}

The results for heterodyne measurements are shown in Fig.~\ref{hetlog}. The
results for the exact calculations agree extremely well with the results for the
continuous approximation over the region where both values have been calculated.
This indicates that the continuous approximation is a very good approximation of
the exact eigenvalue problem.

\begin{figure}
\centering
\includegraphics[width=0.7\textwidth]{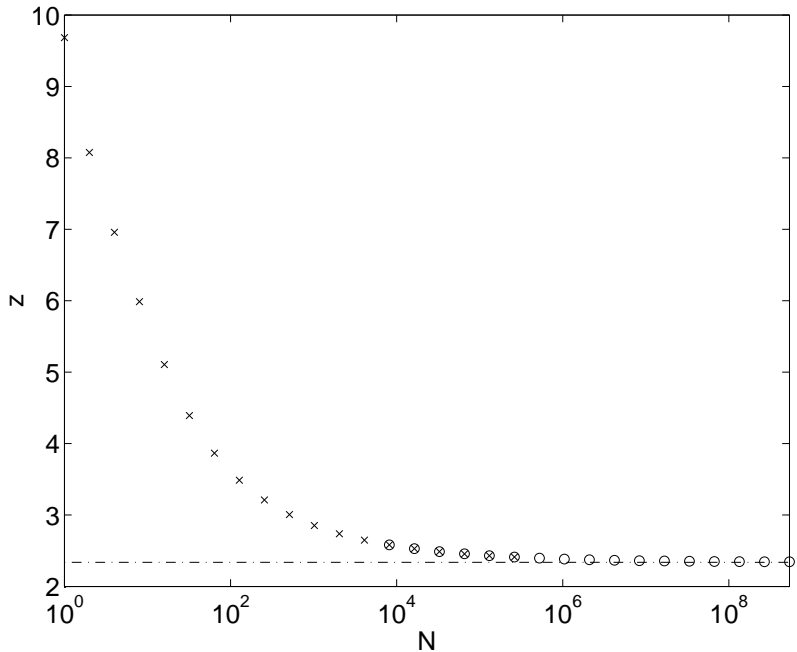}
\caption{The value of $z$ for heterodyne measurements on states with an upper
limit $N$ on the photon number. The exact calculations are shown as the crosses,
the continuous approximation as the circles, and the theoretical asymptotic
value of $|z_1|\approx 2.338107$ is shown as the dash-dotted line.}
\label{hetz}
\end{figure}

The results also agree well with the asymptotic expression (\ref{asymexp}), with
good agreement for photon numbers above 100. In order to better see the
difference between the numeric results and the analytic expression, and in
particular to see how accurate the second term in the analytic expression is,
it is convenient to define the parameter $z$ by
\beq
z=\frac{V(\phi)-2cN^{-p}}{(2cp)^{2/3}N^{-2(1+p)/3}}.
\eeq
From Eq.~(\ref{asymexp}), provided the third term $3c^2 N^{-2p}$ can be ignored,
this should converge to $\st{z_1}\approx 2.338107$.
The values of $z$ for heterodyne measurements are plotted in Fig.~\ref{hetz}.
Again there is extremely good agreement between the values calculated exactly
and those calculated using the continuous approximation. The value of $z$ does
not converge closely to $\st{z_1}$ until a photon number of around $10^6$. In
fact, we do not have 1\% agreement until a photon number of $8 \times 10^6$.

\begin{figure}
\centering
\includegraphics[width=0.7\textwidth]{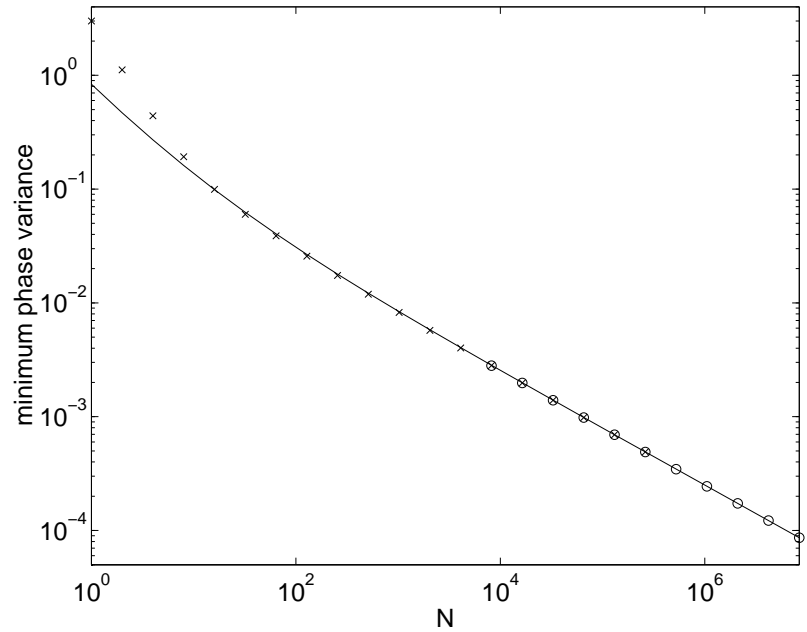}
\caption{The minimum phase variance for mark I measurements on states with
an upper limit $N$ on the photon number. The exact calculations are shown as the
crosses, the continuous approximation as the circles, and the asymptotic
analytic expression as the continuous line.}
\label{markIlog}
\end{figure}

The results for mark I measurements are plotted in Fig.~\ref{markIlog}. The
asymptotic expression plotted here is Eq.~(\ref{markIasym}), which includes the
extra $3c^2 N^{-2p}$ term. There is good agreement between the results and this
asymptotic expression for photon numbers above about 10. There is also good
agreement between the values calculated exactly and those calculated using the
continuous approximation.

\begin{figure}
\centering
\includegraphics[width=0.7\textwidth]{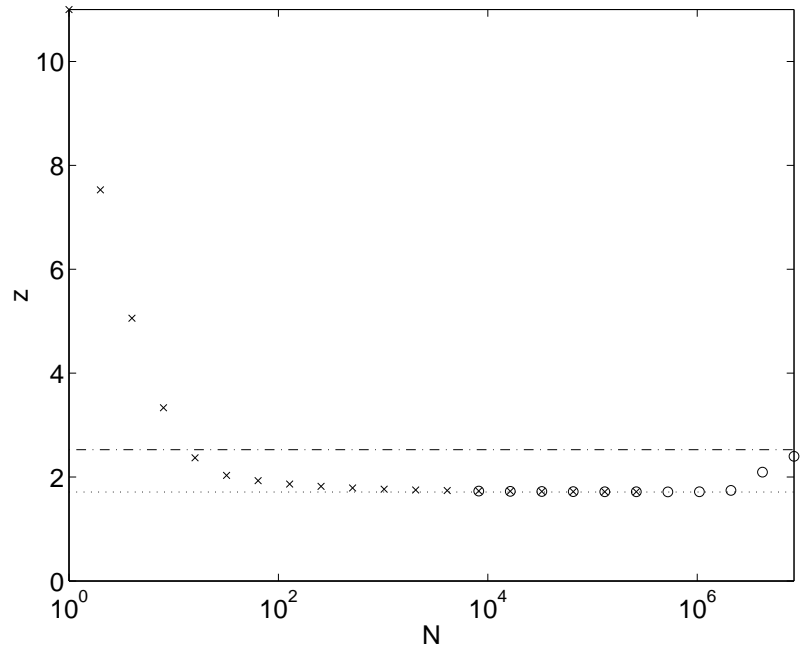}
\caption{The value of $z$ for mark I measurements on states with an upper
limit $N$ on the photon number. The exact calculations are shown as the crosses,
the continuous approximation as the circles, and the theoretical asymptotic value
of $2.525607$ is shown as the dash-dotted line. The asymptotic value taking into
account the second term for $h_{\rm I}(n)$ is shown as the dotted line.}
\label{markIz}
\end{figure}

For mark I measurements the values of $z$ cannot be expected to converge to
$|z_1|$ due to the extra $3c^2 N^{-2p}$ term. From the asymptotic expression
(\ref{markIasym}), the asymptotic value including this term should be
$2.525607$. As can be seen in Fig.~\ref{markIz}, however, $z$ does not
converge to this value. 
The reason for this is that the second term of the asymptotic expression for
$V(\phi_{\rm I})$ is of the same order as the second term in the asymptotic
expression for $h_{\rm I}(n)$ in Eq.~(\ref{aprxhI}). Taking account of
this term, $V(\phi_{\rm I})$ should be
\beq
V(\phi_{\rm I}) \approx \smallfrac 14 N^{-1/2} + 0.428278 \times
N^{-1},
\eeq
and $z$ should converge to approximately $1.713114$. This value is also plotted in
Fig.~\ref{markIz}. As can be seen, the results converge quite accurately to this
value. The results for the largest photon numbers do not agree accurately;
however, this is just due to poor convergence of the numerical technique.

\begin{figure}
\centering
\includegraphics[width=0.7\textwidth]{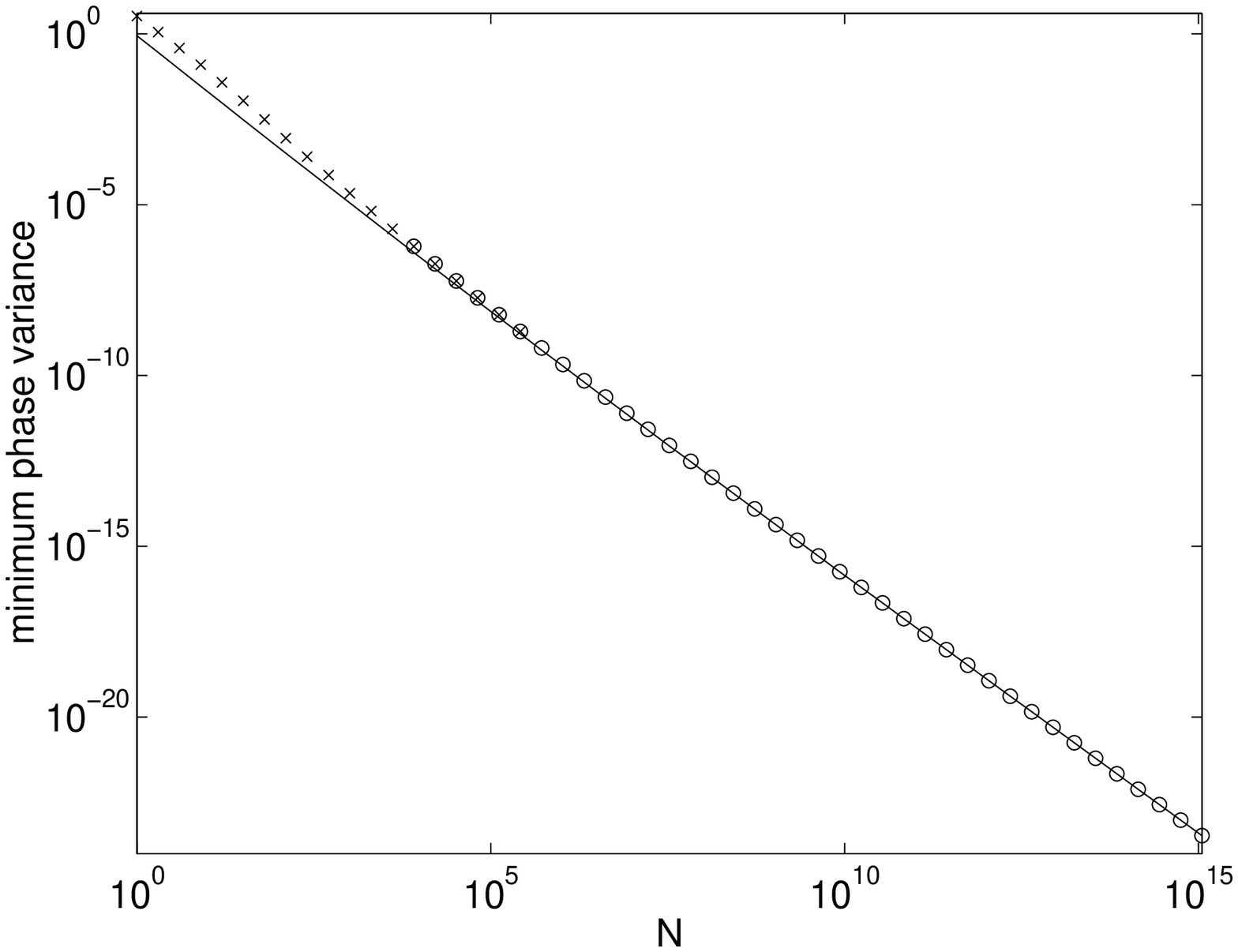}
\caption{The minimum phase variance for mark II measurements on states with
an upper limit $N$ on the photon number. The exact calculations are shown as the
crosses, the continuous approximation as the circles, and the asymptotic
analytic expression as the continuous line.}
\label{markIIlog}
\end{figure}

\begin{figure}
\centering
\includegraphics[width=0.7\textwidth]{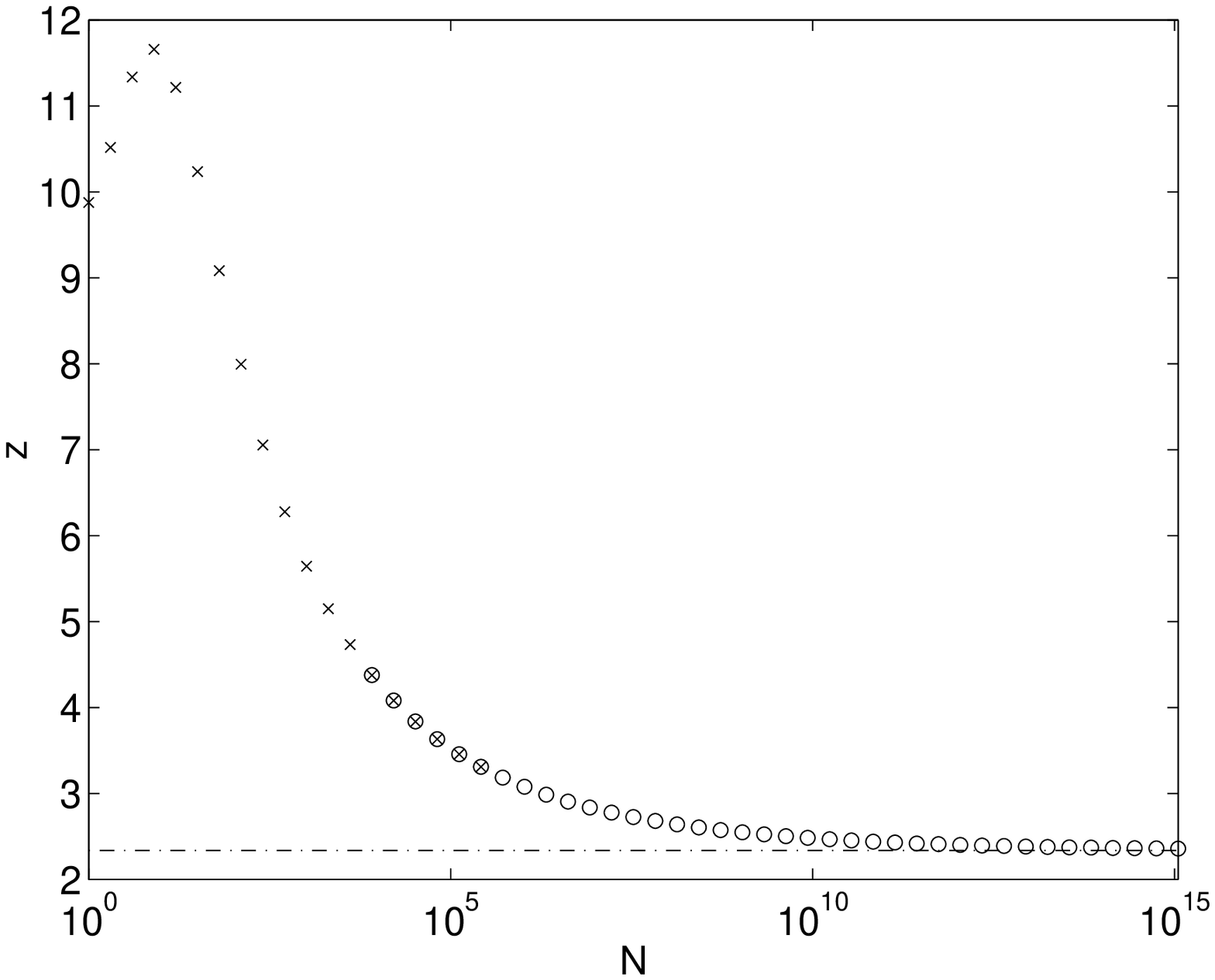}
\caption{The value of $z$ for mark II measurements on states with an upper
limit $N$ on the photon number. The exact calculations are shown as the crosses,
the continuous approximation as the circles, and the theoretical asymptotic value
of $\st {z_1}\approx 2.338107$ is shown as the dash-dotted line.}
\label{markIIz}
\end{figure}

The results for mark II measurements are plotted in Fig.~\ref{markIIlog}. In
this case there is good agreement between the results calculated using the two
different methods, and the results appear to be close to the asymptotic result
for photon numbers above about $10^5$. When we look at the values of $z$ (see
Fig.~\ref{markIIz}), we see that there is again excellent agreement between
the results calculated using the two different methods, but there is good
agreement with the asymptotic value only for photon numbers above $10^{10}$.
In fact, we require a photon number above $5 \times 10^{14}$ in order to have
better than 1\% agreement.

Note that in \cite{semiclass} it is claimed that there should be good
convergence to the asymptotic values for
\beq
N \ge \left( \frac{10^3}{2cp} \right) ^{1/(1-p)}.
\eeq
This means that there should be good agreement with the asymptotic values for
photon numbers above about 4000, 400 and $3\times 10^7$ for heterodyne, mark I
and mark II measurements respectively. The numerical results obtained here show
that, in order to have better than 1\% agreement with the asymptotic results,
the maximum photon numbers should be above about $8 \times 10^6$, 4000 and
$5 \times 10^{14}$ respectively.

These apparent discrepancies of many orders of magnitude are not so great when
considered in terms of the difference in $z$ rather than the photon number. For
a photon number close to 4000 for heterodyne measurements, $z$ is about 13\%
from the asymptotic value, which is not very poor agreement. Similarly for a
photon number around 400 for mark I measurements the value of $z$ is 4\% from
the asymptotic value, and for a photon number around $3\times 10^7$ for mark II
measurements the value of $z$ is 17\% from the asymptotic value. Nevertheless,
these discrepancies are still larger than would be expected.

To understand the reason for this, note that the criterion used in
\cite{semiclass} was that 99.5\% of the solution $\psi(y)$ [as given by
Eq.~(\ref{solnupper})] be confined within the lower half of the interval
$[0,1]$. This criterion means that the approximation that we can ignore the
boundary condition at $y=1$ is very good, but it is a fairly weak criterion for
the approximation that we can linearise $h(n)$ about $n=N$ (equivalent to
$y=0$). A better criterion for this approximation to be accurate is that 99.5\%
of the solution be confined within $y<0.04$. This criterion means that we
require $5b^{-1/3}<0.04$, where $b=2cpN^{2-p}$, and gives
\beq
N > \left( \frac{10^6}{cp} \right)^{1/(2-p)}.
\eeq
With this criterion the minimum photon numbers for heterodyne, mark I and mark
II measurements are $8 \times 10^6$, $6 \times 10^4$ and $10^{14}$ respectively.
With the exception of the mark I result, these values are much closer to the
photon numbers required for 1\% agreement with the asymptotic value of $z$.

\section{Fixed Mean Photon Number}
\label{general}

The next case that I consider is that where, rather than an upper limit being
put on the photon number, the mean photon number is fixed. This case is
rather more complicated, as the method of undetermined multipliers gives two
undetermined constants, so the problem is more complicated than a simple
eigenvalue problem.

\subsection{Canonical Measurements}
\label{meancanon}

The simpler case of canonical measurements with a fixed mean photon number was
solved by Summy and Pegg \cite{SumPeg90}. I will give the derivation here, as it
is simple and indicates the method of solution for the case of general dyne
measurements. I am giving a different but equivalent version of the derivation
for consistency with the other derivations given in this chapter.

We wish to maximise the expectation value of $2\widehat{\cos\phi}$ while
keeping the state normalised and the photon number constant. Using the method
of undetermined multipliers gives the equation
\beq
\left[ \alpha \left( 2 \widehat{\cos \phi} \right) + \beta \hat 1 + \gamma
\hat N \right] \ket{\psi} = 0.
\eeq
Rearranging this gives
\beq
\label{undet}
\left[ \left( 2 \widehat{\cos \phi} \right) - \mu \hat N \right] \ket{\psi}
= \nu \ket{\psi}.
\eeq
This has two unknown constants, and therefore cannot be solved as a
simple eigenvalue problem. Instead what we do is solve it as an eigenvalue
equation for $\nu$ with a fixed value of $\mu$, and the eigenstate corresponding
to the maximum eigenvalue is an optimised state. The mean photon number can then
be found from this state. We can obtain a range of mean photon numbers by
adjusting $\mu$.

Expanding the equation out in terms of the number coefficients of $\ket{\psi}$
gives
\beq
\sum_{n=0}^{\infty} \psi_{n+1} \ket n + \sum_{n=1}^{\infty} \psi_{n-1} \ket{n}-
\sum_{n=0}^{\infty} \mu \psi_n n \ket n = \sum_{n=0}^{\infty} \nu \psi_n \ket n,
\eeq
which implies the recurrence relation
\beq
\psi_{n+1} + \psi_{n-1} - \mu \psi_n n = \nu \psi_n.
\eeq
Similarly to the derivation for general dyne measurements when there is an upper
limit on the photon number, we can use the continuous approximation where we
replace $\psi_n$ with $\psi(n)$, and use
\beq
\psi(n+1)+\psi(n-1) \approx \left[ 2+ \frac{\partial^2}{\partial n^2} \right]
\psi(n).
\eeq

With this approximation the recurrence relation becomes the differential
equation
\beq
\left(-\frac{\partial^2}{\partial n^2} +\mu n \right) \psi(n) = (2-\nu) \psi(n).
\eeq
This equation is exactly equivalent to the differential equation obtained in the
maximum photon number case (\ref{maxNdeq}), and it therefore has the solutions
\beq
\psi_k(n) \propto {\rm Ai} (z_k+\mu^{1/3}n),
\eeq
with the eigenvalues
\beq
\nu_k = 2-\st{z_k}\mu^{2/3}.
\eeq
It is clear that the solution that maximises $\nu$ is again that with $k=1$. In
Ref.~\cite{SumPeg90} the authors say that the solution must have a zero for
$n=-\epsilon$, where $0<\epsilon<1$. The boundary condition for the discrete
equation is $\psi_{-1}=0$, which implies that the solution should have a zero
for $n=-1$. I will take $\psi(0)=0$ here, as the difference only gives higher
order terms to the solution.

It can be shown numerically that
\beq
\ip{z_1+\mu^{1/3}n} \approx -0.779369136819922.
\eeq
For brevity I will call this constant $\ip X$ (for consistency with
\cite{SumPeg90}). Rearranging this gives
\beq
\mu^{1/3}=\frac{\ip X-z_1}{\ip n}.
\eeq
This gives the relation between the mean photon number and $\mu$. The
eigenvalue is then
\beq
\nu_1 = 2-\frac{\st{z_1} (\ip X-z_1)^2}{\ip n ^2}.
\eeq
From Eq.~(\ref{undet}) it is clear that
\beq
\ip{2 \widehat{\cos \phi}} - \mu \ip n = \nu.
\eeq
Substituting the values for $\mu$ and $\nu$ from above gives
\beq
\ip{2 \widehat{\cos \phi}} - \frac{(\ip X-z_1)^3}{\ip n ^2} = 
2-\frac{\st{z_1} (\ip X-z_1)^2}{\ip n ^2}.
\eeq
This simplifies to
\beq
2-\ip{2 \widehat{\cos \phi}} = \frac{-\ip X (\ip X-z_1)^2}{\ip n ^2}
\eeq
Therefore the phase variance is
\beq
V(\phi)=\frac{1.89360591826155}{\nb^2}.
\eeq
This is the result obtained in \cite{SumPeg90}. Therefore we see that the phase
variance for optimum states scales as $\bar n^{-2}$, the same as for the case
with an upper limit on the photon number. Note that it is accurate to
approximate the Holevo phase variance by $2-\ip{2 \widehat{\cos \phi}}$ here,
as the differences will be of order $\nb^{-4}$.

This analytic result has been verified by determining the optimum states
numerically. The eigenvalue problem was solved with various values of $\mu$,
and for each value of $\mu$ the maximum eigenvalue was chosen, and the mean
photon number was determined from the corresponding eigenstate. In order to
make the problem finite, the state coefficients were only considered for photon
numbers up to around 11 times the mean photon number of the state. At this
point the state coefficients had fallen to around $10^{-16}$.

\begin{figure}
\centering
\includegraphics[width=0.7\textwidth]{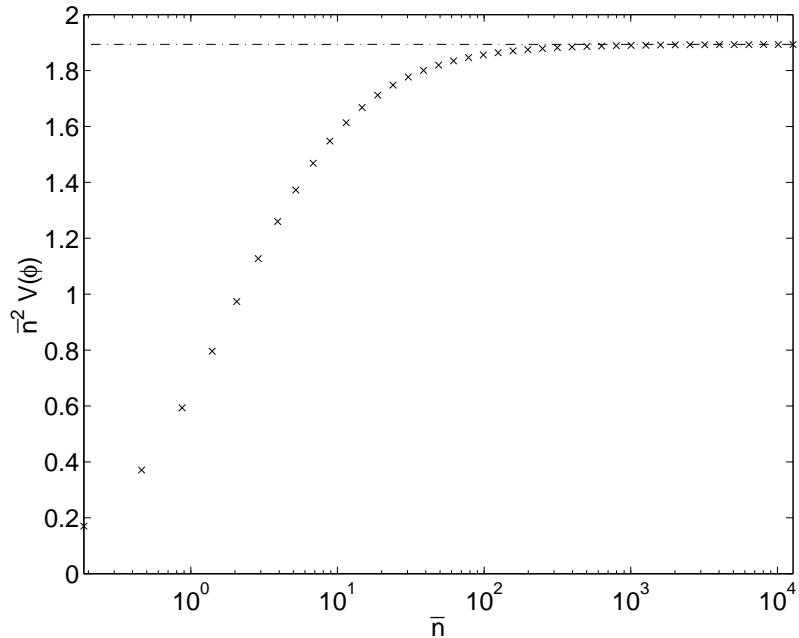}
\caption{The phase variance multiplied by the square of the mean photon number
for states optimised for minimum phase variance with the constraint of fixed
mean photon number. The theoretical asymptotic value of $1.893606$ is shown as
the dash-dotted line.}
\label{canmean}
\end{figure}

The results multiplied by $\nb^2$ are plotted in Fig.~\ref{canmean}. The
results converge rapidly to the asymptotic value, with agreement within 1\% of
the asymptotic value for mean photon numbers above about 250. Note that it
does not make sense to perform calculations with the continuous version of the
eigenvalue equation, as this case was solved exactly.

\subsection{General Dyne Measurements}
\label{gendyne}

Now I will consider the case of optimising for minimum phase variance with a
fixed mean photon number for the case of more general dyne measurements. The
complete derivation is original to this study, and is based on a partial
derivation by Zhong-Xi Zhang (personal communication). This derivation was
published in a brief form in \cite{BerWisZha99}.

Similarly to the case of canonical measurements, the method of
undetermined multipliers gives the equation
\beq
\label{eigeq}
\left[ \left( 2 \widehat{\cos \phi} \right) - \mu \hat N \right] \ket{\psi}
= \nu \ket{\psi}.
\eeq
For the case of more general dyne measurements, we have
\beq
2\widehat{\cos \phi} = \sum_{n=0}^\infty \left[ \ket{n}\bra{n+1}+\ket{n+1}
\bra{n}\right] (1-h(n)).
\eeq
If the state is expressed in the number states basis
\beq
\ket \psi = \sum_{n=0}^\infty \psi_n \ket n,
\eeq
then Eq.~(\ref{eigeq}) can be expanded to obtain
\bqa
\sum_{n=0}^\infty \psi_{n+1} \ket n + \sum_{n=1}^\infty \psi_{n-1} \ket{n}
-\sum_{n=0}^\infty h(n)\psi_{n+1}\ket n-\sum_{n=1}^\infty h(n-1)\psi_{n-1}
\ket{n} \nn \\ - \sum_{n=0}^\infty \mu \psi_n n \ket n = \sum_{n=0}^\infty \nu
\psi_n \ket n.
\eqa
This gives the recurrence relation
\beq
\label{recurr}
\psi_{n+1}+\psi_{n-1}-h(n)\psi_{n+1}-h(n-1)\psi_{n-1}-\mu \psi_n n =
\nu \psi_n .
\eeq

Similarly to the derivation for an upper limit on the photon number, we can
treat $n$ as a continuous variable, and make the approximations
\bqa
\psi(n+1)+\psi(n-1) \!\!\!\! &\approx& \!\!\!\! \left[ 2+ \frac{\partial^2}
{\partial n^2} \right]\psi(n) \nn \\
h(n)\psi(n+1)+h(n-1)\psi(n-1) \!\!\!\! &\approx& \!\!\!\! 2h(n)\psi(n).
\eqa
Using these approximations, Eq.~(\ref{recurr}) becomes
\beq
\label{deq1}
-\frac {d^2 \psi(n)}{dn^2}+\psi(n) \left[2h(n)+\nu+\mu n-2\right] \approx 0.
\eeq
Now define
\beq
f(n)=2h(n)+\nu+\mu n-2.
\eeq
To solve Eq.~(\ref{deq1}), it is convenient to expand $f(n)$ in a Taylor series
around 
\beq
n_0=\left(\frac{\mu}{2cp} \right) ^{-1/(p+1)}.
\eeq
The derivatives of $f(n)$ are
\bqa
 f(n_0)\!\!\!\! &=& \!\!\!\!2h(n_0)+\nu+\mu n_0-2, \\
 f'(n_0)\!\!\!\! &=& \!\!\!\!2h'(n_0)+\mu=0, \\
 f''(n_0)\!\!\!\! &=& \!\!\!\!2h''(n_0)=2cp(p+1)n_0^{-p-2} ,\\
 f'''(n_0)\!\!\!\! &=& \!\!\!\!-2cp(p+1)(p+2)n_0^{-p-3}.
\eqa
Note that this technique requires that the number distribution has its
maximum near $n_0$. This is justified in the derivation in Appendix
\ref{derper}.

Using the Taylor series for $f(n)$, and defining $f_{0}=f(n_{0})$,
$f_{2} = f''(n_{0})/2$ and $f_{3} = f'''(n_{0})/6$, Eq.~(\ref{deq1}) becomes
\beq
\label{eigeq2}
\left\{ -\frac {d^2}{dn^2}+ \left[ (n-n_0)^2 f_2+(n-n_0)^3 f_3
\right] \right\} \psi(n) \approx -f_0 \psi(n).
\eeq
Note that $-f_0=2 - (2h(n_0)+\nu+\mu n_0)$, so the above equation is
equivalent to solving (\ref{eigeq}) as an eigenvalue equation for $\nu$
with a fixed value of $\mu$. Now Eq.~(\ref{eigeq2}) is equivalent to the
time-independent Schr\"odinger's equation with energy eigenvalue
\beq
E = -f_{0},
\eeq
for a perturbed harmonic Hamiltonian $\hat{H}$. We can apply perturbation theory
with
\bqa
\hat{H}\!\!\!\! &=& \!\!\!\!\hat H_0+\hat H_1,\\
\label{zeroth}
\hat H_0\!\!\!\! &=& \!\!\!\!-\frac {d^2}{dn^2}+ \left[ f_{2}(n-n_0)^2\right],\\
\label{perth}
\hat H_1\!\!\!\! &=& \!\!\!\!f_3(n-n_0)^3 .
\eqa

This perturbation theory derivation is fairly lengthy, and the details are
contained in Appendix \ref{derper}. It is shown that the energy eigenvalue of
the unperturbed ground state is $\sqrt{f_2}$, and that the mean photon number
for the perturbed state is
\beq
\label{nbresult}
\bar n \approx n_0 + \frac{p+2}{4\sqrt{cp(p+1)}}n_0^{p/2}.
\eeq
Using this we can find the minimum phase variance based on
\beq
\ip{2 \cos \phi}_{\rm max} = (\nu +\mu \bar n )_{\rm max}.
\eeq
This can be evaluated as
\beq
\nu +\mu \nb = 2-2h(\nb)+f(\nb).
\eeq
Using a Taylor expansion for $f(\nb)$ gives
\beq
\nu +\mu \nb = 2-2h(\nb)+f_0+(\nb-n_0)^2 f_2.
\eeq
Using the result for $\nb$ in Eq.~(\ref{nbresult}), the last term is of order
$\nb^{-2}$, which is small enough to be omitted here.

It is shown in Appendix \ref{derper} that the energy eigenvalue of the state
corresponding to the smallest phase variance is $\sqrt{f_2}$. The correction
found by perturbation theory is of order $\nb^{-2}$, and can be omitted. As
$-f_0$ is the energy eigenvalue, we get
\bqa
\nu +\mu \bar n \!\!\!\! &=& \!\!\!\! 2-\left[2h(\nb)+\sqrt{f_2}\right] \nn \\
\!\!\!\! &=& \!\!\!\! 2-\left[ 2c\bar n^{-p}+\sqrt{cp(p+1)}n_0^{-p/2-1}\right].
\eqa
The Holevo phase variance is then given by
\bqa \label{finalgen}
V(\phi) \!\!\!\! &=& \!\!\!\! [(\nu +\mu \bar n)/2]^{-2} -1 \nn \\
\!\!\!\! & \approx & \!\!\!\! 2c\bar n^{-p}+\sqrt{cp(p+1)}\bar n^{-p/2-1}
+3c^2\bar n^{-2p}.
\eqa
Note that the first term here is the same as the result when an upper limit is
put on the photon number, but the second term scales as a different power of
$\bar n$. Similarly to the case where there is an upper limit on the photon
number, the phase variance should be found in this way rather than using
$V(\phi) \approx 2-(\nu +\mu \bar n)$. This method gives the extra term
$3c^2\bar n^{-2p}$, which is of lower order than the second term
for mark I measurements. This term can be ignored in the other two cases, as it
is of higher order.

Using this relation, the Holevo phase variance for the three cases,
heterodyne, mark I and mark II, should scale as
\bqa
V(\phi_{\rm het}) \!\!\!\! & \approx & \!\!\!\! \smallfrac 14 \bar n^{-1} +
\smallfrac 12 \bar n^{-3/2}, \\
\label{markIas}
V(\phi_{\rm I}) \!\!\!\! & \approx & \!\!\!\! \smallfrac 14 \bar n^{-1/2} + 
\frac 3{64}\nb^{-1} + \frac{\sqrt 3}{4\sqrt 2} \bar n^{-5/4}, \\
V(\phi_{\rm II}) \!\!\!\! & \approx & \!\!\!\! \smallfrac 18 \bar n^{-3/2} +
\frac {\sqrt{15}}8 \bar n^{-7/4}.
\eqa
For mark I measurements there will also be a term of order $\nb^{-1}$ due to
the second term in the expansion for $h_{\rm I}(n)$. Taking this term into
account the variance should be
\beq
\label{cormarI}
V(\phi_{\rm I}) \approx \smallfrac 14 \bar n^{-1/2} -
0.156248 \times \nb^{-1} + \frac{\sqrt 3}{4\sqrt 2} \bar n^{-5/4}.
\eeq

The case of heterodyne detection is of particular interest because it differs
radically from the result claimed by D'Ariano and Paris \cite{DArPar94} of
\beq \label{Dar}
V(\phi_{\rm het}) = \frac{1.00\pm 0.02}{\bar{n}^{1.30\pm 0.02}}.
\eeq
As the quoted errors suggest, this result was obtained entirely numerically, in
contrast to the analytical result obtained here. Judging from the graphs given
in \cite{DArPar94}, the numeric fit was performed for relatively small photon
numbers, only up to about 100. As was found in previous sections, it
generally takes very large photon numbers for the asymptotic scaling to become
evident. In Sec.~\ref{numergen} I present numerical results that show that the
analytic result in Eq.~(\ref{markIas}) is a far better fit than the power law
of D'Ariano and Paris.

\subsection{Numerical Results}
\label{numergen}
These analytic results have been verified by numerically calculating the optimum
states for dyne measurements with a fixed mean photon number. In these
calculations the values of $h(n)$ for heterodyne, mark I and mark II
measurements were calculated using the formulae described in
Sec.~\ref{maxnumer}. The phase variances and photon numbers were determined in
a similar way as in Sec.~\ref{meancanon}. Fixed values of $\mu$ were used, and
for each value of $\mu$ the maximum eigenvalue was found, and the mean photon
number was determined from the corresponding eigenvector.

Similarly to Sec.~\ref{meancanon} a cutoff was used at a photon number
sufficiently above the mean photon number that the state had fallen to around
$10^{-16}$. In addition, calculations were performed using the continuous
approximation in order to obtain results for very large photon numbers. The
method used was similar to that used in Sec.~\ref{maxnumer}. The phase
variances and photon numbers were determined for 512, 1024 and 2048 intervals,
and the result for the continuous case was then estimated by projecting to zero
step size. 

\begin{figure}
\centering
\includegraphics[width=0.7\textwidth]{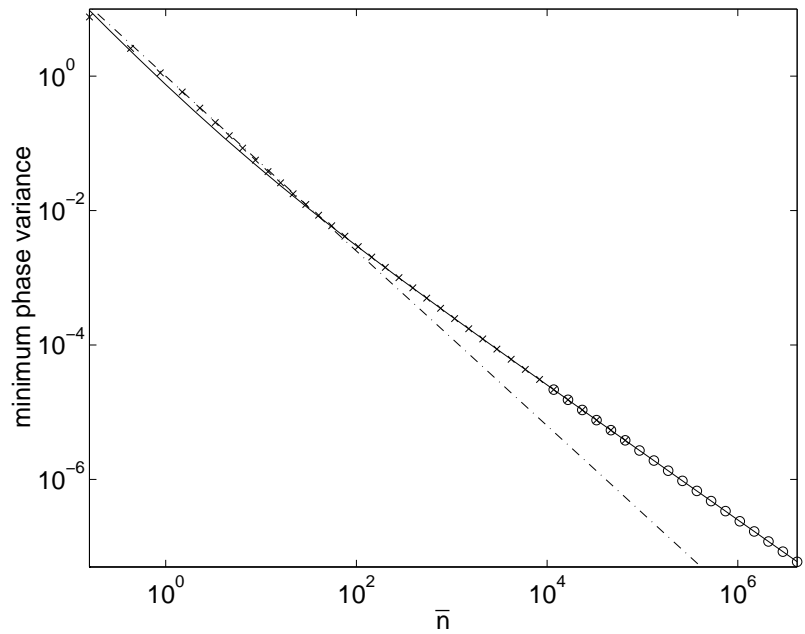}
\caption{The minimum phase variance for heterodyne measurements on states with
a fixed mean photon number. The exact calculations are shown as the crosses, the
continuous approximation as the circles, and the asymptotic analytic expression
as the continuous line. The power law claimed by D'Ariano and Paris for
heterodyne detection is also plotted (dash-dotted line).}
\label{hetlogmean}
\end{figure}

The results for the exact case and continuous approximation for heterodyne
measurements are shown in Fig.~\ref{hetlogmean}. The power law claimed by
D'Ariano and Paris is also shown in this figure. As can be seen, there is
good agreement between the numerically calculated values and the asymptotic
analytic expression for photon numbers above 100. There is also excellent
agreement between the values calculated exactly and those calculated using the
continuous approximation.

Note that the power law of D'Ariano and Paris gives good agreement for photon
numbers below 100, better than the asymptotic analytic expression obtained here.
For photon numbers above 100, however, this power law differs greatly from the
numerically calculated values, demonstrating that this power law is not the
correct asymptotic scaling.

\begin{figure}
\centering
\includegraphics[width=0.7\textwidth]{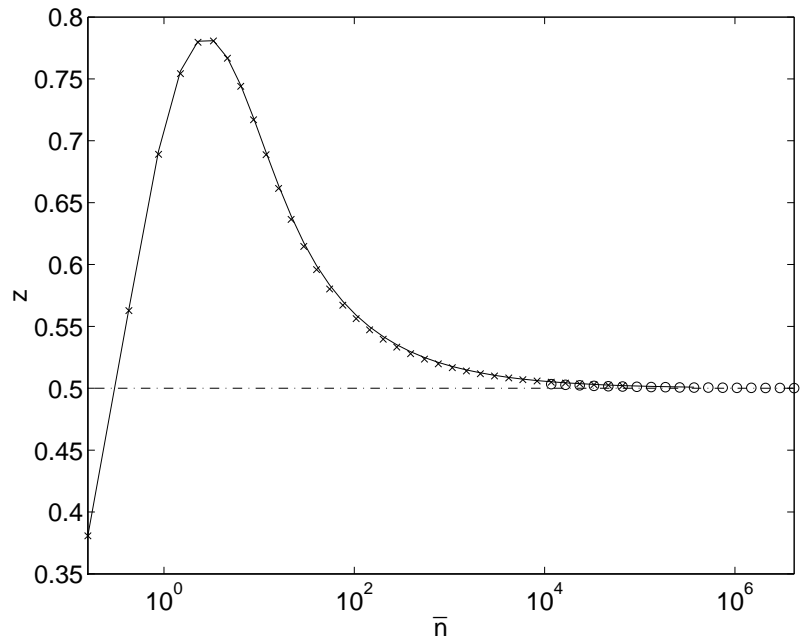}
\caption{The value of $z$ for heterodyne measurements on states with
a fixed mean photon number. The exact calculations are shown as the crosses, the
continuous approximation as the circles, and the asymptotic value of $1/2$
as the dash-dotted line. The results for squeezed states are shown as the
continuous line.}
\label{hetzmean}
\end{figure}

In order to see how accurately the numerical results agree with second term in
the asymptotic limit, we can define the parameter $z$ in a similar way as was
defined in Sec.~\ref{maxnumer}:
\beq
\label{zmean}
z=\frac{V(\phi)-2c\bar n^{-p}}{\bar n^{-p/2-1}}.
\eeq
Provided the term $3c^2 \nb^{-2p}$ can be ignored, the asymptotic value of $z$
should be $\sqrt{cp(p+1)}$. This parameter is plotted in Fig.~\ref{hetzmean}.
Again there is excellent
agreement between the values calculated exactly and the continuous approximation.
There is still a small but significant difference between the numeric values and
the asymptotic limit for a photon number of 100, but for photon numbers above
$10^4$ there is better than 1\% agreement with the asymptotic value.

\begin{figure}
\centering
\includegraphics[width=0.7\textwidth]{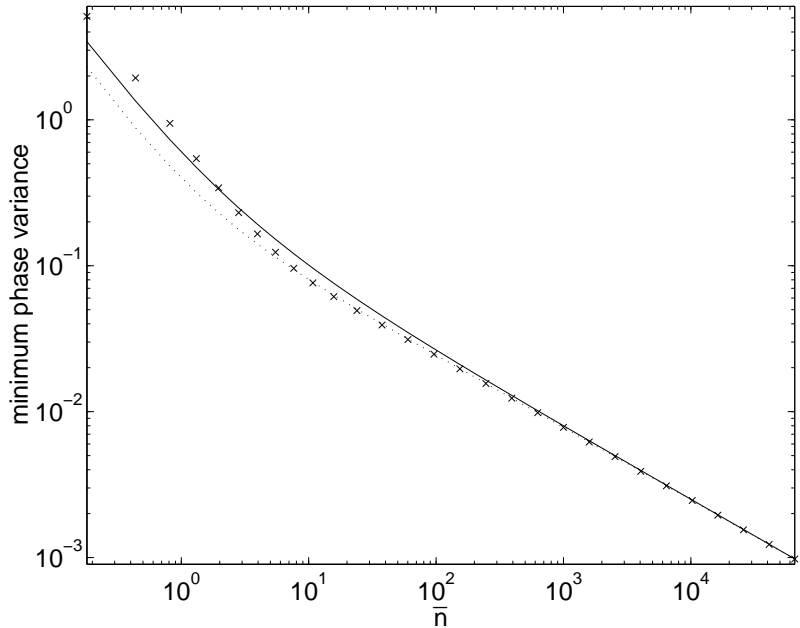}
\caption{The minimum phase variance for mark I measurements on states with a
fixed mean photon number. The exact calculations are shown as the crosses and
the asymptotic analytic expression as the continuous line. The analytic
expression taking account of the second term for $h_{\rm I}(n)$ is shown as the
dotted line.}
\label{mIlogmean}
\end{figure}

The calculated results for mark I measurements are shown in
Fig.~\ref{mIlogmean}. In this case the algorithm for calculating the continuous
approximation did not give convergent results, so only the exact results are
shown. There is good agreement with the asymptotic expression (\ref{markIas})
for photon numbers above about 100. The asymptotic analytic expression taking
account of the second term for $h_{\rm I}(n)$ is also shown in this figure. As
can be seen, the numerical results agree with this expression even more
accurately, with good agreement for photon numbers above about 10.

\begin{figure}
\centering
\includegraphics[width=0.7\textwidth]{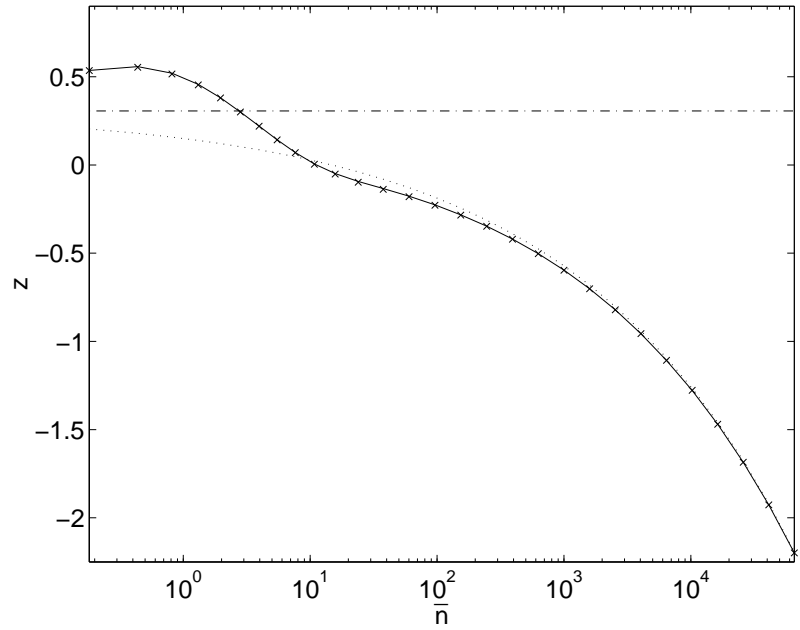}
\caption{The value of $z$ for mark I measurements on states with a fixed mean
photon number. The exact calculations are shown as crosses. The asymptotic
value ignoring the second term in Eq.~(\ref{cormarI}) is shown as the
dash-dotted line, and the asymptotic expression taking this term into account
is plotted as the dotted line. The results for squeezed states are shown as the
continuous line.}
\label{mIzmean}
\end{figure}

For mark I measurements there is a term of order $\nb^{-1}$, so we can not
expect $z$ to converge to $\sqrt{cp(p+1)}$. Using Eq.~(\ref{cormarI}), we see
that $z$ should converge to
\beq
\sqrt{cp(p+1)}-0.156248 \times \nb^{1/4}.
\eeq
The value of $z$ is plotted in Fig.~\ref{mIzmean}, and as can be seen it
converges reasonably accurately to this value, but not to $\sqrt{cp(p+1)}$.

\begin{figure}
\centering
\includegraphics[width=0.7\textwidth]{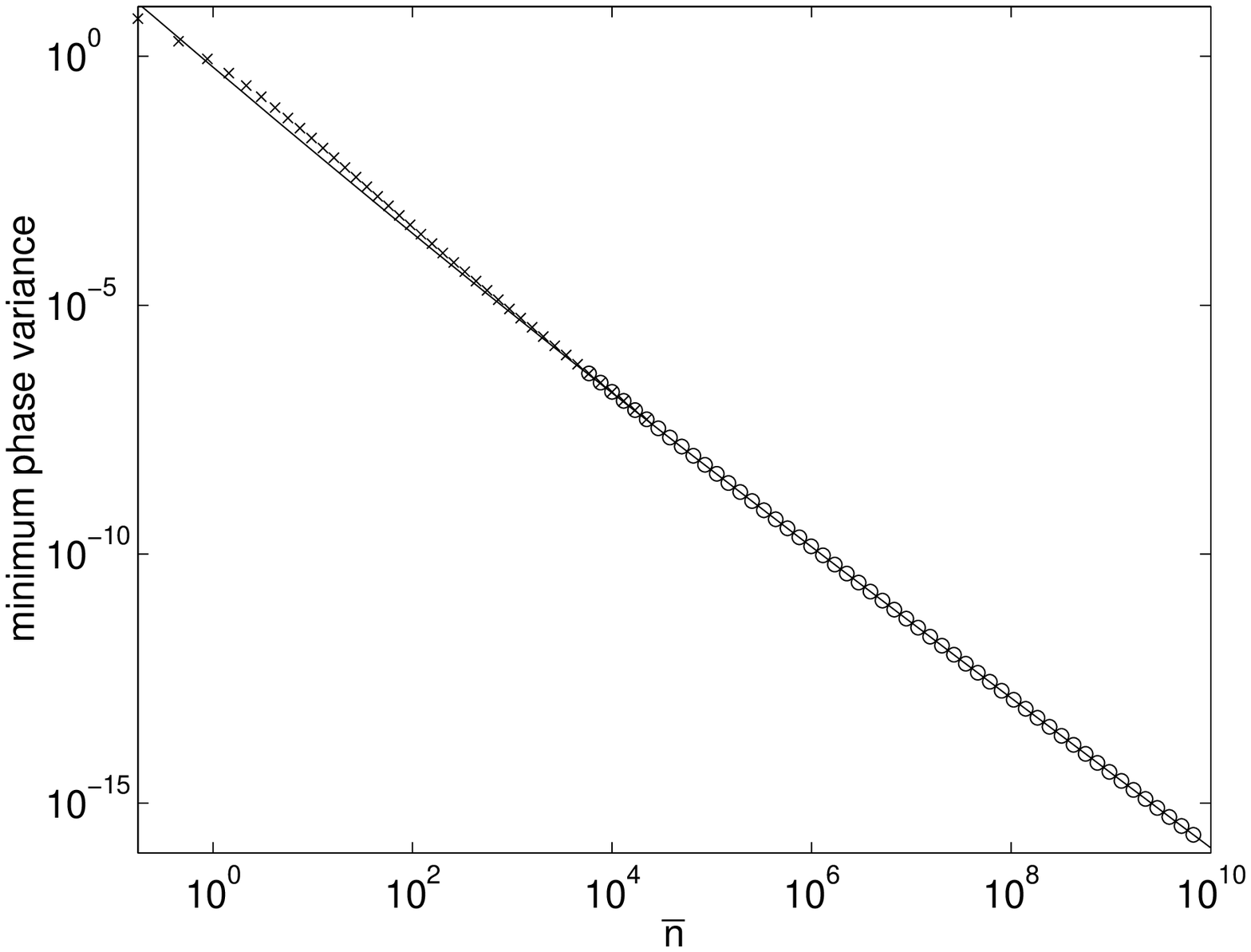}
\caption{The minimum phase variance for mark II measurements on states with
a fixed mean photon number. The exact calculations are shown as the crosses, the
continuous approximation as the circles, and the asymptotic analytic expression
as the continuous line.}
\label{mIIlogmean}
\end{figure}

\begin{figure}
\centering
\includegraphics[width=0.7\textwidth]{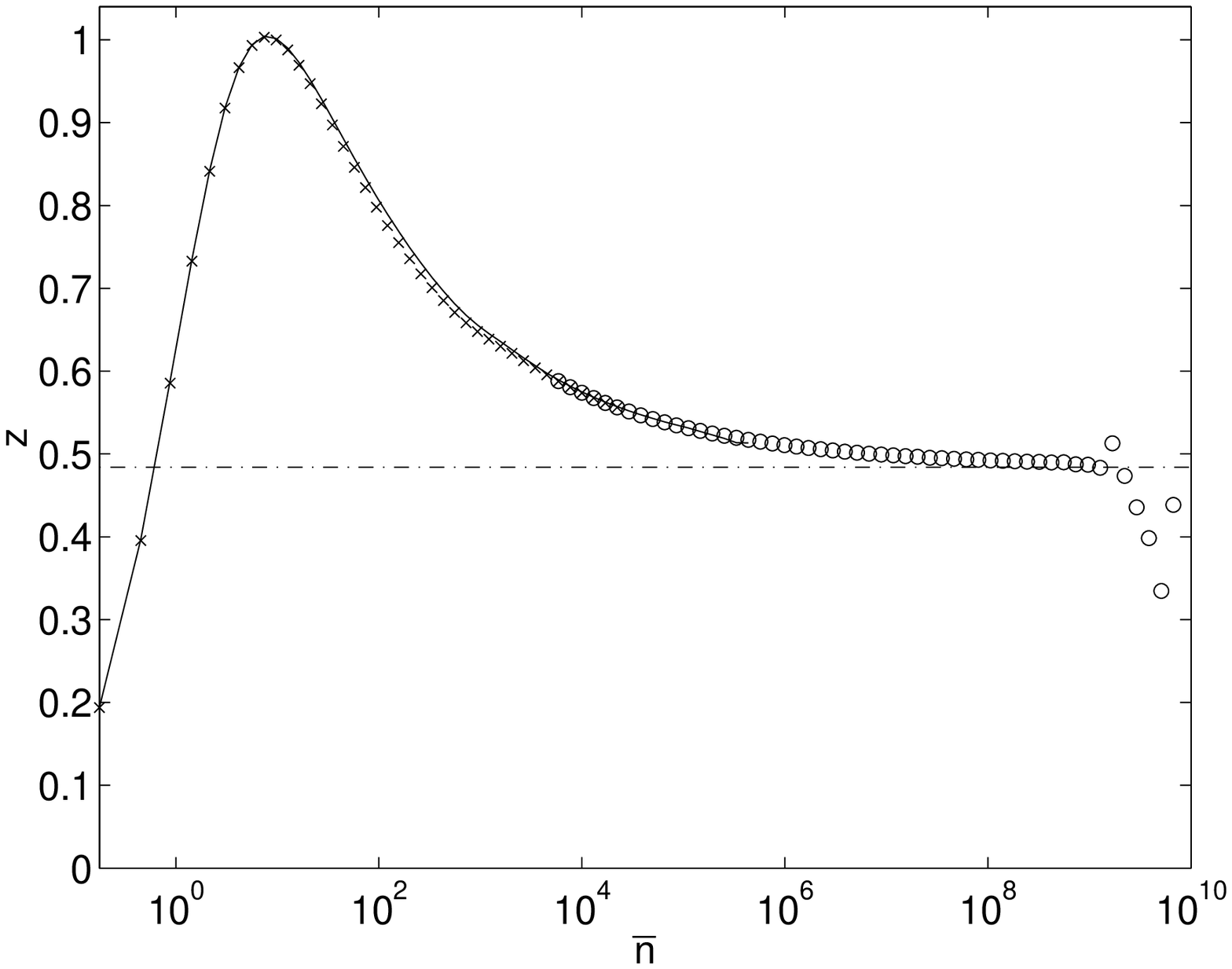}
\caption{The value of $z$ for mark II measurements on states with
a fixed mean photon number. The exact calculations are shown as the crosses, the
continuous approximation as the circles, and the asymptotic value of $\sqrt{15}
/8$ as the dash-dotted line. The results for squeezed states are shown as the
continuous line.}
\label{mIIzmean}
\end{figure}

The results for mark II measurements are shown in Fig.~\ref{mIIlogmean}. There
is very good agreement with the asymptotic analytic expression, and between the
results calculated using the exact method and continuous approximation. If we
plot $z$ (Fig.~\ref{mIIzmean}), we find that the second term does not agree well
until very large photon numbers. In fact, a photon number greater than $7 \times
10^8$ is required to obtain better than 1\% agreement. The agreement is again
poor for the largest photon numbers; however, this is just due to poor
convergence of the numerical technique.

\section{Optimised Squeezed States}
\label{squeezed}

As an alternative to optimising completely general states with a fixed mean
photon number, we can restrict our attention to squeezed states. As mentioned
in the introduction to this chapter, there are three reasons for this:\\
1. Squeezed states are relatively easily generated in the laboratory, whereas
there is no known way of producing general optimised states experimentally.\\
2. Squeezed states can be treated numerically far more easily than general
optimised states.\\
3. It has been found numerically (see Sec.~\ref{numeric}) that the phase
uncertainties of optimised squeezed states are very close to those of optimised
general states, and a partial theoretical explanation can be obtained by the
following analysis.

Squeezed states can be described by just two parameters, the coherent amplitude
and the squeezing parameter. When the mean photon number is fixed there is only
one independent parameter, and the optimisation problem is reduced to function
minimisation in one dimension.

\subsection{Canonical Measurements}
\label{sqzcan}

Again the simplest case is that of canonical measurements. This case was solved
by Collett \cite{collett}, and I will outline the derivation here. The squeezed
states considered are of the form
\beq
\ket{\alpha,\zeta}=\exp(\alpha a^\dagger -\alpha^* a)\exp[(\zeta^*
a^2-\zeta {a^\dagger}^2)/2]\ket 0.
\eeq
The mean photon number for this state is given by
\beq
\bar n = \st{\alpha}^2 + \sinh ^2 \st{\zeta}.
\eeq
Using this relation we can take a fixed value of $\bar n$, and vary $\zeta$. The
value of $\alpha$ can then be determined from $\bar n$ and $\zeta$.

One complication is the phases of $\alpha$ and $\zeta$. Only relative phase is
important, so the phase of $\zeta$ can be considered relative to $\alpha$.
Specifically, if the phase of $\alpha$ is rotated by $\theta$, an equivalent
state is obtained by rotating the phase of $\zeta$ by $2\theta$. We can
therefore take the phase of $\alpha$ to be zero without loss of generality.

All of the following analysis relies on $\zeta$ being real also. We can see that
$\zeta$ should be real if we consider the phase quadrature diagram for the
state. As in the introduction, the general quadrature $\hat X_{\Phi}$ is given
by
\beq
\hat X_{\Phi} = a e^{-i \Phi} + a\dg e^{i \Phi}.
\eeq
The 0 and $\pi/2$ quadratures are therefore
\bqa
\hat X_0 \!\!\!\! &=& \!\!\!\! (a + a\dg ), \\
\hat X_{\pi/2} \!\!\!\! &=& \!\!\!\! -i (a - a\dg ).
\eqa
For squeezed states the expectation values of these operators are
\bqa
\ip{\hat X_0} \!\!\!\! &=& \!\!\!\! 2\st{\alpha} \cos \varphi, \\
\ip{\hat X_{\pi/2}} \!\!\!\! &=& \!\!\!\! 2\st{\alpha} \sin \varphi,
\eqa
where $\varphi$ is the phase of $\alpha$. In addition, the uncertainties of
these quadrature operators are
\bqa
\ip{\Delta \hat X_0^2} \!\!\!\! &=& \!\!\!\! e^{2r} \sin^2 \half\phi_\zeta +
e^{-2r} \cos^2 \half\phi_\zeta, \\
\ip{\Delta \hat X_{\pi/2}^2} \!\!\!\! &=& \!\!\!\! e^{2r} \cos^2
\half\phi_\zeta + e^{-2r} \sin^2 \half\phi_\zeta,
\eqa
where $r$ and $\phi_\zeta$ are the magnitude and phase of $\zeta$.

\begin{figure}
\centering
\includegraphics[width=0.8\textwidth]{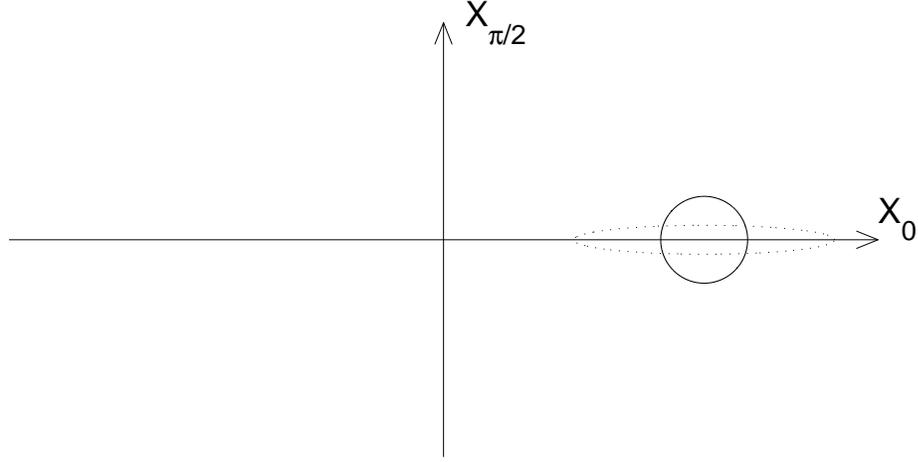}
\caption{Contours of the probability distribution for $X_0$ and
$X_{\pi/2}$. The contour for a coherent state is shown as the continuous
line, and a contour for a squeezed state with $\phi_\zeta = \pi$ is shown as the
dotted line.}
\label{contour}
\end{figure}

A common way of representing coherent and squeezed states is by a contour of the
probability distribution for the measured values $X_0$ and $X_{\pi/2}$, as
in Fig.~\ref{contour}. The contour for a coherent state is a circle; however,
the contour for a squeezed state is an ellipse, with the major axis at an angle
$\half (\phi_\zeta+\pi)$.

If we are considering small deviations from zero phase,
\bqa
\ip{\hat X_0} \!\!\!\! & \approx & \!\!\!\! 2\st{\alpha}, \\
\ip{\hat X_{\pi/2}} \!\!\!\! & \approx & \!\!\!\! 2\st{\alpha} \varphi,
\eqa
so measuring $X_0$ gives information about the photon number, whereas measuring
$X_{\pi/2}$ gives information about the phase. If the actual phase is
zero, then $\phi_\zeta$ should be equal to $\pi$ in order for $X_{\pi/2}$,
and therefore the phase, to have a reduced uncertainty. This means that $\zeta$
should be real and negative in order to have a phase squeezed state.

A complex value of $\zeta$ will mean that the major axis of the ellipse is at an
angle from the horizontal, and this should give a larger phase uncertainty than
if the ellipse is horizontal. This is not a rigorous derivation, and I have
found numerically (see Fig.~\ref{zetapl}) that for very small $\nb$ the optimum
$\zeta$ is positive. Even for these exceptional values of $\nb$, complex
values of $\zeta$ were never found to give a better result. It is therefore
reasonable to assume a real value of $\zeta$ in the following analysis.

In order to determine the phase uncertainty, we wish to determine
\beq
\ip{\widehat{\cos \phi}}=\half \sum_{n=0}^{\infty} \left( \braket{\alpha,\zeta}
{n}\braket{n+1}{\alpha,\zeta}+\braket{\alpha,\zeta}{n+1}\braket{n}{\alpha,\zeta}
\right).
\eeq
Collett determines this sum using the number state representation of squeezed
states as given by \cite{Yuen76}
\beq
\label{sqznumber}
\braket{n}{\alpha,\zeta}=(n!\mu)^{-1/2}(\nu/2\mu)^{n/2}H_n[\beta(2\mu
\nu)^{-1/2}] \exp[-|\beta|^2/2+(\nu^{*}/2\mu)\beta^2],
\eeq
where
\bqa
\mu \!\!\!\! &=& \!\!\!\! \cosh r, \\
\nu \!\!\!\! &=& \!\!\!\! e^{i\phi_\zeta} \sinh r, \\
\beta \!\!\!\! &=& \!\!\!\! \alpha \mu + \alpha^* \nu.
\eqa
The $H_n$ are Hermite polynomials and are given by \cite{Erdelyi53}
\beq
H_n(x)=n!\sum_{m=0}^{\lfloor n/2 \rfloor} \frac{(-1)^m (2x)^{n-2m}}{m!(n-2m)!}.
\eeq
Note that this formula is for the general case of complex $\alpha$ and $\zeta$.

Collett evaluated this for real $\alpha$ and real, negative $\zeta$, and found
\beq
\ip{\widehat{\cos \phi}} \approx \erf(\sqrt{2n_1})\left[ 1-\frac{(\coth r -1)^2}
{32 n_1 \sqrt{2e^{-r}\sinh r}} \left( 1+\frac 1{2n_1}\right) \right],
\eeq
where
\beq
\label{n1def}
n_1 = \half \alpha^2 (1-\tanh r).
\eeq
This in turn gives the phase variance as
\beq
\label{squunc1}
V(\phi) \approx \frac {n_0+1}{4\bar n^2}+ 2 \erfc(\sqrt{2n_0}),
\eeq
where $n_0 = \bar n e^{-2r}$. Taking the derivative with respect to $n_0$ gives
\beq
\frac{\partial}{\partial n_0} V(\phi) \approx \frac 1{4\bar n^2}
-\sqrt{\frac 8{\pi n_0}}e^{-2n_0}.
\eeq
This is zero for
\beq
\frac 1{4\bar n^2}=\sqrt{\frac 8{\pi n_0}}e^{-2n_0},
\eeq
or
\beq
\label{n0prev}
n_0 = \log (4 \bar n) - \smallfrac 14 \log (2\pi n_0).
\eeq
In Ref.~\cite{collett} the factor of $n_0$ is omitted in the second term on the
right hand side. We can make a slightly better approximation by using
$n_0 \approx \log (4 \bar n)$ on the right hand side, so
\beq
\label{n0}
n_0 \approx \log (4 \bar n) - \smallfrac 14 \log \log (4\bar n)
- \smallfrac 14 \log (2\pi).
\eeq
We can also include an arbitrary number of additional terms with iterated logs;
however, this expression gives a very accurate solution, as shown in
Fig.~\ref{v4nmlogn}.

Using the asymptotic expansion of the complementary error function
\beq
\erfc(x) \sim \frac 1{x\sqrt \pi}e^{-x^2} \sum_{k=0}^{\infty} (-1)^k
\frac{1\cdot 3\cdots (2k-1)}{(2x^2)^k},
\eeq
we have, for large $n_0$,
\bqa
\label{badests}
2\erfc(\sqrt{2n_0}) \!\!\!\! & \approx & \!\!\!\! \sqrt{\frac 2{\pi n_0}}
e^{-2n_0} \nn \\ \!\!\!\! &=& \!\!\!\! \frac 1{8\bar n^2}.
\eqa
Using this result and substituting (\ref{n0}) into (\ref{squunc1}) gives the
minimum phase variance as
\beq
\label{squvar2}
V(\phi) \approx \frac{\log \bar n - \smallfrac 14 \log \log (4\bar n)
+ \Delta}{4 \bar n^2},
\eeq
where
\beq
\label{defdelta}
\Delta = \smallfrac 32 + 2\log 2 - \smallfrac 14 \log(2\pi) \approx
2.426825.
\eeq
Therefore we see that the squeezed state optimised for minimum intrinsic phase
variance has a phase variance scaling as $\log \bar n/\bar n^2$. This is not
quite as good as the result for general optimised states, which scale as
$\bar n ^{-2}$. The factor of $\log \bar n$ increases very slowly, however, and
it requires very large photon numbers to produce a significant difference
between the variances of optimised squeezed and general states.

Note that the expression (\ref{squvar2}) for the variance differs from that
given in \cite{collett}. The $\smallfrac 14 \log \log (4\bar n)$ term does not
appear in \cite{collett}, as the approximation $n_0 \sim 1$ is made on the right
hand side of the solution for $n_0$ (\ref{n0prev}). This term increases with
photon number, and is therefore more significant than the $\Delta$ term.

\subsection{Numerical Results for Canonical Measurements}
This approximate analytic result has been tested numerically by determining the optimum
squeezed states for a range of mean photon numbers up to about $8\times 10^7$.
Similarly to the case for optimum general states, the state coefficients were
considered until the point where they had fallen to about $10^{-16}$, where
they did not significantly affect the results.

\begin{figure}
\centering
\includegraphics[width=0.7\textwidth]{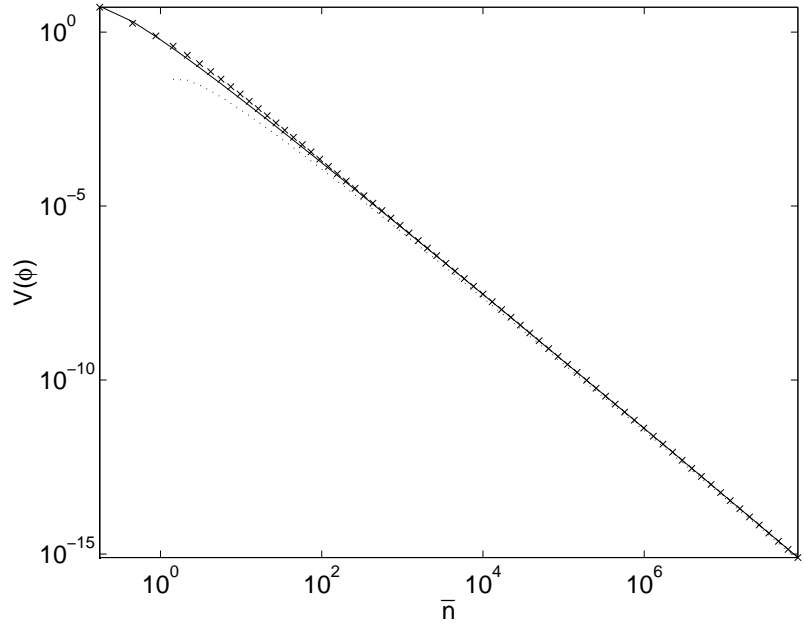}
\caption{The phase variance for optimised squeezed states as a function of mean
photon number. The exact phase variance is shown as the crosses, the
asymptotic expression $(\log \bar n + \Delta)/(4 \bar n^2)$ is shown as the
continuous line, and $\log \bar n/(4 \bar n^2)$ is shown as the dotted line.}
\label{logsqu}
\end{figure}

The phase variances of the minimum uncertainty squeezed states are plotted in
Fig.~\ref{logsqu}. Also shown is the asymptotic expression (with the
$\smallfrac 14 \log \log (4\bar n)$ term omitted for simplicity). There is
good agreement between the calculated values and the analytic expression, with
better than 1\% agreement for photon numbers above $5\times 10^4$. Also shown is
the analytic expression if we omit $\Delta$, and there is much poorer agreement
for the smaller photon numbers in that case. This means that although the
$\Delta$ term is insignificant for large photon numbers, it gives more realistic
results for moderate photon numbers.

\begin{figure}
\centering
\includegraphics[width=0.7\textwidth]{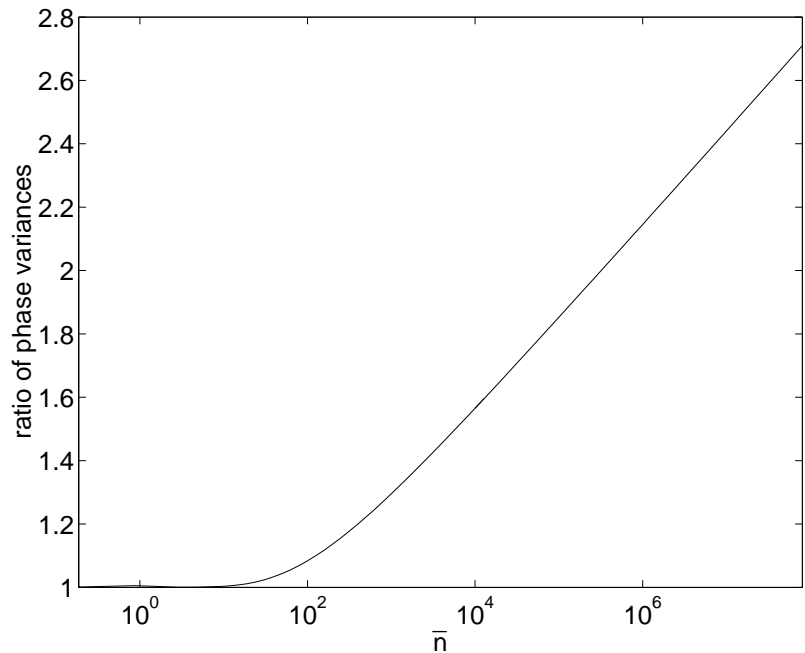}
\caption{The ratio of the phase variance for optimised squeezed states to the
phase variance for optimised general states.}
\label{squzratio}
\end{figure}

To demonstrate how close the phase variance is to that for general states, the
ratio of the phase variance for optimised squeezed states to that for general
states is plotted in Fig.~\ref{squzratio}. For mean photon numbers above about
$10^4$ the phase uncertainty for general states was estimated using the
asymptotic expression, as this is very accurate for large photon numbers.
The phase variances are extremely close for photon numbers up to 10 or 100, and
it takes a photon number of almost a million for the phase variance for
squeezed states to be more than twice that for optimised general states.

\begin{figure}
\centering
\includegraphics[width=0.7\textwidth]{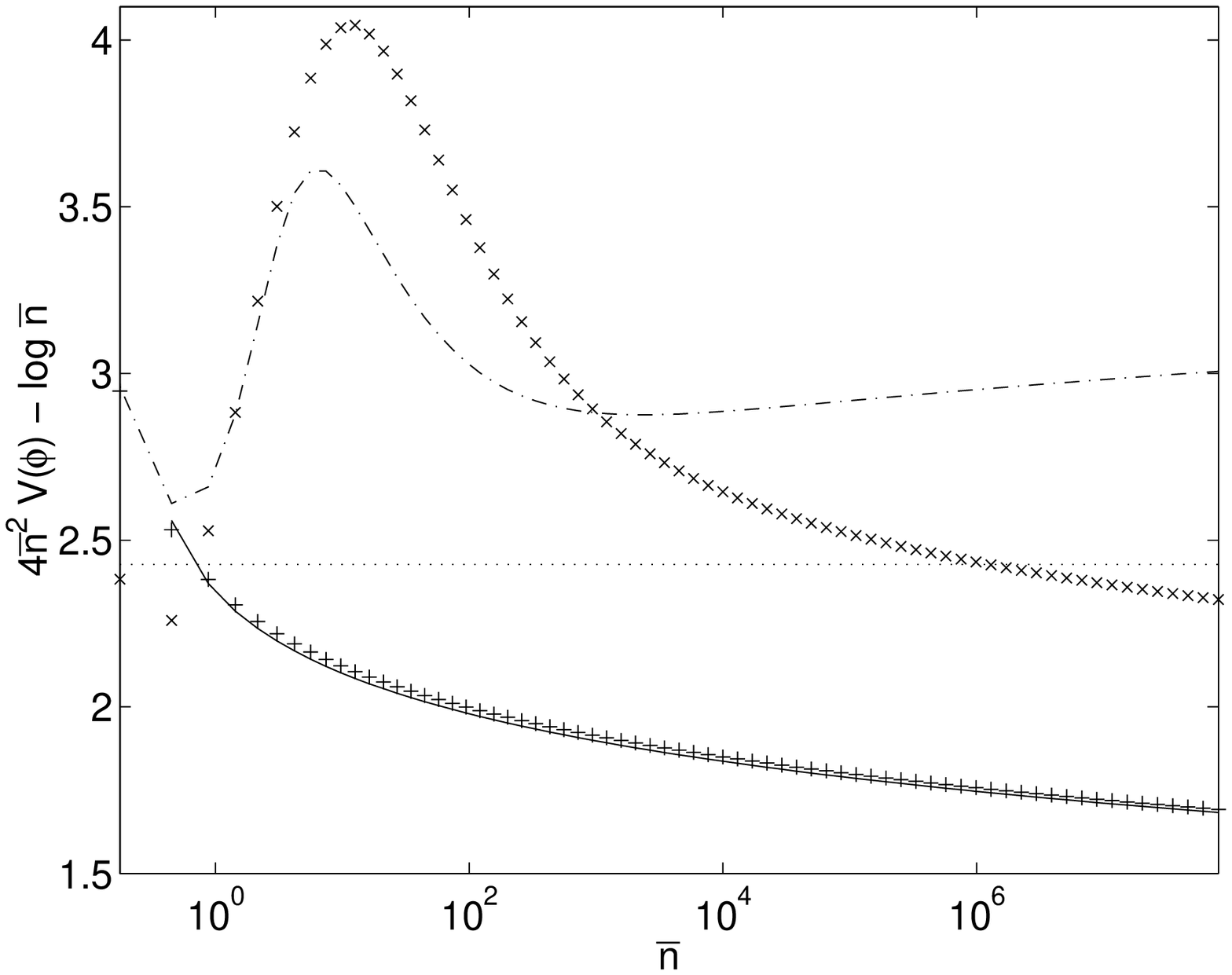}
\caption{The value of $4\bar n^2 V(\phi)-\log\bar n$ as calculated using various
methods. The exact values are shown as crosses, $\Delta$ is shown as the
horizontal dotted line and $- \smallfrac 14 \log \log (4\bar n)+\Delta$ is shown
as the continuous line. The values determined exactly using Eq.~(\ref{squunc1})
are shown as the pluses, and those calculated exactly using the additional
correction are shown as the dash-dotted line.}
\label{v4nmlogn}
\end{figure}

In order to see if the term $\smallfrac 14 \log \log (4\bar n)$ gives a
better estimate of the phase variance, $4\bar n^2 V(\phi)-\log\bar n$ is
plotted in Fig.~\ref{v4nmlogn}. According to the result (\ref{squvar2}), this
should have the asymptotic value $-\smallfrac 14 \log\log(4\bar n)+\Delta$,
rather than $\Delta$. As can be seen, the exact values do not converge to
$\Delta$, but they do not appear to converge to $-\smallfrac 14 \log \log
(4\bar n)+\Delta$ either. The exact values may converge to $-\smallfrac 14 \log
\log (4\bar n)+\Delta$ for much larger photon numbers; however, for the range of
photon numbers shown $\Delta$ gives a better approximation.

This result was checked by calculating the minimum phase variance exactly from
(\ref{squunc1}), and these results are also shown in Fig.~\ref{v4nmlogn}.
There is close agreement with $- \smallfrac 14 \log \log (4\bar n)+\Delta$,
indicating that the asymptotic expression (\ref{squvar2}) is a good
approximation for (\ref{squunc1}). The discrepancy must therefore be due to a
slight inaccuracy in Eq.~(\ref{squunc1}). In Ref.~\cite{collett} Collett
indicates that a correction term for (\ref{squunc1}) is
\beq
2(1-4a)\sqrt{\frac{2n_1}{\pi}}2^{-2n_1},
\eeq
where $a \approx 0.1990726$ and $n_1$ is as defined in Eq.~(\ref{n1def}).
Although the results using this correction agree with the exact values for very
small photon numbers, and reproduce the peak found in the exact results, the
results again do not converge to the exact values for large photon numbers. In
fact the asymptotic expression simply using $\Delta$ is still the most accuate
for photon numbers above about $10^4$. As none of these corrections appear to
give better results, I will omit them in further discussion.

\begin{figure}
\centering
\includegraphics[width=0.7\textwidth]{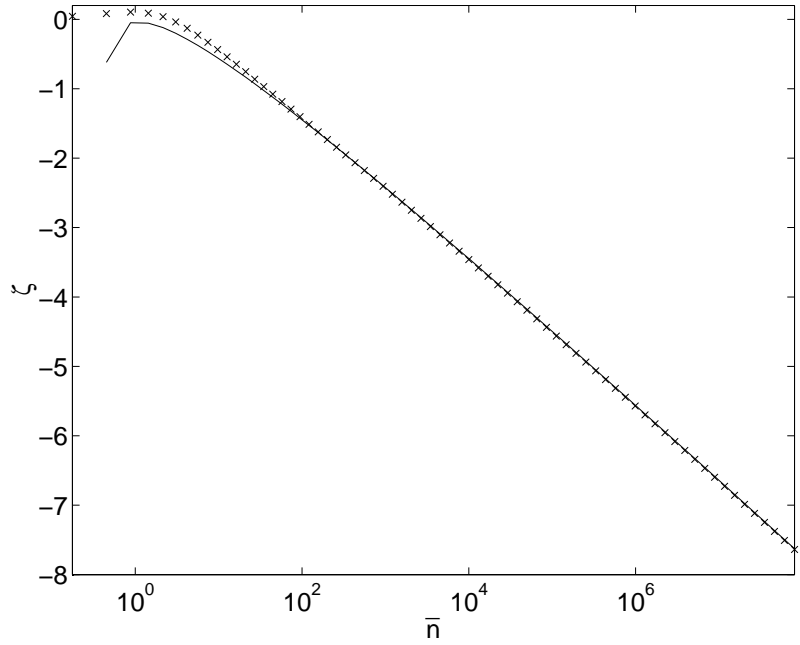}
\caption{The optimum values of $\zeta$ for squeezed states as a function of
$\bar n$. The numerically determined values are shown as crosses, and the
values given by Eq.~(\ref{zetath}) are shown as the continuous line.}
\label{zetapl}
\end{figure}

Another test of the theory is the optimum value of $\zeta$. From
Eq.~(\ref{n0prev}) we have
\beq
n_0 = \bar n e^{2\zeta} \approx  \log (4\bar n) - \smallfrac 14 \log (2\pi),
\eeq
where the factor of $n_0$ has been omitted from the second term. Solving for
$\zeta$ gives
\beq
\label{zetath}
\zeta = \half \log \left( \log (4\bar n) - \smallfrac 14 \log (2\pi) \right)
-\half \log \bar n.
\eeq
The numerically determined optimum values of $\zeta$ and the approximate analytic
expression (\ref{zetath}) are plotted in Fig.~\ref{zetapl}. As can be seen, the
numerically determined values are extremely close to the analytic result for
photon numbers above about 100.

\begin{figure}
\centering
\includegraphics[width=0.5\textwidth]{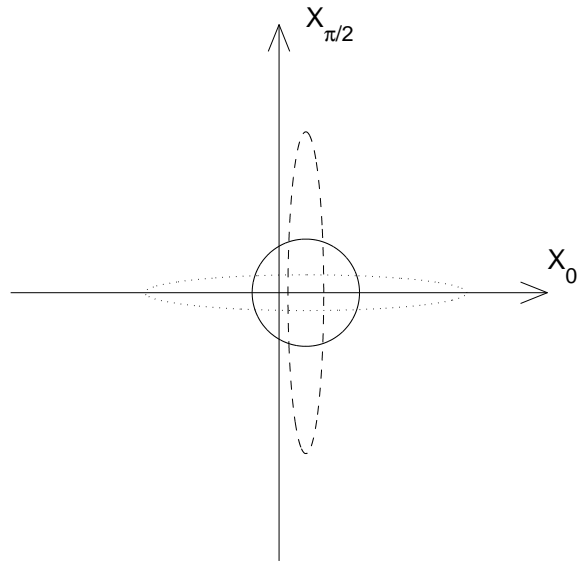}
\caption{Contours of the probability distribution for $X_0$ and
$X_{\pi/2}$. The contour for a coherent state is shown as the continuous
line, a contour for a squeezed state with negative $\zeta$ is shown as the
dotted line, and with positive $\zeta$ as the dashed line.}
\label{contour2}
\end{figure}

Note also that for very small mean photon numbers, below around 2.6, the optimum
value of $\zeta$ is actually positive. This is a strange result, as the theory
given above indicates that $\zeta$ should be negative for reduced phase
uncertainty. The reason for this would seem to be that for cases where the
photon number is very small, the circular error contour for a coherent state,
as shown in Fig.~\ref{contour2}, would overlap the $X_{\pi/2}$ axis. This means
that there would be a significant contribution to the probability distribution
for $\phi=\pm\pi$. A squeezed state with negative $\zeta$ would have an error
contour that extends even further over the $X_{\pi/2}$ axis, resulting in a
larger contribution to the probability distribution near $\phi=\pm\pi$.

If $\zeta$ is positive, however, the major axis of the error ellipse would be
vertical, so the ellipse would avoid overlapping the $X_{\pi/2}$ axis. It is
reasonable that this would result in a smaller phase variance. Note that there
is no corresponding justification for using a complex value of $\zeta$. Complex
values were also tested, but always gave greater phase uncertainties, even for
the very low photon number cases.

\subsection{General Dyne Measurements}
\label{sqzgeneral}
Now I will consider squeezed states optimised for minimum phase variance under
more general dyne measurements. Again we wish to find $\ip{\widehat{\cos\phi}}$,
except now we have
\beq
2 \widehat{\cos \phi} = \sum_{n=0}^{\infty} (1-h(n)) \left[ \ket{n}\bra{n+1}
+\ket{n+1}\bra{n}\right].
\eeq
The measure of the phase variance that I will initially consider is
\beq
V'(\phi) = 2-2\ip{\widehat{\cos \phi}}.
\eeq
I will use this rather than the Holevo variance initially, in order to use the
results for canonical measurements in a straightforward way. This expression is
not a sufficiently accurate estimate of the Holevo variance for the case of
mark I measurements. Therefore, at the end of this derivation, the result for
this measure of the phase variance will be converted to the Holevo variance.
This measure of the phase variance can be used for the intermediate steps, as
minimising this measure of the phase variance is equivalent to minimising the
Holevo phase variance.

Evaluating this measure of the phase variance gives
\bqa
V'(\phi) \!\!\!\! &=& \!\!\!\! 2-\sum_{n=0}^{\infty} (1-h(n)) \left[
\braket{\alpha,\zeta}{n}\braket{n+1}{\alpha,\zeta}
+\braket{\alpha,\zeta}{n+1}\braket{n}{\alpha,\zeta}\right] \nn \\
\!\!\!\! &=& \!\!\!\! 2-\sum_{n=0}^{\infty} \left[\braket{\alpha,\zeta}{n}
\braket{n+1}{\alpha,\zeta}
+\braket{\alpha,\zeta}{n+1}\braket{n}{\alpha,\zeta}\right] \nn \\
&&-\sum_{n=0}^{\infty}h(n)\left[\braket{\alpha,\zeta}{n}\braket{n+1}{\alpha,
\zeta}+\braket{\alpha,\zeta}{n+1}\braket{n}{\alpha,\zeta}\right].
\eqa
Collett \cite{collett} found the first two terms to be approximately
\beq
\frac {n_0+1}{4\nb^2}+2 \erfc(\sqrt{2n_0}),
\eeq
so we therefore only require an expression for the third term. Both $\alpha$ and
$\zeta$ will be taken to be real in this derivation, so the number state
coefficients of the squeezed state are real, and the third term can be expressed
as
\beq
2 \sum_{n=0}^\infty h(n) \braket{\alpha,\zeta}{n}\braket{n+1}{\alpha,\zeta}.
\eeq
Approximating $h(n)$ by $cn^{-p}$, the sum to be evaluated is
\beq
\label{thirdterm}
\sum_{n=0}^\infty cn^{-p} \braket{\alpha,\zeta}{n}\braket{n+1}{\alpha,\zeta}.
\eeq

This expression can be evaluated approximately using
\beq
\label{approxexpect}
\sum_{n=0}^\infty n^{-p} \braket{\alpha,\zeta}{n}\braket{n+1}{\alpha,\zeta}
\approx \ip{n^{-p}}.
\eeq
Expanding this in a series around $n=\nb$ we find
\bqa
\ip{n^{-p}} \!\!\!\! &=& \!\!\!\! \ip{(\nb+\Delta n)^{-p}} \nn \\
\!\!\!\! & \approx & \!\!\!\! \nb^{-p} \left(1+\frac{p(p+1)}2
\frac{\ip{\Delta n^2}}{\nb^2}\right).
\eqa
It is easily shown that for squeezed states
\bqa
\ip{\Delta n^2} \!\!\!\! &=& \!\!\!\! \alpha^2(\mu-\nu)^2+2\mu^2\nu^2, \nn \\
\!\!\!\! & \approx & \!\!\!\! \frac{\nb^2}{n_0},
\eqa
so we find
\beq
\label{sqzresult}
\sum_{n=0}^\infty n^{-p} \braket{\alpha,\zeta}{n}\braket{n+1}{\alpha,\zeta}
\approx \nb^{-p} \left(1+\frac{p(p+1)}{2 n_0}\right).
\eeq
This should be correct to leading order; however, the second order term is not
necessarily correct, as the approximation (\ref{approxexpect}) may only be
correct to first order. This result is derived in a more rigorous way in
Appendix \ref{sqzder}, and it is shown that the second order term is also
correct.

Using this result, the phase uncertainty is given by
\beq
\label{phunc1}
V'(\phi) \approx \frac{n_0+1}{4\nb^2}+2 \erfc(\sqrt{2n_0}) 
+ 2c\nb^{-p}\left[ 1+\frac {p(p+1)}{2n_0} \right].
\eeq
Taking the derivative with respect to $n_0$ gives
\beq
\frac {\partial}{\partial n_0} V'(\phi) \approx \frac {1}{4\nb^2
}-\sqrt{\frac {8}{\pi n_0}}e^{-2n_0}+2c\nb^{-p}\left[\frac{-p
(p+1)}{2n_0^2}\right].
\eeq
As the second term falls exponentially with $n_0$ it can be omitted.
Then we find that for minimum phase variance
\beq
\label{minn0}
n_0 \approx 2\sqrt{cp(p+1)}\nb^{1-p/2}.
\eeq
For $0<p<2$, $n_0$ increases with photon number, but does not increase as
rapidly as $\nb$, in agreement with the assumptions used in the appendix.

Substituting this result into Eq.~(\ref{phunc1}) gives
\beq
V'(\phi) \approx 2c\nb^{-p}+\sqrt{cp(p+1)}\nb^{-p/2-1}.
\eeq
Converting this to the Holevo phase variance gives an additional correction
term:
\beq
V(\phi) \approx 2c\nb^{-p}+\sqrt{cp(p+1)}\nb^{-p/2-1}+3c^2\nb^{-2p}.
\eeq
This correction term is only significant for mark I measurements, where it is
of lower order than the second term. Thus we see that we obtain exactly the
same terms for the phase uncertainty when considering squeezed states as we do
when considering general states.

\subsection{Numerical Results}
\label{numeric}
These results have been tested by numerically determining the optimum squeezed
states for heterodyne and mark I and II measurements. Because the results are
extremely close to those for optimised general states, rather than plotting the
phase variance, I have plotted the phase variance as a ratio to the phase
variance for optimised general states.

\begin{figure}
\centering
\includegraphics[width=0.7\textwidth]{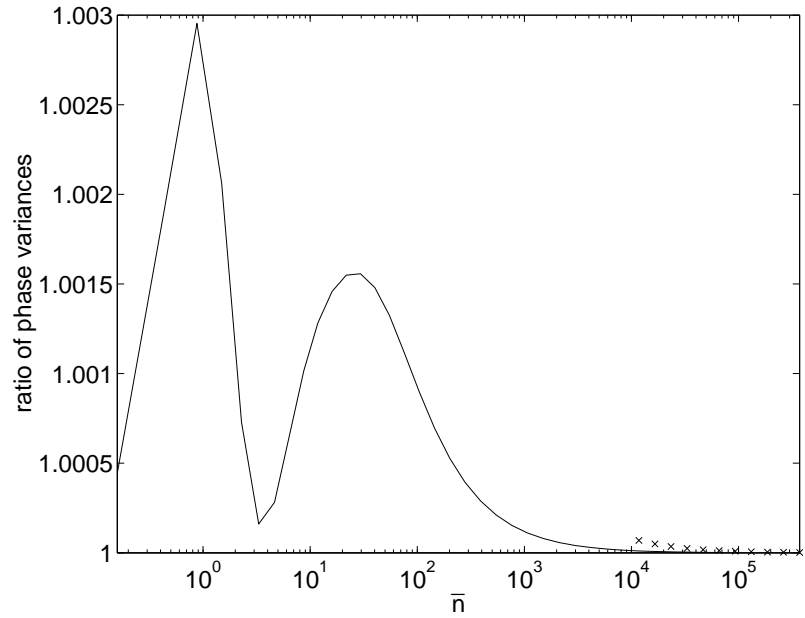}
\caption{The ratio of the phase variance for squeezed states optimised for
minimum phase variance under heterodyne measurements to the phase variance for
optimum general states. The values using the exactly calculated general states
are shown as the continuous line, and the values using the continuous
approximation for the general states are shown as the crosses.}
\label{hetsqz}
\end{figure}

The ratio of the minimum phase variance for squeezed states to that for general
states using heterodyne measurements is plotted in Fig.~\ref{hetsqz}. The
phase variance for optimum squeezed states is never more than 0.3\% above the
phase variance for optimum general states, and for large photon numbers the
phase variances converge. In fact, for some photon numbers the squeezed state
variances are closer to the exact general state variances than those calculated
using the continuous approximation.

\begin{figure}
\centering
\includegraphics[width=0.7\textwidth]{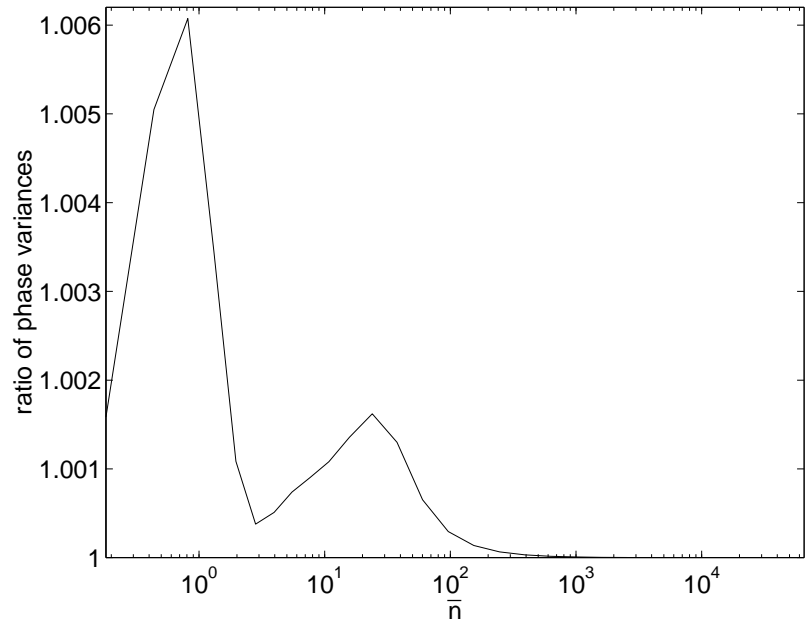}
\caption{The ratio of the phase variance for squeezed states optimised for
minimum phase variance under mark I measurements to the phase variance for
optimum general states.}
\label{mIsqz}
\end{figure}

\begin{figure}
\centering
\includegraphics[width=0.7\textwidth]{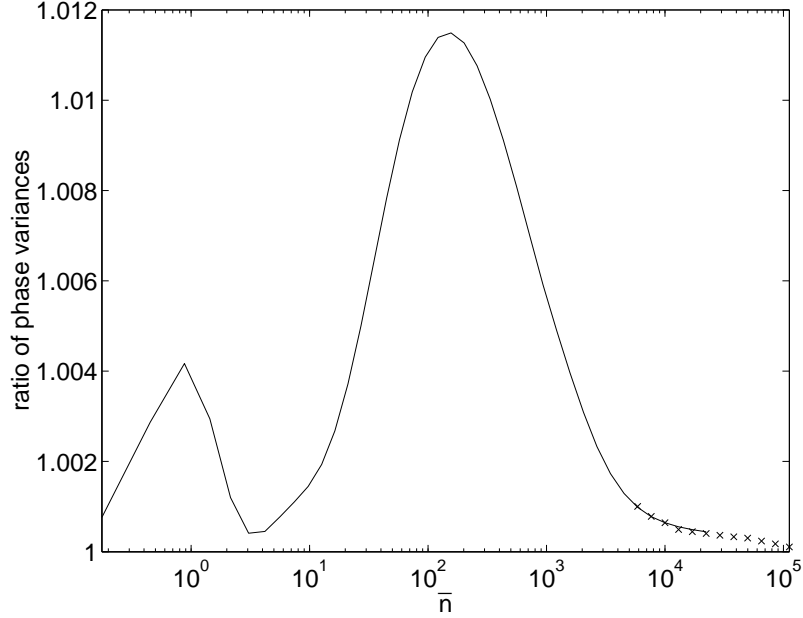}
\caption{The ratio of the phase variance for squeezed states optimised for
minimum phase variance under mark II measurements to the phase variance for
optimum general states. The values using the exactly calculated general states
are shown as the continuous line, and the values using the continuous
approximation for the general states are shown as the crosses.}
\label{mIIsqz}
\end{figure}

The results for mark I measurements are shown in Fig.~\ref{mIsqz}. Here the
squeezed state phase variance is never more than about 0.6\% above the phase
variance for optimum general states, and the phase variances again converge for
large photon numbers. The results for mark II measurements are shown in
Fig.~\ref{mIIsqz}. Here the squeezed state phase variance can be more than 1\%
above the general state phase variance, but the phase variances still converge
for large photon numbers.

Another way of comparing the results for squeezed states and general states is
to plot the $z$ parameter defined by (\ref{zmean}). This is displayed for
heterodyne, mark I and mark II measurements in Figs~\ref{hetzmean},
\ref{mIzmean} and \ref{mIIzmean}. As can be seen, the results for optimum
squeezed states and general states are almost indistinguishable. This means that
the phase variances for squeezed and general states agree to better precision
than just the first terms that I have derived here.

\begin{figure}
\centering
\includegraphics[width=0.7\textwidth]{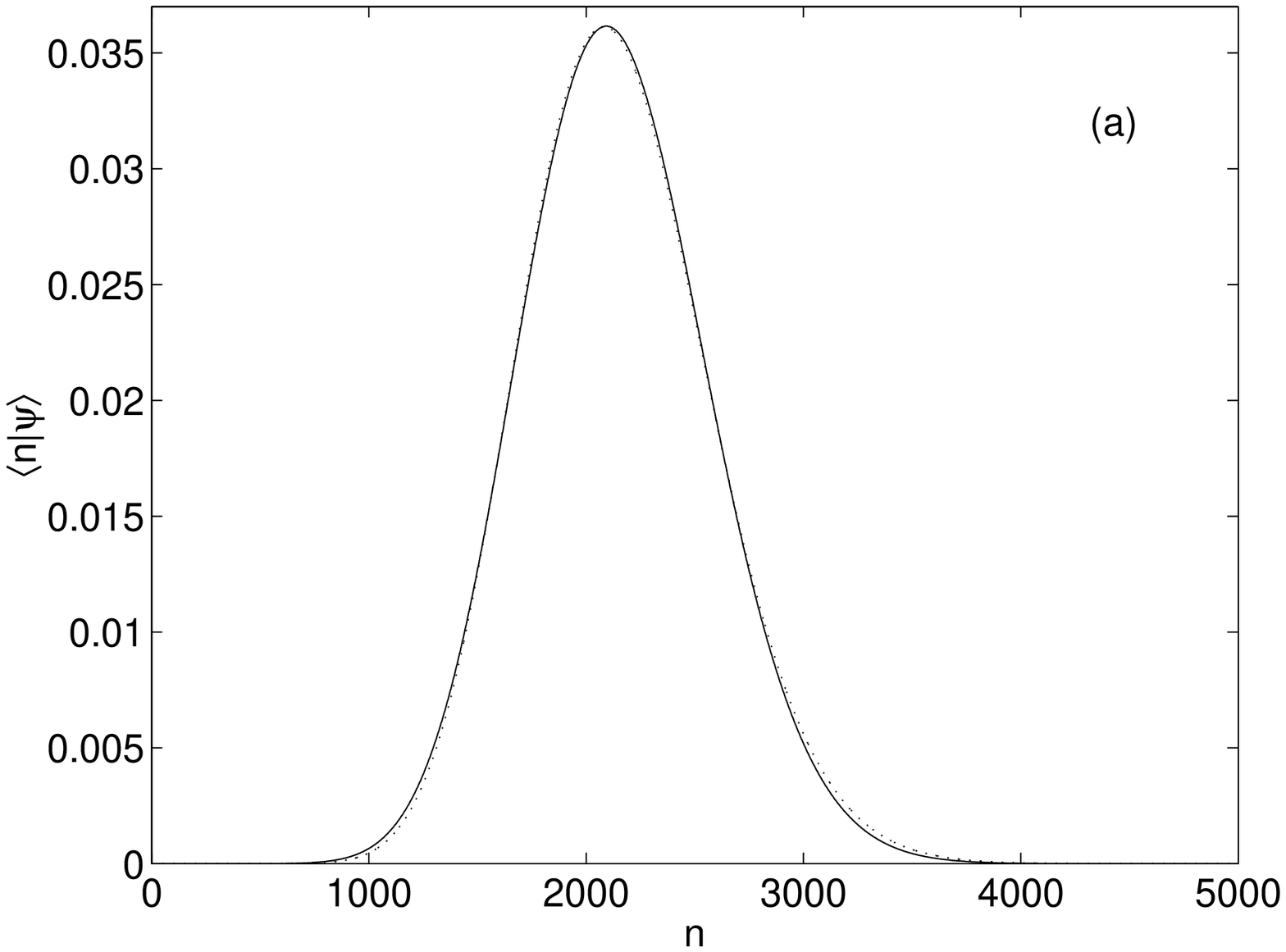}
\includegraphics[width=0.7\textwidth]{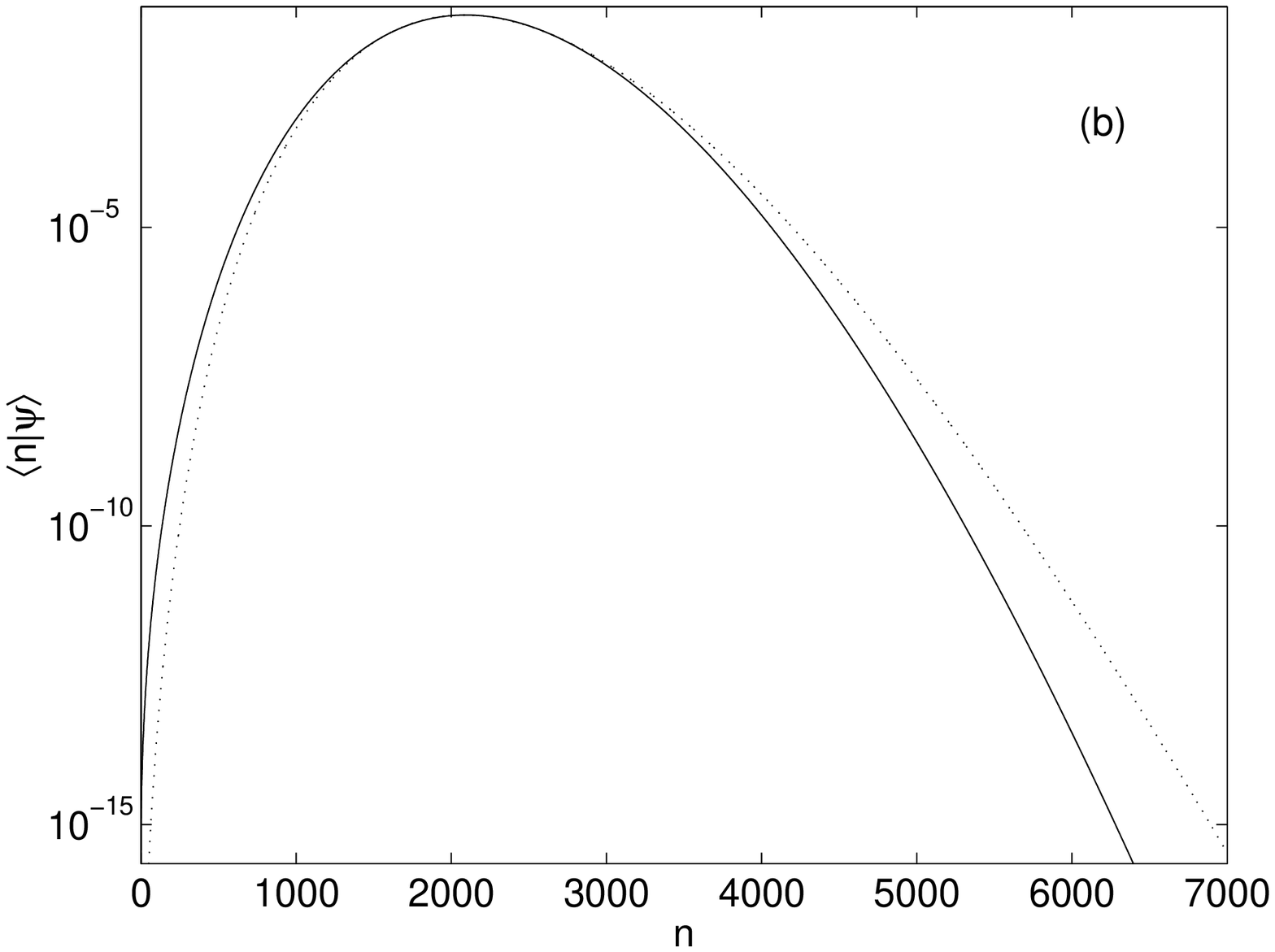}
\caption{The number state coefficients for optimum states for heterodyne
measurements and a mean photon number of about $2116$. The continuous lines are
for the optimised general state with fixed mean photon number, and the dotted lines
are for the optimised squeezed state. Plot (a) is a linear plot, and (b) is a semi-log
plot.}
\label{states}
\end{figure}

In order to compare the squeezed states with the states obtained by general
optimisation in a more direct way, the number state coefficients for the two
cases are plotted in Fig.~\ref{states}(a). This is for the example of heterodyne
measurements, and similar results are obtained for the adaptive measurement
schemes. The two states are extremely close, indicating that the optimum general
states may be converging to squeezed states.

On the other hand, if the state coefficients are plotted logarithmically, there
are large differences between the states [see Fig.~\ref{states}(b)]. Although
the states are fairly close near the peak, the tails of the two states have
different scalings. These differences persist even for very large mean photon
numbers.

\begin{figure}
\centering
\includegraphics[width=0.7\textwidth]{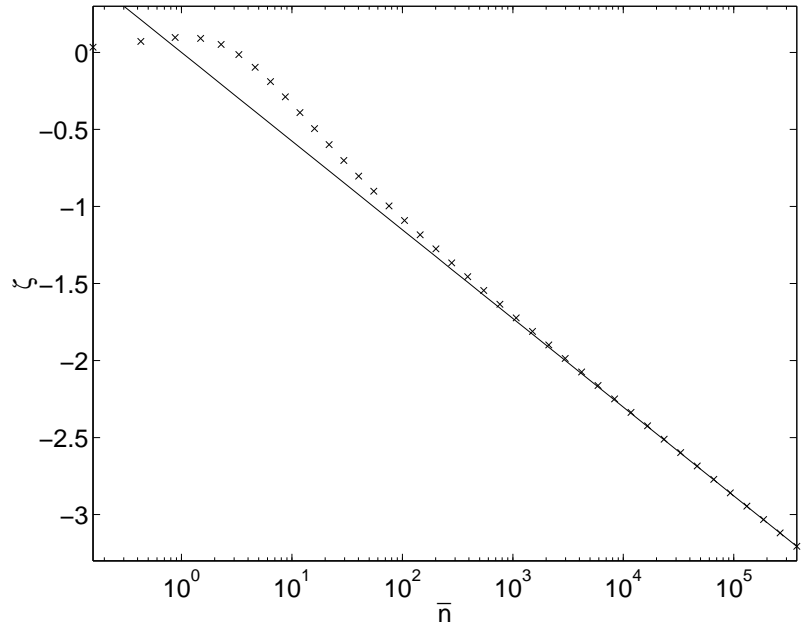}
\caption{The optimum values of $\zeta$ for heterodyne phase measurements on
squeezed states. The numerically determined values are shown as crosses, and the
analytic expression is shown as the continuous line.}
\label{hetzeta}
\end{figure}
\begin{figure}
\centering
\includegraphics[width=0.7\textwidth]{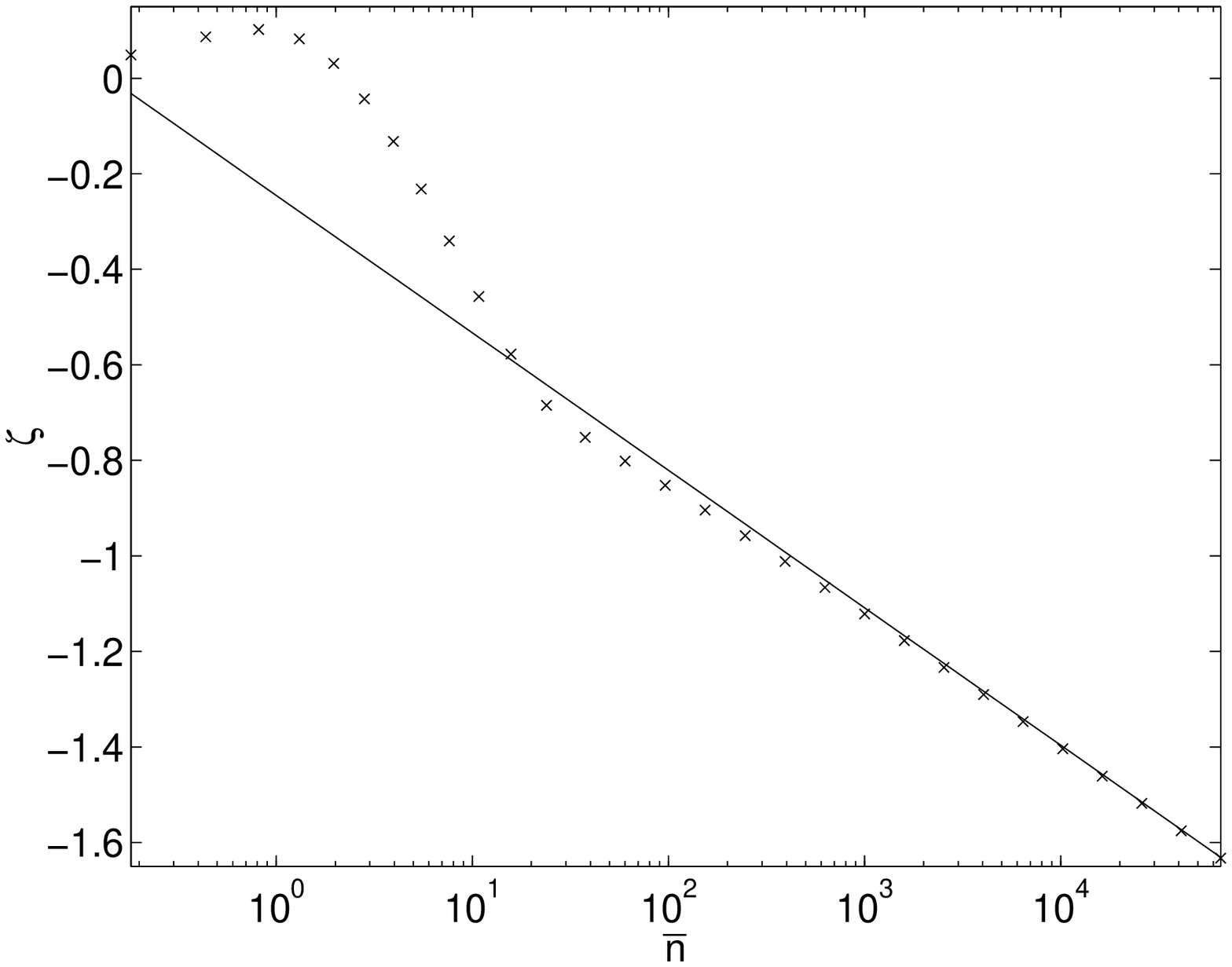}
\caption{The optimum values of $\zeta$ for mark I phase measurements on
squeezed states. The numerically determined values are shown as crosses, and the
analytic expression is shown as the continuous line.}
\label{mIzeta}
\end{figure}
\begin{figure}
\centering
\includegraphics[width=0.7\textwidth]{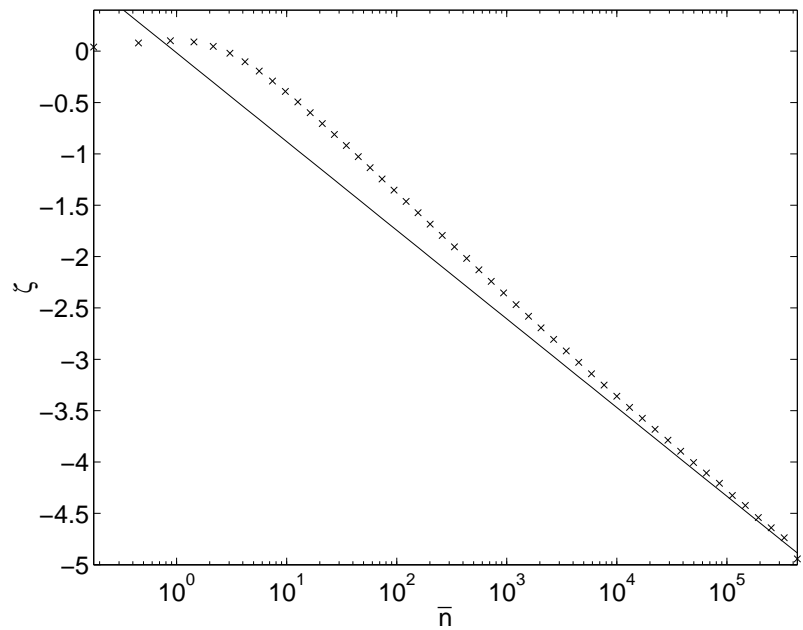}
\caption{The optimum values of $\zeta$ for mark II phase measurements on
squeezed states. The numerically determined values are shown as crosses, and the
analytic expression is shown as the continuous line.}
\label{mIIzeta}
\end{figure}

Another test of the theory is the values of $\zeta$ for the optimum squeezed
states. Using Eq.~(\ref{minn0}) gives the optimum value of $\zeta$ as
\beq
\zeta = -\frac p4 \log \nb + \half \log \left[2\sqrt{cp(p+1)}\right].
\eeq
Using this the optimal values of $\zeta$ for heterodyne, mark I and mark II
measurements should be
\bqa
\zeta_{\rm het}\!\!\!\! &=& \!\!\!\! -\frac 14 \log \nb \\
\zeta_{\rm I}  \!\!\!\! &=& \!\!\!\! -\frac 18 \log \nb
+ \frac 14 \log \frac 38 \\
\zeta_{\rm II} \!\!\!\! &=& \!\!\!\! -\frac 38 \log \nb
+ \frac 14 \log \frac {15}{16}.
\eqa
The numerically found optimum values of $\zeta$ and these asymptotic expressions
are plotted in Figs~\ref{hetzeta}, \ref{mIzeta} and \ref{mIIzeta}. The values
of $\zeta$ converge to the asymptotic analytic expressions in all three cases.
The value of $\zeta$ converges fairly rapidly for heterodyne and mark I
measurements, with agreement within $0.05$ for photon numbers above around
$100$. The convergence is much weaker in the mark II case, and a photon number
around $10^5$ is required for this level of agreement.

%% file: dyne.tex
\setcounter{chapter}{2}
\chapter{Optimum Adaptive Dyne Measurements}
\label{optdyne}
Now I will consider the second area for improvement of dyne measurements: the
measurement technique (as opposed to the input state). This includes both the
local oscillator phase used and the final phase estimate. Recall that for a
state with a mean photon number of $\nb$, heterodyne measurements introduce a
phase variance of about $\frac 14 \nb^{-1}$. This is much larger than the
minimum intrinsic phase variance, which scales as $1.89\times \nb^{-2}$. The
mark II adaptive measurements introduced in Ref.~\cite{semiclass} give a great
improvement on this, introducing a phase variance scaling as
$\frac 18 \nb^{-1.5}$.

This scaling is still far worse than the scaling for minimum intrinsic phase
variance. In \cite{fullquan} it is shown that the minimum introduced phase
variance is the same as for squeezed states optimised for minimum intrinsic
phase variance. As discussed in Sec.~\ref{sqzcan}, this scaling is
$\log \nb /(4\nb^2)$. Recall that this is not quite as good as that for general
optimised states, but it requires very large photon numbers for the difference
to be significant.

In this chapter I describe a feedback scheme that improves on the mark
II adaptive scheme, and is in fact very close to this theoretical limit. It is
therefore very close to being the best possible phase measurement scheme. It is
also possible, under some circumstances, to surpass the theoretical limit. This
is discussed in Sec.~\ref{surpass}.

\section{The Theoretical Limit}
\label{theory}
In order to understand how to attain the theoretical limit, we must first
understand the reason for the theoretical limit. It can be shown
\cite{wise96} that the probability to obtain the results $A,B$ from
an arbitrary (adaptive or non-adaptive) measurement is
\beq
P(A,B)d^{2}A\,d^{2}B = {\rm Tr}[\rho F(A,B)]d^{2}A\,d^{2}B,
\eeq
where $\rho$ is the state of the mode being measured. Here $F(A,B)$ is the POM
(probability operator measure) for the measurement, and is given by
\begin{equation}
F(A,B)=Q(A,B)\ket{\tilde\psi(A,B)}\bra{\tilde\psi(A,B)},
\end{equation}
where $Q(A,B)$ is what the probability distribution $P(A,B)$ would be if
$\rho$ were the vacuum state $\ket{0}\bra{0}$, and
$\ket{\tilde\psi(A,B)}$ is an unnormalised ket defined by
\begin{equation}
\ket{\tilde\psi(A,B)}=\exp(\smallfrac{1}{2} B (a^\dagger)^2-A a^\dagger)\ket{0}.
\end{equation}
This is proportional to a squeezed state \cite{WalMil94}:
\begin{equation}
\exp(\smallfrac{1}{2} B (a^\dagger)^2-A a^\dagger)\ket{0}=\left( 1-|B|^2
\right)^{-\frac 14} \exp\left(\frac {A{{\alpha^{\rm P}}}^*
}{2}\right)\ket{\alpha^{\rm P},\zeta^{\rm P}},
\end{equation}
where
\begin{equation}
\ket{\alpha^{\rm P},\zeta^{\rm P}}=\exp(\alpha^{\rm P} a^\dagger-{\alpha^{\rm P}}^*a)
\exp(\smallfrac 12 {\zeta^{\rm P}}^* a^2-\smallfrac 12 \zeta^{\rm P} (a^\dagger)^2)\ket{0},
\end{equation}
and the squeezing parameters are
\begin{eqnarray}
\label{defineAB1}
\alpha^{\rm P} \!\!\!\! &=& \!\!\!\! \frac{A+BA^*}{1-|B|^2},\\
\label{defineAB2}
\zeta^{\rm P} \!\!\!\! &=& \!\!\!\! -\frac{B \atanh |B|}{|B|}.
\end{eqnarray}
Here I am using the superscript P on the squeezing parameters to indicate that
they are the parameters of the squeezed state in the POM, and not of a physical
state. The squeezing parameters for the input state will be given as $\alpha$
and $\zeta$ with no superscripts. Note that this differs from the notation in
\cite{unpub}, where a superscript $S$ indicates the squeezing parameters for the
input state, and no superscript indicates the squeezing parameters for the
squeezed state in the POM.

In terms of these the POM is given by
\begin{equation}
F(A,B)=Q'(A,B)\ket{\alpha^{\rm P},\zeta^{\rm P}}\bra{\alpha^{\rm P},\zeta^{\rm P}},
\end{equation}
where
\begin{equation}
Q'(A,B)=Q(A,B)\left( 1-|B|^2 \right)^{-\frac 12} \exp\left[ {\rm Re}\left(A
{\alpha^{\rm P}}^*\right)\right].
\end{equation}
If the system state is pure, $\rho = \ket{\psi}\bra{\psi}$ and
the probability distribution is given by
\begin{equation}
\label{POMprob}
P(A,B)=Q'(A,B)\st{\braket{\psi}{{\alpha^{\rm P}},\zeta^{\rm P}}}^2.
\end{equation}
The greatest overlap between the states, and therefore the highest probability,
will be when $\ket{\alpha^{\rm P},\zeta^{\rm P}}$ has the same phase as the input state.
As all the information about the system from the measurement record is
contained in the variables $A$ and $B$, the most probable phase based on the
measurement record is
\bqa
\phi \!\!\!\! &=& \!\!\!\! \arg(\alpha^{\rm P}) \nn \\
\!\!\!\! &=& \!\!\!\! \arg(A+BA^*).
\eqa
If additional information is known about the system, it is possible to obtain a
better phase estimate. This is a far more complicated case, and will be
considered in Sec.~\ref{surpass}.

For an unbiased measurement scheme the probability distribution for this phase
estimate resulting from Eq.~(\ref{POMprob}) depends entirely on the inner
product between the two states, and not on $Q'(A,B)$. To see this, note firstly
that if the measurement is unbiased, the vacuum probability distribution
$Q(A,B)$ will be independent of the phase. Secondly, recall that for the
squeezed state $\ket{\alpha^{\rm P},\zeta^{\rm P}}$, if we rotate the phase of $\alpha^{\rm P}$
by some angle $\theta$, we can obtain an equivalent phase-shifted state by
rotating the phase of $\zeta^{\rm P}$ by $2\theta$. This means that
$\zeta^{\rm P} ({\alpha^{\rm P}}^*)^2$ is independent of the phase. From
Eq.~(\ref{defineAB2}), this means that $B ({\alpha^{\rm P}}^*)^2$ is independent of
the phase. Since
\bqa
A{\alpha^{\rm P}}^* \!\!\!\! &=& \!\!\!\! (\alpha^{\rm P} - B {\alpha^{\rm P}}^*) {\alpha^{\rm P}}^*
\nn \\ \!\!\!\! &=& \!\!\!\! \st{\alpha^{\rm P}}^2 - B ({\alpha^{\rm P}}^*)^2,
\eqa
$A{\alpha^{\rm P}}^*$ and therefore $Q'(A,B)$ are independent of the phase.

Since the probability distribution for the phase estimate depends on the inner
product between the two states, the variance in the measured phase will
approximately be the sum of the intrinsic phase variance and the phase variance
of the squeezed state $\ket{\alpha^{\rm P},\zeta^{\rm P}}$. The maximum overlap between the
states will be when the squeezed state has about the same photon number as the
input state. This means that the theoretical limit to the phase variance that
is introduced by the measurement is the phase variance of the squeezed state
that has the same photon number as the input state and has been optimised for
minimum intrinsic phase variance. Since the phase variance of a squeezed state
optimised for minimum intrinsic phase variance is approximately
$\log\bar n/(4\bar n^2)$ in the limit of large $\bar n$ \cite{collett}, this is
also the limit to the introduced phase variance.

The photon number of the squeezed state at maximum overlap will be mainly
determined by the photon number of the input, but the degree and direction of
squeezing (parametrised by $\zeta^{\rm P}$) will be determined by the multiplying
factor $Q'(A,B)$. The multiplying factor can be expressed as a function of
$\nb^{\rm P}$ and $\xi^{\rm P}$, for which the same symbol $Q'$ will be used, even though it
is a new function $Q'(\nb^{\rm P},\xi^{\rm P})$. Here  $\nb^{\rm P}$ is the mean photon number of
the state $\ket{\alpha^{\rm P},\zeta^{\rm P}}$ (and will be close to the photon number of
the input state), and $\xi^{\rm P} = \zeta^{\rm P} {\alpha^{\rm P}}^* /\alpha^{\rm P}$ is $\zeta^{\rm P}$ with
the phase of $\alpha^{\rm P}$ scaled out. In practice the multiplying
factor tends to be concentrated along a particular line (for example, see
Fig.~\ref{zeta}), effectively giving $\xi^{\rm P}$ as a function of $\nb^{\rm P}$. In order
to obtain the theoretical limit, the measurement scheme must give a multiplying
factor $Q'(\nb^{\rm P},\xi^{\rm P})$ that gives values of $\xi^{\rm P}$ for each $\nb^{\rm P}$
that are the same as for optimised squeezed states.

We can determine the approximate variation of $\xi^{\rm P}$ with $\nb^{\rm P}$ in the
multiplying factor if we can estimate how it varies for measurements on a
coherent state. If the intermediate phase estimates used in the adaptive scheme
are unbiased, it is easy to see that the maximum probability
will be for $B$ real and therefore also $A$ real. These results imply that
\bqa
\alpha^{\rm P} \!\!\!\! & \approx & \!\!\!\! \frac{A(1+B)}{1-B^2} \nn \\
\!\!\!\! &=& \!\!\!\! \frac{A}{1-B}.
\eqa
This means that $B$ should be
\beq
B \approx 1-\frac A{\alpha^{\rm P}} ,
\eeq
so $\zeta^{\rm P}$ should be
\beq
\zeta^{\rm P} \approx - \atanh (1-\frac A{\alpha^{\rm P}}).
\eeq
Note that when $\alpha^{\rm P}$ is real, $\zeta^{\rm P}$ and $\xi^{\rm P}$ are equivalent.
Provided the photon number is large, $B$ should be close to 1 so we can use
the asymptotic approximation of the atanh function, giving
\beq
\zeta^{\rm P} \approx \frac 1 2 \log \frac A{2\alpha^{\rm P}}.
\eeq

The mean photon number for squeezed states is given by
\beq
\nb^{\rm P} = \st{\alpha^{\rm P}}^2 + \sinh^2 \st{\zeta^{\rm P}}.
\eeq
For the states that are considered in this study, $\sinh^2 \st{\zeta^{\rm P}}$ is much
smaller than $\nb^{\rm P}$ for large photon numbers, so $\alpha^{\rm P} \approx
\sqrt{\nb^{\rm P}}$. Using this approximation gives $\zeta^{\rm P}$ as
\beq
\label{defzetaAn}
\zeta^{\rm P} \approx \frac 1 2 \log \frac{A}{2\sqrt{\nb^{\rm P}}}.
\eeq
Since the magnitude of $\zeta^{\rm P}$ is governed by the multiplying factor
$Q'(\nb^{\rm P},\xi^{\rm P})$, this result for $\zeta^{\rm P}$ should hold for more general input
states.

From Sec.~\ref{sqzcan} and Ref.~\cite{collett} the intrinsic phase variance of a
squeezed state is given approximately by
\begin{equation}
\label{pvss}
V(\phi)\approx
\frac{n_0+1}{4\nb^2}+2\erfc (\sqrt{2n_0}),
\end{equation}
where $n_0=\nb e^{2\zeta}$ and $\zeta$ is real. This is minimised for
\begin{equation}
\label{minimn0}
n_0 \approx \log(4\nb)-\frac 14 \log(2\pi).
\end{equation}
Using the result obtained for $\zeta^{\rm P}$ in Eq.~(\ref{defzetaAn}) gives
\beq
e^{2\zeta^{\rm P}} \approx \frac{|A|}{2\sqrt{\nb^{\rm P}}},
\eeq
so
\beq
\label{n0result}
n_0^{\rm P} \approx \frac 1 2 |A| \sqrt{\nb^{\rm P}}.
\eeq
As the optimum value of $n_0^{\rm P}$ is given by Eq.~(\ref{minimn0}), in order for
the measurement to be optimal, $|A|$ should scale with $\nb^{\rm P}$ as
\begin{equation}
|A| \propto \frac{\log \nb^{\rm P}}{\sqrt{\nb^{\rm P}}}.
\end{equation}
Since $\nb^{\rm P} \approx \nb$, we should get the same scaling with the input photon
number. For the case of mark II measurements there is the result that $|A|=1$
\cite{semiclass}, which is why these measurements are not optimal. Note that if
we substitute $|A|=1$ into the expression (\ref{n0result}) to find $n_0^{\rm P}$, and
substitute that into Eq.~(\ref{pvss}), we obtain the correct result for the mark
II introduced phase variance,
\begin{equation}
\Delta V(\phi) \approx \smallfrac 18 {{\nb}}^{-1.5}.
\end{equation}

\section{Improved Feedback}
\label{impfeed}
Now we have the result that for optimal feedback $|A|$ should decrease with
photon number. Therefore, in order to improve the phase measurement scheme, we
want one that gives $|A|<1$. To see in general how this can be achieved,
consider a coherent state with amplitude $\alpha$. The Ito SDE for $|A_v|^2$
will be
\begin{eqnarray}
d|A_v|^2\!\!\!\! &=& \!\!\!\!A_v^*(dA_v)+(dA_v^*)A_v+(dA_v^*)(dA_v) \nn \\
\!\!\!\! &=& \!\!\!\!A_v^* e^{i\Phi(v)}I(v)dv + e^{-i\Phi(v)}I(v)dv A_v+dv \nn\\
\!\!\!\! &=& \!\!\!\!\left[ |A_v| I(v) \left( e^{i\Phi(v)} e^{-i\varphi_v^A}
+ e^{-i\Phi(v)} e^{i\varphi_v^A} \right) +1\right] dv,
\end{eqnarray}
where $\varphi_v^A=\arg A_v$. Usually the local oscillator phase
will be based on a phase estimate $\hat\varphi_v$, so that $\Phi(v)=
\hat\varphi_v+\pi/2$. In terms of this phase estimate the differential
equation becomes
\bqa
d|A_v|^2 \!\!\!\! &=& \!\!\!\!\left[ |A_v| I(v) \left( i e^{i\hat\varphi_v}
e^{-i\varphi_v^A} -i e^{-i\hat\varphi_v} e^{i\varphi_v^A} \right) + 1 \right]
dv \nn \\
\!\!\!\! &=& \!\!\!\!\left[1+ 2|A_v| I(v) \sin(\varphi_v^A-\hat\varphi_v)
\right] dv.
\eqa
Taking the expectation value of $I(v)$ and simplifying gives
\bqa
\ip{I(v)}\!\!\!\! &=& \!\!\!\!\left. \ip{2{\rm Re}\left(\alpha e^{-i\Phi(v)}
\right)dv+dW(v)}\right/ dv \nn \\
\!\!\!\! &=& \!\!\!\!-2|\alpha|\sin(\hat\varphi_v-\varphi),
\eqa
where $\varphi=\arg \alpha$. Using this result, the expectation value for the
increment in $|A_v|^2$ is
\begin{equation}
\label{sines}
\ip{d|A_v|^2}=\left[1- 4|A_v||\alpha|\sin(\hat\varphi_v-\varphi)
\sin(\varphi_v^A-\hat\varphi_v) \right] dv.
\end{equation}
Note that here the average is only over the single increment $dW(v)$, and not
over the different trajectories.

The first term on its own will give $|A|=1$, and in order to get $|A|<1$ the two
sines must have the same sign. This will be the case if the phase estimate is
between the actual phase and the phase of $A_v$. Therefore we would like to use
the phase estimate
\beq
\hat \varphi(v) = \arg(\alpha^{1-\varepsilon(v)} A_v^{\varepsilon(v)}),
\eeq
where $0<\varepsilon<1$. The problem with this phase estimate is that it uses
the actual value of the phase. In order to avoid this problem, the best estimate
of the phase can be used in place of the actual phase. Therefore the phase
estimate that will be considered is
\begin{equation}
\label{phasestimate}
\hat \varphi (v)=\arg(C_v^{1-\varepsilon(v)}A_v^{\varepsilon(v)}),
\end{equation}
where
\beq
C_v = A_v v+B_v A_v^*.
\eeq

It would at first appear that the best value of $\varepsilon$ to use is $1/2$,
as this will make the phase estimate exactly halfway between the best phase
estimate and $\arg A_v$, and this was the initial improvement on mark II
feedback that was tried. This is too simplistic, however, as will be shown.
In order to estimate what the best values of $\varepsilon$ to use are, we can
take the actual phase to be zero, and use the phase estimate
\beq
\hat \varphi (v) = \varepsilon \arg A_v .
\eeq
For simplicity $\varepsilon$ will be taken to be constant. For this value of the
phase estimate, $\varphi_v^A = \hat\varphi_v /\varepsilon$. Substituting this
into Eq.~(\ref{sines}) gives
\begin{equation}
\ip{d|A_v|^2}=\left[1- 4|A_v|\alpha\sin(\hat \varphi_v)
\sin[\hat\varphi_v (1/\varepsilon -1)] \right] dv.
\end{equation}
The absolute value symbols on $\alpha$ have been omitted because the phase has
been taken to be zero. When the phase estimate is close to zero we can use the
linear approximation of the sine function to get
\beq
\ip{d|A_v|^2} = \left[ {1 - 4|A_v|\alpha \hat\varphi_v^2 \left(
1/\varepsilon - 1 \right)} \right] dv.
\eeq

A differential equation for $\hat\varphi_v$ can be obtained in a similar way:
\bqa
  d\hat \varphi _v  \!\!\!\! &=& \!\!\!\! \varepsilon {\rm Im} \left[ {d\ln A_v}
 \right] \nn \\ 
   \!\!\!\! &=& \!\!\!\! \varepsilon {\rm Im} \left[ {\frac{{dA_v }}
{{A_v }} - \frac{{\left( {dA_v } \right)^2 }}
{{2A_v^2 }}} \right] \nn \\ 
   \!\!\!\! &=& \!\!\!\! \varepsilon {\rm Im} \left[ {\frac{{ie^{i\hat
 \varphi_v } I(v)dv}}
{{|A_v|e^{i{{\hat \varphi _v } /\varepsilon }} }} - \frac{{
\left( {ie^{i\hat \varphi _v } I(v)dv} \right)^2 }}
{{2\left( {|A_v|e^{i{{\hat \varphi _v } /\varepsilon }} }
 \right)^2 }}} \right].
\eqa
Note that here $A_v = |A_v|e^{i \hat \varphi _v /\varepsilon}$ has been used.
Continuing the derivation we find
\bqa
  d\hat \varphi _v \!\!\!\! &=& \!\!\!\! \varepsilon {\rm Im} \left[
 {\frac{{ie^{i\hat \varphi _v \left( {1 - 1/\varepsilon } \right)} I\left( v
 \right)dv}}
{{\left| {A_v } \right|}} + \frac{{e^{2i\hat \varphi _v \left( {1 -
 1/\varepsilon } \right)} \left( {I\left( v \right)dv} \right)^2 }}
{{2\left| {A_v } \right|^2 }}} \right] \nn \\ 
   \!\!\!\! &=& \!\!\!\! \frac{\varepsilon }
{{\left| {A_v } \right|}}\left[ {\cos \left[ {\hat \varphi _v \left( {1 -
 1/\varepsilon } \right)} \right]I\left( v \right)dv + \frac{{\sin \left[
 {2\hat \varphi _v \left( {1 - 1/\varepsilon } \right)} \right]dv}}
{{2\left| {A_v } \right|}}} \right] \nn \\ 
   \!\!\!\! &=& \!\!\!\! \frac{\varepsilon }
{{\left| {A_v } \right|}}\left[ {\cos \left[ {\hat \varphi _v \left( {1 -
 1/\varepsilon } \right)} \right]\left[ { - 2\alpha \sin \hat \varphi _v dv +
 dW\left( v \right)} \right] + \frac{{\sin \left[ {2\hat \varphi _v \left( {1 -
 1/\varepsilon } \right)} \right]dv}}
{{2\left| {A_v } \right|}}} \right] \nn \\ 
   \!\!\!\! &=& \!\!\!\! \frac{\varepsilon }
{{\left| {A_v } \right|}}\left[ { - 2\alpha \sin \hat \varphi _v \cos \left[
 {\hat \varphi _v \left( {1 - 1/\varepsilon } \right)} \right]dv + \frac{{\sin
 \left[ {2\hat \varphi _v \left( {1 - 1/\varepsilon } \right)} \right]dv}}
{{2\left| {A_v } \right|}}} 
+ \cos \left[ {\hat \varphi _v \left( {1 - 1/\varepsilon } \right)}
 \right]dW(v) \right] \nn \\ 
   \!\!\!\! &=& \!\!\!\! \frac{\varepsilon }
{{\left| {A_v } \right|}}\left[ { - 2\alpha \hat \varphi _v dv + \frac{{\hat
 \varphi _v \left( {1 - 1/\varepsilon } \right)dv}}
{{\left| {A_v } \right|}} + dW(v)} \right].
\eqa
In the last line the linear approximation for sine and a constant approximation
for cos have been used. Evaluating the increment in the square of the phase
gives
\bqa
\ip{d\hat \varphi_v^2} \!\!\!\! &=& \!\!\!\! \ip{2\hat \varphi _v d\hat
\varphi_v  + \left( d\hat \varphi _v \right)^2} \nn \\
   \!\!\!\! &=& \!\!\!\! \frac{{2\varepsilon }}
{{|A_v|}}\left[ { - 2\alpha \hat \varphi _v^2 dv +
 \frac{{\hat \varphi _v^2 \left( {1 - 1/\varepsilon } \right)dv}}
{{|A_v|}}} \right] + \frac{{\varepsilon ^2 }}
{{|A_v|^2 }}dv \nn \\
 \!\!\!\! &=& \!\!\!\! \frac{\varepsilon }
{{|A_v|}}\left[ { - 4\alpha \hat \varphi _v^2  + \frac{2{\hat
\varphi _v^2 \left( {1 - 1/\varepsilon } \right)}}
{{|A_v|}} + \frac{\varepsilon }
{{|A_v|}}} \right]dv.
\eqa
Thus the expectation values for the increments in $|A_v|^2$ and
$\hat\varphi_v^2$ are:
\bqa
\label{absa}
\ip{d|A_v|^2}  \!\!\!\! &=& \!\!\!\! \left[ {1 - 4\alpha
 |A_v|\hat \varphi _v^2 \left( {1 
/ \varepsilon - 1} \right)} \right]dv, \\
\label{phi2}
\ip{d\hat \varphi _v^2} \!\!\!\! &=& \!\!\!\! \frac{\varepsilon }
{{|A_v|}}\left[ { - 4\alpha \hat \varphi _v^2  + \frac{2{\hat
 \varphi _v^2 \left( {1 - 1/\varepsilon } \right)}}
{{|A_v|}} + \frac{\varepsilon }
{{|A_v|}}} \right]dv.
\eqa

Now an approximate solution for $|A_v|^2$ and $\hat\varphi_v^2$ can be obtained
by solving these as differential equations, ignoring the fact that these are
only the expectation values for the increments. The approximate solution for
large photon number is
\bqa
\st{A_v}^2 \!\!\!\! &=& \!\!\!\! \varepsilon v + O\left(\alpha^{-1} \right), \\
\label{phi2sol}
\hat \varphi _v^2 \!\!\!\! &=& \!\!\!\! \frac 1{{4\alpha }}\sqrt{\frac
{\varepsilon}v} + O\left(\alpha^{-2} \right).
\eqa
To see that these are solutions, we can just substitute them into the above
equations. For Eq.~(\ref{absa}) we have
\bqa
{\rm l.h.s.} \!\!\!\! &=& \!\!\!\! d|A_v|^2 \nn \\ 
\!\!\!\! &=& \!\!\!\! \left[ {\varepsilon  + O\left( {\alpha ^{-1}}\right)}
\right]dv \\ 
{\rm r.h.s.} \!\!\!\! &=& \!\!\!\! \left[ 1-4\alpha \left( \sqrt{\varepsilon v}
+ O\left( \alpha^{-1} \right)\right)\left( \frac{1}{4\alpha}\sqrt{\frac
{\varepsilon}{v}} + O\left( \alpha ^{-2} \right)\right)\left(1/\varepsilon - 1
\right) \right]dv \nn \\
\!\!\!\! &=& \!\!\!\! \left[1 - \left( \sqrt{\varepsilon v} + O\left(
\alpha ^{-1} \right)\right)\left( \sqrt {\frac{\varepsilon}{v}} + O\left(
\alpha ^{-1} \right)\right)\left( 1/\varepsilon - 1 \right) \right]dv \nn \\
\!\!\!\! &=& \!\!\!\! \left[1-\left( \varepsilon + O\left(\alpha^{-1}\right)
\right)\left(1/\varepsilon-1\right)\right]dv \nn \\
\!\!\!\! &=& \!\!\!\! \left[1-\left(1-\varepsilon+O\left(\alpha^{-1}\right)
\right)\right]dv \nn \\
\!\!\!\! &=& \!\!\!\! \left[\varepsilon+O\left(\alpha^{-1}\right)\right]dv.
\eqa
For Eq.~(\ref{phi2}) we find
\bqa
{\rm l.h.s.} \!\!\!\! &=& \!\!\!\! O\left(\alpha^{-1}\right)dv \\
{\rm r.h.s.} \!\!\!\! &=& \!\!\!\! \frac{\varepsilon}{\left(\sqrt{\varepsilon v}
+O\left(\alpha^{-1}\right)\right)}\left[-4\alpha\left(\frac{1}{4\alpha}
\sqrt{\frac{\varepsilon}{v}}+O\left(\alpha^{-2}\right)\right)\right. \nn \\
&&  \left. +\frac{2\left(\frac{1}{4\alpha}\sqrt{\frac{\varepsilon}{v}}+O\left(
\alpha^{-2}\right)\right)\left(1-1/\varepsilon\right)}{\left(\sqrt{\varepsilon
v}+O\left(\alpha^{-1}\right)\right)}+\frac{\varepsilon}{\left(\sqrt{\varepsilon
v}+O\left(\alpha^{-1}\right)\right)}\right]dv \nn \\
\!\!\!\! &=& \!\!\!\! \left(\sqrt{\frac{\varepsilon}{v}}+O\left(\alpha^{-1}
\right)\right)\left[-\left(\sqrt{\frac{\varepsilon}{v}}+O\left(\alpha^{-1}
\right)\right)+O\left(\alpha^{-1}\right)+\left(\sqrt{\frac{\varepsilon}{v}}+
O\left(\alpha^{-1}\right)\right)\right]dv \nn \\
\!\!\!\! &=& \!\!\!\! \left(\sqrt{\frac{\varepsilon}{v}}+O\left(\alpha^{-1}
\right)\right)O\left(\alpha^{-1}\right)dv \nn \\
\!\!\!\! &=& \!\!\!\! O\left(\alpha^{-1}\right)dv.
\eqa
For this equation it is enough that the sides agree to order $\alpha^{-1}$. To
obtain more specific agreement the higher order terms in the solution would be
required.

As a simple check on these results, note that the case $\varepsilon=1$ is
just the standard case and the final phase estimate corresponds to the mark I
phase estimate.  In this case the solutions give to first order
\bqa
\st{A_1}^2 \!\!\!\! &=& \!\!\!\! 1, \\
\hat \varphi _1^2 \!\!\!\! &=& \!\!\!\! \frac 1{4\alpha}.
\eqa
These are exactly the same results as obtained in Ref.~\cite{semiclass}.

The important result is that for the final value of $A_v$, at $v=1$, we have
$|A| \approx \sqrt \varepsilon$. This indicates that we can obtain smaller and
smaller values of $|A|$ simply by using smaller values of $\varepsilon$, and the
minimum is {\it not} for $\varepsilon = 1/2$. To see the reason for this, note
that
\beq
(\varphi_v^A)^2 \approx \frac 1{4\alpha\sqrt{\varepsilon^3 v}}.
\eeq
This means that for smaller values of $\varepsilon$, the value of $\varphi_v^A$
will tend to be greater. This means that the second sine in (\ref{sines}) will
be greater on average, resulting in a smaller value of $|A|$.

For other values of $\varepsilon$ these results have been verified by
numerically performing the stochastic integrals for coherent states. The
numerical technique is very straightforward for coherent states. We simply
replace the infinitesimal interval $dv$ with a finite interval $\delta v$, and
replace the Wiener increment $dW$ with a finite stochastic increment $\delta W$
that has a Gaussian distribution with zero mean and variance $\delta v$. For
these calculations $100 \alpha$ time steps and $2^{10}$ samples were used.

\begin{figure}
\centering
\includegraphics[width=0.7\textwidth]{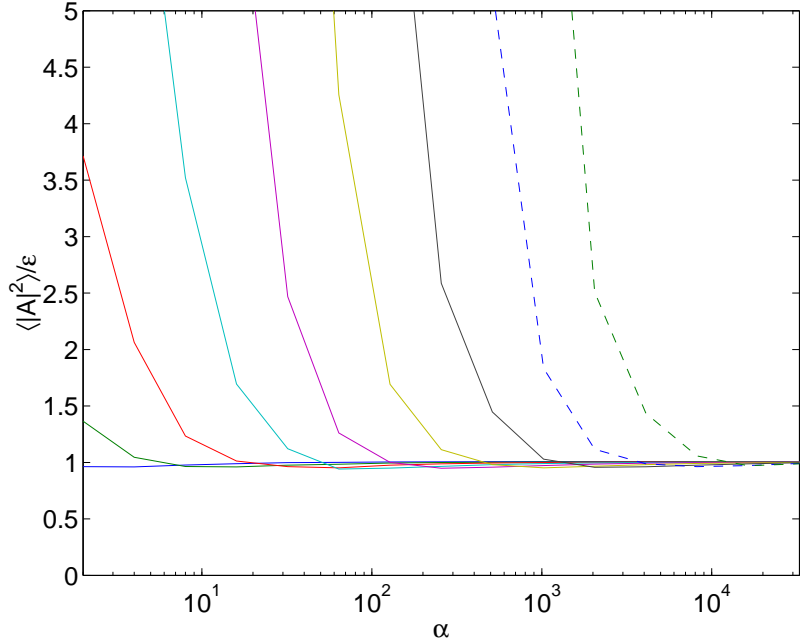}
\caption{The mean values of $|A|^2$ divided by $\varepsilon$ for measurements on
coherent states with $\varepsilon \arg A_v$ feedback. The results for
$\varepsilon = 1/2$ are shown in dark blue, for $\varepsilon = 1/4$ in green,
for $\varepsilon = 1/8$ in red, for $\varepsilon = 1/16$ in light blue, for
$\varepsilon = 1/32$ in purple, for $\varepsilon = 1/64$ in yellow, for
$\varepsilon = 1/128$ in black, for $\varepsilon = 1/256$ as a dashed dark
blue line, and for $\varepsilon = 1/512$ as a dashed green line.}
\label{absA2}
\end{figure}

The results for $|A|^2$ are shown in Fig.~\ref{absA2}. For simplicity, rather
than plotting $|A|^2$, I have shown $|A|^2/\varepsilon$. From the theory above
this should converge to 1. The results for moderate values of $\varepsilon$
agree very well with the theory; however, for smaller values of $\varepsilon$
there is poor agreement unless the photon number is very large. For
$\varepsilon = 1/64$, a value of $\alpha$ above about 512 is required for good
agreement. The photon number required for good agreement with theory
appears to scale roughly as $\varepsilon^{-3}$.

\begin{figure}
\centering
\includegraphics[width=0.7\textwidth]{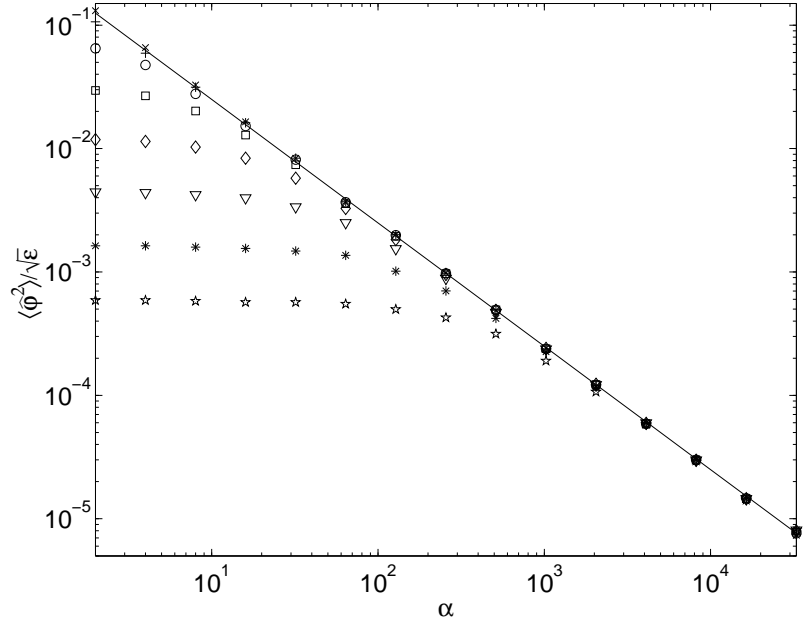}
\caption{The mean values of $\hat \varphi^2$ divided by $\sqrt{\varepsilon}$ for
measurements on coherent states with $\varepsilon \arg A_v$ feedback. The
results for $\varepsilon = 1/2$ are shown as crosses, for $\varepsilon = 1/4$ as
pluses, for $\varepsilon = 1/8$ as circles, for $\varepsilon = 1/16$ as
squares, for $\varepsilon = 1/32$ as diamonds, for $\varepsilon = 1/64$ as
triangles, for $\varepsilon = 1/128$ as asterisks, and for $\varepsilon = 1/256$
as stars. The analytic result of $1/(4\alpha)$ is shown as the continuous
line.}
\label{phi2oneps}
\end{figure}

In Fig.~\ref{phi2oneps} are shown the results for the final variance of the
phase estimate. For simplicity, $\hat \varphi^2/\sqrt{\varepsilon}$ is plotted.
From the theory above this should be approximately $1/(4\alpha)$. Similarly to
the case for $|A|^2$, the results for $\varepsilon = 1/2$ agree very well, but
those for smaller values of $\varepsilon$ require large photon numbers in order
to have good agreement.

\begin{figure}
\centering
\includegraphics[width=0.7\textwidth]{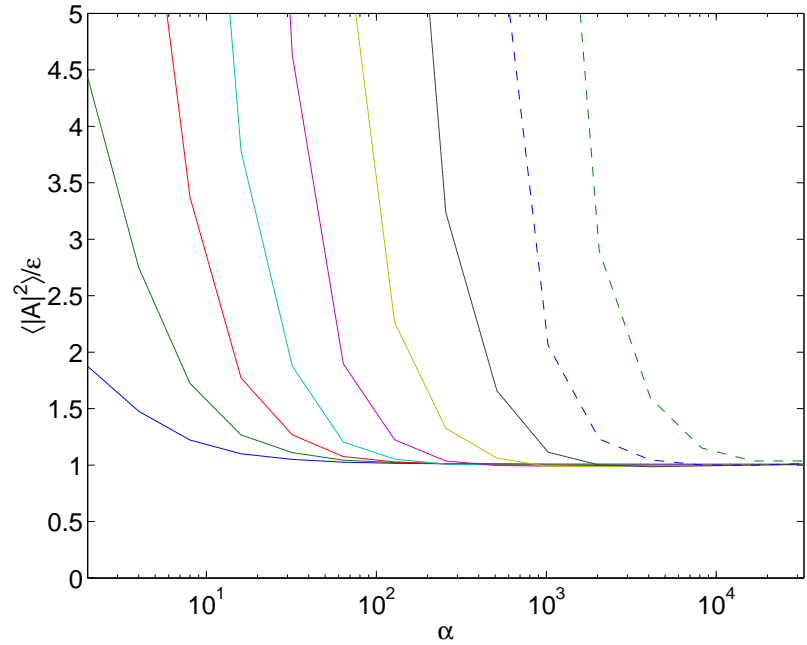}
\caption{The mean values of $|A|^2$ divided by $\varepsilon$ for measurements on
coherent states with $\arg C_v^{1-\varepsilon} A_v^\varepsilon$ feedback. The
results for $\varepsilon = 1/2$ are shown in dark blue, for $\varepsilon = 1/4$
in green, for $\varepsilon = 1/8$ in red, for $\varepsilon = 1/16$ in light
blue, for $\varepsilon = 1/32$ in purple, for $\varepsilon = 1/64$ in yellow,
for $\varepsilon = 1/128$ in black, for $\varepsilon = 1/256$ as a dashed dark
blue line, and for $\varepsilon = 1/512$ as a dashed green line.}
\label{absA2ex}
\end{figure}

The next question is whether this good agreement continues if the best phase
estimate is used in the feedback, rather than the actual phase, as in
Eq.~(\ref{phasestimate}). The calculations were repeated for this feedback, and
the results for $|A|^2$ are shown in Fig.~\ref{absA2ex}. The results are very
similar to the previous case, except that higher photon numbers than before are
required to obtain good agreement in this case. In this case the photon number
required for good agreement with theory appears to scale around
$\varepsilon^{-2}$.

These results indicate that $|A|$ can be reduced to any value required, provided
the photon number is sufficiently large. This implies that the results should be
close to optimum for the feedback given by Eq.~(\ref{phasestimate}), with
$\varepsilon$ given by
\beq
\varepsilon  = \left( 2 \alpha_{\rm opt} e^{2\zeta_{\rm opt}} \right)^2,
\eeq
where $\zeta_{\rm opt}$ is the value of $\zeta$ optimised for minimum intrinsic
phase variance for the photon number $\nb$. Here $\alpha$ has been used rather
than $\sqrt{\nb}$, as it should be closer to $\alpha^{\rm P}$. This is the first
feedback technique that I will discuss.

\section{Simulation Method}
\label{method}
The easiest input states to use for numerical simulations are coherent states,
as they remain coherent with a deterministically decaying amplitude. We simply
have the equations (after the time variation of the amplitude is scaled out)
\bqa
I(v) dv \!\!\!\! &=& \!\!\!\! 2 {\rm Re} (\alpha e^{-i\Phi(v)}) dv + dW(v)
\nn \\ dA_v \!\!\!\! &=& \!\!\!\! e^{i\Phi(v)} I(v) dv \nn \\
dB_v \!\!\!\! &=& \!\!\!\! -e^{2i\Phi(v)} dv,
\eqa
where $\Phi(v) = \hat \varphi_v + \pi/2$, and $\hat \varphi_v$ is obtained by
the feedback technique that we are using. These equations can be integrated
numerically by using a finite increment $\delta v$, and treating $dW(v)$ as a
finite random variable with a normal distribution and variance $\delta v$.

However, in order to estimate the phase variance that is introduced by the
measurement, coherent states would be very inefficient, as the phase variance
would be dominated by the intrinsic phase variance. It is almost as easy (and
much more efficient) to perform calculations on squeezed states, as squeezed
states remain squeezed states under the stochastic evolution, and only the two
squeezing parameters need be kept track of. The best squeezed states to use are
those optimised for minimum intrinsic phase variance. For these states the total
phase variance will be approximately twice the intrinsic phase variance
when the measurements are close to optimal.

In order to determine the stochastic evolution of the state, I will use a
technique similar to that used in Ref.~\cite{wise96}. The master equation for
detection of photons is
\beq
\dot \rho  =  - i\left[ {H,\rho } \right] + \left[ {a\rho a\dg  - \half
\left( {a\dg a\rho  + \rho a\dg a} \right)} \right]dt.
\eeq
For a consistent detection scheme the measurement operators must give this
master equation. In general the master equation is given by
\beq
\rho(t+dt)  = \sum_{n = 0}^N {\Omega _n \rho(t) \Omega _n\dg }.
\eeq
For simple detection the measurement operator is
\beq
\Omega _1  = \sqrt {dt} \; a.
\eeq
In order for this to be consistent with the master equation, the measurement
operator for no detection must be
\beq
\Omega _0  = 1 - \left( {iH + \half a\dg a} \right)dt.
\eeq
To consider measurements where the field is combined with a large amplitude
coherent state that is treated classically, we can use a unitary rearrangement
of the measurement operators. Using that given in \cite{wise96} we obtain
\bqa
\Omega _1 \!\!\!\! &=& \!\!\!\! \sqrt {dt} \left( {a + \gamma } \right) \nn \\
\Omega _0 \!\!\!\! &=& \!\!\!\! 1 - \left[ iH + \half (a\dg+\gamma^*)(a+\gamma)
+\half (\gamma^* a-\gamma a\dg) \right]dt.
\eqa
The operator for no detection can be simplified to
\beq
\Omega _0  = 1 - \left( {iH + \half a\dg a + \gamma ^* a + \half
|\gamma|^2 } \right)dt.
\eeq
When there is a detection the state changes to
\beq
\ket {\psi(t + dt)} = \frac{(a + \gamma)\ket{\psi(t)}}
{\sqrt{\lr{(a\dg+\gamma^*)(a+\gamma)}}}.
\eeq
Now the state after a detection can be multiplied by an arbitrary complex
constant with magnitude 1 without altering the properties of the state.
Therefore this state can be altered to
\beq
\ket {\psi (t + dt)} = \frac{(a e^{-i\Phi} + |\gamma|)\ket{\psi (t)}}
{\sqrt{\lr{(a\dg+\gamma^*)(a+\gamma)}}}.
\eeq
where $\Phi$ is the phase of the local oscillator. This means that the change
in the state is
\beq
\label{evdet}
d\ket {\psi (t)} = \left[ \frac{(a e^{-i\Phi} + |\gamma|)}
{\sqrt{\lr{(a\dg+\gamma^*)(a+\gamma)}}} -1 \right]\ket{\psi (t)}.
\eeq

When there is no detection the unnormalised state changes to
\beq
\ket{\tilde \psi(t + dt)}=\left[ {1 - \left( {iH + \half a\dg a + \gamma^* a +
\half |\gamma|^2 } \right)dt} \right]\ket{\psi(t)}.
\eeq
We find that
\bqa
\braket{\tilde \psi(t + dt)}{\tilde \psi(t + dt)}\!\!\!\! &=& \!\!\!\!
\left\langle \left[ {1 - \left( {-iH+\half a\dg a + \gamma a\dg + \half
|\gamma|^2 } \right)dt}\right]\right. \nn \\
&&\times \left. \left[ {1 - \left( {iH + \half a\dg a + \gamma^* a +
\half |\gamma|^2 } \right)dt} \right] \right \rangle \nn \\ 
\!\!\!\! &=& \!\!\!\! \ip {1 - \left( {a\dg a + \gamma ^* a + \gamma a\dg +
|\gamma|^2 } \right)dt} \nn \\
\!\!\!\! &=& \!\!\!\!1-\left( \lr{a\dg a}+\gamma ^* \lr{a} +
\gamma \lr{a\dg} + |\gamma|^2 \right)dt.
\eqa
Therefore, to obtain a normalised state, we must multiply by a factor of
\beq
\braket{\tilde \psi(t + dt)}{\tilde \psi(t + dt)}^{-1/2} = 1+\half \left(
\lr{a\dg a}+\gamma ^* \lr{a} + \gamma \lr{a\dg} + |\gamma|^2 \right)dt .
\eeq
This means that the normalised state changes to
\beq
\ket{\psi(t + dt)} = \left[ 1+\left( \frac{\lr{a\dg a}}2-\frac{a\dg a}2+
\frac{\lr{\gamma^*a+\gamma a\dg}}2 -\gamma^*a-iH \right)dt\right] \ket{\psi(t)},
\eeq
so the difference in the state is
\beq
\label{evnodet}
d\ket{\psi(t)} = dt \left( \frac{\lr{a\dg a}}2-\frac{a\dg a}2+
\frac{\lr{\gamma^*a+\gamma a\dg}}2 -\gamma^*a-iH \right) \ket{\psi(t)}.
\eeq

The probability of a detection occurring in time $dt$ is
\beq
P_1 = dt\ip{(a\dg+\gamma^*)(a+\gamma)}.
\eeq
In order to determine a stochastic differential equation for the state, we
define a stochastic increment $dN$ which takes the values 0 (for no detection)
and 1 (for detection) and has the expectation value:
\beq
\label{expectation}
E(dN) = dt \ip{(a\dg+\gamma^*)(a+\gamma)}.
\eeq
Using this increment, the SDE for the state taking into account both
alternatives is
\bqa
\label{evboth}
d\ket{\psi(t)} \!\!\!\! &=& \!\!\!\! \left[ dN \left(\frac{(a e^{-i\Phi} +
|\gamma|)} {\sqrt{\lr{(a\dg+\gamma^*)(a+\gamma)}}} -1 \right)\right. \nn \\
\!\!\!\!&&\!\!\!\! \left. +dt \left( \frac{\lr{a\dg a}}2-\frac{a\dg a}2+
\frac{\lr{\gamma^*a+\gamma a\dg}}2 -\gamma^*a-iH \right)\right] \ket{\psi(t)}.
\eqa
When $dN$ takes the value 0, we simply have the increment for no detection
(\ref{evnodet}). When $dN$ takes the value 1, we can ignore the term for no
detection, as it is proportional to $dt$ and is infinitesimal. This means that
we obtain the increment for a detection given by Eq.~(\ref{evdet}).

Note that Eq.~(\ref{evboth}) differs slightly from the result given in
\cite{wise96}. This is because the state for a detection was multiplied by
$e^{-i\Phi}$ here. This is necessary in order to obtain a simple result in the
limit of large $|\gamma|$. To take this limit we can approximate the increment
$dN$ in terms of Wiener increments $dW$:
\beq
dN=\kappa dt+\sqrt{\kappa} dW,
\eeq
where
\beq
\kappa=\ip{(a\dg+\gamma^*)(a+\gamma)}.
\eeq
This approximation is used because it gives the correct expectation value
(\ref{expectation}), as well as $\ip{dN}=\ip{dN^2}$. Expanding the SDE for the
state in terms of this we find
\bqa
d\ket{\psi(t)} \!\!\!\!&=&\!\!\!\! \left[ \left(\kappa dt+\sqrt{\kappa} dW
\right) \left(\frac{(a e^{-i\Phi} + |\gamma|)}{\sqrt{\lr{(a\dg+\gamma^*)(a+
\gamma)}}}-1 \right)\right. \nn \\
&&+ \left. dt \left(\frac{\lr{a\dg a}}2-\frac{a\dg a}2+\frac{\lr{\gamma^*a+
\gamma a\dg}}2 -\gamma^*a-iH\right)\right] \ket{\psi(t)} \nn \\
\!\!\!\!&=&\!\!\!\!\left[\left(-\kappa dt-\sqrt{\kappa} dW+ \sqrt{\kappa}(a
e^{-i\Phi} +|\gamma|)dt+(a e^{-i\Phi} + |\gamma|)dW\right) \right. \nn \\
&&+ \left. dt \left( \frac{\lr{a\dg a}}2-\frac{a\dg a}2+\frac{\lr{\gamma^*a+
\gamma a\dg}}2 -\gamma^*a-iH\right)\right] \ket{\psi(t)}.
\eqa
Expanding $\kappa$ gives
\bqa
\kappa\!\!\!\! &=& \!\!\!\!\ip{\left(a\dg+\gamma^*\right)\left(a+\gamma\right)}
\nn \\ \!\!\!\! &=& \!\!\!\!\ip{a\dg a+a\dg \gamma+\gamma^*a+|\gamma|^2}
\nn \\ \!\!\!\! &=& |\gamma|^2 \left( 1+\frac{2\chi}{|\gamma|}
+\frac{\lr{a\dg a}}{|\gamma|^2} \right),
\eqa
where
\beq
\chi=\half\ip{ae^{-i\Phi}+a\dg e^{i\Phi}}.
\eeq
We can expand $\sqrt{\kappa}$ as
\bqa
\sqrt \kappa\!\!\!\! &=& \!\!\!\!|\gamma|\sqrt{ \left( 1+\frac{2\chi}{|\gamma|}
+\frac{\lr{a\dg a}}{|\gamma|^2} \right)} \nn \\ \!\!\!\! &=&
\!\!\!\!|\gamma|\left( 1+\frac{\chi}{|\gamma|}+\frac{\lr{a\dg a}}
{2|\gamma|^2}- \frac{\chi^2}{2|\gamma|^2}\right).
\eqa
Now using this in the expression for the SDE:
\bqa
d\ket{\psi(t)}\!\!\!\!&=&\!\!\!\!\left[\left(-|\gamma|^2\left( 1+\frac{2
\chi}{|\gamma|}+\frac{\lr{a\dg a}}{|\gamma|^2} \right)dt-|\gamma|\left(
1+\frac{\chi}{|\gamma|}+\frac{\lr{a\dg a}}{2|\gamma|^2}- \frac{\chi^2}
{2|\gamma|^2}\right)dW \right. \right. \nn \\
\!\!\!\!&&\!\!\!\!\left.+|\gamma|\left( 1+\frac{\chi}{|\gamma|}
+\frac{\lr{a\dg a}}{2|\gamma|^2}-\frac{\chi^2}{2|\gamma|^2} \right)
(ae^{-i\Phi}+|\gamma|)dt+(ae^{-i\Phi}+|\gamma|)dW \right) \nn \\
\!\!\!\!&&\!\!\!\!\left.+dt\left(\frac{\lr{a\dg a}}2-\frac{a\dg a}2+\frac{\lr
{\gamma^* a+\gamma a\dg}}2-\gamma^* a -iH\right) \right] \ket{\psi(t)} \nn \\
\!\!\!\!&=&\!\!\!\!\left[\left( -|\gamma|^2 \left(1+\frac{2\chi}{|\gamma|}
+\frac{\lr{a\dg a}}{|\gamma|^2}\right)dt+(ae^{-i\Phi}-\chi)dW \right. \right.
\nn \\
\!\!\!\!&&\!\!\!\!\left.+|\gamma|^2\left(1+\frac{ae^{-i\Phi}}{|\gamma|}
+\frac{\chi}{|\gamma|}+\frac{ae^{-i\Phi}\chi}{|\gamma|^2}
+\frac{\lr{a\dg a}}{2|\gamma|^2}-\frac{\chi^2}{2|\gamma|^2}
\right)dt\right) \nn \\
\!\!\!\!&&\!\!\!\!\left.+dt\left(\frac{\lr{a\dg a}}2-\frac{a\dg a}2+|\gamma|
\chi-|\gamma| ae^{-i\Phi} -iH\right) \right] \ket{\psi(t)} \nn \\
\!\!\!\!&=&\!\!\!\!\left[(ae^{-i\Phi}-\chi)dW-\left(iH+\half a\dg a-ae^{-i\Phi}
\chi+\half\chi^2\right)dt \right]\ket{\psi(t)}.
\eqa

These results are for the case where the field is directly combined with a
local oscillator field, and there is a single photodetector. This was done for
simplicity and for consistency with Ref.~\cite{wise96}. The particular
experimental configuration that we would like to consider, however, is that
shown in Fig.~\ref{dynefig}. Here the field is combined with the local
oscillator field via a 50/50 beam splitter, and there are two photodetectors at
the two outputs of the beam splitter. This case turns out to be entirely
equivalent, as I will now show.

In order to consider this case we can't perform a simple unitary rearrangement
of the measurement operators because we need 3 operators, for detection at one
arm, detection at the other arm, and no detection. As a simple extrapolation
from the previous case we can take the two measurement operators for detection
to be
\bqa
\Omega_+ \!\!\!\! &=& \!\!\!\! \sqrt {\frac{dt}2} (a+\gamma) \nn \\
\Omega_- \!\!\!\! &=& \!\!\!\! \sqrt {\frac{dt}2} (a-\gamma).
\eqa
In order for this measurement scheme to give the correct master equation, the
measurement operator for no detection must be
\beq
\Omega _0  = 1 - \left( {iH + \half a\dg a + \half |\gamma|^2 } \right)dt.
\eeq
The above results for a single photodetector can be simply extended to two
photodetectors. The change in state for a detection at photodetector 1 is
\beq
d\ket{\psi_+(t)} = \left[ \frac{(|\gamma|+ae^{-i\Phi})}
{\sqrt{\lr{(a\dg+\gamma^*)(a+\gamma)}}} - 1 \right]\ket{\psi(t)},
\eeq
and at photodetector 2 is
\beq
d\ket{\psi_-(t)} = \left[ \frac{(|\gamma|-ae^{-i\Phi})}
{\sqrt{\lr{(a\dg-\gamma^*)(a-\gamma)}}} - 1 \right]\ket{\psi(t)}.
\eeq
When there is no detection the unnormalised state changes to
\beq
\ket{\tilde\psi(t+dt)}=\left[1 - \left( iH + \half a\dg a + \half|\gamma|^2
\right)dt\right]\ket{\psi(t)}.
\eeq
We find that
\bqa
\braket{\tilde\psi(t+dt)}{\tilde\psi(t+dt)}\!\!\!\! &=& \!\!\!\!\ip{\left[1 -
\left( {-iH + \half a\dg a + \half |\gamma|^2} \right)dt\right]\left[1 -
\left( {iH + \half a\dg a + \half |\gamma|^2} \right)dt\right]} \nn \\
\!\!\!\! &=& \!\!\!\! \ip{1-(a\dg a+|\gamma|^2)dt} \nn \\
\!\!\!\! &=& \!\!\!\! 1-(\lr{a\dg a}+|\gamma|^2)dt,
\eqa
so the normalising factor is
\beq
\braket{\tilde\psi(t+dt)}{\tilde\psi(t+dt)}^{-1/2}
= 1+\half(\lr{a\dg a}+|\gamma|^2)dt.
\eeq
This means that the normalised state changes to
\beq
\ket{\psi(t+dt)}=\left[1+\left(\frac{\lr{a\dg a}}2-\frac{a\dg a}2-iH
\right)dt\right]\ket{\psi(t)},
\eeq
so the difference in the state is
\beq
d\ket{\psi(t)}=dt\left(\frac{\lr{a\dg a}}2-\frac{a\dg a}2-iH
\right)\ket{\psi(t)}.
\eeq
Then the increment in the state taking into account all three alternatives is
\bqa
d\ket{\psi(t)}\!\!\!\! &=& \!\!\!\!\left[dN_+ \left(\frac{(|\gamma|
+ae^{-i\Phi})}{\sqrt{\lr{(a\dg+\gamma^*)(a+\gamma)}}} - 1\right)+dN_-
\left(\frac{(|\gamma|-ae^{-i\Phi})}{\sqrt{\lr{(a\dg-\gamma^*)(a-\gamma)}}}
- 1\right)\right. \nn \\ &&\left.+dt\left(\frac{\lr{a\dg a}}2-
\frac{a\dg a}2-iH\right)\right]\ket{\psi(t)}.
\eqa

Now rather than defining a single $\kappa$, we can define
\beq
\kappa_\pm=\ip{(a\dg\pm\gamma^*)(a\pm\gamma)}.
\eeq
In terms of this the increment is
\bqa
d\ket{\psi(t)}\!\!\!\! &=& \!\!\!\!\left[dN_+ \left(\frac{(|\gamma|
+ae^{-i\Phi})}{\sqrt{\kappa_+}} - 1\right)+dN_- \left(\frac{(|\gamma|
-ae^{-i\Phi})}{\sqrt{\kappa_-}} - 1\right) \right. \nn \\ &&\left.
+dt\left(\frac{\lr{a\dg a}}2-\frac{a\dg a}2 -iH
\right)\right]\ket{\psi(t)}.
\eqa
For large $|\gamma|$ we can use the approximation
\beq
dN_\pm = \frac{\kappa_\pm}2 dt+\sqrt{\frac{\kappa_\pm}2}dW_\pm.
\eeq
Substituting this in gives
\bqa
d\ket{\psi(t)}\!\!\!\! &=& \!\!\!\!\left[\left(\frac{\kappa_+}2 dt+
\sqrt{\frac{\kappa_+}2}dW_+\right) \left(\frac{(|\gamma|+ae^{-i\Phi})}
{\sqrt{\kappa_+}} - 1\right)\right. \nn \\ &&\left.+\left(\frac{\kappa_-}2 dt+
\sqrt{\frac{\kappa_-}2}dW_-\right)\left(\frac{(|\gamma|-ae^{-i\Phi})}
{\sqrt{\kappa_-}} - 1\right)+dt\left(\frac{\langle a\dg
a \rangle}2-\frac{a\dg a}2-iH \right)\right] \ket{\psi(t)} \nn \\
\!\!\!\! &=& \!\!\!\! \left[ {\frac{{\sqrt{\kappa_+}}}{2}\left( {|\gamma| +
ae^{-i\Phi}}\right)
dt + \frac{1}{{\sqrt 2 }}\left( {|\gamma| + ae^{ - i\Phi } } \right)dW_+
-\frac{{\kappa_+}} {2}dt - \sqrt {\frac{{\kappa_+}}{2}} dW_+} \right. \nn \\
&& + \frac{{\sqrt {\kappa_-}}}{2}\left( {|\gamma| - ae^{-i\Phi}} \right)dt
+\frac{1}{{\sqrt 2 }}\left( {|\gamma| - ae^{ - i\Phi } } \right)dW_-
- \frac{{\kappa_-}}{2}dt - \sqrt {\frac{{\kappa_-}}{2}} dW_ -  \nn \\ 
&& \left. {+dt\left( {\frac{\lr{a\dg a}}{2}
-\frac{{a\dg a}}{2} - iH} \right)} \right]\ket{\psi(t)}.
\eqa
Similarly to the previous case we have
\bqa
\kappa_\pm = |\gamma|^2 \left(1 \pm \frac{2\chi}{|\gamma|} +\frac{\lr{a\dg a}}
{|\gamma|^2} \right),
\eqa
and
\bqa
\sqrt {\kappa_\pm} = |\gamma|\left( 1\pm\frac{\chi}{|\gamma|} +
\frac{\lr{a\dg a}}{2|\gamma|^2}-\frac{\chi^2}{2|\gamma|^2}\right).
\eqa
Substituting this in gives
\bqa
d\ket{\psi(t)} \!\!\!\! &=& \!\!\!\! \left[\frac{|\gamma|^2}2\left(1+\frac{\chi}
{|\gamma|}+\frac{\lr{a\dg a}}{2|\gamma|^2}-\frac{\chi^2}{2|\gamma|^2}\right)
\left(1+\frac{ae^{-i\Phi}}{|\gamma|}\right)dt+\frac {|\gamma|}{\sqrt 2}\left(1+
\frac{ae^{-i\Phi}}{|\gamma|}\right)dW_+\right.\nn \\
&& -\frac{|\gamma|^2}2\left(1+\frac{2\chi}{|\gamma|}+\frac{\lr{a\dg a}}
{|\gamma|^2}\right)dt-\frac{|\gamma|}{\sqrt 2}\left(1+\frac{\chi}{|\gamma|}+
\frac{\lr{a\dg a}}{2|\gamma|^2}-\frac{\chi^2}{2|\gamma|^2}\right)dW_+ \nn \\ 
&& +\frac{|\gamma|^2}2\left(1-\frac{\chi}{|\gamma|}+\frac{\lr{a\dg a}}
{2|\gamma|^2}-\frac{\chi^2}{2|\gamma|^2}\right)\left(1-\frac{ae^{-i\Phi}}
{|\gamma|}\right)dt+\frac{|\gamma|}{\sqrt 2}\left(1-\frac{ae^{-i\Phi}}{|\gamma|}
\right)dW_- \nn \\ 
&& -\frac{|\gamma|^2}2 \left(1-\frac{2\chi}{|\gamma|}+\frac{\lr{a\dg a}}
{|\gamma|^2}\right)dt-\frac{{|\gamma|}}{\sqrt 2}\left(1-\frac{\chi}{|\gamma|}+
\frac{\lr{a\dg a}}{2|\gamma|^2}-\frac{\chi^2}{2|\gamma|^2}\right)dW_- \nn \\
&& \left. +dt\left(\frac{\lr{a\dg a}}2-\frac{a\dg a}2-iH\right)\right]
\ket{\psi(t)} \nn \\
\!\!\!\! &=& \!\!\!\! \left[ \frac{|\gamma|^2}2\left(1+\frac{ae^{-i\Phi}}
{|\gamma|}+\frac{\chi}{|\gamma|}+\frac{a\chi e^{-i\Phi}}{|\gamma|^2}+\frac
{\lr{a\dg a}}{2|\gamma|^2}-\frac{\chi^2}{2|\gamma|^2}\right)dt+\frac{|\gamma|}
{\sqrt 2}\left(\frac{ae^{-i\Phi}}{|\gamma|}\right)dW_+ \right. \nn \\
&& -\frac{|\gamma|^2}2\left(1+\frac{2\chi}{|\gamma|}+\frac{\lr{a\dg a}}
{|\gamma|^2}\right)dt-\frac{|\gamma|}{\sqrt 2}\left(\frac{\chi}{|\gamma|}+
\frac{\lr{a\dg a}}{2|\gamma|^2}-\frac{\chi^2}{2|\gamma|^2}\right)dW_+ \nn \\
&& +\frac{|\gamma|^2}2\left(1-\frac{ae^{-i\Phi}}{|\gamma|}-\frac{\chi}{|\gamma|}
+\frac{a\chi e^{-i\Phi}}{|\gamma|^2}+\frac{\lr{a\dg a}}{2|\gamma|^2}-\frac
{\chi^2}{2|\gamma|^2}\right)dt-\frac{|\gamma|}{\sqrt 2}\left(\frac{a
e^{-i\Phi}}{|\gamma|}\right)dW_- \nn \\
&& -\frac{|\gamma|^2}2\left(1-\frac{2\chi}{|\gamma|}+\frac{\lr{a\dg a}}
{|\gamma|^2}\right)dt-\frac{|\gamma|}{\sqrt 2}\left(-\frac{\chi}{|\gamma|}+
\frac{\lr{a\dg a}}{2|\gamma|^2}-\frac{\chi^2}{2|\gamma|^2}\right)dW_- \nn \\
&& \left. +dt\left(\frac{\lr{a\dg a}}2-\frac{a\dg a}2-iH\right)\right]
\ket{\psi(t)} \nn \\
\!\!\!\! &=& \!\!\!\! \left[|\gamma|^2\left(1+\frac{a\chi e^{-i\Phi}}
{|\gamma|^2}+\frac{\lr{a\dg a}}{2|\gamma|^2}-\frac{\chi^2}{2|\gamma|^2}\right)dt
+\frac{|\gamma|}{\sqrt 2}\left(\frac{ae^{-i\Phi}}{|\gamma|}-\frac{\chi}
{|\gamma|}\right)dW_+ \right. \nn \\
&& -|\gamma|^2 \left(1+\frac{\lr{a\dg a}}{|\gamma|^2}\right)dt-\frac{|\gamma|}
{\sqrt 2}\left(\frac{ae^{-i\Phi}}{|\gamma|}-\frac{\chi}{|\gamma|}\right)dW_-
\nn \\
&& \left. +dt\left(\frac{\lr{a\dg a}}2-\frac{a\dg a}2-iH\right)\right]
\ket{\psi(t)} \nn \\
\!\!\!\! &=& \!\!\!\! \left[|\gamma|^2\left(\frac{a\chi e^{-i\Phi}}{|\gamma|^2}
-\frac{\lr{a\dg a}}{2|\gamma|^2}-\frac{\chi^2}{2|\gamma|^2}\right)dt+\frac 1
{\sqrt 2}(ae^{-i\Phi}-\chi)(dW_+ -dW_-) \right. \nn \\
&& \left. +dt\left(\frac{\lr{a\dg a}}2-\frac{a\dg a}2-iH\right)\right]
\ket{\psi(t)} \nn \\
\!\!\!\! &=& \!\!\!\! \left[ (ae^{-i\Phi}-\chi)\frac{(dW_+-dW_-)}{\sqrt 2}
-\left( iH + \half a\dg a - a\chi e^{-i\Phi}+\half \chi^2 \right)dt
\right]\ket{\psi(t)}.
\eqa
Now we find that the combination of the two stochastic increments is equivalent
to a third
\beq
dW = \frac{\left( dW_+ - dW_- \right)}{\sqrt 2}.
\eeq
In terms of this we find
\beq
d\ket{\psi(t)} = \left[ (ae^{-i\Phi}-\chi)dW-\left(iH+\half a\dg a-a\chi
e^{-i\Phi}+\half \chi^2 \right)dt \right]\ket{\psi(t)} .
\eeq
This case is therefore exactly that same as the case where the
fields are combined directly and there is only one photodetector.

Note also that the POM derived in Ref.~\cite{wise96} was derived for the case
where the signal is combined directly with the local oscillator, and there is
one photodetector. It turns out that this POM also applies to the case with
a 50/50 beam splitter and two photodetectors. This can be shown by performing
the derivation in a similar way to Ref.~\cite{wise96}. In general, we can
consider the stochastic evolution of an unnormalised state vector
\beq
\ket{\tilde \psi _r (t+T)} = \frac{\Omega_r(T)\ket{\psi(t)}}
{\sqrt{\Lambda_r(T)}},
\eeq
where $\Lambda_r(T)$ is the ostensible probability for the result $r$, and the
actual probability is given by
\beq
P_r (T) = \Lambda _r (T) \braket{\tilde\psi_r(t+T)}{\tilde\psi_r(t+T)}.
\eeq
We can use this strategy on the above measurements, with the ostensible
probabilities given by
\bqa
\Lambda_\pm(dt) \!\!\!\! &=& \!\!\!\! \half |\gamma|^2 dt, \\
\Lambda_0(dt) \!\!\!\! &=& \!\!\!\! 1-|\gamma|^2 dt.
\eqa
In this case the increments for the two detection cases are
\beq
  d\ket{\tilde \psi_\pm (t)} = \pm \frac{a}{\gamma}\ket{\psi(t)},
\eeq
and the increment for the case where there is no detection is
\beq
d\ket{\tilde \psi_0(t)} = -dt\left(\half a\dg a + iH \right)\ket{\psi(t)}.
\eeq
Writing out the evolution explicitly
\beq
d\ket{\tilde \psi(t)} = \left[ dN_+(t)\frac a{\gamma}-dN_-(t)\frac a{\gamma}
-dt \left(iH + \half a\dg a\right)\right]\ket{\psi(t)}.
\eeq
Now we can approximate the Poisson process by a Gaussian process:
\beq
dN_\pm = \half|\gamma|^2 dt + \frac{|\gamma|}{\sqrt 2}dW_\pm (t).
\eeq
Using this gives
\bqa
d\ket{\tilde \psi(t)} \!\!\!\! &=& \!\!\!\! \left[ \frac 1{\sqrt 2}dW_+(t)a
e^{-i\Phi}-\frac 1{\sqrt 2}dW_-(t)ae^{-i\Phi}-dt\left(iH+\half a\dg a\right)
\right]\ket{\psi(t)} \nn \\
\!\!\!\! &=& \!\!\!\! \left[ \frac 1{\sqrt 2}ae^{-i\Phi}\left(dW_+(t)-dW_-(t)
\right)-dt\left(iH+\half a\dg a\right)\right]\ket{\psi(t)}.
\eqa
Now replacing $(dW_+(t)-dW_-(t))/\sqrt 2$ with $dW(t)$ as before, we obtain
\beq
d\ket{\tilde \psi(t)} = \left[ ae^{-i\Phi}dW(t)-dt\left(iH+\half a\dg a\right)
\right]\ket{\psi(t)}.
\eeq
Then the recorded photocurrent is given by
\bqa
I(t) \!\!\!\! &=& \!\!\!\! \lim_{\delta t \to 0} \lim_{|\gamma| \to \infty}
\frac{\delta N_+ -\delta N_-}{|\gamma|\delta t} \nn \\
\!\!\!\! &=& \!\!\!\! \lim_{|\gamma| \to \infty}
\frac{\frac{|\gamma|}{\sqrt 2}dW_+(t)-\frac{|\gamma|}
{\sqrt 2 }dW_-(t)}{|\gamma|dt} \nn \\
\!\!\!\! &=& \!\!\!\! \frac{\frac 1{\sqrt 2}\left(dW_+(t)-dW_-(t)\right)}{dt}
\nn \\ \!\!\!\! &=& \!\!\!\! \frac{dW(t)}{dt}.
\eqa
Note that, although the limits are noncommuting when performed on the increments
$\delta N_{\pm}$, they are commuting for Gaussian increments. It is therefore
reasonable to perform the time limit first, as in the second line above.

From this point forward the derivation is identical to that given in
Ref.~\cite{wise96}, and I will therefore not give it explicitly. Note that the
above expression for $I(t)$ is obtained using the same definition as when we are
considering the actual probabilities, and not the ostensible probabilities. The
increments $dN_\pm$ are chosen with the ostensible probabilities here, resulting
in an expression for $I(t)$ that only involves the stochastic part. 

Therefore we find that, similarly to the case for a single detector, the POM for
the parameters $A$ and $B$ defined by
\bqa
A \!\!\!\! &=& \!\!\!\! \int\limits_0^\infty {e^{i\Phi (s)} e^{-s/2}I(s)ds},
\nn \\
B \!\!\!\! &=& \!\!\!\!-\int\limits_0^\infty {e^{2i\Phi (s)} e^{-s} ds},
\eqa
is
\beq
F(A,B)=Q(A,B)\ket{\tilde \psi (A,B)}\bra{\tilde \psi(A,B)},
\eeq
where
\beq
\ket{\tilde \psi(A,B)} = \exp \left(\half B(a\dg)^2 +Aa\dg \right)\ket 0,
\eeq
and $Q(A,B)$ is the ostensible probability distribution. This derivation is for
the case $H=0$, corresponding to a freely damped cavity. From this point forward
I will take $H=0$, rather than continuing to include this term.

Now we can determine the stochastic evolution of a squeezed state. To do this I
use the method of Rigo {\it et al.} \cite{rigo}. Squeezed states obey the
relation
\begin{equation}
\label{rigoreln}
\left( a-B_t^{\rm S} a\dg -A_t^{\rm S} \right) \ket{A_t^{\rm S},B_t^{\rm S}}=0.
\end{equation}
Here the squeezing parameters $A_t^{\rm S}$ and $B_t^{\rm S}$ are defined so that they are
analogous to the parameters $A$ and $B$. They are related to the usual squeezing
parameters by
\bqa
\label{defABS}
B_t^{\rm S}  \!\!\!\! &=& \!\!\!\! - \frac{\zeta_t \tanh\st{\zeta_t}}{\st{\zeta_t}},
\nn \\
A_t^{\rm S}  \!\!\!\! &=& \!\!\!\! \alpha_t - B_t^{\rm S} \left( \alpha_t \right)^*.
\eqa
This is the same as the relation between $A$ and $B$ and the parameters
$\alpha^{\rm P}$ and $\zeta^{\rm P}$ for the squeezed state in the probability distribution.

If the squeezed state remains a squeezed state under the increment $d\ket\psi$,
then it can be shown from Eq.~(\ref{rigoreln}) that
\beq
\left( {a - B_t^{\rm S} a\dg - A_t^{\rm S} }\right) d\ket\psi = \left( {dB_t^{\rm S} a\dg +
dA_t^{\rm S} } \right)\ket \psi,
\eeq
where the Stratonovich formalism has been used.
If we convert the stochastic increment for the state into the Stratonovich form
and substitute it into the above equation, then we can obtain an SDE for the
squeezing parameters. To convert from the Ito to the Stratonovich form we make
the replacement
\beq
XdY \to XdY - \half dXdY
\eeq
Using this on the above SDE gives the Stratonovich increment
\bqa
d\ket{\psi(t)} \!\!\!\! &=& \!\!\!\! \left[(ae^{-i\Phi}-\chi)dW-(\half a\dg a-a
\chi e^{-i\Phi}+\half\chi^2)dt \right. \nn \\
&& \left. -\half\left(a\,d(e^{-i\Phi})-d\chi\right)dW\right]\ket{\psi(t)}
-\half(ae^{-i\Phi}-\chi)dW d\ket{\psi(t)} \nn \\ 
\!\!\!\! &=& \!\!\!\! \left[(ae^{-i\Phi}-\chi)dW-(\half a\dg a-a\chi e^{-i\Phi}+
\half\chi^2)dt \right. \nn \\ && -\half\left(a\,d(e^{-i\Phi})-d\chi\right)dW
\ket{\psi(t)}-\half(ae^{-i\Phi}-\chi)^2 dt\ket{\psi(t)} \nn \\
\!\!\!\! &=& \!\!\!\! \left[(ae^{-i\Phi}-\chi)dW-(\half a\dg a+\half a^2
e^{-2i\Phi}-2a\chi e^{-i\Phi}+\chi^2)dt \right. \nn \\
&& \left. -\half\left(a\,d(e^{-i\Phi})-d\chi\right)dW\right]\ket{\psi(t)}.
\eqa
Here the increments $d(e^{-i\Phi})$ and $d\chi$ have been included because the
phase of the local oscillator can vary stochastically.

In order to determine the SDE for the squeezing parameters, it is convenient to
firstly determine the effect of $(a-B_t^{\rm S} a\dg-A_t^{\rm S})$ on each of the operators
in this equation.
\bqa
(a-B_t^{\rm S} a\dg-A_t^{\rm S})C\ket{B_t^{\rm S},A_t^{\rm S}} \!\!\!\! &=& \!\!\!\! C
(a-B_t^{\rm S} a\dg-A_t^{\rm S})\ket{B_t^{\rm S},A_t^{\rm S}} \nn \\ \!\!\!\! &=& \!\!\!\! 0 \\
(a-B_t^{\rm S} a\dg-A_t^{\rm S})a\ket{B_t^{\rm S},A_t^{\rm S}} \!\!\!\! &=& \!\!\!\!
\left[a(a-B_t^{\rm S} a\dg-A_t^{\rm S})+B_t^{\rm S}\right]\ket{B_t^{\rm S},A_t^{\rm S}} \nn \\
\!\!\!\! &=& \!\!\!\! B_t^{\rm S} \ket{B_t^{\rm S},A_t^{\rm S}} \\
(a-B_t^{\rm S} a\dg-A_t^{\rm S})a^2 \ket{B_t^{\rm S},A_t^{\rm S}} \!\!\!\! &=& \!\!\!\!
\left[a(a-B_t^{\rm S} a\dg-A_t^{\rm S})+B_t^{\rm S}\right]a\ket{B_t^{\rm S},A_t^{\rm S}} \nn \\
\!\!\!\! &=& \!\!\!\! \left\{ a\left[a(a-B_t^{\rm S} a\dg-A_t^{\rm S})+B_t^{\rm S}\right]+aB_t^{\rm S}
\right\}\ket{B_t^{\rm S},A_t^{\rm S}} \nn \\
\!\!\!\! &=& \!\!\!\! 2aB_t^{\rm S} \ket{B_t^{\rm S} ,A_t^{\rm S}} \nn \\
\!\!\!\! &=& \!\!\!\! 2B_t^{\rm S} (B_t^{\rm S} a\dg+A_t^{\rm S})\ket{B_t^{\rm S},A_t^{\rm S}} \\
(a-B_t^{\rm S} a\dg-A_t^{\rm S})a\dg a\ket{B_t^{\rm S},A_t^{\rm S}} \!\!\!\! &=& \!\!\!\!
\left[a\dg(a-B_t^{\rm S} a\dg-A_t^{\rm S})+1\right]a\ket{B_t^{\rm S},A_t^{\rm S}} \nn \\
\!\!\!\! &=& \!\!\!\! \left\{ a\dg \left[a(a-B_t^{\rm S} a\dg-A_t^{\rm S})+B_t^{\rm S}\right]+a
\right\}\ket{B_t^{\rm S},A_t^{\rm S}} \nn \\
\!\!\!\! &=& \!\!\!\! (a\dg B_t^{\rm S}+a)\ket{B_t^{\rm S},A_t^{\rm S}} \nn \\
\!\!\!\! &=& \!\!\!\! (2a\dg B_t^{\rm S}+A_t^{\rm S})\ket{B_t^{\rm S},A_t^{\rm S}}
\eqa
Here $C$ is any scalar constant. Using these results it is straightforward
to show that
\bqa
(a-B_t^{\rm S} a\dg-A_t^{\rm S})d\ket{\psi(t)} \!\!\!\! &=& \!\!\!\! \left\{ \left[-\half
(2a\dg B_t^{\rm S}+A_t^{\rm S})-B_t^{\rm S}(B_t^{\rm S} a\dg+A_t^{\rm S})e^{-2i\Phi}+2B_t^{\rm S}\chi e^{-i\Phi}
\right]dt \right. \nn \\ && \left. +B_t^{\rm S} e^{-i\Phi}dW-\half B_t^{\rm S}d(e^{-i\Phi})
dW \right\}\ket{\psi(t)} \nn \\
\!\!\!\! &=& \!\!\!\! (dB_t^{\rm S} a\dg+dA_t^{\rm S})\ket{\psi(t)}.
\eqa
The Stratonovich SDEs for the squeezing parameters are therefore
\bqa
dB_t^{\rm S} \!\!\!\! &=& \!\!\!\! (-B_t^{\rm S}-(B_t^{\rm S})^2e^{-2i\Phi})dt \nn \\
\!\!\!\! &=& \!\!\!\! -B_t^{\rm S} (1+e^{-2i\Phi}B_t^{\rm S})dt \\
dA_t^{\rm S} \!\!\!\! &=& \!\!\!\! (-\half A_t^{\rm S}-B_t^{\rm S} A_t^{\rm S} e^{-2i\Phi}+2B_t^{\rm S} \chi
e^{-i\Phi})dt+B_t^{\rm S} e^{-i\Phi}dW-\half B_t^{\rm S} d(e^{-i\Phi})dW \nn \\
\!\!\!\! &=& \!\!\!\! \left[-\half A_t^{\rm S}-B_t^{\rm S} A_t^{\rm S} e^{-2i\Phi}+B_t^{\rm S} (\lr a
e^{-2i\Phi}+\lr{a\dg})\right]dt+B_t^{\rm S} e^{-i\Phi}dW-\half B_t^{\rm S} d(e^{-i\Phi})dW
\nn \\
\!\!\!\! &=& \!\!\!\! \left\{ -\half A_t^{\rm S}-B_t^{\rm S} A_t^{\rm S} e^{-2i\Phi}+B_t^{\rm S} \left[
(B_t^{\rm S} \lr{a\dg}+A_t^{\rm S})e^{-2i\Phi}+\lr{a\dg}\right]\right\}dt \nn \\
&&+B_t^{\rm S} e^{-i\Phi}dW-\half B_t^{\rm S} d(e^{-i\Phi})dW \nn \\
\!\!\!\! &=& \!\!\!\! -\half A_t^{\rm S} dt+B_t^{\rm S} \lr{a\dg}(1+e^{-2i\Phi}B_t^{\rm S})dt+
B_t^{\rm S} e^{-i\Phi}dW-\half B_t^{\rm S} d(e^{-i\Phi})dW.
\eqa
In order to convert back to the SDE for the standard (non-scaled) amplitude,
note that
\beq
\alpha_t=\frac{A_t^{\rm S}+B_t^{\rm S}(A_t^{\rm S})^*}{1-|B_t^{\rm S}|^2}.
\eeq
Using this expression, the Stratonovich increment in $\alpha_t$ is
\bqa
d\alpha_t \!\!\!\! &=& \!\!\!\! \frac{dA_t^{\rm S}+B_t^{\rm S} dA_t^{\rm S*}}{1-|B_t^{\rm S}|^2}+
\frac{A_t^{\rm S*} dB_t^{\rm S}}{1-|B_t^{\rm S}|^2}+\left(B_t^{\rm S*} dB_t^{\rm S}+B_t^{\rm S} dB_t^{\rm S*}\right)
\frac{A_t^{\rm S}  + B_t^{\rm S} A_t^{\rm S*}}{\left(1-|B_t^{\rm S}|^2\right)^2} \nn \\
\!\!\!\! &=& \!\!\!\! \left[dA_t^{\rm S}+B_t^{\rm S} dA_t^{\rm S*}+A_t^{\rm S*}dB_t^{\rm S}+\alpha_t
\left(B_t^{\rm S*} dB_t^{\rm S}+B_t^{\rm S} dB_t^{\rm S*}\right)\right]\frac 1{1-|B_t^{\rm S}|^2}.
\eqa
Expanding this gives
\bqa
d\alpha_t \!\!\!\! &=& \!\!\!\! \left\{ -\half A_t^{\rm S} dt + B_t^{\rm S} \lr{a\dg}
\left( 1 + e^{-2i\Phi} B_t^{\rm S} \right)dt + B_t^{\rm S} e^{-i\Phi} dW - \half B_t^{\rm S}
d(e^{-i\Phi})dW + B_t^{\rm S} \left[ -\half A_t^{\rm S*} dt \right. \right. \nn \\
\!\!\!\!&&\!\!\!\! \left. + B_t^{\rm S*} \ip a \left(1+e^{2i\Phi} B_t^{\rm S*}\right)dt
+ B_t^{\rm S*} e^{i\Phi} dW - \half B_t^{\rm S*} d(e^{i\Phi})dW\right] - A_t^{\rm S*} B_t^{\rm S}
\left(1 + e^{-2i\Phi} B_t^{\rm S} \right)dt \nn \\
\!\!\!\!&&\!\!\!\! \left. -\alpha_t \left(B_t^{\rm S*} B_t^{\rm S} \left(1+e^{-2i\Phi}
B_t^{\rm S}\right)dt + B_t^{\rm S} B_t^{\rm S*} \left(1+e^{2i\Phi} B_t^{\rm S*} \right)dt \right)
\right\} \frac 1{1-|B_t^{\rm S}|^2} \nn \\
\!\!\!\! &=& \!\!\!\! -\half\alpha_t dt + \left\{B_t^{\rm S} \alpha_t^* \left(1+
e^{-2i\Phi} B_t^{\rm S} \right)dt + B_t^{\rm S} e^{-i\Phi}dW-\half B_t^{\rm S} d(e^{-i\Phi})dW
\right. \nn \\
\!\!\!\!&&\!\!\!\! +B_t^{\rm S} \left[ B_t^{\rm S*} \alpha_t \left( 1+e^{2i\Phi}B_t^{\rm S*}
\right)dt+B_t^{\rm S*} e^{i\Phi} dW-\half B_t^{\rm S*}d(e^{i\Phi})dW \right]-A_t^{\rm S*}
B_t^{\rm S} \left(1+e^{-2i\Phi}B_t^{\rm S} \right)dt \nn \\
\!\!\!\!&&\!\!\!\! \left. -\alpha_t \left(B_t^{\rm S*} B_t^{\rm S} \left(1+e^{-2i\Phi}
B_t^{\rm S} \right)dt + B_t^{\rm S} B_t^{\rm S*}\left(1+e^{2i\Phi}B_t^{\rm S*}\right)dt \right)
\right\}\frac 1{1-|B_t^{\rm S}|^2} \nn \\
\!\!\!\! &=& \!\!\!\! -\half\alpha_t dt+\left\{ B_t^{\rm S}\left[\alpha_t^*- B_t^{\rm S*}
\alpha_t \right]\left( 1+e^{-2i\Phi} B_t^{\rm S} \right)dt + B_t^{\rm S} e^{-i\Phi} dW +
B_t^{\rm S} B_t^{\rm S*} e^{i\Phi} dW \right. \nn \\
\!\!\!\!&&\!\!\!\! \left. -\half B_t^{\rm S} d\left(e^{-i\Phi} \right)dW-\half B_t^{\rm S}
B_t^{\rm S*} d(e^{i\Phi})dW - A_t^{\rm S*} B_t^{\rm S} \left(1+e^{-2i\Phi}B_t^{\rm S} \right)dt
\right\}\frac 1{1-|B_t^{\rm S}|^2} \nn \\
\!\!\!\! &=& \!\!\!\! -\half\alpha_t dt+\left\{B_t^{\rm S} A_t^{\rm S*}\left(1+e^{-2i\Phi}
B_t^{\rm S} \right)dt+B_t^{\rm S} e^{-i\Phi}dW+B_t^{\rm S} B_t^{\rm S*}e^{i\Phi}dW-\half B_t^{\rm S}
d(e^{-i\Phi})dW \right. \nn \\
\!\!\!\!&&\!\!\!\! \left. -\half B_t^{\rm S} B_t^{\rm S*} d(e^{i\Phi})dW-A_t^{\rm S*} B_t^{\rm S}
\left(1+e^{-2i\Phi}B_t^{\rm S}\right)dt \right\}\frac 1{1-|B_t^{\rm S}|^2} \nn \\
\!\!\!\! &=& \!\!\!\! -\half\alpha_t dt + \frac{B_t^{\rm S} dW}{1-|B_t^{\rm S}|^2}\left(
B_t^{\rm S*} e^{i\Phi}+e^{-i\Phi}\right)-\half\frac{B_t^{\rm S} dW}{1-|B_t^{\rm S}|^2}\left(
B_t^{\rm S*} d(e^{i\Phi})+d(e^{-i\Phi})\right).
\eqa
Converting back to the Ito form of the SDE gives
\beq
d\alpha_t = - \half\alpha_t dt + \frac{B_t^{\rm S} dW}{1-\st{B_t^{\rm S}}^2}\left(B_t^{\rm S*}
e^{i\Phi}+e^{-i\Phi}\right).
\eeq
The SDE for $B_t^{\rm S}$ is unchanged. Note that in converting back to the Ito form,
the extra term involving the increment in the phase of the local oscillator
cancels out.

Now for consistency with Ref.~\cite{wise96} I will take the signal to be
\beq
\label{defineI}
I(t) = \lim_{\delta t \to 0} \lim_{|\gamma| \to \infty} \frac{\delta N_+ -
\delta N_-}{|\gamma|\delta t}.
\eeq
In order to evaluate this, recall that the expression for the photon number
counts at the two photodetectors is
\beq
dN_\pm = \frac{\kappa_\pm}2 dt + \sqrt{\frac{\kappa_\pm}2} dW_\pm,
\eeq
where
\beq
\kappa_\pm = |\gamma|^2 \left( 1 \pm \frac{2\chi}{|\gamma|}
 + \frac{\lr{a\dg a}}{|\gamma|^2} \right),
\eeq
and the expansion of the square root is
\beq
\sqrt{\kappa_\pm}=|\gamma|\left(1\pm\frac{\chi}{|\gamma|}+\frac{\lr{a\dg a}}
{2|\gamma|^2}-\frac{\chi^2}{2|\gamma|^2} \right).
\eeq
Using these expressions gives the photocurrent as
\bqa
I(t)dt \!\!\!\! &=& \!\!\!\! \lim_{|\gamma| \to \infty}\left[\frac{|\gamma|^2}2
\left(1+\frac{2\chi}{|\gamma|}+\frac{\lr{a\dg a}}{|\gamma|^2}\right)dt+
\frac{|\gamma|}{\sqrt 2}\left(1+\frac{\chi}{|\gamma|}+\frac{\lr{a\dg a}}
{2|\gamma|^2}-\frac{\chi^2}{2|\gamma|^2}\right)dW_+ \right. \nn \\ && \left.
-\frac{|\gamma|^2}2 \left(1-\frac{2\chi}{|\gamma|}+\frac{\lr{a\dg a}}
{|\gamma|^2}\right)dt-\frac{|\gamma|}{\sqrt 2}\left(1-\frac{\chi}{|\gamma|}
+\frac{\lr{a\dg a}}{2|\gamma|^2}-\frac{\chi^2}{2|\gamma|^2}\right)dW_-\right]
\frac 1{\st{\gamma}} \nn \\
\!\!\!\! &=& \!\!\!\! \lim_{|\gamma| \to \infty} \left[|\gamma|^2\left(
\frac{2\chi}{|\gamma|}\right)dt+\frac{\st{\gamma}}{\sqrt 2}dW_+ -\frac
{\st{\gamma}}{\sqrt 2}dW_-\right]\frac 1{\st{\gamma}} \nn \\
\!\!\!\! &=& \!\!\!\! 2\chi dt + \frac 1{\sqrt 2}\left(dW_+ -dW_-\right) \nn \\
\!\!\!\! &=& \!\!\!\! 2\chi dt + dW \nn \\
\!\!\!\! &=& \!\!\!\! 2 {\rm Re} \left( \ip a e^{-i\Phi} \right)dt + dW.
\eqa
This result is entirely general and does not depend on the state being coherent
or squeezed. In the case of a squeezed state we have
\beq
\ip a = \alpha_t .
\eeq

Now note that the deterministic part of the SDE for $\alpha_t$ is
\beq
\ip {d\alpha_t} = - \half \alpha_t dt .
\eeq
This means that the deterministic part of the evolution is
\beq
\alpha_t \propto e^{-t/2}.
\eeq
In addition, the parameters $A_t$ and $B_t$ are defined in Ref.~\cite{wise96} by
\bqa
A_t \!\!\!\! &=& \!\!\!\! \int\limits_0^t e^{i\Phi}e^{-s/2}I(s)ds, \\
B_t \!\!\!\! &=& \!\!\!\! - \int\limits_0^t e^{2i\Phi} e^{-s} ds.
\eqa
The exponential factors can be removed by changing the time variable to
\begin{equation}
v=1-e^{-t},
\end{equation}
so that
\bqa
dv \!\!\!\! &=& \!\!\!\! e^{-t} dt \nn \\ 
\!\!\!\! &=& \!\!\!\! (1 - v)dt .
\eqa
This transformation maps the time to the unit interval $[0,1)$. It is also
convenient to define a new scaled coherent amplitude to remove the systematic
variation:
\begin{equation}
\label{scaled}
\alpha_v = \alpha_t e^{t/2}.
\end{equation}
Here the $v$ subscript indicates the scaled amplitude, and the $t$ subscript
indicates the original, unscaled amplitude. Since these are equal to each other
at zero time, there is no ambiguity in the initial amplitude $\alpha$. The
definition of $I(t)$ must also be altered to remove the exponential factors:
\beq
\label{altdefI0}
I(t) = \lim_{\delta t \to 0} \lim_{|\gamma| \to \infty} \frac{\delta N_+ -
\delta N_-}{|\gamma|e^{-t/2}\delta t}.
\eeq
Using this gives the photocurrent as
\beq
I(v)dv = 2{\rm Re}(\alpha_v e^{-i\Phi(v)})dv+dW(v).
\eeq

With these changes of variables, the definitions for $A_v$ and $B_v$ become
\bqa
A_v \!\!\!\! &=& \!\!\!\! \int\limits_0^v e^{i\Phi}I(u)du, \\
B_v \!\!\!\! &=& \!\!\!\! - \int\limits_0^v e^{2i\Phi} du.
\eqa
The SDE for the scaled amplitude $\alpha_v$ is
\bqa
\label{SDEalpha}
d\alpha_v \!\!\!\! &=& \!\!\!\! d\left( {\alpha_t e^{t/2} } \right) \nn \\ 
\!\!\!\! &=& \!\!\!\! \left[ - \half\alpha_t dt + \frac{B_t^{\rm S} dW(t)}
{1 - |B_t^{\rm S}|^2 }\left( B_t^{\rm S*} e^{i\Phi} + e^{-i\Phi} \right)
\right]e^{t/2} + \half\alpha_t e^{t/2} dt \nn \\ 
\!\!\!\! &=& \!\!\!\! e^{t/2} \frac{B_t^{\rm S} dW(t)}{1 - |B_t^{\rm S}|^2 }
\left( B_t^{\rm S*} e^{i\Phi} + e^{-i\Phi} \right) \nn \\
\!\!\!\! &=& \!\!\!\! \frac 1{1-v}\frac{B_v^{\rm S} dW(v)}{1-|B_v^{\rm S}|^2}
\left( B_v^{\rm S*} e^{i\Phi} + e^{-i\Phi} \right).
\eqa
Similarly the SDE for $B_v^{\rm S}$ in terms of the new time variable is
\beq
dB_v^{\rm S} = -\frac{dv}{1-v} B_v^{\rm S} \left( 1 + e^{-2i\Phi} B_v^{\rm S} \right).
\eeq
Initial calculations were performed using these equations, but there is a
further simplification that can be made. The solution for $B_v^{\rm S}$ is
\begin{equation}
\label{Bsoln}
B_v^{\rm S} = \frac {1-v}{(B_0^{\rm S})^{-1}-B^*_v}.
\end{equation}
The only calculations that will be presented in this thesis that were calculated
using numerical integration to find $B_v^{\rm S}$ rather than this solution are those
for larger photon numbers for the constant $\varepsilon$ case where the optimum
value of $\varepsilon$ was found numerically. It was found that performing the
integration rather than using this solution altered the results by less than
$0.5\%$.

It is also possible to consider input states with more general mode-functions in
a similar way \cite{semiclass}. In the general case, the input state has a
time-varying mode function $u(t)$ that is real and positive, and is normalised
so that
\beq
\int\limits_0^T u(t) dt = 1,
\eeq
where $T$ is the total pulse length. The mode function gives the systematic
variation of $\ip a$, i.e.
\beq
\ip a = \alpha_t \propto \sqrt{u(t)},
\eeq
if any stochastic variation is ignored. If the photocurrent is defined as in
(\ref{defineI}), then the definitions for $A_t$ and $B_t$ are
\bqa
A_t \!\!\!\! &=& \!\!\!\! \int\limits_0^t e^{i\Phi(s)} \sqrt{u(s)} I(s)ds , \\
B_t \!\!\!\! &=& \!\!\!\! - \int\limits_0^t e^{2i\Phi(s)} u(s)ds.
\eqa
For the case of squeezed states, the mode-function is
\beq
u(t) = e^{-t}.
\eeq
Using this mode-function these definitions are identical to those used in
\cite{wise96}.

In \cite{semiclass} there is no factor of $\sqrt{u(s)}$ in the definition of
$A_t$. This is because $I(t)$ is defined in a slightly different way. In
\cite{semiclass} it is assumed that the local oscillator is varied with the same
mode-function, so that
\beq
|\gamma| = \sqrt{u(t)}\beta,
\eeq
where $\beta$ is a constant. The photocurrent is then defined as 
\beq
I(t) = \lim_{\delta t \to 0} \lim_{\beta \to \infty} \frac{\delta N_+ -
\delta N_-}{\beta \delta t}.
\eeq
This is equivalent to using
\beq
I(t) = \lim_{\delta t \to 0} \lim_{|\gamma| \to \infty} \frac{\delta N_+ -
\delta N_-}{|\gamma|\delta t} \times \sqrt{u(t)}.
\eeq
As there is a factor of $\sqrt{u(t)}$ here, there is no need for it in the
definition of $A_t$ used in \cite{semiclass}.

The mode-function can be removed by redefining $I(t)$ as 
\beq
I(t) = \lim_{\delta t \to 0} \lim_{|\gamma| \to \infty} \frac{\delta N_+ -
\delta N_-}{|\gamma|\sqrt{u(t)}\delta t},
\eeq
and changing the time variable to
\beq
v = \int\limits_0^t u(s)ds.
\eeq
With these changes of variables, $A_v$ and $B_v$ are now given by
\bqa
A_v \!\!\!\! &=& \!\!\!\! \int\limits_0^v e^{i\Phi}I(u)du, \\
B_v \!\!\!\! &=& \!\!\!\! - \int\limits_0^v e^{2i\Phi} du,
\eqa
and $I(t)$ simplifies to
\beq
I(v) dv = 2 {\rm Re} [\alpha_v e^{-i\Phi(v)}] dv + dW(v),
\eeq
where
\beq
\alpha_v = \alpha_t / \sqrt{u(t)} .
\eeq
Here $\alpha_v$ may vary stochastically, but should have no systematic
variation. Note that for $u(t)=e^{-t}$, these changes of variables are identical
to those used previously to remove the exponential factors.

An additional complication to the numerical technique is that rather than using
the feedback in the form
\beq
\label{simplefeed}
\hat \varphi_v = \arg C_v^{1-\varepsilon(v)} A_v^{\varepsilon(v)},
\eeq
the form used was
\beq
\label{comlxfeed}
\hat \varphi_v = \arg C_v \left( \frac{A_v}{C_v} \right) ^{\varepsilon(v)}.
\eeq
These forms are almost always equivalent; however, for some cases different
results are obtained. If the simple form (\ref{simplefeed}) is used, this tends
to bias the feedback phases towards zero. The reason for this is that fractional
powers generally have multiple solutions, and the program does not necessarily
give the appropriate one.

To demonstrate this, consider for example the case where $\varepsilon = \half$.
If the phases of $A_v$ and $C_v$ are $2\pi /3$ and $-2\pi /3$, then clearly the
phase estimate that we want is halfway between these, at $\hat\varphi_v = \pi$.
The feedback in the form (\ref{simplefeed}) will give
\bqa
\hat\varphi_v \!\!\!\! &=& \!\!\!\! \arg(e^{-i2\pi/3})^{0.5}(e^{i2\pi/3})^{0.5}
\nn \\
\!\!\!\! &=& \!\!\!\! \arg e^{-i\pi/3} e^{i\pi/3} \nn \\
\!\!\!\! &=& \!\!\!\! \arg 1 \nn \\
\!\!\!\! &=& \!\!\!\! 0.
\eqa

What has happened is that the power of $0.5$ has rotated the phases towards
zero, so that multiplying $C_v^{0.5}$ and $A_v^{0.5}$ together has yielded a
phase of zero. On the other hand, if we use the form given by (\ref{comlxfeed}),
we find
\bqa
\hat\varphi_v \!\!\!\! &=& \!\!\!\! \arg e^{-i2\pi/3} (e^{-i2\pi/3})^{0.5}\nn \\
\!\!\!\! &=& \!\!\!\! \arg e^{-i2\pi/3}e^{-i\pi/3} \nn \\
\!\!\!\! &=& \!\!\!\! \arg (-1) \nn \\
\!\!\!\! &=& \!\!\!\! \pi,
\eqa
which is the correct result. I will usually give the feedback
in the simple form (\ref{simplefeed}), even though it is necessary to use
(\ref{comlxfeed}) in actual calculations.

\section{Time Steps}
Before performing the calculations, it is important to estimate the inaccuracy
introduced by the finite time steps. This is necessary to determine how many
time steps are required in order to keep the error due to this factor
negligible. In Ch.~\ref{delays} it is shown that the minimum phase variance
when there is a time {\it delay} of $\tau$ is $\tau/(8\nb)$. It is reasonable
that there should be a similar scaling for the error due to the finite time
step.

In order to estimate the error due to the time step, the phase variance was
determined numerically for mark II measurements on squeezed states optimised for
minimum intrinsic phase variance. Calculations were performed for mean photon
numbers of approximately 122, 432, 1577 and 5877. For each photon number
calculations were performed simultaneously for $2^m$ intervals, with several
different values of $m$. For photon numbers up to 1577, the value of $m$ was
varied from 7 to 16, and for $\nb = 5877$, $m$ was varied from 8 to 17. For
each time delay $2^{14}$ samples were used.

In order to minimise the error between the calculations with different numbers
of intervals, the random numbers for the smaller numbers of intervals were
determined by combining the random numbers for the larger numbers of intervals.
Specifically, for a step size of $\delta v$, $dW$ was replaced in the
integration with $\delta W$, which has a Gaussian distribution with a variance
of $\delta v$. For the integration with twice the step size,
$\delta v' = 2\delta v$, adjacent increments $\delta W$ were added to give the
new increment $\delta W'$, with variance $2\delta v$.

\begin{figure}
\centering
\includegraphics[width=0.7\textwidth]{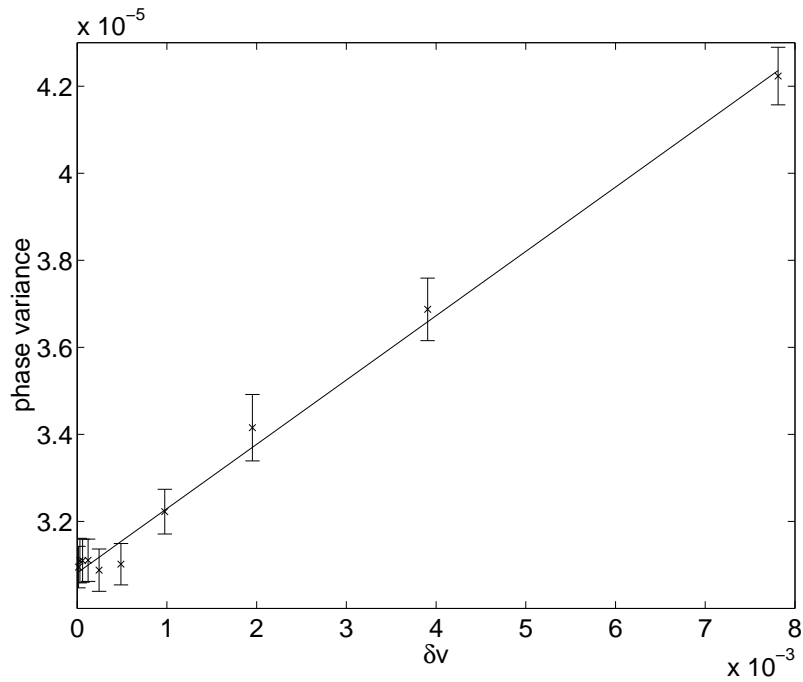}
\caption{The phase variance for mark II measurements on optimised squeezed
states with a mean photon number of 122
with various time step sizes. The crosses are the numerical results, and the
continuous line is that fitted by a linear regression.}
\label{delay20}
\end{figure}

\begin{figure}
\centering
\includegraphics[width=0.7\textwidth]{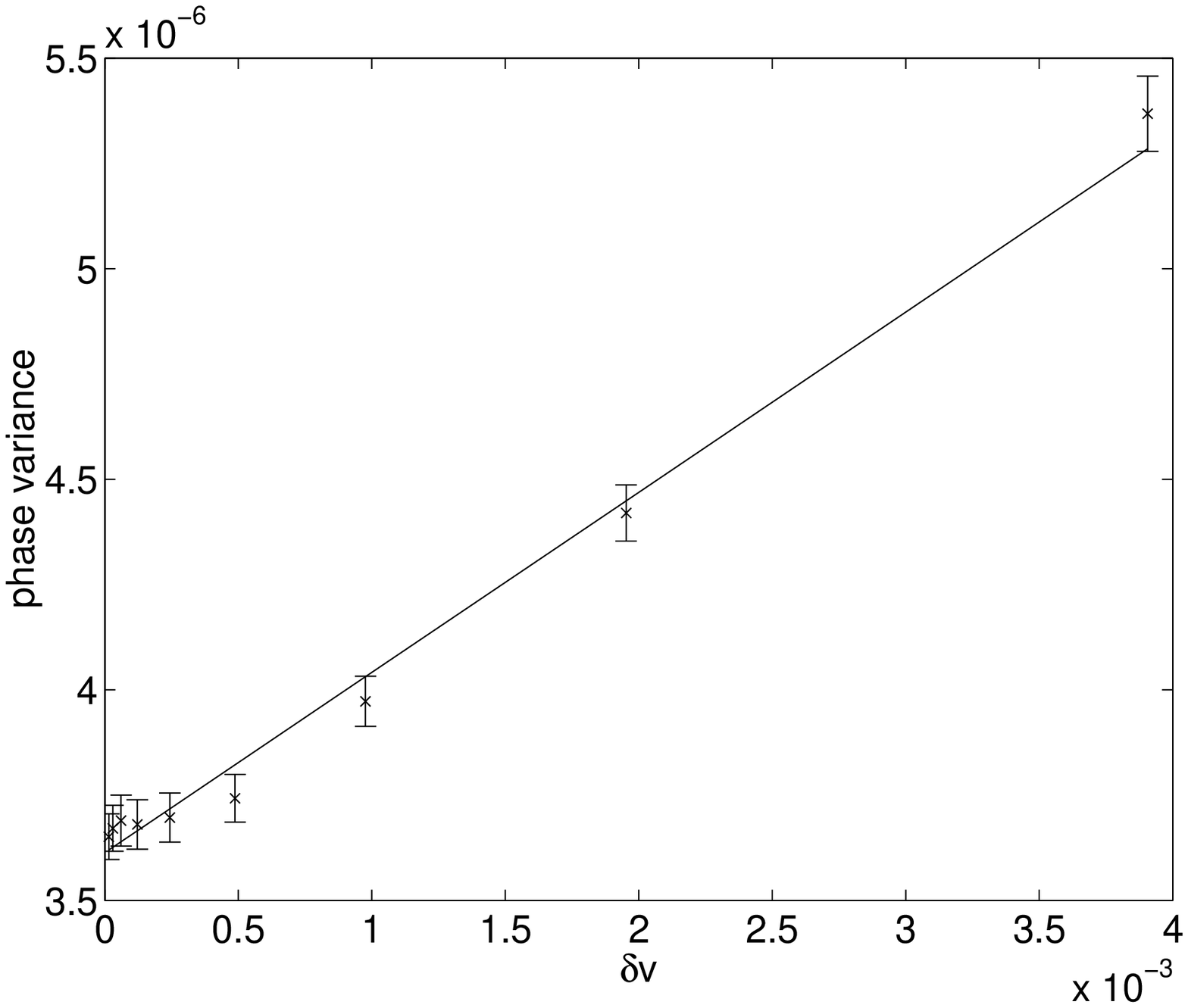}
\caption{The phase variance for mark II measurements on optimised squeezed
states with a mean photon number of 432 with various time step sizes. The
crosses are the numerical results, and the continuous line is that fitted by a
linear regression.}
\label{delay25}
\end{figure}

\begin{figure}
\centering
\includegraphics[width=0.7\textwidth]{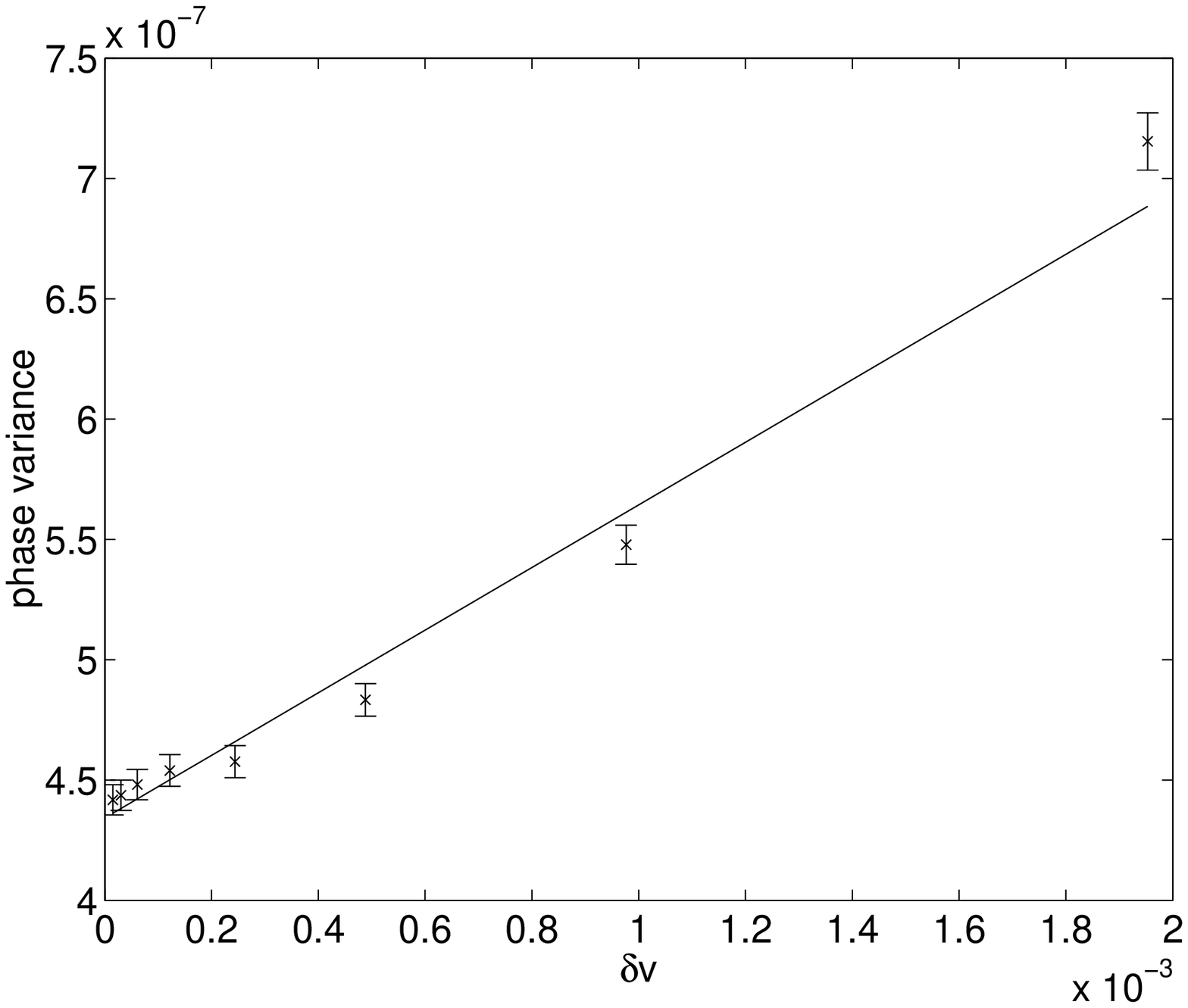}
\caption{The phase variance for mark II measurements on optimised squeezed
states with a mean photon number of 1577 with various time step sizes. The
crosses are the numerical results, and the continuous line is that fitted by a
linear regression.}
\label{delay30}
\end{figure}

\begin{figure}
\centering
\includegraphics[width=0.7\textwidth]{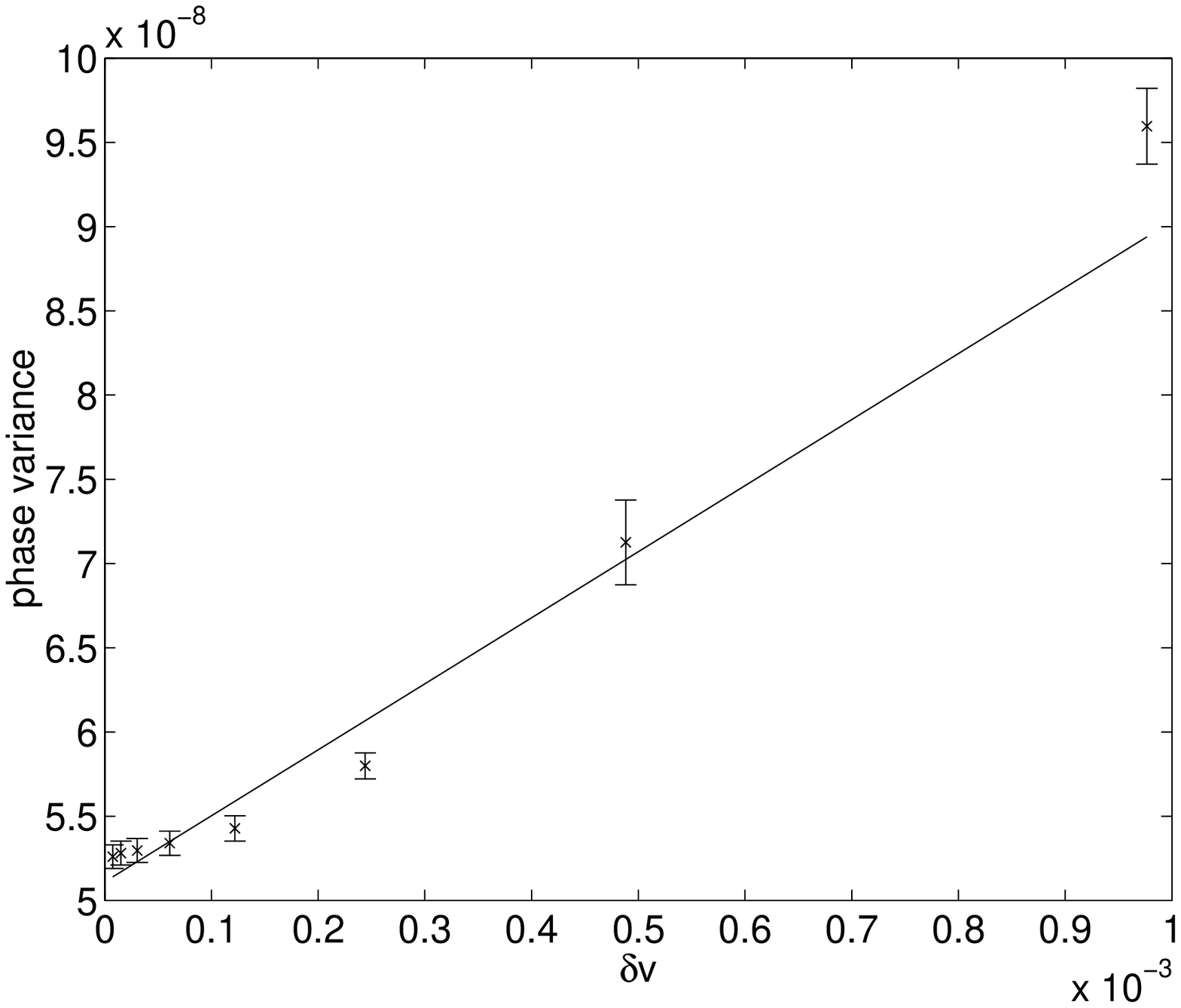}
\caption{The phase variance for mark II measurements on optimised squeezed
states with a mean photon number of 5877 with various time step sizes. The
crosses are the numerical results, and the continuous line is that fitted by a
linear regression.}
\label{delay35}
\end{figure}

The results for mean photon numbers of 122, 432, 1577 and 5877 are shown in
Figs~\ref{delay20}, \ref{delay25}, \ref{delay30} and \ref{delay35},
respectively. In these figures the straight lines shown are based on weighted
linear regressions. The slopes found for each of these graphs were
\bqa
{\rm S_{\ref{delay20}}}\!\!\!\!&=&\!\!\!\! \frac{0.180\pm 0.010}{\nb}, \nn \\
{\rm S_{\ref{delay25}}}\!\!\!\!&=&\!\!\!\! \frac{0.185\pm 0.009}{\nb}, \nn \\
{\rm S_{\ref{delay30}}}\!\!\!\!&=&\!\!\!\! \frac{0.205\pm 0.009}{\nb}, \nn \\
{\rm S_{\ref{delay35}}}\!\!\!\!&=&\!\!\!\! \frac{0.230\pm 0.012}{\nb}.
\eqa
These uncertainties are based on the uncertainties of the individual points. As
the same set of random numbers was used for each time step size, it would be
expected that the uncertainty based on the deviation of the points from the line
would be smaller. This is true for the smaller photon numbers, but not for the
larger photon numbers where the points tend to fit the line poorly.

Note that there tends to be some nonlinearity in the results for the larger
photon numbers. This is most obvious in Fig.~\ref{delay35}, where the variances
for the larger time steps are much higher than would be expected based on a
linear interpolation from the results for small time steps. In addition the
error bars are larger for these time intervals. The reason for this is that
there are a small number of results with large errors in the data sets on which
these points are based. This causes both the variance and the uncertainty in the
variance to be larger. This is also true to a lesser extent for the smaller
photon numbers.

We are only interested in the phase variance for small time steps here, as we
wish to use time steps small enough to keep the error due to the finite step
size below about 1\%. For small step sizes, the effect of this type of
nonlinearity will be that the error is less than that expected based on the
linear fits for larger time steps. For example, if the last two points in
Fig.~\ref{delay35} are omitted, the slope obtained is $0.13/\nb$.

There is a fair amount of variation between the slopes found for the different
photon numbers, and for different ranges of the linear fit. Nevertheless, the
purpose of these calculations is not to find an exact result, but rather to
place an upper limit on the error due to the finite step size. In each case, we
find that
\beq
\label{errorest}
\Delta V(\phi) < \frac{\delta v}{4 \nb}.
\eeq
This means that if we use time steps of
\begin{equation}
\delta v=\frac{\nb V(\phi)}{25},
\end{equation}
where $V(\phi)$ is the total phase variance, the error in the phase variance
should be less than 1\%.

\section{Na\"{\i}ve Constant $\varepsilon$ Feedback}
\label{naive}
The first method that I will consider is that where the values of $\varepsilon$
are the optimum values predicted by the simplified theory:
\beq
\varepsilon  = \left( 2 \alpha_{\rm opt} e^{2\zeta_{\rm opt}} \right)^2.
\eeq
For the results presented here, the states used are squeezed states optimised
for minimum intrinsic phase variance. For these states, the theoretical limit to
the total phase variance is twice the intrinsic phase variance. The variance for
mark II measurements (on general states optimised for minimum phase variance
under these measurements) does not exceed this limit until a photon number of
about 140. This does not contradict the theory, because the theoretical limit is
only expected to be accurate for large photon numbers. Because of this, the
improved phase measurement schemes will only be considered for photon numbers
above about 140.

\begin{figure}
\centering
\includegraphics[width=0.7\textwidth]{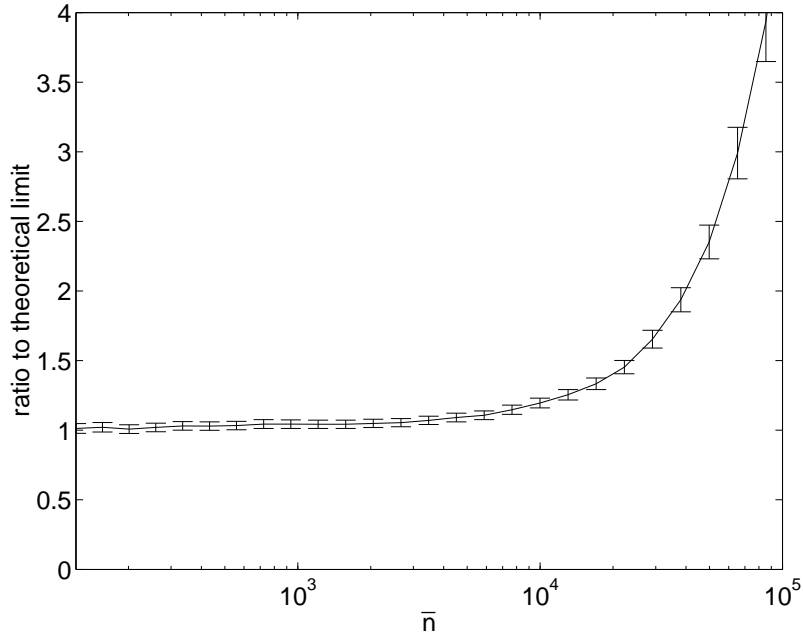}
\caption{The variance for phase measurements using the predicted optimum value
of $\varepsilon$ as a ratio to the theoretical limit.}
\label{naiveth}
\end{figure}

\begin{figure}
\centering
\includegraphics[width=0.7\textwidth]{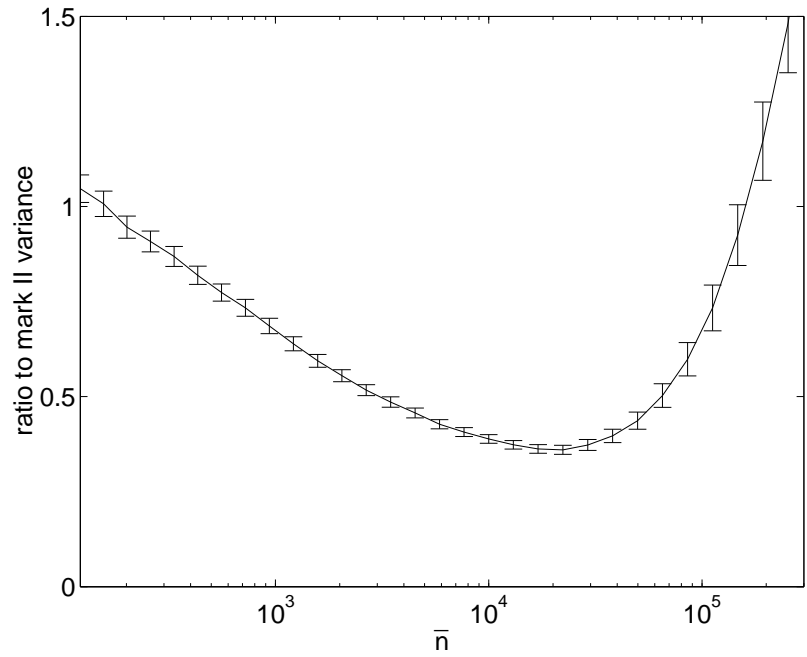}
\caption{The variance for phase measurements using the predicted optimum value
of $\varepsilon$ as a ratio to the phase variance for mark II measurements on
optimum input states.}
\label{naiveII}
\end{figure}

The results using this predicted value of $\varepsilon$ are shown as a ratio to
the theoretical limit (of twice the intrinsic phase variance) in
Fig.~\ref{naiveth}, and as a ratio to the phase variance for mark II
measurements in Fig.~\ref{naiveII}. For the results shown in these figures,
$2^{12}$ samples were used. For moderate photon numbers above 140
this feedback does give an improvement over mark II measurements, and is close
to the theoretical limit. This is true only over a small range of photon
numbers, and for photon numbers over about 5000 the phase variance is
significantly over the theoretical limit. In fact, for photon numbers above
about 160000 this feedback technique gives larger phase variances than the mark
II technique. The best improvement in the phase variance is at a photon number
of about 20000, where the phase variance is about 36\% of the mark II phase
variance.

\begin{figure}
\centering
\includegraphics[width=0.7\textwidth]{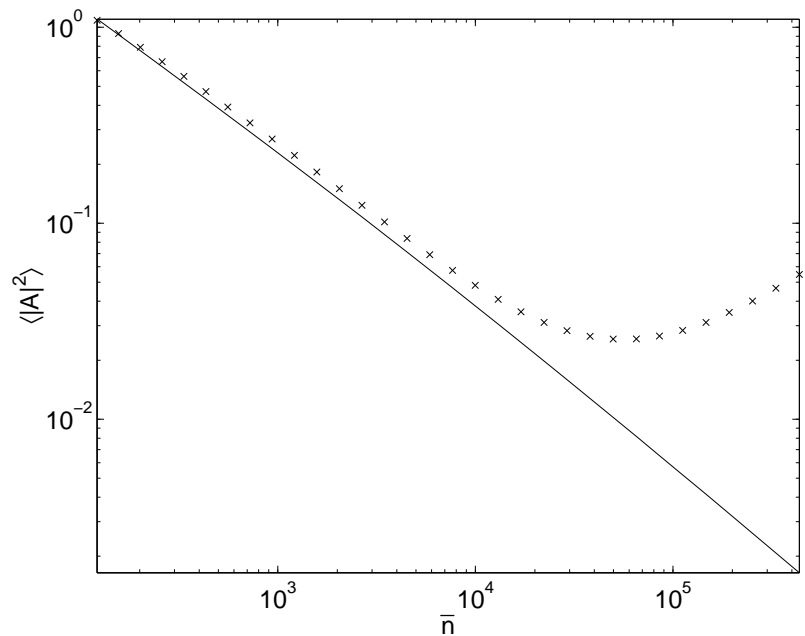}
\caption{The mean values of $|A|^2$ obtained for phase measurements using the
predicted optimum value of $\varepsilon$. The analytic result of
$\ip{|A|^2}=\varepsilon$ is shown as the continuous line and the numerically
obtained values of $\ip{|A|^2}$ are shown as crosses.}
\label{naiveA}
\end{figure}

The reason for these poor results is that the values of $|A|$ given by this
feedback technique are too high. If we plot the mean values of $|A|^2$ (see
Fig.~\ref{naiveA}), we see that although they start out close to $\varepsilon$,
the agreement gradually gets worse and worse. Now recall from the above results
for coherent states that although there is agreement with theory over a large
range of values of $\varepsilon$, for smaller values of $\varepsilon$, larger
photon numbers are required in order to have good agreement with theory.

In fact, the photon number required scales at least as $\varepsilon^{-2}$.
Conversely, we can say that the mimimum value of $\varepsilon$ for good
agreement with the theory scales roughly as $\nb^{-1/2}$. For the predicted
values of $\varepsilon$ above, we find that $\varepsilon$ scales as
$\log^2 \nb /\nb$. Clearly this means that for the larger photon numbers, the
value of $\varepsilon$ will be too small for good agreement with theory.

\section{Optimised Constant $\varepsilon$ Feedback}
\begin{figure}
\centering
\includegraphics[width=0.7\textwidth]{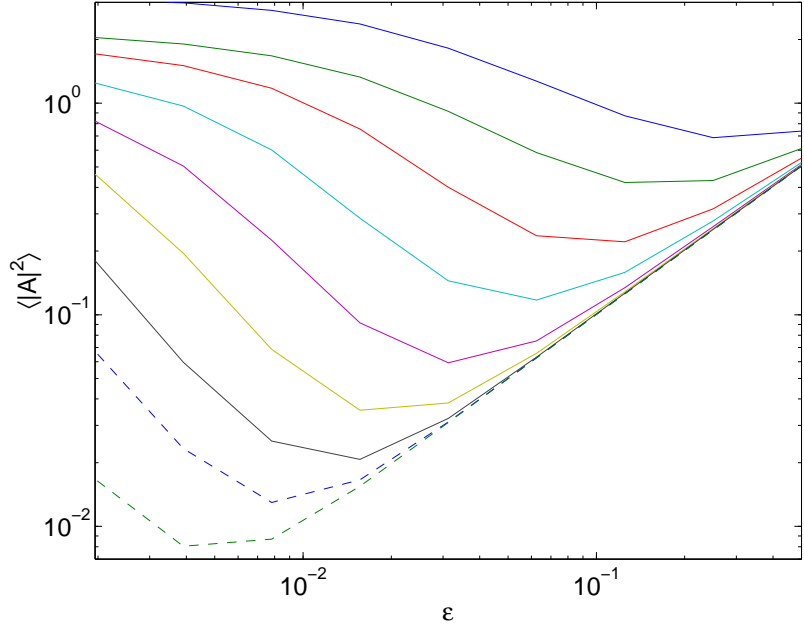}
\caption{The mean values of $|A|^2$ for measurements on coherent states with
$\arg (C_v^{1-\varepsilon} A_v^\varepsilon )$ feedback. The results for
$\alpha = 4$ are shown in dark blue, for $\alpha = 8$ in green, for
$\alpha = 16$ in red, for $\alpha = 32$ in light blue, for $\alpha = 64$ in
purple, for $\alpha = 128$ in yellow, for $\alpha = 256$ in black, for
$\alpha = 512$ as a dashed dark blue line, and for $\alpha = 1024$ as a dashed
green line.}
\label{absA2ab}
\end{figure}

In order to obtain smaller values of $|A|$, and therefore phase variances that
are closer to the theoretical limit, we can use different values of
$\varepsilon$. We would at first expect that reducing $\varepsilon$ would give
smaller values of $|A|$; however, this is not necessarily the case. As
demonstrated in Fig.~\ref{absA2ab}, as the value of $\varepsilon$ is decreased
for a fixed photon number, the value of $|A|^2$ decreases initially, but it
reaches a minimum, and then as $\varepsilon$ is decreased further $|A|^2$
increases. (This figure is for the same data as Fig.~\ref{absA2ex}.)

\begin{figure}
\centering
\includegraphics[width=0.7\textwidth]{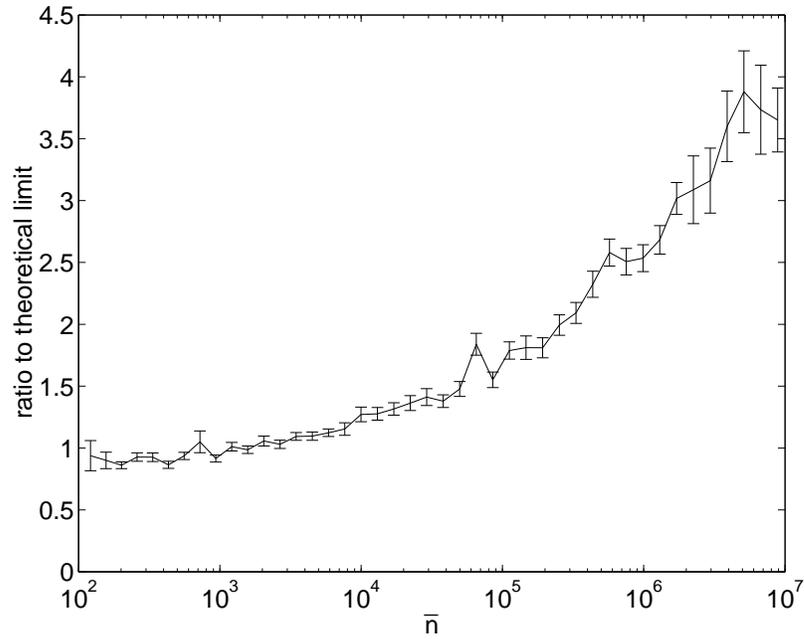}
\caption{The variance for phase measurements using the numerically determined
optimum value of $\varepsilon$ as a ratio to the theoretical limit.}
\label{adaptth}
\end{figure}

\begin{figure}
\centering
\includegraphics[width=0.7\textwidth]{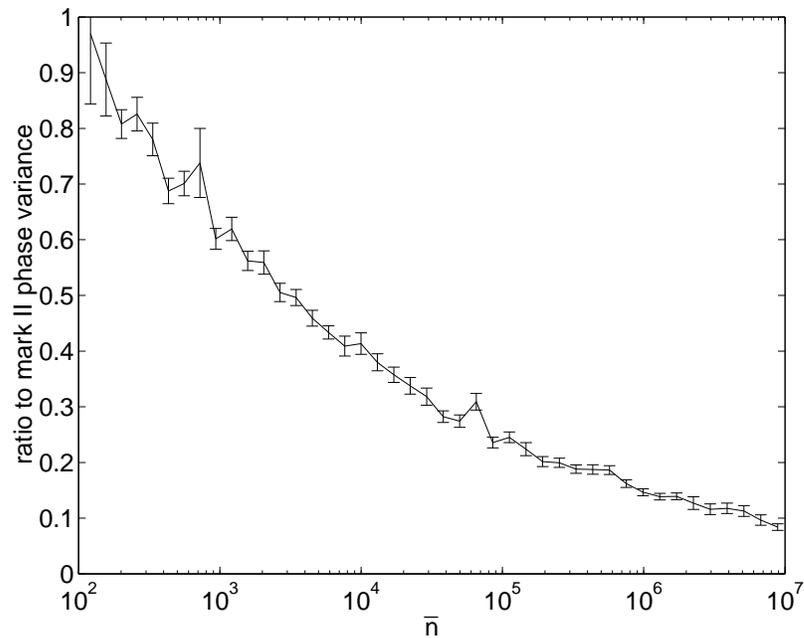}
\caption{The variance for phase measurements using the numerically determined
optimum value of $\varepsilon$ as a ratio to the mark II phase variance.}
\label{adaptII}
\end{figure}

As we are in the region where the approximate theory breaks down, rather than
using a predicted value of $\varepsilon$, it is better to vary $\varepsilon$ to
determine which value gives the smallest phase variance for each mean photon
number. The phase variances obtained by this method are
plotted in Fig.~\ref{adaptth} as a ratio to the theoretical limit, and in
Fig.~\ref{adaptII} as a ratio to the mark II phase variance. For these results
$2^{11}$ samples were used for moderate photon numbers, and $2^9$ samples were
used for photon numbers above $2\times 10^6$.

Again the phase variance is close to the theoretical limit for
photon numbers up to about 5000, but beyond this the phase variance is greater
and greater than the theoretical limit. Unlike the previous case, however,
the phase variance continues to get smaller as compared to the mark II
phase variance. For the maximum photon number calculations have been performed
for, the phase variance is less than 10\% of the mark II phase variance.

\begin{figure}
\centering
\includegraphics[width=0.7\textwidth]{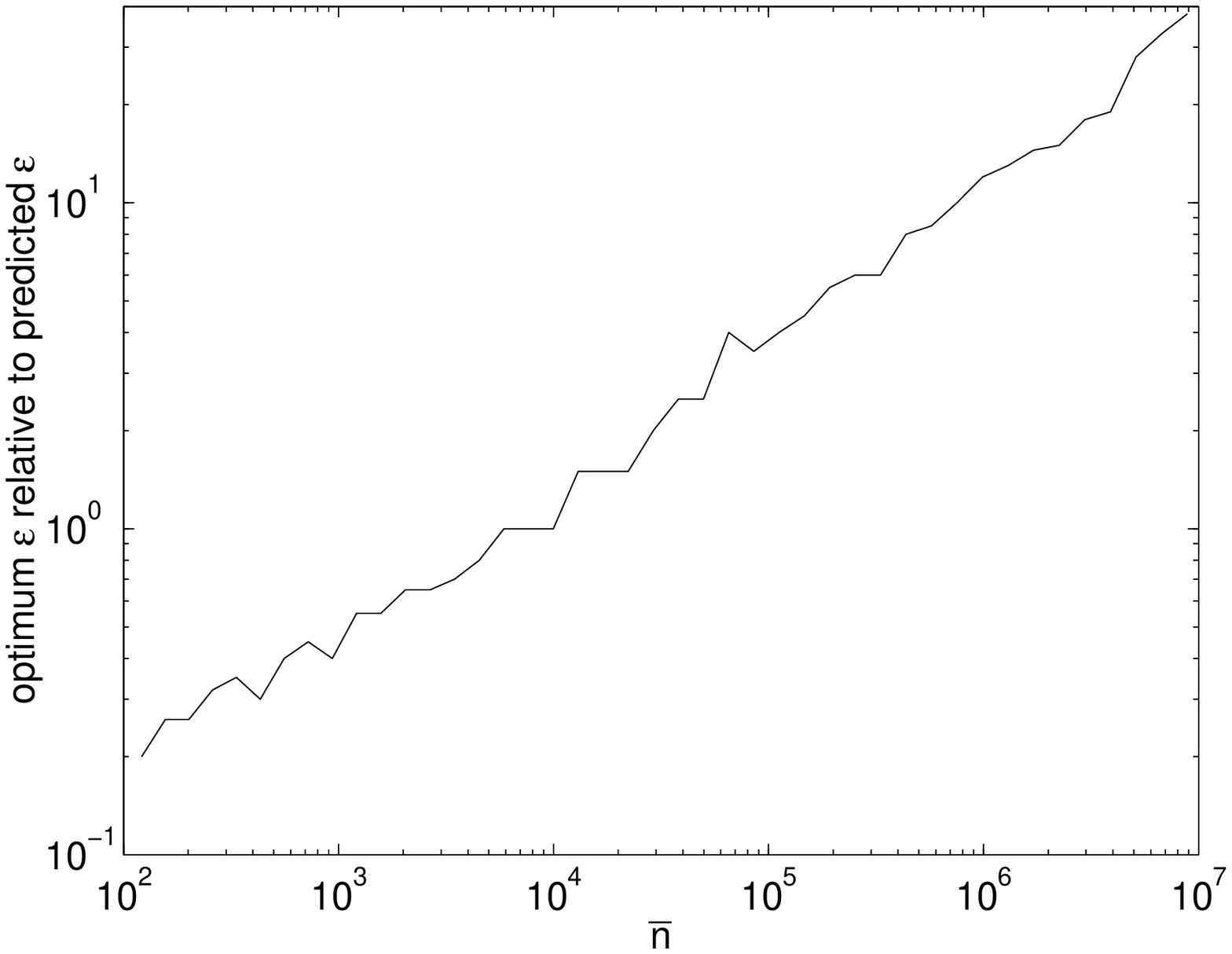}
\caption{The numerically determined optimum values of $\varepsilon$ as ratios
to the analytically predicted optimum values of $\varepsilon$.}
\label{epsilonrel}
\end{figure}

\begin{figure}
\centering
\includegraphics[width=0.7\textwidth]{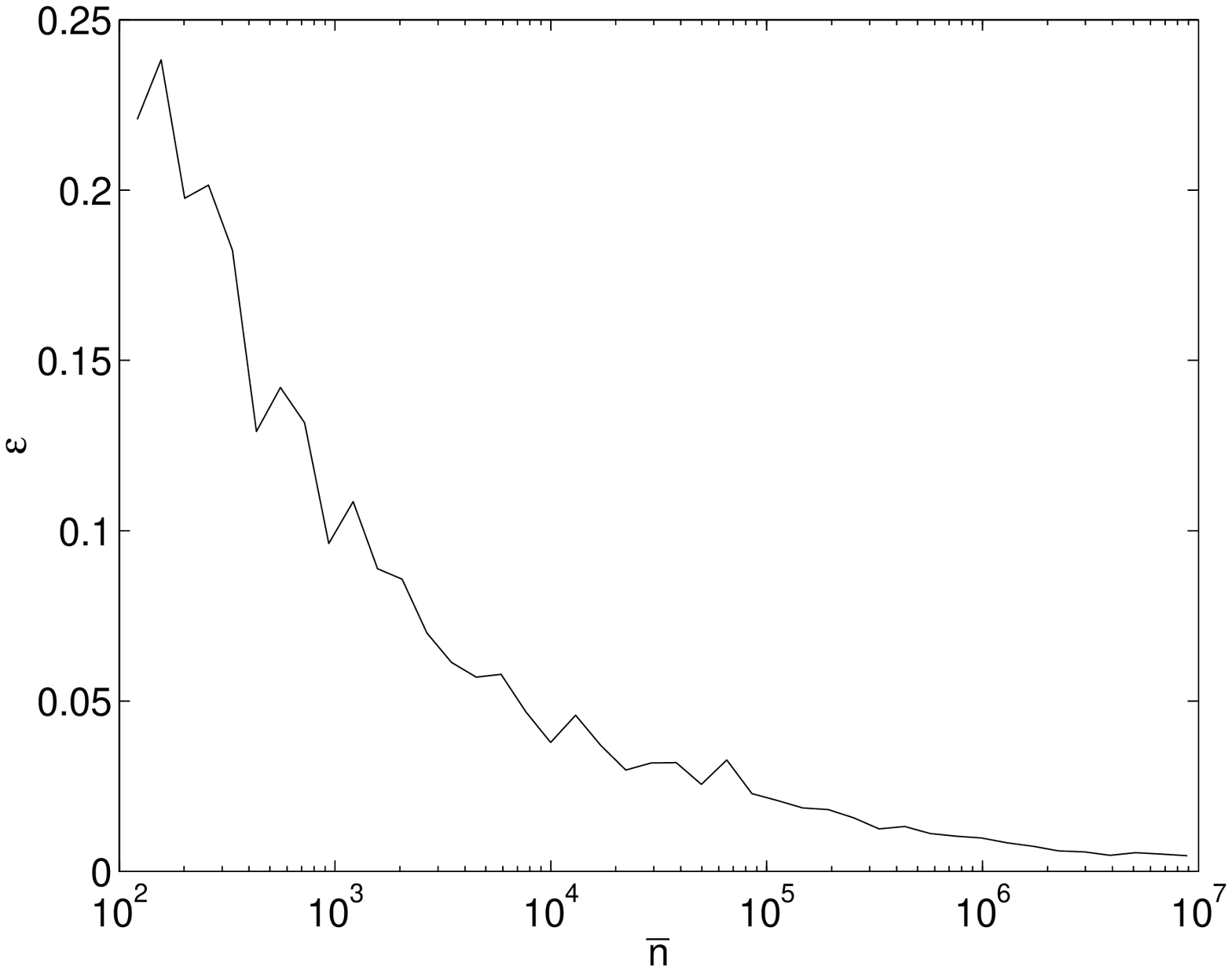}
\caption{The numerically determined optimum values of $\varepsilon$ for
$\arg (C_v^{1-\varepsilon} A_v^\varepsilon )$ feedback.}
\label{epsilon}
\end{figure}

If we plot the ratio of the numerically determined optimum value of
$\varepsilon$ to the analytically predicted value (see Fig.~\ref{epsilonrel}), we find
that for smaller photon numbers the optimum value of $\varepsilon$ is less than
the predicted value. As the photon number is increased, the optimum value of
$\varepsilon$ is greater and greater as a ratio to the predicted value. The
actual value of $\varepsilon$ is still decreasing with photon number, though,
as demonstrated in Fig.~\ref{epsilon}.

\begin{figure}
\centering
\includegraphics[width=0.7\textwidth]{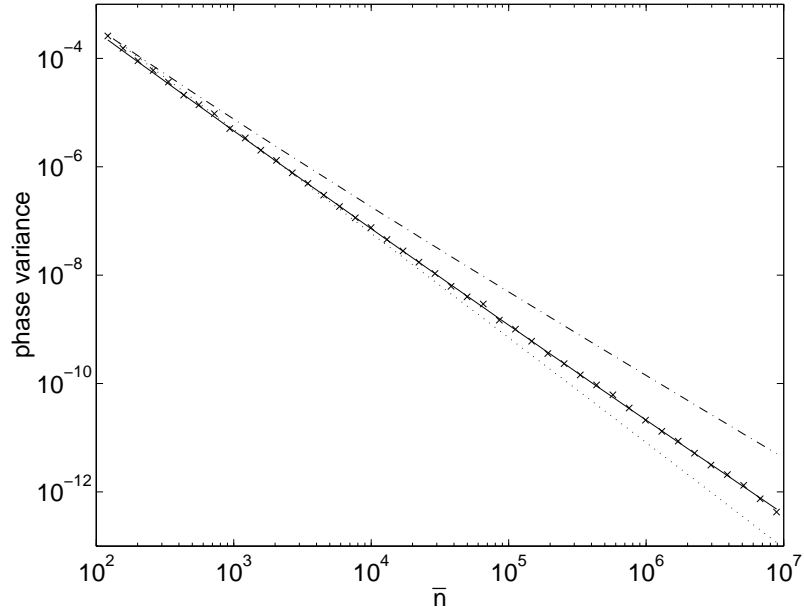}
\caption{The phase variance for $\arg (C_v^{1-\varepsilon} A_v^\varepsilon )$
feedback using the numerically determined optimum values of $\varepsilon$. The
numerical results are shown as crosses and the continuous line is that fitted to
the data. The mark II phase variance (dash-dotted line) and theoretical limit
(dotted line) are also shown.}
\label{constant}
\end{figure}

These results indicate that although this phase feedback scheme is not achieving
the theoretical limit, it is achieving a better scaling than for mark II
measurements. Assuming that $h(n)\approx cn^{-p}$, a numerical fit was performed
on the data to determine $c$ and $p$. Recall from Sec.~\ref{sqzgeneral} that
the approximate phase variance for general measurements on squeezed states is
\beq
V(\phi) \approx V(\phi_{\rm can})
+ 2c\nb^{-p}\left[ 1+\frac {p(p+1)}{2n_0} \right].
\eeq
Rather than simply assuming that the introduced phase variance was $2c\nb^{-p}$,
this expression was used in order to obtain more accurate results. In addition,
only data points for photon numbers above about 10000 were used, as the data for
small photon numbers gave a poor approximation of the asymptotic result. It was
found that the best fit was for $c = 0.095\pm 0.008$ and $p = 1.685\pm 0.007$.
The data and the fitted line along with the mark II case and the theoretical
limit are shown in Fig.~\ref{constant}. As can be seen, the power law is a
very good fit for the data.

Note that, because the value of $\varepsilon$ used depends on the photon number,
this phase measurement scheme can not be described by a single POM. Instead, the
actual POM and $H$ matrix will depend on the photon number. Nevertheless, it is
reasonable to consider these measurements as approximately equivalent to a
measurement scheme with a single POM, and the variation of $h(n)$ as given
above.

\section{Time Dependent $\varepsilon$}
\label{timedepeps}
\subsection{Simple Method}
\label{simplemethod}
In order to improve on this result, the remaining alternative is to vary
$\varepsilon$ during the measurement. As the approximate theory above does not
give any information about how $\varepsilon$ should be varied during the
measurement, trial and error was used initially. It was found that an
improvement over the constant $\varepsilon$ case could be obtained if
$\varepsilon$ was increased linearly during the measurement. Even better
improvements were obtained when $\varepsilon$ was increased proportionally to
$v^2$ or $v^3$, or some higher power of time.

A possible reason why $\varepsilon$ should be increased during the measurement
can be deduced from Eq.~(\ref{sines}). In the case that we are considering
a squeezed state, rather than a coherent state, the phase of $\alpha_v$ is now
time dependent. The phase estimate $\arg C_v$ gives an estimate of the initial
phase, and takes no account of the variation of the phase of the running value
$\alpha_v$. This means that when the variance of $\arg \alpha_v$ is large, the
intermediate phase estimate may not be between the phase of $\alpha_v$ and
$A_v$, even though it is between the phases of $C_v$ and $A_v$.

If the variance of the phase estimate is kept sufficiently above the variance of
$\arg \alpha_v$ during the measurement, then this problem should be corrected.
Therefore a statistical phase feedback scheme was used, where at each time the
value of $\varepsilon$ was chosen such that the variance in the phase estimate
was at or below some fixed multiple of the variance of $\arg \alpha_v$. These
variances were determined numerically from a large number ($2^{12}$) of
simultaneous calculations. The best multiple to use cannot be predicted, and was
determined numerically from these samples.

\begin{figure}
\centering
\includegraphics[width=0.7\textwidth]{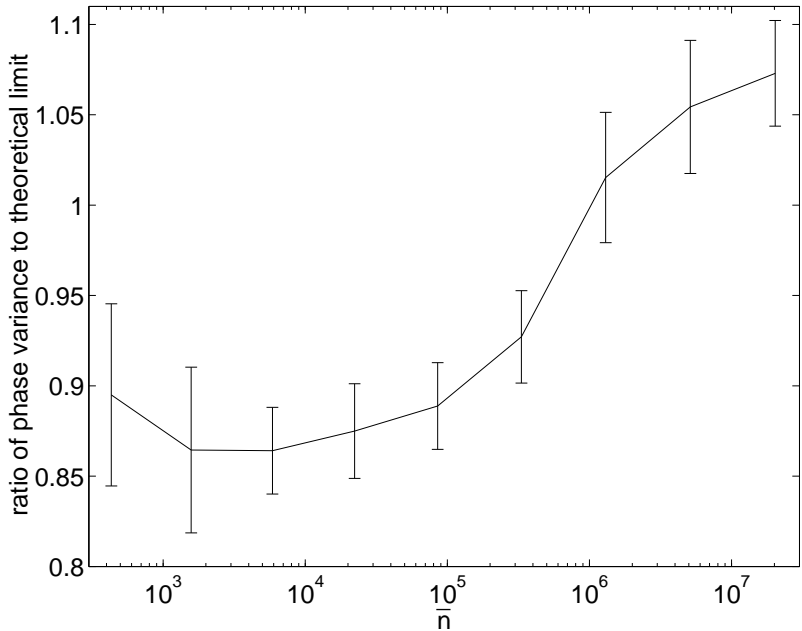}
\caption{The phase variance as a ratio to the theoretical limit for the
feedback scheme where $\varepsilon$ is chosen such that the variance of the
intermediate phase estimate is no less than a fixed multiple of the variance of
$\arg \alpha_v$.}
\label{statistical}
\end{figure}

The results obtained using this method are shown as a ratio to the theoretical
limit of twice the phase variance of an optimised squeezed state in
Fig.~\ref{statistical}. The results are far better than for the
case where $\varepsilon$ is kept constant, with the phase variance being below
or close to the theoretical limit for all the photon numbers tested. Even for
the largest photon number it was feasible to perform calculations for the phase
variance is less than 8\% above the theoretical limit.

\begin{figure}
\centering
\includegraphics[width=0.7\textwidth]{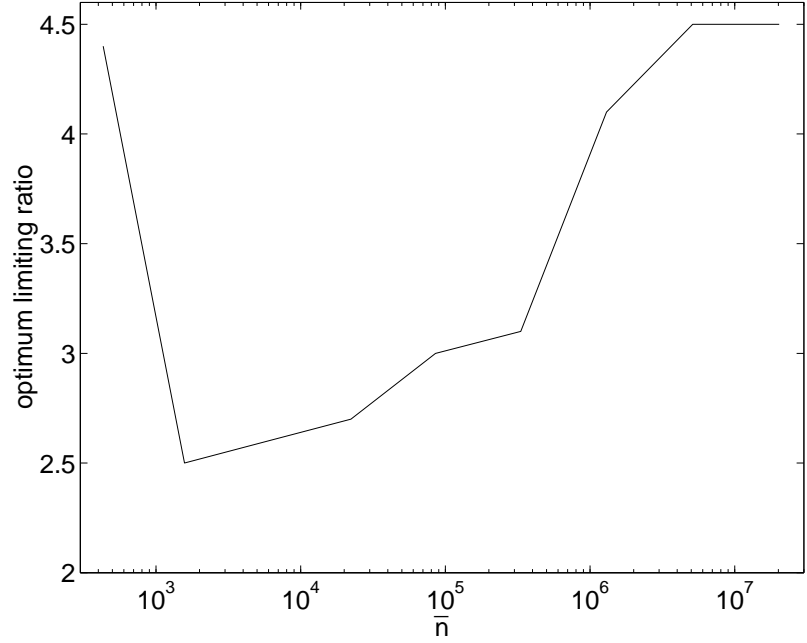}
\caption{The optimum limiting ratio between the variance of the phase estimate
and the variance of $\arg \alpha_v$.}
\label{optratio}
\end{figure}

The optimum limiting ratio between the phase variances turns out to be different
for the different photon numbers. The optimum values obtained are shown in
Fig.~\ref{optratio}. The minimum value is around $2.5$, which is about what
would be expected in order to prevent the phase estimates being outside the
interval between $\arg \alpha_v$ and $\arg A_v$. The optimum limiting ratio can
be much higher, however, and generally increases with photon number.

Another factor is that initially $\arg C_v$ will be a poor phase estimate, and
therefore the phase estimate may not be between the phase of $\alpha_v$ and
$A_v$ if $\varepsilon$ is too small. This would seem to indicate that
$\varepsilon$ should be large initially to take account of this. It was found
that this in fact made the measurement poorer. This indicates that the
situation is not as simple as indicated in the analysis above.

This phase measurement scheme, although it provides measurements at or very
close to the theoretical limit, is unsatisfactory because the values of
$\varepsilon$ are determined statistically from a large number of measurements,
rather than determined from the individual values of $A_v$, $B_v$ and $v$. It is
possible to determine the variation of $\varepsilon$ numerically in this way for
a particular photon number, and use this variation to obtain results close to
the theoretical limit for independent measurements. This requires knowledge of
the photon number, however, and the variation takes a long time to calculate.

In order to determine the values of $\varepsilon$ to use in a simple way, we can
consider the approximate variation of the variance of $\arg \alpha_v$. Taking
the system phase to be zero, and making the approximations that $e^{i\Phi}
\approx i$ and $B_v^{\rm S}$ is approximately real, Eq.~(\ref{SDEalpha}) simplifies to
\beq
d\alpha_v \approx -\frac i{1-v} \frac{B_v^{\rm S} dW}{1+B_v^{\rm S}}.
\eeq
Using the approximation $B_v^{\rm S} \approx 1$, this becomes
\beq
d\alpha_v \approx -\frac i{1-v} \frac{dW}2 .
\eeq
Using this to determine the variation in the phase of $\alpha_v$ gives
\bqa
d\arg \alpha_v \!\!\!\! &=& \!\!\!\! {\rm Im} [d \log \alpha_v] \nn \\
\!\!\!\! &=& \!\!\!\! {\rm Im} \left[ \frac{d \alpha_v}{\alpha_v}-
\frac{(d \alpha_v)^2}{2\alpha_v^2} \right] \nn \\
\!\!\!\! & \approx & \!\!\!\! {\rm Im} \left[ \frac{\frac{-i}{1-v}\frac{dW}2}
{\alpha_v}+\frac{\frac 1{(1-v)^2} \frac {dv}4}{2\alpha_v^2} \right].
\eqa
Assuming that $\alpha_v$ remains close to its initial value $\alpha$, which is
real, this simplifies to
\beq
d\arg \alpha_v \approx \frac{-dW}{2(1-v)\alpha}.
\eeq
Using this, the increment in the expectation value of $(\arg \alpha_v)^2$ is
\beq
d\ip{(\arg \alpha_v)^2} \approx \frac{dv}{4\alpha^2}\frac 1{(1-v)^2}.
\eeq
This gives us
\beq
\ip{(\arg \alpha_v)^2} \approx \frac{1}{4\alpha^2}\frac 1{(1-v)}.
\eeq

During the course of the measurement we wish to keep the variance of the phase
estimate at some constant, $k$, times the variance of the phase of $\alpha_v$.
Therefore the variance in the phase estimate should be
\beq
\ip{\hat \varphi_v^2} \approx \frac{k}{4\alpha^2}\frac 1{(1-v)}.
\eeq
As the simplest approximation, the variance of $\arg A_v$ can be assumed to be
the same as for the mark I case, where it is $1/(4\alpha \sqrt v)$.
Approximating the phase estimate as $\varepsilon(v) \arg A_v$, this gives
\beq
\varepsilon(v)^2\frac 1{4\alpha\sqrt v}\approx\frac{k}{4\alpha^2}\frac 1{(1-v)}.
\eeq
This implies that the value of $\varepsilon(v)$ should be
\beq
\varepsilon(v) \approx \sqrt{\frac k{\alpha}}\sqrt{\frac {\sqrt v}{1-v}}.
\eeq
This approximation is a bit too simplistic, as the variance of $\arg A_v$
increases as $\varepsilon$ is decreased. Alternatively, using the result given
in Eq.~(\ref{phi2sol}), we find
\beq
\varepsilon(v) \approx \left( \frac k{\alpha}\right)^2 \frac v{(1-v)^2}.
\eeq
This result is not necessarily any more accurate than the previous one, however,
as the solution (\ref{phi2sol}) is only accurate for constant $\varepsilon$.

\begin{figure}
\centering
\includegraphics[width=0.7\textwidth]{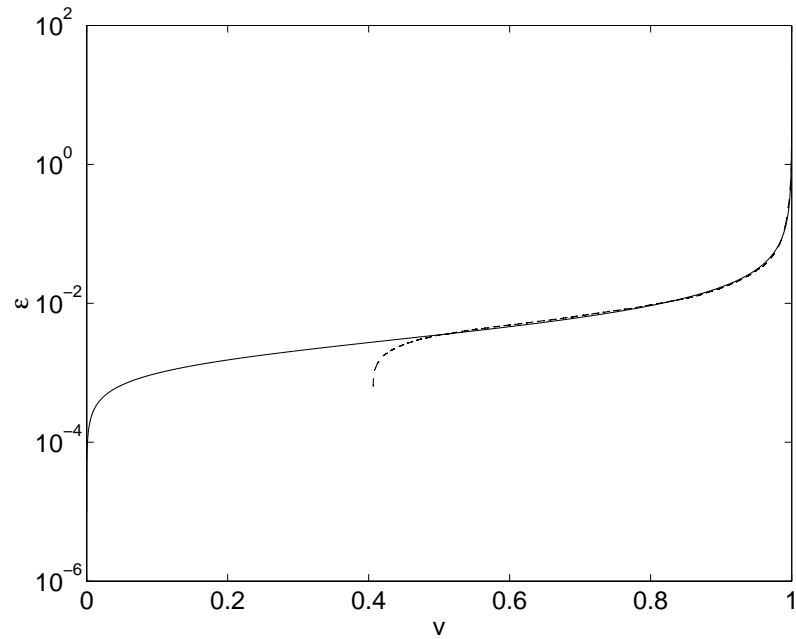}
\caption{The values of $\varepsilon$ found for a mean photon number of 22255 and
a limiting ratio of $2.7$. The dashed line is the numerical results and the
continuous line is an analytic approximation.}
\label{epsilonplot}
\end{figure}

The full differential equations taking into account the time dependent
$\varepsilon$ are extremely difficult to solve, and do not appear to have a
simple solution of the above form. Therefore we will instead consider the
numerical results for the full calculation. The time dependence of
$\varepsilon(v)$ for a mean photon number of about 22255 is shown in
Fig.~\ref{epsilonplot}. The analytic approximation shown was found by trial and
error, and is
\beq
\varepsilon(v) = \frac 1{350}\frac{\sqrt v}{(1-v)^{0.8}}.
\eeq
The dividing factor of 350 here is around twice the value of $\alpha$.

The common features of the above expressions for $\varepsilon$ are a factor of
$v$ to some power in the numerator, and factors of $1-v$ and $\alpha$ to some
power in the denominator. Various expressions with these features were tested,
but the one that gave the best results was
\beq
\varepsilon(v) = \frac 1{\alpha} \sqrt{\frac v{1-v}}.
\eeq
Note that this is very similar to the above analytic approximation for the exact
results, except that $1-v$ is taken to the power of $0.5$ rather than $0.8$ (as
well as a difference of a multiplicative factor).

This expression suffers from the slight drawback that it is dependent on the
value of $\alpha$, rather than the experimentally measured quantities. This is
easily corrected by using the slightly modified expression,
\begin{equation}
\label{uncorfeed}
\varepsilon(v) = \frac {v^2-|B_v|^2}{|C_v|} \sqrt{\frac v {1-v}}.
\end{equation}
Here $|C_v|/(v^2-|B_v|^2)$ is used as an estimator for $\alpha$.

\begin{figure}
\centering
\includegraphics[width=0.7\textwidth]{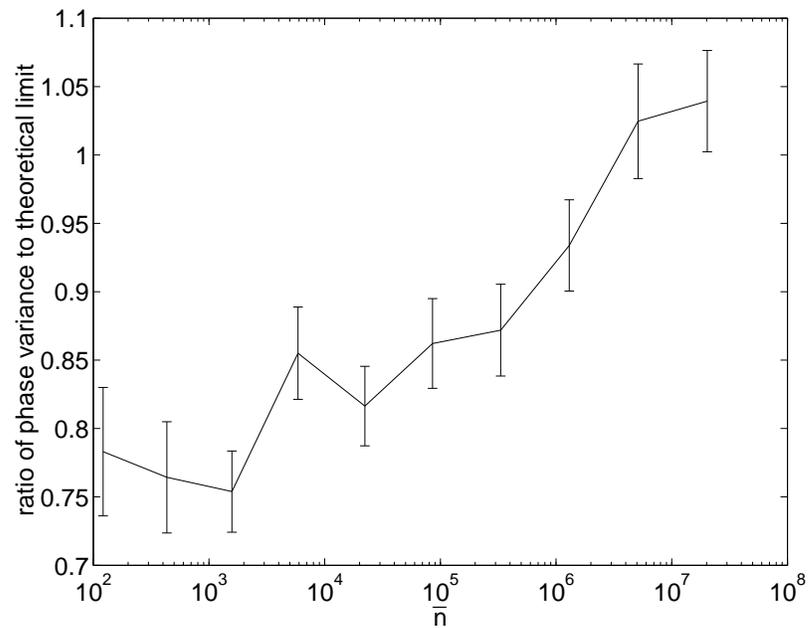}
\caption{Variance for phase measurements with a time dependent $\varepsilon$
given by Eq.~(\ref{uncorfeed}) plotted as a function of the photon number of the
input state. The phase variance is plotted as a ratio to the theoretical minimum
phase variance (i.e.\ twice the intrinsic phase variance).}
\label{ratio}
\end{figure}

The results for this method are shown in Fig.~\ref{ratio} as a ratio to the
theoretical limit. As this figure shows, the results are very close to the
theoretical limit, and even for the largest photon number for which calculations
have been performed the phase uncertainty is only about 4\% above the
theoretical limit. For these results $2^{11}$ samples were used. This sample
size was used for the rest of the results in this chapter, unless otherwise
stated.

If the integration time step is reduced, while keeping the time interval at
which the phase estimates are updated constant, the phase variance converges.
If, however, the phase estimates are updated at smaller and smaller time
intervals then the phase variance does not converge. For example the phase
uncertainty for measurements on an optimised squeezed state with a photon
number of 1577 is $1.54\times 10^{-6}$ if we use the time steps given above.
If we use time steps that are a hundred times smaller, then the phase
variance is $1.93\times 10^{-6}$, and if the time steps are a thousand times
smaller the phase variance is $2.13\times 10^{-6}$. These results indicate that
the phase estimates must be incremented in finite time intervals for this
method to give good results, and the size of the time steps that should be used
depends on the photon number. The phase variance is not strongly dependent on
these time steps, however, and only an order of magnitude estimate of the
photon number would be required.

\subsection{Evaluation of Method}
\label{evaluation}
A problem with determining the phase variance by the method above is that for
highly squeezed states (that are close to optimised for minimum phase
variance), a significant contribution to the phase variance is from low
probability results around $\pi$. In obtaining numerical results the actual
phase variance for the measurement will tend to be underestimated because the
results from around $\pi$ are  obtained too rarely for good statistics. It
would require an extremely large number of samples to estimate this
contribution. However, we can estimate it non-statistically as follows.

\begin{figure}
\centering
\includegraphics[width=0.7\textwidth]{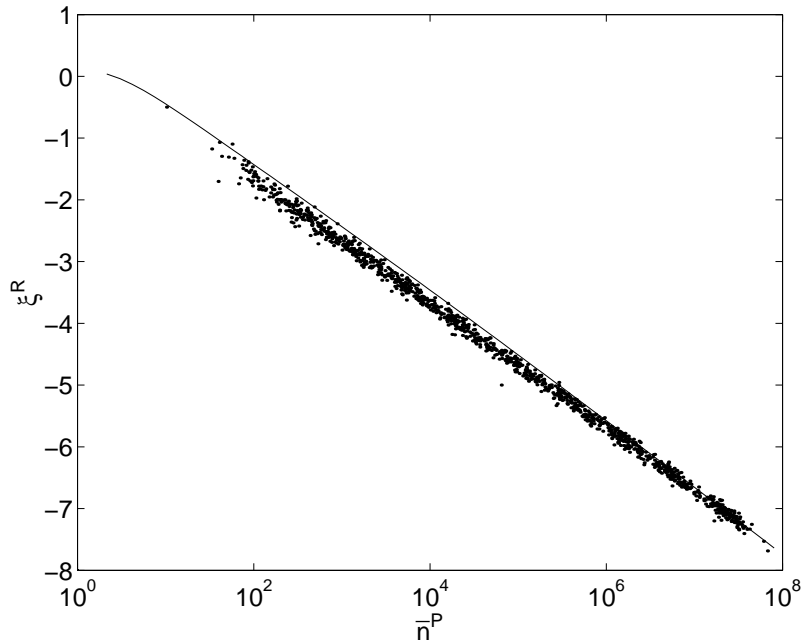}
\caption{Values of $\xi^{\rm R}$ and $\nb^{\rm P}$ (calculated from $A$ and $B$) resulting
from measurements on squeezed states of various mean photon numbers. The
variation of $\xi^{\rm P}$ with $\nb^{\rm P}$ for optimum squeezed states is also
plotted (continuous line).}
\label{zeta}
\end{figure}

Recall that in order to have a measurement that is close to optimum, the
multiplying factor $Q'(\nb^{\rm P},\xi^{\rm P})$ should give values of $\xi^{\rm P}$ for each
$\nb^{\rm P}$ that are close to optimised for minimum phase uncertainty. To test
this for the phase measurement scheme described above, the $\nb^{\rm P}$ and $\xi^{\rm P}$
were determined from the values of $A$ and $B$ from the samples. The resulting
data along with the line for optimum $\xi^{\rm P}$ are plotted in Fig.~\ref{zeta}. The
imaginary part of $\xi^{\rm P}$ should be zero for optimum measurements, and is small
for these results. Therefore in Fig.~\ref{zeta} the real part, $\xi^{\rm R}$, is
plotted. As can be seen, the vast majority of the data points are below the
line, indicating greater squeezing than optimum. (There are more points above
the line for large $\nb$; more will be said about this later.) This means that
if the low probability results around $\pi$ are taken into account the phase
variance for these measurements will be above the theoretical limit.

I will firstly consider the effect of variations in the modulus of $\xi^{\rm P}$,
leaving consideration of error in the phase till later. In order to estimate how
far above the theoretical limit the actual phase variance is, we can make a
quadratic approximation to the expression for the phase variance. From
\cite{collett} the expression for the intrinsic phase variance of a squeezed
state is, for real $\zeta$,
\begin{equation}
V(\phi) \approx \frac{e^{2\zeta}}{4\bar n}+\frac 1{4\bar n^2}+ 2 \erfc
(\sqrt{2\bar n}e^\zeta).
\end{equation}
Taking the derivative with respect to $\zeta$ gives
\begin{equation}
\label{firstder}
\frac d{d\zeta}V(\phi) \approx \frac{e^{2\zeta}}{2\nb}-4e^\zeta
\sqrt{\frac{2\nb}{\pi}}e^{-2\nb e^{2\zeta}}.
\end{equation}
Taking the second derivative we obtain
\beq
\frac {d^2}{d\zeta^2}V(\phi) \approx \frac{e^{2\zeta}}{\nb}-4e^\zeta
\sqrt{\frac{2\bar n}{\pi}}e^{-2\bar n e^{2\zeta}} - 4e^\zeta
\sqrt{\frac{2\bar n}{\pi}}(-4\bar n e^{2\zeta})e^{-2\bar n e^{2\zeta}}.
\eeq
Using the fact that the first derivative (\ref{firstder}) is zero for minimum
phase variance gives
\beq
4e^\zeta\sqrt{\frac{2\nb}{\pi}}e^{-2\nb e^{2\zeta}} = \frac{e^{2\zeta}}{2\nb}.
\eeq
Using this result, the expression for the second derivative simplifies to
\bqa
\frac {d^2}{d\zeta^2}V(\phi) \!\!\!\! &\approx& \!\!\!\! \frac{e^{2\zeta}}{\nb}
-\frac{e^{2\zeta}}{2\nb} + 4\bar n e^{2\zeta} \frac{e^{2\zeta}}{2\nb} \nn \\
\!\!\!\! &=& \!\!\!\! \frac{e^{2\zeta}}{2\nb} + 2e^{4\zeta} \nn \\
\!\!\!\! &=& \!\!\!\! \frac{n_0}{2\bar n^2}\left(1+4n_0\right).
\eqa
This means that for values of $\zeta$ close to optimum the increase in the
phase variance over the optimum value is
\begin{equation}
\Delta V(\phi) \approx (\Delta |\zeta|)^2 \frac{n_0}{4\nb^2}(1+4n_0).
\end{equation}
Here the absolute value of $\zeta$ has been used for greater generality.

The main contribution to the phase uncertainty is $n_0/(4\bar n^2)$, so the
increase in the phase uncertainty as a ratio to the minimum phase uncertainty
is approximately
\begin{equation}
\frac{\Delta V(\phi)}{V_{\rm min}(\phi)} \approx (\Delta|\zeta|)^2 (1+4n_0).
\end{equation}
Note that as the photon number is increased, $n_0$ increases roughly as
$\log \nb$. This means that in order for the percentage increase in the excess
phase uncertainty to remain limited, the error in $|\zeta|$ must decrease with
photon number.

Now I will use the superscript P to indicate specifically the squeezed state
in the POM. Since $|\zeta^{\rm P}|=|\xi^{\rm P}|$, we obtain
\begin{equation}
\Delta V(\phi) \approx (\Delta|\xi^{\rm P}|)^2 \frac{n_0}{4\nb^2}(1+4n_0).
\end{equation}
Here I have used the values of $n_0$ and $\nb$ for the input state, rather than
the squeezed state in the POM. As the input states considered here are optimum
squeezed states, the average values of these quantities for the squeezed state
in the POM should be close to those for the input state. For a more general
input state we would still use the value of $\nb$ for the input state, but the
value of $n_0$ should be chosen as the corresponding value for an optimum
squeezed state with mean photon number $\nb$.

This estimate indicates that the actual phase variance for the measurement
scheme described above can be significantly larger than the intrinsic phase
variance. For example, for a mean photon number of about 332000, the rms
deviation of $|\xi^{\rm P}|$ from the optimum value is only about 0.16, but a
squeezed state with $|\xi^{\rm P}|$ differing this much from optimum will have a
phase variance more than twice the optimum value. This indicates that if the
low probability results around $\pi$ are taken into account, the introduced
phase variance is actually more than twice the theoretical limit.

Now I will estimate the contribution from error in the phase (rather than the
modulus) of $\xi^{\rm P}$. Recall that in attempting to estimate the phase of a state,
we measure quadratures close to $\pi/2$. For the $\pi/2$ quadrature
\beq
\hat X_{\pi/2} = -i (a - a\dg ),
\eeq
the expectation value is
\beq
\ip{\hat X_{\pi/2}} = 2\st{\alpha} \sin \varphi.
\eeq
For small $\varphi$, this means that
\beq
\ip{\hat X_{\pi/2}} \approx 2\st{\alpha} \varphi.
\eeq
Canonical measurements can be considered by assuming that $\alpha$ is known, and
that the $\pi/2$ quadrature is proportional to the phase. This would mean that
a measurement of $\hat X_{\pi/2}$ is equivalent to a direct measurement
of the phase.

For simplicity the system phase will now be taken to be zero. The estimated
phase is then proportional to the measured value of $\hat X_{\pi/2}$, and the
variance in the phase estimate will be proportional to the variance in the
$\pi/2$ quadrature:
\beq
\ip{\Delta \hat X_{\pi/2}^2} \approx 4\nb \ip{\Delta \phi^2}.
\eeq
Here the approximation $\nb \approx |\alpha|^2$ has been used. This is
reasonably accurate for the states that are considered here. Now recall that the
uncertainty in this quadrature for a squeezed state is
\beq
\ip{\Delta \hat X_{\pi/2}^2} = e^{2r} \cos^2
\half\phi_\zeta + e^{-2r} \sin^2 \half\phi_\zeta,
\eeq
where $r$ and $\phi_\zeta$ are the magnitude and phase of $\zeta$.

Therefore the uncertainty in the phase is 
\beq
\ip{\Delta \phi^2} = \frac{e^{2r} \cos^2\half\phi_\zeta + e^{-2r} \sin^2 \half
\phi_\zeta}{4\nb},
\eeq
If the phase of $\zeta$ is close to $\pi$ (so $\zeta$ is close to negative
real), we can make the approximation
\begin{equation}
\ip{\Delta\phi^2}\approx\frac{e^{-2r}+e^{2r}\frac{(\Delta \phi_\zeta)^2}4}
{4\nb},
\end{equation}
where $\Delta\phi_\zeta=\phi_\zeta-\pi$. The first term in the numerator is
identical to the first term for the intrinsic phase variance of a squeezed
state. Clearly the second term is the excess phase variance due to the error in
the phase of $\zeta$. Therefore the extra phase variance due to error in the
phase of $\zeta$ is given by
\begin{equation}
\Delta V(\phi) \approx \frac{(\Delta \arg \zeta)^2}{16 n_0}.
\end{equation}

Note that the error in the phase of $\xi$ is equivalent to the error in the
phase of $\zeta$ above, as the above case is for a system phase of zero. Now
using the superscript P to indicate specifically the squeezed state in the
POM, we have
\begin{equation}
\Delta V(\phi) \approx \frac{(\Delta \arg \xi^{\rm P})^2}{16 n_0}.
\end{equation}
Here we are again using the value of $n_0$ from the input state. Using this
estimate on the example used for the magnitude, with $\nb\approx 332000$, it can
be seen that this is not so much of a problem, with the introduced phase
uncertainty being increased by less than 3\% by this factor.

\subsection{Corrected Method}
\label{improved}
The problem of the large contribution from low probability results around
$\pi$ can be effectively eliminated by using corrections near the end of the
measurement. In order to describe this we must firstly consider the values of
$\alpha^{\rm P}$ and $\zeta^{\rm P}$ for intermediate times. From Ref.~\cite{wise96},
the POM for intermediate times is
\beq
F_v(A_v,B_v) = Q_v(A_v,B_v) \exp(\half B_v {a\dg}^2+A_v a\dg) (1-v)^{a\dg a}
\exp(\half B_v^* a^2+A_v^* a).
\eeq
This is mixed, not pure, and it is therefore not possible to simplify it to an
expression in terms of squeezed states, as is the case at the end of the
measurement. I will therefore define the values of $\alpha_v^{\rm P}$ and $\zeta_v^{\rm P}$
by analogy with the definitions at the end of the measurement, but these will
not actually correspond to squeezing parameters for a squeezed state in the POM.

Recall from the introduction that for a coherent state we have
\beq
v A_v+B_v A_v^* = \alpha \left( v^2-|B_v|^2 \right) + i \left( v \sigma_v - B_v
\sigma_v^* \right),
\eeq
so
\beq
\alpha = \frac{v A_v+B_v A_v^*}{v^2-|B_v|^2} - \frac{i \left( v \sigma_v - B_v
\sigma_v^* \right)}{v^2-|B_v|^2}.
\eeq
This means that $C_v/(v^2-|B_v|^2)$ can be used as an estimator for $\alpha$.
The situation is more complicated for squeezed states, as there are extra
terms due to the stochastic variation of $\alpha$. It is still possible to
use $C_v/(v^2-|B_v|^2)$ as an estimator for $\alpha$, however.
As $\alpha^{\rm P}$ is defined by the value of $C_v/(v^2-|B_v|^2)$ for $v=1$, it is
therefore reasonable to define intermediate values $\alpha_v^{\rm P}$ by
\beq
\label{defalvP}
\alpha_v^{\rm P} = \frac{A_v v+B_v A_v^*}{v^2-|B_v|^2}.
\eeq

For the case of $\zeta_v^{\rm P}$, note that $\zeta^{\rm P}$ is very large when all the
feedback phases are close to each other, and $B$ is correspondingly close to 1.
In order for $\zeta_v^{\rm P}$ to have the same property, the argument of the atanh
function should be close to 1 when the phase estimates are very close together.
As the maximum value $B_v$ can have is $v$, this will be the case if $B_v$ is
divided by $v$. Using this, the definition of $\zeta_v^{\rm P}$ is
\beq
\label{defzevP}
\zeta_v^{\rm P} = -\frac{B_v}{|B_v|} \atanh \left( \frac{|B_v|}v \right).
\eeq

Note that this relation of $\alpha_v^{\rm P}$ and $\zeta_v^{\rm P}$ to $A_v$ and $B_v$ is
{\it not} analogous to the relation of $\alpha_v$ and $\zeta_v$ to $A_v^{\rm S}$ and
$B_v^{\rm S}$. The quantities $A_v^{\rm S}$ and $B_v^{\rm S}$ are defined by analogy to $A_v$ and
$B_v$ at time $v=1$. The time varying values of $A_v^{\rm S}$ and $B_v^{\rm S}$ are found
by using the same definition with the time varying squeezing parameters.

Near the end of the measurement, at each time step the photon number is
estimated from the values of $A_v$ and $B_v$. The estimator used is
\beq
|\alpha_v^{\rm P}|^2 + \sinh^2 |\zeta_v^{\rm P}|.
\eeq
I will call this estimator $\nb_v^{\rm P}$, so that this variable is analogous to
$\nb^{\rm P}$. The optimum value of $\xi_v^{\rm P}$ is then estimated using an asymptotic
formula based on the result in \cite{collett},
\beq
\xi_v^{\rm opt} = - \half \log \left( \frac{\nb_v^{\rm P}}{\log \nb_v^{\rm P} +
\Delta'} \right),
\eeq
where
\beq
\Delta' = 2 \log 2 - \smallfrac 14 \log 2\pi \approx 0.926825.
\eeq
Note that this $\Delta'$ differs from the $\Delta$ defined in
Eq.~(\ref{defdelta}) of Sec.~\ref{sqzcan} by $1.5$. The optimum value of
$\zeta_v^{\rm P}$ will be complex, with a phase dependent on the phase of
$\alpha_v^{\rm P}$.

If $\xi_v^{\rm R}$ (the real part of $\xi_v^{\rm P}$) is too far below this optimum value,
rather than using the feedback phase of Sec.~\ref{simplemethod}, the feedback
phase used is
\begin{equation}
\label{corrfeed}
\Phi(v) = \frac 12 \arg \left[ B_v- v \frac{C_v}{C_v^*} \tanh \left|\xi_v^{\rm
opt}\right|\right].
\end{equation}
Using this feedback phase takes $B_v$ directly towards the optimum value. To
see this, note firstly that the optimum value of $B_v$ is
\beq
B_v^{\rm opt} = v \frac{C_v}{C_v^*} \tanh \left|\xi_v^{\rm opt}\right|.
\eeq
This can be seen by inverting the definition of $\zeta_v^{\rm P}$. The factor of
$C_v/C_v^*$ makes the phase of this zero relative to $\alpha_v^{\rm P}$.

If we then take the exponential of the feedback phase, we find
\beq
e^{2i\Phi} \propto B_v- B_v^{\rm opt},
\eeq
so
\beq
dB_v \propto B_v^{\rm opt} - B_v .
\eeq
Here the proportionality symbol indicates that the phase is equal, but not
necessarily the magnitude. This demonstrates that this feedback phase takes the
value of $B_v$ directly towards the optimum value.

The details of exactly when $\xi_v^{\rm R}$ is considered too far below optimum can
be varied endlessly, but for the results that will be presented here this
alternative phase estimate is used after time $v=0.9$ and when
\begin{equation}
\label{endtime}
|\xi_v^{\rm R}|>|\xi_v^{\rm opt}|e^{\lambda |\alpha_v^{\rm P}|^2(1-v)}.
\end{equation}
Using the exponential multiplying factor means that this alternative feedback is
only used towards the end of the measurement. Only considering the alternative
feedback in the last 10\% of the measurement is necessary for the smaller photon
numbers, where Eq.~(\ref{endtime}) is too weak a restriction. In
Ref.~\cite{unpub} the feedback phase
\begin{equation}
\Phi(v) = \frac 12 \arg \left[\frac{B_v}{|B_v|}-\frac{C_v}{C_v^*}\right],
\end{equation}
was also mentioned. It turns out that this is not necessary, and very good
results can be obtained by using (\ref{corrfeed}) alone.

\begin{figure}
\centering
\includegraphics[width=0.7\textwidth]{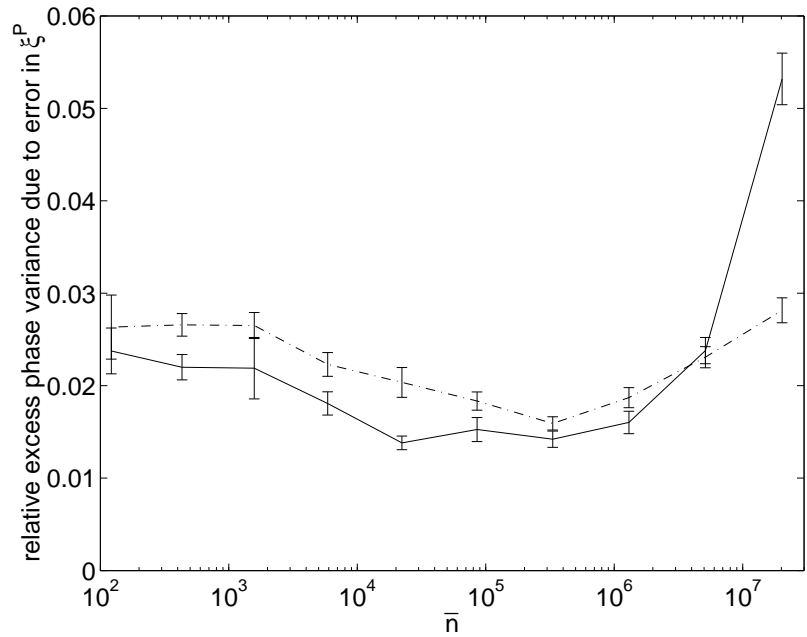}
\caption{Contributions to the phase uncertainty from error in the magnitude of
$\xi^{\rm P}$ (continuous line) and the phase of $\xi^{\rm P}$ (dash-dotted line). No
dividing factors are used, and $\lambda=10^{-3}$. The contributions are plotted
as a ratio to the theoretical minimum introduced phase uncertainty.}
\label{contrib1}
\end{figure}

It was found that good results were obtained for a wide range of photon numbers
when the value of $\lambda$ used was $10^{-3}$. It is also possible to adjust
the value of $\lambda$ individually for the different photon numbers; however,
this only gives marginal improvements. The results for this value of
$\lambda$ are shown in Fig.~\ref{contrib1}. The contribution
due to the error in the phase of $\xi^{\rm P}$ is small for the entire range
considered, below 3\%. The contribution due to error in the magnitude of
$\xi^{\rm P}$ is even smaller for moderate photon numbers, but for the largest photon
numbers it rises dramatically.

The reason for this rise is that the above correction only corrects for values
of $\xi_v^{\rm R}$ that are below optimum, and for the larger photon numbers many of
the uncorrected values of $\xi^{\rm R}$ are above optimum (see Fig.~\ref{zeta}).
In order to make the corrections work well, a dividing factor can be used to
bring the uncorrected values below the line. For the second largest photon
number tested of around $5\times 10^6$, the best results were obtained when the
values of $\varepsilon$ as given by (\ref{uncorfeed}) were divided by $1.1$. For
the largest mean photon number tested, $2\times 10^7$, the best results were
obtained for a dividing factor of $1.2$. The value of $\lambda$ that gave the
best results with these dividing factors was $5\times 10^{-4}$.

\begin{figure}
\centering
\includegraphics[width=0.7\textwidth]{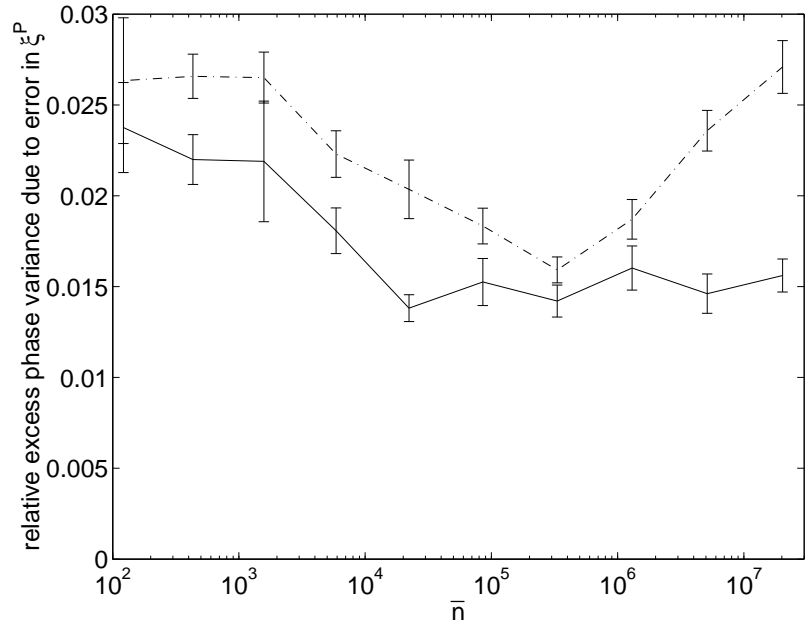}
\caption{Contributions to the phase uncertainty from error in the magnitude of
$\xi^{\rm P}$ (continuous line) and the phase of $\xi^{\rm P}$ (dash-dotted line).
Dividing factors of $1.1$ and $1.2$ are used for photon numbers of
$5\times 10^6$ and $2\times 10^7$, respectively. For these photon numbers
$\lambda=5\times 10^{-4}$, otherwise $\lambda=10^{-3}$. The contributions are
plotted as a ratio to the theoretical minimum introduced phase uncertainty.}
\label{contrib}
\end{figure}

The results using these dividing factors and altered value of $\lambda$ for
the larger photon numbers are shown in Fig.~\ref{contrib}. In this case, the
contribution due to error in the magnitude of $\xi^{\rm P}$ remains small, around
$1.5\%$, for the largest photon numbers. The contribution due to the error in
the phase of $\xi^{\rm P}$ does go up slightly for the largest photon numbers, but
it is still well below 3\%. The total excess phase variance due to error in
$\xi_v^{\rm P}$ does not exceed about 5\% for the entire range of photon numbers
considered.

The indication is that the optimum dividing factor will continue to increase
for photon numbers beyond the maximum for which calculations were performed.
Unfortunately, using these dividing factors means that the photon number must be
known beforehand. Nevertheless, as the dividing factor required increases only
very slowly with the photon number, only a rough, order of magnitude, estimate
of the photon number is required.

With this modified technique the phase variance again does not converge as the
feedback phase is updated in smaller and smaller time intervals. The phase
variance is less dependent on the time step with this technique, however. For
example, for a mean photon number of 1577 the total phase variance for
measurements on an optimised squeezed state only increases by about 9\% as the
time steps are reduced by a factor of 1000. In contrast, the phase variance
increases by a factor of 38\% for the uncorrected technique.

\begin{figure}
\centering
\includegraphics[width=0.7\textwidth]{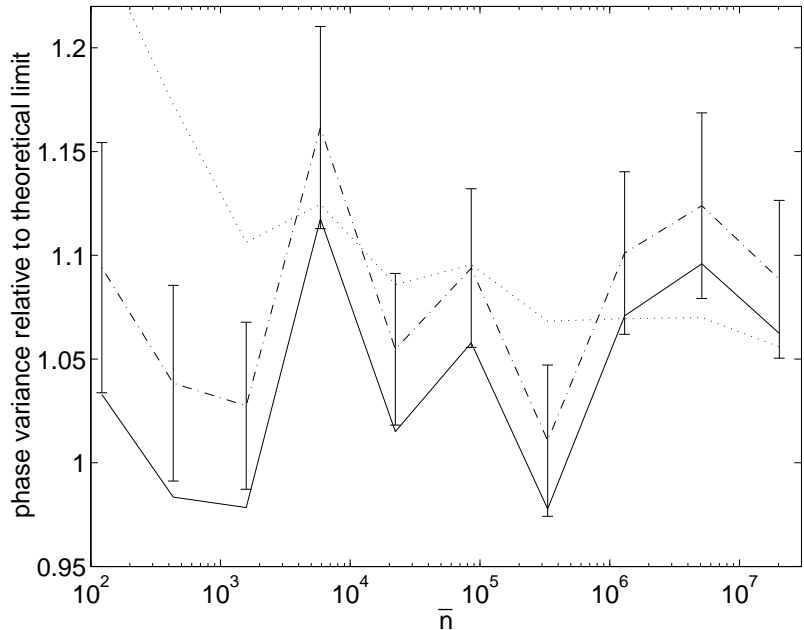}
\caption{The phase variance as estimated using the phase data as a ratio to the
theoretical limit. The continuous line is just the variance of the phase data,
the dash-dotted line is the variance corrected for the low probability results
around $\pm\pi$, and the dotted line is the corrected theoretical limit taking
into account the variation in $\nb^{\rm P}$.}
\label{phsdatcor}
\end{figure}

As an alternative way of evaluating the results we can again consider the
variance of the phase estimates obtained, as shown in Fig.~\ref{phsdatcor}. In
order to take account of the low probability phase results with large error,
we can add a factor of $1/(4\nb)$. The reason for this correction is that, from
Ref.~\cite{collett}, the term $2\erfc(\sqrt{2n_0})$ is the contribution due to
results with large error. From Eq.~(\ref{badests}), for squeezed states near the
theoretical limit this term becomes approximately $1/(8\nb)$. We must add twice
this, as phase measurements near the theoretical limit on optimally squeezed
states have a variance twice the intrinsic variance of the squeezed state.

As can be seen in Fig.~\ref{phsdatcor}, when this correction is included the
results are noticeably above the theoretical limit. The results are on average
around 10\% above the theoretical limit, which is slightly more than would be
expected from the previous analysis. The problem appears to be that the results
where $\nb^{\rm P}$ is small have disproportionately high errors. Even though
$\ip{\nb^{\rm P}} \approx \nb$, we will find that
\beq
\ip{\frac{\log\nb^{\rm P} +\Delta}{4{\nb^{\rm P}}^2}} > \frac{\log\nb +\Delta}{4\nb^2}.
\eeq
This means that the mean phase variance for the states $\ket{\alpha^{\rm P},\zeta^{\rm P}}$
will be higher than that for the state $\ket{\alpha,\zeta}$, even if they are
minimum phase uncertainty squeezed states.

This seems to be an intrinsic problem with these type of measurements, as any
state that has small phase variance will have large uncertainty in the photon
number. It is therefore reasonable to postulate that it is not possible to
reduce the introduced phase variance below the mean value of
$(\log\nb^{\rm P} +\Delta)/(4{\nb^{\rm P}}^2)$, and that this is therefore the actual limit.

When the theoretical limit to the introduced phase variance is corrected based
on the values obtained for $\nb^{\rm P}$, the theoretical limit to the total phase
variance (i.e.\ the intrinsic phase variance plus the limit to the introduced
phase variance), is as shown in Fig.~\ref{phsdatcor}. The phase variance
performs much better when compared with this corrected theoretical limit, and
even for the largest photon numbers is no more than about 5\% above it.

\section{Beyond the Theoretical Limit}
\label{surpass}
The final question that will be addressed in this chapter is whether it is
possible, in some circumstances, to reduce the introduced phase variance
below the theoretical limit. Recall that the theoretical limit is based on the
probability distribution for $A$ and $B$ being given by
\begin{equation}
P(A,B) \propto \st{\braket{\psi}{\alpha^{\rm P},\zeta^{\rm P}}}^2.
\end{equation}
If the phase estimate that is used is $\arg (A+BA^*)$, then this implies that
the introduced variance in this phase estimate is the intrinsic phase variance
of the state $\ket{\alpha^{\rm P},\zeta^{\rm P}}$. If the phase estimate used is {\it not}
$\arg (A+BA^*)$, however, this limit does not apply.

The main example of this is homodyne measurements. For homodyne measurements,
we find that $A+BA^*=0$, so it is not possible to use the phase estimate
$\arg (A+BA^*)$. Instead, we must use a phase estimate that relies on prior
knowledge of the state. For a homodyne measurement the local oscillator phase is
equal to the system phase plus $\pi/2$. We do not need to consider the
photocurrent as a function of time $I(t)$, as the measurement is essentially
just a measurement of the phase quadrature $\hat X_\Phi$, where
\beq
\hat X_\Phi = a e^{-i\Phi} + a \dg e^{i\Phi}.
\eeq
For arbitrary system phase, the expectation value of the $\pi/2$ quadrature is
\beq
\ip{\hat X_{\pi/2}} = 2\st{\alpha}\sin \varphi.
\eeq
This means that, provided $\st{\alpha}$ is known, we can use
\beq
\phi = {\rm asin} \left(\frac{X_{\pi/2}}{2\st{\alpha}}
\right)
\eeq
as a phase estimate. As the asin function is very linear near 0, this means
that these measurements are extremely close to being direct measurements of the
phase.

\subsection{Fitted Phase Estimates}

For the case of adaptive phase measurements, we can consider phase estimates
that are based on fitting the phase to the data. The introduced phase
uncertainty can then be estimated using the techniques of data analysis. If we
consider the photocurrent for a coherent state over a small but finite time
interval $\delta v$, we have
\beq
I(v) \delta v = 2\st{\alpha}\cos(\varphi-\Phi) \delta v + \delta W(v),
\eeq
where $\varphi$ is the system phase, and $\Phi$ is the local oscillator phase.
This is equivalent to a series of data points $y_i = I(v_i) \delta v$, where
\beq
\ip{y_i} = 2\st{\alpha}\cos(\varphi-\Phi) \delta v.
\eeq
If the values of $|\alpha|$ and $\varphi$ are assumed to be $|\alpha^f|$ and
$\phi$ respectively, then the expectation values for the data points are
\beq
\xi_i = 2\st{\alpha^f} \cos(\phi-\Phi) \delta v.
\eeq

I will firstly consider the case where both the magnitude and phase of $\alpha$
are fitted for, then the case where $|\alpha|$ is known, and only the phase
needs to be fitted. To perform the fit, we wish to minimise
\beq
M(\alpha^f) = \sum_i (y_i - \xi_i)^2.
\eeq
Expanding this gives
\beq
M(\alpha^f) = \sum_i \left[ (I(v_i) \delta v)^2
 - 4 I(v) \st{\alpha^f} \cos(\phi -\Phi) \delta v^2
+ 4\st{\alpha^f}^2 \cos^2(\phi -\Phi) \delta v^2 \right].
\eeq
As the $(I(v_i) \delta v)^2$ term does not depend on the fitting values, it
can be omitted. We can also remove a constant factor of $2\delta v$. Then
taking the limit of small $\delta v$ gives
\beq
M'(\alpha^f) = \int\limits_0^1 \left[ -2I(v) \st{\alpha^f}\cos(\phi-\Phi)
+ 2\st{\alpha^f}^2 \cos^2(\phi-\Phi) \right] dv.
\eeq
This can be simplified to an expression in terms of $A$ and $B$:
\beq
\label{solvalp}
M'(\alpha^f) = {\rm Re} \left[ -2A^* \alpha^f+\st{\alpha^f}^2 - B^*(\alpha^f)^2
\right].
\eeq

As this should be minimised for the fitted value of $\alpha$, the derivatives
with respect to the magnitude and phase of $\alpha^f$ should be zero. Thus we
have
\beq
\frac{\partial}{\partial \phi}M'(\alpha^f) = {\rm Im} \left[ -2A^*
\alpha^f - 2B^*(\alpha^f)^2 \right] = 0,
\eeq
and
\beq
\frac{\partial}{\partial \st{\alpha^f}}M'(\alpha^f) = {\rm Re} \left[ -2A^*
e^{i\varphi_f} + 2\st{\alpha^f} - 2B^*\st{\alpha^f}e^{2i\varphi_f} \right] = 0.
\eeq
The solution of these equations is given by
\beq
\alpha^f = \frac{A+BA^*}{1-\st{B}^2}.
\eeq
Thus we find that, if the coherent amplitude is unknown, fitting gives
exactly the same phase estimate $\arg(A+BA^*)$ as has been used previously.

In order to find the uncertainty in the fitted values, we calculate the matrix
$C$, where
\bqa
C_{i1} \!\!\!\! &=& \!\!\!\! \frac{\partial \xi_i}{\partial \phi} \\
C_{i2} \!\!\!\! &=& \!\!\!\! \frac{\partial \xi_i}{\partial |\alpha^f|}.
\eqa
The covariance matrix is then given by
\beq
\sigma^2 M_C^{-1},
\eeq
where $M_C=C^TC$, and $\sigma^2$ is the variance in the individual data points
(which is $\delta v$ in this case). Evaluating $M_C$ gives
\beq
M_C = \left[ {\begin{array}{*{20}c}
   \sum 4\st{\alpha}^2\sin^2(\varphi-\Phi)\delta v^2 &
-\sum 4\st{\alpha}\sin(\varphi-\Phi)\cos(\varphi-\Phi)\delta v^2  \\
-\sum 4\st{\alpha}\sin(\varphi-\Phi)\cos(\varphi-\Phi)\delta v^2 &
\sum 4\cos^2(\varphi-\Phi)\delta v^2  \\
\end{array}} \right].
\eeq
In data analysis we would normally evaluate this using the fitted value of
$\alpha$ rather than the actual value, as the actual value is unknown, and we
wish to estimate the uncertainty in the fitted value. Here, however, we are
interested in the variance in the phase estimates for a given value of $\alpha$,
and it is therefore more useful to express the result in terms of the actual
value of $\alpha$.

Taking the inverse to find the variance in the fitted phase, we find
\beq
{\rm var}(\phi) = \frac{\sum 4\cos^2(\varphi-\Phi)\delta v}
{\left[\sum 4\st{\alpha}^2\sin^2(\varphi-\Phi)\delta v\right] \left[
\sum 4\cos^2(\varphi-\Phi)\delta v \right] - \left[ \sum 4\st{\alpha}
\sin(\varphi-\Phi)\cos(\varphi-\Phi)\delta v \right] ^2},
\eeq
Taking the limit of small $\delta v$, this becomes
\beq
{\rm var}(\phi) = \frac 1{4\st{\alpha}^2} \frac{\int \cos^2(\varphi-
\Phi)dv}{\left[\int \sin^2(\varphi-\Phi)dv\right] \left[\int \cos^2(\varphi
-\Phi)dv \right] - \left[ \int \sin(\varphi-\Phi)\cos(\varphi-\Phi)dv
\right] ^2}.
\eeq
From this expression we can see that smaller phase variances will be obtained
if $\sin(\varphi-\Phi)\cos(\varphi-\Phi)$ has a time averaged value close
to zero. This is possible because $\sin(\varphi-\Phi)\cos(\varphi-\Phi)$
can take negative values. We will also obtain small phase variances if the
time averaged value of $\sin^2(\varphi-\Phi)$ is close to 1. This means that
the local oscillator phase should also be close to $\varphi+\pi/2$. Note that
if we are using a local oscillator phase of $\hat\varphi+\pi/2$, using a better
phase estimate $\hat\varphi$ does not necessarily result in a smaller variance,
as we also want $\sin(\varphi-\Phi)\cos(\varphi-\Phi)$ to average to zero.
This gives an alternative explanation of the result found before that using the
best intermediate phase estimate does not result in the smallest phase
variance.

In order to obtain better phase estimates, we can consider the case where
$\st{\alpha}$ is known, and only the phase is fitted for. Then it is easy to
see that we obtain Eq.~(\ref{solvalp}), except the actual value of
$\st{\alpha}$ is used:
\beq
M'(\phi) = {\rm Re} \left[ -2A^* \st{\alpha}e^{i\phi} +\st{\alpha}^2
- B^*\st{\alpha}^2 e^{2i\phi}\right].
\eeq
As $|\alpha|^2$ does not depend on the fitted phase, it can be omitted, so this
simplifies to
\beq
M'(\phi) = - {\rm Re} \left[ 2A^* \st{\alpha}e^{i\phi}
+ B^*\st{\alpha}^2 e^{2i\phi}\right].
\eeq
Note that this is equivalent to the result obtained in Eq.~(21) of
Ref.~\cite{Wis95c}. To have a minimum, we require
\beq
\frac{\partial}{\partial \phi}M'(\phi) = -{\rm Im} \left[ 2A^*
\st{\alpha}e^{i\phi} + 2B^*\st{\alpha}^2 e^{2i\phi} \right] = 0.
\eeq
Unlike the previous case there is no simple solution in terms of $A$ and $B$.
The solution to this can be found by finding the roots of the fourth order
polynomial
\bqa
\label{fourth}
P(\cos \phi) = 4\st{\alpha}^2\st{B}^2\cos^4 \phi
+ 4\st{\alpha}{\rm Re}(AB^*)\cos^3 \phi
+ (\st{A}^2-4\st{\alpha}^2\st{B}^2)\cos^2 \phi \nn \\
- 2\st{\alpha}({\rm Im}A{\rm Im}B+2{\rm Re}A{\rm Re}B)\cos \phi
+ \st{\alpha}^2({\rm Im}B)^2-({\rm Re}A)^2.
\eqa

The variance in the phase estimate, on the other hand, can be found far more
easily. We simply have
\beq
M_C = \sum 4\st{\alpha}^2\sin^2(\varphi-\Phi)\delta v^2,
\eeq
so the variance in the fitted phase is
\beq
{\rm var}(\phi) = \frac{1}{\sum 4\st{\alpha}^2\sin^2(\varphi-\Phi)
\delta v}.
\eeq
In the limit of small $\delta v$ this becomes
\beq
{\rm var}(\phi) = \frac{1}{4\st{\alpha}^2\int \sin^2(\varphi-\Phi)dv}.
\eeq
This is much simpler than the case where both the amplitude and the phase were
fitted. The smallest phase variance can be obtained by using local oscillator
phases as close as possible to $\varphi+\pi/2$, and there are no extra terms to
complicate the problem.

It is clear that if an accurate estimate of the phase is known beforehand, it
is possible to obtain a phase variance that is extremely close to the intrinsic
phase variance of $1/4\st{\alpha}^2$. If the local oscillator phase is
$\Phi = \hat\varphi+\pi/2$, then the expression for the variance becomes
\bqa
{\rm var}(\phi) \!\!\!\! &=& \!\!\!\! \frac{1}{4\st{\alpha}^2\int
\cos^2(\varphi-\hat\varphi)dv} \nn \\
\!\!\!\! &\approx& \!\!\!\! \frac{1}{4\st{\alpha}^2} + \frac{\int
(\varphi-\hat\varphi)^2 dv}{4\st{\alpha}^2}.
\eqa
This indicates that the introduced phase variance is proportional to the
variance in the intermediate phase estimates, so it should be possible
to reduce the introduced phase variance practically indefinitely by using
better intermediate phase estimates. In particular, it should be possible to
reduce it below the theoretical limit based on $\arg(A+BA^*)$ as the final
phase estimate.

If the intermediate phase estimate is based on the measurement results so far,
then the variance in the phase estimate cannot be smaller than the canonical
phase variance for a coherent state with $\st{\alpha}^2v$ photons,
$1/(4\st{\alpha}^2v)$. We cannot use this directly in the above equation, as
this variance goes to infinity for zero time. Since the average value of
$\cos^2$ for a randomised phase is $1/2$, it is only reasonable to use the
approximation
\beq
\ip{\cos^2(\varphi-\hat\varphi)} \approx 1-\ip{(\varphi-\hat\varphi)^2},
\eeq
when it gives a value less than $1/2$. In order to obtain an approximate
result, we can use this approximation for $v>1/(2\st\alpha^2)$, and use $1/2$
for $v<1/(2\st\alpha^2)$. Then we find
\bqa
\label{veryaprx}
\int\cos^2(\varphi-\hat\varphi)dv \!\!\!\! &\approx& \!\!\!\!
\int\limits_0^{\frac 1{2\st\alpha^2}} \frac{dv}2+\int\limits_{\frac 1
{2\st\alpha^2}}^1\left[1-\frac 1{4\st\alpha^2 v} \right]dv \nn \\
\!\!\!\! &=& \!\!\!\! 1-\frac{1+\log 2\st\alpha^2}{4\st\alpha ^2}.
\eqa
Therefore the phase variance should be approximately
\beq
{\rm var}(\phi) = \frac{1}{4\st{\alpha}^2} +
\frac{1+\log 2\st\alpha^2}{16\st\alpha ^4}.
\eeq
Thus we find that the introduced phase variance scales roughly as
$\log \nb/(16\nb^2)$. This is less than the theoretical limit, which scales as
$\log \nb/(4\nb^2)$.

\begin{figure}
\centering
\includegraphics[width=0.7\textwidth]{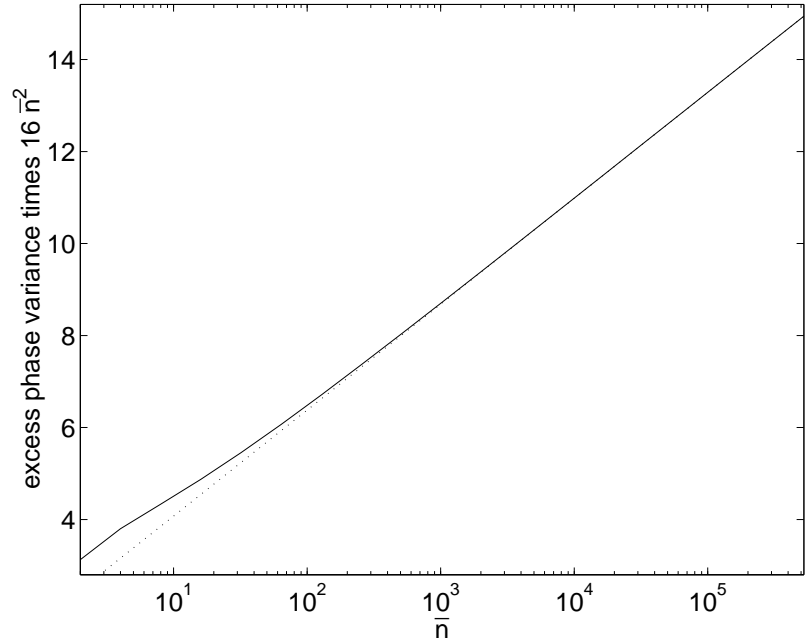}
\caption{The excess phase variance multiplied by $16\nb^2$ for coherent states
calculated using Eq.~(\ref{gamcoh}). The calculated result is shown as the
continuous line, and the dotted line is the fitted expression (\ref{fittedex}).}
\label{exactcos}
\end{figure}

To check the scaling when the calculation is performed exactly, we can use
\beq
\ip{\cos^2 \hat\varphi} = \half (1+{\rm Re}\ip{e^{2i\hat\varphi}}).
\eeq
Here the system phase has been taken to be zero for simplicity. For a coherent
state with amplitude $\alpha \sqrt v$, we have
\beq
\braket{n}{\alpha \sqrt v} = e^{-\alpha ^2 v/2}\frac{(\alpha \sqrt v)^n}
{\sqrt{n!}}
\eeq
This means that
\bqa
\ip{e^{2i\hat\varphi}} \!\!\!\! &=& \!\!\!\! \sum_{n=0}^{\infty} \braket{\alpha
\sqrt v}{n}\braket{n+2}{\alpha \sqrt v} \nn \\
\!\!\!\! &=& \!\!\!\! e^{-\alpha ^2 v}\sum_{n=0}^{\infty}
\frac{(\alpha^2 v)^{n+1}}{\sqrt{n!(n+2)!}}.
\eqa
Now taking the integral of this over time gives
\bqa
\label{gamcoh}
\int\limits_0^1 e^{-\alpha ^2 v}\sum_{n=0}^{\infty}
\frac{(\alpha^2 v)^{n+1}}{\sqrt{n!(n+2)!}}dv \!\!\!\! &=& \!\!\!\!
\frac 1{\alpha^2}\int\limits_0^{\alpha^2} e^{-t}\sum_{n=0}^{\infty}
\frac{t^{n+1}}{\sqrt{n!(n+2)!}}dt \nn \\
\!\!\!\! &=& \!\!\!\! \frac 1{\alpha^2} \sum_{n=0}^{\infty}
\frac{\Gamma(\alpha^2,n+2)}{\sqrt{n!(n+2)!}},
\eqa
where $\Gamma(\alpha^2,n+2)$ is the incomplete gamma function. This can then
be used to evaluate $\ip{\cos^2 \hat\varphi}$ exactly. The results calculated
in this way are plotted in Fig.~\ref{exactcos}.

It was found that the scaling of the introduced phase variance was again
$\log \nb/(16 \nb^2)$ when estimated using this more exact method. There was a
difference of order $\nb^{-2}$ with the previous result, however. It was found
that the introduced phase variance was very close to
\beq
\label{fittedex}
\frac{\log \nb + 1.7733}{16 \nb^2} \nn
\eeq
as compared to
\beq
\frac{\log \nb + 1.693147}{16 \nb^2} \nn
\eeq
using the very approximate method of Eq.~(\ref{veryaprx}).

Note that this result is still based on assuming that the intermediate phase
estimates are as good as canonical. To determine the result for the real case,
where the intermediate phase estimates are based on the preceding data, we must
perform stochastic integrals to determine the result numerically. To determine
the results in this case, rather than performing the integrals separately for
each mean photon number, a continuous calculation with an exponentially
increasing time-step was used. Then at various times, the photon number up to
that time and the integral of $\ip{\cos^2 \hat\varphi}$ were determined.

It is possible to do this in the case of coherent states, because the integral
up to time $v$ with coherent amplitude $\alpha$ is equivalent to the integral
over the unit interval with amplitude $\alpha\sqrt v$. Note also that the
integral of $\ip{\cos^2 \hat\varphi}$ does not need to be calculated
separately, because
\beq
{\rm Re}B_v = \int\limits_0^v \cos^2 \hat\varphi_u du.
\eeq
For this calculation the intermediate phase estimate was the best phase
estimate using the known value of $\st{\alpha}$, as found by solving
Eq.~(\ref{fourth}).

\begin{figure}
\centering
\includegraphics[width=0.7\textwidth]{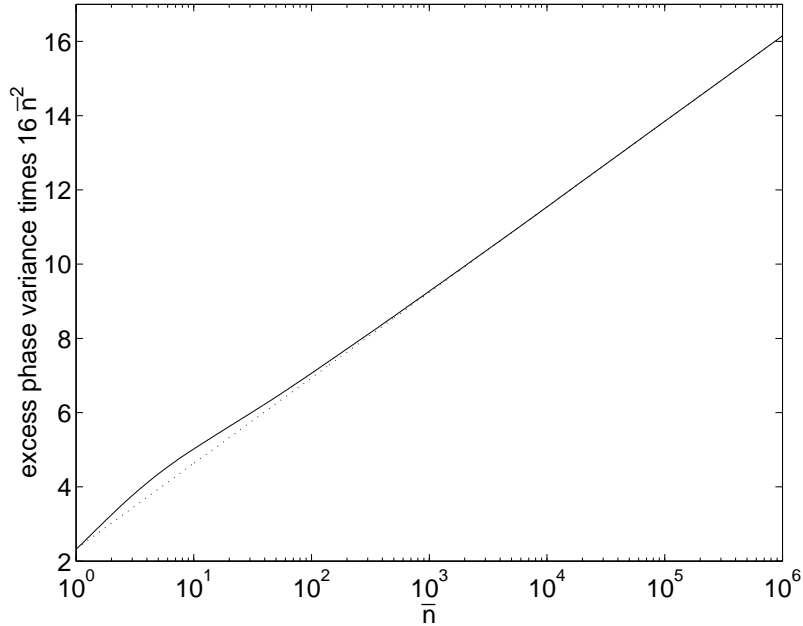}
\caption{The excess phase variance multiplied by $16\nb^2$ for coherent states
as determined using stochastic integrals. The calculated result is shown as the
continuous line, and the dotted line is the fitted expression.}
\label{stochcos}
\end{figure}

This calculation was performed for $2^{14}$ samples, and the results are shown
in Fig.~\ref{stochcos}. The results again scaled as $\log \nb/(16 \nb^2)$, but
the term of order $\nb^{-2}$ was different. The scaling obtained was
approximately
\beq
\frac{\log \nb + 2.4}{16 \nb^2}. \nn
\eeq
As can be expected, this is slightly worse than the result calculated assuming
that the intermediate phase estimates are as good as canonical.

This demonstrates that if the coherent amplitude is known, it is possible to
reduce the introduced phase variance to a factor of 4 below the theoretical
limit (that scales as $\log \nb/(4 \nb^2)$). Unfortunately this is not very
useful in the case of coherent states, as the introduced phase variance is
always (for an adaptive feedback scheme) far less than the intrinsic phase
variance.

If the introduced phase variance remained this small for a squeezed state, then
this would be a very significant result. Unfortunately, the result that the
introduced phase variance should be fairly independent of the input state
breaks down for this type of phase estimate, for similar reasons to why the
theoretical limit does not apply.

The best phase estimate as found by solving Eq.~(\ref{fourth}) is specific
to coherent states, and the best phase estimate for other states will be
different. For the case of squeezed states, this is a difficult calculation
where the evolution of the state for each input phase must be determined based
on the measured values of $I(v)$. Specifically, in the discretised calculation
we are using
\beq
I(v) \delta v = 2{\rm Re}(\alpha_v e^{-i\Phi(v)}) \delta v + \delta W(v).
\eeq
For some assumed system phase, we can determine what value $\delta W(v)$
would have using
\beq
\delta W'(v) = I(v) \delta v - 2{\rm Re}(\alpha'_v e^{-i\Phi(v)}) \delta v,
\eeq
where the primes indicate that these are the values determined based on that
assumed system phase (as opposed to the actual phase). For this value of
$\delta W'(v)$, the evolution of $\alpha'_v$ is then
\beq
\delta\alpha'_v = \frac{1}{1 - v}\frac{{B'}_v^{\rm S} \delta W'(v)} {1 -
\st{{B'}_v^{\rm S}}^2}\left( {B'}_v^{\rm S*} e^{i\Phi} + e^{-i\Phi} \right),
\eeq
where
\beq
{B'}_v^{\rm S} = \frac {1-v}{({B'}_0^{\rm S})^{-1}-B_v^*}.
\eeq
After these values are calculated over the entire time interval $[0,1)$, we
then determine
\beq
\label{sqzmin}
\sum (\delta W'(v))^2.
\eeq
We wish to minimise this in order to find the phase that gives the best fit to
the data.

\begin{figure}
\centering
\includegraphics[width=0.7\textwidth]{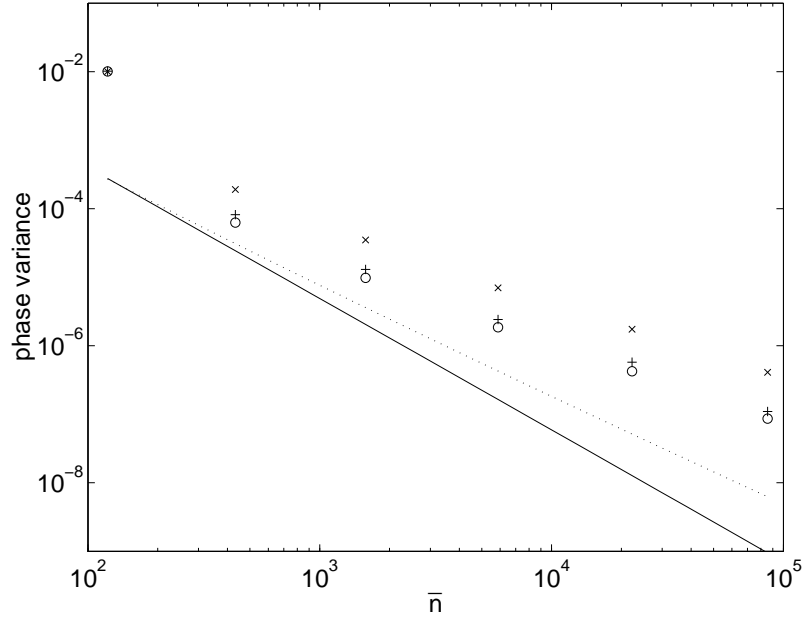}
\caption{The phase variances using $\arg C_v$ feedback on optimally squeezed
input states. The crosses are the results for $\arg C$ phase estimates, the
pluses are those found by solving Eq.~(\ref{fourth}), and the circles are those
for the phase estimates based on numerically minimising Eq.~(\ref{sqzmin}). The
continuous line is the theoretical limit, and the dotted line is the variance
for mark II measurements.}
\label{sqzsol}
\end{figure}

It is feasible to use this method for the final phase estimates; however, it
would make the calculation far too difficult if it was also used for the
intermediate phase estimates. Therefore, for the intermediate phase estimates
$\arg C_v$ was used initially. The results of using this intermediate phase
estimate on optimally squeezed states are plotted in Fig.~\ref{sqzsol}.

If the $\arg C$ phase estimate is used at the end of the measurement as well,
the phase variance is far greater than the theoretical limit. The phase estimate
based on minimising (\ref{sqzmin}) gives a much reduced variance, but it is
still far above the theoretical limit. In fact, it is still worse than the phase
variance for mark II measurements. Also shown in Fig.~\ref{sqzsol} is the phase
estimate based on coherent states, found by solving Eq.~(\ref{fourth}). These
phase estimates give variances that are very close to, but slightly above, the
variances for the fitted phases.

\begin{figure}
\centering
\includegraphics[width=0.7\textwidth]{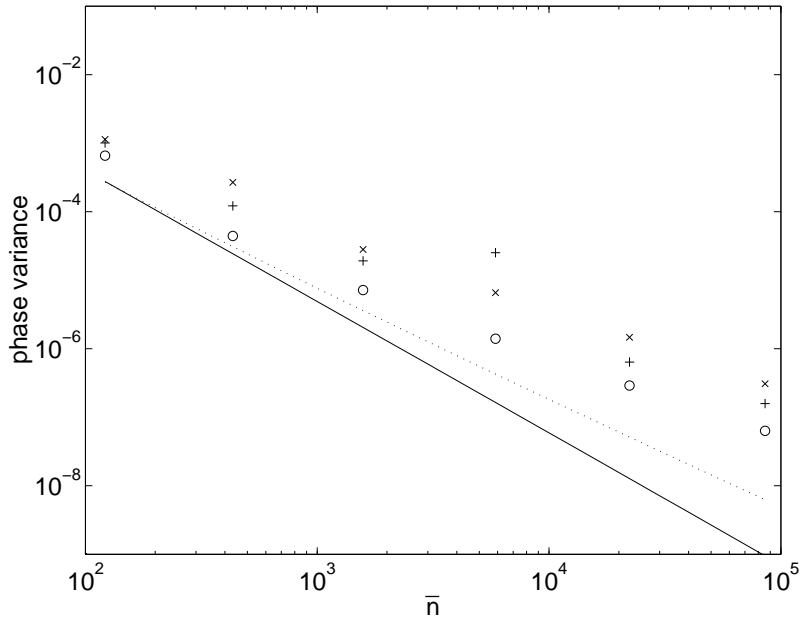}
\caption{The phase variances for measurements on optimally squeezed input states
using phase estimates in the feedback found by solving Eq.~(\ref{fourth}). The
crosses are the results for $\arg C$ phase estimates, the pluses are those
found by solving Eq.~(\ref{fourth}), and the circles are those for the phase
estimates based on numerically minimising Eq.~(\ref{sqzmin}). The continuous
line is the theoretical limit, and the dotted line is the variance for mark II
measurements.}
\label{sqzsolbes}
\end{figure}

As an alternative feedback scheme, we can use the phase estimates based on
coherent states in the feedback. The results using this feedback are shown in
Fig.~\ref{sqzsolbes}. The variances for the fitted phases are slightly less
than those using $\arg C_v$ intermediate phase estimates, but are still far
above the theoretical limit or even the variances for mark II measurements.

\begin{figure}
\centering
\includegraphics[width=0.7\textwidth]{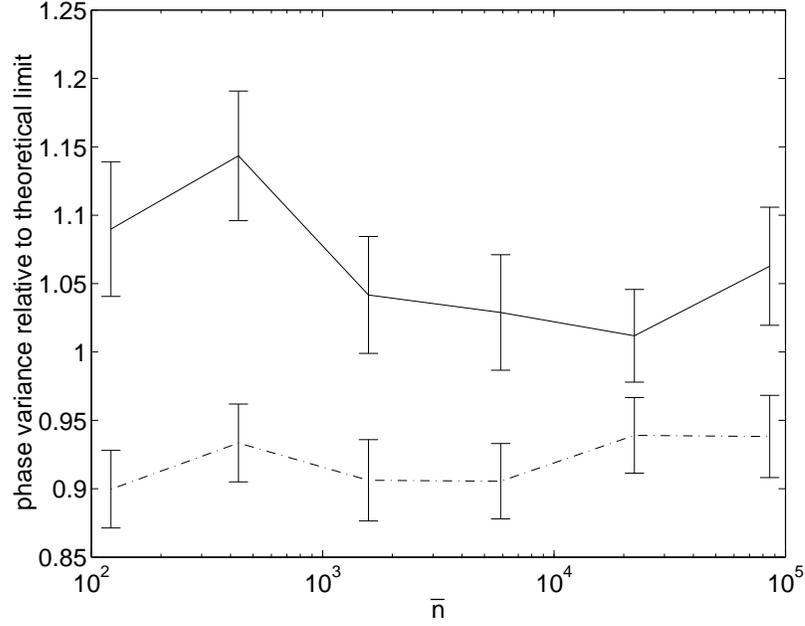}
\caption{The phase variances for measurements on optimally squeezed input states
using intermediate phase estimates as in Sec.~\ref{improved}. The continuous
line is the phase variance for $\arg C$ phase estimates and the dash-dotted line
is the variance for phase estimates based on fitting. All variances are shown as
a ratio to the theoretical limit.}
\label{sqzsolasy}
\end{figure}

It is possible to obtain phase variances close to the theoretical limit if we
use feedback that gives results close to the theoretical limit for the usual
$\arg C$ phase estimates. For example, if we use the corrected feedback as in
Sec.~\ref{improved}, we obtain the results shown in Fig.~\ref{sqzsolasy}. As the
numerical results do not take account of the low probability results with large
errors, a correction factor of $1/(4\nb^2)$ has been added to these results
(as was done for Fig.~\ref{phsdatcor}).

In practice it was found that, for the smaller photon numbers where large error
results were obtained, the fitted phases were still close to the $\arg C$ phase
estimates, so large errors were obtained for the same samples for both phase
estimates. This can also be ensured by only performing the minimisation near
$\arg C$. Then the contribution due to the large error results will be the same
for both cases. This is why the same correction factor of $1/(4\nb^2)$ has been
added to both sets of results in Fig.~\ref{sqzsolasy}.

It is seen in Fig.~\ref{sqzsolasy} that when the corrections are taken into
account, the results for the usual phase estimate are slightly greater than the
theoretical limit, but the results for the fitted phase are slightly less than
the theoretical limit. This demonstrates that it is possible to surpass the
theoretical limit in the case of squeezed states, but only by a very small
margin.

\subsection{Optimal Phase Estimates}
The fitting approach considered in the previous section gives the most probable
phase, but this is not necessarily the same as the phase estimate that minimises
the variance. It is shown in Sec.~\ref{final} that the phase estimate that
minimises the phase variance for interferometric measurements is
\beq
\hat \varphi = \arg \int\limits_{-\pi}^{\pi} e^{i \varphi} P(n_m | \varphi)
d\varphi,
\eeq
where $n_m$ is the measurement record. This derivation is very general, and
should also hold for dyne measurements, where there is a continuous measurement
record ${\bf I}_{[0,v)}$.

Similarly to the case for interferometry, we have [using Eq.~(\ref{genbayes})]
\beq
\label{bayescon}
P({\bf I}_{[0,v)}|\varphi) \propto P(\varphi|{\bf I}_{[0,v)}),
\eeq
so the optimum phase estimate can be expressed as
\bqa
\hat \varphi \!\!\!\! &=& \!\!\!\! \arg \int\limits_{-\pi}^{\pi} e^{i \varphi}
P(\varphi | {\bf I}_{[0,v)}) d\varphi \nn \\
\!\!\!\! &=& \!\!\!\! \arg \ip{e^{i\varphi}},
\eqa
where the average is over the probability distribution for the phase based on
the measurement record.
In order to determine the probability distribution $P({\bf I}_{[0,v)}|\varphi)$,
we need to determine the probability of obtaining the increments $dW(v)$.
Considering the discretised equation, each increment $\delta W(v)$ has a
Gaussian distribution with variance $\delta v$. The probability of obtaining
$\delta W(v)$ is therefore
\beq
P(\delta W(v)) \propto e^{-(\delta W(v))^2/(2\delta v)}.
\eeq
The probability of obtaining the measurement record will therefore be
\beq
P({\bf I}_{[0,v)} | \varphi) \propto
\exp \left( -\sum (\delta W(v))^2/(2\delta v) \right).
\eeq
From Eq.~(\ref{bayescon}), this means that we also have
\beq
P(\varphi|{\bf I}_{[0,v)}) \propto
\exp \left( -\sum (\delta W'(v))^2/(2\delta v) \right).
\eeq
Here the prime on $\delta W'(v)$ indicates that it is calculated from the
measurement record based on an assumed system phase, similarly to the previous
section.

It is obvious that the fitted phase discussed in the previous section is the
most probable phase, as it minimises $\sum (\delta W'(v))^2$ and therefore
maximises $P(\varphi | {\bf I}_{[0,v)})$. To find the optimal phase that
minimises the final phase uncertainty, we need to calculate
\bqa
\hat \varphi \!\!\!\! &=& \!\!\!\! \arg \int\limits_{-\pi}^{\pi} e^{i \varphi}
P(\varphi | {\bf I}_{[0,v)}) d\varphi \nn \\
 \!\!\!\! &=& \!\!\!\! \arg \int\limits_{-\pi}^{\pi} e^{i \varphi} \exp
\left( -\sum (\delta W'(v))^2/(2\delta v) \right) d\varphi .
\eqa
For a coherent state this simplifies to
\beq
\hat \varphi = \arg \int\limits_{-\pi}^{\pi} e^{i \phi} \exp \left( {\rm Re}
\left[ 2A^*\st{\alpha}e^{i\phi} + B^*\st{\alpha}^2 e^{2i\phi}\right]
\right) d\phi .
\eeq
It does not appear to be possible to evaluate this integral analytically, and
it would need to be evaluated numerically.

The calculation is even more difficult in the case of squeezed states. As was
discussed in the previous section, for squeezed states the entire calculation
(i.e.\ determining the time evolution of the state) must be repeated for each
value of $\phi$. For numerical minimisation the number of values of
$\phi$ for which the calculation must be performed is on the order of 10,
which means that the calculations are more lengthy, but not infeasible. For a
numerical integral thousands of function evaluations would be required for an
accurate result, making this method infeasible. For this reason, these optimal
phase estimates were not used for dyne measurements in this study. As will be
seen in Ch.~\ref{interfere}, however, it is much easier to determine the optimal
phase estimates in interferometry.

%% file: delay.tex
\setcounter{chapter}{3}

\chapter{The Effect of Time Delays}
\label{delays}
\section{Introduction}
In practice the adaptive dyne measurements described in the previous chapters
cannot be performed exactly. In any experiment there will be imperfections, for
example calibration errors. In making phase measurements a major source of error
is inefficient photodetectors. This is particularly the case for single
photon photodetectors. These are photodetectors that are designed for
distinguishing between, for example 1 or 2 photons. These photodetectors
currently cannot be made with efficiencies above about 87\% \cite{SingleEff}.

High amplitude photodetectors, on the other hand, can be made with far higher
efficiencies, around 98\% \cite{polzik}. This is the sort of photodetector
required for dyne measurements, due to the large amplitude local oscillator
field. It is fairly straightforward to determine the extra phase uncertainty due
to inefficient photodetectors. In the introduction (and Ref.~\cite{semiclass}),
it is shown that when the photodetector efficiency is $\eta$, the extra phase
variance is approximately
\beq
\frac{1-\eta}{4\eta \nb}.
\eeq

This extra phase variance means that, for large photon numbers, phase feedback
schemes can only reduce the phase variance by a factor of about $1-\eta$. This
will be true for mark II measurements, as well as the more advanced phase
feedback schemes described in the preceding chapter. For current photodetectors,
this extra phase variance is more significant than the introduced phase variance
for mark II measurements for photon numbers above about 1000. Below this the
extra phase variance for mark II measurements is only marginally above the
theoretical limit, and so it is not possible to greatly reduce the phase
variance using more advanced feedback schemes.

Another imperfection that is specific to phase measurements with feedback is the
time delay in the feedback loop. This contribution is more difficult to
estimate. Some highly simplified calculations indicate that the excess phase
variance due to time delays is $\tau/2$ for mark I measurements (where $\tau$ is
the time delay), and $\tau/(2\bar n)$ for mark II measurements \cite{semiclass}.

In Ref.~\cite{delays} I repeated these derivations more rigorously, and this
chapter is based on that paper. While the result for mark I measurements
is reasonably accurate, the perturbation approach is inadequate to obtain a
consistent result for mark II measurements. In Sec.~\ref{thmin} an alternative
derivation is considered that gives the minimum phase variance when there is a
time delay. The phase variance with time delays is evaluated numerically in
Sec.~\ref{result}, and it is shown that for most of the measurement schemes the
phase variance approaches this limit for large time delays.

\section{Perturbation Approach}
\subsection{Mark I}
\label{dtontwo}
Firstly I will estimate the effect of time delays on simplified mark I
measurements in a similar way as in Ref.~\cite{semiclass}, but using
fewer of the simplifications used there. Without a time delay the
stochastic differential equation for the phase estimate is
\begin{eqnarray}
d\hat \varphi_v \!\!\!\! &=& \!\!\!\! \frac{I(v) dv}{\sqrt v} \nn \\
\!\!\!\! &=& \!\!\!\! v^{-1/2} [-2\alpha \sin \hat \varphi_v dv + dW(v)].
\end{eqnarray}
Here the input phase has been taken to be zero for simplicity. For some
time $v_1$ the phase will come to lie near 0, so we can linearise around
$\hat \varphi_v = 0$.  The result, which will be valid for
$v_1 \le v \le 1$ is
\begin{equation}
d\hat \varphi_v = v^{-1/2} [-2\alpha \hat \varphi_v dv + dW(v)].
\label{nodelay}
\end{equation}
We wish to consider the limit of large $\alpha$, so that $v_1$ is small, and
this linearisation is accurate for most of the measurement. Including the time
delay the SDE is
\begin{equation}
d\hat \varphi_v = v^{-1/2} [-2\alpha \hat \varphi_{v-\tau} dv + dW(v)].
\end{equation}
Now the time delay will be treated perturbatively. The solution to the perturbed
equation can be written as
\begin{equation}
\hat\varphi_v=\hat\varphi_v^{(0)}+\alpha\tau\hat\varphi_v^{(1)}
+O(\alpha^2\tau^2).
\end{equation}
Note that for this approximation to be accurate, $\alpha\tau$ must be
small, in addition to $\alpha$ being large.

The zeroth-order term obeys the SDE for no delay (\ref{nodelay}), so the
first-order correction obeys
\begin{equation}
\alpha\tau d\hat \varphi_v^{(1)} = 2\alpha v^{-1/2}(
\hat\varphi_v^{(0)}-\hat\varphi_{v-\tau}^{(0)})dv-2\alpha^2\tau 
v^{-1/2}\hat\varphi_{v-\tau}^{(1)}dv.
\end{equation}
Therefore to first order in $\tau$ we have
\begin{eqnarray}
d\hat \varphi_v^{(1)} \!\!\!\! &=& \!\!\!\! 2 v^{-1/2} d\hat\varphi_v^{(0)}-
2\alpha v^{-1/2}\hat\varphi_v^{(1)}dv \nn \\
\!\!\!\! &=& \!\!\!\! 2 v^{-1/2} \{v^{-1/2} [-2\alpha \hat\varphi_v^{(0)} dv +
dW(v)]\}-2\alpha v^{-1/2}\hat\varphi_v^{(1)}dv \nn \\
\!\!\!\! &=& \!\!\!\! -2\alpha v^{-1/2}\hat\varphi_v^{(1)}dv-\frac {4\alpha}v
\hat\varphi_v^{(0)} dv +\frac 2v dW(v).
\label{firstorder}
\end{eqnarray}
It is straightforward to show that the solution to the zeroth order equation is
\begin{equation}
\label{zerosol}
\hat\varphi_v^{(0)}=e^{4\alpha(\sqrt {v_1}-\sqrt v)}\hat\varphi_{v_1}^{(0)}+
\int\limits_{v_1}^v \frac {e^{4\alpha(\sqrt u-\sqrt v)}} {\sqrt u} dW(u).
\end{equation}
Using this in Eq.~(\ref{firstorder}) and multiplying on both sides by
$e^{4\alpha\sqrt v}$ gives
\begin{equation}
d(e^{4\alpha\sqrt v}\hat \varphi_v^{(1)}) = -\frac{4\alpha}v \left[
e^{4\alpha\sqrt {v_1}}\hat\varphi_{v_1}^{(0)}+\int\limits_{v_1}^v \frac
{e^{4\alpha\sqrt u}} {\sqrt u} dW(u) \right] dv +\frac 2v e^{4\alpha\sqrt v}
dW(v).
\end{equation}
Integrating then gives the solution
\begin{eqnarray}
\hat \varphi_v^{(1)} \!\!\!\! &=& \!\!\!\! e^{4\alpha(\sqrt {v_1}-\sqrt v)}
\hat \varphi_{v_1}^{(1)} -\int\limits_{v_1}^v\frac{4\alpha}s \left[ e^{4
\alpha(\sqrt{v_1}-\sqrt v)}\hat\varphi_{v_1}^{(0)}+\int\limits_{v_1}^s \frac
{e^{4\alpha(\sqrt u-\sqrt v)}} {\sqrt u} dW(u) \right] ds \nn \\
&& +\int\limits_{v_1}^v\frac 2s e^{4\alpha(\sqrt s-\sqrt v)}dW(s).
\end{eqnarray}

In this approximation the mark I phase estimate is given by $\hat\varphi_1=
\hat\varphi_1^{(0)}+\alpha\tau\hat\varphi_1^{(1)}$. To first order in $\tau$
and $\alpha^{-1}$ this has a variance of
\begin{equation}
\ip{\phi_{\rm I}^2}=\frac 1{4\alpha}+2\alpha\tau\ip{\hat\varphi_1^{(0)}
\hat\varphi_1^{(1)}}.
\end{equation}
Here the known variance of $1/(4\alpha)$ of the zeroth order term has been used.
Evaluating the second term on the right hand side we find
\begin{eqnarray}
\ip{\hat\varphi_1^{(0)}\hat\varphi_1^{(1)}} &=& e^{-8\alpha(1-\sqrt{v_1})}\ip{
\hat\varphi_{v_1}^{(0)}\hat\varphi_{v_1}^{(1)}} + 4\alpha\log(v_1)
e^{-8\alpha(1-\sqrt{v_1})}\ip{({\hat\varphi}_{v_1}^{(0)})^2} \nn \\
&&-\int\limits_{v_1}^1 \frac {4\alpha}s \int\limits_{v_1}^s \frac {e
^{8\alpha(\sqrt u-1)}}u du ds +\int\limits_{v_1}^1 \frac {2 e^{8\alpha
(\sqrt u-1)}}{u^{1.5}}du.
\end{eqnarray}
The first two terms decrease exponentially with $\alpha$ and may
therefore be omitted. Exchanging the order of the integrals in the
third term and integrating gives
\begin{equation}
\ip{\hat\varphi_1^{(0)}\hat\varphi_1^{(1)}} = 4\alpha\int\limits_{v_1}^1 \frac
{\log u}u e^{8\alpha(\sqrt u-1)} du + \int\limits_{v_1}^1 \frac
{2 e^{8\alpha(\sqrt u-1)}}{u^{1.5}}du.
\end{equation}
To perform these integrals it is convenient to change variables to
$s=1-\sqrt u$, so $du=-2(1-s)ds$. Then we obtain
\begin{equation}
\ip{\hat\varphi_1^{(0)}\hat\varphi_1^{(1)}} = 16\alpha\int\limits_0^{1-
\sqrt{v_1}}\frac {\log (1-s)} {(1-s)} e^{-8\alpha s} ds +
4\int\limits_0^{1-\sqrt{v_1}} \frac {e^{-8\alpha s}}{(1-s)^2}ds.
\end{equation}
Expanding in a Maclaurin series in $s$ gives
\begin{eqnarray}
\ip{\hat \varphi_1^{(0)}\hat\varphi_1^{(1)}} \!\!\!\! &=& \!\!\!\! -16\alpha
\int\limits_0^{1-\sqrt{v_1}}(s+\frac 32 s^2 +O(s^3)) e^{-8\alpha s} ds
+4\int\limits_0^{1-\sqrt{v_1}} (1+2s+3s^2+O(s^3)){e^{-8\alpha s}}ds \nn \\
\!\!\!\! &=& \!\!\!\!-16\alpha\left[\frac 1{(8\alpha)^2}+\frac 3{(8\alpha)^3}+
O(\alpha^{-4})\right] +4 \left[\frac 1{8\alpha}+\frac 2{(8\alpha)^2}
+ \frac 6{(8\alpha)^3}+O(\alpha^{-4})\right] \nn \\
\!\!\!\! &=& \!\!\!\!\frac 1{4\alpha} + O(\alpha^{-2}).
\end{eqnarray}
Note that the upper bound at $1-\sqrt{v_1}$ has no effect since it
gives a term that decays exponentially with $\alpha$. Using this result, the
total phase variance is
\begin{equation}
\ip{\phi_{\rm I}^2}=\frac 1{4\alpha}+\frac \tau 2.
\end{equation}
This provides a good verification of the result obtained by the highly
simplified method in \cite{semiclass}.

Note that this result is based on continuing to use the intermediate phase
estimate at the end of the measurement. If the phase estimate $\arg A$ is used
at the end of the measurement, the result will be different, and cannot be
predicted using this approach.

\subsection{Mark II}
\label{mark2}
For the mark II case it does not seem to be possible to obtain a consistent
result using this approach. To illustrate this, I will briefly
outline the derivation (the details are in Appendix~\ref{pert}). From
\cite{semiclass}, the mark II phase estimate is effectively a time average of
the mark I phase estimates:
\begin{equation}
\phi_{\rm II} \approx \int\limits_0^1 \hat\varphi_t dt.
\end{equation}
In order to make this consistent with the above theory, we should take the
average only from time $v_1$, then take the limit of small $v_1$. In
perturbation theory the mark II phase estimate is
\begin{equation}
\phi_{\rm II} = \int\limits_{v_1}^1 [\hat\varphi_t^{(0)}+\alpha\tau\hat
\varphi_t^{(1)}] dt.
\end{equation}
Using this expression the variance is
\begin{equation}
\label{markii}
\ip{\phi_{\rm II}^2}=\int\limits_{v_1}^1 dt \int\limits_{v_1}^1 dt' \ip{\hat
\varphi_t^{(0)}\hat\varphi_{t'}^{(0)}} + 2\alpha\tau\int\limits_{v_1}^1 dt
\int\limits_{v_1}^1 dt' \ip{\hat\varphi_t^{(0)} \hat\varphi_{t'}^{(1)}}.
\end{equation}

It is shown in Appendix \ref{pert} that the first term can be simplified to
\beq
\int\limits_{v_1}^1 dt \int\limits_{v_1}^1 dt' \ip{\hat
\varphi_t^{(0)}\hat\varphi_{t'}^{(0)}} \approx \frac 1{4\alpha^2}+
\frac {v_1}{4\alpha^2}\ip{(\hat\varphi_{v_1}^{(0)})^2},
\eeq
which is similar to that obtained in \cite{semiclass}. For the second term,
however, we get
\beq
2\alpha\tau\int\limits_{v_1}^1 dt \int\limits_{v_1}^1 dt'
\ip{\hat\varphi_t^{(0)} \hat\varphi_{t'}^{(1)}}\approx \frac {\tau}{\alpha}
\left( \frac{v_1}2\ip{\hat\varphi_{v_1}^{(0)}\hat\varphi_{v_1}^{(1)}} -
\sqrt{v_1}\ip{(\hat\varphi_{v_1}^{(0)})^2}\right).
\eeq
This is radically different to the result obtained in Ref.~\cite{semiclass}, and
seems to cast some doubt on the simplified theory used there.

Unlike the result for mark I measurements, this result depends entirely on the
conditions at time $v_1$, which are unknown. In order to obtain a usable result,
we would need the initial conditions to give a neglibible contribution, as is
the case for mark I measurements. For this reason I will consider an alternative
approach for estimating the increase in the phase variance due to the time
delay.

\section{Theoretical Minimum}
\label{thmin}
The alternative method of obtaining an estimate for the phase variance with a
time delay is to consider the squeezed state $\ket{\alpha^{\rm P},\zeta^{\rm P}}$ in the
probability distribution. As was explained in Sec.~\ref{theory}, the excess
phase variance due to the measurement scheme is approximately the phase
variance of this squeezed state.

From Ref.~\cite{collett}, the phase variance of a squeezed state is given by
\begin{equation}
\ip{\Delta \phi^2}\approx \frac{n_0+1}{4\nb^2} + 2 {\rm erfc}
\left( \sqrt{2n_0} \right ),
\end{equation}
where $n_0 = \nb e^{2\zeta}$ for real $\zeta$. To determine the excess phase
variance, we would use the values $\nb^{\rm P}$ and $\zeta^{\rm P}$ from the squeezed state
in the POM, rather than those for the input state. The average value of $\nb^{\rm P}$
will be close to the photon number of the input state.

For states that are significantly less squeezed than optimum, the second
term is negligible and we can omit the term of order $(\nb^{\rm P})^{-2}$. Then
this simplifies to
\begin{equation}
\ip{\Delta \phi^2} \approx \frac{e^{2\zeta^{\rm P}}}{4\nb^{\rm P}}.
\end{equation}
Since $\nb^{\rm P}$ will be close to the photon number of the input state, it is
reasonable to replace it with $\nb$.

When there is a delay of $\tau$ in the system, before time $\tau$ we
have no information about the phase of the system to use to adjust the
local oscillator phase. Therefore we must use a heterodyne scheme for this
time period, rapidly varying the local oscillator phase. This means that
$B_{\tau}$ will be equal to zero, and no matter how good the phase estimate
is after time $\tau$, the largest the magnitude of $B_v$ can be made is
$v-\tau$. Then at the end of the measurement, the largest $|B|$ can be is
$1-\tau$, and the largest $|\zeta^{\rm P}|$ can be is $\atanh (1-\tau)$.

The lower limit to the introduced phase variance when there is a time delay of
$\tau$ is therefore
\begin{eqnarray}
\ip{\Delta \phi^2}_{\rm min} \!\!\!\!& \approx &\!\!\!\!
              \frac{e^{-2\atanh (1-\tau)}}{4\bar n}, \nn \\
\!\!\!\!& \approx &\!\!\!\! \frac{\tau}{8\bar n}.
\end{eqnarray}
We can expect that the introduced phase variance will be close to this for
states of small intrinsic phase variance, as there will quickly be very good
phase estimates available for the feedback. In addition, the time delay
must be sufficiently large that the phase variance given by this expression is
significantly above the introduced phase variance for no time delay.

This result obeys the same scaling law as the result given in
\cite{semiclass}, but it is a factor of four times smaller. Note, however,
that the limit condition for the result given in \cite{semiclass} is that
$\tau \alpha$ is small, whereas the above result should only be accurate
when both $\alpha$ and $\tau$ are reasonably large. The result here also
differs in that it is the limit for the total introduced phase variance,
rather than just the extra phase variance due to the time delay.

\section{Numerical Results}
\label{result}
These analytic results were also tested numerically. The numerical
techniques used were similar to those described in Sec.~\ref{method}. Minimum
uncertainty squeezed states were used, with the stochastic differential
equations for the squeezing parameter $\alpha$ as given in Eq.~(\ref{SDEalpha})
and the value of $B_v^S$ as given in Eq.~(\ref{Bsoln}). For all calculations
$2^{20}$ time steps were used, and calculations were performed with time delays
of $2^n$ time steps, where $n$ varies from 0 to 18.

For most of these calculations the same random numbers were used for each time
delay in order to see the difference in the variance due to the time delay more
accurately. If this is not done, the differences between the variances for the
different time delays are primarily due to the variation of the random numbers,
rather than the different time delay. Using the same random numbers makes the
results correlated, so that the differences are primarily due to the time delay.

For the first $2^n$ time steps the local oscillator phase was rotated by
$\pi /2$ each step. For the following time steps the data up to the time
step $2^n$ before the current time step was used. For a delay of $2^0=1$
time steps the data from the previous step was used, corresponding to the
technique for no time delay.

Numerical results for four different phase feedback schemes were obtained: \\
(a) The simplified feedback for mark I and II measurements, where
\begin{equation}
d \hat \varphi_v = \frac{dI(v) dv}{\sqrt v}.
\end{equation}
(b) The unsimplified feedback, where the phase estimate is
\begin{equation}
\hat \varphi(v) = \arg A_v.
\end{equation}
(c) The phase estimate that is intermediate between $\arg A$ and the best
phase estimate
\begin{equation}
\hat \varphi(v) = \arg \left( A_v^\varepsilon C_v^{1-\varepsilon} \right),
\end{equation}
where $\varepsilon$ is a constant. \\
(d) The same as in (c), except that the value of $\varepsilon$ varies with
time as
\begin{equation}
\varepsilon(v) = \frac{v^2-|B_v|^2}{|C_v|}\sqrt{\frac{v}{1-v}}.
\end{equation}

\subsection{Comparison with Perturbative Theory}
\begin{figure}
\centering
\includegraphics[width=0.7\textwidth]{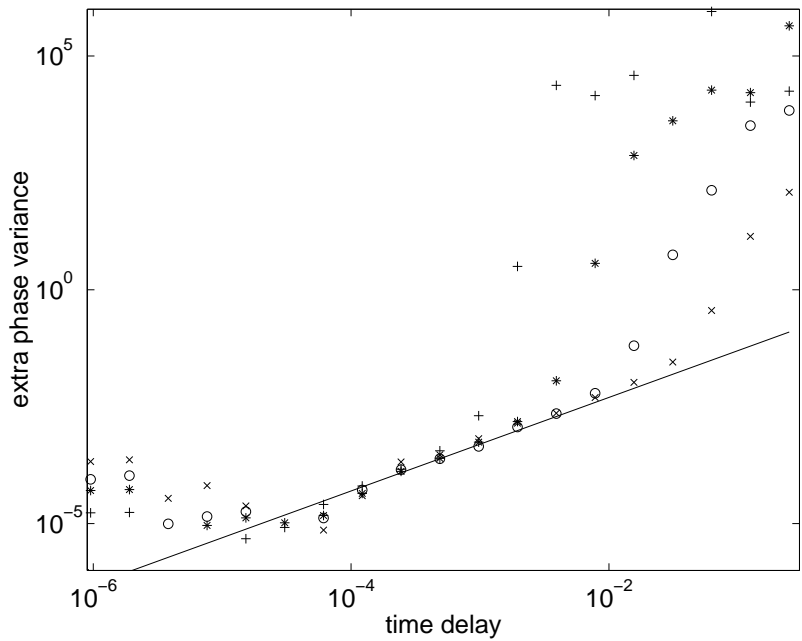}
\caption{The extra phase variance (in the final value of the intermediate
phase estimate for the simplified feedback) due to the time delay plotted
as a function of time delay for four different mean photon numbers. The
data for a mean photon number of 121.590 are shown as crosses, for a
photon number of 1576.55 as circles, for a photon number of 22254.8 as
asterisks and for a photon number of 332067 as pluses. The approximate analytic
result $\tau/2$ is plotted as the continuous line. }
\label{mark0}
\end{figure}

I will firstly consider the case of simplified feedback, and consider
the variance in the final value of the feedback phase, rather than the phase
of $A$ or $C$. This case was examined in Sec.~\ref{dtontwo}, and the
extra phase variance due to the time delay is $\tau /2$ according to that
analysis. The extra phase variance is plotted for four different mean
photon numbers in Fig.~\ref{mark0}. For each of the points shown $2^{13}$
samples were used.

To determine the extra phase variance due to the time delay, an estimate
must be made of the phase variance with no time delay. The most convenient
estimate to use is the minimum variance obtained, as this prevents negative
data points that cannot be plotted on a log-log graph. The minimum variance is
not necessarily that for the smallest time delay, due to the stochastic nature
of the calculations.

The theoretical asymptotic value of $\tau /2$ is also plotted in Fig.~\ref{mark0}.
As can be seen, many of the results are close to the theoretical line for the
intermediate time delays. For small time delays, the extra phase variance
due to the time delay is too small a fraction of the total phase variance
for the results to be accurate. The reason why the results deviate from the
asymptotic result for large time delays is that this result
is for the limit of small $\alpha \tau$. Note also that the results for
larger photon numbers deviate from the asymptotic result for smaller
$\tau$ than the results for smaller photon numbers. This is what can be expected
from this limit condition.

\begin{figure}
\centering
\includegraphics[width=0.7\textwidth]{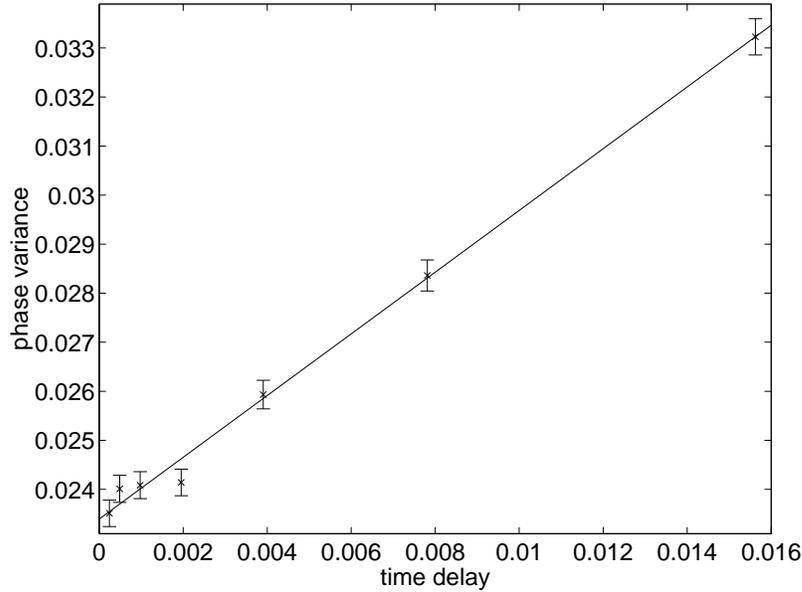}
\caption{The Holevo variance in the final value of the intermediate phase
estimate for a mean photon number of 121.590. The results are shown as the
crosses and the continuous line is that fitted to the data.}
\label{del20}
\end{figure}

\begin{figure}
\centering
\includegraphics[width=0.7\textwidth]{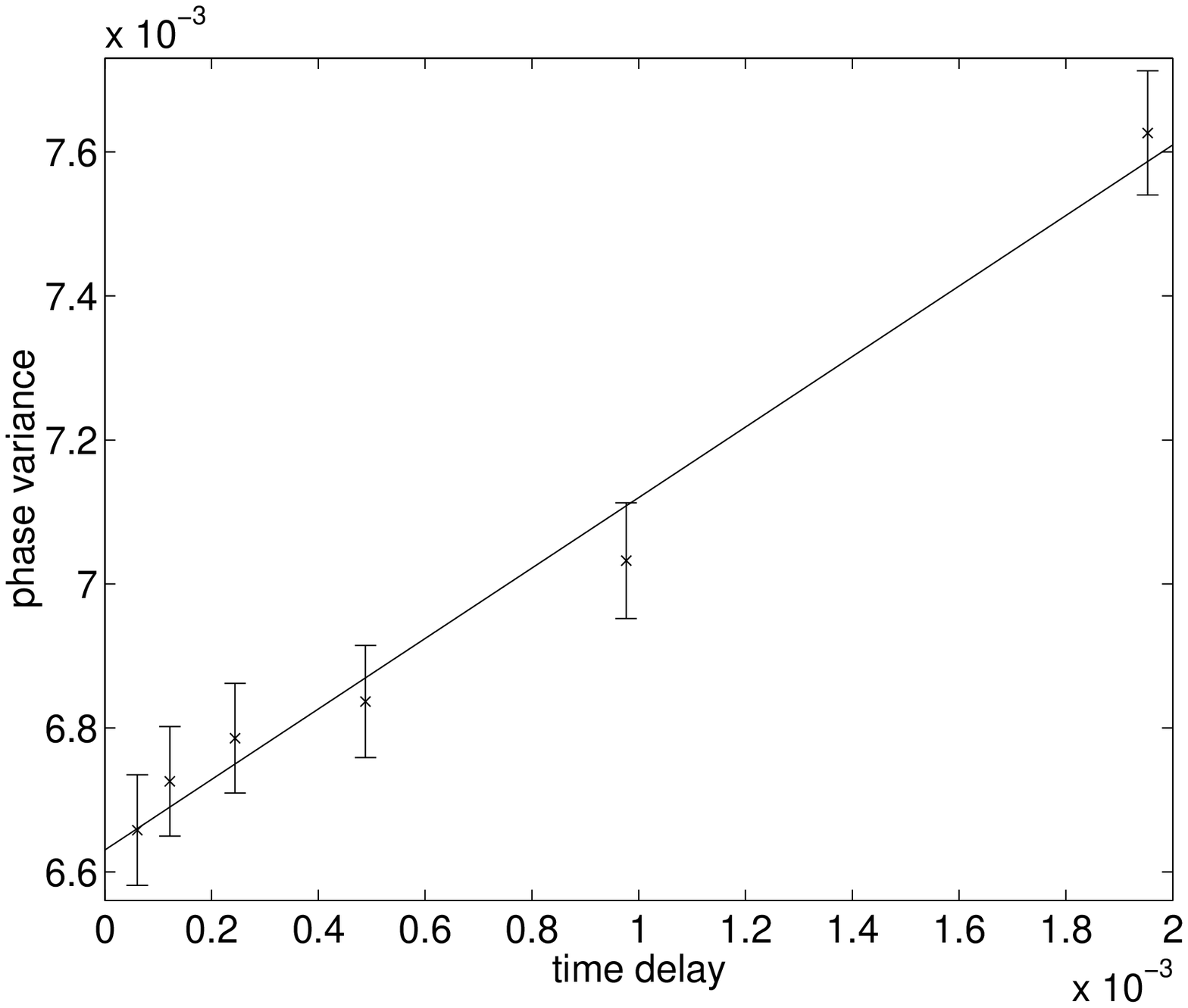}
\caption{The Holevo variance in the final value of the intermediate phase
estimate for a mean photon number of 1576.55. The results are shown as the
crosses and the continuous line is that fitted to the data.}
\label{del30}
\end{figure}

\begin{figure}
\centering
\includegraphics[width=0.7\textwidth]{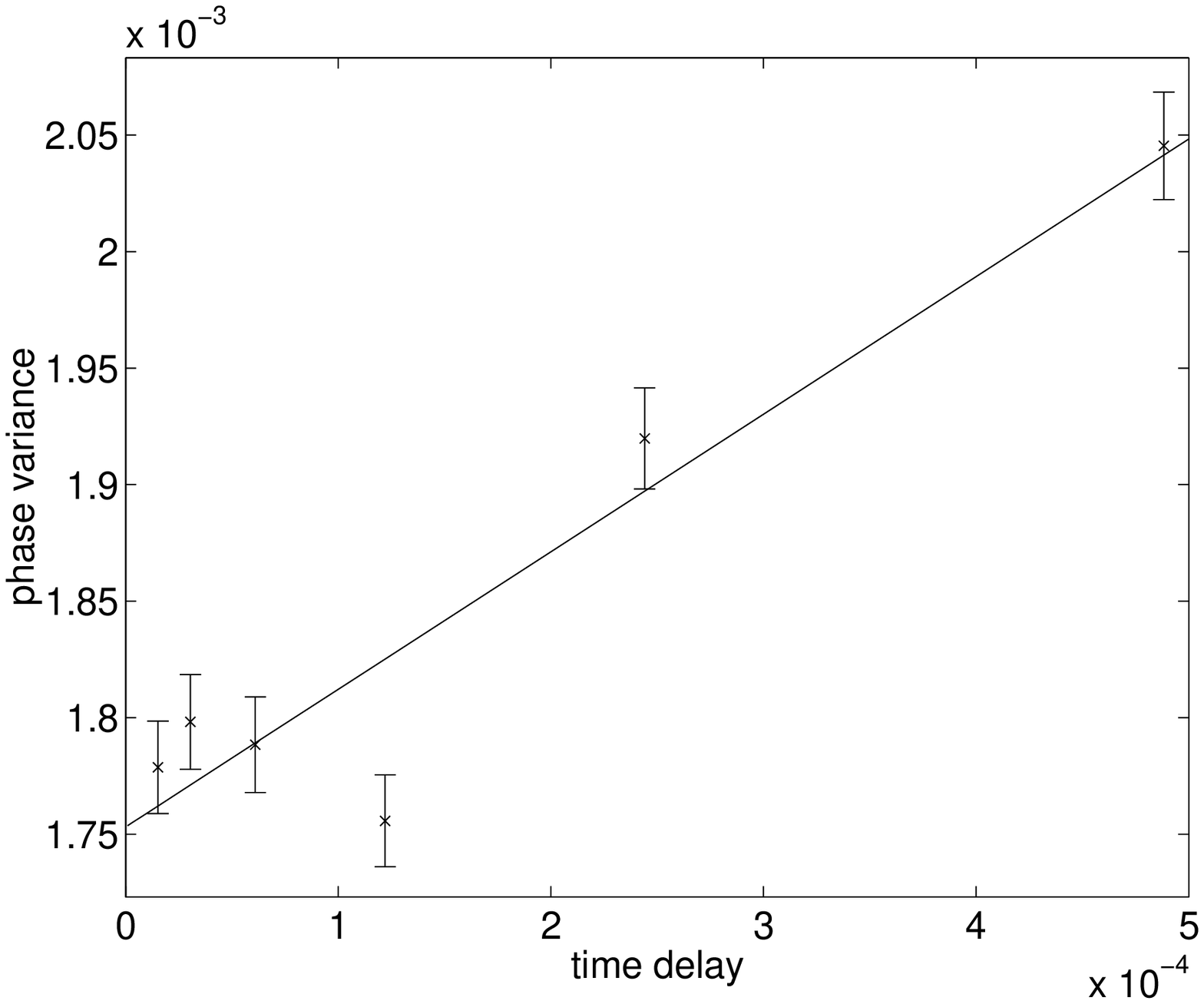}
\caption{The Holevo variance in the final value of the intermediate phase
estimate for a mean photon number of 22254.8. The results are shown as the
crosses and the continuous line is that fitted to the data.}
\label{del40}
\end{figure}

\begin{figure}
\centering
\includegraphics[width=0.7\textwidth]{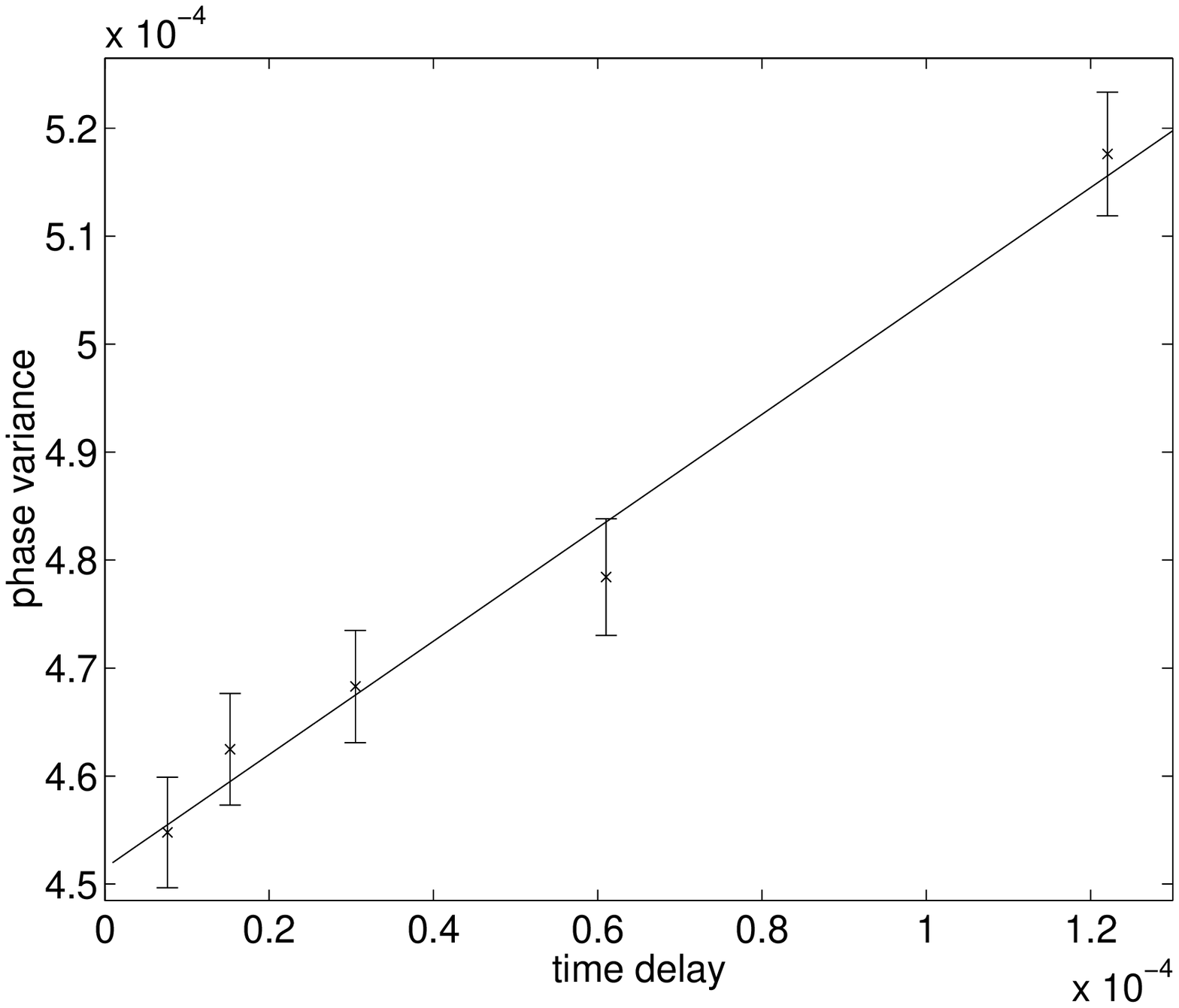}
\caption{The Holevo variance in the final value of the intermediate phase
estimate for a mean photon number of 332067. The results are shown as the
crosses and the continuous line is that fitted to the data.}
\label{del50}
\end{figure}

\begin{figure}
\centering
\includegraphics[width=0.7\textwidth]{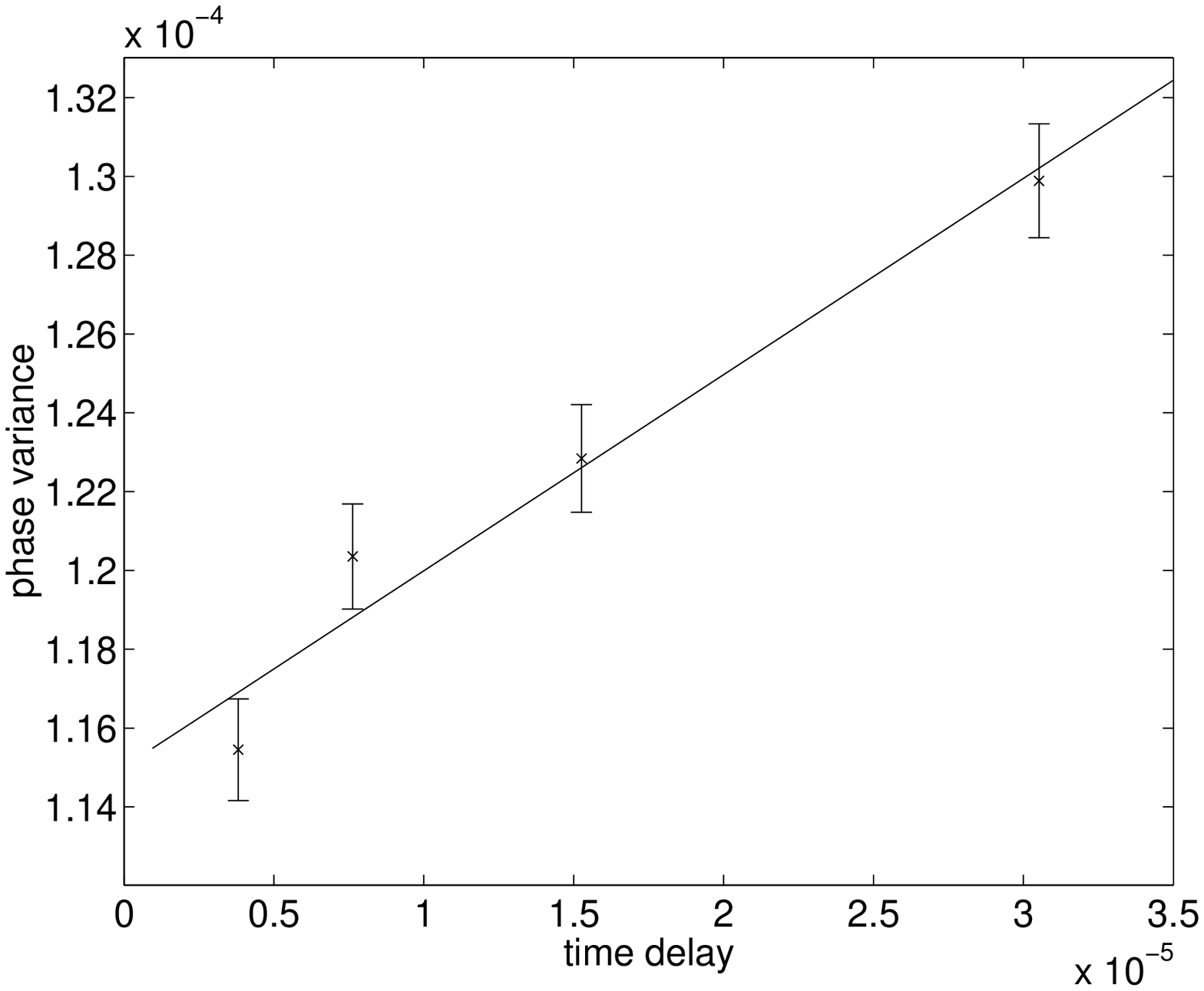}
\caption{The Holevo variance in the final value of the intermediate phase
estimate for a mean photon number of 5122478. The results are shown as the
crosses and the continuous line is that fitted to the data.}
\label{del60}
\end{figure}

It is also possible to use fitting techniques to determine how closely the
numerical results agree with the theoretical asymptotic result of $\tau/2$. The data and
fitted lines for mean photon numbers of 121.590, 1576.55, 22254.8, 332067 and
5122478 are plotted in Figs \ref{del20}, \ref{del30}, \ref{del40}, \ref{del50}
and \ref{del60} respectively. For each of the points shown $2^{14}$ samples were
used. These results were determined using independent random numbers for each
data point. Calculations were also performed using the same random numbers, in
order to reduce the relative error between the data points. (These are the data
points plotted in Fig.~\ref{mark0}.) Unfortunately this method tends to produce
systematic error in the slope, which is not reflected in the uncertainty.
It is therefore better to use independent random numbers but a large number of
samples to estimate the slope.

Some of the initial data points have been omitted in each graph, as these
were too close to each other to be useful. The data points for larger time
delays have also been omitted. These tend to be less accurate, as the
approximation is in the limit of small $\alpha\tau$. More specifically, for most
of these graphs the data points for $\alpha\tau>0.08$ were omitted. The slopes
of the fitted lines found were
\bqa
S_{\ref{del20}} \!\!\!\! &=& \!\!\!\! 0.629 \pm 0.019, \nn \\
S_{\ref{del30}} \!\!\!\! &=& \!\!\!\! 0.490 \pm 0.032, \nn \\
S_{\ref{del40}} \!\!\!\! &=& \!\!\!\! 0.590 \pm 0.098, \nn \\
S_{\ref{del50}} \!\!\!\! &=& \!\!\!\! 0.525 \pm 0.039, \nn \\
S_{\ref{del60}} \!\!\!\! &=& \!\!\!\! 0.498 \pm 0.075. \nn
\eqa
Except for the first result, these results are all consistent with the
theoretical value of 0.5. The third result is higher than 0.5, but also has a
large uncertainty. Note that, due to the linearisation in Eq.~(\ref{nodelay}),
we can only expect the $\tau/2$ result to be accurate in the limit of large
$\alpha$. The reason for the larger slope for the smallest photon number is
likely to be that $\alpha$ is not sufficiently large for the linearisation to
be accurate.

\begin{figure}
\centering
\includegraphics[width=0.7\textwidth]{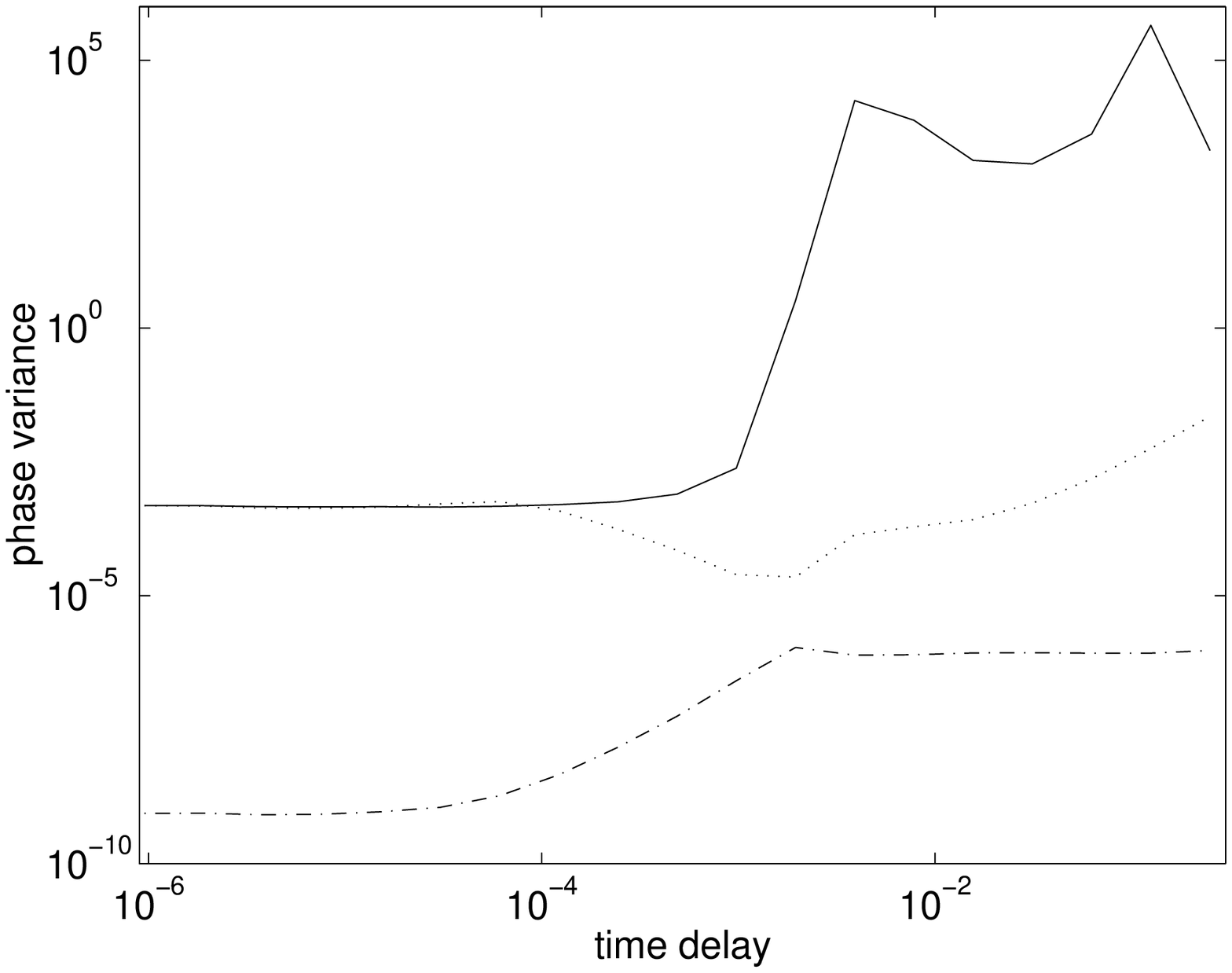}
\caption{The variance of three alternative final phase estimates
for simplified feedback with a time delay plotted as a function of time
delay. The results for the final value of the intermediate phase estimate
are plotted as a continuous line, for $\arg A$ as a dotted line, and for
$\arg C$ as a dash-dotted line. All results are for a photon number of 332067.}
\label{all3}
\end{figure}

As was mentioned above, the result for the additional phase variance
due to the time delay is only valid for the variance in the final
value of the phase estimate, which is not the same as $\arg A$ when there
is a time delay. In Fig.~\ref{all3} I have plotted the variation of the
phase variance with time delay for three alternative final phase estimates,
$\hat \varphi_1$, $\arg A$ and $\arg C$. This is for a photon number of
approximately 332000, and is fairly representative of the results for other
photon numbers. For these results, and the rest of the results in this chapter,
$2^{11}$ samples were used for each data point.

As can be seen, for very small time delays the variances in the $\hat \varphi_1$
and $\arg A$ phase estimates are almost identical. As the time delay is
increased, however, the variance of $\hat \varphi_1$ increases, but the variance
of $\arg A$ {\it decreases}. This is because, as the intermediate phase estimate
gets worse, the value of $|B|$ decreases. This means that $A$ is closer to $C$,
so $\arg A$ is closer to the best phase estimate. Note, however, that the
variance of $\arg A$ rises again, and does not converge to $\arg C$ for large
time delays. This is because $|B|$ does not fall to zero.

\subsection{Comparison with Theoretical Minimum}
\begin{figure}
\centering
\includegraphics[width=0.7\textwidth]{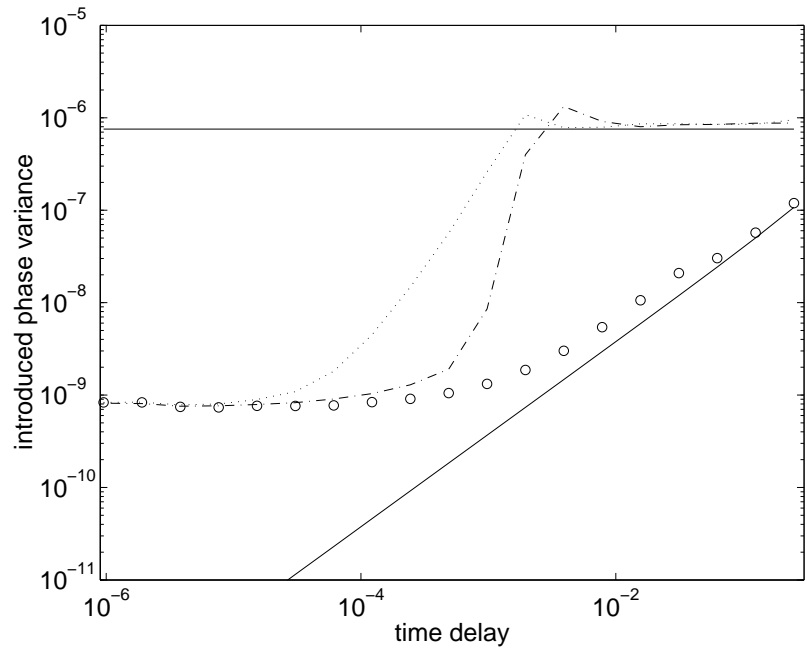}
\caption{The introduced phase variance for three different phase feedback
schemes plotted as a function of time delay. The dotted line is for
simplified feedback, the dash-dotted line is for the corrected simplified
feedback, and the circles are for unsimplified $\arg A_v$ feedback. The
best phase estimate $\arg C$ is used in all three cases. The continuous
horizontal line is the phase variance for heterodyne measurements, and
the continuous diagonal line is the theoretical limit. All results are
for a photon number of 332067.}
\label{limits}
\end{figure}

Lastly the variance in the phase of $C$ will be considered. As was explained
above, the theoretical lower limit to the introduced phase variance is
$\tau/(8\bar n)$. I have plotted the introduced variance in the best phase
estimate $\arg C$ and the theoretical limit in Fig.~\ref{limits}. For
additional accuracy I have plotted
\begin{equation}
\frac{e^{-2\atanh (1-\tau)}}{4\bar n},
\end{equation}
as this will continue to be accurate for time delays that are a large
fraction of 1. This plot is for a photon number of 332000, and similar
results are obtained for other photon numbers. In the case of simplified
feedback, the phase variance is well above the theoretical limit. For large
time delays the phase variance approximately converges to the heterodyne phase
variance, also shown in Fig.~\ref{limits}.

The introduced phase variance for mark II measurements with the unsimplified
$\arg A_v$ feedback is also shown in Fig.~\ref{limits}. The introduced
phase variance for this case increases far more slowly with the time delay,
and for larger time delays it is very close to the theoretical limit. These
results indicate that if there is any significant time delay in the system,
the simplified feedback will give a far worse result than using $\arg A_v$.

It is possible to make a correction to the simplified phase feedback scheme
that improves this result somewhat. Many different alternatives were tried,
and the one that gave the best results was 
\begin{equation}
d \hat \varphi_v = \frac{I(v) dv}{\sqrt {v+\alpha \tau}}.
\end{equation}
This correction is based on the fact that $|A_v|$ is larger than $\sqrt v$
when the phase estimate is worse than $\arg A_v$. (From \cite{semiclass},
the factor of $\sqrt v$ in the simplified feedback comes from a factor of
$|A_v|$.) The results for this correction are also shown in Fig.~\ref{limits}.
The phase variances obtained in this case are significantly below those for
the plain simplified feedback, but are still far above the results for the
unsimplified $\arg A_v$ feedback.

\begin{figure}
\centering
\includegraphics[width=0.7\textwidth]{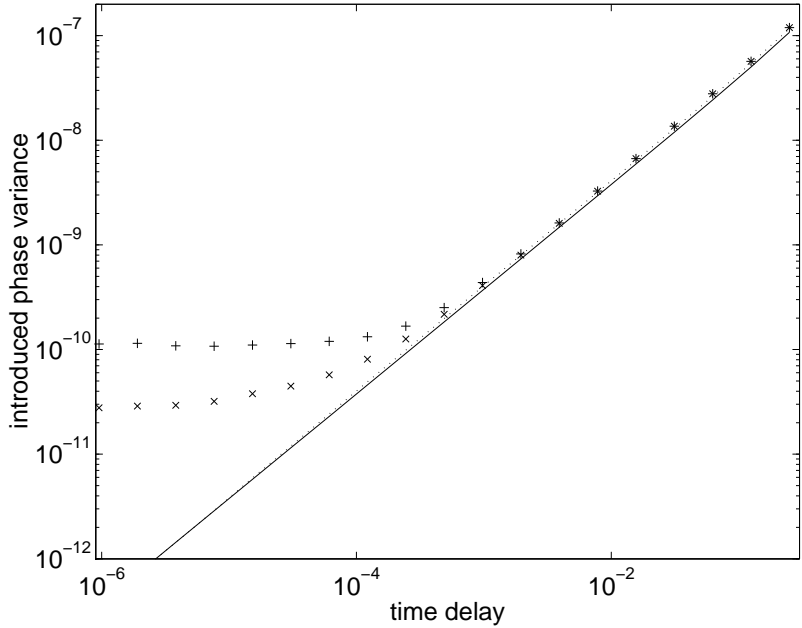}
\caption{The introduced phase variance for better intermediate phase
estimates plotted as a function of time delay. The pluses are for the
constant $\varepsilon$ case and the crosses are for the time dependent
$\varepsilon$ case. The theoretical limit estimated using the mean inverse
photon numbers obtained from the time dependent $\varepsilon$ case is plotted
as the dotted line, and the theoretical limit using the input photon
number is shown as the continuous line. All results are for a photon
number of 332067.}
\label{both50}
\end{figure}

Now we will consider the results for better intermediate phase estimates
that are between $\arg A_v$ and $\arg C_v$. The introduced phase variance
for the constant $\varepsilon$ case and the theoretical limit are shown in
Fig.~\ref{both50}. These results are again for a photon number of about
332000. The results for this case are even closer to the theoretical
limit than those for the mark II case.

The introduced phase variance for the feedback with time-dependent
$\varepsilon$ is also plotted in Fig.~\ref{both50}. The results for this case
converge to the theoretical limit at smaller time delays than for the
constant $\varepsilon$ case. For the larger time delays the results for
the two cases are about the same, slightly above the theoretical limit.

In both cases the phase variance is still noticeably above the theoretical limit
for large time delays, and the values of $B$ obtained are too close to $1-\tau$
to account for this difference. The difference appears to be due to the
approximation that the photon number of the state $\ket{\alpha^{\rm P},\zeta^{\rm P}}$ is
close to the photon number of the input state. The average value of this photon
number is close to the photon number of the input state; however, each individual
value is not necessarily close to $\bar n$. The expression for the introduced
phase variance depends on the inverse of the photon number, and the average of
an inverse is not necessarily equal to the inverse of an average. The general
expression is
\begin{equation}
\ip{\frac 1n} = \frac 1{\ip n}+\frac{\ip{\Delta n^2}}{\ip n^3}+O(\ip n^{-4}).
\end{equation}
In Fig.~\ref{both50} I have also plotted the estimated theoretical limit
based on the average of $1/{\bar n^{\rm P}}$ for the data obtained in the time
dependent $\varepsilon$ case. Specifically, the expression plotted is
\begin{equation}
\frac 14 \ip{\frac 1{\bar n^{\rm P}}}e^{-2 \atanh (1-\tau)}.
\end{equation}
As can be seen, the introduced phase variance converges to this far more
closely than to the limit based on the photon number of the input state.

\begin{figure}
\centering
\includegraphics[width=0.7\textwidth]{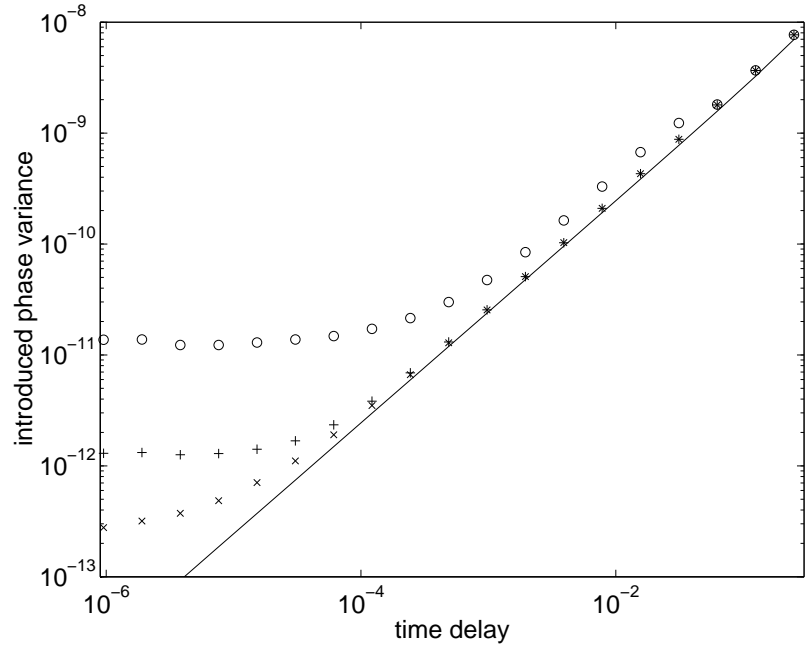}
\caption{The introduced phase variance for three different feedback schemes
plotted as a function of time delay. The circles are for mark II measurements,
the pluses are for the constant $\varepsilon$ case, and the crosses are for
the time dependent $\varepsilon$ case. The theoretical limit estimated using
the input photon number is shown as the continuous line. All results are for a
photon number of 5122478.}
\label{three60}
\end{figure}

For larger photon numbers this factor is not so significant, and the difference
between the results and the theoretical limit is smaller. For example, the
results for the three different feedback schemes (mark II, constant
$\varepsilon$ and time dependent $\varepsilon$) for a photon number of about
$5\times 10^6$ are plotted in Fig.~\ref{three60}.

%% file: intersta.tex
\setcounter{chapter}{4}

\chapter{Optimum Input States for Interferometry}
\label{adaptiveinter}
\section{Introduction}
\begin{figure}
\centering
\includegraphics[width=0.45\textwidth]{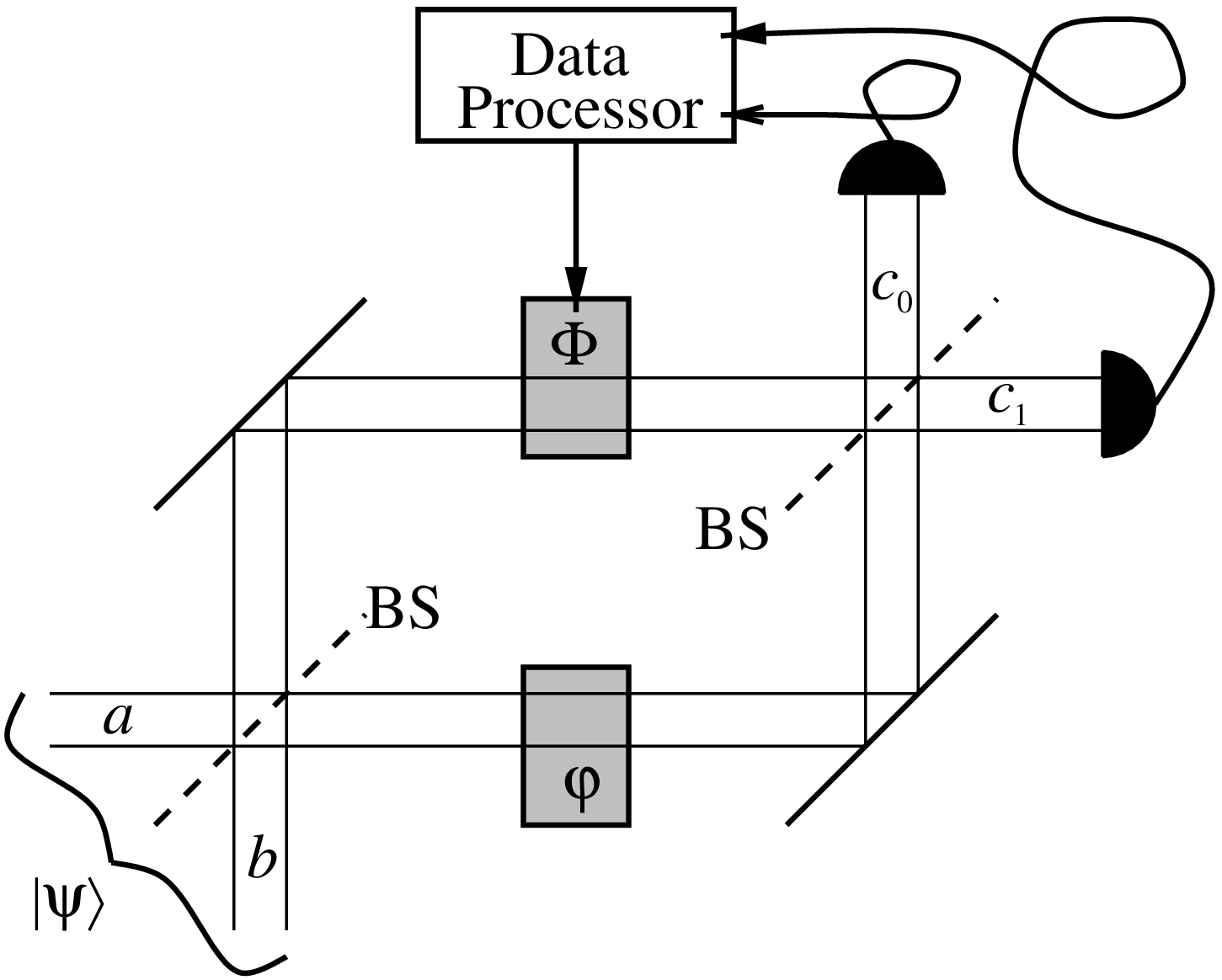}
\caption{The Mach-Zehnder interferometer, with the addition of a controllable
phase $\Phi$ in one arm. The unknown phase to be estimated is $\varphi$. Both
beam splitters (BS) are 50/50. }
\label{diag}
\end{figure}

In chapters \ref{dynesta} to \ref{delays} just a single mode of the
electromagnetic field was considered. The reference phase was provided by a
local oscillator field, which was assumed to be sufficiently large amplitude
that it could be treated classically. The main alternative to this is to
consider two modes, both of which are treated quantum mechanically. Rather than
measuring the absolute phase, we now wish to measure the phase difference
between the two modes. The simplest way of considering this is via the
Mach-Zehnder interferometer, as in Fig.~\ref{diag}.

With the Mach-Zehnder interferometer, an initial two-mode input state
is fed into a beam splitter. The two beams are then subjected to phase shifts of
$\Phi$ and $\varphi$, then recombined at a second beam splitter. The outputs of
this beam splitter are then detected using photodetectors. The counts at the
photodetectors are used to obtain an estimate of the phase difference.

These photodetectors will be assumed to be ideal for this analysis, though
current photodetectors can only achieve a sensitivity of around 87\%
\cite{SingleEff}. The photodetectors required for this need to be able to
distinguish individual photons, in contrast to the large intensity
photodetectors required for dyne detection. This is a more difficult task, and
the efficiencies currently possible are correspondingly lower. In this study,
however, we are concerned with what is possible in principle using foreseeable
technologies. I will therefore be considering only the case of unit efficiency.

In general the interferometer is used to measure the phase difference between
the two arms. Similarly to the single mode case, we wish to add an auxiliary
phase shift $\Phi$ in order to obtain a more accurate phase measurement. For
simplicity I have indicated the phase we wish to measure, $\varphi$, as the
phase shift in one arm in Fig.~\ref{diag}. In the following analysis this
phase can equivalently be taken to be the phase difference between the two arms
in the absence of the auxiliary phase shift $\Phi$.

It is well known that it is possible to obtain a phase uncertainty scaling as
$1/\sqrt N$ (the standard quantum limit) when a photon number state with $N$
photons is input to one port of the interferometer. Several authors
\cite{Caves,Yurke,Holland,SandMil95,SandMil97} have proposed ways of reducing
the phase uncertainty to the Heisenberg limit of $1/N$. Most of these proposals
\cite{Caves,Yurke,Holland} are essentially for detecting small deviations from
some known phase, and therefore only give a $1/N$ scaling for a very small range
of phases. Sanders and Milburn \cite{SandMil95,SandMil97} considered ideal
measurements, that give $1/N$ scaling independent of the phase. Unfortunately
they do not discuss how these measurements can be achieved in practice, and as
will be shown in Ch.~\ref{interfere} it is not possible to perform these
measurements in general, even allowing feedback.

We wish to perform measurements as close as possible to ideal by varying $\Phi$
during the measurement, based on the detections. In general there are three
different areas available for optimisation: \\
1. The initial input state. \\
2. How the feedback phase is changed during the measurement. \\
3. The final phase estimate. \\
In this chapter I discuss the optimum initial input states, and in the next
chapter the feedback phase and final phase estimate are discussed.

\section{Optimum Input States}
\label{OptInpSta}
The input states are most conveniently described using the Schwinger
representation. The operators in this representation are
\bqa
\hat J_x  \!\!\!\! &=& \!\!\!\! (a\dg b+ab\dg)/2, \nn \\
\hat J_y  \!\!\!\! &=& \!\!\!\! (a\dg b-ab\dg)/2i, \nn \\
\hat J_z  \!\!\!\! &=& \!\!\!\! (a\dg a-b\dg b)/2, \nn \\
\hat J^2  \!\!\!\! &=& \!\!\!\! \hat J_x^2 + \hat J_y^2 + \hat J_z^2.
\eqa
The operators $\hat J_x$, $\hat J_y$ and $\hat J_z$ satisfy the commutation
relations for the Lie algebra of SU(2):
\bqa
\left[ \hat J_x, \hat J_y \right] \!\!\!\! &=& \!\!\!\! i\hat J_z, \nn \\
\left[ \hat J_y, \hat J_z \right] \!\!\!\! &=& \!\!\!\! i\hat J_x, \nn \\
\left[ \hat J_z, \hat J_x \right] \!\!\!\! &=& \!\!\!\! i\hat J_y.
\eqa
The operator $\hat J^2$ is the Casimir invariant for this group (i.e.\ it
commutes with all elements). These commutation relations are the same as those
for the operators for components of angular momentum.

I will use the notation $\ket{j\mu}_z$ for the common eigenstate of $\hat J_z$
and $\hat J^2$, with eigenvalues of $\mu$ and $j(j+1)$ respectively. This
corresponds to Fock number eigenstates at the interferometer inputs with photon
numbers in ports $a$ and $b$ of $j+\mu$ and $j-\mu$ respectively. Similarly the
notation $\ket{j\mu}_y$ means the common eigenstate of $\hat J_y$ and $\hat J^2$
with eigenvalues of $\mu$ and $j(j+1)$. This state corresponds to number
eigenstates in the interferometer arms with photon numbers of $j+\mu$ and
$j-\mu$. To see this, note that the annihilation operators for the modes in the
two interferometer arms are
\begin{equation}
(a + ib)/\sqrt{2} ,\;\;\; (ia + b)/\sqrt{2}.
\end{equation}
It is simple to show from this that the operator for the difference in the
photon numbers beween the arms is
\beq
\half (-ia\dg + b\dg)(ia + b) - \half (a\dg - ib\dg)(a + ib) =
(a\dg b-ab\dg)/i,
\eeq
which is twice $\hat J_y$.

Note that the scattering matrix used here for the beam splitter is
\beq
\frac{1}{{\sqrt 2 }}\left[ {\begin{array}{*{20}c}
   1 & i  \\
   i & 1  \\
\end{array}} \right].
\eeq
In contrast, the scattering matrix considered in the case of dyne measurements
was
\beq
\frac{1}{{\sqrt 2 }}\left[ {\begin{array}{*{20}c}
   1 & 1  \\
   -1 & 1  \\
\end{array}} \right].
\eeq
These scattering matrices have been used for consistency with previously
published work. The only important consequence of this difference is that for
dyne measurements, small errors are obtained when the difference between the
signal and local oscillator phases is $\pi/2$, whereas in the interferometer
case the phase difference should be close to zero. This is because, in the
interferometer case, the $i$ in the scattering matrix gives a $\pi/2$ phase
shift.

In order to represent a completely general state we can express it as a sum of
input number states:
\beq
\ket\psi = \sum_{2j = 0}^\infty {\sum_{\mu=-j}^j {\psi_{j\mu}\ket{j\mu}_z}}.
\eeq
Most proposals for reducing the phase uncertainty to the $1/N$ limit consider
only input states with a fixed total photon number of $N=2j$. This restriction
is also applied in this study, as it greatly simplifies the analysis. With this
restriction, the input state can be represented as
\beq
\ket\psi = \sum_{\mu=-j}^j {\psi_\mu \ket{j\mu}_z}.
\eeq

Similarly to the single mode case, the probability distribution for the estimate
of the phase of a two mode input state, $\hat \varphi$, is in general given by
\beq
\label{POM2mode}
P(\hat \varphi)={\rm Tr}\left[ {\rho F(\hat \varphi)} \right],
\eeq
where $F(\hat \varphi)$ is the POM for the measurement. This only depends
on the intrinsic phase of the two mode state, and not the phase shift in the
interferometer. The interferometer transforms the input state $\ket \psi$ to
$\hat I(\varphi)\ket \psi$, where $\varphi$ is the phase in the interferometer
and
\beq
\hat I(\varphi) = \exp \left( -i\varphi \hat J_y \right) .
\eeq
The state matrix therefore transforms to
\beq
\rho' = \hat I(\varphi) \rho \hat I\dg(\varphi).
\eeq
When this transformed state is used in Eq.~(\ref{POM2mode}) the probability
distribution will be dependent on the interferometer phase. This
transformation acts to shift the phase of the state by $\varphi$, so the
probability distribution will depend on the sum of the phase of the input state
and the interferometer phase shift.

We can alternatively include the interferometer in the POM, and use the input
state in Eq.~(\ref{POM2mode}). Then we would transform the POM to
\beq
F'(\hat \varphi) = \hat I\dg(\varphi) F(\hat \varphi) \hat I(\varphi).
\eeq
If the POM describes a shift invariant phase measurement, then this can be
simplified to
\bqa
F'(\hat \varphi) \!\!\!\! &=& \!\!\!\! F(\hat \varphi-\varphi) \nn \\
\!\!\!\! &=& \!\!\!\! F(\phi),
\eqa
where $\phi$ is the error in the phase estimate. Alternatively $\phi$ can be
taken to be the phase estimate when the interferometer phase shift is zero.

Sanders and Milburn \cite{SandMil95,SandMil97} considered what they call
``optimal'' measurements, where the POM is given by
\beq
\label{SMPOM}
F(\phi) = \frac{2j+1}{2\pi}\ket{j\phi}\bra{j\phi},
\eeq
where the $\ket{j\phi}$ are normalised phase states given by
\beq
\ket{j\phi}=(2j+1)^{-1/2} \sum_{\mu=-j}^j e^{i\mu\phi} \ket{j\mu}_y.
\eeq
As I will show below, this POM is equivalent to the ideal or canonical POM for a
single mode. For this reason I will generally use the same terminology for this
POM, and reserve the word ``optimal'' for the best possible measurements that
are realisable using photodetection and feedback.

When expressed in terms of the eigenstates of $\hat J_y$, the canonical POM is
\beq
F(\phi)=\frac 1{2\pi}\sum_{\mu,\nu=-j}^j e^{i(\mu-\nu)\phi}
\ket{j\mu}_y \bra{j\nu}.
\eeq
This is very similar to the POM in the single mode case when there is an upper
limit of $N=2j$ on the photon number. This POM is not given explicitly above;
however, it can be obtained by taking the single mode canonical POM given by
Eq.~(\ref{canPOM}), and limiting the sum to $N$. Explicitly, this POM is
\beq
F^{\rm can}(\phi) = \frac 1{2\pi} \sum_{n,m=0}^N e^{i(n-m)\phi}\ket n \bra m.
\eeq

To show that the interferometer POM of Eq.~(\ref{SMPOM}) is completely
equivalent to this, we can make the change of variables
\bqa
\mu' \!\!\!\! &=& \!\!\!\! \mu + j \nn \\ 
\nu' \!\!\!\! &=& \!\!\!\! \nu + j \nn \\
\ket{\mu '} \!\!\!\! &=& \!\!\!\! \ket{j\mu}_y.
\eqa
With this change in notation, the POM is
\beq
F(\phi)=\frac 1{2\pi} \sum_{\mu',\nu'=0}^N e^{i(\mu'-\nu')
\phi} \ket{\mu'}\bra{\nu'}.
\eeq
This POM is identical to the POM for the single mode case for an upper limit on
the photon number of $N$. This case was considered in Sec.~\ref{simplest}, and
the state that minimises the Holevo phase variance is
\beq
\ket{\psi_{\rm opt}} = \frac 1{\sqrt{j + 1}}\sum\limits_{\mu'=0}^{2j} \sin
\left[ \frac{(\mu'+1)\pi}{2j + 2} \right]\ket{\mu'}.
\eeq
Converting the notation back to the original variables, this state is
\beq
\ket{\psi_{\rm opt}}  = \frac 1{\sqrt{j+1}}\sum_{\mu=-j}^j {\sin \left[
\frac{(\mu+j+1)\pi}{2j+2} \right]\ket{j\mu}_y } .
\eeq
The minimum Holevo phase variance corresponding to this state is
\beq
V(\phi) = \tan ^2 \left( \frac{\pi}{N+2} \right).
\eeq

The states $\ket{j\mu}_y$ correspond to joint number states within the
interferometer, and they do not correspond to input number states in a simple
way. We therefore wish to re-express the optimum states in terms of the input
number states $\ket{j\mu}_z$. To do this, we require the relation from
Ref.~\cite{SandMil95},
\beq
\label{SandResult}
_y \braket{j\mu}{j\nu}_z = e^{i(\pi /2)(\nu-\mu)} I_{\mu \nu }^j (\pi /2),
\eeq
where $I_{\mu \nu }^j (\pi /2)$ are the interferometer matrix elements given by
\beq
I_{\mu\nu}^j(\pi/2) = 2^{-\mu} \left[ \frac{(j-\mu)!}{(j-\nu)!}\frac{(j+\mu)!}
{(j+\nu)!} \right]^{1/2} P_{j-\mu}^{(\mu-\nu,\mu+\nu)} (0),
\eeq
for
\beq
\label{inequal}
\mu  - \nu  \ge 0,\quad \mu  + \nu  \ge 0,
\eeq
where $P_n^{(\alpha,\beta)}$ are the Jacobi polynomials, given by
\beq
P_n^{(\alpha,\beta)} (x) = 2^{-n} \sum_{m=0}^n \binom{n+\alpha}m
\binom{n+\beta}{n-m} (x-1)^{n-m} (x+1)^m .
\eeq
Therefore the explicit expression for determining the values of
$I_{\mu \nu }^j (\pi /2)$ is
\bqa
  I_{\mu \nu }^j \left( {\pi /2} \right) \!\!\!\! &=& \!\!\!\! 2^{-j} \left[
\frac{(j-\mu)!}{(j-\nu)!}\frac{(j+\mu)!}{(j+\nu)!} \right]^{1/2}
\sum_{m = 0}^{j-\mu}\binom{j-\nu}m\binom{j+\nu}{j-\mu-m}(-1)^{j-\mu-m} \nn \\
\!\!\!\! &=& \!\!\!\!
2^{-j} \left[ (j-\mu)!(j+\mu)!(j-\nu)!(j+\nu)! \right]^{1/2} \nn \\
&&\times \sum_{m = 0}^{j-\mu} \left[ m!(j-\nu-m)!(j-\mu-m)!(\mu+\nu+m)!
\right]^{-1} (-1)^{j-\mu-m}.
\eqa
The values of $I_{\mu \nu }^j (\pi /2)$ for values of $\mu$ and $\nu$ that do
not obey the inequalities (\ref{inequal}) can be obtained using the symmetry
relations
\beq
I_{\mu\nu}^j (\theta) = (-1)^{\mu-\nu} I_{\nu\mu}^j (\theta) = I_{-\nu,-\mu}^j
(\theta).
\eeq
Using the result from Eq.~(\ref{SandResult}), the state expressed in terms of
the eigenstates of $\hat J_z$ is
\beq
\ket{\psi_{\rm opt}}=\frac{1}{\sqrt{j + 1}}\sum_{\mu ,\nu = -j}^j {\sin \left[
\frac{(\mu  + j + 1)\pi}{2j + 2} \right]e^{i(\pi /2)(\mu - \nu )}
I_{\mu \nu }^j (\pi /2)\ket{j\nu}_z } .
\eeq

\begin{figure}
\centering
\includegraphics[width=0.7\textwidth]{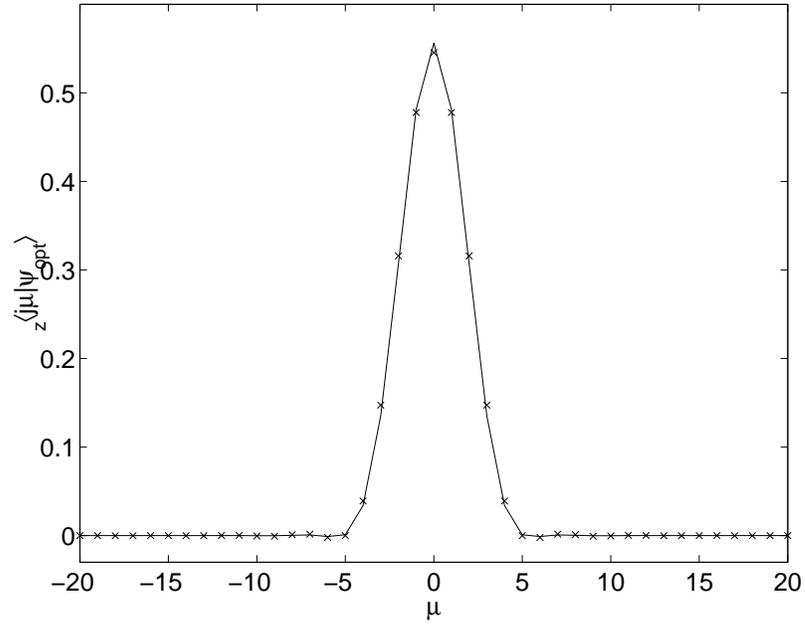}
\caption{The coefficients $_z \braket{j\mu}{\psi_{\rm opt}}$ for the state
optimised for minimum phase variance under canonical measurements. All
coefficients for a photon number of $2j=40$ are shown as the continuous line,
and those near $\mu=0$ for a photon number of 1200 as crosses.}
\label{optstate}
\end{figure}

\begin{figure}
\centering
\includegraphics[width=0.7\textwidth]{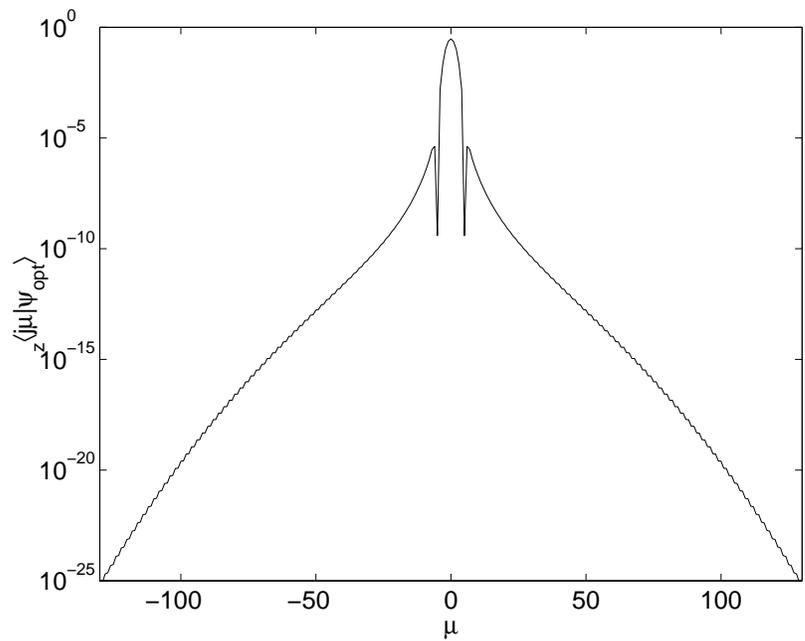}
\caption{The coefficients $_z \braket{j\mu}{\psi_{\rm opt}}$ for larger values
of $\mu$ for the state optimised for minimum phase variance under canonical
measurements for a photon number of 1200.}
\label{optexpon}
\end{figure}

An example of an optimum state for 40 photons calculated using this formula is
shown in Fig.~\ref{optstate}. As can be seen, the only significant
contributions are from 9 or 10 $\hat J_z$ eigenstates near $\mu=0$. In
addition, the distribution near the centre is fairly independent of photon
number, as can be seen by comparing this state with the state for 1200 photons,
also shown in Fig.~\ref{optstate}. For larger values of $\mu$, the
contributions fall approximately exponentially with $\mu$, as can be seen in
Fig.~\ref{optexpon}.

In Ref.~\cite{Yurke} it is shown that it is possible to obtain states similar
to $\ket{j0}_z$ and $\ket{j1}_z$ using a two-mode four-wave mixer. As the
optimum input states have their main contributions from these states and other
$\hat J_z$ eigenstates near $\mu=0$, this suggests that it may be possible to
produce states that are close approximations of the optimum input states using a
suitable modification of the apparatus used in Ref.~\cite{Yurke}. Unfortunately,
rather than producing only the state $\ket{j0}_z$ (for example), the four-wave
mixer produces a superposition of these states with a range of values of $j$.

In \cite{Yurke} the authors state that the value of $j$ can be inferred after
the measurement from the number of photons detected. This would not be
appropriate for the measurements considered in the next chapter, as these rely
on knowing the value of $j$ before the measurement starts. In the limit of large
photon number the spread in the values of $j$ will be small compared to the
mean, so it may be possible to obtain good measurements using the mean value of
$j$. Unfortunately, the measurement scheme considered in the next chapter is
computationally infeasible in the limit of large $j$.

Nevertheless, it should be simpler to produce a small number of $\hat J_z$
eigenstates than the entire range. In order to estimate about how many
$\hat J_z$ eigenstates are required to provide a reasonable approximation of the
optimum states for a given photon number, the coefficients for the optimum
states were determined, and all except a number of coefficients near $\mu=0$
were discarded. These remaining coefficients were then normalised, and the phase
variance of the resulting state was determined.

\begin{figure}
\centering
\includegraphics[width=0.7\textwidth]{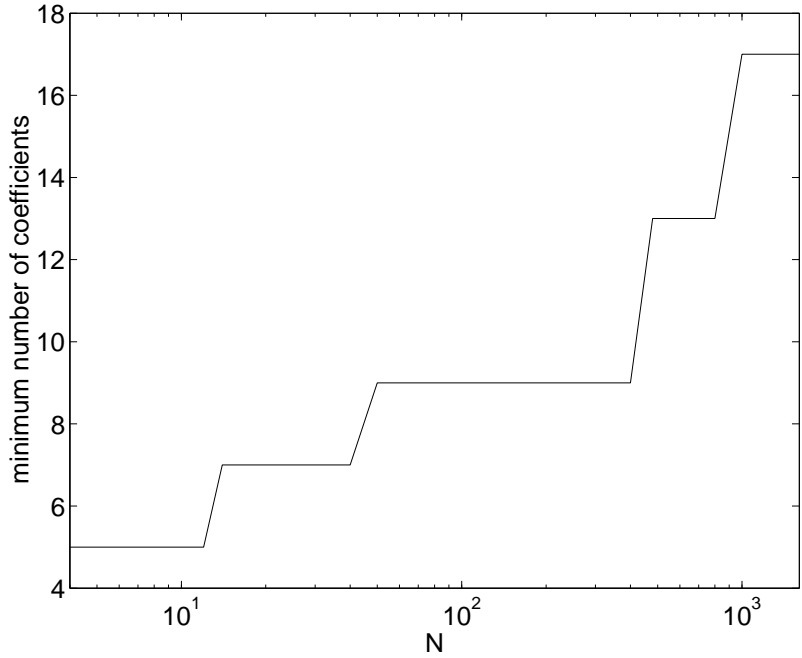}
\caption{The minimum number of $\hat J_z$ eigenstates required to approximate
the optimum state in order to obtain a phase variance less than twice optimum.}
\label{approxno}
\end{figure}

In Fig.~\ref{approxno} I have plotted the number of $\hat J_z$ eigenstates
required to approximate the optimum states for a variety of total photon
numbers.  The criterion used was that the phase variance of the approximate
state be less than twice that of the exact state. For photon numbers up to about
400, only 9 $\hat J_z$ eigenstates are required, but beyond that the number
required increases fairly rapidly, with 17 required for 1600 photons.

\section{Phase Variances for Other States}
\label{others}
I will now consider the phase variances of some other commonly considered states
in order to compare them with the optimum states. These states will be expressed
in terms of the input photon number eigenstates:
\beq
\ket\psi=\sum_{\mu=-j}^j {\psi _\mu  \ket{j\mu}_z }.
\eeq
The probability distribution for the phase for these states will be given by
\bqa
P(\phi) \!\!\!\! &=& \!\!\!\! {\rm Tr}\left[ {\ket\psi\bra\psi F(\phi)}
\right] \nn \\
\!\!\!\! &=& \!\!\!\! \frac{2j + 1}{2\pi}{\rm Tr}\left[ {\ket\psi\braket\psi
{j\phi}\bra{j\phi}} \right] \nn \\
\!\!\!\! &=& \!\!\!\! \frac{2j + 1}{2\pi}\st{\braket{j\phi}\psi}^2.
\eqa
The inner product, when expressed in terms of the coefficients $\psi_\mu$, is
given by
\bqa
\braket{j\phi }\psi\!\!\!\! &=& \!\!\!\!\sum_{\nu=-j}^j {\psi_\nu \braket{j\phi
}{j\nu }_z } \nn \\
\!\!\!\! &=& \!\!\!\! (2j+1)^{-1/2} \sum_{\nu=-j}^j {\psi_\nu \sum_{\mu=-j}^j
{e^{-i\mu \phi} {_y \braket{j\mu }{j\nu }_z} } } \nn \\
\!\!\!\! &=& \!\!\!\! (2j+1)^{-1/2} \sum_{\nu=-j}^j {\psi_\nu \sum_{\mu=-j}^j
{e^{-i\mu \phi} e^{i(\pi /2)(\nu-\mu)} I_{\mu \nu}^j (\pi /2)} }.
\eqa

The state considered in Refs~\cite{Holland,SandMil95,SandMil97} was
$\ket{j0}_z$, the state with equal photon numbers in both input ports. This
state is the biggest contributor to the optimum states, so it is not
unreasonable that this state should have a small phase uncertainty. This state
suffers the drawback that it has equal peaks at 0 and $\pi$, and therefore must
be considered modulo $\pi$ in order to obtain meaningful results. If we add the
state $\ket{j1}_z$ (the next biggest contributor to the optimum state), then we
obtain a state for which we can consider the phase modulo $2\pi$. This state,
$(\ket{j0}_z+\ket{j1}_z)/\sqrt 2$, was considered in Ref.~\cite{Yurke}.

For the state $\ket{j0}_z$, the only non-zero coefficient is $\psi_0=1$, so the
inner product is
\beq
\braket{j\phi}{j0}_z = (2j+1)^{-1/2} \sum_{\mu=-j}^j {e^{-i\mu \left( {\phi+
\pi /2} \right)} I_{\mu 0}^j \left( {\pi /2} \right)}.
\eeq
The probability distribution is then given by
\bqa
\label{exactdist}
  P\left( \phi  \right) \!\!\!\! &=& \!\!\!\! \frac{1}
{{2\pi }}\left| {\sum_{\mu  =  - j}^j {e^{ - i\mu \left( {\phi  + \pi /2}
\right)} I_{\mu 0}^j \left( {\pi /2} \right)} } \right|^2 \nn \\
   \!\!\!\! &=& \!\!\!\! \frac{1}
{{2\pi }}\sum\limits_{\mu ,\nu  =  - j}^j {e^{ - i\left( {\mu  - \nu } \right)
\left( {\phi  + \pi /2} \right)} \left( {I_{\nu 0}^j \left( {\pi /2} \right)}
\right)^ *  I_{\mu 0}^j \left( {\pi /2} \right)}.
\eqa
Because this state has equal peaks at 0 and $\pi$, the usual definition of the
Holevo phase variance will give infinite results. Rather than using the usual
definition, however, we can use a modified definition that is naturally modulo
$\pi$. The definition that I will use is
\beq
\label{holevopi}
V_{\pi}(\phi) = {{\left( \st{\ip{e^{2i\phi}}}^{-2} - 1 \right)} \mathord{\left/
 {\vphantom {{\left( \st{\ip{e^{2i\phi}}}^{-2} - 1 \right)} 4}} \right.
 \kern-\nulldelimiterspace} 4}.
\eeq
From Eq.~(\ref{exactdist}) the value of $\st{\ip{e^{2i\phi}}}$ can be determined
as
\bqa
\label{expectexp}
\st{\ip{e^{2i\phi}}} \!\!\!\! &=& \!\!\!\! \st{\int\limits_{-\pi}^\pi \frac{1}
{2\pi}\sum_{\mu,\nu=-j}^j e^{i(\nu-\mu+2)\phi}e^{i(\nu-\mu)\pi/2} \left(
{I_{\nu 0}^j \left( {\pi /2} \right)}\right)^* I_{\mu 0}^j \left( \pi /2 \right)
d\phi} \nn \\
\!\!\!\! &=& \!\!\!\! \st{\sum_{\mu,\nu=-j}^j {\delta_{\mu ,\nu+2}
\left( I_{\nu 0}^j \left( \pi/2 \right) \right)^* I_{\mu 0}^j (\pi /2)}} \nn \\
\!\!\!\! &=& \!\!\!\! \st{\sum_{\mu=2 - j}^j {\left( {I_{\mu  - 2,0}^j
\left( {\pi /2} \right)} \right)^ *  I_{\mu 0}^j \left( {\pi /2} \right)}}.
\eqa
This can be used in Eq.~(\ref{holevopi}) to determine the phase variance under
this modified definition.

\begin{figure}
\centering
\includegraphics[width=0.7\textwidth]{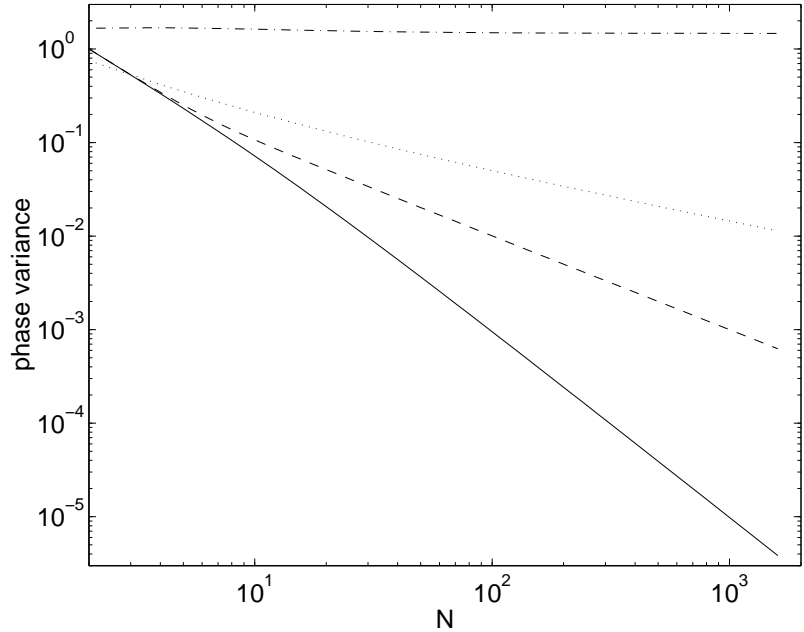}
\caption{The canonical Holevo phase variance versus input photon number $2j$.
The continuous line is for optimum states $\ket{\psi_{\rm opt}}$, the dashed
line is for all photons in one input port $\ket{jj}_z$, the dotted line is for
equal photon numbers in both ports $\ket{j0}_z$, and the dash-dotted line is for
the state $(\ket{j0}_z+\ket{j1}_z)/\sqrt 2$.}
\label{canonvar}
\end{figure}

Similarly, for the state $(\ket{j0}_z+\ket{j1}_z)/\sqrt 2$, we have
$\psi_0=\psi_1=1/\sqrt 2$, so the inner product is
\beq
\bra{j\phi}\left((\ket{j0}_z+\ket{j1}_z)/\sqrt 2 \right) = \frac 1{\sqrt {2
(2j+1)} }
\sum_{\mu=-j}^j {e^{-i\mu\left({\phi+\pi/2}\right)} \left[ {I_{\mu 0}^j (\pi/2)
+ iI_{\mu 1}^j \left( {\pi /2} \right)} \right]}.
\eeq
The probability distribution is therefore
\bqa
P(\phi) \!\!\!\! &=& \!\!\!\! \frac 1{4\pi} \st{\sum_{\mu=-j}^j {e^{-i\mu
(\phi+\pi/2)} \left[ {I_{\mu 0}^j (\pi/2) + iI_{\mu 1}^j (\pi/2)} \right]} }^2
\nn \\
\!\!\!\! &=& \!\!\!\! \frac{1}{4\pi}\sum_{\mu,\nu =-j}^j {e^{i(\nu-\mu)
(\phi+\pi/2)} \left[ I_{\mu 0}^j (\pi/2) + iI_{\mu 1}^j (\pi/2) \right]\left[
I_{\nu 0}^j (\pi/2) + iI_{\nu 1}^j (\pi/2) \right]^* }.
\eqa
Using this expression for the probability distribution we find
\bqa
\st{\ip{e^{i\phi}}}\!\!\!\! &=& \!\!\!\! \st{\int\limits_{-\pi}^\pi \frac 1
{4\pi}\sum_{\mu,\nu=-j}^j e^{i(\nu-\mu+1)(\phi+\pi/2)}e^{-i\pi/2}\left[
I_{\mu 0}^j (\pi/2)+ iI_{\mu 1}^j (\pi/2)\right]\left[I_{\nu 0}^j (\pi/2)
+ iI_{\nu 1}^j (\pi/2) \right]^ * d\phi } \nn \\
\!\!\!\! &=& \!\!\!\! \frac 12 \st{\sum_{\mu,\nu=-j}^j {\delta_{\mu,\nu+1}\left[
{I_{\mu 0}^j (\pi/2) + iI_{\mu 1}^j (\pi/2)} \right]\left[ {I_{\nu 0}^j (\pi/2)
+ iI_{\nu 1}^j \left( {\pi /2} \right)} \right]^* }} \nn \\
\!\!\!\! &=& \!\!\!\! \frac 12 \st{\sum_{\mu=1-j}^j {\left[ {I_{\mu 0}^j (\pi/2)
+ iI_{\mu 1}^j (\pi/2)} \right]\left[ {I_{\mu  - 1,0}^j (\pi/2)
+ iI_{\mu-1,1}^j (\pi/2)} \right]^* }}.
\eqa
This result was used to determine the Holevo phase variance for this state.
The third state that will be considered that does not have a simple solution for
the Holevo phase variance is that with all photons in one port, $\ket{jj}_z$.
For this state the only non-zero coefficient is $\psi_j=1$, so
\beq
\braket{j\phi}{jj}_z = (2j+1)^{-1/2} \sum_{\mu=-j}^j {e^{-i\mu\phi}
e^{i(\pi /2)(j-\mu)} I_{\mu j}^j (\pi /2)}.
\eeq
The probability distribution is therefore
\beq
P(\phi) = \frac 1{2\pi}\sum_{\mu,\nu=-j}^j {e^{i(\nu-\mu)(\phi+\pi /2)} \left(
{I_{\nu j}^j (\pi /2)} \right)^ *  I_{\mu j}^j (\pi /2)},
\eeq
and the value of $\st{\ip{e^{i\phi}}}$ is
\beq
\st{\ip{e^{i\phi}}} = \st{\sum_{\mu=1-j}^j \left(I_{\mu-1,j}^j (\pi/2)\right)^*
I_{\mu j}^j (\pi /2)}.
\eeq
The Holevo phase variance was calculated for the above three states using these
expressions, for photon numbers up to 25600, and the results are shown in
Fig.~\ref{canonvar}. I have also included the analytic expression for the
phase variance for optimised states.

The phase variance for $\ket{j0}_z$ scales down with photon number much more
slowly than the phase variance for optimal states, and in fact even more slowly
than the phase variance for $\ket{jj}_z$, which scales as $N^{-1}$. In fact, for
the range of photon numbers considered the phase variance scales as $N^{-1/2}$.
This would seem to imply a phase uncertainty scaling as $N^{-1/4}$, in dramatic
contrast to the $N^{-1}$ scaling found in \cite{Holland,SandMil95,SandMil97}.
The state $(\ket{j0}_z+\ket{j1}_z)/\sqrt 2$ is even worse, with a phase
variance that does not scale down with photon number at all, and in fact never
falls below 1.

Another unusual feature of the graph is that the phase variance for
$\ket{j0}_z$ is even smaller than that for optimum states for very small photon
numbers. This is not in fact a contradiction, because different measures of the
uncertainty are used for the two states. For very small photon numbers, the
phase probability distribution for optimum states is significant for phases
beyond $\pm \pi/2$. The measure in Eq.~(\ref{holevopi}) that is used for the
phase variance for $\ket{j0}_z$ effectively ignores the distribution beyond
$\pm \pi/2$, so under this measure, this state has a correspondingly lower phase
variance.

\begin{figure}
\centering
\includegraphics[width=0.7\textwidth]{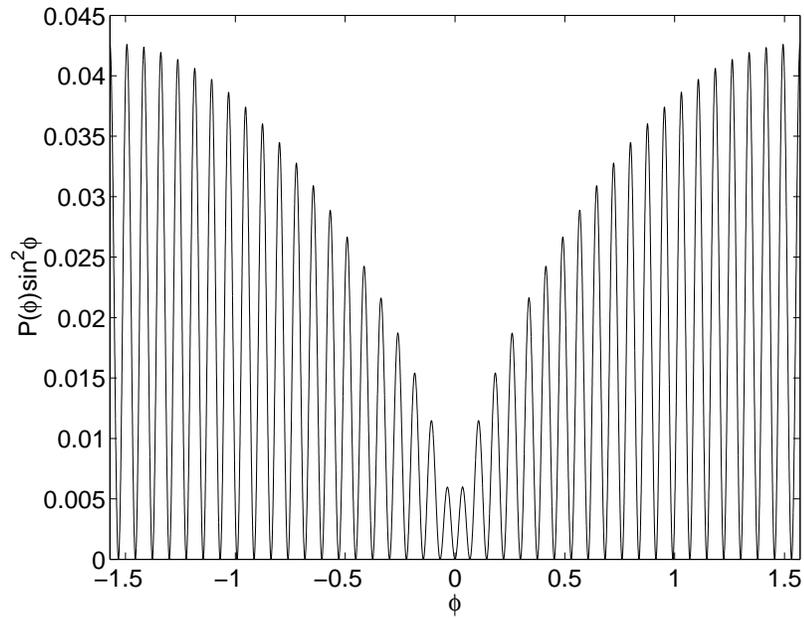}
\caption{The canonical phase probability distribution, multiplied by $\sin^2
\phi$, for the state $\ket{j0}_z$ with $2j=80$ photons.}
\label{sin2phi}
\end{figure}

The reason for the discrepancies between the results obtained here for the
states $\ket{j0}_z$ and $(\ket{j0}_z+\ket{j1}_z)/\sqrt 2$, and those obtained in
\cite{Holland,SandMil95,SandMil97}, is that the results in
\cite{Holland,SandMil95,SandMil97} are all based on the width of the central
peak in the distribution. In contrast, the Holevo phase variance for these
states is primarily due to the tails. To demonstrate this for the state
$\ket{j0}_z$, in Fig.~\ref{sin2phi} I have plotted the phase distribution
multiplied by $\sin^2 \phi$.

\begin{figure}
\centering
\includegraphics[width=0.7\textwidth]{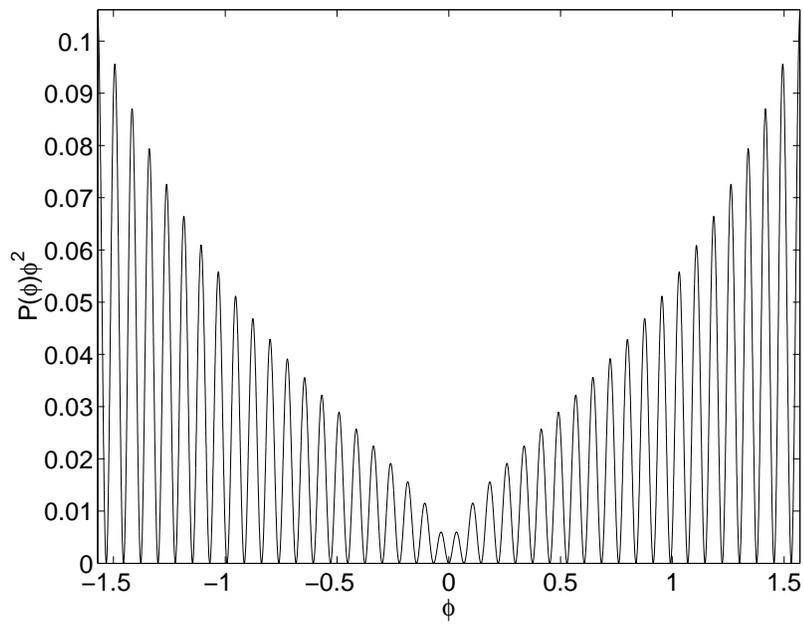}
\caption{The canonical phase probability distribution, multiplied by $\phi^2$,
for the state $\ket{j0}_z$ with $2j=80$ photons.}
\label{Pphi2}
\end{figure}

The reason for multiplying by a factor of $\sin^2 \phi$ is that
\bqa
\label{holevoderive}
V_{\pi}(\phi) \!\!\!\! &=& \!\!\!\! {{\left(\st{\ip{e^{2i\phi}}}^{-2}-1 \right)}
\mathord{\left/ {\vphantom {{\left(\st{\ip{e^{2i\phi}}}^{-2}-1 \right)} 4}}
\right. \kern-\nulldelimiterspace} 4} \nn \\
\!\!\!\! &=& \!\!\!\! {{\left( \ip{\cos (2\phi)}^{-2}-1 \right)}
\mathord{\left/ {\vphantom {{\left( \ip{\cos (2\phi)}^{-2}-1 \right)} 4}}
\right. \kern-\nulldelimiterspace} 4} \nn \\
\!\!\!\! &=& \!\!\!\! {{\left( \ip{1-2\sin ^2 \phi}^{-2}-1 \right)}
\mathord{\left/ {\vphantom {{\left( \ip{1-2\sin ^2 \phi}^{-2}-1 \right)} 4}}
\right. \kern-\nulldelimiterspace} 4} \nn \\
\!\!\!\! &=& \!\!\!\! {{\left[\left(1-2\ip{\sin^2\phi}\right)^{-2}-1\right]}
\mathord{\left/ {\vphantom {{\left[\left(1-2\ip{\sin^2\phi}\right)^{-2}
-1\right]} 4}} \right. \kern-\nulldelimiterspace} 4} \nn \\
\!\!\!\! &\approx& \!\!\!\! {{\left[\left(1+4\ip{\sin^2\phi}\right)-1\right]}
\mathord{\left/ {\vphantom {{\left[\left(1+4\ip{\sin^2\phi}\right)-1\right]}
4}} \right. \kern-\nulldelimiterspace} 4} \nn \\
\!\!\!\! &=& \!\!\!\! \ip{\sin ^2 \phi}.
\eqa
The above approximation is accurate for small phase variance. This derivation
also uses the fact that the phase distribution for this state is unbiased, so
$\ip{e^{2i\phi}}$ is real. Note that in this form the variance is very similar
to the standard variance, $\ip{\phi^2}$. Since $\phi^2 \ge \sin^2 \phi$, the
tails of the distribution are even more significant for the standard variance,
as illustrated in Fig.~\ref{Pphi2}.

\begin{figure}
\centering
\includegraphics[width=0.7\textwidth]{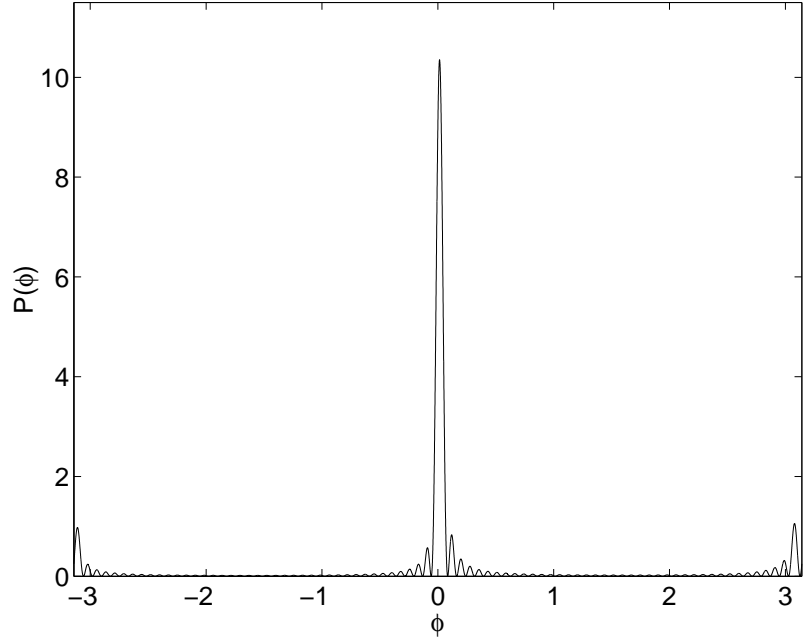}
\caption{The canonical phase probability distribution for
$(\ket{j0}_z+\ket{j1}_z)/\sqrt 2$ for $2j=80$ photons.}
\label{badstate}
\end{figure}

There is a similar problem for the state $(\ket{j0}_z+\ket{j1}_z)/\sqrt 2$. For
this state there are peaks at $\pm \pi$, as shown in Fig.~\ref{badstate}.
Although these peaks are smaller than the main peak at 0, they do not get
smaller with photon number. This means that the Holevo phase variance is almost
entirely due to these peaks, and therefore does not decrease with photon number.
As this state is so poor, I will not consider it further, and restrict attention
to $\ket{j0}_z$ and the optimum states.

The asymptotic expression for the phase probability distribution for
$\ket{j0}_z$ given in \cite{SandMil95} is
\beq
\label{half}
P_j (\phi) = \frac{2j+1}{2\pi}\frac{\left[ \Gamma (3/4) \right]^2}
{2^{3/2}}\frac{\left[ J_{1/4} ({j\phi}) \right]^2 }
{\sqrt {{j\phi}}}.
\eeq
This equation is approximately half what it should be in order to be normalised.
The reason for this is that it approximates the exact distribution over the
interval $[-\pi/2,\pi/2]$, but the exact distribution is normalised over
$[-\pi,\pi]$, and repeats modulo $\pi$. Since Eq.~(\ref{half}) only
approximates the distribution over the region $[-\pi/2,\pi/2]$, its integral
over this region will be approximately $\half$. Therefore the expression that I
will be using is
\beq
\label{full}
P_j (\phi) = \frac{2j+1}{2\pi}\frac{\left[ \Gamma (3/4) \right]^2}{\sqrt 2}
\frac{\left[ J_{1/4} (\st{j\phi}) \right]^2}{\sqrt {\st{j\phi}}}.
\eeq
This expression is correctly normalised in the limit as $j$ goes to infinity. I
have also added absolute value signs so that this expression is correct for
negative values of $\phi$.

Now for large $j\phi$ there is the approximate proportionality \cite{Erdelyi53}
\beq
J_{1/4} ( \st{j\phi}) \propto 1/\sqrt {\st{j\phi}}.
\eeq
This implies that for large $\phi$
\beq
P_j(\phi)\propto\frac 1{\st{\phi}^{3/2}}.
\eeq
This means that we should have
\beq
P_j(\phi)\phi^2\propto \st{\phi}^{1/2}.
\eeq
It can be seen in Fig.~\ref{Pphi2} that the scaling is nothing like this, and
is closer to
\beq
P_j(\phi)\phi^2\propto \st{\phi}.
\eeq

To see where this discrepancy originates, consider the intermediate
approximation made in Ref.~\cite{SandMil95}:
\beq
\braket{j\phi}{j0}_z \approx \frac{(-1)^{j/2}}{\sqrt{2j+1}}\sqrt{\frac{2}
{\pi}} \sum_{\mu=-j/2}^{j/2} {\frac{{e^{ - 2i\mu \phi }}}
{{\left[ j(j+1) -(2\mu)^2 \right]^{1/4}}}}.
\eeq
Using this approximation the probability distribution is
\beq
\label{approxdist}
P(\phi) \approx \frac{2}{\pi^2}\st{\sum_{\mu=-j/2}^{j/2}{\frac{e^{-2i\mu\phi}}
{\left[ j(j+1)-(2\mu)^2 \right]^{1/4}}}}^2.
\eeq
Here I have multiplied by a factor of 2 so the probability distribution is
normalised over the interval $[-\pi/2,\pi/2]$, for consistency with
Eq.~(\ref{full}). In Ref.~\cite{SandMil95} the sum is then approximated by an
integral in order to obtain the Bessel function approximation.

\begin{figure}
\centering
\includegraphics[width=0.7\textwidth]{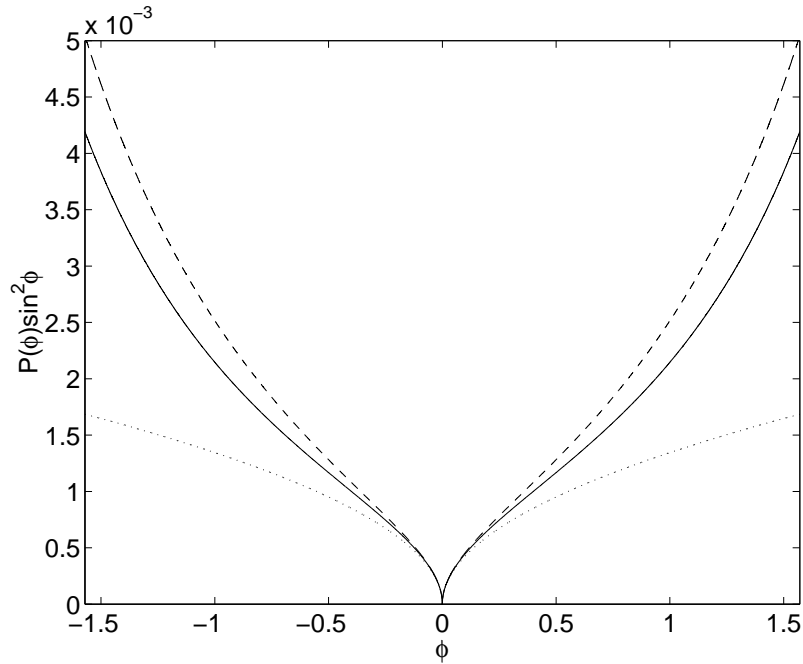}
\caption{The canonical phase probability distribution envelope, multiplied by
$\sin^2 \phi$, for the state $\ket{j0}_z$ with $2j=51200$ photons, calculated
using three different methods. The exact calculation is indicated by the
continuous line, the first approximation (\ref{approxdist}) is shown as the
dashed line, and the Bessel function approximation (\ref{full}) is shown as the
dotted line.}
\label{badagree}
\end{figure}

In order to see where the approximation is deviating from the exact expression,
in Fig.~\ref{badagree} I have plotted the probability distribution multiplied
by $\phi^2$ for the exact expression, the approximation (\ref{approxdist}), and
the Bessel function approximation (\ref{full}). For the exact expression I
have used twice Eq.~(\ref{exactdist}), so that the distribution is normalised
over the interval $[-\pi/2,\pi/2]$. The functions are very rapidly oscillating,
so to make the three curves legible only the peaks are plotted.

Near $\phi=0$ the three expressions for the distribution give very similar
results, but there are large differences in the tails. As can be seen in the
figure, the first approximation (\ref{approxdist}) has tails that are fairly
close to the exact expression, but still noticeably higher. In contrast, the
tails for the Bessel function approximation are much different, and {\it lower}
than the exact expression. These results indicate that it is not primarily the
initial approximation for $I_{\mu 0}^j (\pi /2)$ that is giving the incorrect
scaling for the tails, but the approximation where the sum is approximated by an
integral.

\begin{figure}
\centering
\includegraphics[width=0.7\textwidth]{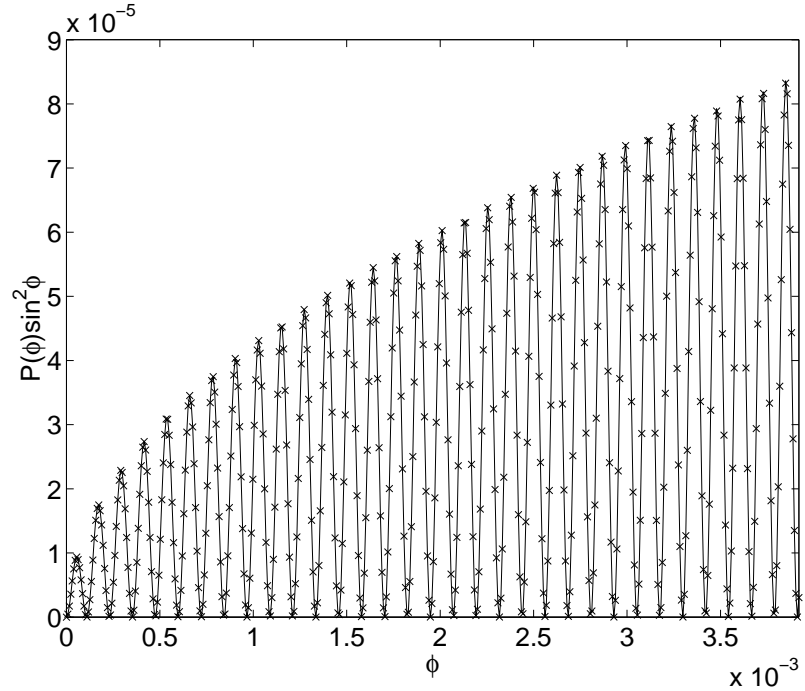}
\caption{The canonical phase probability distribution, multiplied by $\sin^2
\phi$, for the state $\ket{j0}_z$ with $2j=51200$ photons, calculated using two
different methods and restricted to a range of small $\phi$. The exact
calculation is indicated by the continuous line, and the Bessel function
approximation (\ref{full}) is shown as the crosses.}
\label{goodagree}
\end{figure}

On the other hand, if we look at the results close to the centre of the
distribution, we find that there is very good agreement between the curves. To
illustrate this, the distribution near the centre multiplied by $\sin^2 \phi$ is
plotted in Fig.~\ref{goodagree}. For large photon numbers there is good
agreement over a region that is large compared to the central peak, but the
agreement is always poor for phases that are significant compared to $\pi/2$.

In the light of these results, any results based on the Bessel function
approximation should be treated very carefully if they depend on the
distribution for large phases. For example, the Bessel function approximation
can be used to show that the Holevo phase variance should scale as $N^{-1/2}$
(as stated in \cite{short}), but it does not give the correct scaling constant.
From Eq.~(\ref{holevoderive}) the Holevo phase variance is given approximately
by
\beq
\ip {\sin ^2 \phi} = \int\limits_{-\pi /2}^{\pi /2} {P(\phi)\sin^2 \phi d\phi}.
\eeq
Note that this approximation is accurate provided that $\ip {\sin ^2 \phi}$ is
small, but does not depend on the main part of the contribution being from
small $\phi$. Using the Bessel function approximation (\ref{full}), we find
that
\bqa
\ip {\sin ^2 \phi} \!\!\!\! &=& \!\!\!\! \frac{2j+1}{2\pi}\frac{\left[ \Gamma
(3/4) \right]^2}{\sqrt 2}\int\limits_{-\pi/2}^{\pi/2} {\frac{\left[ {J_{1/4}
\left(\st{j\phi}\right)}\right]^2}{\sqrt{\st{j\phi}}}\sin^2 \phi d\phi} \nn \\
\!\!\!\! &=& \!\!\!\! \frac{2j+1}{\pi}\frac{\left[ \Gamma (3/4) \right]^2}
{\sqrt 2}\int\limits_0^{\pi/2} {\frac{\left[ J_{1/4} (j\phi) \right]^2}
{\sqrt{j\phi}}\sin^2 \phi d\phi}.
\eqa
From Ref.~\cite{Erdelyi53}, for large $\phi$ the Bessel function can be
approximated by
\beq
J_{1/4} (j\phi) \approx \sqrt {\frac{2}
{\pi }} \frac{{\sin \left( {j\phi  + \pi /8} \right)}}
{{\sqrt {j\phi } }}.
\eeq
Therefore, since the majority of the contribution to the phase variance is from
large $\phi$, we have
\beq
\ip {\sin ^2 \phi} \approx \frac{{2j + 1}}
{\pi }\frac{{\left[ {\Gamma \left( {3/4} \right)} \right]^2 }}
{{\sqrt 2 }}\int\limits_0^{\pi /2} {\frac{2}
{\pi }\frac{{\sin ^2 \left( {j\phi  + \pi /8} \right)}}
{{\left( {j\phi } \right)^{3/2} }}\sin ^2 \phi d\phi }.
\eeq
Using the average value of $\sin ^2 (j\phi+\pi /8)$ and taking the limit of
large $j$, this becomes
\bqa
\ip{\sin ^2 \phi} \!\!\!\! &\approx& \!\!\!\! \frac{2j}{\pi^2 j^{3/2}}
\frac{\left[ \Gamma (3/4) \right]^2}{\sqrt 2}\int\limits_0^{\pi/2}
{\frac{\sin ^2 \phi}{\phi^{3/2}}d\phi} \nn \\
\!\!\!\! &=& \!\!\!\! \frac{2}{\pi^2}\frac{\left[ \Gamma (3/4) \right]^2}
{\sqrt N}\int\limits_0^{\pi/2} {\frac{\sin^2 \phi}{\phi^{3/2}}d\phi }.
\eqa
This demonstrates that the Bessel function approximation predicts that the
Holevo phase variance should scale as $N^{-1/2}$. The scaling constant is given
by
\bqa
\frac{2}{\pi^2}\left[ \Gamma (3/4) \right]^2 \int\limits_0^{\pi /2}
{\frac{\sin ^2 \phi}{\phi^{3/2}}d\phi} \!\!\!\! &=& \!\!\!\! \frac{2}{\pi^2}
\left[ \Gamma (3/4) \right]^2 \sum\limits_{n=1}^\infty {\frac{(-1)^{n-1}
\pi^{2n} \sqrt {2/\pi}}{(2n)!(4n-1)}} \nn \\
\!\!\!\! &=& \!\!\!\! 0.2845775946062444 \ldots
\eqa
The above expression is easily evaluated by numerical evaluation of the
integral, or by summing the first dozen or so terms of the sum. The scaling
constant obtained is about $0.28$; however, from the numerical results shown in
Fig.~\ref{canonvar}, the actual scaling constant is $0.44$.

Note that for the standard variance, $\ip{\phi^2}$, we are simply replacing
$\sin^2\phi$ with $\phi^2$, so the scaling is again $N^{-1/2}$, except with
a scaling constant of
\bqa
\frac{2}{\pi^2} \left[ \Gamma (3/4) \right]^2 \int\limits_0^{\pi/2}{\phi^{1/2}
d\phi} \!\!\!\! &=& \!\!\!\! \frac{2}{\pi^2}\left[ \Gamma (3/4) \right]^2
\left[ \frac{\phi^{3/2}}{3/2} \right]_0^{\pi /2} \nn \\
\!\!\!\! &=& \!\!\!\! \sqrt{\frac{2}{\pi}} \frac{\left[ \Gamma (3/4) \right]^2}
3 \nn \\
\!\!\!\! &=& \!\!\!\! 0.3993800782451976 \ldots
\eqa
This scaling constant of about 0.40 is again different than the actual scaling
constant of about 0.66 (the results demonstrating this scaling constant will be
discussed below). It is significant that the scaling constants for the Holevo
phase variance and the standard phase variance are different, because it
demonstrates that the Holevo phase variance is not necessarily the same as the
standard phase variance, even in the limit of very sharply peaked distributions.
For the Holevo phase variance to be the same as the standard variance, the phase
distribution must not only be narrowly peaked, but the tails must scale
down rapidly enough that there is no significant contribution to the phase
variance from large $\phi$.

\section{Other Measures of the Phase Uncertainty}

In practice the high tails of the $\ket{j0}_z$ state mean that although most of
the results of phase measurements will have small errors, scaling as $N^{-1}$,
there will always be a significant number of results with large phase errors.
This means that we would need to be very careful analysing results obtained from
this state. For example, if we take the mean of a large number of results
obtained using this state, the error in the mean will scale as $N^{-1/4}$ rather
than $N^{-1}$. In order to obtain results with error scaling as $N^{-1}$, we
would need to use some more sophisticated data analysis technique. In contrast,
because the optimum states derived here have a Holevo phase variance that scales
as $N^{-2}$, we can use all the standard data analysis techniques and still get
an error scaling as $N^{-1}$.

Another issue is that although the phase uncertainty for $\ket{j0}_z$ as
indicated by the Holevo or standard phase variance does not scale as $N^{-1}$,
it does scale as $N^{-1}$ under other measures of uncertainty. The phase
variance is a very stringent measure of uncertainty, and generally gives an
upper limit on other measures of uncertainty. For example, there are the
inequalities \cite{hall00,chebyshev}
\bqa
\sqrt {2\pi e} \Delta \phi \!\!\!\! &\ge& \!\!\!\! L_{\rm H} \ge \sqrt {2\pi e}
L_{\rm F} \nn \\
\frac {\Delta \phi}{\sqrt {1 - C}} \!\!\!\! &\ge& \!\!\!\! L_C,
\eqa
where $\Delta \phi$ is the square root of the variance, $L_{\rm H}$ is the entropic
length, $L_{\rm F}$ is the Fisher length, and $L_C$ is a $C\times 100\%$
confidence interval. These inequalities mean that if $\Delta \phi$ scales as
$N^{-1}$, as is the case for optimum states, then the entropic length, Fisher
length, and confidence intervals must also scale at least as $N^{-1}$.

Two other measures of the phase uncertainty are the reciprocal-of-peak-value,
$L_{\rm rp}$, and the S\"ussman measure, $L_{\rm S}$. Specifically, the
definitions for each of these measures are
\bqa
\label{measures}
  \Delta \phi^2  \!\!\!\! &=& \!\!\!\! \int\limits_{-\pi}^{\pi} {P\left( \phi  \right)\left( {\phi -
 \bar \phi } \right)^2 d\phi } \nn \\
  \log L_{\rm H}  \!\!\!\! &=& \!\!\!\! -\int\limits_{-\pi}^{\pi} {P\left( \phi  \right)\log P
\left( \phi  \right)d\phi } \nn \\
  L_{\rm F}^{ - 2}  \!\!\!\! &=& \!\!\!\! \int\limits_{-\pi}^{\pi} {\left[ {\frac{{dP\left( \phi
\right)}}{{d\phi }}} \right]^2 \frac{1} {{P\left( \phi  \right)}}d\phi } \nn \\
  1 - C \!\!\!\! &=& \!\!\!\! \int\limits_{\bar \phi  - L_C }^{\bar \phi
+ L_C } {P\left( \phi  \right)d\phi } \nn \\
L_{\rm rp}^{-1} \!\!\!\! &=& \!\!\!\! \max \left[ P(\phi) \right] \nn \\
L_{\rm S}^{-1} \!\!\!\! &=& \!\!\!\! \int\limits_{-\pi}^{\pi} {\left[ P(\phi) \right]^2 d\phi }
\eqa
where $\bar \phi$ is the mean phase, defined as
\beq
\bar \phi = \int\limits_{-\pi}^{\pi} P(\phi) \phi d\phi.
\eeq
In addition, there is the usual Holevo phase variance
\beq
V(\phi) = \left| \int\limits_{-\pi}^{\pi} {P(\phi)e^{i\phi} d\phi} \right|^{-2}-1.
\eeq
Most of these measures are discussed in more detail in Ref.~\cite{measures}.

In order to calculate these measures, the full probability distribution for the
measurement scheme is required. In the case of the optimum input state, the
inner product is given by
\bqa
\braket{j\phi }{\psi _{\rm opt}} \!\!\!\! &=& \!\!\!\! \frac{1}{\sqrt {j+1}}
\sum_{\nu=-j}^j {\sin \left[ \frac{\left( {\nu+j+1} \right)\pi}{2j+2} \right]
\braket{j\phi}{j\nu}_y} \nn \\
\!\!\!\! &=& \!\!\!\! \frac{1}{\sqrt {2j+1}}\frac{1}{\sqrt {j+1}}
\sum_{\mu,\nu=-j}^j {\sin \left[ \frac{(\nu+j+1)\pi}{2j+2} \right]e^{-i\mu\phi}
{_y \braket{j\mu}{j\nu}_y} } \nn \\
\!\!\!\! &=& \!\!\!\! \frac{1}{\sqrt {2j+1}}\frac{1}{\sqrt {j+1}}
\sum_{\mu=-j}^j {\sin \left[ \frac{(\mu+j+1)\pi}{2j+2} \right]e^{-i\mu\phi}}.
\eqa
Expressing the sine in terms of exponentials gives a sum that can be evaluated:
\bqa
  \braket{j\phi}{\psi _{\rm opt}} \!\!\!\! &=& \!\!\!\! \frac{1}
{{\sqrt {2j + 1} }}\frac{1}
{{\sqrt {j + 1} }}\sum\limits_{\mu  =  - j}^j {\frac{1}
{{2i}}\left\{ {\exp \left[ {i\frac{{\left( {\mu  + j + 1} \right)\pi }}
{{2j + 2}}} \right] - \exp\left[ { - i\frac{{\left( {\mu  + j + 1} \right)\pi }}
{{2j + 2}}} \right]} \right\}e^{ - i\mu \phi } } \nn \\
\!\!\!\! &=& \!\!\!\! \frac{1}{\sqrt {2j+1}}\frac{1}{\sqrt{j+1}}\frac
{e^{i(j\phi)}}{2i}\left\{ {\sum\limits_{\mu'=0}^{2j} {\exp \left[ i
\frac{(\mu'+1)\pi}{2j+2}-i\mu' \phi \right]} } \right. \nn \\
  && \left. {-\sum\limits_{\mu'= 0}^{2j} {\exp \left[ -i\frac{(\mu'+1)\pi}
{2j+2}-i\mu'\phi \right]} } \right\} \nn \\
   \!\!\!\! &=& \!\!\!\! \frac{1}{\sqrt {2j+1}}\frac{1}{\sqrt {j+1}}\frac
{e^{i(j\phi)}}{2i}\left\{ {\sum\limits_{\mu ' = 0}^N {\exp \left[ {i\mu '\left(
{\frac{\pi}{{N + 2}} - \phi } \right) + i\frac{\pi}
{{N + 2}}} \right]} } \right. \nn \\
  && \left. {-\sum\limits_{\mu'=0}^N {\exp \left[ {-i\mu'\left( {\frac{\pi }
{{N + 2}} + \phi } \right) - i\frac{\pi }
{{N + 2}}} \right]} } \right\}.
\eqa
Using the summation formula
\beq
\sum_{m = 0}^N {a^m } = \frac{1 - a^{N + 1} }{1 - a},
\eeq
this simplifies to
\bqa
  \braket{j\phi}{\psi _{\rm opt}} \!\!\!\! &=& \!\!\!\! \frac{1}
{{\sqrt {2j + 1} }}\frac{1}
{{\sqrt {j + 1} }}\frac{e^{i(j\phi)}}
{{2i}}\left\{ {e^{i\frac{\pi }
{{N + 2}}} \frac{{1 - e^{i\left( {\frac{\pi }
{{N + 2}} - \phi } \right)\left( {N + 1} \right)} }}
{{1 - e^{i\left( {\frac{\pi }
{{N + 2}} - \phi } \right)} }} - e^{ - i\frac{\pi }
{{N + 2}}} \frac{{1 - e^{ - i\left( {\frac{\pi }
{{N + 2}} + \phi } \right)\left( {N + 1} \right)} }}
{{1 - e^{ - i\left( {\frac{\pi }
{{N + 2}} + \phi } \right)} }}} \right\} \nn \\
   \!\!\!\! &=& \!\!\!\! \frac{1}
{{\sqrt {2j + 1} }}\frac{1}
{{\sqrt {j + 1} }}\frac{e^{i(j\phi)}}
{{2i}}\left\{ {\frac{{e^{i\frac{\pi }
{{N + 2}}}  + e^{-i\phi \left( {N + 1} \right)} }}
{{1 - e^{i\left( {\frac{\pi }
{{N + 2}} - \phi } \right)} }} - \frac{{e^{ - i\frac{\pi }
{{N + 2}}}  + e^{ - i\phi \left( {N + 1} \right)} }}
{{1 - e^{ - i\left( {\frac{\pi }
{{N + 2}} + \phi } \right)} }}} \right\}.
\eqa
The probability distribution is therefore
\beq
P(\phi)=\frac 1{4\pi}\frac 1{N+2}
\st{\frac {e^{i\frac{\pi}{N + 2}}+e^{-i\phi(N+1)}}
{1-e^{i\left(\frac{\pi}{N+2}-\phi\right)}}
-\frac{e^{-i\frac{\pi}{N+2}}+e^{-i\phi(N+1)}}
{1-e^{-i\left(\frac{\pi}{N+2}+\phi\right)}}}^2.
\eeq
Simplifying this we find
\bqa
P(\phi) \!\!\!\! &=& \!\!\!\!\frac 1{4\pi}\frac 1{N+2}
\left|\frac { e^{i\frac{\pi}{N+2}}-e^{-i\left(\frac{\pi}{N+2}+\phi(N+2)\right)}
-e^{-i\frac{\pi}{N+2}}+e^{i\left(\frac{\pi}{N+2}-\phi(N+2)\right)}}
{\left(1-e^{i\left(\frac{\pi}{N+2}-\phi\right)}\right)
\left(1-e^{-i\left(\frac{\pi}{N+2}+\phi\right)}\right)}\right|^2 \nn \\
\!\!\!\! &=& \!\!\!\!\frac 1{4\pi}\frac 1{N+2}
\left|\frac { \left(e^{i\frac{\pi}{N+2}}-e^{-i\frac{\pi}{N+2}}\right)
\left(1+e^{-i\phi(N+2)}\right)}
{\left(1-e^{i\left(\frac{\pi}{N+2}-\phi\right)}\right)
\left(1-e^{-i\left(\frac{\pi}{N+2}+\phi\right)}\right)}\right|^2 \nn \\
\!\!\!\! &=& \!\!\!\!\frac 1{2\pi}\frac 1{N+2}
\frac { \sin^2\left(\frac{\pi}{N+2}\right)
\left(1+\cos \phi(N+2)\right)}
{\left(1-\cos\left(\frac{\pi}{N+2}-\phi\right)\right)
\left(1-\cos\left(\frac{\pi}{N+2}+\phi\right)\right)} \nn \\
\label{analopt} \!\!\!\! &=& \!\!\!\!\frac 1{2\pi}\frac 1{N+2}
\frac { \sin^2\left(\frac{\pi}{N+2}\right)
\left(1+\cos \phi(N+2)\right)}
{\left(\cos\phi-\cos\left(\frac{\pi}{N+2}\right)\right)^2}.
\eqa
This simple result for the probability distribution can be used to determine the
phase uncertainty under each of the measures in (\ref{measures}) using numerical
integrals.

For the state with equal numbers of photons in each port, the probability
distribution is given by Eq.~(\ref{exactdist}). This expression cannot be
evaluated to a simple expression like Eq.~(\ref{analopt}), and the full sum must
be used in order to treat the distribution exactly. This becomes prohibitively
time consuming if the integrals must be performed numerically; however, all of
the integrals can be evaluated using sums, except for the entropic length.
To see this, note that the probability distribution can be expressed as 
\bqa
P(\phi) \!\!\!\! &=& \!\!\!\! \frac 1{2\pi}\sum_{k=-2j}^{2j}{P_k e^{ik\phi}},
\eqa
where
\beq
P_k = \sum_{\mu,\nu=-j}^j {e^{-i(\mu-\nu)(\pi/2)} \left( I_{\nu 0}^j (\pi/2)
\right)^* I_{\mu 0}^j (\pi/2)\delta_{k,\nu-\mu}}.
\eeq
Recall that the modified Holevo phase variance for this state can be evaluated
using (\ref{holevopi}) and (\ref{expectexp}). For the other measures of the
phase uncertainty the integrals will be performed over the interval
$[-\pi/2,\pi/2]$, so the above distribution must be multiplied by a factor
of 2 to be correctly normalised. The modified probability distribution is
therefore
\beq
P(\phi)=\frac 1{\pi}\sum_{k=-2j}^{2j}{P_k e^{ik\phi}}.
\eeq
In addition, it can be shown that all the $P_k$ are real, and $P_k=0$ for odd
$k$. For the standard phase variance, it is simpler to express the probability
distribution as
\beq
P(\phi)=\frac 1{\pi}P_0+\frac 2{\pi}\sum_{k=2}^{2j}{P_k\cos (k\phi)}.
\eeq
The standard phase variance can then be determined using
\beq
\Delta \phi^2=\frac 1{\pi}P_0 \int\limits_{ - \pi /2}^{\pi /2} {\phi ^2 d\phi }+
\frac{2}{\pi}\sum\limits_{k=2}^{2j} {P_k \int\limits_{-\pi/2}^{\pi/2}
{\cos(k\phi)\phi^2 d\phi}}.
\eeq
Evaluating this we find that
\bqa
  \Delta \phi^2  \!\!\!\! &=& \!\!\!\! \frac{1}
{\pi }P_0 \frac{2}
{3}\left( {\frac{\pi }
{2}} \right)^3  + \frac{2}
{\pi }\sum\limits_{k = 2}^{2j} {P_k \frac{\pi }
{2}\frac{{\left( { - 1} \right)^{k/2} }}
{{\left( {k/2} \right)^2 }}} \nn \\
   \!\!\!\! &=& \!\!\!\! \frac{{\pi ^2 }}
{{12}}P_0  + \sum\limits_{k = 2}^{2j} {P_k \left( {\frac{2}
{k}} \right)^2 \left( { - 1} \right)^{k/2} }.
\eqa
This provides a simple formula to determine the standard phase variance, once
the coefficients $P_k$ have been determined.

Next, to determine the Fisher information, note that for the state with equal
photon numbers in both ports, the probability distribution can be expressed as
\beq
P(\phi)=\psi^2(\phi),
\eeq
with
\beq
\psi(\phi) = \frac {i^j}{\sqrt{\pi}} \sum_{\mu=-j}^j e^{-i\mu(\phi+\pi/2)}
I_{\mu 0}^j (\pi/2).
\eeq
Here the factor of $i^j$ ensures that this is always real, so the absolute value
sign is not required. Note also that a factor of $1/\sqrt{\pi}$ (rather than
$1/\sqrt{2\pi}$) is required for the probability distribution to be normalised
over the interval $[-\pi/2,\pi/2]$. In terms of this function the Fisher
information is
\bqa
L_{\rm F}^{-2} \!\!\!\! &=& \!\!\!\! \int\limits_{-\pi/2}^{\pi/2} \left[
\frac{d\psi^2(\phi)}{d\phi} \right]^2 \frac 1{\psi^2(\phi)} d\phi \nn \\
\!\!\!\! &=& \!\!\!\! \int\limits_{-\pi/2}^{\pi/2} \left[ 2 \psi(\phi)
\frac{d\psi(\phi)}{d\phi} \right]^2 \frac 1{\psi^2(\phi)} d\phi \nn \\
\!\!\!\! &=& \!\!\!\! 4 \int\limits_{-\pi/2}^{\pi/2} \left[ \frac{d\psi(\phi)}
{d\phi}\right]^2 d\phi.
\eqa
Evaluating this gives
\bqa
L_{\rm F}^{-2} \!\!\!\! &=& \!\!\!\! 4(-1)^j \int\limits_{-\pi/2}^{\pi/2} \left[
\frac 1{\sqrt{\pi}}\sum_{\mu=-j}^j (-i\mu) e^{-i\mu(\phi+\pi/2)} I_{\mu 0}^j
(\pi/2)\right]^2 d\phi \nn \\
\!\!\!\! &=& \!\!\!\! \frac 4{\pi}(-1)^j \int\limits_{-\pi/2}^{\pi/2}
\sum_{\mu,\nu=-j}^j \mu (-\nu) e^{-i(\mu+\nu)(\phi+\pi/2)} I_{\mu 0}^j (\pi/2)
I_{\nu 0}^j (\pi/2) d\phi \nn \\
\!\!\!\! &=& \!\!\!\! 4(-1)^j \sum_{\mu=-j}^j \mu^2 I_{\mu 0}^j (\pi/2)
I_{-\mu 0}^j (\pi/2) d\phi \nn \\ \!\!\!\! &=& \!\!\!\!
4 \sum_{\mu=-j}^j \mu^2 \left[I_{\mu 0}^j(\pi/2)\right]^2 d\phi.
\eqa
In the last line the result that $I_{-\mu 0}^j (\pi/2) = (-1)^j I_{\mu 0}^j
(\pi/2)$ has been used. This expression gives a simple method for calculating
the Fisher length.

Next, in order to evaluate the confidence interval we wish to perform the
integral
\beq
1-C = \int\limits_{-L_C}^{L_C} P(\phi) d\phi.
\eeq
This is easily evaluated as
\bqa
1-C \!\!\!\! &=& \!\!\!\! \int\limits_{-L_C}^{L_C} \left[ \frac {P_0}{\pi} +
\frac 2{\pi} \sum_{k=2}^{2j} P_k \cos(k\phi) \right] d\phi \nn \\
\!\!\!\! &=& \!\!\!\! \frac {2L_C P_0}{\pi} + \frac 4{\pi} \sum_{k=2}^{2j}
\frac {P_k}k \sin(kL_c).
\eqa
This expression can then be used to find $L_C$ for a given $C$ numerically.

The evaluation of the reciprocal-of-peak-value is simple, as it merely requires
the evaluation of the probability distribution at $\phi=0$. Lastly, the
S\"ussman measure can be evaluated using
\bqa
L_{\rm S}^{-1} \!\!\!\! &=& \!\!\!\! \int\limits_{-\pi/2}^{\pi/2} \frac 1{\pi^2}
\sum_{k,k'=-2j}^{2j} P_k P_{k'} e^{i(k+k')\phi} d\phi \nn \\
\!\!\!\! &=& \!\!\!\! \frac 1{\pi} \sum_{k,k'=-2j}^{2j} P_k^2 .
\eqa

\begin{figure}[tb]
\centering
\includegraphics[width=0.7\textwidth]{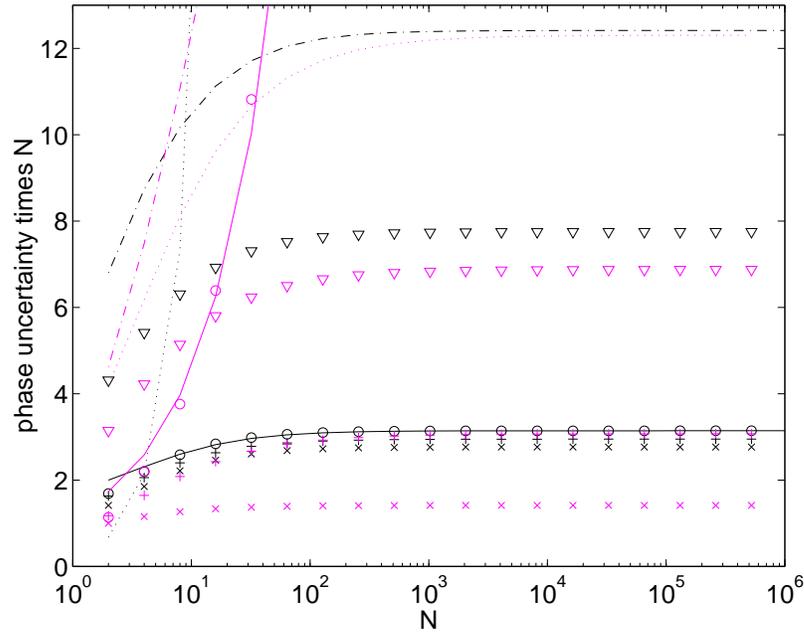}
\caption{The phase uncertainty of the $\ket{j0}_z$ state (purple) and states
optimised for minimum Holevo phase variance (black) under several measures
multiplied by $N$. The square root of the Holevo phase variance is shown as the
continuous lines, the square root of the standard variance as the
circles, the inverse-of-maximal-value as the triangles, the S\"ussman
measure as the dotted lines, the entropic length as the
dash-dotted lines, the Fisher length as the crosses, and the 67\%
confidence interval is shown as the pluses.}
\label{meas}
\end{figure}

\begin{figure}[tb]
\centering
\includegraphics[width=0.7\textwidth]{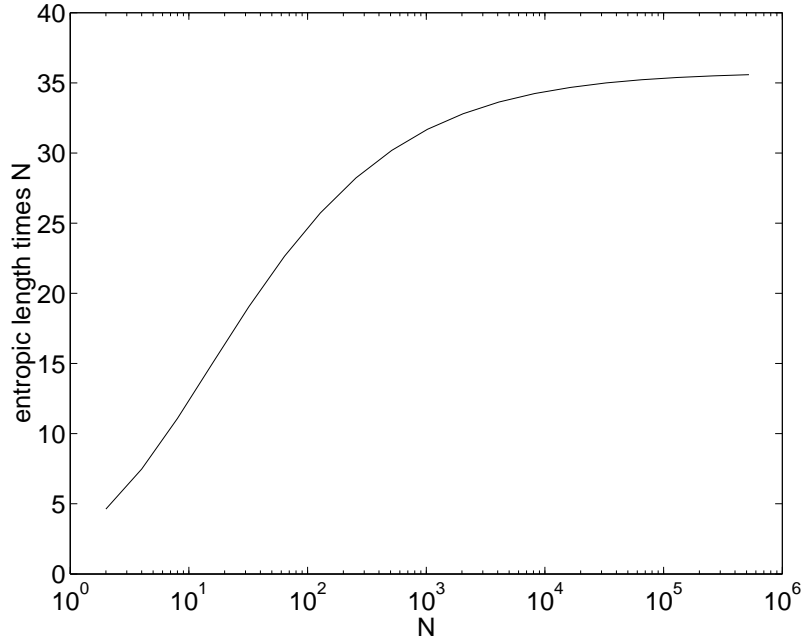}
\caption{The entropic length for the $\ket{j0}_z$ state multiplied by $N$ on a
larger scale.}
\label{entropic}
\end{figure}

These methods were used to calculate each of the above measures of uncertainty
for the optimum input state and $\ket{j0}_z$ for a large range of photon
numbers, and the results multiplied by $N$ are plotted in Fig.~\ref{meas}.
It is clear from this plot that the asymptotic
values for most of these measures are good approximations for photon numbers of
order 100 or greater. The results for the Holevo and standard variance for
$\ket{j0}_z$ do not converge, as these measures of the phase uncertainty scale
as $N^{-1/4}$. The results for the entropic length for $\ket{j0}_z$ do not
appear to converge at this scale; however, if we increase the scale of the plot,
as in Fig.~\ref{entropic}, we find that the results converge to a large value
for very large photon number.

It can be seen from Fig.~\ref{meas} that the optimal state is actually worse
than the $\ket{j0}_{z}$ state as evaluated using the reciprocal peak
$L_{\rm rp}$ or the Fisher length $L_{F}$, but it is better under all the other
measures of phase uncertainty. The S\"ussman measure and the entropic length are
smaller for $\ket{j0}_{z}$ for smaller photon numbers, but the asymptotic values
are larger.

It can be argued that the reciprocal peak and Fisher length do not give very
meaningful estimates of the phase uncertainty in the case of $\ket{j0}_{z}$.
The reciprocal peak gives good scaling because the probability distribution
has a high peak, but it does not take the tails into account at all. As an
extreme example of where this happens, consider the probability distribution
given by
\beq
P(\phi) = \frac {1-\lambda}{2\pi} + \lambda\delta(\phi),
\eeq
where $\lambda \ll 1$. Under the reciprocal peak measure, the phase uncertainty
is zero; however, it is clear that the actual phase uncertainty is very large.

The Fisher length is small when the probability distribution has a large first
derivative. This means that the Fisher length gives a meaningful estimate of
the uncertainty when the probability distribution has a smooth peak, as the
larger the derivative is, the sharper the peak. The Fisher length is not so
meaningful when the probability distribution oscillates, as is the case for
$\ket{j0}_{z}$. The small Fisher length is then just due to the rapid
oscillations, rather than a single narrow peak. As an extreme example of this,
consider
\beq
P(\phi) = \frac{\cos^2(n \phi)}{\pi},
\eeq
where $n \gg 1$. The Fisher length will be very small for this distribution;
however, it is clear that the actual phase uncertainty is very large.
In view of these considerations, it appears that the reciprocal peak and Fisher
length give misleadingly small estimates of the phase uncertainty for the
$\ket{j0}_{z}$ state. The better scaling using these measures should not be
taken to imply that $\ket{j0}_{z}$ states are better than the optimal states.

Using the results plotted in Fig.~\ref{meas} we can determine the scaling
constants for each of these measures for the two states, and the results are
listed in Table \ref{table:meastab}. In this table $\Delta \phi_{\rm H}$ denotes the
square root of the Holevo phase variance. Note that for $\ket{\psi_{\rm opt}}$
the square roots of both the standard variance and the Holevo variance scale as
$\pi/N$, in agreement with the scaling predicted analytically. For $\ket{j0}_{z}$, the
standard and Holevo variance scale as
\bqa
\Delta \phi^2 \!\!\!\! &\approx& \!\!\!\! \frac {0.6573863}{\sqrt N} \\
V(\phi) \!\!\!\! &\approx& \!\!\!\! \frac {0.4395}{\sqrt N}.
\eqa
These are different from each other, and also from the values obtained
previously using the Bessel function approximation.

\begin{table}[tbp]
\centering
\caption{The scaling constants for each of the measures of phase uncertainty
for the state optimised for minimum Holevo phase variance and $\ket{j0}_{z}$.}
\begin{tabular}{||c|c|c||} \hline \hline
Measure          & $\ket{\psi_{\rm opt}}$ & $\ket{j0}_{z}$  \\ \hline
$N\Delta \phi$   &    $3.14159265359$ &    $0.8107937 \times N^{3/4}$ \\ \hline
$N\Delta \phi_{\rm H}$ &    $3.1415927$     &    $0.66292 \times N^{3/4}$   \\ \hline
$NL_{\rm rp}$    &    $7.7515691701$  &    $6.87519$      \\ \hline
$NL_{\rm S}$           &   $10.710529485$   &   $12.30505$      \\ \hline
$NL_{\rm H}$           &   $12.414819836$   &   $35.79$         \\ \hline
$NL_{2/3}$       &    $2.9481552495$  &    $3.07129$      \\ \hline
$NL_{\rm F}$           &    $2.76615948$    &    $1.4142136$    \\ \hline \hline
\end{tabular}
\label{table:meastab}
\end{table}

It is interesting to note that the coefficients found here for the
confidence interval ($3.071$) and the Fisher length ($1.414$) for the state
$\ket{j0}_{z}$ differ from those found in \cite{SandMil97}, of $3.36$ and $2$
respectively. In order to see the reason for this difference we can use the
asymptotic approximation (\ref{full}). In the large $j$ limit the $+1$ can be
ignored, so this becomes
\beq
\label{analaprox}
P_j (\phi) = \frac j{\pi}\frac{\Gamma^2(3/4)}{\sqrt 2}
\frac{\left[ J_{1/4} (\st{j\phi}) \right]^2 }{\sqrt {\st{j\phi}}}.
\eeq
Therefore the $2/3$ confidence interval can be determined as
\bqa
\frac 13 \!\!\!\! &=& \!\!\!\! \int\limits_0^{L_{2/3}} \frac j{\pi}
\frac{\Gamma^2(3/4)}{\sqrt 2}\frac{J_{1/4}^2(j\phi)}{\sqrt{j\phi}}d\phi \nn \\
\!\!\!\! &=& \!\!\!\! \frac 1{\pi}\frac{\Gamma^2(3/4)}{\sqrt 2}
\int\limits_0^{jL_{2/3}} \frac{J_{1/4}^2(x)}{\sqrt{x}}dx
\eqa
where $x=j\phi$. Rather than using the asymptotic approximation of the Bessel
function, as used in \cite{SandMil97}, it is fairly straightforward to
numerically evaluate the integral and use Newton's method to find $jL_{2/3}$.
This gives $jL_{2/3} \approx 1.53564794820384$, so $N L_{2/3} \approx
3.07129589640767$. This agrees with the value found from the exact calculations
to six significant digits. Thus we see that the difference in the scaling
constant is due to the asymptotic approximation for the Bessel function used in
\cite{SandMil97}.

For the Fisher length, if we perform the derivation in the same way as in
\cite{SandMil97}, we find
\bqa
\frac 1{P_j (\phi)}\left[ \frac{d}{d\phi} P_j (\phi) \right]^2 \!\!\!\! &=&
\!\!\!\! \frac j{\pi}\frac{\Gamma^2(3/4)}{\sqrt 2} \frac{\sqrt{j\phi}}
{J_{1/4}^2(j\phi)} \left[ \frac{d}{d\phi} \frac{J_{1/4}^2(j\phi)}
{\sqrt{j\phi}} \right]^2 \nn \\
\!\!\!\! &=& \!\!\!\! \frac j{\pi}\frac{\Gamma^2(3/4)}{\sqrt 2}
4j^2 \frac{J_{5/4}^2(j\phi)}{\sqrt{j\phi}}.
\eqa
So the Fisher information is
\bqa
F_j \!\!\!\! &=& \!\!\!\! 2\int\limits_0^{\pi/2} \frac j{\pi}\frac{\Gamma^2
(3/4)}{\sqrt 2}4j^2 \frac{J_{5/4}^2(j\phi)}{\sqrt{j\phi}} d\phi \nn \\
\!\!\!\! &\approx& \!\!\!\! \frac {j^2}{\pi}\frac{\Gamma^2(3/4)}{\sqrt 2}8
\int\limits_0^{\infty} \frac{J_{5/4}^2(x)}{\sqrt{x}} dx \nn \\
\!\!\!\! &=& \!\!\!\! \frac {j^2}{\pi}\frac{\Gamma^2(3/4)}{\sqrt 2}8
\frac{\pi}{\sqrt 8 \Gamma^2(3/4)} \nn \\
\!\!\!\! &=& \!\!\!\! 2j^2.
\eqa
This is twice what was found in \cite{SandMil97}, and using this the Fisher
length is approximately $L_{\rm F} \approx \sqrt 2/N$. The reason for this factor of
two is that the approximation used for the probability distribution in
\cite{SandMil97} is not correctly normalised. Here the normalised version has
been used, which is twice that given in \cite{SandMil97}.

Note that this approximate analytic result agrees very precisely with the
result found using the exact calculations. We can also determine the scaling
constants for the other measures of phase variance using the analytic
approximation (\ref{analaprox}). The scaling constants for the standard and
Holevo phase variances were found above. The reciprocal peak can be evaluated as
\bqa
L_{\rm rp} \!\!\!\! &=& \!\!\!\! \frac {\pi}j \frac{\sqrt 2}{\Gamma^2(3/4)}
\lim_{\phi \to 0} \frac{\sqrt{j\phi}}{J_{1/4}^2(j\phi)} \nn \\
\!\!\!\! &=& \!\!\!\! \frac{2\pi}j
\left[\frac{\Gamma(5/4)}{\Gamma(3/4)}\right]^2.
\eqa
The scaling is therefore
\beq
L_{\rm rp} \approx \frac 1N 4\pi\left[\frac{\Gamma(5/4)}{\Gamma(3/4)}\right]^2.
\eeq
This scaling constant is approximately $6.87518581802037$, agreeing very
accurately with that found from the exact calculations.

The S\"ussman measure and entropic length can also be determined by numerical
integrals. The scaling constants found, along with the scaling constants found
above for the other measures, are given in Table \ref{table:meastaban}. Note
that there is very good agreement with the results based on the exactly
calculated data in these two cases as well. The numerical integral for the
entropic length is particularly difficult to calculate, as the probability
distribution falls very slowly with the phase, and the integral must be
calculated to very large values of $j\phi$. For this reason only 7 significant
digits could be found.

\begin{table}[tbp]
\centering
\caption{The scaling constants for each of the measures of phase uncertainty
for the state optimised for minimum Holevo phase
variance and $\ket{j0}_{z}$ as determined using the asymptotic approximations.}
\begin{tabular}{||c|c|c||} \hline \hline
Measure       & $\ket{\psi_{\rm opt}}$ & $\ket{j0}_{z}$  \\ \hline
$N\Delta \phi$   & $\pi=3.14159265359265\ldots$ &
$0.631965250820959 \times N^{3/4}$ \\ \hline
$N\Delta \phi_{\rm H}$ & $\pi=3.14159265359265\ldots$ &
$0.533458147005221 \times N^{3/4}$ \\ \hline
$NL_{\rm rp}$    & $\pi^3/4 = 7.75156917007495\ldots$ & $4\pi\left[
\frac{\Gamma(5/4)}{\Gamma(3/4)}\right]^2 = 6.87518581802037\ldots$ \\ \hline
$NL_{\rm S}$           & $10.7105294850660$     & $12.305050002393$    \\ \hline
$NL_{\rm H}$           & $12.4148198362985$     & $35.78817$           \\ \hline
$NL_{2/3}$       &  $2.94815524951393$    &  $3.07129589640767$  \\ \hline
$NL_{\rm F}$           &  $2.7661594839$        &
$\sqrt{2} = 1.41421356237309\ldots$ \\ \hline \hline
\end{tabular}
\label{table:meastaban}
\end{table}

It is also possible to determine the scaling constants for the state optimised
for minimum intrinsic phase variance using an asymptotic approximation. It is
easy to see that in the limit of large $N$ and small $\phi$,
Eq.~(\ref{analopt}) becomes
\bqa
P(\phi) \!\!\!\! & \approx & \!\!\!\! \frac 1{2\pi}\frac 1{N+2}\frac { \left(
\frac{\pi}{N+2}\right)^2 \left(1+\cos \phi(N+2)\right)}
{\left[\half\phi^2-\half\left(\frac{\pi}{N+2}\right)^2\right]^2} \nn \\
\!\!\!\! & \approx & \!\!\!\! 2\pi N\frac {1+\cos N\phi}
{\left[(N\phi)^2-\pi^2\right]^2}.
\eqa
In the second line it has been assumed that $N+2 \approx N$ in the limit of
large photon number.

Note that the asymptotic approximations for both of these states have the
general form
\beq
P(\phi)=N f(N\phi).
\eeq
For asymptotic functions of this form, we obtain $1/N$ scaling for each of the
measures of phase uncertainty, provided that the integrals are bounded. To see
this, note that each of the measures is, in terms of the asymptotic function,
\bqa
(N\Delta \phi)^2 \!\!\!\! &=& \!\!\!\! \int\limits_{-N\pi}^{N\pi} f(x)(x-\bar x)^2 dx \nn \\
\log NL_{\rm H} \!\!\!\! &=& \!\!\!\! -\int\limits_{-N\pi}^{N\pi} f(x) \log f(x) dx \nn \\
(NL_{\rm F})^{-2} \!\!\!\! &=& \!\!\!\! \int\limits_{-N\pi}^{N\pi} \frac{[f'(x)]^2}{f(x)}dx \nn \\
1-C \!\!\!\! &=& \!\!\!\! \int\limits_{\bar x-NL_C}^{\bar x+NL_C}
f(x) dx \nn \\
(NL_{\rm rp})^{-1} \!\!\!\! &=& \!\!\!\! \max f(x) \nn \\
(NL_{\rm S})^{-1} \!\!\!\! &=& \!\!\!\! \int\limits_{-N\pi}^{N\pi} f^2(x) dx.
\eqa
In the asymptotic limit the Holevo variance is given by the same expression as
the standard variance. For the standard or Holevo variance in the case of the
$\ket{j0}_{z}$ state, the integral is not bounded, and must be considered up to
the limit $\phi=\pi/2$. That is why $1/N$ scaling is not obtained in that case.

For the asymptotic expression for the optimal input states, the scaling constant
for the variance is already known to be $\pi$, using the analytic result for
the Holevo phase variance. The scaling constant for the reciprocal-peak measure
of the phase uncertainty is easily determined analytically as $\pi^3/4$. For the
other measures, the scaling constants can be determined using numerical
integrals, and the results are as shown in Table \ref{table:meastaban}.
Comparing the results in Tables \ref{table:meastab} and \ref{table:meastaban},
it can be seen that there is excellent agreement between the scaling constants
for all of the measures.

We therefore find that the asymptotic expressions accurately give the scaling
constants for all the cases considered, except the Holevo or standard variance
for the $\ket{j0}_z$ state. For these cases the main contribution is from
significant values of $\phi$, where the asymptotic expression is not accurate.
The scaling constants are higher for $\ket{j0}_z$ than the optimal state for
all of the measures except the reciprocal-of-peak and Fisher length. As
explained above, these measures appear to give unrealistically low estimates of
the phase uncertainty for the $\ket{j0}_z$ state.

Similarly the Holevo and standard variances give unrealistically high estimates
of the phase uncertainty for $\ket{j0}_z$. A more accurate way of comparing the
two states is through confidence intervals. For the $2/3$ confidence interval,
the scaling constant is only slightly higher for $\ket{j0}_z$ than the optimal
state. The difference is far more pronounced if we consider a higher probability
confidence interval. For a 90\% confidence interval, the scaling constant is
only a little higher for the optimal state, at about $4.88$, but it is $37.69$
for $\ket{j0}_z$.

Probably the most accurate way of comparing the two distributions is through the
entropic length, as this corresponds exactly with the phase information
contained in the distribution. The scaling constant for the entropic length is
almost three times higher for the $\ket{j0}_z$ state than for the optimal state.
These results demonstrate that although the phase uncertainty scales as $N^{-1}$
for $\ket{j0}_z$, it is not as good as the optimal state.

%% file: interfer.tex
\setcounter{chapter}{5}

\chapter{Optimum Adaptive Interferometry}
\label{interfere}
Now that the input states have been fully considered, the next factor to
consider is the optimum way of measuring these states. There are basically two
components to this: the feedback phase that is used during the measurement, and
the phase estimate that is used at the end of the measurement. It has been shown
\cite{semiclass,unpub,fullquan} that in the single mode case it is possible to
make very good phase measurements by using feedback to an auxiliary phase shift.
In order to apply the same principle here, I consider adaptive measurements
where the phase to be measured, $\varphi$, is in one arm of the interferometer,
and a known phase shift, $\Phi$, is introduced into the other arm of the
interferometer, as in Fig.~\ref{diag}.

The initial feedback scheme that will be considered is that where the introduced
phase shift is adjusted in order to minimise the variance in the phase estimate
after the next photodetection. In order to evaluate this feedback scheme, the
optimal phase estimates are required; these are derived in Sec.~\ref{final}. It
is also possible to select the feedback phases in order to minimise the final
variance, using numerical minimisation techniques.

For these adaptive schemes to work, the feedback that adjusts $\Phi$ must act
much faster than the average time between photon arrivals. For simplicity I
make the assumption that the feedback is arbitrarily fast, which simply means
that the phase $\Phi$ is always assumed to have been changed before the next
detection occurs. It is the ability to change $\Phi$ during the passage of a
single (two-mode) pulse that makes photon counting measurements more general
than a measurement of the output $\hat{J}_{z}$ considered in
\cite{Yurke,Holland}.

As mentioned in the previous chapter, it is assumed that the photodetectors
have unit efficiency. In addition, only single detection events are considered,
rather than multiple detections. Physically, simultaneous detections correspond to
individual detections that are too close together to be resolved by the apparatus.
Here it will be assumed that the apparatus is arbitrarily fast, so that all the
detections can be resolved, and the feedback phase updated between detections.
This is not necessarily realistic for current technology, particularly for
larger photon numbers. However, the purpose of this study is to examine how
far phase measurements can, in principle, be improved, rather than to examine
the limits of current technology.

\section{Preliminary Theory}

To describe the measurement, it is convenient to denote the result $u$ from
the $m$th detection as $u_m$ (which is 0 or 1 depending on which output the
photon is detected in), and the measurement record up to and including the
$m$th detection as the binary string $n_m \equiv u_m \ldots u_2 u_1$. The input
state after $m$ detections will be a function of the measurement record and
$\varphi$, and is denoted as $\ket{\psi(n_m,\varphi)}$. In the calculations
these states are not normalised (except for the initial state), in order to
express the state as a power series in $e^{i\varphi}$.

The time between detections does not give any information, and is not included
in the measurement record. To see this, note that the measurement operator for
no detection is
\beq
\Omega_0 = 1-\half (a\dg a+b\dg b)dt,
\eeq
where $a$ and $b$ are the annihilation operators for the two input modes. As
this does not depend on the phase, the probability distribution for the phase
is unchanged between detections. In addition, we consider only states with a
fixed total photon number $N$. This means that, after $m$ detections, the system
will be in an eigenstate of $a\dg a+b\dg b$ with eigenvalue $N-m$. Therefore the
measurement operator $\Omega_0$ does not produce any system evolution. As
neither the probability distribution for the phase nor the system state changes
between detections, the time between detections may be ignored.

Now the evolution produced by detections will be considered.
After the first beam splitter the operators for the two beams are (in
the Heisenberg picture)
\begin{equation}
(a + ib)/\sqrt{2} ,\;\;\; (ia + b)/\sqrt{2}.
\end{equation}
The two beams are then subjected to phase shifts of $\varphi$ and $\Phi$, so
the operators become
\begin{equation}
e^{i\varphi}(a + ib)/\sqrt{2} ,\;\;\;
e^{i\Phi}(ia + b)/\sqrt{2}.
\end{equation}
Lastly, the effect of the second beam splitter is the same as the first,
giving the two operators
\begin{eqnarray}
ie^{i(\varphi+\Phi)/2}(a\sin {\theta} + b\cos {\theta}),
\nonumber \\
ie^{i(\varphi+\Phi)/2}(a\cos {\theta} - b\sin {\theta}),
\end{eqnarray}
where $\theta=(\varphi - \Phi)/2$. Ignoring the unimportant initial phase
factors, these can be represented as the operators $\hat{c}_0, \hat{c}_1$,
where
\begin{equation}
\label{detectop}
\hat{c}_{u}=a\sin \left( \theta + u\pi/2 \right) +b\cos \left(\theta+u\pi/2
\right).
\end{equation}

The input state is then determined by the initial condition
\beq
\ket{\psi(n_0,\varphi)}=\ket{\psi},
\eeq
where $n_0$ is the empty string, and the recurrence relation
\beq
\label{recur}
\ket{\psi(u_m n_{m-1},\varphi)}=\hat c_{u_m}(\varphi)
\frac{\ket{\psi(n_{m-1},\varphi)}}{\sqrt{2j+1-m}}.
\eeq
These states are not normalised, and their norm represents the probability for
the measurement record $n_m$ given $\varphi$. That is,
\beq
P(n_m|\varphi) = \braket{\psi(n_m,\varphi)}{\psi(n_m,\varphi)}.
\eeq

An arbitrary input state with $N=2j$ photons can be expressed as a sum of
$\hat J_z$ eigenstates
\beq
\ket{\psi} = \sum_{\mu=-j}^{j} \psi_{\mu}\ket{j \mu}_z.
\eeq
The state after $m$ detections will be (for $m\ge 1$) a function of $\varphi$.
It will be denoted as follows:
\beq
\ket{\psi(n_m,\varphi)} = \sum_{\mu=-j+m/2}^{j-m/2}
\psi_{\mu,m}(n_{m},\varphi) \ket{j-m/2, \mu}_z .
\eeq
Using the recurrence relation (\ref{recur}), it can be shown that the functional
form of $\psi_{\mu,m}(n_{m},\varphi)$ is always
\beq
\psi_{\mu,m}(n_{m},\varphi) = \sum_{k=-m/2}^{m/2}
\psi_{\mu,m,k}(n_{m})e^{i k \varphi}.
\eeq
The recurrence relation for the coefficients $\psi_{\mu,m,k}$ is
\begin{eqnarray}
\label{explicit}
\psi_{\mu,m,k}(n_{m}) \!\!\!\! &=& \!\!\!\! \frac{e^{-i(\Phi_m -u_m\pi)/2}}
{2\sqrt{2j-m+1}} \left[ s_-\psi_{\mu-\frac{1}{2},m-1,k-\frac{1}{2}}(n_{m-1})
-is_+ \psi_{\mu+\frac{1}{2},m-1,k-\frac{1}{2}}(n_{m-1})\right] \nn \\
\!\!\!\! && \!\!\!\!+\frac{e^{i(\Phi_m -u_m\pi)/2}}{2\sqrt{2j-m+1}}\left[ s_-
\psi_{\mu-\frac 12,m-1,k+\frac{1}{2}}(n_{m-1})+is_+
\psi_{\mu+\frac{1}{2},m-1,k+\frac{1}{2}}(n_{m-1})\right],
\end{eqnarray}
where
\begin{equation}
s_\pm = \sqrt{j-\frac{m}{2} \pm \mu+1}.
\end{equation}

We need to express the system state in this form for two main reasons. Firstly
the feedback scheme should not depend on the actual value of the phase, only on
the measurement record. To determine the feedback phase, the probability
distribution for the system phase is required. This can be determined from the
variation of the state with the unknown system phase. This requires the above
coefficients, as repeating the calculation for each individual value of the
phase would be too inefficient. Secondly, we can perform the entire calculation
independently of the system phase, and average over the system phase at the end
of the measurement. This allows us to take account exactly of the full range of
input phases.

The probability distribution for the unknown phase can be determined using
Bayes' theorem
\beq
P(\varphi|n_m) = \frac{P(\varphi)P(n_m|\varphi)}{P(n_m)}.
\eeq
The probability distribution for the phase at the start of the measurement,
$P(\varphi)$, will be flat, as it is assumed that there is no prior knowledge
about the phase. The divisor $P(n_m)$ is independent of the phase, and
therefore only provides a normalising factor to the phase distribution. Ignoring
these terms gives
\beq
P(\varphi|n_m) \propto P(n_m|\varphi),
\eeq
and therefore
\begin{equation}
\label{prob}
P(\varphi|n_m) \propto \braket{\psi(n_m,\varphi)}{\psi(n_m,\varphi)}.
\end{equation}
Note that this result is equivalent to Eq.~(\ref{genbayes}) for the case of
continuous measurements. A similar Bayesian approach to interferometry has been
considered before, and used in analysing experimental results \cite{Hra96}.
However, this was done only with non-adaptive measurements and with all
particles incident on one port.

\section{Optimum Phase Estimates}
\label{final}
Before describing the phase feedback technique, I will firstly describe how
to select the final phase estimate. This is necessary because it is not
possible to determine the phase variance produced by a phase feedback technique
without specific final phase estimates. The best estimate to use is
that which minimises the Holevo variance in the final phase estimate. This can
be determined by summing over the $2^{2j}$ combinations of results, then
averaging over the input phase.

For greater generality, the optimum phase estimate after $m$ detections will be
considered, rather than specifically the phase estimate at the end of the
measurement. The result for the final phase estimate can then be obtained by
substituting $m=2j$. First, summing over the combinations of results gives the
probability distribution for the error in the phase estimate as
\begin{equation}
\label{starter}
P(\phi|\varphi) = \sum_{[n_m]=0}^{2^m-1} P(n_m|\varphi)
\delta(\phi-(\hat\varphi(n_m)-\varphi)),
\end{equation}
where the square brackets in $[n_m]$ denote the numerical value of this
binary string interpreted as a binary number, and $\hat\varphi(n_m)$ is the
final phase estimate.

Next we wish to average over the system phase and the initial feedback phase.
For a feedback scheme to be unbiased (in that it treats all input phases
equivalently), the initial feedback phase should be chosen at random, as there
is no information to base this phase on. This phase should therefore be averaged
over in order to determine the average probability distribution. In determining
the coefficients $\psi_{\mu,m,k}(n_{m})$, a specific initial feedback phase must
be chosen, so at first it might appear that this phase cannot be averaged over.
Note, however, that all the successive feedback phases are relative to the
initial feedback phase. If the initial feedback phase is altered by some amount
$\Delta \Phi$, then all the successive feedback phases will be altered by the
same amount (for the same detection results). This is because, if the feedback
scheme is unbiased, the initial feedback phase provides a reference phase.

Therefore it is only the difference between the system phase and the initial
feedback phase that is significant, and we need only average over one of them.
I will take the initial feedback phase to be zero, which is equivalent to
measuring all phases relative to the initial feedback phase, and average over
the system phase. Performing this average gives
\begin{eqnarray}
\label{probdist}
P(\phi) \!\!\!\! &=& \!\!\!\! \int\limits_{-\pi}^{\pi} d\varphi \frac{1}{2\pi}
\sum_{[n_m]=0}^{2^m-1}P(n_m|\varphi)\delta(\phi- (\hat\varphi(n_m) - \varphi)),
\nn \\ \!\!\!\! &=& \!\!\!\! \frac{1}{2\pi}\sum_{[n_m]=0}^{2^m-1}
P(n_m|\hat\varphi(n_m) - \phi).
\end{eqnarray}

The exact phase variance for the measurement scheme can be determined from this
probability distribution. Evaluating $\ip{e^{i\phi}}$ gives
\begin{eqnarray}
\label{singleopt}
\ip{e^{i\phi}} \!\!\!\! &=& \!\!\!\! \int\limits_{-\pi}^{\pi} d\phi e^{i\phi}
\frac 1{2\pi}\sum_{[n_m]=0}^{2^m-1} P(n_m|\hat\varphi(n_m)-\phi) \nn \\
\!\!\!\! &=& \!\!\!\! \frac{1}{2\pi}\sum_{[n_m]=0}^{2^m-1}
e^{i\hat\varphi(n_m)} \int\limits_{-\pi}^{\pi} d\phi e^{-i(\hat\varphi (n_m)-
\phi)}P(n_m|\hat\varphi(n_m) - \phi) \nn \\
\!\!\!\! &=& \!\!\!\! \frac{1}{2\pi}\sum_{[n_m]=0}^{2^m-1}
e^{i\hat\varphi(n_m)} \int\limits_{-\pi}^{\pi} dx e^{-ix} P(n_m|x). 
\end{eqnarray}
In order to minimise the Holevo phase variance, we wish to maximise
$\st{\ip{e^{i\phi}}}$. A phase estimate that maximises this is
\begin{eqnarray} \label{estimate}
\hat\varphi(n_m) \!\!\!\! &=& \!\!\!\! \arg\int\limits_{-\pi}^{\pi} e^{i\varphi}
P(n_m|\varphi) d\varphi \nn \\
\!\!\!\! &=& \!\!\!\! \arg\int\limits_{-\pi}^{\pi} e^{i\varphi}
\braket{\psi(n_m,\varphi)}{\psi(n_m,\varphi)} d\varphi \nn \\
\!\!\!\! &=& \!\!\!\! \arg\int\limits_{-\pi}^{\pi} e^{i\varphi} P(\varphi|n_m)
d\varphi \nn \\ \!\!\!\! &=& \!\!\!\! \arg \ip{e^{i\varphi}},
\end{eqnarray}
where the expectation value is determined from the probability distribution for
the phase based on the measurement record.
For the specific case of the phase estimate at the end of the measurement, this
can be calculated as
\beq
\label{specest}
\hat\varphi(n_{2j})= \arg \sum_{k=-j}^{j-1} \psi_{0,2j,k} \psi_{0,2j,k+1}^*.
\eeq

Unfortunately there is a slight ambiguity here, as the same Holevo phase
variance will be obtained if a constant is added to these phase estimates. This
would make the probability distribution biased, and as discussed in the
introduction, a better way of evaluating the variance when the measurements may
be biased is
\beq
V_0(\phi)=\left[{\rm Re}{\ip{e^{i\phi}}}\right]^{-2}-1.
\eeq
Note that $e^{i\phi}$ is used here, rather than $e^{i(\phi-\varphi)}$, because
$\phi$ is the deviation from the system phase. Minimising this estimate of the
phase variance is equivalent to maximising ${\rm Re}{\ip{e^{i\phi}}}$. If any
constant is added to the phase estimates given by Eq.~(\ref{estimate}), the
value of $\st{\ip{e^{i\phi}}}$ will be the same, but the value of
${\rm Re}{\ip{e^{i\phi}}}$ will be smaller, as $\ip{e^{i\phi}}$ is complex.
Therefore the phase estimates given by Eq.~(\ref{estimate}) are the unique
solution that maximises ${\rm Re}{\ip{e^{i\phi}}}$.

With these phase estimates, the exact variance after $m$ detections can be
determined using the simple expression
\beq
\label{exactint}
\st{\ip{e^{i\phi}}}=\frac{1}{2\pi}\sum_{[n_m]=0}^{2^m-1} \left| \int
\limits_{-\pi}^{\pi}
e^{i\varphi} \braket{\psi(n_m,\varphi)} {\psi(n_m,\varphi)} d\varphi \right|.
\eeq
As these phase estimates are unbiased, the absolute value will be used, rather
than the real part. The final variance can be determined using
\beq
\label{exact}
\st{\ip{e^{i\phi}}}=\sum_{[n_{2j}]=0}^{2^{2j}-1} \left| \sum_{k=-j}^{j-1}
\psi_{0,2j,k} \psi_{0,2j,k+1}^* \right|.
\eeq
This means that during calculations, after each sequence of measurements, the
phase estimate and contribution to $\st{\ip{e^{i\phi}}}$ can be determined
as the phase and magnitude respectively of
\beq
\sum_{k=-j}^{j-1} \psi_{0,2j,k} \psi_{0,2j,k+1}^*.
\eeq

\section{The Feedback Technique}
\label{feedback}
In order to find the optimum phase feedback technique, we need to find the
feedback phase $\Phi_m$ for each measurement record $n_{m-1}$ that maximises
the value of $\st{\ip{e^{i\phi}}}$ as determined using Eq.~(\ref{exact})
(and therefore minimises the Holevo phase variance). This is, in general, a very
difficult problem, and I will initially consider a much simpler feedback
technique that can be determined analytically.

Rather than choosing each feedback phase to minimise the final phase variance
at the end of the measurement, we can choose the feedback phase that minimises
the phase variance after the next detection. This will mean that the last
feedback phases are optimal, but not necessarily the intermediate feedback
phases. This is because they minimise the intermediate phase variances, not the
final phase variance.

In order to see how to choose the phase estimate, recall that the value of
$\st{\ip{e^{i\phi}}}$ after the $m$th detection is given by
Eq.~(\ref{exactint}). For the feedback phase before the $m$th detection,
$\Phi_m$, we need only consider the part of the sum for the measurement record
$n_{m-1}$, as this is the only part of the sum that is affected by the feedback
phase. We must still sum over the $m$th detection result, as this is still
unknown. The expression to be maximised is therefore
\beq
\label{maxim}
M(\Phi_{m})= \sum_{u_{m}=0,1}\left| \int\limits_{-\pi}^{\pi} \braket{\psi(u_{m}
n_{m-1},\varphi)}{\psi(u_{m}n_{m-1},\varphi)} e^{i\varphi} d \varphi\right|.
\eeq

In order to use this expression we require $\braket{\psi(u_{m}n_{m-1},\varphi)}
{\psi(u_{m}n_{m-1},\varphi)}$ explicitly in terms of $u_m$. It is
straightforward to show from Eq.~(\ref{explicit}) that
\begin{eqnarray}
\label{statechange}
\braket{\psi(u_{m}n_{m-1},\varphi)}{\psi(u_{m}n_{m-1},\varphi)} = \frac 12
[\braket{\psi(n_{m-1},\varphi)}{\psi(n_{m-1},\varphi)}
\nn \\ +\lambda_{m-1}(\varphi)e^{i\varphi}e^{-i(\Phi_m-u_m\pi)}
+ \lambda_{m-1}^*(\varphi)e^{-i\varphi}e^{i(\Phi_m-u_m\pi)}],
\end{eqnarray}
where $\lambda_{m-1}(\varphi)$ is defined by
\beq
\lambda_{m-1}(\varphi) = \sum_{n=-m+1}^{m-1}\lambda_{m-1,n}e^{in\varphi},
\eeq
where
\beq
\lambda_{m-1,n} = -\frac{\xi_{m-1,n} + i \zeta_{m-1,n}}{2j-m+1},
\eeq
and where
\bqa
\xi_{m-1,n} \!\!\!\! &=& \!\!\!\! \sum_{k,k^\prime=-\frac{m-1}{2}}^{\frac{m-1}
2}\sum_{\mu=-j+\frac{m-1}2}^{j-\frac{m-1}2}\mu 
\psi_{\mu,m-1,k}\psi_{\mu,m-1,k'}^* \delta_{n,k-k'}, \\
\zeta_{m-1,n} \!\!\!\! &=& \!\!\!\! \sum_{k,k^\prime=-\frac{m-1}{2}}^
{\frac{m-1}{2}} \sum_{\mu=-j+\frac m2}^{j-\frac m2}
\frac{s_+s_-}2 (\psi_{\mu-\frac 12,m-1,k}\psi_{\mu+\frac 12,m-1,k'}^*
+\psi_{\mu+\frac 12,m-1,k} \psi_{\mu-\frac 12,m-1,k'}^* )\delta_{n,k-k'}. \nn\\
\eqa

Using this result it is possible to show that
\begin{eqnarray}
M(\Phi_{m}) \!\!\!\! &=& \!\!\!\! \sum_{u_{m}=0,1}\left|\int\limits_{-\pi}^{\pi}
\frac{1}{2}[\braket{\psi(n_{m-1},\varphi)}{\psi(n_{m-1},\varphi)}
+\lambda_{m-1}(\varphi)e^{i\varphi}e^{-i(\Phi_m-u_m\pi)} \right. \nn \\
&&\left. + \lambda_{m-1}^*(\varphi)e^{-i\varphi}e^{i(\Phi_m-
u_m\pi)}]e^{i\varphi} d \varphi\right| \nn \\
\!\!\!\! &=& \!\!\!\! \frac 12 \sum_{u_{m}=0,1} \left| \int\limits_{-\pi}^{\pi}
\left[\sum_{\mu=-j+\frac{m-1}2}^{j-\frac{m-1}2}\sum_{k,k'=-\frac{m-1}2}^{\frac
{m-1}2}\psi_{\mu,m-1,k}\psi_{\mu,m-1,k'}^*e^{i(k-k'+1)\varphi} \right. \right.
\nn \\
&& \!\!\!\!\!\!\!\! \left. \left. +e^{-i(\Phi_m-u_m\pi)} \sum_{n=1-m}^{m-1}
\lambda_{m-1,n}e^{i (n+2) \varphi} +e^{i(\Phi_m-u_m\pi)} \sum_{n=1-m}^{m-1}
\lambda_{m-1,n}^* e^{-i n \varphi} \right] d\varphi \right| \nn \\
\!\!\!\! &=& \!\!\!\! \pi \sum_{u_{m}=0,1} \left| \sum_{\mu=-j+\frac{m-1}2}^
{j-\frac{m-1}2}\sum_{k=-\frac{m-1}2}^{\frac{m-3}2}\psi_{\mu,m-1,k}
\psi_{\mu,m-1,k+1}^* \right. \nn \\ && \left. +\lambda_{m-1,-2}
e^{-i(\Phi_m-u_m\pi)}+\lambda_{m-1,0}^* e^{i(\Phi_m-u_m\pi)} \right|.
\end{eqnarray}
This can be expressed in the form
\begin{equation}
\label{simple}
M(\Phi_{m}) = \left| a+b e^{-i\Phi_m}+c e^{i\Phi_m}\right|+
\left| a-b e^{-i\Phi_m}-c e^{i\Phi_m}\right|,
\end{equation}
where
\begin{eqnarray}
a\!\!\!\! &=& \!\!\!\! \pi \sum_{\mu=-j+\frac{m-1}{2}}^{j-\frac{m-1}{2}}
\sum_{k=-\frac{m-1}{2}}^{\frac{m-3}{2}}
\psi_{\mu,m-1,k} \psi_{\mu,m-1,k+1}^*, \\
b\!\!\!\! &=& \!\!\!\! \pi \lambda_{m-1,-2}, \\
c\!\!\!\! &=& \!\!\!\! \pi \lambda_{m-1,0}^*.
\end{eqnarray}
There is an analytic solution for the $\Phi_m$ that maximises $M(\Phi_{m})$.
This solution gives three phases, $\Phi_0$ and $\Phi_\pm$, and the phase that
is optimal must be found by substituting into Eq.~(\ref{simple}). These phases
are given by
\begin{equation}
\Phi_0=\arg (b a^*-c^* a),
\end{equation}
and
\begin{equation}
\Phi_\pm = \arg \sqrt{\frac {c_2 \pm \sqrt{c_2^2+|c_1|^2}}{c_1}},
\end{equation}
where
\begin{eqnarray}
c_1\!\!\!\! &=& \!\!\!\!(a^* c)^2-(ab^*)^2+4(|b|^2-|c|^2)b^* c, \\
c_2\!\!\!\! &=& \!\!\!\!-2i {\rm Im} (a^2 b^* c^*).
\end{eqnarray}
Note that $M(\Phi)=M(\Phi+\pi)$ so that in addition to the solution found by
this method there will be another differing by $\pi$. It does not matter which
of these is chosen; it simply reverses the significance of the two alternative
detection results $u_m=0$ and 1.

As mentioned above, the initial feedback phase $\Phi_1$ should be chosen at
random, as there is no information to base it on. At each following step, we
determine the optimal feedback phase by the method described above, then
determine the evolution of the state for that feedback phase. This process
continues until all photons have been counted. The measurement record is then
the binary string $n_{2j} = u_{2j} \ldots u_2 u_1$ and the result is a
posterior distribution $P(\varphi|n_{2j})$ that is proportional to
$\braket{\psi(n_{2j},\varphi)}{\psi(n_{2j},\varphi)}$ and is characterised by
the $2j+1$ numbers $\psi_{0,2j,k}(n_{2j})$.

\section{Stochastic Method}
\label{stoch}
For a moderate number of photons it is possible to determine the exact phase
variance by systematically determining the evolution of the state for each
combination of measurement results and using Eq.~(\ref{exact}). As the
number of possible measurement records increases as $2^{2j}$, the exact variance
can be determined only for moderate photon numbers (up to 20 or 30). For larger
photon numbers it is necessary to determine the phase distribution
stochastically.

The initial feedback phase is selected at random, and the results of the
detections are selected by their probability of occurring. In order to
determine the probabilities, a specific system phase must be selected. For
simplicity, the system phase was taken to be zero in the results that will be
presented here. This leads to no loss of generality, as the initial feedback
phases were selected at random.

In order to determine the probabilities, it is simplest to determine the phase
dependent state coefficients $\psi_{\mu,m}(n_m,\varphi)$. The state coefficients
will change as
\bqa
\label{stochrec}
\psi_{\mu,m}(n_m,\varphi) \!\!\!\!&=&\!\!\!\! 
s_+ \sin \Theta_m \psi_{\mu+\frac 12,m-1}(n_{m-1},\varphi)
+s_- \cos\Theta_m \psi_{\mu-\frac 12,m-1}(n_{m-1},\varphi),
\eqa
where
\begin{equation}
\Theta_m=\frac{\varphi-\Phi_m+u_m \pi}2.
\end{equation}
The probability of obtaining the result $u_m$ is given by
\begin{eqnarray}
&& P(u_m|\varphi,n_{m-1})=\frac 12+\frac{e^{iu_m\pi}}
{2j-m+1}\left[ -\cos (\varphi-\Phi_m) \sum_{\mu=-j+\frac{m-1}2}^
{j-\frac{m-1}2}\mu \left|\psi_{\mu,m-1}(n_{m-1},\varphi)\right|^2 \right. \nn \\
&& \left. +\sin (\varphi-\Phi_m) \sum_{\mu=-j+\frac m2}^
{j-\frac m2} 
s_+s_- 
{\rm Re} \left(\psi_{\mu+\frac 12,m-1}
(n_{m-1},\varphi)\psi_{\mu-\frac 12,m-1}^*(n_{m-1},\varphi)\right)\right].
\end{eqnarray}
Here it has been assumed that the coefficients $\psi_{\mu,m-1}(n_{m-1},\varphi)$
are normalised. The recurrence relation (\ref{stochrec}) does not give the
normalised coefficients. The coefficients determined from Eq.~(\ref{stochrec})
should be normalised in a separate calculation.

The detection results were chosen with probabilities determined using
$\varphi=0$ and these formulae, and the final phase estimates were determined
using Eq.~(\ref{specest}). For the ensemble $\{\phi_{\mu}\}_{\mu=0}^M$ of
$M$ final phase estimates the Holevo phase variance was estimated by
\beq
\left[{\rm Re}\left(M^{-1} \sum_{n=1}^M e^{i\phi_n}\right)\right]^{-2}-1.
\eeq

\section{Modulo $\pi$ States}
In the case of the state with equal photon numbers in both ports, the above
techniques will not apply exactly, as the phase distribution repeats modulo
$\pi$. This means, for example, that $\ip{e^{i\phi}}=0$. In the previous
chapter the measure of the phase variance used for this state was
\beq
V_{\pi}(\phi)=(\st{\ip{e^{2i\phi}}}^{-2}-1)/4.
\eeq
Most of the above analysis must also be modified to consider the phase modulo
$\pi$.

Firstly the distribution (\ref{starter}) should be averaged over the interval
$[-\pi/2,\pi/2]$ rather than $[-\pi,\pi]$. This gives
\beq
P(\phi) = \frac 1{\pi}\sum_{[n_m]=0}^{2^m-1}P(n_m|\hat\varphi(n_m) - \phi).
\eeq
Now, rather than evaluating ${\ip{e^{i\phi}}}$, we wish to find
${\ip{e^{2i\phi}}}$. This can be evaluated as
\begin{eqnarray}
{\ip{e^{2i\phi}}} \!\!\!\! &=& \!\!\!\! \left| \int\limits_{-\pi/2}^{\pi/2}
d\phi e^{2i\phi} \frac 1{\pi}\sum_{[n_m]=0}^{2^m-1} P(n_m|\hat\varphi(n_m)-\phi)
\right| \nn \\ \!\!\!\! &=& \!\!\!\! \left| \frac{1}{\pi}\sum_{[n_m]=0}^{2^m-1}
e^{2i\hat\varphi(n_m)} \int\limits_{-\pi/2}^{\pi/2} dx e^{-2ix}
\braket{\psi(n_m,x)}{\psi(n_m,x)} \right|.
\end{eqnarray}

This means that the optimal phase estimate is
\beq
\hat\varphi(n_m) = \half \arg\int\limits_{-\pi/2}^{\pi/2} e^{2i\varphi}
\braket{\psi(n_m,\varphi)} {\psi(n_m,\varphi)} d\varphi.
\eeq
This can be evaluated at the end of the measurement using
\beq
\hat\varphi(n_{2j})=\half\arg\sum_{k=-j}^{j-2} \psi_{0,2j,k} \psi_{0,2j,k+2}^*,
\eeq
and the final variance is calculated using
\beq
\st{\ip{e^{2i\phi}}}=\sum_{[n_{2j}]=0}^{2^{2j}-1} \left| \sum_{k=-j}^{j-2}
\psi_{0,2j,k} \psi_{0,2j,k+2}^* \right|.
\eeq
Now instead of Eq.~(\ref{maxim}), the expression to be maximised is
\beq
M(\Phi_m)= \sum_{u_m=0,1}\left| \int\limits_{-\pi/2}^{\pi/2} \braket{\psi(u_{m}
n_{m-1},\varphi)}{\psi(u_{m}n_{m-1},\varphi)} e^{2i\varphi} d \varphi\right|.
\eeq
Using Eq.~(\ref{statechange}), this becomes
\begin{eqnarray}
M(\Phi_{m}) \!\!\!\! &=& \!\!\!\! \sum_{u_{m}=0,1} \left|
\int\limits_{-\pi/2}^{\pi/2}\frac{1}{2}[\braket{\psi(n_{m-1},\varphi)}
{\psi(n_{m-1},\varphi)}+\lambda_{m-1}(\varphi)e^{i\varphi}e^{-i(\Phi_m-u_m\pi)}
\right. \nn \\
&&\left. + \lambda_{m-1}^*(\varphi)e^{-i\varphi}e^{i(\Phi_m-
u_m\pi)}]e^{2i\varphi} d \varphi\right| \nn \\
\!\!\!\! &=& \!\!\!\! \frac 12 \sum_{u_{m}=0,1} \left|
\int\limits_{-\pi/2}^{\pi/2}\left[\sum_{\mu=-j+\frac{m-1}2}^{j-\frac{m-1}2}
\sum_{k,k'=-\frac{m-1}2}^{\frac{m-1}2}\psi_{\mu,m-1,k} \psi_{\mu,m-1,k'}^*
e^{i(k-k'+2)\varphi} \right. \right. \nn \\
&& \!\!\!\!\!\!\!\! \left. \left. +e^{-i(\Phi_m-u_m\pi)} \sum_{n=1-m}^{m-1}
\lambda_{m-1,n}e^{i (n+3) \varphi} +e^{i(\Phi_m-u_m\pi)} \sum_{n=1-m}^{m-1}
\lambda_{m-1,n}^* e^{-i(n-1)\varphi} \right] d\varphi \right| \nn \\
\!\!\!\! &=& \!\!\!\! \frac {\pi}2 \sum_{u_{m}=0,1} \left| \sum_{\mu=-j+
\frac{m-1}2}^{j-\frac{m-1}2}\sum_{k=-\frac{m-1}2}^{\frac{m-5}2}\psi_{\mu,m-1,k}
\psi_{\mu,m-1,k+2}^* \right. \nn \\ && \left. +\lambda_{m-1,-3}
e^{-i(\Phi_m-u_m\pi)}+\lambda_{m-1,1}^* e^{i(\Phi_m-u_m\pi)} \right|.
\end{eqnarray}
Thus $a$, $b$ and $c$ are given by
\begin{eqnarray}
a\!\!\!\!&=&\!\!\!\! \frac {\pi}2 \sum_{\mu=-j+\frac{m-1}{2}}^{j-\frac{m-1}{2}}
\sum_{k=-\frac{m-1}{2}}^{\frac{m-5}{2}}
\psi_{\mu,m-1,k} \psi_{\mu,m-1,k+2}^*, \\
b\!\!\!\! &=& \!\!\!\! \frac {\pi}2 \lambda_{m-1,-3}, \\
c\!\!\!\! &=& \!\!\!\! \frac {\pi}2 \lambda_{m-1,1}^*.
\end{eqnarray}
The last equation that must be modified for this case is the formula for
estimating the phase variance from the data, which becomes
\beq
\frac 14 \left\{ \left[ {\rm Re} \left( M^{-1}
\sum_{n=1}^M e^{2i\phi_n}\right)\right]^{-2}-1 \right\} .
\eeq

\section{Results}
\label{results}
\begin{figure}
\centering
\includegraphics[width=0.7\textwidth]{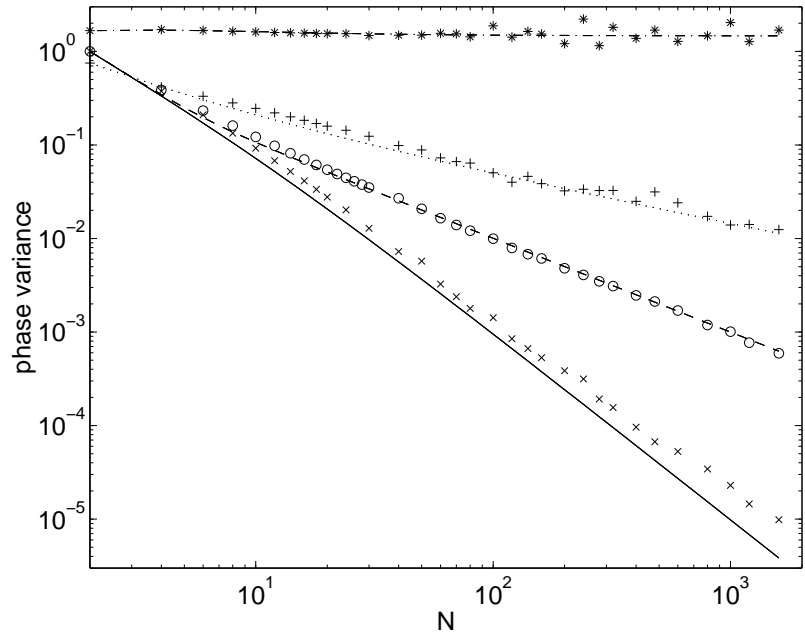}
\caption{Variances in the phase estimate versus input photon number $2j$. The
lines are exact results for canonical measurements on optimal states
$\ket{\psi_{\rm opt}}$ (continuous line), on states with all photons incident
on one input port $\ket{jj}_{z}$ (dashed line), on states with equal photon
numbers incident on both input ports $\ket{j0}_{z}$ (dotted line), and the
state $(\ket{j0}_{z}+\ket{j1}_{z})/\sqrt{2}$ (dash-dotted line). The crosses are
the numerical results for the adaptive phase measurement scheme on
$\ket{\psi_{\rm opt}}$, the circles are those on $\ket{jj}_{z}$, the pluses
are those on $\ket{j0}_z$, and the asterisks are those on a
$(\ket{j0}_{z}+\ket{j1}_{z})/\sqrt{2}$ input state. All variances for the
$\ket{j0}_{z}$ state are for phase modulo $\pi$.}
\label{three}
\end{figure}

The results of using this adaptive phase measurement scheme on the four
alternative input states are shown in Fig.~\ref{three}. The phase variances
for states up to $N=20$ (or $N=30$ for $\ket{jj}_z$) were determined exactly
using Eq.~(\ref{exact}), whereas those for larger photon numbers were
determined using the stochastic method described in Sec.~\ref{stoch}. The sample
sizes used were about $2^{15}$ for the smallest photon numbers, down to $2^{10}$
for the larger photon numbers. The phase variances are very close to the phase
variances for canonical measurements for all of these input states.

\begin{figure}
\centering
\includegraphics[width=0.7\textwidth]{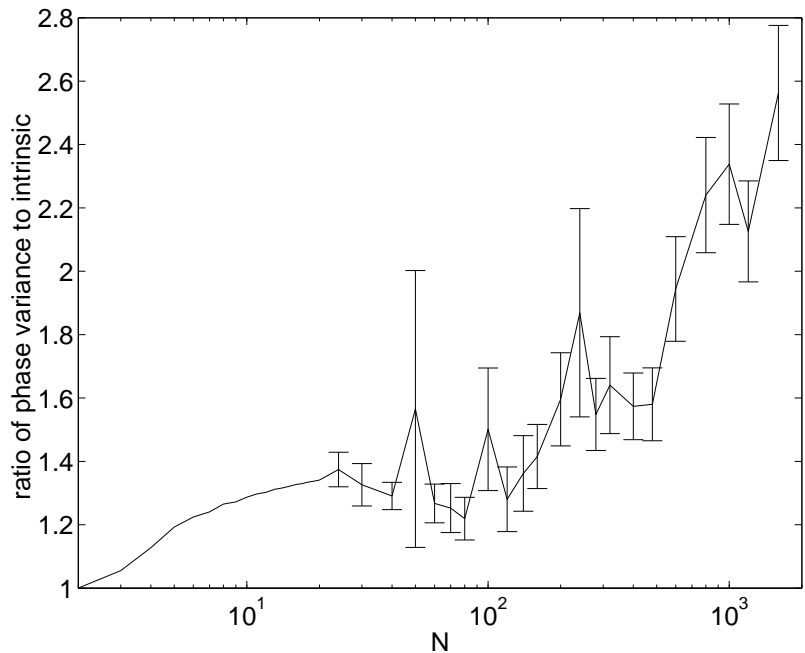}
\caption{The ratio of the phase variance using the feedback scheme of
Sec.~\ref{feedback} to the canonical phase variance for optimal input states.}
\label{ratioopt}
\end{figure}

For the optimal input states described in Ch.~\ref{adaptiveinter}, the
scaling is close to $1/N^2$, but the phase variances do differ relatively more
from the canonical values for larger photon numbers. If we plot the phase
variances as a ratio to the canonical phase variance (see Fig.~\ref{ratioopt}),
we find that the ratio of the phase variance to the canonical phase variance
increases fairly regularly with photon number. This ratio possibly increases
proportional to $\log N$. In that case the introduced phase variance would
increase as $\log N/N^2$, as is the case for the optimal single mode phase
measurements considered in Ch.~\ref{optdyne}.

\begin{figure}
\centering
\includegraphics[width=0.7\textwidth]{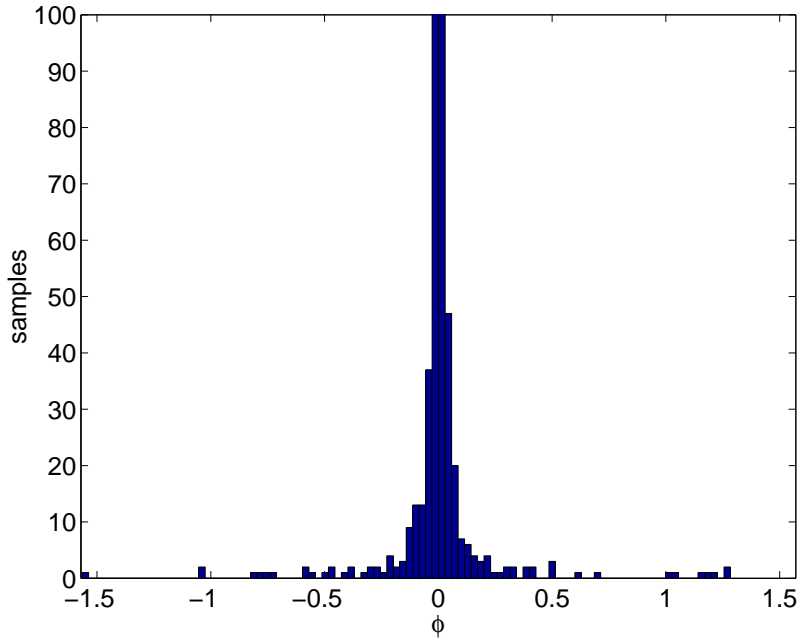}
\caption{The phase distribution resulting from using adaptive phase
measurements on an input state of $\ket{j0}_z$ for 800 photons. The
vertical axis has been cut off at 100 (the peak count is almost 500)
to show the tails more clearly.}
\label{resdist}
\end{figure}

\begin{figure}
\centering
\includegraphics[width=0.7\textwidth]{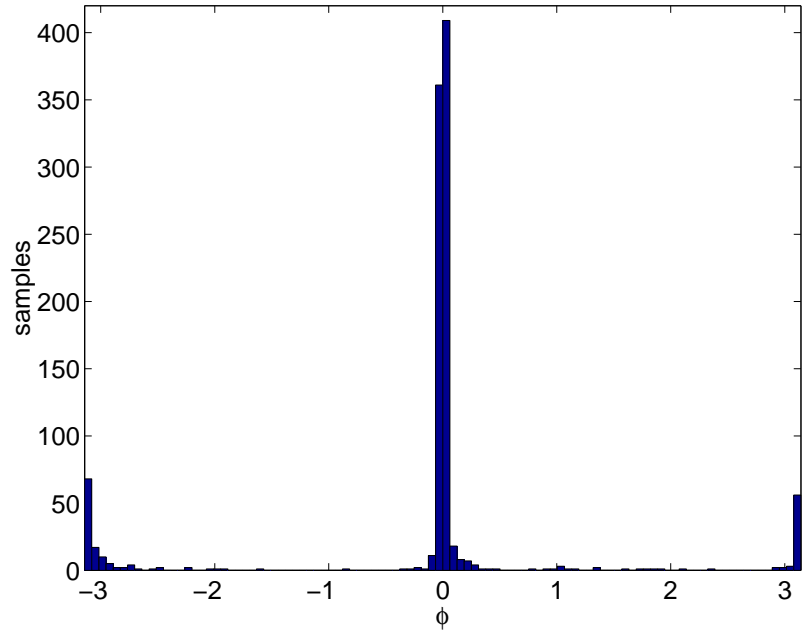}
\caption{The phase distribution resulting from using adaptive phase measurements
on an input state of $(\ket{j0}_{z}+\ket{j1}_{z})/\sqrt{2}$ for 800 photons.}
\label{xxxdist}
\end{figure}

For $\ket{j0}_z$ the variances are very close to those for canonical
measurements, scaling as $1/N^{1/2}$. If we look at the distribution of the
phases resulting from these measurements, we find that there is a sharp peak,
but a significant number of results with large error that produce the large
variance (see Fig.~\ref{resdist}). Similarly, for the case of the state
$(\ket{j0}_{z}+\ket{j1}_{z})/\sqrt{2}$, there are a large number of results at
$\pm \pi$, as seen in Fig.~\ref{xxxdist}. As for the canonical distribution,
this is why the phase variance does not decrease significantly with photon
number.

Now recall that although the phase uncertainty of the $\ket{j0}_z$ state as
measured by the square root of the variance does not scale as $N^{-1}$, other
measures of the phase uncertainty do scale as $N^{-1}$. This indicates that if
the data is analysed in a way corresponding to one of these other measures, the
uncertainty should scale down more rapidly with photon number. I will consider
$2/3$ confidence intervals, as the other measures do not make sense for discrete
data.

\begin{figure}
\centering
\includegraphics[width=0.7\textwidth]{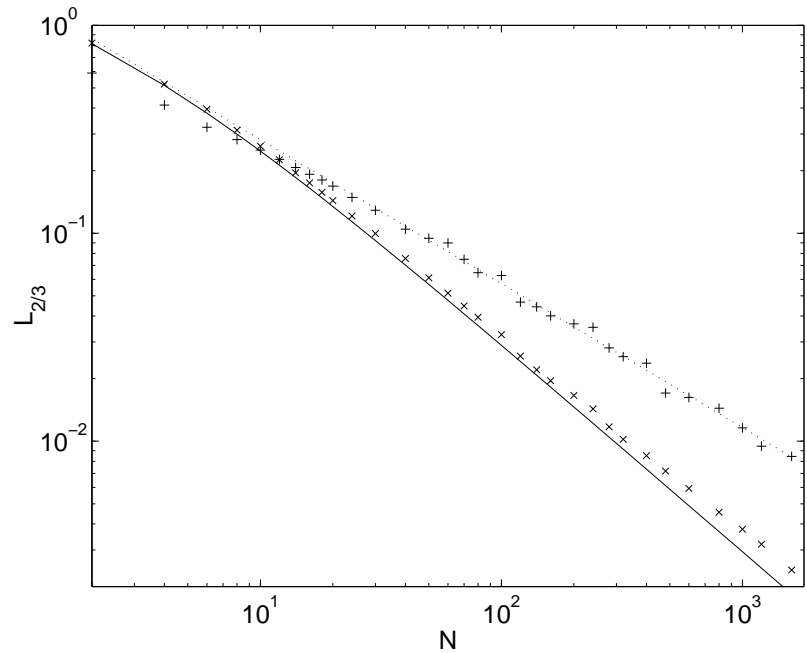}
\caption{The phase uncertainty resulting from using adaptive phase measurements
as measured using $2/3$ confidence intervals. The pluses are the numerical
results for an input state of $\ket{j0}_z$ and the crosses are those for
optimal states. The continuous line is the confidence interval for the
canonical distribution for $\ket{\psi_{\rm opt}}$, and the dotted line is the
 function fitted to the $\ket{j0}_z$ data.}
\label{confidence}
\end{figure}

The phase uncertainty for adaptive measurements on the $\ket{j0}_z$ state as
measured by the $2/3$ confidence interval is plotted in Fig.~\ref{confidence}.
A power law of the form $cN^{-p}$ was fitted to this data set (for photon numbers
of 20 and over), and it was found that the best fit was for $c=1.39 \pm 0.08$
and $p=0.69 \pm 0.01$. This means that the scaling is not as good as
the $N^{-1}$ scaling for the canonical phase distribution, but it is still far
better than the $N^{-1/4}$ scaling indicated by the variance.

In contrast, the confidence interval for measurements on the optimal state (also
shown in Fig.~\ref{confidence}) scales very close to $N^{-1}$. In fact for the
range of photon number considered the confidence interval for the measurements
is never more than about 30\% above the canonical confidence interval. This
means that although the $\ket{j0}_z$ state has $2/3$ confidence intervals for
the canonical distribution that are close to those for the optimal state, the
confidence intervals for the measurements are far worse.

As mentioned in the previous chapter, the uncertainty in the mean phase will
scale as $N^{-1/4}$ rather than $N^{-1}$ for $\ket{j0}_z$. According to the
Central Limit Theorem, the probability distribution for the mean of a large
number of measurements will be close to a normal distribution, even if the
probability distribution for each individual sample is far from a normal
distribution. In addition, the variance in the mean is the variance for an
individual sample divided by the number of samples. As the probability
distribution for the mean is approximately Gaussian, this means that the
uncertainty for any other measure will have the same scaling as the square root
of the variance, $N^{-1/4}$.

For example, for a photon number of 1600 and a sample size of $2^{10}$, the
uncertainty in the mean phase as indicated by the standard error (the standard
deviation divided by the square root of the number of samples) is
$4.0\times 10^{-3}$. The mean phase is also around this value, at
$3.5\times 10^{-3}$. On the other hand the median is much closer to zero, at
about $5.9\times 10^{-5}$.

In order to find the best phase estimate based on the data, we would need to
multiply together the probability distributions for the phase from each of the
samples. Similarly to the case for the individual samples, the optimal phase
estimate would then be given by $\half \arg\ip{e^{2i\phi}}$, where the average
is based on that distribution. The problem with this method is that it is
extremely computationally intensive. For the example given, we would need to
multiply together 1024 sums, each with 3200 terms, resulting in a total of more
than $3.2$ million terms.

I have also considered phase measurements using two other measurement schemes.
The first is a non-adaptive phase measurement introduced in Ref.~\cite{short}
and defined by
\beq \label{nonadapt}
\Phi_m = \Phi_0 + \frac{m\pi}{N}.
\eeq
This is analogous to heterodyne detection \cite{fullquan} on a single mode, in
that the phase $\Phi$ equally weights all relevant values over the course of
the measurement. The second scheme is a simple adaptive feedback scheme using a
running estimate of the phase:
\begin{eqnarray} \label{est}
\Phi_m \!\!\!\! &=& \!\!\!\! \arg \ip{e^{i\phi}} \nn \\
\!\!\!\! &=& \!\!\!\! \arg \sum_{\mu=-j+\frac{m-1}{2}}^{j-\frac{m-1}{2}}
\sum_{k=-\frac{m-1}{2}}^{\frac{m-3}{2}}
\psi_{\mu,m-1,k} \psi_{\mu,m-1,k+1}^*.
\end{eqnarray}
This is motivated by the relative success of the analogous simple feedback
scheme \cite{fullquan} for phase measurement of a single mode. The difference
here is that there is no $\pi/2$ term. For interferometry,
small phase uncertainties are obtained when the phase difference between the
arms is close to zero \cite{BondShap,Yurke,Holland}, rather than $\pi/2$. As
mentioned in the previous chapter, the reason for this difference is that a
different scattering matrix for the beam splitters has been assumed for
interferometry than for dyne measurements. 

\begin{figure}
\centering
\includegraphics[width=0.7\textwidth]{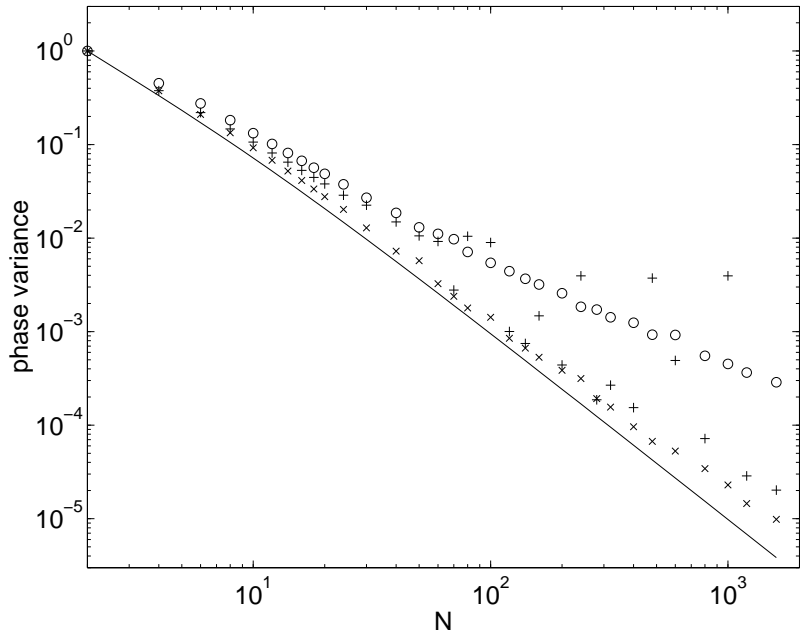}
\caption{The Holevo phase variance for optimal input states under various
measurement schemes. The canonical phase variance is shown as the continuous
line, the results for the adaptive measurement scheme of Sec.~\ref{feedback} as
crosses, the non-adaptive measurement scheme of Eq.~(\ref{nonadapt}) as
circles, and the feedback scheme of Eq.~(\ref{est}) as pluses.}
\label{schemes}
\end{figure}

The results of using these two measurement schemes, as well as the adaptive
measurement scheme of Sec.~\ref{feedback}, on the optimal input states, are
shown in Fig.~\ref{schemes}. The non-adaptive measurement scheme is far inferior
to the adaptive measurement scheme of Sec.~\ref{feedback}, and the variance
scales as $N^{-1}$. The simple adaptive feedback scheme also gives poor results.
Although most of the phase results for this feedback scheme have small error,
there are a small number of results with very large error. This also means that
the results shown in Fig.~\ref{schemes} are fairly erratic, as the results for
which a large error sample was obtained have much larger phase variance.

\begin{figure}
\centering
\includegraphics[width=0.7\textwidth]{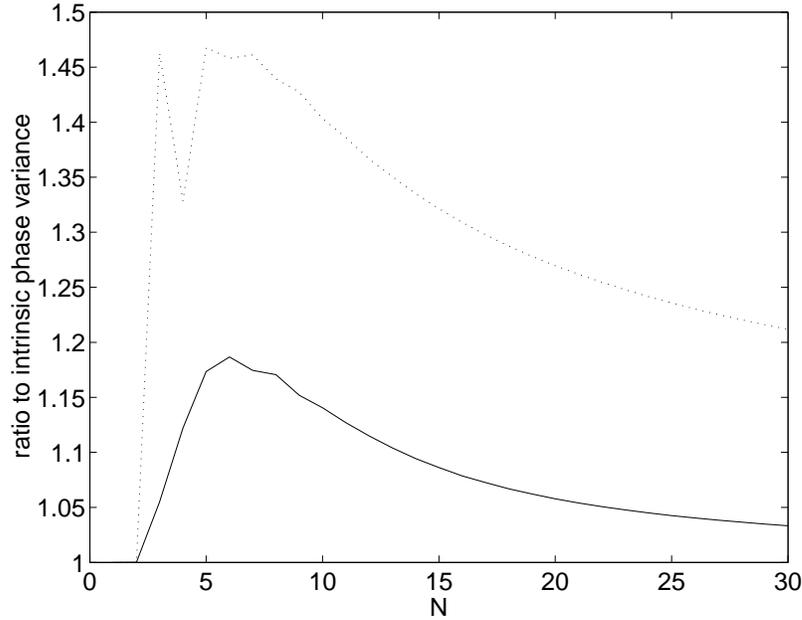}
\caption{The exact phase variance for $\ket{jj}_z$ input states under two
different measurement schemes as a ratio to the canonical phase variance. The
results for the adaptive measurement scheme of Sec.~\ref{feedback} are shown as
the continuous line, and the non-adaptive measurement scheme of
Eq.~(\ref{nonadapt}) as the dotted line.}
\label{schemes2}
\end{figure}

I also considered the non-adaptive measurement scheme on the state with all
photons in one port. The exact results for that case for $N$ up to 30 are
shown in Fig.~\ref{schemes2}. The phase variance is not much more than the
canonical phase variance, about 20\% more for $N=30$ and still decreasing.
This demonstrates that for this state there is relatively little improvement in
using a more advanced feedback scheme for larger photon numbers. The biggest
improvement is a reduction in the variance of about 24\% for $N=3$.

\section{Optimal Feedback}
\label{optimal}

The next question is whether the adaptive measurement technique described above
is optimal. Note firstly that the initial feedback phase has no effect, because
it is effectively averaged over by averaging over the system phase. Secondly the
last feedback phase is always optimal, as was noted above. This means that for
states with 1 or 2 photons the measurement technique must be optimal. In fact,
for the states considered here the phase variance was equal to the canonical
phase variance for 1 or 2 photons.

To see if the phase variance was equal to canonical for arbitrary states, the
complete range of possible states was considered. For 1 photon the state can be
expressed as
\beq
\ket \psi = \sum_{\mu = -1/2}^{1/2} \psi_\mu \ket{1/2,\mu}_y.
\eeq
There are therefore two coefficients, $\psi_{\pm 1/2}$, which can in general
take complex values. The magnitude and phase of these coefficients give four
real numbers that can be varied for these states. There is the restriction,
however, that the states must be normalised, which removes one degree of
freedom.

Also the absolute phase of the coefficients is irrelevant. That is, changing the
phase of both coefficients $\psi_{\pm 1/2}$ by the same amount gives an
equivalent state. In addition, the relative phases of $\psi_{\pm 1/2}$ are
irrelevant. This is because adding a phase difference between the coefficients
simply gives an equivalent input state with a phase shift. These considerations
mean that only real $\psi_{\pm 1/2}$ need be considered, and there is therefore
only one degree of freedom.

\begin{figure}
\centering
\includegraphics[width=0.7\textwidth]{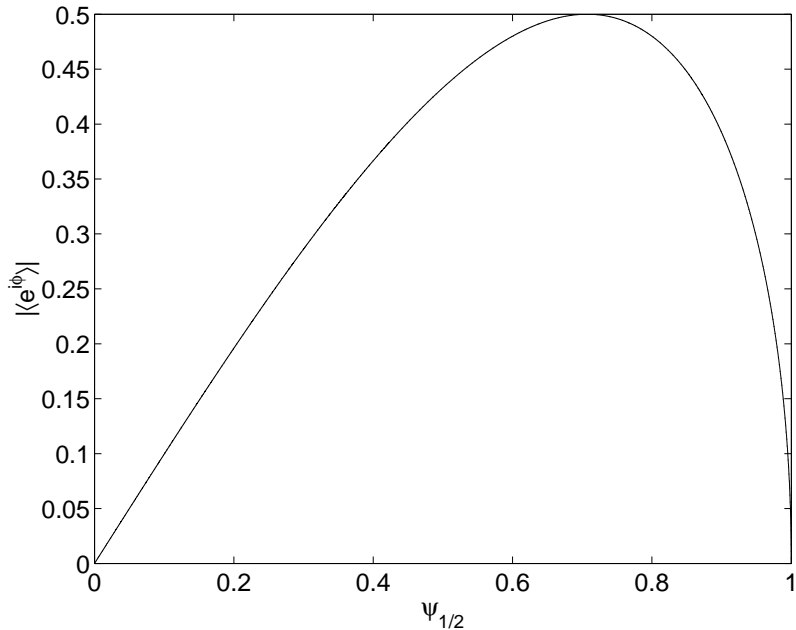}
\caption{The value of $|\lr{e^{i\phi}}|$ for a 1 photon input state as a
function of $\psi_{1/2}$. The canonical variance is shown as the continuous line
and the value for the measurements is shown as the dotted line (these lines
overlap).}
\label{canproof}
\end{figure}

The value of $\psi_{1/2}$ was varied in 10000 steps from 0 to 1 (see
Fig.~\ref{canproof}), and it was found that the phase variance obtained was
identical to the canonical phase variance for the entire range. Note that this
case is independent of the feedback scheme, as there is only the initial
feedback phase, which has no effect. This means that the variance is the same
as the canonical variance, regardless of the input state or feedback scheme.

For the case of 2 photons the state has three coefficients. Taking account of
the magnitude and phase this gives 6 real numbers that can be varied. The same
considerations as for the 1 photon case apply, leaving only 3 degrees
of freedom. We can vary the magnitude of two of the coefficients independently,
and the phase of one.

The magnitudes of $\psi_{-1}$ and $\psi_1$ and the phase of $\psi_0$ were
varied over the complete range with 100 steps in each of the variables, and it
was found that the phase variance obtained was identical to the canonical phase
variance. This demonstrates that the phase variance is as good as canonical for
2 photons, independent of the input state. This case is not independent of the
feedback scheme, and if any other feedback phase is used the feedback is not
quite as good as canonical. These results were also obtained when the states
were selected completely at random.

For the case of optimal input states with 3 or 4 photons, it was found that it
is not possible to decrease the variance by altering the intermediate feedback
slightly, so showing that the feedback technique is locally optimal for the
phase variance. For more than 4 photons it is possible to reduce the phase
variance by varying the intermediate feedback phases, and so the feedback is
not optimal.

In order to show that the feedback is globally optimal for 3 or 4 photons, it
is necessary to test the entire phase range. Only optimal input states will be
considered here, and the more general case will be considered later. There are
three factors that reduce the number of phases that need be varied. The first
two are as noted above: the first feedback phase has no effect (and so may be
ignored), and the last feedback phase is always optimal (and so need not be
varied). The third is that the contribution to the phase variance for a sequence
of detections is independent of the first detection result. This is because
changing the first feedback phase by $\pi$ reverses the significance of the
first detection results, and the first feedback phase is arbitrary.

\begin{figure}
\centering
\includegraphics[width=0.7\textwidth]{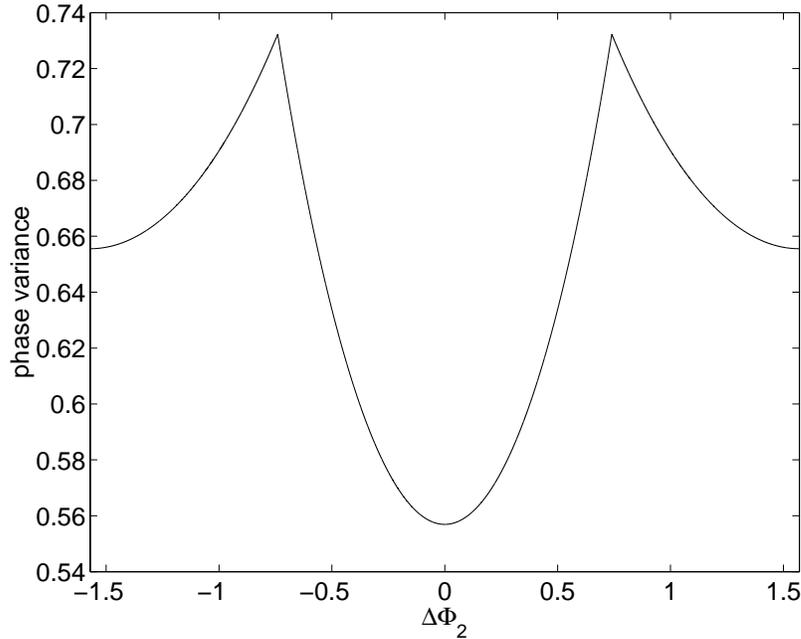}
\caption{The phase variance for the 3 photon optimum input state
as a function of the second feedback phase. The phase given here is relative to
the second feedback phase given by the feedback scheme of Sec.~\ref{feedback}.
The other feedback phases are as given by this feedback scheme.}
\label{proof}
\end{figure}

The consequence of these three factors is that for 3 photons the variation of
only one feedback phase needs to be considered, and for 4 photons the variation
of three feedback phases needs to be considered. The phase variance for the 3
photon case as the second feedback phase is varied from its value for the
feedback technique of Sec.~\ref{feedback} is shown in Fig.~\ref{proof}. This
figure shows that this feedback technique is globally optimal for 3 photons.
Since the phase variance is above the canonical phase variance in this case,
this demonstrates that it is not possible to perform canonical measurements
using photodetection and feedback alone. For 4 photons the second feedback phase
and two third feedback phases must be varied. This case was tested with 100
steps in each of the three variables, and it was found that the feedback
technique is globally optimal in this case also.

\begin{figure}
\centering
\includegraphics[width=0.7\textwidth]{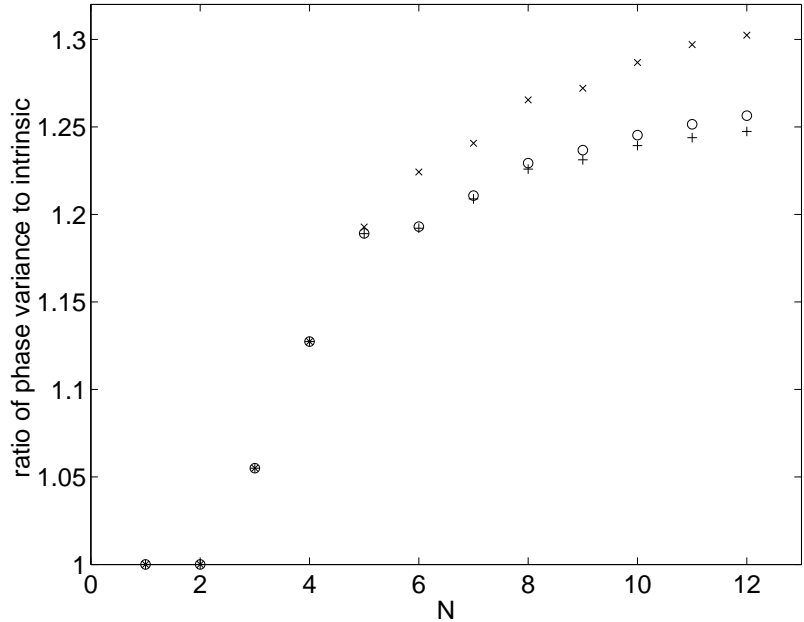}
\caption{The phase variance for the feedback scheme of Sec.~\ref{feedback} and
two numerically optimised feedback schemes as ratios to the minimum intrinsic
phase variance. The variance for the feedback scheme of Sec.~\ref{feedback} is
shown as the crosses, the case where the feedback alone is numerically
optimised is shown as circles, and the case where both the state and the
feedback are numerically optimised is shown as pluses.}
\label{optimum}
\end{figure}

In order to see how far the phase variance could be improved for photon numbers
above 4, the feedback phases were optimised using function minimisation
techniques, and the results are shown in Fig.~\ref{optimum}. Unfortunately
the number of feedback phases increases exponentially with the photon number,
making this technique infeasible very rapidly, and therefore only results up to
$N=12$ are shown. As can be seen, this optimisation only gives minor
improvements in the phase variance, with the maximum reduction in the phase
variance being about 3.5\%.

Note that the input states used here give minimum canonical phase variance,
but as the phase variances obtained for these feedback schemes differ from the
canonical phase variance (above 2 photons), these input states are not
necessarily optimum for these measurements. Therefore numerical optimisations
were performed where both the input state and the feedback were optimised for,
and the results are also shown in Fig.~\ref{optimum}.

It was found that the phase variance was reduced below the results where only
the feedback was optimised for, even for the case with 3 photons. In the case
with 3 photons, the improvement is only about $0.0005\%$, which is not visible
on the graph. The improvements for the larger photon numbers are slightly
larger, almost $1\%$ for the largest photon number calculations have been
performed for. These improvements are still very minor, and much less than the
improvements obtained by solving for the feedback.

\begin{figure}
\centering
\includegraphics[width=0.7\textwidth]{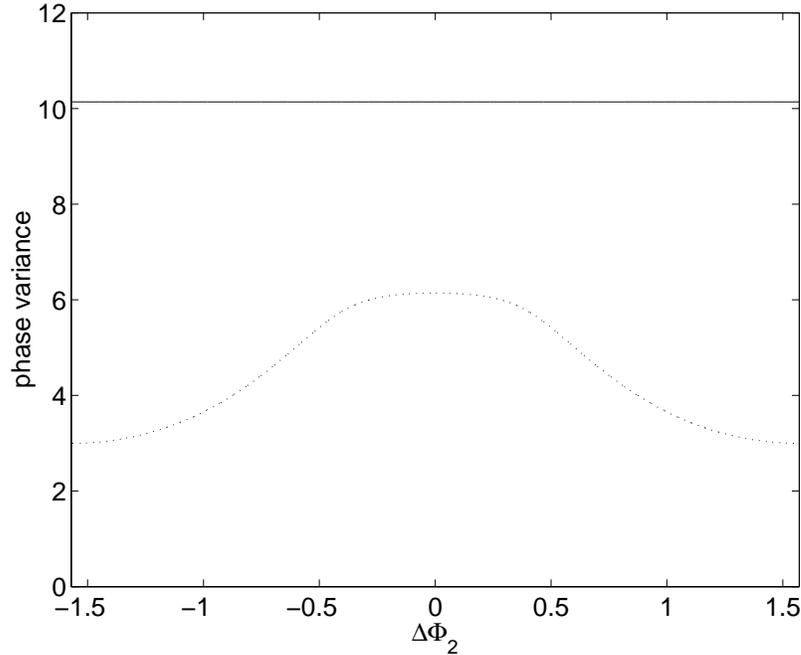}
\caption{The phase variance for the 3 photon input state of
Eq.~(\ref{except}) as a function of the second feedback phase. The phase given
here is relative to the second feedback phase given by the feedback scheme of
Sec.~\ref{feedback}. The other feedback phases are as given by this feedback
scheme. The continuous line is the canonical phase variance, and the dotted
line is the variance for the measurements.}
\label{exception}
\end{figure}

For 3 or 4 photons it was found that the feedback scheme of Sec.~\ref{feedback}
is still optimum for the numerically optimised states. For more general
states, however, this is not the case. For example, for the 3 photon state
given by
\beq
\label{except}
\ket \psi = \frac 1{\sqrt 3} \left( \ket{3/2,-3/2}_z + \ket{3/2,1/2}_z +
\ket{3/2,3/2}_z \right),
\eeq
the variation of the phase variance with the second feedback phase is as given
in Fig.~\ref{exception}. The phase variance here is much smaller when the
feedback phase is altered by $\pi$. In fact, the variance is a maximum when this
feedback phase is at the value given by the feedback scheme of
Sec.~\ref{feedback}. Not only this, but the variance is less than the canonical
variance for the entire range.

This is a strange result, as we would normally expect that it is impossible for
the phase variance for the measurements to be smaller than the canonical phase
variance. To understand the reason for this, consider the expression for
$\ip{e^{i\phi}}$. From Eq.~(\ref{expectgen}) of the introduction, for
measurements of a single mode we have
\beq
\ip{e^{i\phi}} = \sum_{n=0}^\infty \braket{\psi}{n}\braket{n+1}{\psi}H_{n,n+1}.
\eeq
The corresponding expression for interferometric measurements is
\beq
\ip{e^{i\phi}} = \sum_{\mu=-j}^{j-1} \braket{\psi}{j\mu}_y
\braket{j,\mu+1}{\psi}H_{\mu,\mu+1}.
\eeq
For canonical measurements all of the $H_{\mu,\mu+1}$ are equal to one. We would
normally expect that any smaller value of $H_{\mu,\mu+1}$ would lead to a
smaller value of $\st{\ip{e^{i\phi}}}$ (and a larger Holevo phase variance).
This is not necessarily the case, however. For example, if all the values of
$_y\braket{j\mu}{\psi}$ are positive except for one negative value for
$\mu=\mu_1$, then $\braket{\psi}{j\mu_1}_y\braket{j_1,\mu_1+1}{\psi}$ will be
negative. In that case, it is obvious that a smaller value of
$H_{\mu_1,\mu_1+1}$ will lead to a smaller variance.

In general, the POM with $H_{nm}=1$ gives the smallest variance only for states
where the phases of $_y\braket{j\mu}{\psi}$ vary as
\beq
\label{linphase}
\arg {_y\braket{j\mu}{\psi}} = \mu\varphi+\phi_0.
\eeq
As is discussed in Ref.~\cite{Hel76}, if the phases of $_y\braket{j\mu}{\psi}$
have some more arbitrary variation
\beq
\arg {_y\braket{j\mu}{\psi}} = \phi_\mu,
\eeq
then the optimum POM (for the error in the phase) is given by 
\beq
\label{optimPOM}
F(\phi) = \frac 1{2\pi} \sum_{\mu,\nu=-j}^j e^{i(\mu-\nu)\phi}
\ket{j\mu}_y\bra{j\nu} e^{i(\phi_\mu-\phi_\nu)}.
\eeq
I will call this the corrected canonical POM, to avoid confusion with the
optimum measurements based on feedback. With this POM we find
\beq
\ip{e^{i\phi}} = \sum_{\mu=-j}^{j-1} \st{\braket{\psi}{j\mu}_y
\braket{j,\mu+1}{\psi}}.
\eeq
Using this expression, the variance for the state of Eq.~(\ref{except}) is
approximately $0.6863$, which is much smaller than the variance for the
measurements. This is possible because, although the $_z\braket{j\mu}{\psi}$
are positive, the $_y\braket{j\mu}{\psi}$ are complex.

For most of the states considered in this study, the phases of
$_y\braket{j\mu}{\psi}$ vary linearly as in Eq.~(\ref{linphase}), so using the
corrected canonical POM does not change the results. The only exceptions [apart
from the state of Eq.~(\ref{except})] are: \\
1. The state $(\ket{j0}_z+\ket{j1}_z)/\sqrt 2$. This has a slightly smaller
variance for corrected canonical measurements than for the usual canonical
measurements. The results are not qualitatively changed, however, as the phase
variance still remains on the order of 1 for very large photon numbers, due to
the large peak at $\pm\pi$. This state is therefore still too poor to be useful.
\\
2. Arbitrary 2 photon states. For any 2 photon state where the phases of
$_y\braket{j\mu}{\psi}$ do not vary linearly, although the feedback scheme of
Sec.~\ref{feedback} gives a variance equal to the canonical variance, it does
not give a variance as small as that for corrected canonical measurements. For
other 2 photon states and for 1 photon states there is no change.

%% file: continue.tex
\setcounter{chapter}{6}

\chapter{Continuous Phase Measurements}
\label{continue}
Now I will consider the case of continuous adaptive measurements, where we are
continuously measuring a phase that is varying. This case is closer to what is
usually done in practice, where rather than measuring a single, fixed phase of a
pulse, a signal is transmitted in the varying phase of a continuous beam, and we
wish to measure that varying phase.

\section{Continuous Dyne Measurements on Coherent States}

Firstly I will consider the case of continuous measurements on a single mode
field. For this case it is simplest to consider coherent states.
For continuous coherent states, the coherent amplitude $\alpha$ has
a constant magnitude, but varying phase. As in the single-shot case, a
quadrature of the field is measured by combining the mode to be measured with a
large amplitude local oscillator field that is treated classically. The two
fields at the two output ports of the beam splitter are given by
\beq
b_\pm (t) = \frac 1{\sqrt 2}\left( a \pm \gamma \right),
\eeq
where $a$ is the operator for the mode to be measured and $\gamma$ is the
amplitude of the local oscillator field. The instantaneous rate of
photodetection at each photodetector is
\bqa
\ip{b_\pm\dg(t)b_\pm(t)} \!\!\!\! &=& \!\!\!\! \frac 12\ip{\left(a\dg\pm\gamma^*
\right)\left(a\pm\gamma\right)} \nn \\
\!\!\!\! &=& \!\!\!\! \frac 12\left(|\alpha|^2+|\gamma|^2 \pm 2{\rm Re}\left(
\alpha \gamma^* \right) \right).
\eqa
The signal of interest is the difference between the two photocurrents at the
two detectors. The number of photocounts at each of the detectors in the time
interval $[t,t+\delta t)$ will be denoted by $\delta N_\pm (t)$.
I will use the usual definition of the signal photocurrent
\beq
\label{defphoto}
I(t) = \lim_{\delta t \to 0 } \lim_{|\gamma| \to \infty }
\frac{\delta N_+(t)-\delta N_-(t)}{|\gamma| \delta t}.
\eeq
The expectation values of the increments $\delta N_\pm (t)$ in the infinitesimal
limit are
\bqa
\ip{dN_\pm(t)} \!\!\!\! &=& \!\!\!\! \ip{b_\pm\dg(t)b_\pm(t)} dt \nn \\
\ip{dN_\pm^2(t)} \!\!\!\! &=& \!\!\!\! \ip{dN_\pm(t)}.
\eqa
For large $|\gamma|$, $\delta N_\pm (t)$ can be approximated by the Gaussian
increments $\delta W_\pm (t)$:
\beq
\delta N_\pm(t) \approx \kappa_\pm\delta t+\sqrt{\kappa_\pm}\delta W_\pm(t),
\eeq
where
\beq
\kappa_\pm = \frac 12 \left(|\alpha|^2+|\gamma|^2\pm 2{\rm Re}\left(\alpha
\gamma^*\right)\right).
\eeq
Using this in the definition of the photocurrent (\ref{defphoto}) gives
\bqa
I(t) \!\!\!\! &=& \!\!\!\! \lim_{\delta t \to 0} \lim_{|\gamma|\to\infty}
\frac{\delta N_+(t)-\delta N_-(t)}{|\gamma| \delta t} \nn \\
\!\!\!\! &=& \!\!\!\! \lim_{\delta t \to 0}\frac{2{\rm Re}\left(\alpha
e^{-i\Phi(t)}\right)\delta t + \frac 1{\sqrt 2}\left(\delta W_+(t)-\delta W_-(t)
\right)}{\delta t} \nn \\
\!\!\!\! &=& \!\!\!\! 2{\rm Re} \left( \alpha e^{-i\Phi(t)}\right)
+ \frac{dW(t)}{dt},
\eqa
where $\Phi(t)$ is the phase of $\gamma$. The result is therefore
\beq
I(t)dt = 2{\rm Re} \left( \alpha e^{-i\Phi(t)} \right)dt + dW(t).
\eeq
This is identical to the result in the case of single-shot measurements, except
in this case the time is not scaled to the unit interval. 

In making adaptive phase measurements the phase of the local oscillator is
usually taken to be
\beq
\Phi (t) = \hat \varphi (t) + \pi/2,
\eeq
where $\hat \varphi (t)$ is some estimate of the system phase $\varphi(t)$. The
only case where the feedback phase is not based on this (for dyne measurements)
is for the corrections to the close-to-optimal measurements considered in
Ch.~\ref{optdyne}. With this, the signal becomes
\beq
I(t)dt = 2\st{\alpha} \sin \left( {\varphi (t) - \hat \varphi (t)} \right)dt +
dW(t).
\eeq

\subsection{Linear Approximation}

Provided that the estimated system phase is sufficiently close to the actual
system phase, we can make the linear approximation
\beq
\label{linearapprox}
I(t)dt = 2\st{\alpha}\left( {\varphi (t) - \hat \varphi (t)} \right)dt + dW(t).
\eeq
Rearranging this equation gives
\bqa
2\st{\alpha}\hat \varphi (t)dt + I(t)dt \!\!\!\! &=& \!\!\!\! 2\st{\alpha}
\varphi (t)dt + dW(t) \nn \\
\hat \varphi (t) + \frac{I(t)}{2\st{\alpha}} \!\!\!\! &=& \!\!\!\! \varphi(t)+
\frac{dW(t)}{2\st{\alpha}dt},
\eqa
which gives
\beq
\ip {\hat\varphi(t) + \frac{I(t)}{2|\alpha|}} = \varphi(t).
\eeq
Therefore, under the linear approximation, each data point $I(t)$
can be used to obtain an independent estimate of the phase.

Now I will denote the best phase estimate based on all the data up to time
$t$ by $\Theta(t)$. I will also denote the best phase estimate based on the
data in the infinitesimal time interval $[t,t+dt)$ by $\theta(t)$. Note that
these are the {\it best} phase estimates, in contrast to the phase estimate used
in the feedback $\hat\varphi(t)$. The phase estimates $\theta(t)$ are given by
\beq
\theta(t)=\hat\varphi(t) + \frac{I(t)}{2|\alpha|},
\eeq
which has the expectation value $\varphi(t)$ and the variance
\bqa
\label{bigvar}
\ip{\left( \theta(t) - \varphi(t) \right)^2}
\!\!\!\! &=& \!\!\!\! \ip{\left( {\frac{dW(t)}
{2\st{\alpha}dt}} \right)^2 } \nn \\
\!\!\!\! &=& \!\!\!\! \frac 1{4\st{\alpha}^2 dt}.
\eqa
Here the simple definition of the variance has been used, rather than the Holevo
phase variance, because we are using the linear approximation.

If we were considering a measurement over the time interval $[0,1]$, during
which there are $\nb=\st \alpha ^2$ photons from the signal mode, then we would
be effectively averaging over $1/dt$ estimates of the phase, each with
a variance given by Eq.~(\ref{bigvar}). As usual with averages, the variance
in the mean is the variance of each data point divided by the number of data
points. This means that the variance in the final phase estimate is
\bqa
  \frac{{{1 \mathord{\left/
 {\vphantom {1 {\left( {4\left| \alpha  \right|^2 dt} \right)}}} \right.
 \kern-\nulldelimiterspace} {\left( {4\left| \alpha  \right|^2 dt} \right)}}}}
{{{1 \mathord{\left/
 {\vphantom {1 {dt}}} \right.
 \kern-\nulldelimiterspace} {dt}}}} \!\!\!\! &=& \!\!\!\! \frac{1}
{{4\left| \alpha  \right|^2 }} \nn \\
   \!\!\!\! &=& \!\!\!\! \frac 1{4\nb},
\eqa
which is the standard result for a coherent state under adaptive phase
measurements.

In this case we want to consider a measurement that is continued indefinitely.
If the system phase is not varied, then the variance in the phase estimate will
just go down indefinitely. In that case, however, no information is transmitted
beyond a single real number. What we wish to do is vary the system phase in
order to transmit information, as is the case for FM radio.

Next is the question of how the system phase $\varphi(t)$ will be varied.
The simplest way of varying the system phase is to vary it stochastically via
Wiener increments. I will therefore take the variation in the system phase to
be
\beq
\label{varyphase}
\varphi (t+dt) = \varphi (t) + \kappa dW'(t).
\eeq
This Wiener increment is independent from that used previously for the
photocurrent, as indicated by the prime.

It is clear that the most recent data will be least affected by the
variation in the system phase, and the data from further and further back will
be less and less accurate. To quantify this, note that when the current system
phase $\varphi(t)$ is used as a reference, the variance in the system phase
at some previous time $t'$ is $\kappa^2 (t-t')$. As the phase estimate
$\theta(t')$ based on the time interval $[t',t'+dt')$ is an estimate of the
system phase at time $t'$, when it is considered relative to the current system
phase its variance will be increased by $\kappa^2 (t-t')$.
Therefore the total variance in each individual phase estimate $\theta(t')$ is
\beq
\label{badvar}
\frac{1}
{{4\left| \alpha  \right|^2 dt}} + \kappa^2 ( t - t' ).
\eeq

In order to determine the current best phase estimate $\Theta(t)$, we would like
to form a weighted average of each of the individual phase estimates
$\theta(t')$. If this is done in the usual way for weighted averages, using
the variance given by (\ref{badvar}), the results obtained are ridiculous, as
the contribution to the variance from the variation in the system phase is
infinitesimal as compared to the quantum noise. The problem is that the
variation in the system phase makes the error in the phase estimates
$\theta(t')$ correlated, so the contribution from the variation in the system
phase is infinitesimal over infinitesimal time intervals, but becomes
significant over finite time intervals. This means that the usual method of
performing weighted averages will not work.

As an alternative approach, we can consider just the weighting of the latest
phase estimate $\theta(t)$ as compared to the phase estimate from all the
previous data $\Theta(t)$. In this case $\theta(t)$ has no variance from
the variation in the system phase, and so it is uncorrelated with the previous
phase estimate and we can use a weighted average in the usual way.

The equilibrium value of the variance of $\Theta(t)$, with all the individual
phase estimates correctly weighted, will be denoted by $\Delta\Theta^2$. After a
time $dt$ the phase variance of $\Theta(t)$ with respect to the new system phase
$\varphi(t+dt)$, i.e.
\beq
\ip {\left( {\Theta (t) - \varphi (t+dt)} \right)^2 },
\eeq
will be $\Delta\Theta^2+\kappa^2dt$. The variance in the phase estimate from the
latest time interval, $\theta(t)$, will be given by Eq.~(\ref{bigvar}). If
we take a weighted average of $\Theta(t)$ and $\theta(t)$, then the
contributions from each of the phase estimates from the individual time
intervals should be correctly weighted, and the variance in the weighted
average should be the equilibrium value, $\Delta\Theta^2$. This implies that
\beq
{\frac{1}{{\Delta\Theta ^2 + \kappa^2 dt}} + 4\st{\alpha}^2 dt} =
\frac{1}{\Delta\Theta ^2}.
\eeq
This expression can be used to determine the equilibrium value of the phase
variance. Solving for $\Delta\Theta^2$ gives
\beq
\label{derive22}
\Delta\Theta ^2  = \frac{\kappa} {{2\left| \alpha  \right|}}.
\eeq
Thus we find that for continuous measurements the phase variance scales as
$1/|\alpha|$ rather than $1/|\alpha|^2$.

Showing explicitly how the weighted average is performed,
\beq
\Theta (t+dt) = \frac{{\left( {4\st{\alpha}^2 dt} \right)\theta(t) + \frac{1}
{{\Delta\Theta ^2 + \kappa^2 dt}}\Theta (t)}}{{{1 /{\Delta\Theta ^2 }}}}.
\eeq
Simplifying this gives
\beq
\Theta \left( {t + dt} \right) = \left( {2\st{\alpha}\kappa\,dt} \right)
\theta(t)+ \left( {1 - 2\st{\alpha}\kappa\,dt} \right)\Theta \left( t \right).
\eeq
In terms of the increment in the phase estimate this is
\beq
d\Theta (t) + \left( {2\st{\alpha} \kappa dt} \right)\Theta \left( t \right) =
\left( {2\st{\alpha} \kappa dt} \right)\theta(t).
\eeq
Solving this we find that
\bqa
d\Theta(t)e^{2\st{\alpha}\kappa t}+\left( 2\st{\alpha}\kappa\,dt \right)\Theta
(t)e^{2\st{\alpha}\kappa t} \!\!\!\! &=& \!\!\!\! \left( 2\st{\alpha}\kappa\,dt
\right)\theta(t)e^{2\st{\alpha}\kappa t} \nn \\
d\left(\Theta(t)e^{2\st{\alpha}\kappa t}\right) \!\!\!\! &=& \!\!\!\! \left(
2\st{\alpha}\kappa \right)\theta(t)e^{2\st{\alpha}\kappa t} dt \nn \\
\Theta(t) \!\!\!\! &=& \!\!\!\! 2\st{\alpha}\kappa\int\limits_{-\infty}^t
{\theta (s)e^{2\st{\alpha}\kappa(s-t)} ds}.
\eqa
Therefore this method corresponds to a simple negative exponential scaling of
the weighting.

We can also consider a more general negative exponential scaling given by
\beq
\label{negativeexp}
\Theta (t) = \chi \int\limits_{-\infty}^t {\theta(s)e^{\chi (s-t)} ds}.
\eeq
Note that with this more general scaling, $\Theta(t)$ is no longer necessarily
the best phase estimate. For most of the remainder of this chapter, $\Theta(t)$
will be used in this more general sense, rather than as specifically the best
phase estimate. The best phase estimate will be found by finding the optimum
value of $\chi$. Taking the derivative of this expression with respect to time
gives
\bqa
d\left( {\Theta (t)e^{\chi t} } \right) \!\!\!\! &=& \!\!\!\! \chi \theta(t)
e^{\chi t} dt \nn \\
d\Theta (t) \!\!\!\! &=& \!\!\!\! \chi dt\left(\theta(t) - \Theta (t)\right)
\nn \\ \Theta (t+dt) \!\!\!\! &=& \!\!\!\! \chi dt\theta(t)
+ \left(1 - \chi dt \right)\Theta (t).
\eqa
This means that this method is again a weighted average, except with a weighting
that is not optimum. If we find the variance of both sides of this equation we
obtain
\bqa
{\rm var} \left( {\Theta (t+dt)} \right) \!\!\!\! &=& \!\!\!\! \left( {\chi dt}
\right)^2 {\rm var} \left({\theta(t)} \right) + \left( {1 - \chi dt} \right)^2
{\rm var} \left(\Theta(t)\right) \nn \\
\Delta\Theta^2  \!\!\!\! &=& \!\!\!\! \frac{{\left( {\chi dt} \right)^2 }}
{{4\left| \alpha  \right|^2 dt}} + \left( {1 - 2\chi dt} \right)\left(\Delta
\Theta^2 + \kappa^2 dt \right) \nn \\
\Delta\Theta^2  \!\!\!\! &=& \!\!\!\! \frac{{\chi ^2 dt}}{{4\left| \alpha
\right|^2 }} + \Delta\Theta^2 + \kappa^2 dt - 2\chi dt\Delta\Theta^2.
\eqa
Solving for $\Delta\Theta^2$ gives
\beq
\label{sigma2}
\Delta\Theta^2 = \frac{\chi}{8\st{\alpha}^2}
+ \frac{\kappa^2}{2\chi}.
\eeq
The optimum value of $\chi$ can be verified from this equation. Taking the
derivative with respect to $\chi$ gives
\beq
\frac{\partial }
{{\partial \chi }}\left( {\Delta\Theta^2 } \right) = \frac{1}
{8\left| \alpha  \right|^2} - \frac{\kappa^2}{2\chi^2}.
\eeq
For the variance to be minimised this must be zero, so
\bqa
\frac{1}{8\left| \alpha  \right|^2} \!\!\!\! &=& \!\!\!\! \frac{\kappa^2}
{2\chi^2} \nn \\
\chi  \!\!\!\! &=& \!\!\!\! 2\st{\alpha} \kappa.
\eqa
This is the exponential constant which was found directly. Substituting this
value of $\chi$ into the expression for the variance gives
\bqa
\Delta\Theta^2 \!\!\!\! &=& \!\!\!\! \frac{2\st{\alpha}\kappa}{8\st{\alpha}^2}+
\frac{\kappa^2}{4\st{\alpha}\kappa} \nn \\
\!\!\!\! &=& \!\!\!\! \frac{\kappa}{2\st{\alpha}},
\eqa
which is the result found for the phase variance in Eq.~(\ref{derive22}).

\subsection{Exact Case}
\label{exactcase}
The results of the previous section are all using the linear approximation
(\ref{linearapprox}). Although this approximation is very useful for obtaining
the asymptotic value of the variance, it does not directly tell us what to do
in the exact case. In the exact case for single-shot measurements (see
Sec.~\ref{introadapt}), rather than forming independent phase estimates from
each time interval $dt$ and then averaging them, we determine $A_v$ and $B_v$,
and the phase estimate is given by
\beq
\label{phaseest}
\Theta (v) = \arg \left( {vA_v  + B_v A_v^* } \right).
\eeq
Therefore the average phase estimate with exponential weighting does not make
much sense in this case. In addition the intermediate phase estimate must be
considered. An alternative approach is to use an exponential weighting in
determining $A_v$ and $B_v$, and then use these to determine the phase estimate.
Specifically, I will replace the definitions of $A_v$ and $B_v$,
\bqa
A_v \!\!\!\! &=& \!\!\!\! \int\limits_0^v {e^{i\Phi } I(u)du} \nn \\
B_v \!\!\!\! &=& \!\!\!\!  - \int\limits_0^v {e^{2i\Phi } du},
\eqa
by
\bqa
\label{defineAB}
A_t \!\!\!\! &=& \!\!\!\! \int\limits_{-\infty}^t e^{\chi(u-t)}
e^{i\Phi } I\left( u \right)du \nn \\
B_t \!\!\!\! &=& \!\!\!\! - \int\limits_{-\infty}^t e^{\chi(u-t)}e^{2i\Phi} du.
\eqa
Then $\arg A_t$ can still be used as the intermediate phase estimate. I will
not consider any better intermediate phase estimates here, as these only
give very small improvements over the mark II case for coherent states. To find
a phase estimate to use for $\Theta(t)$, we can use a similar approach to that
used in Ref.~\cite{semiclass}. If the system phase is constant, then we find
\bqa
\label{expandA}
A_t \!\!\!\! &=& \!\!\!\! \int\limits_{-\infty}^t e^{\chi(u-t)}e^{i\Phi} I(u)
du \nn \\
\!\!\!\! &=& \!\!\!\! \int\limits_{-\infty}^t e^{\chi(u-t)}e^{i\Phi}\left[
\left(\alpha e^{-i\Phi} + \alpha^* e^{i\Phi} \right)du + dW(u) \right] \nn \\
\!\!\!\! &=& \!\!\!\! \alpha \int\limits_{-\infty}^t e^{\chi(u-t)} du +\alpha^*
\int\limits_{-\infty}^t e^{\chi(u-t)} e^{2i\Phi} du + \int\limits_{-\infty}^t
e^{\chi(u-t)} e^{i\Phi} dW(u) \nn \\
\!\!\!\! &=& \!\!\!\! \frac{\alpha}{\chi} - \alpha^* B_t + i\sigma_t.
\eqa
where
\beq
\sigma_t  = \int\limits_{-\infty}^t e^{\chi(u-t)} e^{i\left(\Phi-\pi/2 \right)}
dW(u).
\eeq
This result is analogous to the result (\ref{expandA0}) for the case of
single-shot measurements, except with $v$ replaced with $1/\chi$.
Note that from this derivation it naturally emerges that we should use the same
exponential scaling for $B_t$ as for $A_t$. From Eq.~(\ref{expandA}) it can be
shown that
\beq
A_t + \chi B_t A_t^* = \alpha \left(\frac{1}{\chi}- \chi \st{B_t}^2 \right) +
i\sigma_t - i\chi B_t \sigma_t^*.
\eeq
This means that
\beq
\ip{A_t + \chi B_t A_t^* } \approx \alpha \left( \frac{1}{\chi}-\chi\st{B_t}^2
\right).
\eeq
Similarly to the single-shot case this is not necessarily exact if the local
oscillator phase is dependent on the measurement record, but it should still be
approximately true. Therefore the phase estimate that will be used here is
\beq
\Theta(t) = \arg (A_t+\chi B_tA_t^*).
\eeq
Note that the factor of $\chi$ makes sense, as the time over which
the data is used is effectively $1/\chi$. Similarly to the single-shot case, I
will define the variable $C_t = A_t + \chi B_t A_t^*$,
so $\Theta(t)=\arg C_t$. The above derivation is not exact if the
system phase is not constant; however, $\arg C_t$ should still be a good
estimator for the phase.

A differential equation for the feedback phase can be determined in a similar
way as in Ref.~\cite{semiclass}. Using Eq.~(\ref{defineAB}), we can determine
the increment in $A_t$:
\bqa
e^{\chi t} A_t \!\!\!\! &=& \!\!\!\! \int\limits_{-\infty }^t {e^{\chi u}
e^{i\Phi } I\left( u \right)du} \nn \\
d\left( {e^{\chi t} A_t } \right) \!\!\!\! &=& \!\!\!\! e^{\chi t} e^{i\Phi}
I(t)dt \nn \\
dA_t \!\!\!\! &=& \!\!\!\! e^{i\Phi} I(t)dt - \chi A_t dt.
\eqa
Taking the local oscillator phase to be
\beq
\Phi (t) = \arg A_t  + \frac{\pi}2,
\eeq
as in the case of mark II measurements, we find that
\beq
dA_t  = i\frac{{A_t }}
{{\left| {A_t } \right|}}I(t)dt - \chi A_t dt.
\eeq
So the magnitude of $A_t$ varies as
\bqa
d\st{A_t}^2 \!\!\!\! &=& \!\!\!\! A_t^* \left( dA_t \right) + \left( dA_t^*
\right)A_t + \left( dA_t^* \right)\left( dA_t \right) \nn \\
\!\!\!\! &=& \!\!\!\! A_t^* \left( i\frac{A_t}{\st{A_t}}I(t)dt - \chi A_t dt
\right) + \left( -i\frac{A_t^*}{\st{A_t}}I(t)dt - \chi A_t^* dt \right)A_t + dt
\nn \\
\!\!\!\! &=& \!\!\!\! \left(1 - 2\chi \st{A_t}^2 \right)dt.
\eqa
This demonstrates that, rather than increasing linearly as in the standard case,
$\st {A_t}$ increases up to an equilibrium value given by
\beq
\left| {A_t } \right|^2  = \frac 1{2\chi}.
\eeq

Using this result, the increment in the feedback phase is
\bqa
d\Phi(t) \!\!\!\! &=& \!\!\!\! {\rm Im} \left[ d\ln A_t \right] \nn \\
\!\!\!\! &=& \!\!\!\! {\rm Im}\left[\frac{dA_t}{A_t}-\frac{\left(dA_t\right)^2}
{2A_t^2} \right] \nn \\
\!\!\!\! &=& \!\!\!\! {\rm Im}\left[\frac{i\frac{A_t}{\st{A_t}}I(t)dt - \chi
A_t dt}{A_t} + \frac{\frac{A_t^2}{\st{A_t}^2}dt}{2A_t^2} \right] \nn \\
\!\!\!\! &=& \!\!\!\! \frac{I(t)dt}{\st{A_t}} \nn \\
\!\!\!\! &=& \!\!\!\! \sqrt{2\chi} I(t)dt.
\eqa
Therefore the feedback for this case is actually much simpler than for the
single-shot case. The feedback phase just changes linearly with the signal, and
there is no $\sqrt t$ scaling as there is in the standard case.

Using this result gives the stochastic differential equation for the phase
estimate $\hat \varphi (t)$ as
\bqa
d\hat\varphi(t) \!\!\!\! &=& \!\!\!\! \sqrt{2\chi} \left[I(t)dt\right] \nn \\
\!\!\!\! &=& \!\!\!\! \sqrt{2\chi} \left[2\st{\alpha}\sin (\varphi(t) - \hat
\varphi (t) )dt + dW(t) \right].
\eqa
Unlike the standard case, no change of variables is required, as there is no
factor of $\sqrt t$. Making a linear approximation gives
\beq
d\hat \varphi(t) = \sqrt{2\chi} \left[ 2\st{\alpha}(\varphi(t) - \hat \varphi
(t))dt + dW(t) \right].
\eeq
Rearranging this gives
\bqa
d\hat \varphi (t) + 2\st{\alpha}\sqrt{2\chi} \hat \varphi(t)dt
\!\!\!\! &=& \!\!\!\! 2\st{\alpha}\sqrt{2\chi} \varphi(t)dt + \sqrt{2\chi} dW
(t) \nn \\
d\left[e^{2\st{\alpha}\sqrt{2\chi}t} \hat \varphi(t)\right]
\!\!\!\! &=& \!\!\!\! e^{2\st{\alpha}\sqrt{2\chi}t} \left[2\st{\alpha}
\sqrt{2\chi} \varphi (t)dt + \sqrt{2\chi} dW(t) \right]. \nn
\eqa
Integrating then gives the solution as
\beq
\hat \varphi (t) = \sqrt{2\chi} \int\limits_{-\infty}^t e^{2\st{\alpha}
\sqrt{2\chi}(u-t)} \left[ 2\st{\alpha}\varphi(u)du + dW(u) \right].
\eeq

If the phase is measured relative to the current system phase, then
\beq
\varphi(u) = -\kappa\int\limits_u^t dW'(s).
\eeq
Using this, the solution for the phase estimate is
\beq
\label{phestsol}
\hat \varphi(t) = \sqrt{2\chi} \int\limits_{-\infty}^t e^{2\st{\alpha}
\sqrt{2\chi}(u-t)} \left[dW(u)-2\st{\alpha}\kappa\int\limits_u^t dW'(s) du
\right].
\eeq
The variance in this phase estimate will be
\bqa
\label{bigderiv}
\ip{\hat \varphi^2 (t)} \!\!\!\! &=& \!\!\!\! 2\chi\ip{\int\limits_{-\infty}^t
du_1 \int\limits_{-\infty }^t du_2 e^{2\st{\alpha}\sqrt{2\chi} (u_1+u_2-2t)}
4\st{\alpha}^2 \kappa^2 \int\limits_{u_1 }^t \int\limits_{u_2}^t dW'(s_2)
dW'(s_1)} \nn \\ 
&& + 2\chi \ip{\int\limits_{-\infty}^t \int\limits_{-\infty}^t e^{2\st{\alpha}
\sqrt{2\chi}(u_1+u_2-2t)} dW(u_2) dW(u_1)} \nn \\ 
\!\!\!\! &=& \!\!\!\! 8\chi \st{\alpha}^2 \kappa^2 \int\limits_{-\infty}^t du_1
\int\limits_{-\infty}^t du_2 e^{2\st{\alpha}\sqrt{2\chi}(u_1+u_2-2t)}
\int\limits_{\max (u_1,u_2)}^t ds + 2\chi \int\limits_{-\infty}^t
e^{4\st{\alpha}\sqrt{2\chi}(u-t)} du \nn \\ 
\!\!\!\! &=& \!\!\!\! 8\chi \st{\alpha}^2 \kappa^2 \int\limits_{-\infty}^t du_1
\int\limits_{-\infty}^{u_1} du_2 e^{2\st{\alpha}\sqrt{2\chi}(u_1+u_2-2t)}
\int\limits_{u_1}^t ds \nn \\
&& + 8\chi \st{\alpha}^2 \kappa^2 \int\limits_{-\infty}^t du_1
\int\limits_{u_1}^t du_2 e^{2\st{\alpha}\sqrt{2\chi}(u_1+u_2-2t)}
\int\limits_{u_2}^t ds + 2\chi \left[\frac{e^{4\st{\alpha}\sqrt{2\chi}(u-t)}}
{4\st{\alpha}\sqrt{2\chi}}\right]_{-\infty}^t \nn \\
\!\!\!\! &=& \!\!\!\! 8\chi \st{\alpha}^2 \kappa^2 \int\limits_{-\infty}^t du_1
\int\limits_{-\infty}^{u_1} e^{2\st{\alpha}\sqrt{2\chi}(u_1+u_2-2t)}(t-u_1)du_2
\nn \\ 
&& + 8\chi \st{\alpha}^2 \kappa^2 \int\limits_{-\infty}^t du_1
\int\limits_{u_1}^t e^{2\st{\alpha}\sqrt{2\chi}(u_1+u_2-2t)}(t-u_2)du_2+
\frac{\sqrt{2\chi}} {4\st{\alpha}} \nn \\ 
\!\!\!\! &=& \!\!\!\! 16\chi \st{\alpha}^2 \kappa^2 \int\limits_{-\infty}^t du_1
\int\limits_{-\infty}^{u_1} e^{2\st{\alpha}\sqrt{2\chi}(u_1+u_2-2t)}(t-u_1)du_2
+\frac{\sqrt{2\chi}} {4\st{\alpha}} \nn \\
\!\!\!\! &=& \!\!\!\! \frac{16\chi \st{\alpha}^2 \kappa^2}{2\st{\alpha}
\sqrt{2\chi}} \int\limits_{-\infty}^t e^{4\st{\alpha}\sqrt{2\chi}(u_1-t)}(t-u_1)
du_1+\frac{\sqrt{2\chi}} {4\st{\alpha}} \nn \\
\!\!\!\! &=& \!\!\!\! \frac{16\chi \st{\alpha}^2 \kappa^2}{2\st{\alpha}
\sqrt{2\chi}\left(4\st{\alpha}\sqrt{2\chi}\right)^2}+\frac{\sqrt{2\chi}}
{4\st{\alpha}} \nn \\
\!\!\!\! &=& \!\!\!\! \frac{\kappa^2}{4\st{\alpha}\sqrt{2\chi}}+
\frac{\sqrt{2\chi}}{4\st{\alpha}}.
\eqa
This result for the variance of the intermediate phase estimate is quite
different from that for the variance of $\Theta(t)$ given in Eq.~(\ref{sigma2}).
Of particular interest is the fact that the contribution due to the variation in
the system phase is not simply $\kappa^2/(2\chi)$, as is usually the case.

At first it would seem that the method used to find the mark II phase
variance in \cite{semiclass} would be applicable here also. It turns out that
this method is not useable, because when the system phase varies it is not
possible to separate the different terms as in Eq.~(\ref{expandA}). A more
promising way is to use a method similar to that in Eq.~(\ref{bigderiv}). The
phase estimate can be simplified to
\bqa
\Theta(t) \!\!\!\! &=& \!\!\!\! \arg\left( A_t + \chi B_t A_t^* \right) \nn \\
\!\!\!\! &=& \!\!\!\! \arg A_t + \arg\left( 1+\chi B_t A_t^*/A_t \right) \nn \\
\!\!\!\! &=& \!\!\!\! \hat \varphi(t) + \arg\left( 1+\chi
e^{-2i\hat \varphi(t)} B_t \right).
\eqa
Expressing $B_v$ as an integral gives
\beq
\Theta(t) = \hat\varphi(t) + \arg \left( 1+\chi e^{-2i\hat\varphi(t)}
\int\limits_{-\infty}^t e^{\chi (u-t)} e^{2i\hat\varphi(u)} du \right).
\eeq
Expanding the exponentials to first order we get
\bqa
\Theta(t) \!\!\!\! & \approx & \!\!\!\! \hat \varphi(t) + \arg \left( 1+\chi
\left( 1-2i\hat\varphi(t) \right)\int\limits_{-\infty}^t e^{\chi (u-t)} \left(
1+2i\hat\varphi(u) \right)du \right) \nn \\
\!\!\!\! &=& \!\!\!\! \hat \varphi(t) + \arg \left[ 1 + \left( 1-
2i\hat \varphi(t) \right) \left( 1 + 2i\chi \int\limits_{-\infty}^t
e^{\chi (u-t)} \hat \varphi(u)du \right) \right] \nn \\
\!\!\!\! & \approx & \!\!\!\! \hat \varphi(t) + \arg \left( 2-2i\hat \varphi(t)
+2i\chi \int\limits_{-\infty}^t e^{\chi (u-t)} \hat \varphi(u)du \right) \nn \\
\!\!\!\! &=& \!\!\!\! \hat \varphi(t) + \arg \left( 1-i\hat \varphi(t)
+i\chi \int\limits_{-\infty}^t e^{\chi (u-t)} \hat \varphi(u)du \right) \nn \\
\!\!\!\! & \approx & \!\!\!\! \hat \varphi(t) - \hat \varphi(t) + \chi
\int\limits_{-\infty}^t e^{\chi (u-t)} \hat \varphi(u)du \nn \\
\!\!\!\! &=& \!\!\!\! \chi \int\limits_{-\infty}^t \hat \varphi(u)
e^{\chi (u-t)} du.
\eqa
This demonstrates that the mark II phase estimate is approximately a weighted
average of the intermediate phase estimates, just as in the standard case it is
approximately a normal average. Note also the similarity of this result to the
result for the linear case (\ref{negativeexp}). Unfortunately the simple
technique used in the linear case cannot be applied here. The problem here
is that the phase estimates $\hat \varphi(t)$ are based on the previous data,
not just on the data from the infinitesimal time interval $[t,t+dt)$. This
means that $\hat \varphi(t)$ is not independent of $\Theta(t)$, and we
therefore cannot use the simple techniques based on weighted averages.

Using the result (\ref{phestsol}) for the intermediate phase estimate gives
\beq
\Theta(t) \approx \sqrt{2\chi^3} \int\limits_{-\infty }^t e^{\chi(v-t)} \left[
\int\limits_{-\infty}^v e^{2\st{\alpha}\sqrt{2\chi}(u-v)} \left(dW(u)
-2\st{\alpha}\kappa \int\limits_u^t dW'(s) du \right) \right]dv.
\eeq
Note that the integral for the system phase variation is taken up to time $t$,
rather than the time of the intermediate phase estimate. This is because the
phase is measured relative to the current system phase, rather than the system
phase at the time of that intermediate phase estimate.

In order to determine the variance, we need to evaluate
\bqa
\label{evalbig}
&&\!\!\!\!\!\!\!\! \ip{\Theta^2(t)}=2\chi^3\ip{\int\limits_{-\infty}^t dv_1
\int\limits_{-\infty}^t dv_2 e^{\chi (v_1+v_2-2t)} \int\limits_{-\infty}^{v_1}
dW(u_1)\int\limits_{-\infty}^{v_2}dW(u_2)e^{2\st{\alpha}\sqrt{2\chi}(u_1+u_2
-v_1-v_2)}} \nn \\
&&\!\!\!\!\!\!\!\!\!\!\!\! + 8\chi^3 \st{\alpha}^2 \kappa^2 \ip{\int
\limits_{-\infty}^t \!\! dv_1 \int\limits_{-\infty}^t \!\! dv_2e^{\chi (v_1+v_2-
2t)}\int\limits_{-\infty}^{v_1} \!\! du_1 \int\limits_{-\infty}^{v_2} \!\! du_2
e^{2\st{\alpha}\sqrt{2\chi} (u_1+u_2-v_1-v_2)} \int\limits_{u_1}^t \!\! dW'(s_1)
\int\limits_{u_2}^t \!\! dW'(s_2)}. \nn \\
\eqa
This is a lengthy calculation, and is performed in Appendix \ref{dercont}. The
result found is
\beq
\label{sameresult}
\ip{\Theta^2 (t)} \approx \frac{\chi}{8\st{\alpha}^2} + \frac{\kappa^2}{2\chi}.
\eeq
This is exactly the same result as for the simplified linear case
(\ref{sigma2}), and the minimum phase variance is
\beq
\ip{\Theta^2 (t)}_{\min} \approx \frac \kappa {2\st{\alpha}}
\eeq
for
\beq
\chi = 2\st{\alpha}\kappa.
\eeq

\section{Continuous Heterodyne Measurements}

In order to determine how much of an improvement feedback gives for continuous
measurements, I will compare it with the case of continuous heterodyne
measurements. For heterodyne measurements on a pulsed coherent state, the
introduced phase variance is equal to the intrinsic phase variance. This
indicates that the first term in Eq.~(\ref{sameresult}) should be double for the
heterodyne case, so the phase variance is
\beq
\label{phvar79}
\ip{\Theta^2(t)} \approx \frac{\chi}{4\st \alpha ^2} + \frac{\kappa^2}{2\chi}.
\eeq

This can be shown more rigorously using a similar technique to that used in
\cite{semiclass}. Expanding $A_v$ gives
\bqa
A_t \!\!\!\! &=& \!\!\!\! \int\limits_{-\infty}^t e^{\chi(u-t)} e^{i\Phi(u)}
I(u)du \nn \\
\!\!\!\! &=& \!\!\!\! \int\limits_{-\infty}^t e^{\chi(u-t)} e^{i\Phi(u)}
\left[ \left( \alpha e^{-i\Phi} + \alpha^* e^{i\Phi} \right)du + dW(u) \right]
\nn \\
\!\!\!\! &=& \!\!\!\! \st{\alpha}\int\limits_{-\infty}^t e^{\chi(u-t)}
e^{i\varphi(u)} du + \st{\alpha}\int\limits_{-\infty}^t e^{\chi(u-t)}
e^{2i\Phi(u) - i\varphi(u)} du + \int\limits_{-\infty}^t e^{\chi(u-t)}
e^{i\Phi(u)} dW(u).
\eqa
For the heterodyne case, the local oscillator phase $\Phi(t)$ varies very
rapidly, so the second term above will be very small. This means that $A_t$
simplifies to
\beq
A_t = \st \alpha \int\limits_{-\infty}^t e^{\chi(u-t)} e^{i\varphi(u)} du +
i\sigma_t.
\eeq
Since $B_v$ is negligible, the phase estimate $\Theta(t)$ simplifies to
\beq
\Theta(t) = \arg A_v.
\eeq
As above, the phase will be measured relative to the current system phase. In
the limit of small phase variance, the system phase does not vary significantly
during the time $1/\chi$, so we can take the linear approximation, giving
\bqa
A_t \!\!\!\! & \approx & \!\!\!\! \st{\alpha}\int\limits_{-\infty}^t
e^{\chi(u-t)} (1+i\varphi(u))du + i\sigma _t \nn \\
\!\!\!\! &=& \!\!\!\! \frac{\st{\alpha}}{\chi} + i\st{\alpha}
\int\limits_{-\infty}^t e^{\chi(u-t)}\varphi(u)du + i\sigma_t .
\eqa
Using this the phase estimate is
\bqa
\Theta(t) \!\!\!\! &=& \!\!\!\! {\rm Im} \left[ \log \left( \frac{\st{\alpha}}
{\chi} + i\st{\alpha}\int\limits_{-\infty}^t e^{\chi (u-t)} \varphi(u)du +
i\sigma _t \right) \right] \nn \\
\!\!\!\! &=& \!\!\!\! {\rm Im} \left[\log\left( 1+i\chi \int\limits_{-\infty}^t
e^{\chi(u-t)} \varphi(u)du + i\chi \sigma_t/\st{\alpha} \right) \right] \nn \\
\!\!\!\! & \approx & \!\!\!\! {\rm Im} \left[ i\chi \int\limits_{-\infty}^t
e^{\chi(u-t)} \varphi(u)du + i\chi \sigma_t/\st{\alpha} \right].
\eqa
In the last line the linear approximation has again been used. Further
evaluating this gives
\bqa
\Theta(t) \!\!\!\! &=& \!\!\!\! \chi \int\limits_{-\infty}^t e^{\chi(u-t)}
\varphi(u)du + \frac{\chi}{2\st{\alpha}}(\sigma_t + \sigma_t^*) \nn \\
\!\!\!\! &=& \!\!\!\! - \kappa\chi \int\limits_{-\infty}^t du\,e^{\chi(u-t)}
\int\limits_u^t dW'(s) + \frac{\chi}{2\st{\alpha}}(\sigma_t+\sigma_t^*).
\eqa
The variance is therefore
\beq
\ip{\Theta^2(t)} = \kappa^2 \chi ^2 \ip{\int\limits_{-\infty}^t du_1
\int\limits_{-\infty}^t du_2 e^{\chi(u_1+u_2-2t)} \int\limits_{u_1}^t dW'(s_1)
\int\limits_{u_2}^t dW'(s_2)} + \frac{\chi ^2}{4\st{\alpha}^2} \ip{(\sigma_t
+\sigma_t^*)^2}.
\eeq
The first term here can be evaluated to give
\bqa
\label{bigder37}
&&\ip{\int\limits_{-\infty}^t du_1 \int\limits_{-\infty}^t
du_2 e^{\chi(u_1+u_2-2t)} \int\limits_{u_1}^t dW'(s_1)\int\limits_{u_2}^t
dW'(s_2)} \nn \\
\!\!\!\! &=& \!\!\!\! \int\limits_{-\infty}^t ds_1 \int\limits_{-\infty}^t ds_2
e^{\chi (s_1+s_2-2t)} \int\limits_{\max (s_1,s_2)}^t du \nn \\
\!\!\!\! &=& \!\!\!\! \int\limits_{-\infty}^t ds_1 \int\limits_{-\infty}^{s_1}
ds_2 e^{\chi(s_1+s_2-2t)} \int\limits_{s_1}^t du + \int\limits_{-\infty}^t ds_1
\int\limits_{s_1}^t ds_2 e^{\chi(s_1+s_2-2t)} \int\limits_{s_2}^t du \nn \\
\!\!\!\! &=& \!\!\!\! \int\limits_{- \infty }^t ds_1 \int\limits_{-\infty}^{s_1}
ds_2 e^{\chi(s_1+s_2-2t)}(t-s_1) + \int\limits_{-\infty}^t ds_1
\int\limits_{s_1}^t ds_2 e^{\chi(s_1+s_2-2t)} (t-s_2) \nn \\
\!\!\!\! &=& \!\!\!\! 2\int\limits_{- \infty }^t ds_1
\int\limits_{-\infty}^{s_1} ds_2 e^{\chi(s_1+s_2-2t)}(t-s_1) \nn \\
\!\!\!\! &=& \!\!\!\! \frac 2\chi\int\limits_{-\infty}^t ds_1
e^{2\chi(s_1-t)}(t-s_1) \nn \\
\!\!\!\! &=& \!\!\!\! \frac 1{2\chi^3}.
\eqa
In addition, it is easy to show that
\beq
\ip{\sigma_t^2} = -\int\limits_{-\infty}^t e^{2\chi(u-t)}e^{2i\Phi(u)}du.
\eeq
Note that $\ip{\sigma_t^2}\ne B_t$, so these variables are not completely
analogous to those defined in the single-shot case. Nevertheless, as $\Phi$ is
rotating rapidly in the heterodyne case, we should still find that
$\ip{\sigma_t^2}\approx 0$. Similarly, evaluating $\lr{\st{\sigma_t}^2}$ gives
\bqa
\ip{\st{\sigma_t}^2} \!\!\!\! &=& \!\!\!\! \int\limits_{-\infty}^t
e^{\chi(u_1+u_2-2t)} e^{i(\Phi(u_1)-\Phi(u_2))} dW(u_1)dW(u_2) \nn \\
\!\!\!\! &=& \!\!\!\! \int\limits_{-\infty}^t e^{2\chi(u-t)} du \nn \\
\!\!\!\! &=& \!\!\!\! \frac 1{2\chi}.
\eqa
Using these results, as well as Eq.~(\ref{bigder37}), the variance is
\bqa
\ip{\Theta^2(t)}\!\!\!\! &=& \!\!\!\! \frac{\kappa^2}{2\chi} + \frac{\chi^2}
{4\st{\alpha}^2} \ip{\sigma_t^2 + 2\st{\sigma_t}^2 + \sigma_t^{*2}} \nn \\
\!\!\!\! &=& \!\!\!\! \frac{\kappa^2}{2\chi} + \frac{\chi}{4\st{\alpha}^2}.
\eqa
This shows that Eq.~(\ref{phvar79}) is correct. Taking the derivative
of Eq.~(\ref{phvar79}) gives
\beq
\frac{\partial}{\partial \chi}\ip{\Theta^2(t)} = \frac 1{4\st{\alpha}^2}
- \frac{\kappa^2}{2\chi^2}.
\eeq
Therefore the variance is minimised by
\beq
\chi  = \sqrt 2 \kappa\st \alpha ,
\eeq
and the minimum variance is
\beq
\ip{\Theta^2(t)}_{\min} = \frac{\kappa}{\sqrt 2 \st{\alpha}}.
\eeq
The minimum phase variance is therefore $\sqrt 2$ times the minimum phase
variance for the adaptive case.

\section{Results for Continuous Dyne Measurements}

In order to verify these approximate analytic results, the equilibrium phase variance was
determined numerically for a variety of parameters. Although it at first appears
that there are three parameters that should be varied, $\st \alpha$, $\kappa$
and $\chi$, these parameters are not completely independent, and we need only
consider variation in two parameters. To see this, consider the equations for
this system:
\bqa
\label{unscaled}
I(t)dt \!\!\!\! &=& \!\!\!\! 2\st{\alpha}\sin \left(\varphi(t)-\hat \varphi(t)
\right)dt + dW(t) \nn \\
\varphi (t+dt) \!\!\!\! &=& \!\!\!\! \varphi(t) + \kappa dW'(t) \nn \\
A_t \!\!\!\! &=& \!\!\!\! \int\limits_{-\infty}^t e^{\chi(u-t)} e^{i\Phi}
I(u)du \nn \\
B_t \!\!\!\! &=& \!\!\!\! - \int\limits_{-\infty}^t e^{\chi(u-t)} e^{2i\Phi} du
\nn \\
\Theta(t) \!\!\!\! &=& \!\!\!\! \arg(A_t+\chi B_t A_t^*) .
\eqa
Consider a change in the time variable
\beq
t'=t/\lambda^2.
\eeq
For this change in the time variable, the variables $I(t)$, $A_t$ and $B_t$
should be scaled to
\bqa
I'(t') \!\!\!\! &=& \!\!\!\! \lambda I(t') \nn \\
A'_{t'} \!\!\!\! &=& \!\!\!\! A_{t'} / \lambda \nn \\
B'_{t'} \!\!\!\! &=& \!\!\!\! B_{t'} / \lambda ^2 .
\eqa
This can be done because these are merely intermediate variables. With these
substitutions the equations become
\bqa
I'(t')dt' \!\!\!\! &=& \!\!\!\! 2\lambda \st{\alpha}\sin \left( \varphi(t')
-\hat \varphi(t') \right)dt' + dW(t') \nn \\
\varphi(t'+dt') \!\!\!\! &=& \!\!\!\! \varphi(t') + \lambda\kappa dW'(t') \nn \\
A'_{t'} \!\!\!\! &=& \!\!\!\! \int\limits_{-\infty}^{t'} e^{\lambda^2\chi(u-t')}
e^{i\Phi} I'(u)du \nn \\
B'_{t'} \!\!\!\! &=& \!\!\!\! -\int\limits_{-\infty}^{t'}e^{\lambda^2\chi(u-t')}
e^{2i\Phi} du \nn \\
\Theta(t') \!\!\!\! &=& \!\!\!\! \arg(A'_{t'}+\lambda^2\chi B'_{t'}
{A'_{t'}}^* ).
\eqa
This time scaling is therefore equivalent to making the substitutions
\bqa
\st \alpha &\to& \lambda \st \alpha \nn \\
\kappa &\to& \lambda \kappa \nn \\ 
\chi  &\to& \lambda^2 \chi.
\eqa
If $\lambda=1/\st\alpha$, then $\st\alpha$ is replaced in the equations with
1, $\kappa$ is replaced with $\kappa/\st\alpha$, and $\chi$ is replaced with
$\chi/\st\alpha^2$. This means that the results depend only on the ratios
$\kappa/\st\alpha$ and $\chi/\st\alpha^2$. If the individual values of $\kappa$,
$\st\alpha$ and $\chi$ are varied while keeping these ratios the same, then the
results will not change. It is therefore convenient to define the variables
\bqa
{\rm K} \!\!\!\! &=& \!\!\!\! \frac \kappa {\st\alpha} \\
{\rm X} \!\!\!\! &=& \!\!\!\! \frac \chi {\st\alpha^2}.
\eqa
Note that these variables are equal to the values of $\kappa$ and $\chi$ if
$\st\alpha$ is equal to 1. In terms of these variables, the minimum phase
variance for adaptive measurements is ${\rm K}/2$, and the optimum value of X is
$2{\rm K}$.

The value of ${\rm K}$ was varied from 1 down to $2\times 10^{-19}$. For each
value of ${\rm K}$, ${\rm X}$ was varied from a quarter to four times its
optimum value of $2{\rm K}$. The time steps used were
\beq
\Delta t = \frac{1}{10^3 \rm X}.
\eeq
For these calculations 1024 simultaneous integrations were performed and the
variance was sampled repeatedly. The integrations were taken up to time
$10/{\rm X}$, in order for the variance to reach its equilibrium value, then
the variance was sampled at time intervals of $1/{\rm X}$ up until time
$100/{\rm X}$.

\begin{figure}
\centering
\includegraphics[width=0.7\textwidth]{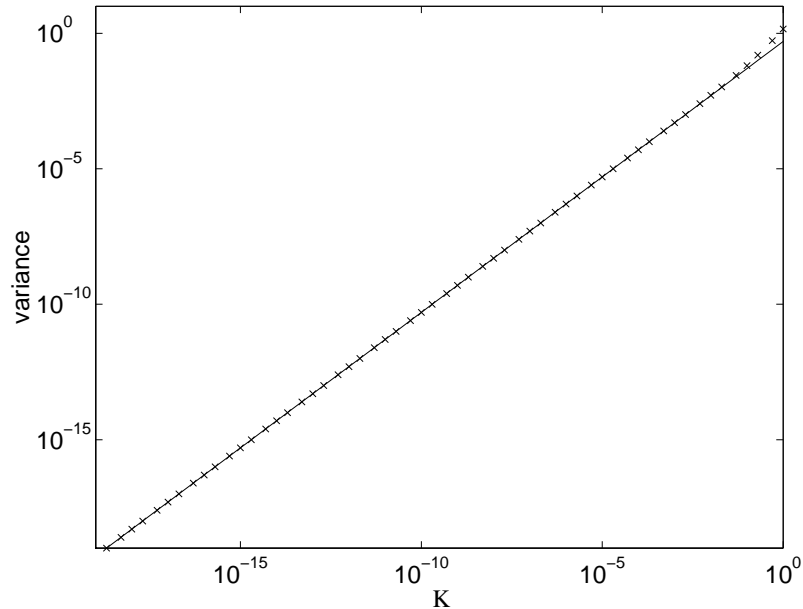}
\caption{The phase variance for continuous adaptive measurements for
${\rm X}=2{\rm K}$. The numerical results are shown as crosses and the
approximate analytic relation ${\rm K}/2$ is shown as the continuous line.}
\label{chi2k}
\end{figure}

\begin{figure}
\centering
\includegraphics[width=0.7\textwidth]{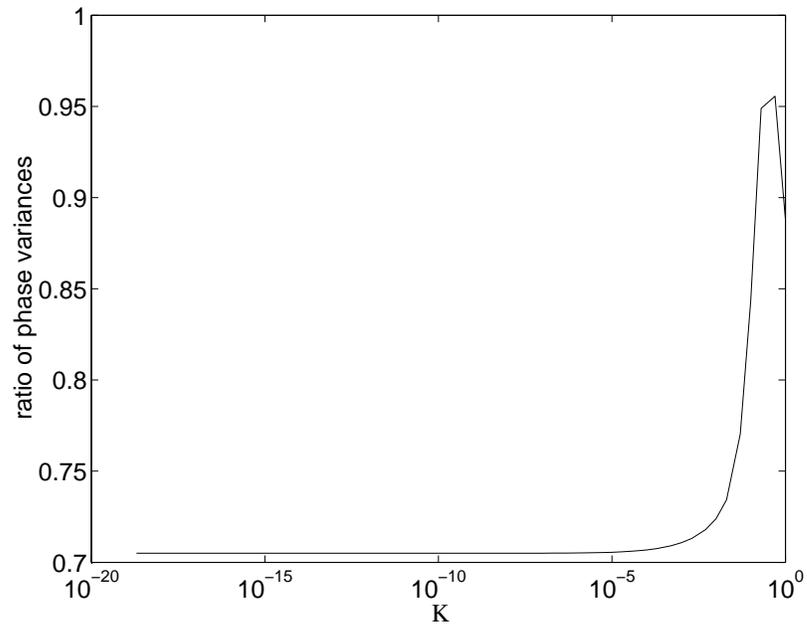}
\caption{The ratio of the minimum phase variance for continuous adaptive
measurements to the minimum phase variance for continuous heterodyne phase
measurements.}
\label{chir2k}
\end{figure}

The results for ${\rm X}=2{\rm K}$ are plotted in Fig.~\ref{chi2k}. The
variances for ${\rm K}=1$ to $5\times 10^{-7}$ are the Holevo variances, and for
below $5\times 10^{-7}$ are the standard variances. As can be seen, the results
are very close to the analytic expression. To show the improvement over
heterodyne measurements, the ratio of the minimum phase variance for adaptive
measurements to the minimum phase variance for heterodyne measurements (with
${\rm X}=\sqrt 2{\rm K}$) is plotted in Fig.~\ref{chir2k}. The ratio is close
to 1 for large K, but for smaller K the ratio gets closer and
closer to $1/\sqrt 2$.

\begin{figure}
\centering
\includegraphics[width=0.7\textwidth]{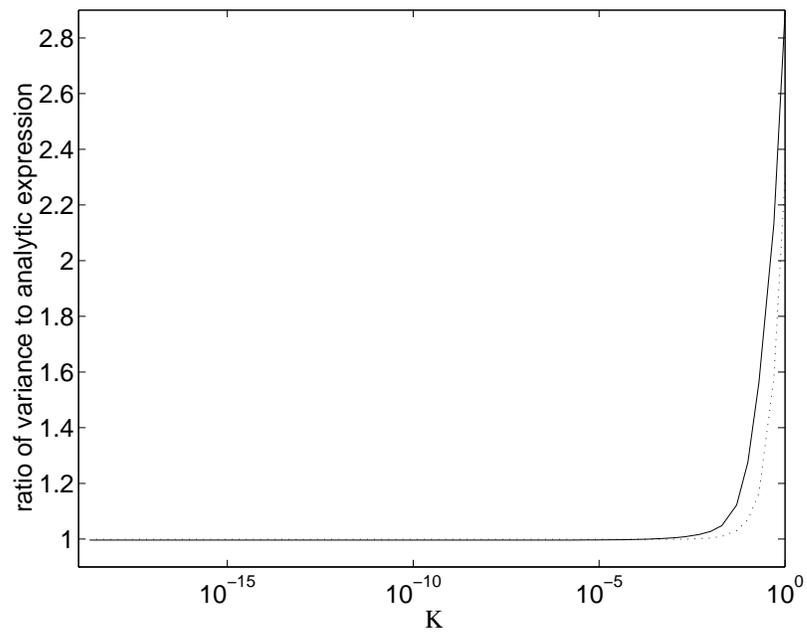}
\caption{The ratio between the numerically obtained phase variance and the
analytic expression ${\rm K}/2$ for adaptive measurements (continuous line),
and ${\rm K}/\sqrt 2$ for heterodyne measurements (dotted line).}
\label{ratioboth}
\end{figure}

In order to see the differences from the analytic expression more
clearly, the ratios of the variances to the analytic values for adaptive and
heterodyne measurements are plotted in Fig.~\ref{ratioboth}. As can be seen
the phase variances do differ significantly from the analytic values for
large K, but the agreement is extremely good for small K. This can be
expected, because the approximations made are for the limit of large $\st
\alpha$, which is equivalent to small K. Also the agreement at large
K is slightly better for heterodyne measurements than for adaptive
measurements.

\begin{figure}
\centering
\includegraphics[width=0.7\textwidth]{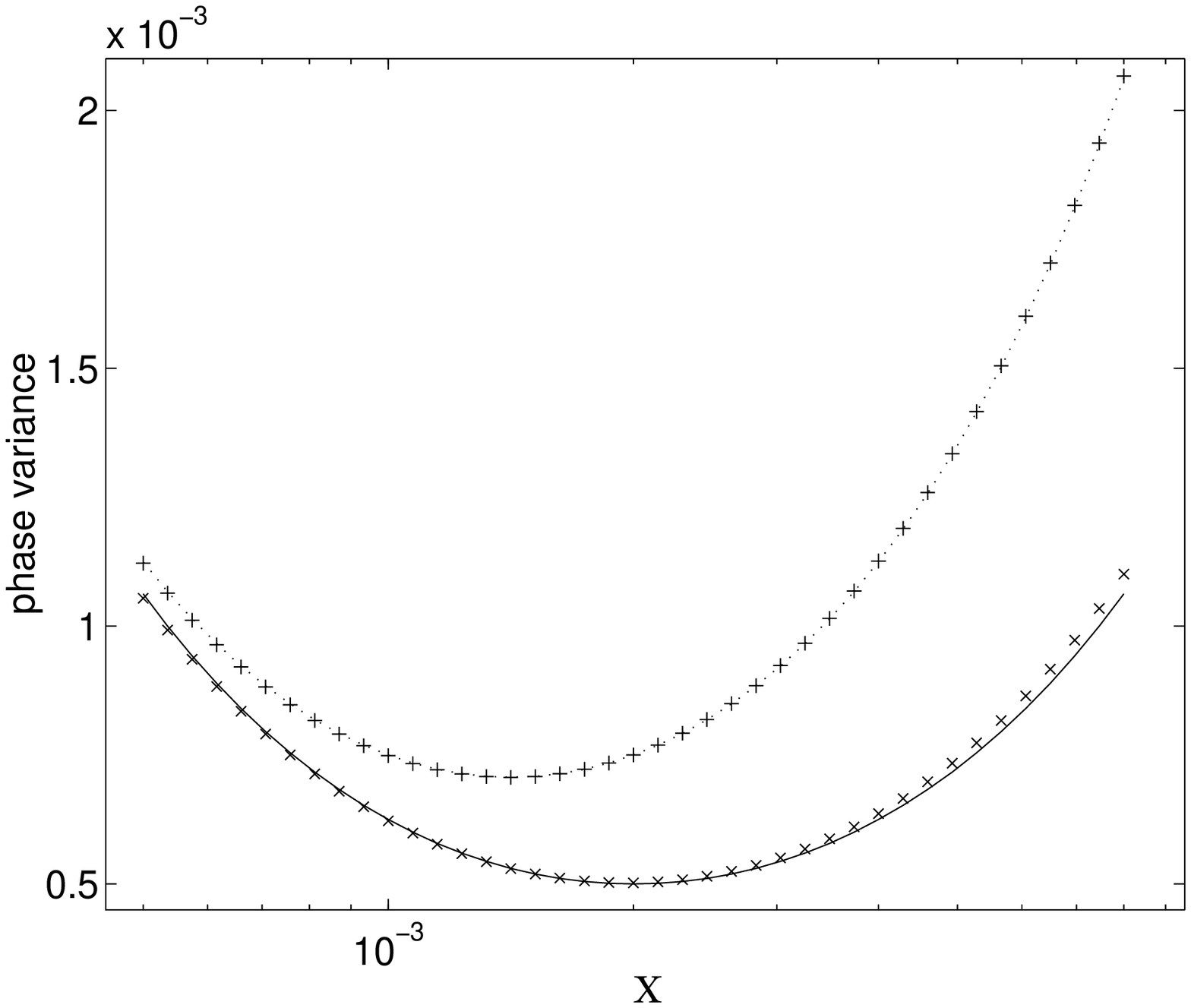}
\caption{The phase variance as a function of X for ${\rm K}=0.001$. The
numerical results for adaptive and heterodyne measurements are shown as the
crosses and pluses respectively and the approximate analytic results for adaptive
and heterodyne measurements are shown as the continuous line and dotted line
respectively.}
\label{bothpoor}
\end{figure}

\begin{figure}
\centering
\includegraphics[width=0.7\textwidth]{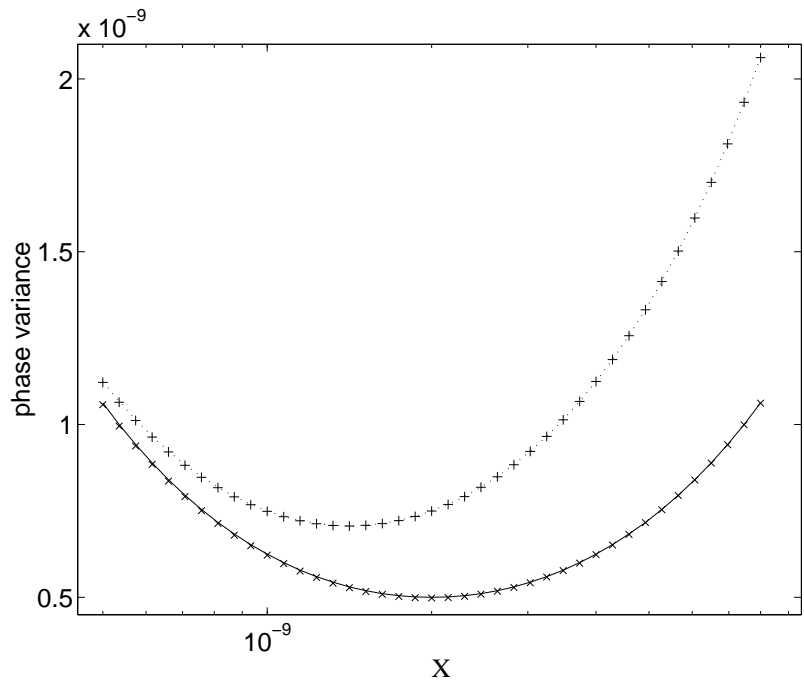}
\caption{The phase variance as a function of X for ${\rm K}=10^{-9}$. The
numerical results for adaptive and heterodyne measurements are shown as the
crosses and pluses respectively and the approximate analytic results for adaptive and
heterodyne measurements are shown as the continuous line and dotted line
respectively.}
\label{bothgood}
\end{figure}

Alternatively we can plot the phase variance as a function of X for fixed
K. In Fig.~\ref{bothpoor} I have shown the phase variance as a
function of X for ${\rm K}=0.001$ for adaptive and heterodyne measurements.
The numerical results agree reasonably closely with the analytic values,
although there is a noticeable difference for adaptive measurements for the
larger values of X. Note that the minimum phase variance for adaptive
measurements is at ${\rm X}=2{\rm K}$, and the minimum phase variance for heterodyne
measurements is larger and at a smaller value of X. When the value of
K is reduced further, as in Fig.~\ref{bothgood}, the numerical results
agree even more closely with the analytic values.

\section{Continuous Squeezed State Measurements}
\label{contsqz}
It is also possible to consider dyne measurements on continuous squeezed states.
At first it appears that it does not make sense to consider squeezed states
in the continuous case. This is because, in the single-shot case, the reduced
variance is due to the back-action of the measurement on the state. The
photocurrent is given by
\beq
I(t)dt=2{\rm Re}(\ip{a}e^{-i\Phi(t)})dt + dW(t).
\eeq
Here there is no factor multiplying $dW$, so the variances in the individual
values of $I(t)$ are not reduced for a squeezed state. Instead, the reduced
phase variance comes from the variation of $\ip{a}$ during the measurement.
If we are considering continuous measurements, then the state should remain
constant during the measurement. This means that $\ip{a}$ is constant, and
the phase variance cannot be reduced by this factor.

For the state to remain constant, we must consider a squeezed state produced by
a driven parametric oscillator in the limit that the decay time of the cavity is
extremely short. This limit must be taken before the limit $\delta t \to 0$
in the definition for the photocurrent. In this limit, the changes in the state
over the time scale of the decay constant result in reduced noise, but these
changes do not persist over the time interval $\delta t$, which is much larger
than the decay time.

The net result of this is that when we take the limit $\delta t \to 0$ the
photocurrent is given by
\beq
\label{photosqz}
I(t)dt=2{\rm Re}(\alpha e^{-i\Phi(t)})dt + \sqrt{e^{-2r}
\cos^2(\Phi-\phi_\zeta/2)+e^{2r}\sin^2(\Phi-\phi_\zeta/2)} dW(t),
\eeq
where $\alpha$ is the amplitude of the squeezed state, and $r$ and $\phi_\zeta$
are the magnitude and direction of the squeezing. Therefore in this
case, rather than the uncertainty being reduced by variation in $\ip a$, it
is reduced by a multiplying factor for $dW$.

For reduced phase uncertainty, the phase of the squeezing should be $\phi_\zeta
=2\varphi + \pi$, where $\varphi$ is the system phase. If we are using
feedback given by
\beq
\Phi = \hat\varphi + \pi/2,
\eeq
where $\hat\varphi$ is an estimate of the phase, then the photocurrent can be
expressed as
\beq
I(t)dt=2|\alpha| \sin(\varphi-\hat\varphi)dt + \sqrt{e^{-2r}\cos^2
(\hat\varphi-\varphi)+e^{2r}\sin^2(\hat\varphi-\varphi)} dW(t).
\eeq
It is clear that if the intermediate phase estimate used is very close to the
system phase, then the factor multiplying $dW$ will be close to $e^{-r}$ and
will be at a minimum. The better the intermediate phase estimate is, the smaller
this multiplying factor will be. If the intermediate phase estimate is not
perfect, it is clear that increasing the squeezing past a certain level will not
reduce the multiplying factor. This is because the $e^{2r}$ term will start to
dominate.

It is possible to estimate the optimum squeezing and the minimum phase variance
under these measurements using the linear approximation. In this approximation,
the variance in the individual phase estimates $\theta(t)$ is equal to
\beq
\frac{e^{-2r}\cos^2(\hat\varphi-\varphi)+e^{2r}\sin^2(\hat\varphi-\varphi)}
{4\st{\alpha}^2 dt}.
\eeq
It is clear that the minimum phase variance (in this approximation) will be
obtained when the best phase estimates are used for $\hat\varphi$. It is
therefore reasonable to use the phase estimates $\Theta(t)$ for $\hat\varphi$.
These will be the best phase estimates when the correct value of $\chi$ is used.
As the variance of these estimates is $\Delta\Theta^2$, we obtain
\beq
\ip{e^{-2r}\cos^2(\hat\varphi-\varphi)+e^{2r}\sin^2(\hat\varphi-\varphi)}
\approx e^{-2r} + e^{2r} \Delta\Theta^2.
\eeq
This approximation will be true for small phase variances and large squeezing.
Following the same derivation as for the coherent state case, the only
difference is the multiplying factor, so we obtain
\beq
\label{simpler1}
\Delta\Theta^2 = \frac{\chi}{8\st{\alpha}^2}\left(e^{-2r} + e^{2r}
\Delta\Theta^2\right)+\frac{\kappa^2}{2\chi}.
\eeq
Solving this for $\Delta\Theta^2$ gives
\beq
\Delta\Theta^2 = \frac{\frac{\chi e^{-2r}}{8\st{\alpha}^2} + \frac{\kappa^2}
{2\chi}}{1-\frac{\chi e^{2r}}{8\st{\alpha}^2}}.
\eeq

This expression has two independent variables, $\chi$ and $r$, that can be
varied in order to find the minimum phase variance. Taking the derivative of
Eq.~(\ref{simpler1}) with respect to $\chi$ gives
\beq
\frac{\partial\Delta\Theta^2}{\partial\chi} = \frac{1}{8\st{\alpha}^2}
\left(e^{-2r}+e^{2r} \Delta\Theta^2\right)+\frac{\chi e^{2r}}{8\st{\alpha}^2}
\frac{\partial\Delta\Theta^2}{\partial\chi}-\frac{\kappa^2}{2\chi^2}.
\eeq
Since the derivative is zero for the minimum this gives
\beq
0 = \frac{1}{8\st{\alpha}^2}\left(e^{-2r} + e^{2r} \Delta\Theta^2\right)
-\frac{\kappa^2}{2\chi^2}.
\eeq
Together with Eq.~(\ref{simpler1}), this gives
\beq
\label{opchi}
\chi = \frac{\kappa^2}{\Delta\Theta^2}.
\eeq
Substituting this into Eq.~(\ref{simpler1}) gives
\beq
\label{simpler2}
\Delta\Theta^2 = \frac{\kappa^2}{4\st{\alpha}^2}\left( e^{2r} + \frac{e^{-2r}}
{\Delta\Theta^2} \right).
\eeq
Taking the derivative of this with respect to $r$ gives
\beq
\frac{\partial\Delta\Theta^2}{\partial r} = \frac{\kappa^2}{2\st{\alpha}^2}
\left(e^{2r} - \frac{e^{-2r}}{\Delta\Theta^2} - \frac{e^{-2r}}
{(\Delta\Theta^2)^2}\frac{\partial\Delta\Theta^2}{\partial r}\right).
\eeq
Since the derivative is zero at the minimum, this becomes
\beq
0 = \frac{\kappa^2}{2\st{\alpha}^2}\left(e^{2r}-\frac{e^{-2r}}{\Delta\Theta^2}
\right).
\eeq
This can be solved to give
\beq
\label{e2r}
e^{-4r} = \Delta\Theta^2.
\eeq
Substituting this back into Eq.~(\ref{simpler2}) gives the phase variance as
\beq
\label{therval}
\Delta\Theta^2 = \left( \frac{\kappa}{\sqrt 2\st{\alpha}} \right)^{4/3}.
\eeq

Thus we see that even for an arbitrarily squeezed state, the best scaling we
can obtain for the phase variance is $|\alpha|^{-4/3}$, as compared to
$|\alpha|^{-1}$ for a coherent state. This difference is less than for
single-shot measurements, where optimum squeezed states scale as almost
$\nb^{-2}$, as compared to $\nb^{-1}$ for coherent states.

\section{Results for Squeezed States}

The results for the continuous squeezed state case were obtained by a similar
method as for the coherent state case. Similarly to the case for coherent
states, only variation in the variables K and X was considered, rather than
in all three variables $\st\alpha$, $\kappa$ and $\chi$. The step sizes used
were
\beq
\Delta t = \frac{1}{10^3 \rm X}.
\eeq
The integrations were taken up to time $30/{\rm X}$, then the variance was
sampled every time step until time $130/{\rm X}$. The integration was performed
using the photocurrent given in Eq.~(\ref{photosqz}), except with the time
scaled such that $\st\alpha$ was not required (similarly to the coherent state
case). The phase of the squeezing was assumed at all times to be in the correct
direction for reduced phase variance, i.e.\ negative with respect to the coherent
amplitude.

It was found that when the phase estimate $\arg C_t$ was used in the
feedback, very poor results were obtained. This is a similar result to the case
for single-shot measurements, where using $\arg C_v$ feedback results in large
phase variances. This is because, when the intermediate phase estimates are
extremely good, the results do not distinguish easily between the real system
phase and the system phase plus $\pi$. The means that many of the results are
out by $\pi$, resulting in a large overall phase variance.

In order to avoid this problem, rather than using $\arg C_t$ in the feedback, an
intermediate phase estimate given by
\beq
\hat\varphi (t) = \arg (C_t^{1-\varepsilon} A_t^{\varepsilon}),
\eeq
was used. Note that this is similar to the phase estimate used to obtain
phase measurements close to optimum in the single-shot case. Here a constant
$\varepsilon$ was used, as the system state is not changing like in the
single-shot case.

For each value of K there are three variables that can be altered to
minimise the phase variance: X, $r$ and $\varepsilon$. It is not
calculationally feasible to consider a range of values for all three variables.
Rather than considering a range of values, the values of the variables were
varied in order to find the values that gave the minimum phase variance.

\begin{figure}
\centering
\includegraphics[width=0.7\textwidth]{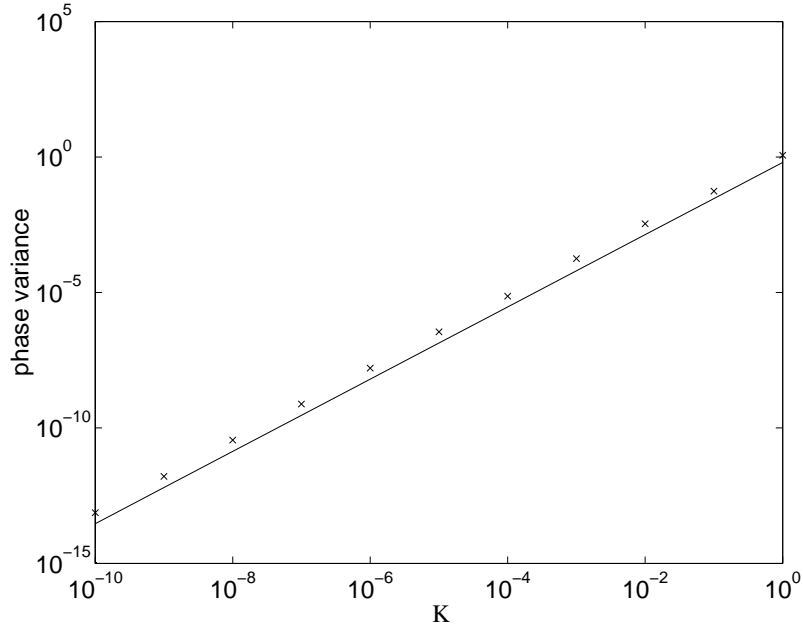}
\caption{The phase variance as a function of K for continuous squeezed
states. The continuous line is the theoretical analytic relation, and the crosses
are the numerical results.}
\label{mincontsqz}
\end{figure}

\begin{figure}
\centering
\includegraphics[width=0.7\textwidth]{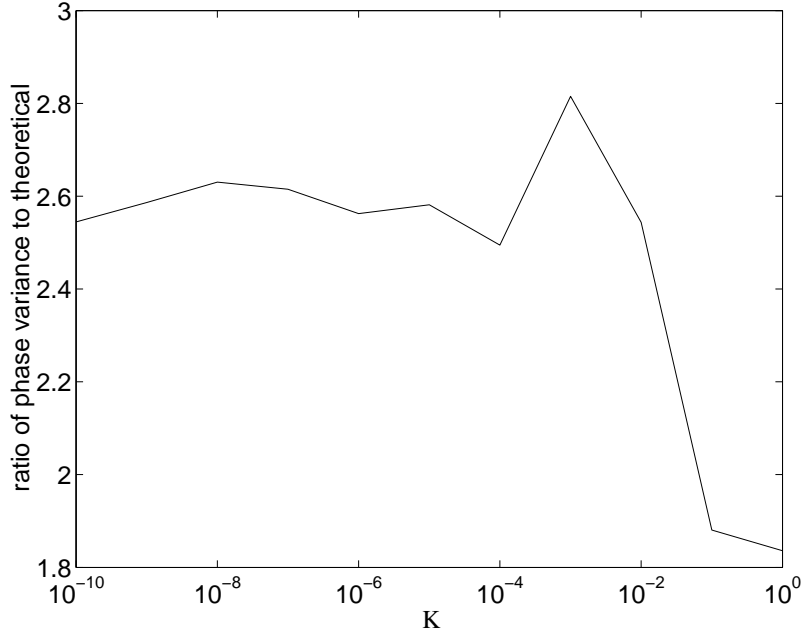}
\caption{The ratio of the numerically obtained phase variance to the
analytic relation as a function of K for continuous squeezed states.}
\label{mincontrat}
\end{figure}

The minimum phase variances obtained by this method are plotted as a function
of K in Fig.~\ref{mincontsqz}. The values given by the approximate analytic
expression (\ref{therval}) are also shown in this figure. The numerical results
are higher than the analytic expression, but for small K they have the same
scaling. If we plot the ratio of the numerical results to the analytic values
as in Fig.~\ref{mincontrat}, we find that for the smallest values of K
the ratio levels off at about $2.6$. Thus we see that the stochastic results
also give a scaling of $\st{\alpha}^{-4/3}$.

Now note that, from Eqs~(\ref{e2r}) and (\ref{therval}), the optimum value of
$e^{-2r}$ should be
\bqa
e^{-2r} \!\!\!\! &=& \!\!\!\! \left( \frac{\kappa}{\sqrt 2\st{\alpha}}
\right)^{2/3} \nn \\
\!\!\!\! &=& \!\!\!\! \left( \frac{\rm K}{\sqrt 2} \right)^{2/3}.
\eqa
Similarly, from Eqs~(\ref{opchi}) and (\ref{therval}), the optimum value of
$\chi$ should be
\beq
\chi = \left( 2\st{\alpha}^2\kappa \right)^{2/3}.
\eeq
The corresponding optimum value of X is
\beq
{\rm X} = \left( 2{\rm K} \right)^{2/3}.
\eeq
This indicates that the optimum values for both $e^{-2r}$ and X should
scale as ${\rm K}^{2/3}$.

\begin{figure}
\centering
\includegraphics[width=0.7\textwidth]{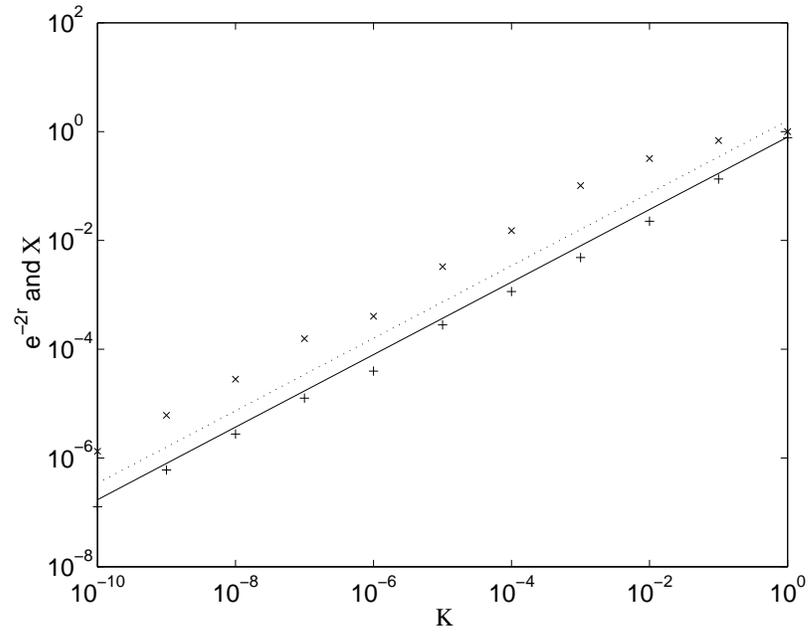}
\caption{The optimum values of $e^{-2r}$ and X for measurements on
continuous squeezed states. The numerically found values of $e^{-2r}$ are
plotted as crosses, and the approximate analytic expression as a continuous line. 
The numerically found values of X are plotted as pluses, and the
approximate analytic expression as a dotted line.}
\label{rccontsqz}
\end{figure}

The numerically obtained optimum values of $e^{-2r}$ and X, as well as
these analytic expressions, are plotted in Fig.~\ref{rccontsqz}. Similarly
to the case for the phase variance, the scaling is the same as that predicted
analytically, but the scaling constants are different. For the case of $e^{-2r}$,
the optimum values are about 8 times those analytically predicted, whereas the
values of X are around a third of those analytically predicted.

\begin{figure}
\centering
\includegraphics[width=0.7\textwidth]{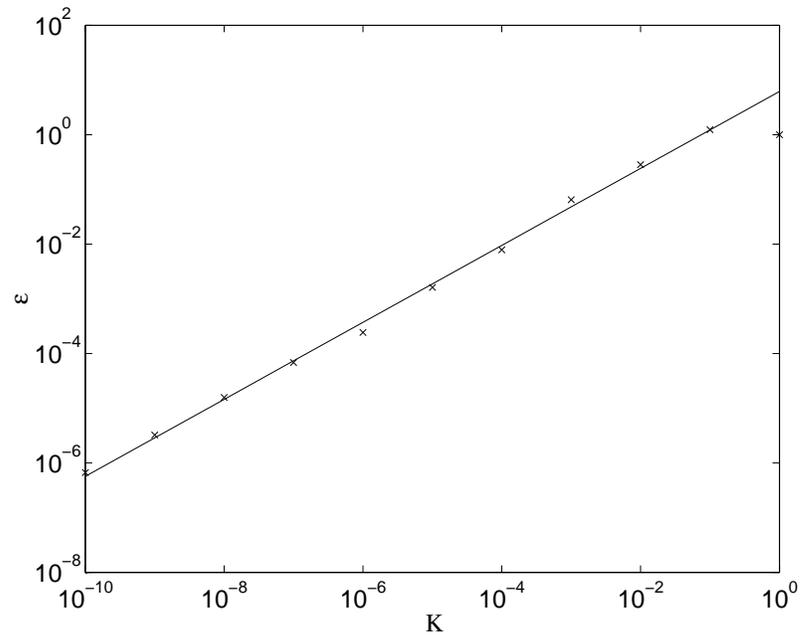}
\caption{The optimum values of $\varepsilon$ for measurements on continuous
squeezed states. The crosses are the numerically found values, and the
continuous line is the expression fitted to the data.}
\label{epcontsqz}
\end{figure}

For the case of $\varepsilon$ there is no analytic prediction for the optimum
value. The numerically obtained values are shown in Fig.~\ref{epcontsqz}, and
as can be seen $\varepsilon$ decreases in a regular way with $\kappa$. A power
law was fitted to these values (for ${\rm K}<1$), and the power found was
$0.70\pm 0.01$. This is very similar to the ${\rm K}^{2/3}$ scaling found for
$e^{-2r}$ and X.

A problem with these results is that they do not take
account of the low probability results with large error. Unfortunately it is
not possible to take account of these results in an analytic way as in the case
of single shot measurements on squeezed states. The best that can be done is to
use a very large number of samples. For these calculations the phase variance
was sampled every time step, resulting in about $10^8$ samples. Unfortunately
the phase estimates from adjacent time steps are strongly correlated, so there
is only around $10^5$ independent samples.

Despite the large number of samples used, there were generally no large error
samples for the results shown. The optimum parameters found usually gave the
minimum variance because varying the parameters beyond these values, such that
the variance would be smaller according to the linear theory, resulted in
samples with large error.

\section{Continuous Interferometric Measurements}

Now I will consider the case of continuous measurements for two-mode
interferometry.  In this case we have a Mach-Zehnder interferometer as in
Ch.~\ref{adaptiveinter}, and are attempting to continuously track the phase
in one arm, and control the phase in the other arm in order to obtain the best
possible estimate of the phase. The optimum states discussed in
Ch.~\ref{adaptiveinter} make little sense in this context, so instead I will
consider the state with all the photons in one arm. This state can be
generalised to the continuous case by considering a state with $N$ photons per
unit time.

This case is essentially semiclassical, and the detections can be considered
independently. Therefore, consider a single photon incident on port $a$, so the
state is $\ket{\half,\half}_z$. Upon detection the unnormalised state changes as 
\beq
\ket {\psi(u,\varphi)} = \hat c_{u} (\varphi)\ket{\psi}.
\eeq
Using Eq.~(\ref{detectop}) for the detection operator, the probability of
detecting the photon in detector $u$ is given by
\beq
\label{probdect}
\braket{\psi(u,\varphi)}{\psi(u,\varphi)} = \sin^2 \left(
\frac {\varphi-\Phi+u\pi}2 \right).
\eeq
Using Bayes' theorem, the probability distribution for the system phase after
the detection is proportional to this probability times the initial phase
distribution.

The probability distribution for the phase based on $m$ of these independent
detections, $P(\varphi |n_m)$, can be expressed as
\bqa
P(\varphi |n_m) = \sum_{k=-m}^m P_{mk} (n_m)e^{ik\varphi}.
\eqa
It is straightforward to show from Eq.~(\ref{probdect}) that the coefficients
$P_{mk}(n_m)$ can be determined by
\beq
\label{probrecurse0}
P_{mk} (n_m) \propto P_{m-1,k} (n_{m-1}) - \half e^{-i(\Phi_m-u_m\pi)}
P_{m-1,k-1} (n_{m-1}) - \half e^{i(\Phi_m-u_m\pi)} P_{m-1,k+1} (n_{m-1}).
\eeq
The normalisation condition on the probability distribution becomes
\beq
P_{m0} (n_m) = 1.
\eeq
The normalised probability distribution can be obtained by a simple addition to
the recursion relation:
\beq
\label{probrecurse}
P_{mk} (n_m) = \frac{P_{m-1,k} (n_{m-1}) - \half e^{-i(\Phi_m-u_m\pi)}
P_{m-1,k-1} (n_{m-1}) - \half e^{i(\Phi_m-u_m\pi)} P_{m-1,k+1} (n_{m-1})}
{P_{m0} (n_m)}.
\eeq

I will consider the same variation in the system phase as in the case of dyne
measurements,
\beq
\varphi (t+dt) = \varphi (t) + \kappa dW(t).
\eeq
When the phase varies in time, the time between detections is important. For a
photon flux of $N$, the probability of a photodetection in time $dt$ is $Ndt$.
The probability distribution for the time between detections is given by
\beq
\label{timedist}
P_{\rm P} (t)dt = Ne^{-Nt} dt.
\eeq
In the results that will be presented here, the time between detections,
$\Delta t$, was determined according to this probability distribution.

Now in order to determine the effect of this phase diffusion on the probability
distribution between detections, we must first consider the effect over some
very small time interval $\delta t$. This is necessary because the probability
distribution for the change in the system phase over time $\Delta t$ does not
go to zero for $\Delta \varphi = \pm \pi$. This means that the probability
distribution will not be exactly Gaussian, due to the overlap.
In contrast, if we look at a very small time interval $\delta t$, the change in
the phase will have a normal distribution with a variance of
$\kappa^2 \delta t$. Explicitly the probability distribution is
\beq
\label{phaseprobdist}
P_{\rm G} (\Delta \varphi)d(\Delta \varphi) = \frac{1}{\kappa \sqrt{2\pi\delta
t}} e^{-\Delta \varphi^2 /(2\kappa^2 \delta t)} d(\Delta \varphi).
\eeq

The probability distribution for the phase after time $\delta t$ will be the
convolution of the initial probability distribution with the Gaussian described
by Eq.~(\ref{phaseprobdist}). Evaluating this convolution gives
\bqa
P'(\varphi |n_m) \!\!\!\! &=& \!\!\!\! \int\limits_{-\pi }^\pi P(\varphi-\theta
|n_m)P_{\rm G} (\theta)d\theta \nn \\
\!\!\!\! &=& \!\!\!\! \int\limits_{-\pi }^\pi \sum\limits_{k=-m/2}^{m/2} P_{mk}
(n_m)e^{ik(\varphi-\theta)} P_{\rm G} (\theta)d\theta \nn \\
\!\!\!\! &=& \!\!\!\! \sum_{k=-m/2}^{m/2} P_{mk} (n_m)e^{ik\varphi}
\int\limits_{-\pi}^\pi e^{-ik\theta} P_{\rm G} (\theta)d\theta.
\eqa
As $\delta t$ is assumed to be small, $\kappa^2 \delta t \ll 1$, so
\beq
\int\limits_{-\pi }^\pi e^{-ik\theta} P_{\rm G} (\theta)d\theta =
e^{-k^2 \kappa ^2 \delta t/2} .
\eeq
The effect of the variation of the system phase on the probability distribution
is therefore
\beq
P^{\delta t}(\varphi |n_m ) = \sum\limits_{k=-m/2}^{m/2} P_{mk}(n_m)e^{-k^2
\kappa ^2 \delta t/ 2} e^{ik\varphi}.
\eeq
This shows that the coefficients for the probability distribution are just
multiplied by a Gaussian:
\beq
P^{\delta t}_{mk} (n_m) = P_{mk} (n_m)e^{-k^2 \kappa^2 \delta t/2}.
\eeq
This result is related to the usual result for convolutions and Fourier
transforms.

In order to take account of the effect of the phase diffusion on the probability
distribution over some significant time interval $\Delta t$, this time interval
can be divided into $M$ small time intervals $\delta t$. Then we find
\bqa
\label{narrower}
P^{\Delta t}_{mk} (n_m) \!\!\!\! &=& \!\!\!\! P_{mk} (n_m) \prod_{j=1}^M
e^{-k^2 \kappa^2 \delta t/2} \nn \\
\!\!\!\! &=& \!\!\!\! P_{mk} (n_m)
e^{-k^2 \kappa^2 \sum_{j=1}^M \delta t/2} \nn \\
\!\!\!\! &=& \!\!\!\! P_{mk} (n_m) e^{-k^2 \kappa^2 \Delta t/2}.
\eqa
This result can be used to take account of the variation of the system phase
very simply in the probability distribution.

As time passes the effect of Eq.~(\ref{probrecurse}) is to broaden the
distribution of probability coefficients in $k$, corresponding to a smaller
variance in the phase distribution. In contrast the Gaussian term in
Eq.~(\ref{narrower}) tends to narrow the distribution of probability
coefficients, corresponding to a greater phase variance. The initially broad
phase distribution narrows until an approximate equilibrium is reached, where
the two effects cancel each other out.

The derivation for the optimum phase estimate for the single-shot case given in
Eq.~(\ref{singleopt}) is general enough to hold in this case also. Therefore the
optimal phase estimate is given by
\bqa
\hat \varphi  \!\!\!\! &=& \!\!\!\! \arg \ip {e^{i\varphi}} \nn \\
\!\!\!\! &=& \!\!\!\! \arg P_{m,-1} (n_m).
\eqa
In addition, much of the reasoning for the feedback phase still holds.
The phase variance after the next detection can be minimised by
minimising the value of
\beq
M(\Phi_{m}) = \sum_{u_m=0,1} \st{\int\limits_{-\pi}^{\pi} P(n_m|\varphi)
e^{i\varphi} d\varphi}.
\eeq
The values of $P(n_m|\varphi)$ can be obtained, except for a normalising
constant that is common to $u_m=0$ and 1, by using Eq.~(\ref{probrecurse0}).
This means that we can express $M(\Phi_{m})$ as in Eq.~(\ref{simple}) with the
parameters $a$, $b$ and $c$ given by
\bqa
a \!\!\!\! &=& \!\!\!\! P_{m-1,-1} (n_{m-1}) \nn \\
b \!\!\!\! &=& \!\!\!\! \half P_{m-1,-2} (n_{m-1}) \nn \\
c \!\!\!\! &=& \!\!\!\! \half P_{m-1,0} (n_{m-1}).
\eqa
These values of $a$, $b$ and $c$ can be used to determine the feedback phase as
in Sec.~\ref{feedback}.

The phase uncertainty at equilibrium can be estimated using a similar approach
as was used for the single mode case. Let us assume that the equilibrium
variance in the best estimate for the system phase is $\Delta\Theta^2$. After
time $\Delta t$, the variance in this phase estimate with respect to the
new system phase, $\varphi(t+\Delta t)$, will be $\Delta\Theta^2+\kappa^2\Delta
t$. In the equilibrium case this increase in the variance should, on average, be
balanced by the decrease due to the next detection.

We now wish to estimate the equilibrium variance based on a weighted average
with the previous best phase estimate, and a phase estimate from the new
detection. If we use the actual variance for a phase estimate based on a single
detection, then we do not get accurate results. This is because the variance for
a single detection is large, so the weighted average does not accurately
correspond to the exact theory. In order to make the theory based on weighted
averages accurate, we need to assume an {\it effective} variance for the
single detection, that is different from the actual variance.

In the case where there is no variation in the system phase, the phase variance
after $N$ detections is approximately $1/N$ (see the results given in Secs
\ref{others} and \ref{results}). Denoting the variance after $N$
detections as $\Delta\Theta_N^2$, and the effective phase variance from a new
detection as $\Delta\Theta_1^2$, the weighted average gives
\bqa
\frac{1}{\Delta\Theta_N^2} + \frac{1}{\Delta\Theta_1^2} = \frac{1}{\Delta
\Theta_{N+1}^2} \nn \\ N + \frac{1}{\Delta\Theta_1^2 } = N+1.
\eqa
Therefore, in order to take account of a single detection via a weighted
average, it can be assumed to be equivalent to a phase estimate with a variance
of 1. This is, in fact, equal to the variance as estimated using
$\lr{2(1-\cos\varphi)}$.

Applying this to the case with a varying system phase gives
\beq
\frac{1}{\Delta\Theta^2  + \kappa^2 \Delta t} + 1 = \frac{1}{\Delta\Theta^2}
\eeq
Simplifying this to solve for $\Delta\Theta^2$, we find
\bqa
\frac{1}{1+\kappa^2 \Delta t/\Delta\Theta^2} + \Delta\Theta^2 \!\!\!\! &=&
\!\!\!\! 1 \nn \\ 1 - \kappa ^2 \Delta t/\Delta\Theta ^2 + \Delta\Theta^2
\!\!\!\! &\approx& \!\!\!\! 1 \nn \\ \kappa ^2 \Delta t/\Delta\Theta ^2
\!\!\!\! &\approx& \!\!\!\! \Delta\Theta^2 .
\eqa
Thus the approximate value of the variance is
\beq
\Delta\Theta ^2 \approx \kappa \sqrt {\Delta t}.
\eeq
On average the time between detections is $1/N$, so the approximate value of the
variance should be
\beq
\Delta\Theta ^2 \approx \kappa / \sqrt N.
\eeq

\section{Results for Continuous Interferometric Measurements}

In order to verify this result, the equilibrium phase variance was
determined numerically for a variety of parameters. In this case there are only
two parameters, $\kappa$ and $N$. In the case of dyne measurements there was the
additional parameter $\chi$ describing how the latest results were weighted as
compared to the previous results. In this case we do not have that parameter, as
the phase estimates are not determined in that way.

Similarly to the case of dyne measurements, these two parameters are related.
Here the average time between detections is $1/N$, and the mean squared change
in the system phase after time $\Delta t$ is $\kappa^2 \Delta t$. In the above
theory the parameters $\kappa$ and $\Delta t$ only appear through the product
$\kappa^2 \Delta t$. It is therefore the value of $\kappa^2 /N$ that matters,
and if $\kappa$ and $N$ are varied while keeping this product the same, the
results will not change. It is therefore convenient to define
\beq
{\rm K} = \frac \kappa {\sqrt N},
\eeq
similarly to the case for continuous dyne measurements.

The calculations
were run for $10^5$ detections (or $2\times 10^5$ for the minimum value of
K), and the phase error was sampled every detection after $10/{\rm K}$
detections. This was done 100 times for each value of K. The equilibrium
phase variance was determined in this way for the nearly optimum feedback scheme
and in addition for a nonadaptive feedback scheme given by
\beq
\Phi_m = \Phi_0  + m {\rm K} \pi ,
\eeq
where $\Phi_0$ is a random initial phase. When the value of K was 1 or
more this was modified to
\beq
\Phi_m = \Phi_0 + m\pi/2,
\eeq
to prevent $\Phi_m$ being constant (modulo $\pi$). This is equivalent to the
non-adaptive feedback in the single-shot case given by Eq.~(\ref{nonadapt}), and
is analogous to heterodyne feedback for dyne measurements. The reason for the
factor of K is that the effective number of detections used for the phase
estimate is $1/{\rm K}$. To see this, note that the phase variance is
approximately K.

A minor problem with continuous adaptive measurements is that the number of
probability coefficients $P_{mk}(n_m)$ needed to determine the probability
distribution for the phase rises indefinitely with the number of detections. The
narrowing effect of the varying system phase, however, means that the
probability coefficients fall approximately exponentially with $k$. The
probability distribution can therefore be approximated very accurately by
keeping only a certain number of coefficients. For the results presented
here all probability coefficients with a magnitude above about $10^{-20}$ were
used.

\begin{figure}
\centering
\includegraphics[width=0.7\textwidth]{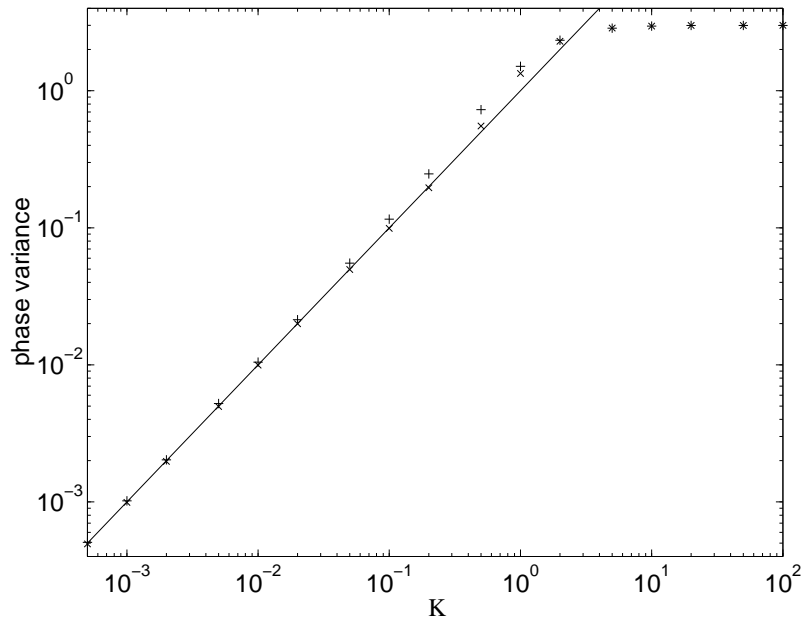}
\caption{The phase variance as a function of K. The numerical
results for adaptive and nonadaptive measurements are shown as the crosses and
pluses respectively and the analytic result is shown as the continuous
line.}
\label{runintlog}
\end{figure}

All phase variances for the two feedback schemes are plotted in
Fig.~\ref{runintlog}. Note that, from the previous section, the analytically
predicted value of the phase variance is K. As can be seen, the results for both
cases are very close to this analytic result for the smaller values of K. For
values of K close to 1 the results for the nonadaptive scheme are noticeably
above the analytically predicted values. For large values of K above 1 the
variance converges to 3 for both the feedback schemes. This is what can be
expected, as the system phase is randomised between detections. This means that
the measurements are equivalent to phase measurements with a single photon, for
which the Holevo phase variance is 3. The feedback has no effect, as there is
no information on which to base it on.

\begin{figure}
\centering
\includegraphics[width=0.7\textwidth]{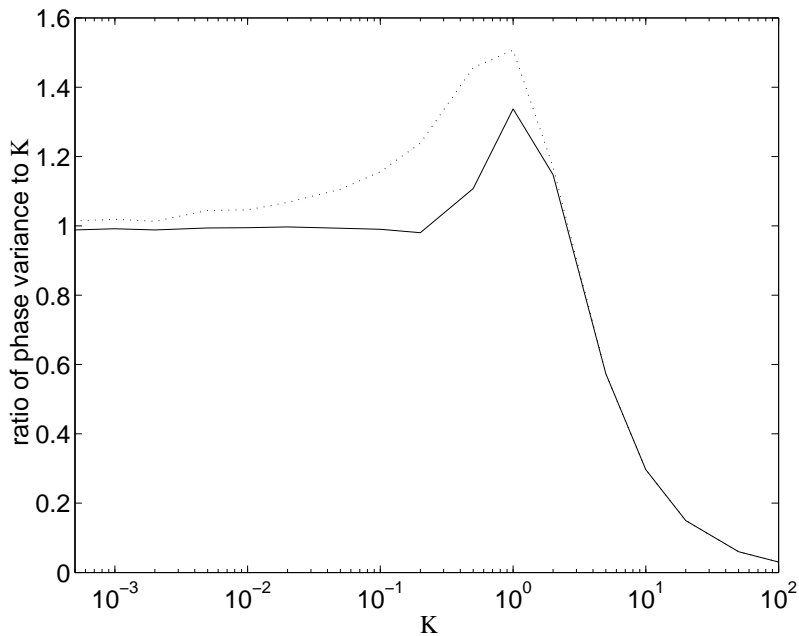}
\caption{The phase variance as a ratio to the approximate analytic result of K.
The results for adaptive and nonadaptive measurements are shown as the
continuous line and dotted line respectively.}
\label{runintrat}
\end{figure}

To see the differences more clearly, the phase variances are plotted as ratios
to the analytic results in Fig.~\ref{runintrat}. As can be seen, the results
for both measurement schemes are very close to the analytic result for very
small values of K, but differ increasingly for larger values. The
adaptive scheme gives phase variances that are very close to, and slightly
below, the analytically predicted values for smaller values of K. In contrast
the results for nonadaptive measurements are all above the analytically
predicted values (for ${\rm K}\le 1$). For large values of K the variance for
both schemes is below K, as the variance is converging to 3.

These results show that there is a small improvement in using an adaptive
scheme over a non-adaptive scheme; however, when the system phase is varying
slowly there will be very little improvement. This can be expected from the
results in the single shot case, where there was very little difference between
the results for the adaptive and non-adaptive measurements when all photons were
in one port. This means that we do not even get an improvement by a factor of
$\sqrt 2$, as we do in the case of continuous dyne measurements on a coherent
state. If the system phase is fluctuating rapidly, we can get a significant
improvement in the accuracy of the phase measurements. The maximum improvement
is about 24\% for ${\rm K}=0.5$.

%% file: conc.tex
\setcounter{chapter}{7}
\chapter{Conclusions}
\label{conc}
The major results of this study can be described very succinctly: I have
found the optimal input states and adaptive measurement schemes for both dyne
and interferometric phase measurements.

\section{Input States for Dyne Measurements}
In Ch.~\ref{dynesta} the problem of optimum input states for dyne measurements
was considered. Rather than just considering optimisation for minimum intrinsic
phase variance, the problem of optimisation for minimum phase variance under
more general measurements where $h(n)=cn^{-p}$ was considered. There are three
different types of constraints on the states that were considered: an upper
limit on the photon number, a fixed mean photon number, and squeezed states with
fixed mean photon number. The optimum states for canonical measurements with
these three constraints on the states have been determined previous to this
study, as has the result for general measurements with an upper limit on the
photon number. In Ch.~\ref{dynesta} the set of results was completed with
analytic results for fixed mean photon number and optimum squeezed states under
general measurements. The complete set of results are summarised in
Table~\ref{table:opttab}.

\begin{table}[tbp]
\centering
\caption{The asymptotic formulae for minimum Holevo phase variance for the
canonical distribution and general measurements under three different
constraints on the states. The results in bold are those that are original to
this study.}
\begin{tabular}{||c|c|c||} \hline \hline
                      & Canonical & General measurements \\ \hline
Maximum photon number & $ \frac {\pi^2}{N^2}$ &
$ 2cN^{-p} + \st{z_1}(2cp)^{2/3}N^{-2(1+p)/3}{\bf +3c^2N^{-2p}} $ \\ \hline
Mean photon number    & $ \frac{1.893606}{\nb^2} $ &
$ {\bf 2c\bar n^{-p}+\sqrt{cp(p+1)}\bar n^{-p/2-1}+3c^2\nb^{-2p}} $ \\ \hline
Squeezed states       & $ \frac{\log \bar n + \Delta}{4 \bar n^2} $ &
$ {\bf 2c\bar n^{-p}+\sqrt{cp(p+1)}\bar n^{-p/2-1}+3c^2\nb^{-2p}} $ \\ \hline
\hline
\end{tabular}
\label{table:opttab}
\end{table}

All of these results (except the exact analytic result for canonical
measurements with an upper limit on the photon number) were verified by
exact numerical calculation of the optimum states. For the case of canonical
measurements on squeezed states it was found that the exact results did not
agree with the approximate analytic result to order $\nb^{-2}$. In addition,
corrections to this analytic expression derived from the results in
\cite{collett} did not produce any better agreement.

For the states with an upper limit on the photon number optimised for minimum
phase variance under general measurements, it was found that although there was
good agreement with the analytic results, larger photon numbers were required
for good agreement than was claimed in \cite{semiclass}. This is because in
\cite{semiclass} the photon numbers required for good agreement were estimated
from the approximation made in omitting a boundary condition. The main
approximation made is actually the linearisation of the equations, which means
that the photon numbers were underestimated in \cite{semiclass}.

An additional correction term of $3c^2N^{-2p}$ was found for this case. This
term is specific to the Holevo phase variance, and is only significant for
mark I measurements. For mark I measurements this term is required in order to
obtain agreement to second order with the numerical results. Similar terms were
found for the other two constraints upon the state.

For the case of general measurements on states with fixed mean photon number,
it was found that the results for both the general case and the squeezed state
case agreed well with the analytic result. In addition the results for these
two cases agreed very closely with each other; more closely than with the
analytic approximation. In fact, when the states were compared, it was found
that they were extremely close, although there were small systematic
differences, particularly in the tails of the distribution.

These results mean
that it should be possible to experimentally produce very good approximations of
the states that are optimised for minimum phase variance under the various
measurement schemes by using squeezed states. This is an advantage in
theoretical work also, as squeezed states are more easily treated numerically.
Only the two squeezing parameters need be kept track of, rather than the entire
state, resulting in much faster calculation times. The exception to this is
canonical measurements, for which the results for general and squeezed states
have different scalings. The ratio between the two results only scales as
$\log \nb$, however, so the difference becomes significant only for very large
photon numbers.

\section{Optimum Dyne Measurements}
The next issue considered was the problem of how to make optimum measurements
of these states. Prior to this study it was known that the minimum phase
variance introduced by an arbitrary dyne measurement scheme was that of an
optimised squeezed state, with variance scaling as $\log \nb/\nb^2$. The best
adaptive dyne measurement scheme previously considered, the mark II scheme,
only gave scaling of $\nb^{-1.5}$. In this study I have found that it is
possible to obtain results that are very close to the theoretical limit
using a feedback phase estimate of
\beq
\hat\varphi_v = \arg {C_v^{1-\varepsilon(v)} A_v^{\varepsilon(v)}},
\eeq
with
\beq
\varepsilon(v) = \frac {v^2-|B_v|^2}{|C_v|} \sqrt{\frac v {1-v}}.
\eeq

If this is used without corrections, the variance as estimated from the variance
of the samples is very close to the theoretical limit. Unfortunately this method
of estimating the variance does not take account of very low probability results
with large phase error. This contribution can be estimated from the values of
$A$ and $B$ at the end of the measurement, and it is quite significant for this
uncorrected phase feedback scheme.

It is possible to correct the feedback scheme in order to minimise this problem.
It was found that very good results were obtained by using a correction that
brings the value of $B_v$ directly towards the estimated optimum value. This is
used only at the very end of the measurement, and then only if $\st{B_v}$ is too
far above optimum. Using this correction, the introduced phase variance as
estimated from the values of $A$ and $B$ is only about 4\% above the optimum
value, even for the largest photon number considered.

It was found that the phase variance of the samples when this corrected feedback
was used was slightly worse than that estimated from the values of $A$ and
$B$. This appears to be because of the large scatter in the photon number of the
squeezed states in the POM. The data points with smaller photon number give a
disproportionately large contribution to the phase variance, as the phase
variance for optimum squeezed states rises very rapidly as the photon number is
reduced. This is unavoidable, as any state with a small phase variance will have
a large uncertainty in the photon number. This is the case both for the input
state and the squeezed state in the POM. It is therefore reasonable to claim
that the corrected phase measurement scheme is about 4\% above what is
theoretically possible.

Under some circumstances, it is actually possible to surpass the theoretical
limit based on the POM. This is possible if the input state is known accurately.
For coherent states, it is possible to reduce the introduced phase variance to
about a quarter of the theoretical limit. Unfortunately, for states with reduced
phase uncertainty, this level of improvement does not appear to be possible. The
improvement found for squeezed states was only around 10\%.

\section{Time Delays}
All the adaptive feedback schemes considered are adversely affected by time
delays. This was considered previous to this study in Ref.~\cite{semiclass},
which predicted the excess phase variance for mark I and II measurements in a
very approximate way. In this study the excess phase variance due to time delays
was determined in a far more rigorous way, and verified by numerical
calculations.

In Ref.~\cite{semiclass} it was predicted that the excess phase variance for
mark I measurements was $\tau/2$. In this study it was found that this is the
correct result, provided the phase estimate at the end of the measurement is
the final value of the running phase estimate. This was found by repeating the
derivation of Ref.~\cite{semiclass} in a more rigorous way, and the same result
was obtained. Fitting to the numerical results gave scaling constants close to
the theoretical value of $\half$, particularly for the larger photon numbers.
There was some discrepancy for the smallest photon number considered, which can
be expected as the approximation is in the limit of large $\alpha$.

If, rather than using the final value of the intermediate phase estimate as the
final phase estimate, $\arg A$ is used, the phase variance actually decreases
with the time delay. This is because the intermediate phase estimates have a
larger variance, resulting in the mark I measurements being closer to
heterodyne measurements.

It was found that the scaling of the excess phase variance for mark II
measurements predicted in \cite{semiclass} was correct, though the scaling
constant was incorrect. When the derivation of Ref.~\cite{semiclass} was
repeated more rigorously, it was found that completely different terms were
obtained, casting some doubt on the method used in Ref.~\cite{semiclass}.
The terms found here are unusable, as they depend explicitly on the initial
conditions of the integration.

In order to avoid this problem, an alternative derivation was considered,
that gives a lower limit to the total introduced phase variance with a given
time delay, rather than the approximate excess phase variance due to the time
delay. This result was approximately $\tau/(8\nb)$, as opposed to the introduced
variance of $\tau/(2\nb)$ found in Ref.~\cite{semiclass}. This result can be
expected to give the introduced phase variance accurately if both the photon
number and time delay are relatively large. In contrast, the result found in
Ref.~\cite{semiclass} was for the limit of small $\alpha\tau$.

Numerically it was found that the introduced phase variance converges to this
theoretical limit in the three different cases with unsimplified feedback
considered. For the case of simplified feedback, however, the phase variance was
far higher. For larger time delays, rather than converging to the theoretical
lower limit, the phase variance converged to the variance for heterodyne
measurements. It was found to be possible to correct the feedback to improve on
these results, but the variance was still much higher than for
unsimplified feedback, and did not converge to the theoretical limit.

This means that, if there is any significant time delay in the system, the
simplified analog feedback will give far worse results than the exact,
unsimplified feedback. This makes the more sophisticated feedback schemes
considered in Ch.~\ref{optdyne} more attractive, as one of the main
advantages of the mark II measurement scheme was that it allowed the simplified
feedback to be used.

\section{Input States for Interferometry}
In Ch.~\ref{adaptiveinter} optimum input states for interferometry were
considered. When the total photon number is fixed, this problem is equivalent to
the dyne case with an upper limit on the photon number. This problem
has a simple solution, and there is a correspondingly simple solution in the
case of interferometry. This state has a canonical phase variance scaling as
$N^{-2}$, as compared to $N^{-1}$ when all the photons are incident on one port.

This state requires contributions from a large number of input photon
eigenstates; however, it requires significant contributions from only about 10.
This means that although the complete state will be very difficult to produce,
it should be less difficult to produce close approximations of it.
Unfortunately, as the photon number is increased, more and more of these
eigenstates are required to give a variance that is close to that for the exact
state.

An alternative state that has been considered in previous work, is one where
equal photon numbers are incident on each input port. This state is the photon
number eigenstate that has the highest contribution to the optimum state, and
can therefore be considered to be an approximation of it. Unfortunately this
state has a canonical phase variance that scales as $N^{-1/2}$. This indicates
that the phase uncertainty should scale as $N^{-1/4}$. This is an
extraordinary result considering that previous work indicated that this state
should give a phase uncertainty scaling as $N^{-1}$.

This discrepancy was shown to be due to the tails of the phase distribution,
which give the main contribution to the phase variance. The previous work, in
contrast, only considered the central peak of the distribution. To compare these
two states more thoroughly, these states were each evaluated under several
different measures of the uncertainty. It was found that the phase uncertainty
for the $\ket{j0}_z$ state scaled as $N^{-1}$ for all of the measures of the
phase uncertainty except the square root of the variance (standard and Holevo).

The optimal states gave smaller scaling constants under all of the measures
except two: the inverse-of-maximal-value and the Fisher length. There are
reasons to consider the results given by these measures as being slightly
misleading, however. In practice, the $\ket{j0}_z$ state will give results with
small errors most of the time, but a small number of results with very large
errors. This means that results obtained using this state must be carefully
analysed to avoid problems due to the results with large error. For example, if
we simply take the mean of a number of results obtained using this state, the
uncertainty in this mean will scale as $N^{-1/4}$ rather than $N^{-1}$.

\section{Optimum Interferometric Measurements}
In order to make interferometric phase measurements with the minimum possible
phase uncertainty, the other two areas to optimise are the feedback phases used
and the final phase estimate. It was shown in Ch.~\ref{interfere} that it is
reasonably simple to determine the final phase estimate that gives the minimum
uncertainty. Determining the optimum feedback phases to use is a far more
difficult problem.

Solving this problem for $N$ photons is a minimisation problem with
approximately $2^{N-2}$ independent variables. It is possible to solve it
numerically; however, the rapidly increasing number of variables means that this
is only feasible for photon numbers up to about 12. Even then there is a small
possibility of a better result for a significantly different combination of
numbers, as the numerical technique finds a local minimum, but does not prove
that it is a global minimum.

This numerical minimisation was performed for input states optimised for minimum
phase variance under canonical measurements. It is also possible to solve
simultaneously for the feedback phases and the input state. This is only a
slightly more difficult problem, as the number of state coefficients increases
linearly with $N$. Improvements over the optimum input states were found for
every photon number above 2; however, these improvements were only very small,
less than 1\%.

An alternative approach is to, rather than trying to determine the feedback
phase that will minimise the final phase estimate, determine the phase that
will simply minimise the phase uncertainty after the next detection. This
approach was used in Ch.~\ref{interfere}, and this feedback scheme gave
results that were the same as optimum for up to 4 photons with optimum input
states. For more than 4 photons the phase variances obtained were only slightly
higher than for the numerically optimised feedback. For the maximum photon
number for which the optimum feedback was determined, 12, the increase was only
about 3.6\%.

This feedback scheme was tested on optimum input states for higher photon
numbers up to 1600, and it was found that the phase variance was only slightly
higher than the canonical phase variance. That means that this feedback scheme
must be very close to optimum for these larger photon numbers as well.
Unfortunately, the ratio of the phase variance under this feedback scheme to the
canonical phase variance increases fairly systematically with photon number,
indicating that the introduced phase variance is not quite scaling as $N^{-2}$.
It is quite likely that the scaling is $\log N/N^2$, as for dyne measurements,
though it would require calculations at much larger photon numbers to confirm
this.

For input states other than the optimum input state, this feedback scheme gave
variances almost indistinguishable from the canonical phase variance. Of
particular interest is the $\ket{j0}_z$ state, for which the canonical phase
uncertainty scales as $N^{-1}$ for measures other than the square root of the
variance. For this state, the $2/3$ confidence interval scales approximately as
$N^{-0.69}$ under this measurement scheme. This is not as good as the $N^{-1}$
scaling for the canonical distribution, but it is an improvement on the result
if all photons are in one port. Note, however, that this measurement scheme
was designed with the aim of minimising the phase variance, not any other
measure of phase uncertainty. This means that the feedback phases and phase
estimates are not necessarily close to optimum for minimising the confidence
interval. Minimising the confidence interval would require quite a different
approach.

Alternative phase feedback schemes were also considered in Ch.~\ref{interfere}.
A nonadaptive scheme was considered, where rather than using information from
the detections to determine the feedback phase, the feedback phase was merely
varied linearly. This is equivalent to heterodyne measurements for the single
mode case, in that all values of the feedback phase are used with roughly equal
probability. This scheme gave a phase variance scaling as $N^{-1}$ for optimum
input states, similarly to the heterodyne scheme for single mode measurements.
When used on the state with all photons in one port (which has a canonical phase
variance scaling as $N^{-1}$) there was only a marginal increase in the phase
variance over the adaptive scheme.

An alternative adaptive feedback scheme based on phase estimates was found to
give far poorer results than the feedback scheme of Sec.~\ref{feedback}.
Although most of the results had small error, it gave a small proportion of
results with very large error, resulting in a high phase variance. On average
the phase variance was between that for the feedback scheme of
Sec.~\ref{feedback} and the nonadaptive scheme.

For 1 or 2 photons, the phase variances obtained by the feedback scheme of
Sec.~\ref{feedback} were identical to the phase variances for canonical
measurements, for any input state. This means that, provided we have no more
than 2 photons, it is possible to perform measurements as good as canonical. For
3 or 4 photons, however, the phase variance was higher than canonical. As the
feedback phases are optimum for these photon numbers, this means that it is not
possible to make canonical measurements in general, even with perfect
photodetectors and the best possible feedback phases. This result was checked by
evaluating the entire range of feedback phases, in order to demonstrate that the
feedback phases used gave a global minimum to the phase variance.

If arbitrary states are considered, this feedback scheme is no longer optimum
for 3 or 4 photons. In addition, it is possible to obtain phase variances that
are below canonical. This is not a contradiction, as the variance is not
necessarily minimised by having all the $H_{nm}$ equal to 1. For some states the
optimal POM should have complex values of $H_{nm}$. Using this corrected
canonical POM, the feedback scheme of Sec.~\ref{feedback} no longer gives
variances as small as canonical for all 2 photon states.

\section{Continuous Feedback}
The last area considered in this study was the problem of continuous phase
measurements, where the phase is being varied and the aim is to follow this
variation with the minimum possible excess uncertainty. For this problem, we
cannot use the optimum input states considered in other chapters, as these are
based on single-shot measurements with a limited number of photons. Instead,
I considered a continuous coherent state for dyne measurements, and a
state with all photons in one port for the interferometric case. These two cases
are extremely similar, with both having canonical phase variances proportional
to $N^{-1}$. In addition, continuous squeezed states for dyne measurements were
considered. In all cases the phase variation considered was Gaussian
diffusion.

In the case of dyne measurements it was found that good results were obtained
using $\arg A_t$ feedback, similarly to mark I and II single-shot measurements.
In the continuous case, the feedback simplifies to a very simple form, where the
feedback phase is adjusted proportional to the photocurrent. This form is even
simpler than for the single-shot case. When the correct proportionality
constant is selected, a minimum equilibrium phase variance is found that is
proportional to $\alpha^{-1}$. This is much poorer scaling than the case where
the system phase is constant, where the variance scales as $\alpha^{-2}$ (or
$\nb^{-1}$). This is because, as the photon number is increased, data
from a shorter time is used (to reduce the contribution to the variance from
the varying system phase), with the result that the effective photon number that
is used increases proportionally to $\alpha$ rather than $\alpha^2$.

When heterodyne feedback is used, rather than adaptive feedback, the phase
variance is increased by a factor of $\sqrt 2$. This is less than the
difference for the single-shot case for a similar reason as the scaling is
different: when a more accurate measurement can be made, the time interval from
which data is used is reduced. It was also found that the numerical results
matched these theoretical predictions very accurately, particularly for the
results with smaller phase variances.

For the case of dyne measurements on continuous squeezed states, the situation
is considerably more complicated. Rather than just a single constant that must
be varied to find the minimum phase variance there are three. Nevertheless, it
is still possible to obtain an approximate analytic result. It was found that
the minimum phase variance should vary as $\alpha^{-4/3}$. This is only slightly
better scaling than the case for coherent states.

Numerically it was found that the scaling of the minimum phase variance was very
close to the $\alpha^{-4/3}$ scaling found analytically. The numerical results
were well above the analytic result, however, on average more than twice.
This appears to be due to the contribution from large phase error results,
which were ignored in the analytic treatment. Numerically it was found that
when the parameters were varied beyond the optimal values found, the variance
increased due to these large error results.

The case for interferometry is more difficult to treat, as it does not give a
simple result for the feedback. The feedback used was based on minimising the
variance after the next detection, similarly to the single-shot case.
Nevertheless, it was found that it is possible to determine an approximate
theory that agrees reasonably well the numerical results. Similarly to the
dyne case with a coherent state, the phase variance is proportional to
$N^{-1/2}$, where $N$ is the photon flux. When a linearly changing feedback
phase was used (analogous to the heterodyne scheme), it was found that the phase
variance is above that for the adaptive feedback, but the difference is only
small, particularly for the smaller phase variances. This is as can be expected,
as the difference is very small for large photon numbers in the single-shot
case.

\section{Questions for Future Research}
Although the main aims of this project, finding the optimum states and
measurement schemes for phase measurements, have been achieved, this project has
also raised a number of unanswered questions that are possible future directions
for research.

\subsection{Optimum Dyne Measurements}
In Ch.~\ref{optdyne} a feedback scheme was found that produced dyne measurements
extremely close to the theoretical limit. Although there are good qualitative
reasons for expecting a feedback scheme of this type to be close to optimal,
there is no rigorous justification for why this scheme works. This question is
only of theoretical interest, as it would only tell us why a feedback scheme
that has already been found works, rather than leading to any better phase
measurements.

Another question is raised by the results showing that the theoretical limit
does not hold when different final phase estimates are used. An alternative
limit was found in the case of coherent states, but there is no corresponding
result for the case of squeezed states or more general states. When the
feedback that gives results close to the theoretical limit for the usual
$\arg C$ phase estimates was used, it was found that it is only possible to
improve the results slightly using better phase estimates. It is conceivable,
though unlikely, that it is possible to do better using some other feedback
scheme.

A more promising way of improving upon the theoretical limit would be to use
nonlinear elements. The basic reason why there is an introduced phase variance
is because the phase is not measured directly. The best we can do is to measure
a quadrature of the phase, which is proportional to the sine of the phase. The
introduced phase variance is due to the fact that the sine function is
nonlinear. In principle, it should be possible to compensate for this
nonlinearity using nonlinear optics.

Another possible area for further research is to consider alternative measures
of the phase uncertainty, as was done for interferometric measurements. For
the Holevo phase variance, it is necessary to take account of very low
probability, large error results in order to obtain an accurate estimate of the
variance. This is unnecessary for other measures of the uncertainty with less
emphasis on the tails. Nevertheless, I do not anticipate that this would give
qualitatively different results.

\subsection{Optimum Interferometry}
There are a number of promising areas for future study in the area of
interferometry. One is the question of whether the theoretical limit for dyne
measurements also holds for interferometric measurements. There is a great deal
of similarity between interferometric measurements and dyne measurements, as in
both cases we are measuring the phase difference between two modes combined at a
beam splitter. In principle it should be possible to take the large amplitude
limit in a similar way as for dyne measurements, in order to get a similar
theoretical limit.

The main problem with taking this limit is that we cannot assume that one mode
is of much larger amplitude than the other, as we can for dyne measurements.
Some other complications are that there will be quantum correlations between the
modes for interferometry, and the final phase estimate that is being used is not
$\arg C$. These factors mean that a theoretical limit would be very difficult to
find, and it need not be the same as for the case of dyne measurements.

Another area that can be considered is feedback where the feedback phase is
selected so as to minimise the phase variance two or more detections in advance,
rather than just one. This should improve on the feedback scheme that minimises
the variance after the next detection. In addition, if only very minor
improvements were obtained for feedback that minimises two or three detections
ahead, then this would indicate that the feedback is very close to optimum.

Unfortunately this approach would be very computationally intensive, as there
does not appear to be any analytic solution for the feedback phase. The feedback
phases would have to be determined numerically, making the calculations far more
time consuming, though not as much as for the case where the feedback phases are
chosen to minimise the final phase variance.

Another promising direction is using other measures of the phase uncertainty.
In particular it would be interesting to optimise the measurements for minimum
entropic length, as this would correspond to maximum information. It is
conceivable that minimising the entropic length one detection in advance would
also minimise the final entropic length. This problem would also be very
computationally intensive, as it would require numerical integrals to evaluate
the entropic length.

A more difficult problem is performing measurements that have zero error
probability, as in Ref.~\cite{Shapiro}. The measurements considered in
Ref.~\cite{Shapiro} are not physically possible, but it may be possible to
perform measurements with zero error probability using feedback. The states
considered in Ref.~\cite{Shapiro} do not have a fixed total photon number in the
two modes, as was assumed in this study, so this theory would have to be
substantially modified in order to consider these states.

I will lastly mention that it should be possible to generalise the theory
considered in this study to phase measurements with an arbitrary number of
modes. These are discussed in Ref.~\cite{multi}, and here it is the phase
differences between the modes that we wish to measure. This is a difficult
problem, as for example it does not seem to be possible to generalise the
optimal states for two-mode interferometry to this case in a simple way.

%% file: derivs.tex
\chapter{Longer Derivations}
In this appendix I give some of the longer derivations that are too lengthy
to present in the main text.

\section{Perturbation Theory for Optimum Dyne States}
\label{derper}

We wish to solve Eq.~(\ref{eigeq2}) of Sec.~\ref{gendyne} by perturbation
theory, with the unperturbed Hamiltonian and perturbation term given by
Eqs~(\ref{zeroth}) and (\ref{perth}) respectively. The unperturbed solution is
\beq
\psi_j^{(0)}(n) = \frac {f_{2}^{1/8}}{\sqrt{\pi^{1/2} 2^j j!}}  
\exp\left[-\rt{f_{2}}(n-n_0)^2/2 \right] H_j\left[ f_{2}^{1/4}(n-n_0)\right],
\eeq
where $H_j$ are Hermite polynomials. This is the standard result for the
harmonic potential, and is easily derived from the properties of Hermite
polynomials given in \cite{Erdelyi53}. Similarly to the other derivations given
in Ch.~\ref{dynesta}, the boundary condition is $\psi(0)=0$. The unperturbed
solution, and the more accurate perturbed solutions below, will approximately
obey this boundary condition when $\rt{f_{2}}n_0^{2} \gg 1$. For this to be the
case, $p<2$ is required, in addition to $n_0\gg 1$.

The energy eigenvalues are
\beq
E_j^{(0)}=(2j+1)\sqrt{f_{2}}.
\eeq
From perturbation theory we then have
\bqa
E_j \!\!\!\! & \approx & \!\!\!\! E_j^{(0)}+\bra{j}\hat H_1\ket{j}+\sum_{k\neq 
j}\frac{\left|\bra{k}\hat H_1\ket{j}\right|^2}{E_j^{(0)}-E_k^{(0)}}, \\
\psi_j(n) \!\!\!\! & \approx & \!\!\!\! \psi_j^{(0)}(n)+\sum_{k\neq j}
\frac{\bra{k}\hat H_1\ket{j}}{E_j^{(0)}-E_k^{(0)}}\psi_k^{(0)}(n),
\eqa
where $\ket{j}$ is the state corresponding to $\psi_j^{(0)}(n)$. In terms
of number states
\beq
\ket{j}=\sum_{n=0}^\infty \psi_j^{(0)}(n)\ket{n}.
\eeq
We can rewrite the perturbation as $\hat H_1=b\xi^3$, where
\bqa
b\!\!\!\! &=& \!\!\!\!-\frac{p+2}3 \left[cp(p+1)\right]^{1/4}n_0^{-p/4-3/2},\\
\xi\!\!\!\! &=& \!\!\!\!f_{2}^{1/4}(n-n_0).
\eqa
In terms of $\xi$ the unperturbed eigenstates are
\beq
\psi_j^{(0)}(\xi) = \frac 1{\sqrt{\pi^{1/2} 2^j j!}} e^{-\xi^2/2}H_j (\xi).
\eeq
The factor of $f_2^{1/8}$ has been omitted here, so these states are normalised
when integrated with respect to $\xi$.

The lowest energy eigenvalue and eigenstate, corresponding to the maximum value
of $\nu$ and therefore $\ip{\cos \phi}$, can be expressed as
\bqa
\label{lowpert}
E_0 \!\!\!\! & \approx & \!\!\!\! E_0^{(0)}+b \bra{0} \xi ^3 \ket{0}-
\frac{b^2}{2\sqrt{f_{2}}}
\sum_{k=1}^\infty\frac{\left|\bra{k}\hat \xi^3\ket{0}\right|^2}k ,\\
\psi_0(n) \!\!\!\! & \approx & \!\!\!\! \psi_0^{(0)}(n)-\frac {b}{2\sqrt{f_{2}}}
\sum_{k=1}^\infty\frac{\bra{k}\hat \xi^3\ket{0}}k \psi_k^{(0)}(n).
\eqa
Here $\hat \xi$ is the operator that transforms the state $\ket j$ in the same
way as $\xi$ transforms the function $\psi_j^{(0)}(\xi)$.

The first four Hermite polynomials are
\bqa
H_0(\xi) \!\!\!\! &=& \!\!\!\! 1, \\
H_1(\xi) \!\!\!\! &=& \!\!\!\! 2\xi, \\
H_2(\xi) \!\!\!\! &=& \!\!\!\! 4\xi^2 -2, \\
H_3(\xi) \!\!\!\! &=& \!\!\!\! 8\xi^3 -12\xi.
\eqa
Therefore the first four unperturbed eigenstates are
\bqa
\psi_0^{(0)}(\xi) \!\!\!\! &=& \!\!\!\! \pi^{-1/4} e^{-\xi^2/2}, \\
\psi_1^{(0)}(\xi) \!\!\!\! &=& \!\!\!\! \pi^{-1/4} e^{-\xi^2/2} \sqrt 2 \xi, \\
\psi_2^{(0)}(\xi) \!\!\!\! &=& \!\!\!\! \pi^{-1/4} e^{-\xi^2/2} \left(\sqrt 2
\xi^2-\frac 1{\sqrt 2}\right), \\
\psi_3^{(0)}(\xi) \!\!\!\! &=& \!\!\!\! \pi^{-1/4} e^{-\xi^2/2} \left(\frac 2
{\sqrt 3}\xi^3-\sqrt 3\xi \right).
\eqa
It is therefore easy to show that
\beq
\xi^3 \psi_0^{(0)}(\xi) = \frac 3{2\sqrt 2}\psi_1^{(0)}(\xi) +
\frac{\sqrt 3}2 \psi_3^{(0)}(\xi),
\eeq
or, in the alternative notation
\beq
\hat \xi^3 \ket 0 = \frac 3{2\sqrt 2}\ket 1 + \frac{\sqrt 3}2 \ket 3.
\eeq

From this it is evident that the only non-zero terms in the sums in
(\ref{lowpert}) are $\bra{1}\xi^3\ket{0}=3/(2\sqrt 2)$ and
$\bra{3}\xi^3\ket{0}={\sqrt 3}/2$. This then gives the lowest energy eigenvalue
and eigenstate as
\bqa
E_0 \!\!\!\! & \approx & \!\!\!\! E_0^{(0)}-b^2 \frac{11}{16}f_{2}^{-1/2} ,\\
\psi_0(n) \!\!\!\! & \approx & \!\!\!\! \psi_0^{(0)}(n)-\frac {b}
{4\sqrt{2f_{2}}}\left[3\psi_1^{(0)}+\sqrt\frac23 \psi_3^{(0)}\right].
\eqa
The correction term to the eigenvalue is of order $n_0^{-2}$, which is
sufficiently small to be omitted. The eigenstate can alternatively be expressed
as
\beq
\ket{\psi} \approx \ket 0 -\frac {3b}{4\sqrt{2f_{2}}}\ket 1
-\frac {b}{4\sqrt{3f_{2}}}\ket 3 .
\eeq

Therefore the expectation value of the photon number is
\beq
\ip{n} = \left(\bra 0 - \frac{3b}{4\sqrt{2f_{2}}}\bra 1 - 
\frac {b}{4\sqrt{3f_{2}}}\bra 3 \right)\left[\frac {\hat \xi}{(f_{2})^{1/4}}
+n_0 \right]\left(\ket 0 -\frac{3b}{4 \sqrt{2f_2}}\ket 1-
\frac{b}{4\sqrt{3f_{2}}}\ket 3 \right)
\eeq
In order to evaluate this, note first that we wish to keep terms only up to
first order in the perturbation, so we can simplify this to
\bqa
\bar n \!\!\!\! & \approx & \!\!\!\! n_0 + \left(\bra 0 - \frac{3b}
{4\sqrt{2f_{2}}}\bra 1 - \frac {b}{4\sqrt{3f_{2}}}\bra 3 \right)\left[\frac 
{\hat \xi}{(f_{2})^{1/4}}\right]\ket 0
+\bra 0 \left[\frac {\hat \xi}{(f_{2})^{1/4}}\right]\left(-\frac{3b}{4
\sqrt{2f_2}}\ket 1-\frac{b}{4\sqrt{3f_{2}}}\ket 3 \right) \nn \\
\!\!\!\! &=& \!\!\!\! n_0 + \left( \bra 0 - \frac{3b}{2\sqrt{2f_{2}}}
\bra 1 - \frac {b}{2\sqrt{3f_{2}}}\bra 3 \right)\left[\frac {\hat \xi}
{(f_{2})^{1/4}}\right]\ket 0.
\eqa

From the above listing of the eigenstates it can be seen that
\beq
\xi \psi_0^{(0)}(\xi) = \frac 1{\sqrt 2} \psi_1^{(0)}(\xi),
\eeq
or
\beq
\hat \xi \ket 0 = \frac 1{\sqrt 2} \ket 1.
\eeq
Therefore the only non-zero matrix element above is $\bra 1 \hat \xi \ket 0$.
Using this gives
\bqa
\bar n \!\!\!\! & \approx & \!\!\!\! n_0 - \frac {3b}{4(f_2)^{3/4}} \nn \\
\!\!\!\! &=& \!\!\!\! n_0 + \frac{p+2}{4\sqrt{cp(p+1)}}n_0^{p/2}.
\eqa
As we are assuming $p<2$,
\beq
n_0 \approx \bar n\left[1-\frac{p+2}{4\sqrt{cp(p+1)}}\bar n^{p/2-1}
\right],
\eeq
so the mean photon number is close to $n_0$, justifying the expansion
around $n_0$.

\section{Derivation for Optimum Squeezed States}
\label{sqzder}

Next the result (\ref{sqzresult}) of Sec.~\ref{sqzgeneral} will be derived in a
more rigorous way. Recall that the number state coefficients for squeezed states
are given by an expression that depends on Hermite polynomials
(\ref{sqznumber}). Hermite polynomials satisfy the recursion relation
\cite{Erdelyi53}
\beq
\label{Recurse}
H_{n+1}(x)-2xH_n(x)+2nH_{n-1}(x)=0.
\eeq
This can be used to derive the recursion relation for number state coefficients:
\beq
\braket{n+1}{\alpha,\zeta}\mu\sqrt{n+1}-\braket{n}{\alpha,\zeta}\beta
+\braket{n-1}{\alpha,\zeta}\nu\sqrt{n}=0.
\eeq
Rearranging this and squaring gives
\beq
|\braket{n+1}{\alpha,\zeta}|^2\mu^2(n+1)=|\braket{n}{\alpha,\zeta}|^2
\beta^2+|\braket{n-1}{\alpha,\zeta}|^2\nu^2n-2\braket{\alpha,\zeta}{n}
\braket{n-1}{\alpha,\zeta}\beta\nu\sqrt{n}.
\eeq
Multiplying this by $n^k$ and summing gives
\bqa
\sum_{n=1}^{\infty} |\braket{n+1}{\alpha,\zeta}|^2\mu^2(n+1)n^k \!\!\!\! &=&
\!\!\!\! \sum_{n=1}^{\infty} |\braket{n}{\alpha,\zeta}|^2\beta^2n^k+
\sum_{n=1}^{\infty} |\braket{n-1}{\alpha,\zeta}|^2\nu^2n^{k+1} \nn \\
&& -\sum_{n=1}^{\infty} 2\braket{\alpha,\zeta}{n} \braket{n-1}{\alpha,\zeta}
\beta\nu n^{k+1/2} \nn \\
2\beta\nu \sum_{n=1}^{\infty} n^{k+1/2} \braket{\alpha,\zeta}{n}
\braket{n-1}{\alpha,\zeta} \!\!\!\! &=& \!\!\!\!
\beta^2 \sum_{n=1}^{\infty} n^k |\braket{n}{\alpha,\zeta}|^2+
\nu^2 \sum_{n=1}^{\infty} n^{k+1} |\braket{n-1}{\alpha,\zeta}|^2 \nn \\
&& -\mu^2 \sum_{n=1}^{\infty} (n+1)n^k |\braket{n+1}{\alpha,\zeta}|^2 \nn \\
2\beta\nu \sum_{n=0}^{\infty} (n+1)^{k+1/2} \braket{\alpha,\zeta}{n}
\braket{n+1}{\alpha,\zeta} \!\!\!\! &=& \!\!\!\!
\beta^2 \sum_{n=1}^{\infty} n^k |\braket{n}{\alpha,\zeta}|^2+
\nu^2 \sum_{n=0}^{\infty} (n+1)^{k+1} |\braket{n}{\alpha,\zeta}|^2 \nn \\
&& -\mu^2 \sum_{n=2}^{\infty} n(n-1)^k |\braket{n}{\alpha,\zeta}|^2
\eqa
If $k>0$, then the sums in the first and third terms on the right hand side can
be extended to $n=0$, giving
\beq
\label{kg0}
2\beta\nu\sum_{n=0}^\infty (n+1)^{k+1/2} \braket{\alpha,\zeta}{n}\braket{n+1}
{\alpha,\zeta} = \beta^2\ip{n^k}+\nu^2\ip{(n+1)^{k+1}}-\mu^2\ip{n(n-1)^k}.
\eeq

In the following we wish to take $k = \half -p$, so $k \le 0$. In this case the
sums cannot be extended to zero, and in fact the additional terms would be
infinite. In the following expansions, however, only the behaviour of the state
near $n=\bar n$ is considered, and the contribution from $n \sim 0$ is
negligible. It is therefore reasonable to use Eq.~(\ref{kg0}) as a basis
for the approximate expansions, despite the fact that some of the terms would be
infinite if they were worked out exactly.

Taking $k = \half -p$ and considering the deviation from the mean photon number
gives
\begin{eqnarray}
2\beta\nu\sum_{n=0}^\infty (n+1)^{-p}\braket{\alpha,\zeta}{n}\braket{n+1}
{\alpha,\zeta}\!\!\!\! & \approx & \!\!\!\!\beta^2\ip{(\nb+\Delta n)^{-(p+1/2)}}
+\nu^2\ip{(\nb+1+\Delta n)^{-(p-1/2)}} \nn \\ -\mu^2\ip{(\nb-1+\Delta n)^
{-(p-1/2)}} \!\!\!\!&-&\!\!\!\! \mu^2\ip{(\nb-1+\Delta n)^{-(p+1/2)}}.
\eqa
Expanding this in a series in $\Delta n$ gives
\begin{eqnarray}
\label{series}
2\beta\nu\sum_{n=0}^\infty (n+1)^{-p} \braket{\alpha,\zeta}{n}\braket{n+1}
{\alpha,\zeta} \approx \sum_{j=0}^\infty \frac {(-1)^j\ip{\Delta n^j}}{j!}
\left\{ \frac{(p+j-1/2)!}{(p-1/2)!}\left(\beta^2\nb^{-(p+j+1/2)} \right. \right.
\nn \\ \left. \left. -\mu^2 (\nb-1)^{-(p+j+1/2)}\right)
+\frac{(p+j-3/2)!}{(p-3/2 )!}\left[\nu^2
(\nb+1)^{-(p+j-1/2)}-\mu^2(\nb-1)^{-(p+j-1/2)}\right]\right\}.
\eqa

Now we have an expression that can be used to evaluate Eq.~(\ref{thirdterm}).
Here the approximation that will be used for $h(n)$ is $c(n+1)^{-p}$ rather than
$cn^{-p}$. This is reasonable, as the difference is of order $\nb^{-(p+1)}$,
which is of higher order than will be considered here. Using the first three
terms of the sum in Eq.~(\ref{series}) gives
\bqa
&& 2\beta\nu\sum_{n=0}^\infty (n+1)^{-p} \braket{\alpha,\zeta}{n}\braket{n+1}
{\alpha,\zeta} \approx \beta^2\nb^{-(p+1/2)} -\mu^2(\nb-1)^{-(p+1/2)}+\nu^2
(\nb+1)^{-(p-1/2)} \nn \\ && -\mu^2(\nb-1)^{-(p-1/2)}+ \frac{\ip{\Delta n^2}}{2}
\left\{ (p+3/2)(p+1/2) \left(\beta^2 \nb^{-(p+5/2)}-\mu^2 (\nb-1)^{-(p+5/2)}
\right) \right. \nn \\ && \left. +(p+1/2)(p-1/2)\left[\nu^2(\nb+1)^{-(p+3/2)}
-\mu^2(\nb-1)^{-(p+3/2)}\right]\right\}
\eqa
Note that the $j=1$ term of the sum is zero, as $\ip{\Delta n}=0$. Using
$\ip{\Delta n^2}=\alpha^2(\mu-\nu)^2+2\mu^2\nu^2$ this equation becomes
\begin{eqnarray}
\label{first3}
&& 2\beta\nu\sum_{n=0}^\infty (n+1)^{-p}\braket{\alpha,\zeta}{n}\braket
{n+1}{\alpha,\zeta} \approx \beta^2\nb^{-(p+1/2)}-\mu^2(\nb-1)^{-(p+1/2)}
+\nu^2(\nb+1)^{-(p-1/2)} \nn \\ && -\mu^2(\nb-1)^{-(p-1/2)}
 + \left[\frac {\alpha^2(\mu-\nu)^2}{2}+\mu^2\nu^2\right]
\left\{ \left(p^{2}+2p+\frac 34\right)\left(\beta^2\nb^{-(p+5/2)} \right.\right.
\nn \\ &&\left. \left. -\mu^2(\nb-1)^{-(p+5/2)}\right)+\left(p^{2}-\frac 14
\right)\left[\nu^2(\nb+1)^{-(p+3/2)}-\mu^2(\nb-1)^{-(p+3/2)}\right]\right\}.
\eqa

At this stage the main problem is to determine which terms should be kept. This
depends on how $n_0$ scales with $\bar n$. Recall that if the state is optimised
for minimum intrinsic phase uncertainty, then $n_0\propto\log(\nb)$. That
scaling cannot be assumed in this case; however, it is possible to make some
general assumptions about the scaling of $n_0$. Firstly $n_0$ should increase
with $\nb$, so in order to obtain two orders in the approximation we will keep
terms up to leading order divided by $n_0$.

Secondly $n_0$ should not increase as rapidly as $\nb$, as the squeezing should
increase as the photon number is increased. This means that $n_0/\nb^2$ should
be higher order than $1/n_0$, and we can therefore omit terms of leading order
times $n_0/\nb^2$. We will, however, include terms of leading order times
$n_0 /\nb$, as we cannot assume that these are higher order. We will not obtain
any terms of leading order times $(n_0 /\nb)^2$ or any higher power, so we do
not need to consider these terms.

The third and fourth terms in (\ref{first3}) appear to be of order
$\nb^{-(p-3/2)}/n_0$. We can partially cancel these terms, however, by making
the expansion
\beq
(\nb \pm 1)^{-(p-1/2)} = \nb^{-(p-1/2)}\left[ 1 \mp \frac{(p-\half)}{\nb} 
+\frac{(p+\half)(p-\half)}{2\nb^2} + O(\nb^{-3}) \right].
\eeq
This means that
\bqa
\nu^2(\nb+1)^{-(p-1/2)}-\mu^2(\nb-1)^{-(p-1/2)} \!\!\!\! & \approx & \!\!\!\!
(\nu^2-\mu^2)\nb^{-(p-1/2)}\left[ 1+\frac{(p+\half)(p-\half)}{2\nb^2}\right]
\nn \\ && - (\nu^2+\mu^2)(p-\half) \nb^{-(p+1/2)}.
\eqa
Now $\nu^2-\mu^2=-1$, so together these terms are of order $\nb^{-(p-1/2)}$.
This is the leading order, and we will now omit terms known to be higher order
than $\nb^{-(p-1/2)}/n_0$. This means that we will use
\beq
\nu^2(\nb+1)^{-(p-1/2)}-\mu^2(\nb-1)^{-(p-1/2)} \approx -\nb^{-(p-1/2)}
- (\nu^2+\mu^2)(p-\half) \nb^{-(p+1/2)}.
\eeq

Similarly for the terms on the third line of (\ref{first3}) there is the
expansion
\beq
(\nb \pm 1)^{-(p+3/2)}=\nb^{-(p+3/2)}\left[ 1 \mp \frac{(p+\smallfrac 32)}{\nb}
+\frac{(p+\smallfrac 52)(p+\smallfrac 32)}{2\nb^2} + O(\nb^{-3}) \right],
\eeq
so
\beq
\nu^2(\nb+1)^{-(p+3/2)}-\mu^2(\nb-1)^{-(p+3/2)} \approx 
-\nb^{-(p+3/2)} - (\nu^2+\mu^2)(p+\smallfrac 32) \nb^{-(p+5/2)}.
\eeq
Using this result, and omitting all terms known to be of higher order than
$\nb^{-(p-1/2)}/n_0$, Eq.~(\ref{first3}) simplifies to
\bqa
&& 2\beta\nu\sum_{n=0}^\infty (n+1)^{-p}\braket{\alpha,\zeta}{n}\braket
{n+1}{\alpha,\zeta} \approx \beta^2\nb^{-(p+1/2)}-\mu^2(\nb-1)^{-(p+1/2)}
-\nb^{-(p-1/2)} \nn \\ &&- (\nu^2+\mu^2)(p-\half) \nb^{-(p+1/2)}
 + \left[\frac {\alpha^2(\mu-\nu)^2}{2}\right]
\left(p^{2}-\frac 14 \right)\left[ -\nb^{-(p+3/2)} \right].
\eqa

Now using $\beta^2 \approx n_0$, $\mu^2 \approx \nu^2 \approx \nb /(4n_0)$ and
$\alpha (\mu - \nu)^2 \approx \nb^2/n_0$, this simplifies to
\beq
2\beta\nu\sum_{n=0}^\infty (n+1)^{-p}\braket{\alpha,\zeta}{n}\braket
{n+1}{\alpha,\zeta} \approx - \nb^{-(p-1/2)}\left\{ 1 - \frac{n_0}{\nb}
+\frac {1}{2n_0}\left[ p(p+1)-\frac 14 \right] \right\}.
\eeq
Now we can expand $2\beta\nu$ to give
\bqa
2\beta\nu \!\!\!\! &=& \!\!\!\! -\bar n^{1/2}\sqrt{ 1- \frac {\nu^2}{\nb}}
\left(1- \frac {n_0}{\bar n} \right) \nn \\
\!\!\!\! & \approx & \!\!\!\! -\bar n^{1/2}\sqrt{1- \frac 1{4n_0}}
\left(1- \frac {n_0}{\nb} \right).
\eqa
Using this gives
\bqa
\left(1- \frac {n_0}{\bar n} \right)\sum_{n=0}^\infty (n+1)^{-p}
\braket{\alpha,\zeta}{n}\braket{n+1}{\alpha,\zeta} \!\!\!\! & \approx & \!\!\!\!
\nb^{-p}\frac{\left\{ 1-\frac {n_0}{\nb}+\frac {1}{2n_0}\left[
p(p+1)-\frac 14 \right] \right\}}{\sqrt{1- \frac 1{4n_0}}} \nn \\
\!\!\!\! & \approx & \!\!\!\! \nb^{-p}\left\{ 1-\frac {n_0}{\nb}
+\frac {p(p+1)}{2n_0} \right\}.
\eqa
The two terms of $n_0/\nb$ just cancel, giving the simple result
\beq
\label{appresult}
\sum_{n=0}^\infty (n+1)^{-p} \braket{\alpha,\zeta}{n}\braket{n+1}
{\alpha,\zeta} \approx \nb^{-p}\left[ 1+\frac {p(p+1)}{2n_0} \right].
\eeq
This result is identical to that obtained simply using $\ip{n^{-p}}$.

\section{Perturbation Theory for Mark II Measurements}
\label{pert}

From Sec.~\ref{mark2}, the phase variance for mark II measurements with a
time delay is, according to perturbation theory
\beq
\label{m2pert}
\ip{\phi_{\rm II}^2}=\int_{v_1}^1 dt \int_{v_1}^1 dt' \ip{\hat\varphi_t^{(0)}
\hat\varphi_{t'}^{(0)}} + 2\alpha\tau\int_{v_1}^1 dt \int_{v_1}^1 dt'
\ip{\hat\varphi_t^{(0)} \hat\varphi_{t'}^{(1)}}.
\eeq
Using the result for $\hat\varphi_t^{(0)}$ given in Eq.~(\ref{zerosol}), the
first term can be evaluated as
\bqa
\label{twoterms}
&& \!\!\!\!\!\!\!\!\!\!\!\!\!\!\!\! \int_{v_1}^1 dt \int_{v_1}^1 dt'\left\langle
\left[e^{4\alpha(\sqrt{v_1}-\sqrt t)}\hat\varphi_{v_1}^{(0)}+\int_{v_1}^t \frac
{e^{4\alpha(\sqrt u-\sqrt t)}}{\sqrt u} dW(u) \right] \right. \nn \\ && \left.
\times\left[e^{4\alpha(\sqrt{v_1}-\sqrt{t'})}\hat\varphi_{v_1}^{(0)}+\int_{v_1}^
{t'} \frac {e^{4\alpha(\sqrt u-\sqrt{t'})}}{\sqrt u} dW(u) \right]\right\rangle
\nn \\ \!\!\!\! &=& \!\!\!\!
\left[\int_{v_1}^1 e^{4\alpha(\sqrt {v_1}-\sqrt t)}dt\right]^2\ip{(\hat
\varphi_{v_1}^{(0)})^2}+\int_{v_1}^1 dt\int_{v_1}^1 dt'\int_{v_1}
^{\min(t,t')}du\frac{e^{4\alpha(2\sqrt u-\sqrt t-\sqrt{t'})}}u \nn \\ \!\!\!\!
&=& \!\!\!\! \left[\int_{v_1}^1 e^{4\alpha(\sqrt {v_1}-\sqrt t)}dt\right]^2
\ip{(\hat\varphi_{v_1}^{(0)})^2}+2\int_{v_1}^1 dt \int_{v_1}^t dt'
\int_{v_1}^{t'} du \frac{e^{4\alpha(2\sqrt u-\sqrt t-\sqrt{t'})}}u.
\eqa
Considering the first term, this simplifies to
\bqa
&&\left[ 2\int_0^{1-\sqrt{v_1}}(\sqrt{v_1}+s)e^{-4\alpha s} ds \right]^2
\ip{(\hat\varphi_{v_1}^{(0)})^2} \nn \\
&& = \left[ \frac{\sqrt{v_1}}{2\alpha}+\frac 1{8\alpha^2} \right]^2
\ip{(\hat\varphi_{v_1}^{(0)})^2} \nn \\
&& = \left[ \frac{v_1}{4\alpha^2}+\frac{\sqrt{v_1}}{8\alpha^3}+
\frac 1{64\alpha^4} \right] \ip{(\hat\varphi_{v_1}^{(0)})^2}.
\eqa
The term that we take to be leading order here depends on how we take the
limits. If we consider the limit of large $\alpha$ with a fixed value of $v_1$,
then the first term is largest. On the other hand, if we consider the limit of
small $v_1$ the third term is the largest. As the variance should be small at
time $v_1$, a large number of detections should have been made by this time.
This means that we should have $\alpha^2 v_1 \gg 1$. In this limit, the first
term above is largest.

Now consider the second term in Eq.~(\ref{twoterms}). Changing variables to
$s=\sqrt{t'}-\sqrt u$ gives
\bqa
\!\!\!\! && \!\!\!\! 2\int_{v_1}^1 dt \int_{v_1}^t dt' \int_{\sqrt{t'}-
\sqrt{v_1}}^0 [-2(\sqrt{t'}-s)ds] e^{4\alpha(\sqrt{t'}-\sqrt t)}
\frac{e^{-8\alpha s}}{(\sqrt{t'}-s)^2} \nn \\ \!\!\!\! &=& \!\!\!\!
4\int_{v_1}^1 dt \int_{v_1}^t dt' e^{4\alpha(\sqrt{t'}-\sqrt t)}
\int_0^{\sqrt{t'}-\sqrt{v_1}} ds \frac{e^{-8\alpha s}}{\sqrt{t'}-s}.
\eqa
Performing an expansion in $s$,
\bqa
\!\!\!\! && \!\!\!\!4\int_{v_1}^1 dt \int_{v_1}^t dt' e^{4\alpha(\sqrt{t'}-
\sqrt t)}\int_0^{\sqrt{t'}-\sqrt{v_1}} \left(\frac 1 {\sqrt{t'}} +\frac s {t'}+
O(s^2)\right)e^{-8\alpha s} ds \nn \\ \!\!\!\! &=& \!\!\!\!
4\int_{v_1}^1 dt \int_{v_1}^t dt' e^{4\alpha(\sqrt{t'}-\sqrt t)}
\left(\frac 1 {\sqrt{t'}8\alpha} +\frac 1 {t'(8\alpha)^2}+O(\alpha^
{-3})\right).
\eqa
Next we substitute $s=\sqrt t-\sqrt{t'}$ so that $dt'=-2(\sqrt t-s)ds$. This
gives
\bqa
\!\!\!\! && \!\!\!\! 8\int_{v_1}^1 dt \int_0^{\sqrt t-\sqrt{v_1}}
e^{-4\alpha s} \left(\frac 1 {8\alpha} + \frac 1{(\sqrt t-s)(8\alpha)^2}+
O(\alpha^{-3})\right)ds.
\eqa
Again expanding in a series we obtain
\bqa
\!\!\!\! && \!\!\!\! 8\int_{v_1}^1 dt \int_0^{\sqrt t-\sqrt{v_1}}
e^{-4\alpha s}\left( \frac 1{8\alpha} +\frac 1{\sqrt t (8\alpha)^2}+\frac s
{t(8\alpha)^2}+O(s^2 \alpha^2)+O(\alpha^{-3}) \right) ds \nn \\ \!\!\!\! &=&
\!\!\!\! 8\int_{v_1}^1 \left(\frac 1{(8\alpha)(4\alpha)}+O(\alpha^{-3})\right)
dt \nn \\ \!\!\!\! &=& \!\!\!\! \frac {1-v_1}{4\alpha^2}+O(\alpha^{-3}).
\eqa
In the limit of small $v_1$ this simplifies to $1/(4\alpha^2)$. Note that the
higher order terms diverge if the limit $v_1 \to 0$ is taken for fixed $\alpha$.
However, as we are taking $v_1$ such that $\alpha^2 v_1 \gg 1$, they will be
smaller than the term given here.

Thus we find that the first term of Eq.~(\ref{m2pert}) is
\beq
\int_{v_1}^1 dt \int_{v_1}^1 dt' \ip{\hat\varphi_t^{(0)}\hat\varphi_{t'}^{(0)}}
=\frac 1{4\alpha^2}+\frac {v_1}{4\alpha^2}\ip{(\hat\varphi_{v_1}^{(0)})^2}
+O(\alpha^{-3}).
\eeq

Next consider the second term in Eq.~(\ref{m2pert}). Evaluating this gives
\bqa
\label{verynasty}
\!\!\!\!&&\!\!\!\!\int_{v_1}^1 dt \int_{v_1}^1 dt'\left\langle\left[e^{4\alpha
(2\sqrt{v_1}-\sqrt t-\sqrt{t'})}\hat\varphi_{v_1}^{(0)}\hat\varphi_{v_1}^{(1)}
-\int_{v_1}^{t'}\frac {4\alpha}se^{4\alpha(2\sqrt{v_1}-\sqrt t-\sqrt{t'})}
(\hat\varphi_{v_1}^{(0)})^2 ds \right. \right. \nn \\
&& -\int_{v_1}^{t'}\frac{4\alpha}s ds\int_{v_1}^t\int_{v_1}^s \frac
{e^{4\alpha(\sqrt v-\sqrt t)}}{\sqrt v}dW(v)\frac {e^{4\alpha(\sqrt u-
\sqrt{t'})}}{\sqrt u}dW(u) \nn \\ &&\left.\left. +\int_{v_1}^{t}\int_{v_1}^{t'}
\frac{e^{4\alpha(\sqrt v-\sqrt t)}}{\sqrt v}dW(v)\frac 2u e^{4\alpha(\sqrt u-
\sqrt{t'})}dW(u)\right]\right\rangle \nn \\ \!\!\!\! &=&
\!\!\!\! \int_{v_1}^1 dt \int_{v_1}^1 dt' e^{4\alpha(2\sqrt{v_1}-\sqrt t-
\sqrt{t'})}\ip{\hat\varphi_{v_1}^{(0)}\hat\varphi_{v_1}^{(1)}} \nn \\
&&+\int_{v_1}^1 dt \int_{v_1}^1 dt'\left[4\alpha(\log{v_1}-\log{t'})
e^{4\alpha(2\sqrt{v_1}-\sqrt t-\sqrt{t'})}\ip{(\hat\varphi_{v_1}^{(0)})^2}
\right. \nn \\ &&\left. -\int_{v_1}^{t'}\frac{4\alpha}s ds\int_{v_1}^{\min(t,s)}
\frac {e^{4\alpha(2\sqrt u-\sqrt t-\sqrt{t'})}}u du 
+2\int_{v_1}^{\min(t,t')}\frac{e^{4\alpha(2\sqrt u-\sqrt t-\sqrt{t'})}}
{u^{1.5}} du\right].
\eqa
The first term simplifies to
\beq
\left[ \frac{v_1}{4\alpha^2}+\frac{\sqrt{v_1}}{8\alpha^3}+
\frac 1{64\alpha^4} \right]
\ip{\hat\varphi_{v_1}^{(0)}\hat\varphi_{v_1}^{(1)}}.
\eeq
As was discussed above, the first term here will be largest when we take $v_1$
such that $\alpha^2 v_1 \gg 1$.

Now consider the second term in Eq.~(\ref{verynasty}). Taking the integral
over $t$ this simplifies to
\beq
\left(2\sqrt{v_1}+\frac 1{2\alpha}\right) \int_{v_1}^1 (\log{v_1}-\log{t'})
e^{4\alpha(\sqrt{v_1}-\sqrt{t'})}\ip{(\hat\varphi_{v_1}^{(0)})^2} dt'.
\eeq
Now we make the substitution $s=\sqrt{t'}-\sqrt{v_1}$ so $dt'=2(s+\sqrt{v_1})
ds$, giving
\bqa
&& \!\!\!\!\!\!\!\! -\left(2\sqrt{v_1}+\frac 1{2\alpha}\right)
\int_0^{1-\sqrt{v_1}}\log\left(\frac{(s+\sqrt{v_1})^2}{v_1}\right)e^{-4\alpha s}
\ip{(\hat\varphi_{v_1}^{(0)})^2} [2(s+\sqrt{v_1})ds] \nn \\
\!\!\!\! &=& \!\!\!\! -\left(8\sqrt{v_1}+\frac 2{\alpha}\right)
\int_0^{1-\sqrt{v_1}}\log\left(1+\frac s{\sqrt{v_1}}\right)e^{-4\alpha s}
\ip{(\hat\varphi_{v_1}^{(0)})^2} [(s+\sqrt{v_1})ds].
\eqa
Again, there is a slight problem, as the result obtained depends on how the
limits are taken. The contributions to this integral will only be significant
when $s$ is less than about $1/\alpha$. As we should have $\alpha^2 v_1 \gg 1$,
this means we should have $s/\sqrt{v_1} \ll 1$. In this limit, we obtain
\bqa
\!\!\!\! && \!\!\!\! -\left(8\sqrt{v_1}+\frac 2{\alpha}\right)
\int_0^{1-\sqrt{v_1}}\left(s+\frac {s^2}{2 \sqrt{v_1}}+O(s^3)\right)
e^{-4\alpha s}\ip{(\hat\varphi_{v_1}^{(0)})^2} ds \nn \\
\!\!\!\! &=& \!\!\!\!
-\left(8\sqrt{v_1}+\frac 2{\alpha}\right) \left(\frac 1{(4\alpha)^2}+\frac 1
{\sqrt{v_1}(4\alpha)^3}+O(\alpha^{-4})\right)\ip{(\hat\varphi_{v_1}^{(0)})^2}
\nn \\ \!\!\!\! &=& \!\!\!\! -\left(\frac {\sqrt{v_1}}{2\alpha^2}+\frac 1
{4\alpha^3}+O(\alpha^{-4})\right)\ip{(\hat\varphi_{v_1}^{(0)})^2}.
\eqa

Next we will consider the third term. To treat the minimum we must split the
integral into three parts:
\bqa
&&\!\!\!\!\!\!-\int_{v_1}^1 dt \int_{v_1}^1 dt' \int_{v_1}^{t'}\frac{4\alpha}v
dv \int_{v_1}^{\min(t,v)}\frac {e^{4\alpha(2\sqrt u-\sqrt t-\sqrt{t'})}}u du
\nn \\
&&\!\!\!\!\!\!\!\! =-\int_{v_1}^1 dt \int_{v_1}^t dt' \int_{v_1}^{t'}
\frac{4\alpha}v dv \int_{v_1}^v\frac {e^{4\alpha(2\sqrt u-\sqrt t-\sqrt{t'})}}
u du \nn \\
&&-\int_{v_1}^1 dt \int_t^1 dt' \int_t^{t'}\frac{4\alpha}v dv \int_{v_1}^t
\frac {e^{4\alpha(2\sqrt u-\sqrt t-\sqrt{t'})}}u du \nn \\
&&-\int_{v_1}^1 dt \int_t^1 dt' \int_{v_1}^t\frac{4\alpha}v dv \int_{v_1}^v
\frac {e^{4\alpha(2\sqrt u-\sqrt t-\sqrt{t'})}}u du \nn \\
&&\!\!\!\!\!\!\!\! =-2\int_{v_1}^1 dt \int_{v_1}^t dt' \int_{v_1}^{t'}
\frac{4\alpha}v dv \int_{v_1}^v\frac {e^{4\alpha(2\sqrt u-\sqrt t-\sqrt{t'})}}
u du \quad\quad {\rm (a)}\nn \\
&&-\int_{v_1}^1 dt \int_t^1 dt' \int_t^{t'}\frac{4\alpha}v dv \int_{v_1}^t
\frac {e^{4\alpha(2\sqrt u-\sqrt t-\sqrt{t'})}}u du. \quad\quad {\rm (b)}
\eqa
Considering term (a), this simplifies to
\bqa
&&-2\int_{v_1}^1 dt \int_{v_1}^t dt' \int_{v_1}^{t'} \frac{4\alpha}v dv
\int_0^{\sqrt{v}-\sqrt{v_1}}\frac {e^{-8\alpha s}}{\sqrt{v}-s}
e^{4\alpha(2\sqrt v-\sqrt t-\sqrt{t'})} ds \nn \\
&&= -2\int_{v_1}^1 dt \int_{v_1}^t dt' \int_{v_1}^{t'} \frac{8\alpha}v dv
\int_0^{\sqrt{v}-\sqrt{v_1}}\left(\frac 1{\sqrt{v}}+\frac sv +O(s^2) \right)
e^{-8\alpha s}e^{4\alpha(2\sqrt v-\sqrt t-\sqrt{t'})} ds \nn \\
&&= -2\int_{v_1}^1 dt \int_{v_1}^t dt' \int_{v_1}^{t'} \frac{8\alpha}v dv
\left(\frac 1{8\alpha\sqrt{v}}+\frac s{(8\alpha)^2 v} +O(\alpha^{-3}) \right)
e^{4\alpha(2\sqrt v-\sqrt t-\sqrt{t'})} \nn \\
&&= -4\int_{v_1}^1 dt \int_{v_1}^t dt' \int_{0}^{\sqrt{t'}-\sqrt{v_1}}
\left(\frac 1{(\sqrt{t'}-s)^2}+\frac 1{(8\alpha)(\sqrt{t'}-s)^3}+O(\alpha^{-3})
\right) e^{4\alpha(\sqrt{t'}-\sqrt t)}e^{-8\alpha s} ds \nn \\
&&= -4\int_{v_1}^1 dt \int_{v_1}^t dt' \int_{0}^{\sqrt{t'}-\sqrt{v_1}}
\left(\frac 1{t'}+\frac {2s}{t'^{1.5}}+\frac 1{(8\alpha)t'^{1.5}}+O(\alpha^{-3})
\right) e^{4\alpha(\sqrt{t'}-\sqrt t)}e^{-8\alpha s} ds \nn \\
&&= -\frac 1{2\alpha}\int_{v_1}^1 dt \int_{v_1}^t dt' \left(\frac 1{t'}+\frac 3
{(8\alpha)t'^{1.5}}+O(\alpha^{-3})\right)e^{4\alpha(\sqrt{t'}-\sqrt t)} \nn \\
&&= -\frac 1{\alpha}\int_{v_1}^1 dt \int_0^{\sqrt{t}-\sqrt{v_1}} \left(\frac
1{\sqrt{t}-s}+\frac 3{(8\alpha)(\sqrt{t}-s)^2}+O(\alpha^{-2})\right)
e^{-4\alpha s} ds \nn \\
&&= -\frac 1{\alpha}\int_{v_1}^1 dt \int_0^{\sqrt{t}-\sqrt{v_1}} \left(\frac 1
{\sqrt{t}}+\frac st+\frac 3{8\alpha t}+O(\alpha^{-2})\right)e^{-4\alpha s}ds
\nn \\
&&= -\frac 1{4\alpha^2}\int_{v_1}^1 dt \left(\frac 1{\sqrt{t}}+\frac 5{8\alpha
t}+O(\alpha^{-2})\right).
\eqa
Term (b) simplifies to
\bqa
&&-4\alpha\int_{v_1}^1 dt \int_t^1 dt' [\log{t'}-\log{t}] \int_{v_1}^t
\frac {e^{4\alpha(2\sqrt u-\sqrt t-\sqrt{t'})}}u du \nn \\
&&=-8\alpha\int_{v_1}^1 dt \int_t^1 dt' [\log{t'}-\log{t}] \int_0^{\sqrt{t}-
\sqrt{v_1}}\frac{e^{-8\alpha s}}{\sqrt{t}-s}e^{4\alpha(\sqrt t-\sqrt{t'})} ds
\nn \\
&&=-8\alpha\int_{v_1}^1 dt \int_t^1 dt' [\log{t'}-\log{t}] \int_0^{\sqrt{t}-
\sqrt{v_1}}\left(\frac 1{\sqrt{t}}+\frac st\right)e^{-8\alpha s}
e^{4\alpha(\sqrt t-\sqrt{t'})} ds \nn \\
&&=-\int_{v_1}^1 dt \int_t^1 dt' [\log{t'}-\log{t}]\left(\frac 1{\sqrt{t}}+
\frac 1{8\alpha t}+O(s^2)\right)e^{4\alpha(\sqrt t-\sqrt{t'})} \nn \\
&&=-4\int_{v_1}^1 dt \left(\frac 1{\sqrt{t}}+\frac 1{8\alpha t}+O(\alpha^{-2})
\right)\int_0^{1+\sqrt{t}} \log\left(1+\frac s{\sqrt{t}}+O(\alpha^{-2}\right)
(\sqrt{t}+s) e^{-4\alpha s} ds \nn\\
&&=-4\int_{v_1}^1 dt \left(\frac 1{\sqrt{t}}+\frac 1{8\alpha t}+O(\alpha^{-2})
\right)\int_0^{1+\sqrt{t}} \left(s+\frac{s^2}{2\sqrt t}+O(s^3)\right)
e^{-4\alpha s} ds \nn \\
&&=-4\int_{v_1}^1 dt \left(\frac 1{\sqrt{t}}+\frac 1{8\alpha t}+O(\alpha^{-2})
\right)\left(\frac 1{(4\alpha)^2}+\frac 1{(4\alpha)^3 \sqrt t}+O(\alpha^{-4})
\right) \nn \\
&&=-\frac 1{4\alpha^2}\int_{v_1}^1 \left(\frac 1{\sqrt{t}}+\frac 3{8\alpha t}+
O(\alpha^{-2})\right) dt.
\eqa
Therefore the third term of Eq.~(\ref{verynasty}) simplifies to
\beq
-\frac 1{2\alpha^2}\int_{v_1}^1 \left(\frac 1{\sqrt{t}}+\frac 1{2\alpha t}+
O(\alpha^{-2})\right) dt.
\eeq
Next considering the fourth term of Eq.~(\ref{verynasty}), we obtain
\bqa
&& \!\!\!\!\!\!\!\! 2\int_{v_1}^1 dt \int_{v_1}^t dt' \int_{v_1}^{t'}
\frac{e^{4\alpha(2\sqrt u-\sqrt t-\sqrt{t'})}}{u^{1.5}}du 
+2\int_{v_1}^1 dt \int_t^1 dt' \int_{v_1}^t
\frac{e^{4\alpha(2\sqrt u-\sqrt t-\sqrt{t'})}}{u^{1.5}}du \nn \\
&&=4\int_{v_1}^1 dt \int_t^1 dt' \int_{v_1}^t
\frac{e^{4\alpha(2\sqrt u-\sqrt t-\sqrt{t'})}}{u^{1.5}}du \nn \\
&&=4\int_{v_1}^1 dt \left(\frac{\sqrt t}{2\alpha}+\frac 1{8
\alpha^2}\right) \int_{v_1}^t \frac{e^{8\alpha(\sqrt u-\sqrt t)}}
{u^{1.5}}du \nn \\
&&=\frac 4{\alpha}\int_{v_1}^1 dt \left(\sqrt t+\frac 1{4\alpha}\right)
\int_0^{\sqrt t-\sqrt{v_1}} \frac{e^{-8\alpha s}}{(\sqrt t-s)^2}ds \nn \\
&&=\frac 4{\alpha}\int_{v_1}^1 dt \left(\sqrt t+\frac 1{2\alpha}\right)
\int_0^{\sqrt t-\sqrt{v_1}} e^{-8\alpha s}\left(\frac 1t+\frac {2s}{t^{1.5}}+
O(s^2)\right)ds \nn \\
&&=\frac 4{\alpha}\int_{v_1}^1 dt \left(\sqrt t+\frac 1{4\alpha}\right)\left(
\frac 1{8\alpha t}+\frac 2{(8\alpha)^2t^{1.5}}+O(\alpha^{-3})\right) \nn \\
&&=\frac 1{2\alpha^2}\int_{v_1}^1 dt \left(\frac 1{\sqrt t}+\frac 1{2\alpha t}+
O(\alpha^{-2})\right).
\eqa
The third and fourth terms therefore cancel to order $\alpha^{-3}$. This means
that
\beq
\int_{v_1}^1 dt \int_{v_1}^1 dt' \ip{\hat\varphi_t^{(0)}\hat\varphi_{t'}^{(1)}}
\approx \frac{v_1}{4\alpha^2}\ip{\hat\varphi_{v_1}^{(0)}\hat\varphi_{v_1}^{(1)}}
-\frac{\sqrt{v_1}}{2\alpha^2}\ip{(\hat\varphi_{v_1}^{(0)})^2}.
\eeq
Therefore we find that Eq.~(\ref{m2pert}) simplifies to
\beq
\ip{\phi_{\rm II}^2} \approx \frac 1{4\alpha^2}+\frac {v_1}{4\alpha^2}
\ip{(\hat\varphi_{v_1}^{(0)})^2} + \frac {\tau}{\alpha} \left( \frac{v_1}2
\ip{\hat\varphi_{v_1}^{(0)}\hat\varphi_{v_1}^{(1)}} - \sqrt{v_1}
\ip{(\hat\varphi_{v_1}^{(0)})^2}\right).
\eeq

\section{Derivations for Continuous Measurements}
\label{dercont}

Evaluating Eq.~(\ref{evalbig}) of Sec.~\ref{exactcase} for the phase
variance under continuous measurements, we find
\bqa
\label{bigder64}
&&\!\!\!\!\!\!\!\! \ip{\Theta^2(t)}=2\chi^3\ip{\int\limits_{-\infty}^t dv_1
\int\limits_{-\infty}^t dv_2 e^{\chi (v_1+v_2-2t)} \int\limits_{-\infty}^{v_1}
dW(u_1)\int\limits_{-\infty}^{v_2}dW(u_2)e^{2\st{\alpha}\sqrt{2\chi}(u_1+u_2
-v_1-v_2)}} \nn \\
&&\!\!\!\!\!\!\!\! + 8\chi^3 \st{\alpha}^2 \kappa^2 \ip{\int\limits_{-\infty}^t \!\! dv_1
\int\limits_{-\infty}^t \!\! dv_2 e^{\chi (v_1+v_2-2t)} \int\limits_{-\infty}^{v_1}
\!\! du_1 \int\limits_{-\infty}^{v_2} \!\! du_2 e^{2\st{\alpha}\sqrt{2\chi} (u_1+u_2-v_1
-v_2)} \int\limits_{u_1}^t \!\! dW'(s_1)\int\limits_{u_2}^t \!\! dW'(s_2)} \nn \\
&&~~~~~~ =2\chi^3 \int\limits_{-\infty}^t dv_1 \int\limits_{-\infty}^t
dv_2 e^{\chi(v_1+v_2-2t)} \int\limits_{-\infty}^{\min(v_1,v_2)} du\,
e^{2\st{\alpha}\sqrt{2\chi}(2u-v_1-v_2)} \nn \\
&&\!\!\!\!\!\!\!\! +8\chi^3 \st{\alpha}^2 \kappa^2 \int\limits_{-\infty }^t
dv_1 \int\limits_{-\infty}^t dv_2 e^{\chi(v_1+v_2-2t)}
\int\limits_{-\infty}^{v_1} du_1 \int\limits_{-\infty}^{v_2} du_2
e^{2\st{\alpha}\sqrt{2\chi}(u_1+u_2-v_1-v_2)} \int\limits_{\max(u_1,u_2)}^t ds.
\eqa
Considering the first term first, we have
\bqa
\label{result66}
&&\!\!\!\!\!\!\!\!  2\chi ^3 \int\limits_{-\infty}^t dv_1
\int\limits_{-\infty}^t dv_2 e^{\chi(v_1+v_2-2t)}
\int\limits_{-\infty}^{\min(v_1,v_2)} du e^{2\st{\alpha}
\sqrt{2\chi}(2u-v_1-v_2)} \nn \\
\!\!\!\! &=& \!\!\!\! 2\chi^3 \int\limits_{-\infty}^t dv_1
\int\limits_{-\infty}^{v_1} dv_2 e^{\chi(v_1+v_2-2t)}
\int\limits_{-\infty}^{v_2} du e^{2\st{\alpha}\sqrt{2\chi}(2u-v_1-v_2)} \nn \\
&& + 2\chi^3 \int\limits_{-\infty}^t dv_1 \int\limits_{v_1}^t dv_2
e^{\chi(v_1+v_2-2t)} \int\limits_{-\infty}^{v_1} du e^{2\st{\alpha}\sqrt{2\chi}
(2u-v_1-v_2)}  \nn \\
\!\!\!\! &=& \!\!\!\! 4\chi^3 \int\limits_{-\infty}^t dv_1
\int\limits_{-\infty}^{v_1} dv_2 e^{\chi(v_1+v_2-2t)}
\int\limits_{-\infty}^{v_2} du e^{2\st{\alpha}\sqrt{2\chi}(2u-v_1-v_2)} \nn \\
\!\!\!\! &=& \!\!\!\! \frac{\chi^3}{\st{\alpha}\sqrt{2\chi}}
\int\limits_{-\infty}^t dv_1 \int\limits_{-\infty}^{v_1} dv_2 e^{\chi(v_1+v_2
-2t)} e^{2\st{\alpha}\sqrt{2\chi}(v_2-v_1)}  \nn \\
\!\!\!\! &=& \!\!\!\! \frac{\chi^3}{\st{\alpha}\sqrt{2\chi}}
\int\limits_{-\infty}^t \frac{e^{2\chi(v_1-t)}}{\chi+2\st{\alpha}\sqrt{2\chi}}
dv_1 \nn \\
\!\!\!\! &=& \!\!\!\! \frac{\chi^3}{\st{\alpha}(2\chi)^{3/2}
(\chi+2\st{\alpha}\sqrt{2\chi}) } \nn \\
\!\!\!\! &\approx& \!\!\!\! \frac{\chi}{8\st{\alpha}^2}.
\eqa
For the approximation in the last line, it has been assumed that $\st \alpha \gg
\sqrt \chi$.

Now considering the second term in (\ref{bigder64}), we obtain
\bqa
&&\!\!\!\!\!\!\!\! 8\chi^3 \st{\alpha}^2 \kappa^2 \int\limits_{-\infty}^t dv_1
\int\limits_{-\infty}^t dv_2 e^{\chi(v_1+v_2-2t)} \int\limits_{-\infty}^{v_1}
du_1 \int\limits_{-\infty}^{v_2} du_2 e^{2\st{\alpha}\sqrt{2\chi}(u_1+u_2-v_1-
v_2)} \int\limits_{\max(u_1,u_2)}^t ds \nn \\
&&\!\!\!\!\!\!\!\! =8\chi^3 \st{\alpha}^2 \kappa^2 \int\limits_{-\infty}^t dv_1
\int\limits_{-\infty}^t dv_2 e^{\chi(v_1+v_2-2t)} \int\limits_{-\infty}^{v_1}
du_1 \int\limits_{-\infty}^{\min(v_2,u_1)} du_2 e^{2\st{\alpha}\sqrt{2\chi}
(u_1+u_2-v_1-v_2)} \int\limits_{u_1}^t ds \nn \\
&&\!\!\!\!\!\!\!\! +8\chi^3 \st{\alpha}^2 \kappa^2 \int\limits_{-\infty}^t dv_1
\int\limits_{-\infty}^t dv_2 e^{\chi(v_1+v_2-2t)}
\int\limits_{-\infty}^{\min(v_1,v_2)} du_1 \int\limits_{u_1}^{v_2} du_2
e^{2\st{\alpha}\sqrt{2\chi}(u_1+u_2-v_1-v_2)} \int\limits_{u_2}^t ds \nn \\
&&\!\!\!\!\!\!\!\! =8\chi^3 \st{\alpha}^2 \kappa^2 \int\limits_{-\infty}^t dv_1
\int\limits_{-\infty}^{v_1} du_1 \int\limits_{-\infty}^{u_1} dv_2
e^{\chi(v_1+v_2-2t)} \int\limits_{-\infty}^{v_2} du_2
e^{2\st{\alpha}\sqrt{2\chi}(u_1+u_2-v_1-v_2)}(t-u_1) \quad {\rm (a)} \nn \\
&&\!\!\!\!\!\!\!\! +8\chi^3 \st{\alpha}^2 \kappa^2 \int\limits_{-\infty}^t dv_1
\int\limits_{-\infty}^{v_1} du_1 \int\limits_{u_1}^t dv_2 e^{\chi(v_1+v_2-2t)}
\int\limits_{-\infty}^{u_1} du_2 e^{2\st{\alpha}\sqrt{2\chi}(u_1+u_2-v_1-v_2)}
(t-u_1) \quad \;\;\,{\rm (b)} \nn \\
&&\!\!\!\!\!\!\!\! +8\chi^3 \st{\alpha}^2 \kappa^2 \int\limits_{-\infty}^t dv_1
\int\limits_{-\infty}^{v_1} dv_2 e^{\chi(v_1+v_2-2t)}
\int\limits_{-\infty}^{v_2} du_1 \int\limits_{u_1}^{v_2} du_2
e^{2\st{\alpha}\sqrt{2\chi}(u_1+u_2-v_1-v_2)}(t-u_2) \quad \;\;\;{\rm (c)}\nn\\
&&\!\!\!\!\!\!\!\! +8\chi^3 \st{\alpha}^2 \kappa^2 \int\limits_{-\infty}^t dv_1
\int\limits_{v_1}^t dv_2 e^{\chi(v_1+v_2-2t)} \int\limits_{-\infty}^{v_1} du_1
\int\limits_{u_1}^{v_2} du_2 e^{2\st{\alpha}\sqrt{2\chi}(u_1+u_2-v_1-v_2)}
(t-u_2) \quad \;\;\;\;{\rm (d)} \nn \\
\eqa
Due to the difficult nature of the integration limits, we have obtained four
terms. Evaluating each of these terms in turn, we firstly find for (a)
\bqa
\label{result69}
&& 8\chi^3 \st{\alpha}^2 \kappa^2 \int\limits_{-\infty}^t dv_1
\int\limits_{-\infty}^{v_1} du_1 \int\limits_{-\infty}^{u_1} dv_2
e^{\chi(v_1+v_2-2t)} \int\limits_{-\infty}^{v_2} du_2
e^{2\st{\alpha}\sqrt{2\chi}(u_1+u_2-v_1-v_2)}(t-u_1) \nn \\
&& =\frac{8\chi^3 \st{\alpha}^2 \kappa^2}{2\st{\alpha}\sqrt{2\chi}}
\int\limits_{-\infty}^t dv_1 \int\limits_{-\infty}^{v_1} du_1
\int\limits_{-\infty}^{u_1} dv_2 e^{\chi(v_1+v_2-2t)}
e^{2\st{\alpha}\sqrt{2\chi}(u_1-v_1)}(t-u_1) \nn \\
&& =\frac{4\chi^2 \st{\alpha}\kappa^2}{\sqrt{2\chi}}\int\limits_{-\infty}^t
dv_1 \int\limits_{-\infty}^{v_1} du_1 e^{\chi(v_1+u_1-2t)}
e^{2\st{\alpha}\sqrt{2\chi}(u_1-v_1)}(t-u_1) \nn \\
&& =\frac{4\chi^2 \st{\alpha}\kappa^2}{\sqrt{2\chi}\left(\chi+2\st{\alpha}
\sqrt{2\chi}\right)}\int\limits_{-\infty}^t dv_1 e^{2\chi(v_1-t)}\left(t+
\frac{1}{\chi+2\st{\alpha}\sqrt{2\chi}}-v_1\right) \nn \\
&& =\frac{2\chi\st{\alpha}\kappa^2}{\sqrt{2\chi}\left(\chi+2\st{\alpha}
\sqrt{2\chi}\right)}\left(\frac{1}{\chi+2\st{\alpha}\sqrt{2\chi}}+ \frac{1}
{2\chi} \right) \nn \\
&& \approx \frac{\kappa^2}{4\chi}.
\eqa
The approximation in the last line is accurate in the limit $\st \alpha \gg
\sqrt\chi$. Turning to the next term, (b), we obtain
\bqa
&& 8\chi^3 \st{\alpha}^2 \kappa^2 \int\limits_{-\infty}^t dv_1
\int\limits_{-\infty}^{v_1} du_1 \int\limits_{u_1}^t dv_2 e^{\chi(v_1+v_2-2t)}
\int\limits_{-\infty}^{u_1} du_2 e^{2\st{\alpha}\sqrt{2\chi}(u_1+u_2-v_1-v_2)}
(t-u_1) \nn \\
&& = \frac{8\chi^3 \st{\alpha}^2 \kappa^2}{2\st{\alpha}\sqrt{2\chi}}
\int\limits_{-\infty}^t dv_1 \int\limits_{-\infty}^{v_1} du_1
\int\limits_{u_1}^t dv_2 e^{\chi(v_1+v_2-2t)} e^{2\st{\alpha}\sqrt{2\chi}
(2u_1-v_1-v_2)}(t-u_1) \nn \\
&& = \frac{4\chi ^3 \st{\alpha}\kappa^2}{\sqrt{2\chi}}\int\limits_{-\infty}^t
dv_1 \int\limits_{-\infty}^{v_1} du_1 e^{\chi(v_1-2t)}
e^{2\st{\alpha}\sqrt{2\chi}(2u_1-v_1)} \frac{\left(e^{(\chi-2\st{\alpha}
\sqrt{2\chi})t}-e^{(\chi-2\st{\alpha}\sqrt{2\chi})u_1} \right)}
{\chi-2\st{\alpha}\sqrt{2\chi}} (t-u_1) \nn \\
&& = \frac{\chi^3 \kappa^2}{2\chi(\chi-2\st{\alpha}\sqrt{2\chi})}
\int\limits_{-\infty}^t dv_1 e^{(\chi+2\st{\alpha}\sqrt{2\chi})(v_1-t)}\left(
t-v_1+\frac{1}{4\st{\alpha}\sqrt{2\chi}} \right) \nn \\
&& - \frac{4\chi^3 \st{\alpha}\kappa^2}{\sqrt{2\chi}(\chi^2-8\st{\alpha}^2\chi)}
\int\limits_{-\infty}^t dv_1 e^{2\chi(v_1-t)}\left(t-v_1+\frac{1}{\chi+
2\st{\alpha}\sqrt{2\chi}} \right) \nn \\
&& = \frac{\chi^3 \kappa^2}{2\chi(\chi^2-8\st{\alpha}^2 \chi)}\left(\frac{1}
{4\st{\alpha}\sqrt{2\chi}}+\frac{1}{\chi+2\st{\alpha}\sqrt{2\chi}}\right)\nn \\
&& - \frac{2\chi^2 \st{\alpha}\kappa^2}{\sqrt{2\chi}(\chi^2-8\st{\alpha}^2\chi)}
\left(\frac{1}{\chi+2\st{\alpha}\sqrt{2\chi}}+\frac{1}{2\chi}\right) \nn \\
&& \approx \frac{\kappa^2}{8\st{\alpha}\sqrt{2\chi}}.
\eqa
Again this is accurate in the limit $\st \alpha \gg \sqrt \chi$. This
term is higher order than the result obtained for (a), and so can be ignored.

Now looking at the third term (c),
\bqa
&& 8\chi^3 \st{\alpha}^2 \kappa^2 \int\limits_{-\infty}^t dv_1
\int\limits_{-\infty}^{v_1} dv_2 e^{\chi(v_1+v_2-2t)}\int\limits_{-\infty}^{v_2}
du_1 \int\limits_{u_1}^{v_2} du_2 e^{2\st{\alpha}\sqrt{2\chi}(u_1+u_2-v_1-v_2)}
(t-u_2) \nn \\
&& = \frac{4\chi^3 \st{\alpha}\kappa^2}{\sqrt{2\chi}}\int\limits_{-\infty}^t
dv_1 \int\limits_{-\infty}^{v_1} dv_2 e^{\chi(v_1+v_2-2t)}
\int\limits_{-\infty}^{v_2} du_1 e^{2\st{\alpha}\sqrt{2\chi}(u_1-v_1)}\left(t-
v_2+\frac{1}{2\st{\alpha}\sqrt{2\chi}} \right) \nn \\
&& -\frac{4\chi^3 \st{\alpha}\kappa^2}{\sqrt{2\chi}}\int\limits_{-\infty}^t
dv_1 \int\limits_{-\infty}^{v_1} dv_2 e^{\chi(v_1+v_2-2t)}
\int\limits_{-\infty}^{v_2} du_1 e^{2\st{\alpha}\sqrt{2\chi}(2u_1-v_1-v_2)}
\left(t-u_1+\frac{1}{2\st{\alpha}\sqrt{2\chi}} \right) \nn \\
&& = \chi^2 \kappa^2\int\limits_{-\infty}^t dv_1 \int\limits_{-\infty}^{v_1}
dv_2 e^{\chi(v_1+v_2-2t)} e^{2\st{\alpha}\sqrt{2\chi}(v_2-v_1)}\left(t-v_2+
\frac{1}{2\st{\alpha}\sqrt{2\chi}} \right) \nn \\ && -\frac{\chi^2 \kappa^2}{2}
\int\limits_{-\infty}^t dv_1 \int\limits_{-\infty}^{v_1} dv_2
e^{\chi(v_1+v_2-2t)} e^{2\st{\alpha}\sqrt{2\chi}(v_2-v_1)}\left(t-v_2+\frac{3}
{4\st{\alpha}\sqrt{2\chi}} \right) \nn \\
&& = \frac{\chi^2 \kappa^2}{2}\int\limits_{-\infty}^t dv_1
\int\limits_{-\infty}^{v_1} dv_2 e^{\chi(v_1+v_2-2t)}
e^{2\st{\alpha}\sqrt{2\chi}(v_2-v_1)}\left(t-v_2+\frac{1}{4\st{\alpha}
\sqrt{2\chi}} \right) \nn \\
&& = \frac{\chi^2 \kappa^2}{2\left( \chi+2\st{\alpha}\sqrt{2\chi} \right)}
\int\limits_{-\infty}^t dv_1 e^{2\chi(v_1-t)}\left(t-v_1+\frac{1}{4\st{\alpha}
\sqrt{2\chi}} + \frac{1}{\chi+2\st{\alpha}\sqrt{2\chi}} \right) \nn \\
&& = \frac{\chi \kappa^2}{4\left(\chi+2\st{\alpha}\sqrt{2\chi} \right)}
\left(\frac{1}{4\st{\alpha}\sqrt{2\chi}} + \frac{1}{\chi+2\st{\alpha}
\sqrt{2\chi}} + \frac{1}{2\chi} \right) \nn \\
&& \approx \frac{\kappa^2}{16\st{\alpha}\sqrt{2\chi}}.
\eqa
This is again higher order than term (a), and so can be ignored. Now looking at
the final term, (d), we find
\bqa
&& 8\chi^3 \st{\alpha}^2 \kappa^2 \int\limits_{-\infty}^t dv_1
\int\limits_{v_1}^t dv_2 e^{\chi(v_1+v_2-2t)} \int\limits_{-\infty}^{v_1} du_1
\int\limits_{u_1}^{v_2} du_2 e^{2\st{\alpha}\sqrt{2\chi}(u_1+u_2-v_1-v_2)}
(t-u_2) \nn \\
&& = \frac{4\chi^3 \st{\alpha}\kappa^2}{\sqrt{2\chi}}\int\limits_{-\infty}^t
dv_1 \int\limits_{v_1}^t dv_2 e^{\chi(v_1+v_2-2t)} \int\limits_{-\infty}^{v_1}
du_1 e^{2\st{\alpha}\sqrt{2\chi}(u_1-v_1)} \left(t-v_2+\frac{1}{2\st{\alpha}
\sqrt{2\chi}} \right) \nn \\
&& - \frac{4\chi^3 \st{\alpha}\kappa^2}{\sqrt{2\chi}}\int\limits_{-\infty}^t
dv_1 \int\limits_{v_1}^t dv_2 e^{\chi(v_1+v_2-2t)} \int\limits_{-\infty}^{v_1}
du_1 e^{2\st{\alpha}\sqrt{2\chi}(2u_1-v_1-v_2)} \left(t-u_1+\frac{1}
{2\st{\alpha}\sqrt{2\chi}} \right) \nn \\
&& = \chi^2 \kappa^2 \int\limits_{-\infty}^t dv_1 \int\limits_{v_1}^t dv_2
e^{\chi(v_1+v_2-2t)}\left(t-v_2+\frac{1}{2\st{\alpha}\sqrt{2\chi}}\right) \nn \\
&& - \frac{\chi^2 \kappa^2}2 \int\limits_{-\infty}^t dv_1 \int\limits_{v_1}^t
dv_2 e^{\chi(v_1+v_2-2t)} e^{2\st{\alpha}\sqrt{2\chi}(v_1-v_2)}\left(t-v_1+
\frac{3}{4\st{\alpha}\sqrt{2\chi}} \right) \nn \\
&& = \chi \kappa^2 \int\limits_{-\infty}^t dv_1 e^{\chi(v_1-t)} \left(\frac{1}
{2\st{\alpha}\sqrt{2\chi}} + \frac{1}{\chi} \right) - \chi \kappa^2
\int\limits_{-\infty}^t dv_1 e^{2\chi(v_1-t)} \left(t-v_1+\frac{1}{2\st{\alpha}
\sqrt{2\chi}} + \frac{1}{\chi} \right) \nn \\
&& - \frac{\chi^2 \kappa^2}{2\left(\chi-2\st{\alpha}\sqrt{2\chi}\right)}
\int\limits_{-\infty}^t dv_1 e^{\left(\chi+2\st{\alpha}\sqrt{2\chi} \right)
(v_1-t)}\left(t-v_1+\frac{3}{4\st{\alpha}\sqrt{2\chi}} \right) \nn \\
&& + \frac{\chi^2 \kappa^2}{2\left(\chi-2\st{\alpha}\sqrt{2\chi}\right)}
\int\limits_{-\infty}^t dv_1 e^{2\chi(v_2-t)}\left(t-v_1+\frac{3}{4\st{\alpha}
\sqrt{2\chi}} \right) \nn \\
&& = \frac{\kappa^2}{4\chi}+\frac{\kappa^2}{4\st{\alpha}\sqrt{2\chi}} -
\frac{\chi\kappa^2}{2(\chi-8\st{\alpha}^2)}\left(\frac{3}{4\st{\alpha}
\sqrt{2\chi}} + \frac{1}{\chi+2\st{\alpha}\sqrt{2\chi}} \right) \nn \\
&&   + \frac{{\chi \kappa^2 }}
{{4\left( {\chi  - 2\st{\alpha}\sqrt {2\chi } } \right)}}\left( {\frac{3}
{{4\st{\alpha}\sqrt {2\chi } }} + \frac{1}
{{2\chi }}} \right).
\eqa
For $\st \alpha \gg \sqrt\chi$, the only significant term here is
\beq \frac{\kappa^2}{4\chi}.\eeq
Using this, together with the results from Eqs (\ref{result66}) and
(\ref{result69}), the total phase variance is
\beq
\ip{\Theta^2(t)} \approx \frac{\chi} {8\st{\alpha}^2} + \frac{\kappa^2}{2\chi}.
\eeq